\documentstyle[12pt]{report}

\makeatletter
\@addtoreset{equation}{subsection}
\@addtoreset{section}{part}
\makeatother
\renewcommand{\theequation}{\thesubsection.
\arabic{equation}}
\renewcommand{\thesection}{\thepart.\arabic{section}}
\newtheorem{Theorem}{Theorem}[section]
\newtheorem{Definition}{Definition}[section]
\newtheorem{Lemma}{Lemma}[section]
\newtheorem{Corollary}{Corollary}[section]

\def\be{\begin{equation}}
\def\ee{\end{equation}}
\def\ba{\begin{eqnarray}}
\def\ea{\end{eqnarray}}
\def\a{{\cal A}}
\def\ab{\overline{\a}}
\def\ac{\a^\Cl}

\def\g{{\cal G}}
\def\gb{\overline{\g}}
\def\gc{\g^\Cl}

\def\ag{\a/\g}
\def\agb{\overline{\ag}}

\def\abgb{\ab/\gb}
\def\abc{\ab^\Cl}
\def\agbc{\agb^\Cl}
\def\Nl{{\mathchoice
{\setbox0=\hbox{$\displaystyle\rm N$}\hbox{\hbox to0pt
{\kern0.4\wd0\vrule height0.9\ht0\hss}\box0}}
{\setbox0=\hbox{$\textstyle\rm N$}\hbox{\hbox to0pt
{\kern0.4\wd0\vrule height0.9\ht0\hss}\box0}}
{\setbox0=\hbox{$\scriptstyle\rm N$}\hbox{\hbox to0pt
{\kern0.4\wd0\vrule height0.9\ht0\hss}\box0}}
{\setbox0=\hbox{$\scriptscriptstyle\rm N$}\hbox{\hbox to0pt
{\kern0.4\wd0\vrule height0.9\ht0\hss}\box0}}}}
\def\Zl{{\mathchoice
{\setbox0=\hbox{$\displaystyle\rm Z$}\hbox{\hbox to0pt
{\kern0.4\wd0\vrule height0.9\ht0\hss}\box0}}
{\setbox0=\hbox{$\textstyle\rm Z$}\hbox{\hbox to0pt
{\kern0.4\wd0\vrule height0.9\ht0\hss}\box0}}
{\setbox0=\hbox{$\scriptstyle\rm Z$}\hbox{\hbox to0pt
{\kern0.4\wd0\vrule height0.9\ht0\hss}\box0}}
{\setbox0=\hbox{$\scriptscriptstyle\rm Z$}\hbox{\hbox to0pt
{\kern0.4\wd0\vrule height0.9\ht0\hss}\box0}}}}
\def\Ql{{\mathchoice
{\setbox0=\hbox{$\displaystyle\rm Q$}\hbox{\hbox to0pt
{\kern0.4\wd0\vrule height0.9\ht0\hss}\box0}}
{\setbox0=\hbox{$\textstyle\rm Q$}\hbox{\hbox to0pt
{\kern0.4\wd0\vrule height0.9\ht0\hss}\box0}}
{\setbox0=\hbox{$\scriptstyle\rm Q$}\hbox{\hbox to0pt
{\kern0.4\wd0\vrule height0.9\ht0\hss}\box0}}
{\setbox0=\hbox{$\scriptscriptstyle\rm Q$}\hbox{\hbox to0pt
{\kern0.4\wd0\vrule height0.9\ht0\hss}\box0}}}}
\def\Rl{{\mathchoice
{\setbox0=\hbox{$\displaystyle\rm R$}\hbox{\hbox to0pt
{\kern0.4\wd0\vrule height0.9\ht0\hss}\box0}}
{\setbox0=\hbox{$\textstyle\rm R$}\hbox{\hbox to0pt
{\kern0.4\wd0\vrule height0.9\ht0\hss}\box0}}
{\setbox0=\hbox{$\scriptstyle\rm R$}\hbox{\hbox to0pt
{\kern0.4\wd0\vrule height0.9\ht0\hss}\box0}}
{\setbox0=\hbox{$\scriptscriptstyle\rm R$}\hbox{\hbox to0pt
{\kern0.4\wd0\vrule height0.9\ht0\hss}\box0}}}}
\def\Cl{{\mathchoice
{\setbox0=\hbox{$\displaystyle\rm C$}\hbox{\hbox to0pt
{\kern0.4\wd0\vrule height0.9\ht0\hss}\box0}}
{\setbox0=\hbox{$\textstyle\rm C$}\hbox{\hbox to0pt
{\kern0.4\wd0\vrule height0.9\ht0\hss}\box0}}
{\setbox0=\hbox{$\scriptstyle\rm C$}\hbox{\hbox to0pt
{\kern0.4\wd0\vrule height0.9\ht0\hss}\box0}}
{\setbox0=\hbox{$\scriptscriptstyle\rm C$}\hbox{\hbox to0pt
{\kern0.4\wd0\vrule height0.9\ht0\hss}\box0}}}}
\def\Co{{\mathchoice
{\setbox0=\hbox{$\displaystyle\rm C$}\hbox{\hbox to0pt
{\kern0.4\wd0\vrule height0.9\ht0\hss}\box0}}
{\setbox0=\hbox{$\textstyle\rm C$}\hbox{\hbox to0pt
{\kern0.4\wd0\vrule height0.9\ht0\hss}\box0}}
{\setbox0=\hbox{$\scriptstyle\rm C$}\hbox{\hbox to0pt
{\kern0.4\wd0\vrule height0.9\ht0\hss}\box0}}
{\setbox0=\hbox{$\scriptscriptstyle\rm C$}\hbox{\hbox to0pt
{\kern0.4\wd0\vrule height0.9\ht0\hss}\box0}}}}
\def\Hl{{\mathchoice
{\setbox0=\hbox{$\displaystyle\rm H$}\hbox{\hbox to0pt
{\kern0.4\wd0\vrule height0.9\ht0\hss}\box0}}
{\setbox0=\hbox{$\textstyle\rm H$}\hbox{\hbox to0pt
{\kern0.4\wd0\vrule height0.9\ht0\hss}\box0}}
{\setbox0=\hbox{$\scriptstyle\rm H$}\hbox{\hbox to0pt
{\kern0.4\wd0\vrule height0.9\ht0\hss}\box0}}
{\setbox0=\hbox{$\scriptscriptstyle\rm H$}\hbox{\hbox to0pt
{\kern0.4\wd0\vrule height0.9\ht0\hss}\box0}}}}
\def\Ol{{\mathchoice
{\setbox0=\hbox{$\displaystyle\rm O$}\hbox{\hbox to0pt
{\kern0.4\wd0\vrule height0.9\ht0\hss}\box0}}
{\setbox0=\hbox{$\textstyle\rm O$}\hbox{\hbox to0pt
{\kern0.4\wd0\vrule height0.9\ht0\hss}\box0}}
{\setbox0=\hbox{$\scriptstyle\rm O$}\hbox{\hbox to0pt
{\kern0.4\wd0\vrule height0.9\ht0\hss}\box0}}
{\setbox0=\hbox{$\scriptscriptstyle\rm O$}\hbox{\hbox to0pt
{\kern0.4\wd0\vrule height0.9\ht0\hss}\box0}}}}

\def\dprime{{\prime\prime}}

\title{Introduction\\ to\\ Modern Canonical\\ Quantum General Relativity}
\author{T. Thiemann\thanks{thiemann@aei.mpg.de} \\
       MPI f. Gravitationsphysik, Albert-Einstein-Institut, \\
           Am M\"uhlenberg 1, 14476 Golm near Potsdam, Germany}
\date{{\small Preprint AEI-2001-119}}

\begin{document}



\setcounter{page}{1}

\maketitle

\begin{abstract}
This is an introduction to the by now fifteen years old research field of
canonical quantum general relativity, sometimes called
``loop quantum gravity".
The term ``modern" in the title refers to the fact
that the quantum theory is based on formulating classical general relativity
as a theory of connections rather than metrics as compared to
in original version due to Arnowitt, Deser and Misner.

Canonical quantum general relativity is an attempt to define a mathematically
rigorous,
non-perturbative, background independent theory of Lorentzian quantum gravity
in four spacetime dimensions in the continuum. As such it differs considerably
from perturbative ans\"atze. It provides a {\it unified} theory of
all interactions in the sense that all interactions transform under a
common gauge group: The four-dimensional diffeomorphism group
of the underlying differental manifold which is maximally broken in
perturbative approaches. The approach
is minimal in that one simply analyzes the logical consequences of
combining the principles of general relativity with the principles of
quantum mechanics. As a consequence, no extra dimensions, no
corresponding Kaluza-Klein compactifications, no supersymmetry and its
associated phenomenology -- compatible spontaneous breaking at low energies,
seem to be necessary. On the other hand, no explanation for the particle 
content and the dimension of the universe is provided by the theory.

The requirement to preserve
background independence has lead to new, fascinating mathematical
structures which
one does not see in perturbative approaches, e.g. a fundamental discreteness
of spacetime
seems to be a prediction of the theory which is a first substantial
evidence for a
theory in which the gravitational field acts as a natural UV cut-off.

An effort has been made to provide a self-contained exposition at the
appropriate level of rigour which at the same time is accessible to graduate
students with only basic
knowledge of general relativity and quantum field theory on Minkowski space.
To be
useful, not all facets of the field have been covered, however, guides to
further reading
and a detailed bibliography is included. This report is submitted to the
on-line journal
``Living Reviews" and is thus subject to being updated on at least a
bi-annual basis.
\end{abstract}

\setcounter{page}{2}

\cleardoublepage

\tableofcontents


\cleardoublepage

\section*{Introduction}
\label{s1}
\addcontentsline{toc}{chapter}{\numberline{}Introduction}

This report tries to give a status overview of the field of modern
canonical quantum general relativity, sometimes called
``loop quantum gravity".  The term ``modern" accounts for the fact that
this is a ``connection dynamics" formulation of Einstein's
theory, rather than the original
``geometrodynamics" formulation due to Arnowitt, Deser and Misner.
As this is an  article submitted to the on-line journal ``Living Reviews´´,
the report will be updated on an at least bi-annual basis.

The field of modern canonical quantum general relativity was born in 1986 and
since then an order of 500 research papers closely related to the subject
have been published. Pivotal structures of the theory are scattered over an
order of 100 research papers, reports, proceedings and books, issues which
were believed to be essential initially turned out to be
negligible later on and vice versa. These facts make it hard for the
beginner to go
quickly to the frontier of original research. The present report aims at
giving beginners a guideline, a kind of {\it geodesic} through the literature
suggesting in which order to read a minimal number of papers
in order to understand the foundations. The report should be accessible to
students at the
graduate level (say European students in their sixth or seventh
semester) with only basic prior knowledge of general relativity and
quantum field theory on Minkowski space. However, the target audience
does not consist of the light hearted readers who wish to get just some
rough idea
of what the field is all about, those readers are advised to study the
excellent review articles \cite{126,0,10f}. Rather, we typically have in
mind the serious student who wants to gain thorough understanding of the
foundations and to pass quickly to the frontier of active research.
As financial support over an extended period of time is a serious issue for
almost all graduate students in the world, time is pressing and it is
important
that one does not waste too much time in learning the basics. We have
therefore decided to select only a minimal amount of material, just enough
in order to reach a firm understanding of the foundations,
which on the other hand is presented in great detail so that one does not
have to read much besides this report for this purpose. Since some of
the mathematics that is needed maybe unfamiliar to the younger students
we have also included a large mathematical ``appendix" that serves to
fill in some of the necessary background.
Remembering too freshly still our own experience how annoying and
time consuming it can be to collect papers, to compare and streamline
notations, to adapt numerical coefficient conventions etc. this should
also help to avoid unnecessary confusions and time delays.\\
\\
The number of people working in the field of canonical quantum general
relativity is
of the order of $10^2$ (including students and post-docs) which is
unfortunately
quite small, when compared, for instance, with the size of the string theory
community (of the order $10^3$) and it is one of the
aims of this report to attract more young researchers to dive into this
alternative, fascinating research
subject. Here is a, to the best knowledge of the author, complete list of 
locations where research in canonical
quantum general relativity (and related) is, at least partly, currently
performed {\it and funded} together with the names of main contact 
persons in alphabetical order and their
main current research directions. The locations are listed alphabetically by
continent and within one continent geographically by country from north to
south (p$=$permanent position, n$=$non-permanent position, we only list
post-docs among the non-permanent staff):
\begin{itemize}
\item[A)] {\bf Central America}
\begin{itemize}
\item[1.] {\it Universidad Autonoma Metropolitana Itztapalapa,
Mexico City, Mexico}

Hugo Morales-Tecotl (p): semiclassical quantum gravity
\item[2.] {\it Universidad Nacionale Autonoma de Mexico, Mexico City,
Mexico}

Alexandro Corichi (p):
semiclassical quantum gravity, isolated horizon quantum black holes;\\
Michael Ryan (p) : mini(midi)superspace models;\\
Jose´-Antonio Zapata (p) : semiclassical quantum gravity, spin foams
\end{itemize}
\item[B)] {\bf North America}
\begin{itemize}
\item[3.] {\it University of Alberta, Edmonton, Canada}

Stephen Fairhurst (n): isolated horizon quantum black holes;\\
Viquar Husain (p): mini(midi)superspace and integrable models;\\
Don Page (p): mini(midi)superspace models
\item[4.] {\it Perimeter Institute for Theoretical Physics and University of
Waterloo, Waterloo, Ontario, Canada}

Olaf Dreyer (n): isolated horizon quantum black holes;\\
Fotini Markopoulou (p): causal spin foams, renormalization;\\
Lee Smolin (p): connections between loop quantum gravity and string theory,
causal quantum dynamics, spin foam models;\\
O. Winkler (n): semi-classical quantum gravity
\item[5.] {\it Syracuse University, Syracuse, NY, USA}

Donald Marolf (p): refined algebraic quantization, general relativistic
aspects of string theory
\item[6.] {\it Center for Gravitational Physics and Geometry, The Pennsylvania
State University at University Park, PA, USA}

Abhay Ashtekar (p): isolated
horizon quantum black holes, semi-classical quantum gravity;\\
Martin Bojowald (n): mini(midi)superspace models, quantum dynamics;\\
Amit Ghosh (n): isolated horizon black holes;\\
Roger Penrose\footnote{Distinguished Visiting Professor from
the University of Oxford, Oxford, UK} (p): twistor theory, fundamental
issues;\\
Alexandro Perez (n): spin foam models
\item[7.] {\it University of Utah, Salt Lake City, UT, USA}

Christopher Beetle (n): covariant formulation;\\
Karel Kucha\v{r} (p): mini -- and midisuperspace models, covariant
formulation
\item[8.] {\it University of Maryland, College Park, MD, USA}

Ted Jacobson (p): classical actions, (quantum) black hole physics
\item[9.] {\it Kansas State Unviversity, Kansas City, KS, USA}

Louis Crane (p): state sum models, spin foam models;\\
David Yetter(p): state sum models, spin foam models
\item[10.] {\it Utah State University, Logan, UT, USA}

Charles Torre (p): mini(midi)superspace models, fundamental issues
\item[11.] {\it University of California, Riverside, CA, USA}

John Baez (p):
isolated horizon quantum black holes, spin foam models
\item[12.] {\it University of California and Institute of Theoretical
Physics, Santa Barbara, CA, USA}

Jim Hartle (p): connection between consistent histories -- and canonical
approach;\\
Kirill Krasnov (n) : isolated horizon quantum black holes, spin foam
models, aspects of string theory
\item[13.] {\it University of Mississippi, Oxford, MS, USA}

Luca Bombelli (p): semiclassical quantum gravity
\item[14.] {\it Lousiana State University, Baton Rouge, LA, USA}

Jorge Pullin (p):
quantum dynamics, Dirac observables, semiclassical quantum gravity
\end{itemize}
\item[C)] {\bf South America}
\begin{itemize}
\item[15.] {\it Universidad de la Republica, Montevideo, Uruguay}

Rodolfo Gambini (p): quantum dynamics, Dirac observables;\\
Jorge Griego (p): quantum dynamics;\\
Michael Reisenberger (p): dynamical lattice formulations,
spin foam models
\item[16.] {\it Centro de Estudios Cientificos, Valdivia, Chile}

Claudio Teitelboim: mini(midi)superspace models, quantization 
of gauge systems;\\
Andres Gomberoff (p): refined algebraic quantization, general
relativistic aspects of string theory
\end{itemize}
\item[D)] {\bf Asia}
\begin{itemize}
\item[17.] {\it Raman Research Institute, Bangalore, India}

Joseph Samuel (p): classical formulation;\\
Madhavan Varadarajan (p): semiclassical quantum gravity
\end{itemize}
\item[E)] {\bf Europe}
\begin{itemize}
\item[18.] {\it The Niels Bohr Institute, Copenhagen, Denmark}

Jan Ambjorn (p): path integral formulation (dynamical triangulations);\\
Matthias Arnsdorf (n): semiclassical quantum gravity, connections
with string theory
\item[19.] {\it University of Nottingham, Nottingham, UK}

John Barrett (p): spin foam models, state sum models;\\
Jorma Louko (p): mini(midi)superspace models, general relativistic
aspects of string theory
\item[20.] {\it Albert -- Einstein -- Institut, Golm near Potsdam, Germany}

Hermann Nicolai (p): supergravity, superstring theory, connections between
canonical quantum gravity and M -- Theory;\\
Thomas Thiemann (p): quantum dynamics, semi-classical analysis
\item[21.] {\it University of Warsaw, Warsaw, Poland}

Jurek Lewandowski (p):
isolated horizon quantum black holes, semi-classical
quantum gravity
\item[22.] {\it Utrecht University and Spinoza Institute, Utrecht, 
The Netherlands}

Gerhard 't Hooft (p): quantum black hole physics, fundamental issues;\\
Renate Loll (p): quantum geometry, path integral formulation (dynamical
triangulations)
\item[23.] {\it Cambridge University, Cambridge, UK}

Ruth Williams (p): spin foam models, state sum models
\item[24.] {\it Imperial College, London, UK}

Chris Isham (p): fundamental issues (topos theory)
%
\item[25.] {\it Universit\'e Libre de Bruxelles, Bruxelles, Belgium}

Marc Henneaux (p): BRST analysis, quantization of gauge systems, 
general relativistic aspects of supergravity and string theory
\item[26.] {\it Ecole Normale Sup\'erieure, Paris, France}
 
Bernard Julia (p): canonical quantization of supergravity theories,
Noether charges

\item[27.] {\it Universit\"at Wien, Wien, Austria}

Peter Aichelburg (p): general relativistic aspects of supergravity and 
string theory
\item[28.] {\it Universit\"at Bern, Bern, Switzerland}

Peter Haj\'i\v{c}ek (p): midi(mini)superspace models

\item[29.] {\it Ecole Normale Sup\'erieur, Lyon, France}

Laurent Freidel (p) : spin foam models
\item[30.] {\it Universita di Torino, Torino, Italy}

Jeanette Nelson (p): mini(midi)superspace models,
Regge calculus
\item[31.] {\it Universita di Parma, Parma, Italy}

Roberto de Pietri (n) : quantum dynamics, spin foam models
\item[32.] {\it Universit\'e de Marseille, Luminy, France}

Carlo Rovelli (p):
quantum dynamics, Dirac observables, spin foam models
\item[33.] {\it Instituto de Matematicas y Fisica Fundamental, Madrid, Spain}

Guillermo  Mena-Magu\'an (p): mini(midi)superspace models,
Euclidean versus Lorentzian formulation
\item[34.] {\it Universidad Europea, Madrid, Spain}

Fernando Barbero (p): classical actions
\item[35.] {\it Instituto Superior Tecnico, Lisboa, Portugal}

Jos\'e Mour\~ao (p): mathematical framework
\item[36.] {\it Universidad do Algarve, Faro, Portugal}

Nenad Manojlovic (p): mini(midi)superspaces, integrable models
\end{itemize}
\end{itemize}
All these places are definitely worthwhile applying to for
graduate studies or post-doc positions. Notice that we have listed
only those researchers that are at least partly involved or interested in 
quantum general
relativity research and only those aspects of their work that touch on 
quantum gravity. For 
instance, the Albert -- Einstein -- Institute is currently the
largest (by number of members and budget) institute in the world that
focusses on gravitational physics,
consisting of altogether four divisions: Quantum Gravity and Unified
Theories with focus on M -- Theory (director: Hermann Nicolai),
Astrophysics (director: Bernard
Schutz), Mathematical General Relativity (director: Gerhard Huisken),
Gravitational Wave Detection (director: Karsten Danzmann).
Institutes of similar sizes are the Center for Gravitational
Physics and Geometry, the Perimeter Institute for Theoretical
Physics and the Institute for Theoretical Physics at Santa Barbara.
Unfortunately the University of Pittsburgh, Pittsburgh, PA, USA (Ted 
Newman (p): null surface formulation, fundamental issues) no longer 
appears in the above list since quantum gravitational research is no 
longer funded there.\\ 
\\
This report is organized as follows :\\
\\
Before diving into the subject, the next subsection motivates the search for a
quantum theory of gravity, points out what the essential problems are that
one has to deal with, lists the possible approaches and their respective
strengths and weaknesses and finally motivates our choice to study
canonical quantum general relativity. We also list our notation and
conventions.\\
\\
We then approach the main text of the report which is subdivided into three
parts:\\

The first part contains the foundations of the theory, that is, results
which are physically and mathematically robust. Thus we will study in detail
a) the classical canonical formulation of general relativity in terms of
connection variables, b) the general programme of canonical quantization,
c) the application of this programme to general relativity in terms of
connections and the resulting Hilbert space structures, d) the proof that
the Hilbert space found implements the correct quantum kinematics (that is,
it represents the correct commutation relations and supports the
kinematical constraints of the theory) and finally e) the kinematical
geometrical operators which measure for instance areas of (coordinate)
surfaces. We also sketch how spectral properties of these kinematical
operators extend to their physical (dynamical) counterparts in the
presence of matter.

The second part discusses the main, current research directions within
canonical quantum general relativity and describes their respective status.
These results are less robust and still, at least partially, in flow.
Thus we outline in detail a) the implementation of the quantum dynamics
or {\bf Quantum Einstein Equantions} (also known by ``Wheeler -- DeWitt
Equation"), b) the coupling of standard quantum matter, c) the
semiclassical analysis necessary in order to verify whether the
theory constructed is indeed a quantization of general relativity, d)
the path integral formulation of the theory (also called spin foam
formulation), e) quantum black hole physics and finally f) the
possible links between string theory and canonical quantum general
relativity. This part closes with a section in which we list
a selected number of open and fascinating research problems.

Finally, in the third part we provide some mainly mathematical background
material. Thus we give elementary but fairly extensive introductions to
elements of
a) the Dirac algorithm for dealing with theories with constraints,
b) the theory of fibre bundles, c) general topology, d) Gel'fand
theory for Abelean $C^\ast-$algebras, e) measure theory, f) the GNS
construction and g) refined algebraic quantization (RAQ).\\
\\
\\
\\
{\it\bf Acknowledgements}\\
\\
\\
My thanks go to Theresa Velden, for a long time managing director
of the on-line journal ``Living Reviews", for continuously encouraging me
to finally finish this review.


\newpage

\subsection*{Defining Quantum Gravity}
\addcontentsline{toc}{section}{\numberline{}Defining Quantum Gravity}
\label{s1.1}

In the first subsection of this section we explain why the problem of
quantum gravity cannot be ignored in nowadays physics, even though the
available accelerator energies lie way beyond the Planck scale.
Then we define what a quantum theory
of gravity and all interactions is widely expected to achieve and
point out the two main directions of research divided into the
perturbative and non-perturbative approaches. In the
third subsection we describe these approaches in more detail and finally
in the fourth motivate
our choice to do canonical quantum general relativity as opposed to other
approaches.

\subsubsection*{Why Quantum Gravity in the 21st Century ?}
\addcontentsline{toc}{subsection}{\numberline{}
Why Quantum Gravity in the 21st Century ?}
\label{s1.1.1}

It is often argued that quantum gravity is not relevant for the physics
of this century because in our most powerful accelerator, the LHC to be
working in 2005, we obtain energies of the order of a few $10^3$ GeV
while the energy scale at which quantum gravity is believed to become important
is the Planck energy of $10^{19}$ GeV. While that is true, it is false that
nature does not equip us with particles of energies much beyond the TeV
scale, there are astrophysical particles with energy of a fist stroke and the
next generation of particle microscopes is therefore not going to be
built on the
surface of Earth any more but in its orbit. Moreover, as we will
describe in this
report in more detail, even with TeV energy scales it might be possible to see
quantum gravity effects in the close future.

But even apart from these purely experimental considerations, there are
good theoretical reasons for studying quantum gravity. To see why,
let us summarize our current understanding of the fundamental interactions :\\
Ashamingly,
the only quantum fields that we fully understand to date in four dimensions
are {\it free quantum fields on four-dimensional Minkowski
space}. Formulated more provocatively:\\
\\
{\bf In four dimensions we only understand an (infinite)
collection of uncoupled harmonic oscillators on Minkowski space} !\\
\\
In order to leave the domain of these rather trivial and unphysical quantum
field theories, physicists have developed two techniques : perturbation
theory and quantum field theory on curved backgrounds. This means the
following : \\
With respect to accelerator experiments, the most important processes are
scattering amplitudes between particles. One can {\it formally} write down a
unitary operator
that accounts for the scattering interaction between particles and which maps
between the well-understood free quantum field Hilbert spaces in the far
past and future.  Famously, by Haags theorem \cite{62}, whenver that operator
is really unitary, there is no interaction and if it is not unitary, then it
is ill-defined.
In fact, one can only define the operator perturbatively by writing down the
formal power expansion in terms of the generator of the would-be unitary
transformation between the free quantum field theory Hilbert spaces. The
resulting
series is divergent order by order but if the theory is ``renormalizable"
then one can make these orders artificially finite by a regularization and
renormalization procedure with, however, no control on convergence
of the resulting series. Despite these
drawbacks, this recipe has worked very well so far, at least for the
electroweak interaction.

Until now, all we have said applies only to free (or perturbatively
interacting)
quantum fields on Minkowski spacetime for which the so-called Wightman
axioms \cite{62} can be verified. 
Let us summarize them for the case of a scalar field
in $(D+1)-$dimensional Minkowski space:

\newpage

\begin{itemize}
\item[{\bf W1}] {\it Representation}\\
There exists a unitary and continuous representation
$U:\;\mbox{\boldmath ${\cal P}$\unboldmath}\to {\cal B}({\cal H})$
of the {\bf Poincar\'e group} $\mbox{\boldmath ${\cal P}$\unboldmath}$
on a Hilbert space ${\cal H}$.
\item[{\bf W2}] {\it Spectral Condition}\\
The momentum operators $\mbox{\boldmath $P^\mu$\unboldmath}$ have
spectrum in the forward lightcone:\\
$\mbox{\boldmath $\eta_{\mu\nu} P^\mu P^\nu \le 0;\;P^0\ge 0$\unboldmath}$.
\item[{\bf W3}] {\it Vacuum}\\
There is a unique {\bf Poincar\'e} invariant vacuum state
$U(\mbox{\boldmath $p$\unboldmath})
\mbox{\boldmath $\Omega$\unboldmath}
=\mbox{\boldmath $\Omega$\unboldmath}$ for all
$\mbox{\boldmath $p$\unboldmath}
\in \mbox{\boldmath ${\cal P}$\unboldmath}$.
\item[{\bf W4}] {\it Covariance}\\
Consider the smeared field operator valued tempered distributions
$\phi(f)=\int_{\mbox{\boldmath $\Rl^{D+1}$\unboldmath}}
d^{D+1}x \phi(x) f(x)$ where
$f\in {\cal S}(\mbox{\boldmath $\Rl^{D+1}$\unboldmath})$
is a test function of rapid decrease.
Then finite linear combinations of the form
$\phi(f_1)..\phi(f_N)\mbox{\boldmath $\Omega$\unboldmath}$
lie dense in $\cal H$ (that is, $\mbox{\boldmath $\Omega$\unboldmath}$
is a cyclic vector) and $U(p)\phi(f) U(p)^{-1}=\phi(f\circ p)$
for any 
$\mbox{\boldmath $p$\unboldmath}\in \mbox{\boldmath ${\cal P}$\unboldmath}$.
\item[{\bf W5}] {\it Locality (Causality)}\\
Suppose that the supports (the set of points where a function
is different from zero) of $f,f'$ are {\it spacelike
separated} (that is, the points of their supports cannot be
connected by a non-spacelike curve) then
$\phi(f),\phi(f')=0$.
\end{itemize}
The most important objects in this list are those that are highlighted in
bold face letters: The fixed, non-dynamical Minkowski background metric
$\mbox{\boldmath $\eta$\unboldmath}$ with its well-defined causal
structure, its
Poincar\'e
symmetry group $\mbox{\boldmath ${\cal P}$\unboldmath}$, the associated
representation $U(\mbox{\boldmath $p$\unboldmath})$ of its
elements, the invariant vacuum state
$\mbox{\boldmath $\Omega$\unboldmath}$ and finally the fixed, non-dynamical
topological, differentiable manifold
$\mbox{\boldmath $\Rl^{D+1}$\unboldmath}$. 
Thus the Wightman axioms assume the existence of a
non-dynamical, Minkowski background metric which implies that we have a
preferred notion of causality (or locality) and its symmetry group, the
Poincar\'e group from which one builds the usual Fock Hilbert spaces of the
free fields. We see that the whole structure
of the theory is heavily based on the existence of these objects which
come with a fixed, non-dynamical background metric on a fixed,
non-dynamical topological and differentiable manifold.

For a general background spacetime, things are already under much less control:
We still have a notion of causality
(locality) but generically no symmetry group any longer and thus there is no
obvious generalization of the Wightman axioms and no
natural perturbative Fock Hilbert space any longer. These obstacles can
partly be overcome by the
methods of algebraic quantum field theory \cite{62a} and the so-called
microlocal analysis \cite{62a1} (in which the locality axiom is taken care of
pointwise rather than globally) which recently have
also been employed to develop perturbation theory on arbitrary background
spacetimes \cite{62b} by invoking the
mathematically more rigorous implementation of the renormalization programme
developed by Epstein and Glaser in which no divergent expressions ever
appear at least order by order (see, e.g., \cite{62c}). \\
\\
{\it However, the whole framework of ordinary quantum field
theory
breaks down once we make the gravitational field (and the differentiable
manifold) dynamical, once there is no background metric any longer} !\\
\\
Combining these issues, one can say that we have a working understanding of
scattering processes between elementary particles in arbitrary spacetimes as
long as the backreaction of matter on geometry can be neglected and that
the coupling constant between non-gravitational interactions is small enough
(with QCD being an important exception) since then
the classical Einstein equation, which says that curvature of geometry is
proportional to the stress energy of matter, can be approximately solved by
neglecting matter altogether. Thus,
for this set-up, it seems fully sufficient to have only a classical theory of
general relativity and perturbative quantum field theory on curved
spacetimes.

From a fundamental point of view, however, this state of affairs is
unsatisfactory for many reasons among which we have the following:
\begin{itemize}
\item[i)] {\it Classical Geometry -- Quantum Matter Inconsistency}\\
At a fundamental level, the backreaction of matter on geometry  
cannot be neglected. Namely, geometry couples to matter through
{\it Einstein's equations}
\\
\\
\fbox{\fbox{\parbox{12cm}{
\mbox{\boldmath$R_{\mu\nu}-\frac{1}{2} R\cdot g_{\mu\nu}=\kappa
T_{\mu\nu}[g]$\unboldmath}}}}
\\
\\
and since matter undelies the rules of quantum mechanics, the right hand 
side of this equation, the stress-energy tensor $T_{\mu\nu}[g]$, becomes 
an operator. One has tried to keep geometry classical while matter
is quantum mechanical by replacing $T_{\mu\nu}[g]$ by the Minkowski 
vacuum expectation
value $<\hat{T}_{\mu\nu}[\eta]>$ but the solution of this equation 
will give $g\not=\eta$ which one then has to feed back into the 
definition of the vacuum expectation value etc. The resulting 
iteration does not converge in general. Thus, such a procedure is also
inconsistent whence we {\it must quantize the gravitational field as well}.
This leads to the {\it Quantum Einstein Equations}
\\
\\
\fbox{\fbox{\parbox{12cm}{
\mbox{\boldmath$\hat{R}_{\mu\nu}-\frac{1}{2} \hat{R}\cdot 
\hat{g}_{\mu\nu}=\kappa \hat{T}_{\mu\nu}[\hat{g}]$\unboldmath}}}}
\\
\\
Of course, this equation is only formal at this point and must 
be embedded in an appropriate Hilbert space context.
\item[ii)] {\it Inherent Classical Geometry Inconsistency}\\
Even without quantum theory at all Einstein's field equations predict
spacetime singularities (black holes, big bang singularities etc.) at which
the equations become meaningless. In a truly fundamental theory, there is no
room for such breakdowns and it is suspected by many
that the theory cures itself upon quantization in analogy to the Hydrogenium
atom whose stability is classically a miracle (the electron should
fall into the nucleus after a finite time lapse due to
emission of Bremsstrahlung) but is easily explained by quantum theory.
\item[iii)] {\it Inherent Quantum Matter Inconsistency}\\
As outlined above, perturbative quantum field theory on curved
spacetimes is itself also ill-defined due to its UV (short distance)
singularities which can be cured only with an ad hoc recipe order by
order which lacks a fundamental explanation, moreover, the perturbation
series is presumably divergent. Besides that, the usually infinite vacuum
energies being usually neglected in such a procedure
contribute to the cosmological constant and should
have a large gravitational backreaction effect. That such energy subtractions
are quite significant is maybe best demonstrated by the Casimir
effect. Now, since general relativity possesses a fundamental length
scale, the Planck length, it has been argued ever since that gravitation
plus matter should give a finite quantum theory since gravitation
provides the necessary, built-in, short distance cut-off.
\item[iv)] {\it Perturbative Quantum Geometry Inconsistency}\\
Given the fact that perturbation theory works reasonably well if the
coupling constant is small for the non-gravitational interactions on a
background metric it is natural to try whether the methods of
quantum field theory on curved spacetime work as well for the
gravitational field. Roughly, the procedure is to write the
dynamical metric tensor as $g=\eta+h$ where $\eta$ is the Minkowski metric
and $h$ is the deviation of $g$ from it. One
arrives at a formal, infinite series with finite radius of convergence
which becomes meaningless if the fluctuations are large. Although the
naive power counting argument implies that general relativity so defined
is a non-renormalizable theory it was hoped
that due to cancellations of divergencies the perturbation theory could be
actually finite. However, that this hope was unjustified was shown in
\cite{6} where calculations demonstrated the appearance of divergencies
at the two-loop-level, which suggests that at every order of
perturbation theory one must introduce new coupling constants which the
classical theory did not know about and one loses predictability.

It is well-known that the (locally) supersymmetric extension
of a given
non-supersymmetric field theory usually improves the ultra-violet convergence
of the resulting theory as compared to the original one due to fermionic
cancellations \cite{9}. It was therefore natural to hope that quantized
supergravity might be finite. However, in \cite{10} a serious argument
against the expected cancellation of perturbative divergences was raised
and recently even the again popular (due to its M-Theory context) most
complicated 11D
``last hope" supergravity theory was shown not to have the magical
cancellation property \cite{10.1}.

Summarizing, although a definite proof is still missing up to date
(mainly due to the
highly complicated algebraic structure of the Feynman rules for quantized
supergravity) it is today widely believed that perturbative quantum field
theory approaches to quantum gravity are meaningless.
\end{itemize}
The upshot of these considerations is that our understanding of quantum field
theory and therefore fundamental physics is quite limited unless one
quantizes the gravitational field as well. Being very sharply critical
one could say:\\
\\
{\bf The current situation in fundamental physics can be compared with
the one at the end of the nineteenth century: While one had a successful
theory of electromagnetism, one could not explain the stability of atoms.
One did not need to worry about this from a practical point of view
since atomic length scales could not be resolved at that time but from a
fundamental point of view, Maxwell's theory was incomplete. The discovery
of the mechanism for this stability, quantum mechanics, revolutionized
not only physics. Still today
we have no thorough understanding for the stability of nature (an experiment
that everybody can repeat by looking out of the window) and it is
similarly expected that the more complete theory of quantum gravity will
radically change our view of the world. That is, considering the metric as a
quantum operator will bring us beyond standard model physics even
without the discovery of new forces, particles or extra dimensions}.

\newpage

\subsubsection*{The Role of Background Independence}
\addcontentsline{toc}{subsection}{\numberline{}
The Role of Background Independence}
\label{s1.1.2}

The twentieth century has dramatically changed our understanding of nature :
It revealed that
physics is based on two profound principles : quantum mechanics and general
relativity. Both principles revolutionize two pivotal structures of
Newtonian physics : First, the determinism of Newton's equations of motion
evaporates at a fundamental level, rather dynamics is reigned by
probabilities underlying the Heisenberg uncertainty obstruction. Secondly,
the notion of absolute time and space has to be corrected, space and time
and distances between points of the spacetime manifold, that is,
the metric, become themselves dynamical, geometry it is no longer just an
observer. The usual Minkowski metric ceases to be
a distinguished, externally prescribed, background structure.
Rather, the laws of physics are {\it background independent}, mathematically
expressed by the classical Einstein equations which are {\it generally
(or four-diffeomorphism) covariant}. As we have argued, it is this new
element of {\it background independence} brought in with Einstein's
theory of gravity which completely changes our present understanding of
quantum field theory.

A satisfactory physical theory must combine both of these fundamental
principles, quantum mechanics and general relativity, in a consistent
way and will be called ``Quantum Gravity". However, the quantization of the
gravitational field has turned out to be one of the most challenging
unsolved problems in theoretical and mathematical physics. Although
numerous proposals towards a quantization have been made since
the birth of general relativity and quantum theory, none of them can be
called successful so far. This is in sharp contrast to what we see
with respect to the other three interactions whose description has culminated
in the so-called standard model of matter, in particular, the spectacular
success
of perturbative quantum electrodynamics whose theoretical predictions
could be verified {\it to all digits within the experimental error bars}
until today.

Today we do not have
a theory of quantum gravity, what we have is :\\
1) The Standard Model, a quantum theory of the non-gravitational interactions
(electromagnetic, weak and strong) or {\it matter} which, however, completely
ignores general relativity.\\
2) Classical General Relativity or {\it geometry}, which is a background
independent
theory of all interactions but completely ignores quantum mechanics.

What is so special about the gravitational force that it persists its
quantization for about seventy years already ? As outlined in the previous
subsection, {\it the answer is
simply that today we only know how to do Quantum Field Theory (QFT) on fixed
background metrics}. The whole formalism of ordinary QFT
heavily relies on this background structure and collapses
to nothing when it is missing. It is already much more difficult
to formulate a QFT on a non-Minkowski (curved) background but it seems to
become
a completely hopeless task when the metric is a dynamical, even fluctuating
quantum field itself. This underlines once more the source of our current
problem of quantizing gravity : {\it We have to learn how to do QFT on a
differential manifold (or something even more rudimentary, not even
relying on a fixed topological, differentiable manifold) rather than a
spacetime}.

In order to proceed, today a high energy physicist has the choice between the
following two, extreme approaches :\\
Either the {\it particle physicist's}, who
prefers to take over the well-established mathematical machinery from QFT
on a background at the price of dropping background independence altogether
to begin with and then tries to find the true background independent
theory by summing the perturbation series (summing over all possible 
backgrounds). Or the {\it quantum
geometer's}, who believes that background independence lies at the heart of
the solution to the problem and pays the price to have to invent
mathematical tools that go beyond the framework of ordinary QFT right
from the beginning. Both approaches
try to unravel the truly deep features that are unique to Einstein's theory
associated with background independence from different ends.

The particle physicist's language is perturbation theory, that is,
one writes the quantum metric operator as a sum consisting of a
background piece and a perturbation piece around it, the graviton,
thus obtaining a graviton QFT on a Minkowski background. {\it We see that
perturbation theory, by its very definition, breaks background 
independence and diffeomorphism invariance at every finite order of 
perturbation theory}. Thus one can restore background independence 
only by summing up the entire perturbation series which is of course not
easy.
Not surprisingly, as already mentioned,
applying this programme to Einstein's theory itself results in a mathematical
desaster, a so-called non-renormalizable theory without any predictive power.
In order to employ perturbation theory, it seems that one has to go to
string theory which, however, requires the introduction of new additional
structures that Einstein's classical theory did not know about:
supersymmetry, extra dimensions and an infinite tower of new and very
heavy particles next to the graviton. This is a fascinating but
extremely drastic modification of
general relativity and one must be careful not to be in conflict
with phenomenology
as superparticles, Kaluza Klein modes from the dimensional reduction and
those heavy particles have not been observed until today. On the other hand,
string theory has a good chance to be a unified theory of the perturbative 
aspects
of all interactions in the sense that all interactions follow from a
common object, the string, thereby explaining the particle content of the
world.

The quantum geometer's language is a non-perturbative one, keeping
background independence as a guiding principle at every stage of the
construction of the theory, resulting in mathematical structures
drastically different form the ones of ordinary QFT on a background
metric. One takes Einstein's theory absolutely seriously, uses only
the principles of general relativity and quantum mechanics and lets the
theory build itself, driven by mathematical consistency. Any theory meeting
these standards will be called {\it Quantum General Relativity (QGR)}.
Since QGR does not modify the matter content of the known interactions,
QGR is therefore not in conflict with phenomenology but also it cannot
explain the particle content so far. However, it tries to unify
all interactions in a different sense: all interactions must transform
under a common gauge group, the four-dimensional diffeomorphism group
which on the other hand is mlmost completely broken in perturbative
approaches.

Let us remark that
even without specifying further details, any QGR theory is a promising
candidate for a theory that is free from two divergences
of the so-called perturbation series of Feynman diagrammes
common to all perturbative QFT's on a background metric:
(1) Each term in
the series diverges due to the ultraviolet (UV) divergences of the
theory which one can cure for renormalizable theories, such as string theory,
through so-called
renormalization techniques and (2) the series of these renormalized, finite
terms diverges, one says the theory is not finite. The first, UV, problem
has a chance to be absent
in a background independent theory for a simple but profound reason:
In order to to say that a momentum becomes large
one must refer to a background metric with respect to which it
is measured, but there simply is no background metric in the
theory. The second, convergence, problem of the series might be void
as well since there are simply no Feynman diagrammes ! Thus, the mere
existence of a consistent background independent quantum gravity theory
could imply a finite quantum theory of all interactions.


\subsubsection*{Approaches to Quantum Gravity}
\addcontentsline{toc}{subsection}{\numberline{}
Approaches to Quantum Gravity}

\label{s1.1.3}

The aim of the previous subsection was to convince the reader that
background independence is, maybe, {\bf The Key Feature} of quantum gravity
to be dealt with. No matter how one deals with this issue, whether
one starts from a perturbative ($=$ background dependent) or from a
non-perturbative ($=$ background independent) platform, one has to
invent something drastically new in order to quantize the gravitational
field. We will now explain these approaches in more detail, listed
in decreasing numbers of researchers working in the respective fields.
\begin{itemize}
\item[1)] {\it Perturbative Approach : String Theory}\\
The only known consistent perturbative approach to quantum gravity is string
theory which has good chances to be a theory that unifies all interactions.
String Theory \cite{11} is not a field theory in the ordinary sense of the
word.
Originally, it was a two-dimensional field theory of world-sheets
embedded into a
fixed, D-dimensional pseudo-Riemannian manifold $(M,g)$ of Lorentzian
signature which is
to be thought of as the spacetime of the physical world. The Lagrangean of
the theory is a kind of non-linear $\sigma-$model Lagrangean for the
associated embedding variables $X$ (and their supersymmetric partners in
case of the superstring). If one perturbes $g(X)=\eta+h(X)$ as above and
keeps only the lowest order in $X$ one obtains a free field theory in
two dimensions which, however, is consistent only when $D=26$ (bosonic
string) or $D=10$ (superstring) respectively. Strings propagating in those
dimensions are called critical strings, non-critiical strings exist
but have so far not played a significant role due to phenomenological 
reasons. Remarkably, the 
mass spectrum of the particle-like excitations of the closed worldsheet
theory contain a massless
spin-two particle which one interprets as the graviton. Until recently,
the superstring
was favoured since only there it was believed to be possible to get rid
off an unstable
tachyonic vaccum state by the GSO projection. However,
one recently also tries to construct stable bosonic string theories
\cite{12c}.

Moreover, if one
incorporates the higher order terms $h(X)$ of the string action, sufficient
for one loop corrections, into
the associated path integral one finds a consistent quantum theory up to
one loop only if the background metric satisfies the Einstein equations.
These are the most powerful outcomes of the theory : although one started
out with a fixed background metric, the background is not arbitrary
but has to satisfy the Einstein equations up to higher loop
corrections indicating that the one-loop effective action for the low
energy quantum field theory in those $D$ dimensions is Einstein's theory
plus corrections. Finally, at least the type II superstring theories are
are one-loop and, possibly, to all orders, {\it finite}. String theorists
therefore argue to have found candidates for a consistent theory of
quantum gravity with the
additional advantage that they do not contain any free parameters
(like those of the standard model) except for the string tension.

These facts are very impressive, however, some cautionary remarks
are appropriate:
\begin{itemize}
\item {\it Vacuum Degeneracy}\\
Dimension $D+1=10,26$ is not the dimension of everyday physics so that one
has to argue that the extra $D-3$ dimensions are ``tiny" in the
Kaluza-Klein sense although nobody knows the mechanism responsible for
this ``spontaneous compactification". According to \cite{12b}
there exists an order of $10^4$ consistent, distinct Calabi-Yau
compactifications (other compactifications such as toroidal 
ones seem to be inconsistent with phenomenology)
each of which has an order of $10^2$ free, continuous parameters (moduli)
like the vacuum expectation value of the Higgs field in the standard
model. For each compactification of each of the five string theories in
$D=10$ dimensions and for each choice of the moduli one obtains a
distinct low energy effective theory. 
This is
clearly not what one expects from a theory that aims to unify all the
interactions, the 18 (or more for massive neutrinos) free, continuous
parameters of the standard model have been replaced by $10^2$ continuous
plus at least $10^4$ discrete ones.

This vacuum degeneracy problem is not cured by the M-Theory
interpretation of string theory but it is conceptually simplified if certain
conjectures are indeed correct : String theorists
believe (bearing on an impressively huge number of successful checks) that 
so-called
T (or target space) and S (or strong -- weak coupling) duality transformations
between all these string theories exist which suggest that we
do not have $10^4$ unrelated $10^2$-dimensional moduli spaces but that
rather these $10^2$-dimensional manifolds intersect in singular,
lower dimensional submanifolds corresponding to certain singular moduli
configurations.
This typically happens when certain masses vanish or certain couplings
diverge or vanish (in string theory the coupling is related to the vacuum
expectation value of the dilaton field). Crucial in this picture are
so-called D-branes, higher dimensional objects additional to strings
which behave like solitons (``magnetic monopoles") in the electric
description of a string theory and like fundamental objects (``electric
degrees of freedom") in the S-Dual description of the same string
theory, much like the electric -- magnetic duality of Maxwell theory
under which strong and weak coupling are exchanged. Further relations
between different string theories are obtained by compactifying them in
one way and decompactifying them in another way, called a T -- duality
transformation. The resulting picture is that there exists only one
theory which has all these compactification limits just described,
called M -- Theory. Curiously, M -- Theory is an 11D theory whose low
energy limit is 11D supergravity and whose weak coupling limit is type
IIA superstring theory (obtained by one of these singular limits
since the size of the 11th compactified
dimension is related to the string coupling again). Since 11D
supergravity is also the low energy limit of the 11D supermembrane, some
string theorists interpret M-Theory as the quantized 11D supermembrane
(see, e.g., \cite{12} and references therein).
\item {\it Phenomenology Match}\\
Until today, no conclusive proof exists that for any of the
compactifications described above we obtain a low energy effective theory
which is experimentally consistent with the data
that we have for the standard model \cite{12d} although one seems
to get at least rather close. The challenge in
string phenomenology is to consistently and spontaneously break
supersymmetry in order to get rid off the so far
non-observed superpartners. There is also an
infinite tower of very massive (of the order of the Planck mass and
higher) excitations of the string but these are too heavy to be 
observable. More interesting are the Kaluza Klein modes whose masses 
are inverse proportional to the compactification radii and which have 
recently given rise to speculations about ``sub-mm-range" gravitational forces
\cite{12e} which one must make consistent with observation also.
\item {\it Fundamental Description}\\
Even before the M -- Theory revolution, string theory has always been a
theory without Lagrangean description, S -- Matrix element computations
have been guided by conformal invariance but there is no ``interaction
Hamiltonian", string theory is a first quantized theory.
Second quantization of string theory, called string field theory
\cite{12f}, has so far not attracted as much attention as it possibly
deserves. However, a fascinating possibility is that 
the 11D supermembrane, and thus M -- Theory, is an already second quantized 
theory \cite{12g}. 
\item {\it Background Dependence}\\
As mentioned above, string theory is best understood as a free 2D field
theory propagating on a 10D Minkowski target space plus perturbative
corrections for scattering matrix computations.
This is a heavily background dependent description, issues like the
action of the 10D diffeomorphism group, the fundamental symmetry of
Einstein's action,  or the backreaction of matter on geometry, cannot be
asked.
Perturbative string theory, as far as quantum gravity is
concerned, can describe graviton scattering in a background spacetime,
however, most problems require a non-perturbative description
when the backreaction can no longer be ignored, such as scattering
at quantum black holes. As a first step in that direction,
recently stringy black holes have been discussed \cite{12h}. Here one 
uses so-called BPS D-brane configurations which are so special that
one can do a perturbative calculation and extend it to the
non-perturbative regime since the results are
protected against non-perturbative corrections due to supersymmetry.
So far 
this works only for extremely charged, supersymmetric black holes which are
astrophysically not very realistic. But still these developments are 
certainly a move in the right direction since they use for the first
time non-perturbative ideas in a crucial way and have 
been celebrated as one of the triumphes of string theory.
%
%
\end{itemize}
\item[2)] {\it Non-Perturbative Approaches}\\
The non-perturbative approaches to quantum gravity can be grouped into
the following five main categories.
\begin{itemize}
\item[2a)] {\it Canonical Quantum General Relativity}\\
If one wanted to give a definition of this theory then one
could say the following:\\
\\
\fbox{\fbox{\parbox{12cm}{{\bf Canonical Quantum General
Relativity is an attempt to construct a mathematically rigorous,
non-perturbative, background independent quantum theory
of four-dimensional, Lorentzian general relativity plus all
known matter in the continuum.}}}}
\\
\\
This is the oldest approach and goes back to the pioneering work
by Dirac \cite{1} started in the 40's and was further developed
especially by Wheeler and DeWitt \cite{2} in the 60's. The idea of this
approach is to apply the Legendre transform to the Einstein-Hilbert
action by splitting spacetime into space and time and to cast it into
Hamiltonian form. The resulting ``Hamiltonian" $H$ is actually a so-called
Hamiltonian constraint, that is, a Hamiltonian density which is
constrained to vanish by the equations of motion. A Hamiltonian constraint
must occur in any theory that, like general relativity, is invariant under
local reparameterizations of time. According to Dirac's theory of the
quantization of constrained Hamiltonian systems, one is now supposed to
impose the vanishing of the quantization $\hat{H}$ of the
Hamiltonian constraint $H$ as a condition on states $\psi$ in a suitable
Hilbert space $\cal H$, that is, formally
\\
\\
\fbox{\fbox{\parbox{12cm}{
\mbox{\boldmath$\hat{H}\psi=0$\unboldmath}}}}
\\
\\
This is the famous Wheeler-DeWitt equation or
{\bf Quantum-Einstein-Equation} of canonical
quantum gravity and resembles a Schr\"odinger equation, only that the
familiar $\partial\psi/\partial t$ term is missing, one of several
occurences of the ``absence or problem of time" in this approach
(see, e.g., \cite{4} and references therein).\\
Since the status of this programme is the subject of the present review
we will not go too much into details here. The successes of the
theory are a mathematically rigorous framework, manifest background
independence, a manifestly non-perturbative language, an inherent
notion of quantum discreteness of spacetime which is {\it derived} rather
than postulated, certain UV finiteness results, a promising path
integral formulation (spin foams) and finally a consistent formulation of
quantum black hole physics.

The following issues are at the moment unresolved within this
approach:
\begin{itemize}
\item {\it Tremendously Nonlinear Structure}\\
The Wheeler-DeWitt operator is, in the so-called ADM formulation,
a functional differential operator of second order of the worst kind,
namely with non-polynomial, not even analytic (in the basic
configuration variables) coefficients. To even define such an operator
rigorously has been a major problem for more than 60 years. What should
be a suitable Hilbert space that carries such an operator ? It is known
that a Fock Hilbert space is not able to support it.
Moreover, the structure
of the solution space is expectedly very complicated. Thus we see
that one meets a great deal of mathematical problems before one
can even start addressing physical questions. As we will describe in this
report there has been a huge amount of progress in this direction
since the introduction of new canonical variables due to
Ashtekar \cite{5} in 1986. However, the physics of the
Wheeler-DeWitt operator is still only poorly understood.
\item {\it Loss of Manifest Four-Dimensional Diffeomorphism Covariance}\\
Due to the split of spacetime into space and time the
treatment of spatial and time diffeomorphisms is somewhat
different and the original four-dimensional covariance of the
theory is no longer manifest. Classically one can prove
(and we will in fact do that later on) that four-dimensional
diffeomorphism covariance is encoded in a precise sense into
the canonical formalism, although it is deeply hidden.
In quantum theory the proper implementation of the
diffeomorphism group is the question whether the so-called
Dirac algebra (of which the Wheeler -- DeWitt operator is an element)
has an anomaly or not and which at the moment has no conclusive answer.

Let us clarify an issue that comes up often in debates between quantum 
geometers and string theorists:\\ 
What one means by $(D+1)$-dimensional covariance
in string theory on a Minkowski target space is just $(D+1)-$dimensional
{\it Poincar\'e} covariance but not {\it Diffeomorphism} covariance.
Clearly the Poincar\'e group is a subgroup of the diffeomorphism group
(for asymptotically flat spacetimes) of measure zero and the rest
of the diffeomorphism group, which is in fact the symmetry group of
Einstein's theory, is completely broken in string theory. In canonical
QGR at least the huge spatial subgroup of the diffeomorphism group is 
manifestly without anomalies and possibly the remaining part of the 
diffeomorphism group as well.
\item {\it Interpretational (Conceptual) Issues}\\
Once one has found the solutions of the Quantum Einstein
Equations one must find a complete set of Dirac observables
(operators that leave the space of solutions invariant)
which is an impossible task to achieve even in classical
general relativity. One must therefore find suitable
approximation methods which is a development which has just
recently started. However, even if one would have found those
(approximate) operators, which would be in some sense even
time independent and therefore extremely non-local, one would need to 
deparameterize the theory, that is, one must find an explanation for 
the local dynamics in our world. There are some proposals for dealing with this
issue but there is no rigorous framework available at
the moment.
\item {\it Classical Limit}\\
As we will see, our Hilbert space is of a new (background independent)
kind, operators are regulated in a non-standard (background independent)
way. It is therefore no longer clear that the theory that has been
constructed so far indeed has general relativity as its classical
limit. Again, semiclassical analysis has just been launched recently.
\end{itemize}
\item[2b)] {\it Continuum Functional Integral Approach}\\
Here one tries to give meaning to the sum over histories of $e^{-S_E}$
where $S_E$ denotes the Euclidean Einstein-Hilbert action \cite{7}.
It is extremely hard to do the path integral and
apart from semi-classical approximations and steepest descent
methods in simplified models with a finite number of degrees of freedom
one could not get very far within this framework yet.
There are at least the two following reasons for this :\\
1) The action functional $S_E$ is unbounded from below. Therefore the
path integral is badly divergent from the outset and although rather
sophisticated proposals have been made of how to improve the convergence
properties, none of them has been fully successful to the best knowledge of
the author.\\
2) The Euclidean field theory underlying the functional integral and
the quantum theory of fields propagating on a Minkowski background
are related by Wick rotating the Schwinger functions of the former into
the Wightman functions of the latter (see, e.g., \cite{8}).
However, in the case of quantum gravity the metric itself becomes 
dynamical and is being integrated over, therefore the concept 
of Wick rotation becomes ill-defined. In other words, there is no
guarantee that the Euclidean path integral even has any relevance for the
quantum field theory underlying the Lorentzian
Einstein-Hilbert action. \\
On the other hand, the functional integral approach has motivated the
consistent histories approach to the quantum mechanics of closed
systems (cosmologies) \cite{8a} which in many senses is superiour
over the Copenhagen interpretation.
\item[2c)] {\it Lattice Quantum Gravity}\\
This approach can be subdivided into two main streams (see 
\cite{9a} for a review):\\
a) Regge Calculus \cite{10a}. Here one introduces a fixed triangulation
of spacetime and integrates with a certain measure over the lengths of
the links of this triangulation. The continuum limit is reached by
refining the triangulation.\\
b) Dynamical Triangulations \cite{10b}. Here one takes the opposite point
of view and keeps the lengths of the links fixed but sums over
all triangulations. The continuum limit is reached by taking the 
link length to zero.\\
In both approaches one has to look for critical points (second order 
phase transitions). An issue in both approaches is the 
choice of the correct measure. Although there is no guideline, 
it is widely believed that the dependence on the measure is weak 
due to universality in the statistical mechanical sense. The reason for 
the possibility that the path integral exists although the Euclidean 
action is unbounded from below is that the configurations with large 
negative action have low volume (measure) so that ``entropy wins over
energy". Especially in the field of dynamical triangulations there 
has been a major breakthrough recently \cite{9c}: The convergence of the 
partition function could be established {\it analytically} in two dimensions
(the action is basically a cosmological constant term)
and the relation between the Lorentzian and Euclidean theory becomes 
transparent. This opens the possibility that similar results hold in 
higher dimensions, in particular, it seems as if the Lorentzian theory
is much better behaved than the Euclidean theory because one has to sum
over fewer configurations (those that are compatible with quantum causality).
There are also promising new results concerning a non-perturbative Wick 
rotation \cite{9d}.

What is still missing within this approach (in more than two dimensions), 
as in any path integral 
approach for quantum gravity that has been established so far, is a clear 
physical interpretation of the expectation values of observables as 
transition amplitudes in a given Hilbert space. A possible way out
could be proposed if one were able to establish reflection positivity of 
the measure, see \cite{8}. 
\item[2d)] {\it Non-Orthodox Approaches}\\
Approaches belonging to this approach start by questioning standard
quantum field theory at an even more elementary level. Namely, if the
ideas about spacetime foam (discrete structure of spacetime) are indeed true
then {\it one should not even start formulating quantum field theory on a
differentiable manifold} but rather something intrinsically discrete.
Maybe we even have to question the foundations of quantum mechanics
and to depart from a purely binary logic. 
To this category belong the {\it Non-Commutative Geometry} by Alain
Connes \cite{8b} also considered recently by string theorists \cite{8c}, the
{\it Topos Theory} by Chris Isham \cite{10c},
the {\it Twistor Theory} by Roger Penrose \cite{10d}, the {\it Causal Set
Programme} by Raphael Sorkin \cite{10e} and finally the Consistent History
approach due to Gell-Mann and Hartle \cite{8a} which recently has picked 
up some new momentum in terms of the history phase space due to Isham 
et al and Kucha\v{r} et al \cite{10e1}.\\
These approaches are, maybe, the most radical reformulations of fundamental
physics but they are also the most difficult ones because the contact with
standard quantum field theory is very small. Consequently, these programmes
are in some sense ``farthest" from observation and are consequently least
developed so far. However, the ideas spelled out in these programmes 
could well reappear in the former approaches as well once the latter have 
reached a sufficiently high degree of maturity.
\end{itemize}
All of these four nonperturbative programmes are mutually loosely 
connected : Roughly, the operator
formulation of the canonical approach is equivalent to the continuous
path integral formulation through some kind of Feynman-Kac formula
a concrete implementation of which are the so-called spin-foam models to
be mentioned later,
lattice quantum gravity is a discretization of the path integral formulation
and both the canonical and the lattice approach seem to hint at discrete
structures on which the non-orthodox programmes are based.

Finally, every non-perturbative programme better contains a sector 
which is well described by perturbation theory and therefore string theory
which then provides an interface between the two big research streams.
\end{itemize}
This ends our survey of the existent quantum gravity programmes.
By far most of the people, of the order of $10^3$, work in string theory,
followed by the canonical programme 
with an order of $10^2$ scientists, then the lattice quantum gravity --,
functional integral --
and the non-commutative programme with an order of $10^1$ researchers and
finally the non-orthodox programmes except for the non-commutative one
with an order of $10^0$ physicists.

\subsubsection*{Motivation for Canonical Quantum General Relativity}
\addcontentsline{toc}{subsection}{\numberline{}
Motivation for Canonical Quantum General Relativity}

\label{s1.1.4}

We close this section by
motivating our choice to follow the non-perturbative, canonical approach.
Of course, our discussion cannot be entirely objective.
\begin{itemize}
\item[I)] {\it Non-Perturbative versus Perturbative}


Our preference for a nonperturbative approach is twofold:\\ 
\\
The first 
reason is certainly a matter of taste, a preference for a certain 
methodology: 
Try to combine the two fundamental principles, general relativity and 
quantum mechganics {\it with no additional structure}, explore the 
logical consequences and push the framework until success or until
there is a contradiction (inconsistency) either within the theory or with 
the experiment. In the latter case, examine the reason for failure and 
try to modify theory appropriately. The reason for not allowing 
additional structure (principle of minimality) is that unless we only use 
structures which have been confirmed to be a property of nature then we are 
standing in front 
of an ocean of possible new theories which a priori could be equally 
relevant. In a sense we are saying that if 
gravity cannot be quantized perturbatively without extra structures such 
as necessary in string theory, then one should try a nonperturbative 
approach. If that still fails then maybe we find out why and 
exactly which extra 
structures are necessary rather than guessing them. Such a methodology has 
proved to be very successful in the history of science.\\
\\
The second reason, however, is maybe more serious: 
{\it It is not at all true that perturbation theory is always a 
good approximation in a non-empty neighbourhood of the expansion point}.
To quote an example from \cite{10f}, consider the harmonic oscillator
Hamiltonian $H=p^2+\omega^2 q^2$ and let us treat the potential
$V=\omega^2 q^2$ as an interaction Hamiltonian perturbing the free
Hamiltonian $H_0=p^2$ at least for low frequencies $\omega$. The exact
spectrum of $H$ is discrete while that of $H_0$ is continuous. The point is
now that {\it one is never going to see, for no value of the 
``coupling constant" $\omega>0$, the
discreteness of the unperturbed
Hamiltonian by doing perturbation theory and thus one completely misses the
correct physics} !

Finally, borrowing from \cite{10g}, let us exhibit a calculation which
demonstrates the {\it regularizing mechanism of a non-perturbative treatment
of general gravity taking its very non-linear nature very serious}.\\
Consider the self-energy of a bare point charge $e_0$ with rest mass 
$m_0$ due to static electromagnetic and gravitational interaction.
From the point of view of Newtonian physics, this energy is of the form
($\hbar=c=1$, the bare Newton's constant is denoted by $G_0$)
$$
m(r)=m_0+e_0^2/r-G_0 m_0^2/r
$$
and diverges as $r\to 0$ unless $e_0,m_0, G_0$ are fine tuned. However, 
general relativity
tells us that all of the mass of the charge, that is rest mass plus 
field energy within a shell of radius $r$ couples
to the gravitational field which is why above equation should be replaced by
$$
m(r)=m_0+e_0^2/r-G_0 m(r)^2/r
$$
which can be solved for
$$
m(r)=\frac{r}{2G_0}[-1+\sqrt{1+\frac{4G_0}{r}(m_0+\frac{e_0^2}{r})}]
$$
Notice that now the bare mass $m(r=0)=e_0/\sqrt{G_0}$ is {\it finite without 
fine tuning. Moreover,
the result is non-analytical in Newton's constant $G_0$ and is not accessible
by perturbation theory, in particular, the bare mass is independent of the rest
mass } ! Of course, this calculation should not be taken too seriously since
e.g. no quantum effects have been brought in, it merely serves to illustrate
our point that general relativity could serve as natural  regulator of
field theory divergences. (However, a proper general relativistic 
treatment (ADM mass of the Reissner-Nordstr{\o}m solution) \cite{10g}
can be performed, see also for more details).

These arguments can be summarized by saying that there is a good chance that
perturbative quantum gravity {\it completely misses the point} although,
of course, there is no proof ! 
\item[II)] {\it Canonical versus other Non-Perturbative Approaches}

Here our motivation is definitely just a matter of taste, that is,
we take a practical viewpoint:

Path integrals
have the advantage that they are manifestly four-dimensionally 
diffeomorphism invariant but their huge disadvantage is that they are
awfully hard to compute analytically, even in quantum mechanics. While 
numerical methods 
will certainly enter the canonical approach as well in the close future 
one gets farther with analytical methods. However, it should be stressed 
that path integrals and canonical methods are very closely related and 
usually one can derive one from the other through some kind of 
Feynman -- Kac formula.

The non-orthodox approaches have the
advantage of starting from a discrete/non-commutative spacetime structure
from scratch while in canonical quantum gravity one begins with a smooth
spacetime manifold and obtains discrete structures as a derived concept only
which is logically less clean: The true theory is the quantum theory and 
if the world is discrete one should not begin with smooth structures at all.
Our viewpoint is here that, besides the fact that again 
the canonical approach is more minimalistic, at some stage in 
the development of 
the theory there must be a quantum leap and in the final reformulation 
of the theory everything is just combinatorical. This can actually 
be done in 2+1 gravity as we will describe later on ! 
\end{itemize}

\newpage

\subsection*{Notation and Conventions}
\addcontentsline{toc}{section}{\numberline{}
Notation and Conventions}
\label{s1.2}

You can tell whether a high energy physicist is a particle physicist or
a quantum geometer from the index notation that she or he uses.
We obviously use here the quantum geometer's (that is, latin letters
from the beginning of the alphabet are tensorial while those from the
middle are Lie algebra indices), the particle physicists's is often just
opposite.\\
\\
G$=6.67\cdot 10^{-11} m^3 kg^{-1} s^{-2}$: Newton's constant\\
$\kappa=8\pi$G$/c^3$: Gravitational coupling constant  \\
$\ell_p=\sqrt{\hbar \kappa}=10^{-33}$ cm: Planck length\\
$m_p=\sqrt{\hbar/\kappa}/c=10^{19}$GeV: Planck mass\\
$Q$: Yang-Mills coupling constant\\
$M,\;\dim(M)=D+1$: Spacetime manifold\\
$\sigma,\; \dim(\sigma)=D$: Abstract spatial manifold\\
$\Sigma$: Spatial manifold embedded into $M$\\
$G$: Compact gauge group\\
Lie$(G)$: Lie algebra\\
$N-1$: Rank of gauge group\\
$\mu,\nu,\rho,..=0,1,..,D$: Tensorial spacetime indices\\
$a,b,c,..=1,..,D$: Tensorial spatial indices\\
$\epsilon_{a_1..a_D}$: Levi-Civita totally skew tensor density\\
$g_{\mu\nu}$: Spacetime metric tensor\\
$q_{ab}$: Spatial (intrinsic) metric tensor of $\sigma$\\
$K_{ab}$: Extrinsic curvature of $\sigma$\\
$R$: Curvature tensor\\
$\underline{h}$: Group elements for general $G$\\
$\underline{h}_{mn},\; m,n,o,..=1,..,N$: Matrix elements for general $G$\\
$I,J,K,..=1,2,..,\dim(G)$: Lie algebra indices for general $G$\\
$\underline{\tau}_I$: Lie algebra generators for general $G$\\
$k_{IJ}
=-\mbox{tr}(\underline{\tau}_I\underline{\tau}_J)/N:=\delta_{IJ}$:
Cartan-Killing metric for $G$\\
$[\underline{\tau}_I,\underline{\tau}_J]=
2 f_{IJ}\;^K\underline{\tau}_K$:
Structure constants for $G$\\
$\underline{\pi}(\underline{h})$:
(Irreducible) representations for
general $G$\\
$h$: Group elements for $SU(2)$\\
$h_{AB},\;A,B,C,.. =1,2$: Matrix elements for $SU(2)$\\
$i,j,k,..=1,2,3$: Lie algebra indices for $SU(2)$ \\
$\tau_i$: Lie algebra generators for $su(2)$\\
$k_{ij}=\delta_{ij}$: Cartan-Killing metric for $SU(2)$\\
$f_{ij}\;^k=\epsilon_{ijk}$: Structure constants for $SU(2)$\\
$\pi_j(h)$: (Irreducible) representations for $SU(2)$ with spin $j$\\
$\underline{A}$: Connection on $G$-bundle over $\sigma$\\
$\underline{A}_a^I$: Pull-back of $\underline{A}$ to $\sigma$ by local
section\\
$g$: gauge transformation or element of complexification of $G$\\
$P$: Principal $G-$bundle\\
$A$: Connection on $SU(2)$-bundle over $\sigma$\\
$A_a^i$: Pull-back of $A$ to $\sigma$ by local section\\
$\ast\underline{E}$: $(D-1)$-form covector bundle associated to the
$G$-bundle under the adjoint representation\\
$\ast\underline{E}_{a_1..,a_{D-1}}^I=:
k^{IJ}\epsilon_{a_1..,a_D}\underline{E}^{a_D}_J$:
Pull-back of $\ast\underline{E}$ to $\sigma$ by local section\\
$\ast E$: $(D-1)$-form covector bundle associated to the $SU(2)$-bundle
under the adjoint representation\\
$\ast E_{a_1..,a_{D-1}}^i=:k^{ij}\epsilon_{a_1..,a_D} E^{a_D}_j$:
Pull-back of $\ast E$ to $\sigma$ by local section\\
$E^a_j:=\epsilon^{a_1 .. a_{D-1}} (\ast E)^k_{a_1..a_{D-1}} k_{jk}/((D-1)!)$:
``Electric fields"\\
$e$: One-form covector bundle associated to the $SU(2)$-bundle under the
defining representation ($D$-bein)\\
$e_a^i$: Pull-back of $e$ to $\sigma$ by local section\\
$\Gamma_a^i$: Pull-back by local section of $SU(2)$ spin connection
over $\sigma$\\
$\cal M$: Phase space\\
$\cal E$: Banach manifold or space of smooth electric fields\\
$T_{(a_1..a_n)}:=\frac{1}{n!}\sum_{\iota\in S_n}
T_{a_{\iota(1)}..a_{\iota(n)}}$:
Symmetrization of indices\\
$T_{[a_1..a_n]}:=\frac{1}{n!}\sum_{\iota\in S_n}
\mbox{sgn}(\iota)\;T_{a_{\iota(1)}..a_{\iota(n)}}$:
Antisymmetrization of indices\\
$\a$: Space of smooth connections or abstract algebra\\
$\g$: Space of smooth gauge transformations\\
$\ab$: Space of distributional connections\\
$\ac$: Space of smooth complex connections\\
$\gc$: Space of smooth complex gauge transformations\\
$\gb$: Space of distributional gauge transformations\\
$\ag$: Space of smooth connections modulo smooth gauge transformations\\
$\abgb$: Space of distributional connections modulo distributional gauge
transformations\\
$\agb$: Space of distributional gauge equivalence classes of connections\\
$\abc$: Space of distributional complex connections\\
$\agbc$: Space of distributional complex gauge equivalence
classes of connections\\
$\cal C$: Set of piecewise analytic curves or classical configuration
space\\
$\overline{{\cal C}}$: Quantum configuration space\\
$\cal P$: Set of piecewise analytic paths\\
$\cal Q$: Set of piecewise analytic closed and basepointed paths\\
$\cal L$: Set of tame subgroupoids of $\cal P$ or general label set\\
$\cal S$: Set of tame subgroups of $\cal Q$ (hoop group) or
set of spin network labels\\
$s$: spin-net$=$ spin-network label (spin-net)\\
$\Gamma^\omega_0$: Set of piecewise analytic, compactly supported graphs\\
$\Gamma^\omega_\sigma$: Set of piecewise analytic, countably infinite 
graphs\\ 
$c$: Piecewise analytic curve\\
$p$: Piecewise analytic path\\
$e$: Entire analytic path (edge)\\
$\alpha$: Entire analytic closed path (hoop)\\
$\gamma$: Piecewise analytic graph\\
$v$: vertex of a graph\\
$E(\gamma)$: Set of edges of $\gamma$\\
$V(\gamma)$: Set of vertices of $\gamma$\\
$h_p(A)=A(p)$: holonomy of $A$ along $p$\\
$\prec$: abstract partial order\\
$\Omega$: vector state or symplectic structure or curvature two-form\\
$F$: pull-back to $\sigma$ of $2\Omega$ by a local section\\
$\omega$: general state\\
$\cal H$: general Hilbert space\\
Cyl: Space of cylindrical functions\\
$\cal D$: dense subspace of $\cal H$ equipped with a stronger topology\\
${\cal D}'$: topological dual of $\cal D$\\
${\cal D}^\ast$: algebraic dual of $\cal D$\\
${\cal H}^0=L_2(\ab,d\mu_0)$: Uniform measure $L_2$ space\\
${\cal H}^\otimes$: Infinite Tensor Product extension of ${\cal H}^0$\\
Cyl$_l$: Restriction of Cyl to functions cylindrical over $\gamma$\\
$[.],(.)$: Equivalence classes\\
Diff$^\omega(\sigma)$: Group of analytic diffeomorphisms of $\sigma$\\
$\varphi$: analytic diffeomorphism

\newpage

~~~~~~~~~~~\\
{\huge\bf DISCLAIMER}\\
\\
Like every review also this one is necessarily incomplete. There are
many more fascinating results which we could not possibly also describe
due to lack of space and time.

The selection of material is certainly
biased by the author's own taste and by the fact that we wanted to keep
this review at the same time compact and complete. This is
not meant, at all, as a quality evaluation. We apologize to all those
researchers whose results were, in their mind, not (or not sufficiently) 
highlighted and
hope that the extensive bibliography displays a reasonably fair and
objective
overview of the available literature.

In that respect, let us mention that
until four years ago a complete and nicely structured literature list
\cite{202} was available
which unfortunately has not been updated since then. It would be
nice if some good soul would find the time and strength to fill in
the huge amount of literature that has piled up over the last four years.

\newpage

\cleardoublepage

\part{Foundations}
\label{p1}


\section{Classical Hamiltonian Formulation of General Relativity and the
Programme of Canonical Quantization}
\label{s2}

In this section we focus on the classical Hamiltonian formulation of
general relativity. First we repeat the most important steps that have
lead to the metrical formulation due to Arnowitt, Deser and Misner
\cite{16}. Then we introduce the main ideas behind the programme
of canonical quantization and summarize the status of that programme
when applied to general relativity in the ADM formulation. As we will see,
not much progress could be achieved in that formulation which has
motivated Ashtekar to look for a different one, more suitable to quantization.
We introduce this connection formulation in the final subsection of this
section.

\subsection{The ADM Formulation}
\label{s2.1}

The major reference for this subsection is definitely the beautiful textebook
by Wald (especially appendix E and chapter 10) and by Hawking\&Ellis
\cite{14}.

The object of interest is the Einstein -- Hilbert action for metric tensor
fields $g_{\mu\nu}$ of Lorentzian ($s=-1$) or Euclidean ($s=+1$) signature
which propagate on a $(D+1)$-dimensional manifold $M$
\be \label{2.1}
S=\frac{1}{\kappa}\int_M d^{D+1}X \sqrt{|\det(g)|}R^{(D+1)} \;.
\ee
In this article we will be mostly concerned with $s=-1,D=3$ but since
the subsequent derivations can be done without extra effort we will be more
general here.
Our signature convention is ``mostly plus´´, that is, $(-,+,..,+)$ or
$(+,+,..,+)$ in the Lorentzian or Euclidean case respectively so that
timelike vectors have negative norm in the Lorentzian case..
Here $\mu,\nu,\rho,..=0,1,..,D$ are indices for the components of spacetime
tensors and $X^\mu$ are the coordinates of $M$ in local
trivializations. $R^{(D+1)}$ is the curvature scalar
associated with $g_{\mu\nu}$ and $\kappa=8\pi$G where G is Newton's
constant (in units where $c=1$). The definition of the Riemann
curvature tensor is in terms of one-forms given by
\be \label{2.2}
[\nabla_\mu,\nabla_\nu]u_\rho=^{(D+1)}R_{\mu\nu\rho}\;^\sigma \; u_\sigma
\ee
where $\nabla$ denotes the unique, torsion-free, metric-compatible,
covariant differential associated with $g_{\mu\nu}$.
To make the action principle
corresponding to (\ref{2.1}) well-defined one has, in general, to add
boundary terms which we avoid by assuming that $M$ is spatially compact
without boundary. The
more general case can be treated similarly, see e.g. \cite{19},
but would unnecessarily complicate the analysis.

In order to cast (\ref{2.1}) into canonical form
one makes the assumption that $M$ has the special topology $M=\Rl
\times \sigma$ where $\sigma$ is a fixed three-dimensional,
compact manifold without boundary. By a theorem due to Geroch
\cite{19a},
if the spacetime is globally hyperbolic (existence of Cauchy surfaces
in accordance with the determinism of classical physics)
then it is necessarily of this kind of topology. Therefore, for classical
physics our assumptions about the topology of $M$ seems to be no
restriction at all, at least in the Lorentzian signature case. In
quantum gravity, however, different kinds of topologies
and, in particular, {\it topology changes} are conceivable. Our philosophy
will be first to construct the quantum theory of the gravitational field
based on the classical assumption that $M=\Rl\times \sigma$ and then
to {\it lift this restriction} in the quantum theory. A concrete proposal
for such a lifting which naturally suggests itself in our
formulation will be given later on, we are even able to allow for certain
classes of signature changes ! Notice that {\it none} of the approaches
listed in the introduction, except for the path integral approach which,
however, is mathematically poorly defined in more than two spacetime
dimensions, knows how to deal with topology changes from first principles,
there exist only ad hoc prescriptions.

Having made this assumption, one knows that $M$ foliates into
hypersurfaces $\Sigma_t:=X_t(\sigma)$, that is, for each fixed $t\in \Rl$
we have an embedding (a globally injective immersion)
$X_t\; : \; \sigma\to M$ defined by
$X_t(x):=X(t,x)$ where $x^a,\;a,b,c,..=1,2,3$ are local coordinates of
$\sigma$. Likewise we have a diffeomorphism $X\;:\;\Rl\times\sigma\mapsto
M;\;(t,x)\mapsto X(t,x):=X_t(x)$, in other words, a one-parameter family
of embeddings is equivalent to a diffeomorphism. We would like to use
these special diffeomorphisms in order to give a $D+1$ (space and time)
decomposition of the action (\ref{2.1}). Now, since the action (\ref{2.1}) is
invariant under all diffeomorphisms (changes of the coordinate system) of $M$
the families of embeddings $X_t$ are not specified by it and
we allow them to completely arbitrary (a
precise characterization of these ``embedding diffeomorphisms" as
compared to Diff$(M)$ can be found in \cite{15}).
A useful parameterization of the embedding and its arbitrariness can be
given through its deformation vector field
\be \label{2.3}
T^\mu(X):=(\frac{\partial X^\mu(t,x)}{\partial t})_{|X=X(x,t)}
=:N(X)n^\mu(X)+N^\mu(X)
\ee
Here $n^\mu$ is a unit normal vector to $\Sigma_t$, that is,
$g_{\mu\nu}n^\mu n^\nu=s$ and $N^\mu$ is tangential,
,$g_{\mu\nu}n^\mu X^\nu_{,a}=0$.
The coefficients of proportionality $N$ and $N^\mu$ respectively are called
{\it lapse} function and {\it shift} vector field respectively. Notice that
implicitly information about the metric $g_{\mu\nu}$ has been invoked
into (\ref{2.3}), namely we are only dealing with spacelike embeddings and
metrics of the above specified signature. The lapse is nowhere vanishing
since for a foliation $T$ must be timelike everywhere. Moreover, we take
it to be {\it positive} everywhere as we want a future directed foliation
(negative sign would give a past directed one and mixed sign would not
give a foliation at all since then necessarily the leaves of the foliation
would intersect).\\
We need one more property of $n$ :
By the inverse function theorem, the surface $\Sigma_t$ can be
defined by an equation of the form $f(X)=t=const.$. Thus,
$0=\lim_{\epsilon\to 0} [f(X_t(x+\epsilon b)-f(X_t(x))]/\epsilon=
b^a X^\mu_{,a} (f_{,\mu})_{X=X_t(x)}$ for any tangential vector $b$ of
$\sigma$ in $x$. It follows that up to normalization the normal
vector is proportional to an exact one-form, $n_\mu=F f_{,\mu}$
or, in the language of forms, $n=n_\mu dX^\mu= F df$. Actually,
this fact is an easy corollary from Frobenius' theorem (the surfaces
$\Sigma_t$ are the integral manifolds of the distribution
$v:\;M\mapsto T(M);\;X\mapsto V_X(n)=\{v\in T_X(M);
i_v(n)=0\}\subset T_X(M)$).

Let us forget about the foliation for a moment and just suppose that
we are given a hypersurface $\sigma$ embedded into $M$ via
the embedding $X$. Let $n$ be its unit normal vector field and
$\Sigma=X(\sigma)$ its image.
We now have the choice to work either on $\sigma$ or on $\Sigma$
when developing the tensor calculus of so-called spatial tensor fields.
To work
on $\Sigma$ has the advantage that we can compare spatial tensor
fields with arbitrary tensor tensor fields {\it restricted to $\Sigma$}
because they are both tensor fields on a subset of $M$.
Moreover, once we have developed tensor calculus on $\Sigma$
we immediately have the one on $\sigma$ by just pulling back
(covariant) tensor fields on $\Sigma$ to $\sigma$ via the embedding,
see below.

Consider then the following tensor fields, called the
{\it first and second fundamental form} of $\Sigma$
\be \label{2.4}
q_{\mu\nu}:=g_{\mu\nu}-s n_\mu n_\nu \mbox{ and }
K_{\mu\nu}:=q_\mu^\rho q_\nu^\sigma \nabla_\rho n_\sigma
\ee
where all indices are moved with respect to $g_{\mu\nu}$. Notice
that both tensors in \ref{2.4}, are ``spatial", i.e.
they vanish when either of their indices is contracted  with $n^\mu$.
A crucial property of $K_{\mu\nu}$ is its symmetry : We have
$K_{[\mu\nu]}=q_\mu^\rho q_\nu^\sigma( (\nabla_{[\rho} \ln(F) )
n_{\sigma]}+ F \nabla_{[\mu}\nabla{\nu]}f)=0$ since $\nabla$ is
torsion free. The square brackets denote antisymmetrization defined
as an idempotent operation. From this fact one derives another
useful differential geometric identity by employing the relation between the
covariant differential and the Lie derivative :
\ba \label{2.5}
2 K_{\mu\nu} &=& q_\mu^\rho q_\nu^\sigma (2\nabla_{(\rho} n_{\sigma)})
\nonumber\\
&=& q_\mu^\rho q_\nu^\sigma ({\cal L}_n g)_{\rho\sigma}=
q_\mu^\rho q_\nu^\sigma ({\cal L}_n q+s{\cal L}_n n\otimes n)_{\rho\sigma}
\nonumber\\
&=& q_\mu^\rho q_\nu^\sigma ({\cal L}_n q)_{\rho\sigma}
=({\cal L}_n q)_{\mu\nu}
\ea
since $n^\mu {\cal L}_n q_{\mu\nu}=-q_{\mu\nu} [n,n]^\mu=0$. Using
$n^\mu=(T^\mu-N^\mu)/N$ we can write (\ref{2.5}) in the form
\be \label{2.6}
2 K_{\mu\nu}
=\frac{1}{N}({\cal L}_{T-N} q)_{\mu\nu}-2n^\rho q_{\rho(\mu} \ln(N)_{,\nu)}
=\frac{1}{N}({\cal L}_{T-N} q)_{\mu\nu}
\ee

Next we would like to construct a covariant differential associated with
the metric $q_{\mu\nu}$. We would like to stress that this metric
is non-degenerate as a bijection between {\it spatial tensors only} and
not as a metric between arbitrary tensors defined on $\Sigma$.
Recall that, by definition, a differential $\nabla$ is said to be
covariant with
respect to a metric $g$ (of any signature) on a manifold $M$
if it is 1) metric compatible, $\nabla g=0$ and 2) torsion free,
$[\nabla_\mu,\nabla_\nu]f=0\;\forall C^\infty(M)$. According to a
classical theorem, these two conditions fix $\nabla$ uniquely in terms
of the Christoffel connection which in turn is defined
by its action on one-forms through
$\nabla_\mu u_\nu:=\partial_\mu u_\nu-\Gamma^\rho_{\mu\nu} u_\rho$.
Since the tensor $q$ is a metric of Euclidean signature on $\Sigma$
we can thus apply these two conditions to $q$ and we are looking for a
covariant differential $D$ on spatial tensors only such that
1) $D_\mu q_{\nu\rho}=0$ and 2) $D_{[\mu} D_{\nu]} f=0$ for scalars
$f$. Of course, the operator $D$ should preserve the
set of spatial tensor fields. It is easy to verify that $D_\mu f
:=q_\mu^\nu \nabla_\nu \tilde{f}$ and
$D_\mu  u_\nu:=q_\mu^\rho q_\nu^\sigma
\nabla_\rho \tilde{u}_\sigma$ for $u_\mu n^\nu=0$ and extended to arbitrary
tensors by linearity and Leibniz' rule, does the job and thus, by the above
mentioned theorem, is the unique choice. Here, $\tilde{f}$ and $\tilde{u}$
denote arbitrary smooth extensions of $f$ and $u$ respectively into a
neighbourhood of $\Sigma$ in $M$, necessary in order to perform the
$\nabla$ operation. The covariant differential is independent of that extension
as derivatives not tangential to $\Sigma$ are projected out by the
$q$ tensor (go into a local, adapted system of coordinates to see this)
and we will drop the tilde again. One can convince
oneself that the action of $D$ on arbitrary spatial tensors is then given by
acting with $\nabla$ in the usual way followed by spatial projection of all
appearing indices including the one with respect to which the derivative
was taken.

We now ask what the Riemann curvature $R^{(D)}_{\mu\nu\rho}\;^\sigma$
of $D$ is in terms of that of $\nabla$. To answer this question we
need the second covariant differential of a spatial co-vector
$u_\rho$ which when carefully using the definition of $D$ is given by
\ba \label{2.7}
D_\mu D_\nu u_\rho
&=& q_\mu^{\mu'} q_\nu^{\nu'} q_\rho^{\rho'}
\nabla_{\mu'} D_{\nu'} u_{\rho'}\nonumber\\
&=& q_\mu^{\mu'} q_\nu^{\nu'} q_\rho^{\rho'}
\nabla_{\mu'} q_{\nu'}^{\nu^{\prime\prime}} q_{\rho'}^{\rho^{\prime\prime}}
\nabla_{\nu^{\prime\prime}} u_{\rho^{\prime\prime}}
\ea
The outer derivative hits either a $q$ tensor or $\nabla u$
the latter of which will give rise to a curvature term.
Consider then the $\nabla q$ terms.\\
Since $\nabla$
is $g$ compatible we have $\nabla q=s\nabla n\otimes n=
s[(\nabla n)\otimes n+n\otimes (\nabla n)]$´. Since all of these
terms are contracted with $q$ tensors and $q$ annihilates $n$,
the only terms that survive are proportional to terms of either
the form
$$
(\nabla_{\mu'} n_{\nu'}) (n^{\rho^{\prime\prime}}
(\nabla_{\nu^{\prime\prime}} u_{\rho^{\prime\prime}})
=-(\nabla_{\mu'} n_{\nu'}) (\nabla_{\nu^{\prime\prime}}
n^{\rho^{\prime\prime}}) u_{\rho^{\prime\prime}}
$$
where $n^\mu u_\mu=0\Rightarrow \nabla_\nu (n^\mu u_\mu)=0$
was exploited, or of the form
$(\nabla_{\mu'} n_{\nu'}) (\nabla_n u_{\rho'})$.
Concluding, the only terms that survive from
$\nabla q$ terms can be transformed terms into proportional to
$\nabla n\otimes \nabla n$ or $\nabla n\otimes \nabla_n u$
where the $\nabla n$ factors, since contracted with $q$ tensors,
can be traded for extrinsic curvature terms
(use $u_\mu=q_\mu^\nu u_\nu$ to do that).\\
It turns out that the terms proportional to $\nabla_n u$ cancel each other
when computing the antisymmetrized second $D$ derivative of $u$
due to the symmetry of $K$ and we are thus left with the famous
{\it Gauss equation}
\ba \label{2.8}
R^{(D)}_{\mu\nu\rho}\;^\sigma u_\sigma: &=& 2D_{[\mu} D_{\nu]} u_\rho
\nonumber\\
&=& [2s K_{\rho[\mu} K_{\nu]}^\sigma+
q_\mu^{\mu'} q_\nu^{\nu'} q_\rho^{\rho'} q_{\sigma'}^\sigma
R^{(D+1)}_{\mu'\nu'\rho'}\;^{\sigma'}] u_\sigma
\nonumber\\
R^{(D)}_{\mu\nu\rho\sigma}
&=& 2s K_{\rho[\mu} K_{\nu]\sigma}+
q_\mu^{\mu'} q_\nu^{\nu'} q_\rho^{\rho'} q_\sigma^{\sigma'}
R^{(D+1)}_{\mu'\nu'\rho'\sigma'}
\ea
Using this general formula we can specialize to the Riemann curvature
scalar which is our ultimate concern in view of the Einstein-Hilbert action.
Employing the standard abbreviations $K:=K_{\mu\nu}  q^{\mu\nu}$ and
$K^{\mu\nu}=q^{mu\rho} q^{\nu\sigma} K_{\nu\sigma}$ (notice that
indices for spatial tensors can be moved either with $q$ or with $g$)
we obtain
\ba \label{2.9}
R^{(D)} &=& R^{(D)}_{\mu\nu\rho\sigma} q^{\mu\rho} q^{\nu\sigma}
\nonumber\\
&=&
s [K^2-K_{\mu\nu} K^{\mu\nu}]+
q^{\mu\rho} q^{\nu\sigma} R^{(D+1)}_{\mu\nu\rho\sigma}
\ea
Equation (\ref{2.9}) is not yet quite what we want since it is not yet purely
expressed in terms of  $R^{(D+1)}$ alone. However, we can eliminate
the second term in (\ref{2.9}) by using $g=q+s n\otimes n$ and the definition
of curvature
$R^{(D+1)}_{\mu\nu\rho\sigma} n^\sigma=
2\nabla{[\mu} \nabla_{\nu]} n_\rho$ as follows
\ba \label{2.11}
R^{(D+1)} &=& R^{(D+1)}_{\mu\nu\rho\sigma} g^{\mu\rho} g^{\nu\sigma}
\nonumber\\
&=&
q^{\mu\rho} q^{\nu\sigma} R^{(D+1)}_{\mu\nu\rho\sigma}
+2s q^{\rho\mu} n^\nu[\nabla_\mu,\nabla_\nu] n_\rho
\nonumber \\&=&
q^{\mu\rho} q^{\nu\sigma} R^{(D+1)}_{\mu\nu\rho\sigma}
+2s n^\nu[\nabla_\mu,\nabla_\nu] n_\nu
\ea
where in the first step we used the antisymmetry of the Riemann tensor
to eliminate the term quartic in $n$ and in the second step we used again
$q=g-s n\otimes n$ and the antisymmetry in the $\mu\nu$ indices.
Now
$$
n^\nu ([\nabla_\mu, \nabla_\nu] n^\mu)=
 -(\nabla_\mu n^\nu)( \nabla_\nu n^\mu)=
 +(\nabla_\mu n^\mu)( \nabla_\nu n^\nu)
+\nabla_\mu (n^\nu \nabla_\nu n^\mu-n^\mu \nabla_\nu n^\nu)
 $$
and using $\nabla_\mu s=2n^\nu \nabla_\mu n_\nu=0$ we have
\ba \label{2.12}
\nabla_\mu n^\mu &=& g^{\mu\nu} \nabla_\nu n^\mu =
q^{\mu\nu} \nabla_\nu n^\mu =K \\
&& (\nabla_\mu n^\nu)(\nabla_\nu n^\mu)
=g^{\nu\sigma} g^{\rho\mu} (\nabla_\mu n_\sigma)(\nabla_\nu n_\rho)
=q^{\nu\sigma} q^{\rho\mu} (\nabla_\mu n_\sigma)(\nabla_\nu n_\rho)
=K_{\mu\nu} K^{\mu\nu} \nonumber
\ea
Combining (\ref{2.9}), (\ref{2.11}) and (\ref{2.12}) we obtain the
{\it Codacci equation}
\be \label{2.13}
R^{(D+1)}=R^{(D)}-s[K_{\mu\nu} K^{\mu\nu}-K^2]
+2s\nabla_\mu (n^\nu \nabla_\nu n^\mu-n^\mu \nabla_\nu n^\nu)
\ee
Inserting this differential geometric identity back into the action,
the third term in (\ref{2.13}) is a total differential which we drop
for the time being as one can rederive it later on when making the
variational principle well-defined.

At this point it is useful to pull back various quantities to $\sigma$.
Consider the $D$ spatial vextor fields on $\Sigma_t$ defined by
\be \label{2.14}
X^\mu_a(X):=X^\mu_{,a}(x,t)_{|X(x,t)=X}
\ee
Then we have due to $n_\mu X^\mu_a=0$ that
\be \label{2.15}
q_{ab}(t,x):=(X^\mu_{,a} X^\nu_{,b} q_{\mu\nu})(X(x,t))
=g_{\mu\nu}(X(t,x))X^\mu_{,a}(t,x) X^\nu_{,b}(t,x)
\ee
and
\be \label{2.16}
K_{ab}(t,x):=(X^\mu_{,a} X^\nu_{,b} K_{\mu\nu})(X(x,t))
=(X^\mu_{,a} X^\nu_{,b}\nabla_\mu n_\nu)(t,x)
\ee
Using $q_{ab}$ and its inverse
$q^{ab}=\epsilon^{a a_1 .. a_{D-1}} \epsilon^{b b_1 .. b_{D-1}}
q_{a_1 b_1}..q_{a_{D-1} b_{D-1}}/[\det((q_{cd}))\;(D-1)!]$ we can
express $q_{\mu\nu}, q^{\mu\nu}, q_\mu^\nu$ as
\ba \label{2.17}
q^{\mu\nu}(X)&=&[q^{ab}(x,t) X^\mu_{,a} X^\mu_{,b}](x,t)_{|X(x,t)=X}
\nonumber\\
q_\mu^\nu(X) &=& g_{\mu\rho}(X) q^{\rho\nu}(X)
\nonumber\\
q_{\mu\nu}(X) &=& g_{\nu\rho}(X) q_\mu^\rho(X)
\ea
To verify that this coincides with our previous defintion $q=g-s n\otimes n$
it is sufficient to check the matrix elements in the basis given by the
vector fields $n, X_a$. Since for both definitions $n$ is annihilated we just
need to verify that (\ref{2.17}) when contracted with $X_a \otimes X_b$
reproduces (\ref{2.15}) which is indeed the case.

Next we define $N(x,t):=N(X(x,t)),
\vec{N}^a(x,t):=q^{ab}(x,t) (X^\mu_b g_{\mu\nu} N^\nu)(X(x,t))$. Then it is
easy to verify that
\be \label{2.18}
K_{ab}(x,t)=\frac{1}{2N}(\dot{q}_{ab}-({\cal L}_{\vec{N}} q)_{ab})(x,t)
\ee
We can now pull back the expressions quadratic in $K_{\mu\nu}$
that appear in (\ref{2.13}) using (\ref{2.17}) and find
\ba \label{2.19}
K(x,t) &=& (q^{\mu\nu} K_{\mu\nu})(X(x,t))=(q^{ab} K_{ab})(x,t) \nonumber\\
(K_{\mu\nu} K^{\mu\nu})(x,t) &=& (K_{\mu\nu} K_{\rho\sigma}
q^{\mu\rho} q^{\nu\sigma})(X(x,t))=(K_{ab} K_{cd} q^{ac} q^{bd})(x,t)
\ea
Likewise we can pull back the curvature scalar $R^{(D)}$. We have
\be \label{2.20}
R^{(D)}(x,t)=(R^{(D)}_{\mu\nu\rho\sigma} q^{\mu\rho}q^{\nu\sigma})(X(x,t))
(R^{(D)}_{\mu\nu\rho\sigma}  X^\mu_a X^\nu_b X^\rho_c X^\sigma_d)(X(x,t))
q^{ac}(x,t) q^{bd}(x,t)
\ee
We would like to show that this expression equals the curvature scalar
$R$ as defined in terms of the Christoffel connection for $q_{ab}$. To see this
it is sufficient to compute $(X^\mu_a D_\mu f)(X(x,t))=\partial_a f(X(x,t))=:
(D_a f)(x,t)$ with $f(x,t):=F(X(x,t))$ and with
$u_a(x,t):=(X^\mu_a u_\mu)(X(x,t)),\;u^a(x,t)=q^{ab}(x,t) u_b(x,t)$
\ba \label{2.21}
(D_a u_b)(x,t) &:=& (X^\mu_a X^\nu_b D_\mu u_\nu)(X(x,t))
\nonumber\\
&=& X^\mu_{,a}(x,t) X^\nu_{,b}(x,t)(\nabla_\mu u_nu)(X(x,t))
\nonumber\\
&=&
(\partial_a u_b)(x,t)-X^\mu_{,ab} u_\mu(X(x,t))
\nonumber\\
&& -u^c(x,t)
\Gamma^{(D+1)}_{\rho\mu\nu}(X(x,t))
X^\rho_{,c}(x,t) (X^\mu_{,a}(x,t) X^\nu_{,b}(x,t)
\nonumber\\ &=&
(\partial_a u_b)(x,t)-\Gamma^{(D)}_{cab}(x,t) u^c(x,t)
\ea
where in the last step we have used the explicit expressions of
the Christoffel connections $\Gamma^{(D+1)}$ and $\Gamma^{(D)}$ in terms of
$g_{\mu\nu}$ and $q_{ab}$ respectively. Now since every tensor field $W$
is a linear combination of tensor products of one forms and since $D_\mu$
satisfies the Leibniz rule we easily find
$(X^\mu_a X^\nu_b .. D_\mu W_{\nu..})(X(x,t))=:(D_a W_{b..})(x,t)$ where
now $D_a$ denotes the uniqe torsion-free covariant differential associated with
$q_{ab}$ and $W_{a..}$ is the pull-back of $W_{\mu..}$. In particular, we have
$X^\mu_a X^\nu_b X^\rho_c D_\mu D_\nu u_\rho=
D_a X^\mu_b X^\nu_c D_\mu u_\rho=D_a D_b u_c$ from which our assertion
follows since
\ba \label{2.22}
(R_{abcd} u_d)(x,t):=([D_a,D_b]u_c)(x,t) &=&
(X^\mu_a X^\nu_b X^\rho_c [D_\mu,D_\nu]u_\rho)(X(x,t))
\\
&=&(X^\mu_a X^\nu_b X^\rho_c X^\sigma_d R^{(D)}_{\mu\nu\rho\sigma})(X(x,t))
u^d(x,t) \nonumber
\ea
From now on we will move indices with the metric $q_{ab}$ only.

One now expresses the line element in the new system of coordinates $x,t$
using the quantities $q_{ab},N,N^a$ (we refrain from displaying the
arguments of the components of the metric)
\ba \label{2.23}
ds^2&=&g_{\mu\nu}dX^\mu\otimes dX^\nu \\
&=& g_{\mu\nu}(X(t,x))
[X^\mu_{,t}dt+X^\mu_{,a} dx^a]\otimes
[X^\nu_{,t}dt+X^\nu_{,b} dx^b]\nonumber\\
&=& g_{\mu\nu}(X(t,x))
[N n^\mu dt+X^\mu_{,a} (dx^a+N^a dt)]\otimes
[N n^\nu dt+X^\nu_{,b} (dx^b+N^b dt)]\nonumber\\
&=& [s N^2+q_{ab}N^a N^b]dt\otimes dt
+q_{ab}N^b [dt\otimes dx^a+dx^a\otimes dt]
+q_{ab} dx^a\otimes dx^b \nonumber
\ea
and reads off the components $g_{tt}, g_{ta}, g_{ab}$ of $X^\ast g$
in this frame.
Since the volume form $\Omega(X):=\sqrt{|\det(g)|}d^{D+1}X$ is covariant,
i.e.,
$(X^\ast \Omega)(x,t)=\sqrt{|\det(X\ast g)|}dt d^Dx$ we just
need to
compute $\det(X^\ast g)=s N^2 \det(q_{ab})$ in order to finally cast the action
(\ref{2.1}) into $D+1$ form. The result is (dropping the total differential
in (\ref{2.13}))
\be \label{2.24}
S=\frac{1}{\kappa}\int_\Rl dt\int_\sigma d^Dx
\sqrt{\det(q)}|N| (R-s[K_{ab} K^{ab}-(K_a^a)^2])
\ee
We could drop the absolute sign for $N$ in (\ref{2.24}) since we took $N$
positive but we will keep it for the moment to see what happens if we
allow arbitrary sign.
Notice that (\ref{2.24}) {\it vanishes identically} for $D=1$, indeed
in two spacetime dimensions the Einstein action is proportional to a
topological
charge, the so-called Euler characteristic of $M$ and in what
follows we concentrate on $D>1$.

We now wish to cast this action into canonical form, that is, we would like
to perform the Legendre transform from the Lagrangean density appearing
in (\ref{2.24}) to the corresponding Hamiltonian density.
The action (\ref{2.24}) depends on the velocities $\dot{q}_{ab}$
of $q_{ab}$ but not on those of $N$ and $N^a$. Therefore we obtain for
the conjugate momenta (use (\ref{2.18}) and the fact that $R$ does not
contain time derivatives)
\ba \label{2.25}
\frac{1}{\kappa}P^{ab}(t,x)&:=&\frac{\delta S}{\delta \dot{q}_{ab}(t,x)}=
-s\frac{|N|}{N\kappa}\sqrt{\det(q)}[K^{ab}-q^{ab}(K_c^c)]
\nonumber\\
\Pi(t,x)&:=& \frac{\delta S}{\delta \dot{N}(t,x)}=0
\nonumber\\
\Pi_a(t,x)&:=& \frac{\delta S}{\delta \dot{N}^a(t,x)}=0
\ea
The Lagrangean in (\ref{2.24}) is therefore a {\it singular} Lagrangean,
one cannot solve all velocities for momenta \cite{18}. We can solve
$\dot{q}_{ab}$ in terms of $q_{ab},N,N^a$ and $P^{ab}$ using (\ref{2.18})
but this is not possible for $\dot{N},\dot{N}^a$, rather we have the
so-called {\it primary constraints}
\be \label{2.26}
C(t,x):=\Pi(t,x)=0 \mbox{ and } C^a(t,x):=\Pi^a(t,x)=0
\ee
The Hamiltonian treatment of systems with constraints has been developed
by Dirac \cite{3} to which a short introduction is given in
section \ref{sd}.
According to that theory, we are supposed to introduce Lagrange multiplier
fields $\lambda(t,x),\lambda_a(t,x)$ for the primary constraints and
to perform the Legendre transform as usual with respect to the remaining
velocities which can be solved for. We have
\ba \label{2.27}
\dot{q}_{ab} &=& 2N K_{ab}+({\cal L}_{\vec{N}} q)_{ab}\nonumber\\
\dot{q}_{ab} P^{ab} &=&
({\cal L}_{\vec{N}} q)_{ab} P^{ab}
-2s|N|\sqrt{\det(q)}[K_{ab} K^{ab}-K^2]
\nonumber\\
P_{ab}P^{ab} &=& \det(q) (K_{ab} K^{ab}+(D-2)K^2)\nonumber\\
P^2 &:=& (P_a^a)^2=(1-D)^2\det(q) K^2
\ea
and by means of these formulae we obtain the canonical form of the
action (\ref{2.24})
\ba \label{2.28}
\kappa S &=& \int_\Rl dt\int_\sigma d^Dx \{\dot{q}_{ab} P^{ab}+\dot{N} \Pi
+\dot{N}^a \Pi_a
\\
&& -[\dot{q}_{ab}(P,q,N,\vec{N}) P^{ab}+\lambda C+\lambda^a C_a
-\sqrt{\det(q)}|N| (R-s[K_{ab} K^{ab}-K^2])(P,q,N,\vec{N})] \}
\nonumber\\
&=& \int_\Rl dt\int_\sigma d^Dx \{\dot{q}_{ab} P^{ab}+\dot{N} \Pi
+\dot{N}^a \Pi_a
\nonumber\\
&& -[({\cal L}_{\vec{N}} q)_{ab} P^{ab}+\lambda C+\lambda^a C_a
-\sqrt{\det(q)}|N|(R+s[K_{ab} K^{ab}-K^2])(P,q,N,\vec{N})] \}
\nonumber\\
&=& \int_\Rl dt\int_\sigma d^Dx \{\dot{q}_{ab}
P^{ab}+\dot{N}\Pi+\dot{N}^a \Pi_a \nonumber\\
&& -[({\cal L}_{\vec{N}} q)_{ab} P^{ab}+\lambda C+\lambda^a C_a
+|N|(-\frac{s}{\sqrt{\det(q)}}[P_{ab} P^{ab}-\frac{1}{D-1}P^2]
-\sqrt{\det(q)}R)] \} \nonumber
\ea
Upon performing a spatial integration by parts (whose boundary term
vanishes since $\partial\sigma=\emptyset$) one can cast it into the following
more compact form
\be \label{2.29}
S=\frac{1}{\kappa}
\int_\Rl dt\int_\sigma d^Dx \{\dot{q}_{ab} P^{ab}+\dot{N} \Pi+\dot{N}^a \Pi_a
-[\lambda C+\lambda^a C_a+N^a H_a+|N| H]\}
\ee
where
\ba \label{2.30}
H_a &:=& -2 q_{ac}D_b P^{bc} \nonumber\\
H& :=& -(\frac{s}{\sqrt{\det(q)}}
[q_{ac}q_{bd}-\frac{1}{D-1}q_{ab} q_{cd}]P^{ab}P^{cd}
+\sqrt{\det(q)}R)
\ea
are called the {\it (spatial) Diffeomorphism constraint} and {\it
Hamiltonian
constraint} respectively, for reasons which we will derive below.

The geometrical meaning of these quantities is as follows :\\
At fixed $t$ the fields $(q_{ab}(t,x),N^a(t,x),N(t,x);
P^{ab}(x,t),\Pi_a(t,x),\Pi(t,x))$ label
points (configuration;canonically conjugate momenta) in an infinite
dimensional phase space $\cal M$ (or symplectic manifold). Strictly
speaking, we should now specify on what Banach space this manifold is
modelled \cite{17}, however, we will be brief here as we are primarily
not interested in the metric formulation of this section but rather in
the connection formulation of the next section where we will give more
details. For the purpose of this subsection it is sufficient to say that
we can choose the model space to be the direct product of the space
$T_2(\sigma)\times T_1(\sigma)\times T_0(\sigma)$
of smooth symmetric covariant tensor fileds of rank $2,1,0$ on $\sigma$
respectively and the space
$\tilde{T}^2(\sigma)\times\tilde{T}^1(\sigma)\times\tilde{T}^0(\sigma)$ of
smooth symmetric contravariant tensor density fields of weight one and
of rank $2,1,0$ on $\sigma$ respectively, equipped with some Sobolev norm.
(The precise
functional analytic description is somewhat more complicated in case that
$\sigma$ is unbounded with boundary but can also be treated). In particular,
one shows that the action (\ref{2.29}) is differentiable in this topology.

The phase space carries the strong \cite{17}
symplectic structure $\Omega$ or Poisson bracket
\be \label{2.31}
\{P(f^2),F_2(q)\}=\kappa F_2(f^2),\;
\{\vec{\Pi}(\vec{f}^1),\vec{F}_1(\vec{N})\}=
\kappa \vec{F}_1(\vec{f}^1),
\;\{\Pi(f),F(N)\}=\kappa F(f)
\ee
(all other brackets vanishing)
where we have defined the following pairing, invariant under
diffeomorphisms of $\sigma$, e.g.
\be \label{2.32}
\tilde{T}^2(\sigma)\times T_2(\sigma)\to \Rl;\;
(F_2,f^2)\to F^2(f_2):=\int_\sigma d^Dx F_2^{ab}(x) f^2_{ab}(x)
\ee
and similar for the other fields.
Physicists use the following short-hand notation for (\ref{2.31})
\be \label{2.33}
\{P^{ab}(t,x),q_{cd}(t,x')\}=\kappa\delta^a_{(c}\delta^b_{d)} \delta^{(D)}(x,y)
\;.
\ee
In the language of symplectic geometry, the first term in the action
(\ref{2.29}) is a symplectic potential for the symplectic structure
(\ref{2.31}). We now turn to the meaning of the term in the square bracket
in (\ref{2.29}), that is, the ``Hamiltonian"
\be \label{2.34}
\kappa\mbox{{\bf H}}:=
\int_\sigma d^Dx [\lambda C+\lambda^a C_a+N^a H_a+|N| H]
=:\vec{C}(\vec{\lambda})+C(\lambda)+\vec{H}(\vec{N})+H(|N|)
\ee
of the action and the associated equations of motion.

The variation of the action with respect to the Lagrange multiplier fields
$\vec{\lambda}, \lambda$ reproduces the primary constraints (\ref{2.26}).
If the dynamics of the system is to be consistent, then these constraints must
be preserved under the evolution of the system, that is, we should have e.g.
$\dot{C}(t,x):=\{\mbox{{\bf H}},C(t,x)\}=0$ for all $x\in\sigma$, or
equivalently,
$\dot{C}(f):=\{\mbox{{\bf H}},C(f)\}=0$ for all ($t$-independent)
smearing fields
$f\in T_0(\sigma)$. However, we do not get zero but rather
\be \label{2.35}
\{\vec{C}(\vec{f}),\mbox{{\bf H}}\}=\vec{H}(\vec{f})
\mbox{ and } \{C(f),\mbox{{\bf H}}\}=H((\frac{N}{|N|} f)
\ee
which is supposed to vanish for all $f,\vec{f}$. Thus, consistency of the
equations of motion ask us to impose the
{\it secondary constraints}
\ba \label{2.36}
H(x,t)=0 \mbox{ and } H_a(x,t)=0
\ea
for all $x\in\sigma$.
Since these two functions appear next to the $C,C_a$ in (\ref{2.34}), in
general
relativity the ``Hamiltonian" is constrained to vanish ! General
relativity is an
example of a so-called constrained Hamiltonian system with no true
Hamiltonian. The reason for this will become evident in a moment.

Now one might worry that imposing consistency of the secondary constraints
under evolution results in tertiary constraints etc., but fortunately,
this is not the case.
Consider the smeared quantities $H(f),\vec{H}(\vec{f})$ where, e.g.,
$\vec{H}(\vec{N}):=\int_\sigma d^3x N^a V_a$ (notice that indeed $H,\Pi$ and
$H_a,\Pi_a$ are, respectively, scalar and
co-vector densities of weight one on $\sigma$).
Then we obtain
\ba \label{2.37}
&&
\{\mbox{{\bf H}},\vec{H}(\vec{f})\}=\vec{H}({\cal L}_{\vec{N}} \vec{f})-
H({\cal L}_{\vec{f}} |N|)
\nonumber\\
&& \{\mbox{{\bf H}},H(f)\}=H({\cal L}_{\vec{N}} f)
+\vec{H}(\vec{N}(|N|,f,q))
\ea
where $\vec{N}(f,f',q)^a=q^{ab}(f f'_{,b}-f' f_{,b})$. Equations (\ref{2.37})
are equivalent to the {\it Dirac algebra} \cite{3}
\ba \label{2.38}
\{\vec{H}(\vec{f}),\vec{H}(\vec{f}')\}&=& \kappa
\vec{H}({\cal L}_{\vec{f}}\vec{f}') \nonumber\\
\{\vec{H}(\vec{f}),H(f))\}&=&\kappa H({\cal L}_{\vec{f}}f)\nonumber\\
\{H(f),H(f'))\}&=&\kappa \vec{H}(\vec{N}(f,f',q))
\ea
also called the {\it hypersurface deformation algebra}. The meaning of
(\ref{2.35},\ref{2.38}) is that the constraint surface $\overline{{\cal M}}$ of
$\cal M$, the submanifold of $\cal M$ where the constraints hold,
is preserved under the motions generated by the constraints.
In the terminology of Dirac \cite{3}, all constraints are of first
class (determine coisotropic constraint
submanifolds \cite{18} of $\cal M$) rather than of second class (determine
symplectic constraint submanifolds \cite{18} of $\cal M$).

It remains to study the equations of motion of the canonical coordinates on
the phase space themselves. Since $C=\Pi,C_a=\Pi_a$ it remains to study
those of $N,N^a,q_{ab},P^{ab}$. For shift and lapse we obtain
$\dot{N}^a=\lambda^a,\dot{N}=\lambda$. Since $\lambda^a,\lambda$ are
{\it arbitrary, unspecified functions} we see that also the trajectory of
lapse and shift is completely arbitrary.
Moreover, the equations of motion of $q_{ab}, P^{ab}$ are
completely unaffected by the term $\vec{C}(\vec{\lambda})+C(\lambda)$
in {\bf H}. It is therefore completely straightforward to solve the equations of
motion as far as $N,N^a,\Pi,\Pi_a$ are concerned : Simply treat
$N,N^a$ as  {\it Lagrange multipliers} and drop all terms proportional to
$C, C_a$ from the action (\ref{2.29}). The result is the reduced action
\be \label{2.39}
S=\frac{1}{\kappa}
\int_\Rl dt\int_\sigma d^Dx \{\dot{q}_{ab} P^{ab}-[N^a H_a+|N| H]\}
\ee
called the {\it Arnowitt -- Deser -- Misner action} \cite{16}. It is
straightforward
to check that as far as $q_{ab},P^{ab}$ are concerned, the actions
(\ref{2.29}) and (\ref{2.39}) are completely equivalent.

The equations of motion of $q_{ab},P^{ab}$ then finally allow us
to interpret the motions that the constraints generate on
$\cal M$ geometrically. Since the reduced Hamiltonian (using the same
symbol as before)
\be \label{2.40}
\mbox{{\bf H}}=\frac{1}{\kappa} \int_\sigma d^Dx [N^a H_a+|N| H]
\ee
is a linear combination of constraints we obtain the equations of motion
once we know the Hamiltonian flow of the functions $\vec{H}(\vec{f}), H(f)$
for any $\vec{f},f$ separately. Denoting, for any function $J$ on $\cal M$,
\be \label{2.41}
\delta_{\vec{f}} J:=\{\vec{H}(\vec{f}),J\} \mbox{ and }
\delta_f J:=\{H(f),J\}
\ee
it is easiest to begin with the corresponding equations for $J=F_2(q)$
since upon integration by parts we have
$\vec{H}(\vec{f})=\int d^Dx P^{ab} ({\cal L}_{\vec{f}} q)_{ab}$ so that
both constraint
functions are simple polynomials in $P^{ab}$ not involving their derivatives.
We then readily find
\ba \label{3.42}
\delta_{\vec{f}} F_2(q) &=& \kappa F_2({\cal L}_{\vec{f}} q)
\nonumber\\
\delta_f  F_2(q) &=& -2s\kappa  \int_\sigma d^Dx
\frac{P_{ab}-P q_{ab}/(D-1)}{\sqrt{\det(q)}}
\ea
Using the relations (\ref{2.25}), (\ref{2.18}) the second identity in
(\ref{3.42}) can be written as
$$
\delta_{|N|} q_{ab}=2 N \kappa K_{ab}=
\kappa(\dot{q}_{ab}-({\cal L}_{\vec{N}} q)_{ab})
$$
In order to interprete this quantity, notice that the components of $n_\mu$
in the frame $t,x^a$ are given by $n_t=n_\mu X^\mu_{,t}=sN,\;n_a=
n_\mu X^\mu_{,a}=0$. In order to compute the contravariant components
$n^\mu$ in that frame we need the corresponding contravariant
metric components. From (\ref{2.23}) we find the covariant components
to be $g_{tt}=s N^2+q_{ab} N^a N^b,g_{ta}=q_{ab} N^b,g_{ab}=q_{ab}$ so
that the inverse metric has components $g^{tt}=s/N^2, g^{ta}=-sN^a/N^2,
g^{ab}=q^{ab}+sN^a N^b/N^2$. Thus $n^t=1/N, n^a=-N^a/N$ and since
$q_{at}=q_{tt}=0$ we finally obtain
\be \label{2.43}
\delta_{|N|} F_2(q)=\kappa F_2({\cal L}_{Nn} q)
\ee
which of course we guessed immediately from the $D+1$ dimensional identiy
(\ref{2.6}). Concluding, as far as $q_{ab}$ is concerned, $H_a$ generates
{\it on all of $\cal M$}
diffeomorphisms of $M$ that preserve $\Sigma_t$ while $H$ generates
diffeomorphisms of $M$ orthogonal to $\Sigma_t$.

The corresponding computation for $P(f^2)$ is harder by an order of magnitude
due to the curvature term involved in $H$ and due to the fact that the
identity corresponding to ({\ref{2.43}) holds {\it only on shell}, that is, when
the (vacuum) Einstein equations $G^{(D+1)}_{\mu\nu}:=
R^{(D+1)}_{\mu\nu}-\frac{g_{\mu\nu}}{2} R^{(D+1)}=0$ hold.
The variation with respect to
$\vec{H}(\vec{f})=-\int_\sigma d^Dx q_{ab} ({\cal L}_{\vec{f}}  P)^{ab}$
(notice that $P^{ab}$ carries density weight one to verify this identity)
is still easy and yields the expected result
\be \label{2.44}
\delta_{\vec{f}} P(f^2) = \kappa ({\cal L}_{\vec{f}} P)(f^2)
\ee
We will now describe the essential steps for the analog of (\ref{2.43}).
The ambitious reader who wants to fill in the missing steps should expect
to perform at least one Din A4 page of calculation in between each of the
subsequent formulae. \\
We start from formula (\ref{2.30}). Then
\ba \label{2.44a}
\{H(|N|),P^{ab}\}&=&\frac{\delta H(|N|)}{\delta q_{ab}}
\nonumber\\
&=&\frac{s|N|}{\sqrt{\det(q)}}[2(P^{ac} P^b_c-P^{ab} P/(D-1))
-\frac{q^{ab}}{2}(P^{cd} P_{cd}-P^2/(D-1))]
\nonumber\\
&& +\frac{\delta}{\delta q_{ab}}\int d^Dx |N| \sqrt{\det(q)} R
\ea
where the second term comes from the $\sqrt{\det(q)}^{-1}$ factor and
we used the well-known formula $\delta \det(q)=\det(q) q^{ab} \delta q_{ab}$.
To perform the remaining variation in (\ref{2.44a}) we write
$$
\delta \sqrt{\det(q)} R=[\delta\sqrt{\det(q)}] R
+\sqrt{\det(q)}[\delta q^{ab}] R_{ab}
+\sqrt{\det(q)} q^{ab}[\delta R_{ab}]
$$
use $\delta\delta^a_b=\delta[q^{ac} q_{cb}]=0$ in the second variation
and can simplify (\ref{2.44a})
\ba \label{2.45}
\{H(|N|),P^{ab}\}&=&
\frac{2s|N|}{\sqrt{\det(q)}}[2(P^{ac} P^b_c-P^{ab} P/(D-1)]
+\frac{q^{ab}|N| H}{2}+|N| \sqrt{\det(q)} (q^{ab} R- R^{ab})
\nonumber\\
&& +\int d^Dx |N| \sqrt{\det(q)} q^{cd} \frac{\delta}{\delta q_{ab}} R_{cd}
\ea
The final variation is the most difficult one since $R_{cd}$ contains second
derivatives of $q_{ab}$. Using the explicit expression of $R_{abcd}$ in terms
of the Christoffel connection $\Gamma^c_{ab}$ and observing that, while
the connection itself is not a tensor, its variation in fact {\it is} a tensor,
we find after careful use of the definition of  the covariant derivative
\be \label{2.46}
q^{cd} \delta R_{cd}=q^{cd} [-D_c\delta\Gamma^e_{ed}+D_e \delta
\Gamma^e_{cd}]
\ee
We now use the explicit expression of $\Gamma^a_{bc}$ in terms of
$q_{ab}$ and find
\be \label{2.47}
\delta \Gamma^a_{bc}=\frac{q^{ad}}{2}[D_c\delta q_{bd}
+D_b\delta q_{cd}-D_d\delta q_{bc}]
\ee
Next we insert (\ref{2.46}) and (\ref{2.47}) into the integral appearing in
(\ref{2.45}) and integrate by parts two times using the fact that for
the divergence of a vector $v^a$ we have
$\sqrt{\det(q)} D_a v^a=D_a (\sqrt{\det(q)} v^a)=\partial_a (\sqrt{\det(q)} v^a)$
(no boundary terms due to $\partial\sigma=\emptyset$) and find
\ba \label{2.48}
&& \int d^Dx |N| \sqrt{\det(q)} q^{cd} \delta R_{cd}
=\int d^Dx \sqrt{\det(q)}
q^{cd} [(D_c |N|) \delta\Gamma^e_{ed}-(D_e |N|) \delta \Gamma^e_{cd} ]
\nonumber\\
&=&\int d^Dx \sqrt{\det(q)} q^{cd} q^{ef}
[(D_c |N|) (D_d\delta q_{ef})-(D_e |N|) (D_c\delta q_{df})]
\nonumber\\
&=& \int d^Dx \sqrt{\det(q)}
[-(D_c D^c|N|) q^{ab} + (D^a D^b |N|)]\delta q_{ab}
\ea
Collecting all contributions we obtain the desired result
\ba \label{2.49}
&&\{H(|N|),P^{ab}\}=
\frac{2s|N|}{\sqrt{\det(q)}}[2(P^{ac} P^b_c-P^{ab} P/(D-1)]
+\frac{q^{ab}|N| H}{2}
\nonumber\\
&& + |N| \sqrt{\det(q)} (q^{ab} R- R^{ab})
+\sqrt{\det(q)}[-(D_c D^c|N|) q^{ab}-(D^a D^b |N|)]
\ea
which does not look at all as ${\cal L}_{Nn} P^{ab}$ !

In order to compute ${\cal L}_{Nn} P^{ab}$ we need
an identity for ${\cal L}_{Nn} K_{\mu\nu}=N{\cal L}_n K_{\mu\nu}$
which we now derive. Using the definition of the Lie derivative in terms
of the covariant derivative $\nabla_\mu$ and using $g=q+s n\otimes n$
one finds first of all
\be \label{2.50}
{\cal L}_n K_{\mu\nu}=-K K_{\mu\nu} +2K_{\rho\mu} K^\rho_\nu
+[\nabla_\rho(n^\rho K_{\mu\nu})+2s K_{\rho(\mu} n_{\nu)} \nabla_n n^\rho]
\ee
Using the Gauss equation (\ref{2.8}) we find for the Ricci tensor
$R^{(D)}_{\mu\nu}$ the following equation (use again $g=q+s n\otimes n$
and the defintion of curvature as $R=[\nabla,\nabla]$)
\be \label{2.51}
R^{(D+1)}_{\rho\sigma}q^\rho_\mu q^\sigma_\nu-R^{(D)}_{\mu\nu}
=s[-K_{\mu\nu} K +K_{\mu\rho} K_\nu^\rho
+q^\rho_\mu q^\sigma_\nu n^\lambda [\nabla_\rho,\nabla_\lambda] n_\sigma]
\ee
We claim that the term in square brackets on the right hand side of
(\ref{2.50}) equals $(-s)$ times the sum of the
left hand side of (\ref{2.51}) and the term
$-s (D_\mu D_\nu N)/N$. In order to prove this we manipulate the commutator
of covariant derivatives appearing in (\ref{2.51}) making use of the definition
of the extrinsic curvature. One finds
\ba \label{2.52}
&&
q^\rho_\mu q^\sigma_\nu n^\lambda [\nabla_\rho,\nabla_\lambda] n_\sigma]
\nonumber\\
&=&
q^\rho_\mu q^\sigma_\nu n^\lambda (\nabla_\rho \nabla_\lambda n_\sigma)
+K K_{\mu\nu}-\nabla_\rho(n^\rho K_{\mu\nu})
\nonumber\\
&& -s(\nabla_n n^\rho) n_\nu K_{\mu\rho} -s(\nabla_n(n_\mu n^\rho))
(\nabla_\rho n_\nu)
\ea
Using this identity we find for the sum of
the term in square brackets on the right hand side of
(\ref{2.50}) and  $s$ times the sum of the
right hand side of (\ref{2.51}) the expression (dropping the obvious
cancellations) \ba \label{2.53}
&&
K_{\mu\rho} K_\nu^\rho
+q^\rho_\mu q^\sigma_\nu n^\lambda (\nabla_\rho \nabla_\lambda n_\sigma)
+s[K_{\rho\nu} n_\mu (\nabla_n n^\rho)-(\nabla_n (n_\mu n^\rho))
(\nabla_\rho n_\nu)]
\nonumber\\
&=&
K_{\mu\rho} K_\nu^\rho
+q^\rho_\mu q^\sigma_\nu n^\lambda (\nabla_\rho \nabla_\lambda n_\sigma)
+s[n_\mu (\nabla_n n^\rho)\{q_\rho^\sigma-\delta_\rho^\sigma\}
(\nabla_\sigma n_\nu)
-(\nabla_n n\mu) (\nabla_n n_\nu)]
\nonumber\\
&=&
K_{\mu\rho} K_\nu^\rho
+q^\rho_\mu q^\sigma_\nu (\nabla_\rho \nabla_n n_\sigma)
-q^\rho_\mu q^\sigma_\nu  (\nabla_\rho n^\lambda)( \nabla_\lambda n_\sigma)
-s (\nabla_n n\mu) (\nabla_n n_\nu)
\nonumber\\
&=&
+q^\rho_\mu q^\sigma_\nu (\nabla_\rho \nabla_n n_\sigma)
-s (\nabla_n n\mu) (\nabla_n n_\nu)
\ea
where in the second step it has been used that the curly bracket vanishes
since it is proportional
to $n_\rho$ and contracted with the spatial  vector $\nabla_n n^\rho$,
in the third step we moved $\nabla^\lambda$ inside a covariant derivative and
picked up a correction term and in the fourth step one realizes that
this correction term is just the negative of the first term using that
$K_{\mu\nu}=q_\mu^\rho\nabla_\rho n_\nu$. Our claim is equivalent to showing
that the last line of (\ref{2.53}) is indeed given by $-s (D_\mu D_\nu N)/N$.\\
To see this notice that if the surface $\Sigma_t$ is defined by
$t(X)=t=const.$ then $1=T^\mu\nabla_\mu t$. Since $\nabla_\mu t$ is orthogonal
to $\Sigma_t$ we have $n_\mu=s N \nabla_\mu t$ as one verifies by contracting
with $T^\mu$ and thus $N=1/(\nabla_n t)$. Thus
\ba \label{2.54}
D_\mu N &=&- N^2 D_\mu (\nabla_n t)=-N^2 q_\mu^\nu n^\rho
(\nabla_\rho\nabla_\nu t)
\nonumber\\
&=& -s N (\nabla_n n_\nu )
= -s N  \nabla_n n_\mu
\ea
where in the first step we interchanged the second derivative due to torsion
freeness and could pull $n^\rho$ out of the second derivative
because the correction term is proportional to $n_\rho \nabla n^\rho=0$
and in the second we have pulled in a factor of $N$, observed that the
correction
is annihilated by the projection, used once more $s N \nabla t=n$
and finally used that $\nabla_n n_\nu$ is already spatial. The second
derivative then gives simply
\ba \label{2.55}
D_\mu D_\nu N &=&
= -s (D_\mu N)  \nabla_n n_\nu
-s N  q_\mu^\rho q_\nu^\sigma \nabla_\rho \nabla_n n_\sigma
\nonumber\\
&=&  N (\nabla_n n_\mu ) (\nabla_n n_\nu)
-s N  q_\mu^\rho q_\nu^\sigma \nabla_\rho \nabla_n n_\sigma
\ea
which is indeed $N$ times (\ref{2.53}) as claimed. Notice that in
(\ref{2.55}) we cannot replace $N$ by $|N|$ if $N$ is not everywhere
positive so {\it the interpretation that we are driving at would not hold
if we would not set $N=|N|$ everywhere}. It is at this point that
we must take $N$ positive in all that follows. \\
We have thus established
the key result
\ba \label{2.56}
{\cal L}_{Nn} K_{\mu\nu} &=& N(-K K_{\mu\nu} +2K_{\rho\mu} K^\rho_\nu)
-s[D_\mu D_\nu N
\nonumber\\
&&
+N(R^{(D+1)}_{\rho\sigma}q^\rho_\mu q^\sigma_\nu-R^{(D)}_{\mu\nu})]
\ea

In order to finish the calculation for ${\cal L}_{Nn} P^{\mu\nu}$
we need to know ${\cal L}_{Nn} \sqrt{\det(q)},
{\cal L}_{Nn} q^{\mu\nu}$. So far we have defined $\det(q)$ in the
ADM frame only, its generalization to an arbirtrary frame is given by
\be \label{2.57}
\det((q_{\mu\nu})(X)):=\frac{1}{D !}
[(\nabla_{\mu_0}t)(X)\epsilon^{\mu_0..\mu_D}]
[(\nabla_{\nu_0}t)(X)\epsilon^{\nu_0..\nu_D}]
q_{\mu_1\nu_1}(X)..q_{\mu_D\nu_D}(X)
\ee
as one can check by specializing to the ADM coordinates $X^\mu=t,x^a$.
Here $\epsilon^{\mu_0..\mu_D}$ is the metric independent, totally skew
Levi-Civita tensor density of weight one.
One can verify that with this definition we have $\det(g)=s N^2\det(q)$
by simply expanding $g=q+s n\otimes n$. It is important to see that
${\cal L}_T \nabla_\mu t={\cal L}_{N} \nabla_\mu t=0$ from which then
follows immediately that
\be \label{2.58}
{\cal L}_{Nn} \sqrt{\det(q)}
=\frac{1}{2}\sqrt{\det(q)} q^{\mu\nu} {\cal L}_{Nn} q_{\mu\nu}
=N\sqrt{\det(q)} K
\ee
where (\ref{2.6}) has been used. Finally, using once more (\ref{2.54})
we find indeed
\be \label{2.59}
{\cal L}_{Nn} q^{\mu\nu}=-q^{mu\rho} q^{\nu\sigma}
{\cal L}_{Nn} q_{\rho\sigma}=-2N K^{\mu\nu}
\ee
We are now in position to compute the Lie derivative of
$P^{\mu\nu}=-s\sqrt{\det(q)}[q^{mu\rho} q^{\nu\sigma}
-q^{mu\nu} q^{\rho\sigma}]K_{\rho\sigma}$. Putting all six contributions
carefully together and comparing with (\ref{2.49}) one finds the
non-trivial result
\ba \label{2.60}
\{H(N),P^{\mu\nu}\}&=&\frac{q^{\mu\nu} N H}{2}-N\sqrt{\det(q)}
[q^{mu\rho} q^{\nu\sigma}-q^{mu\nu} q^{\rho\sigma}]R^{(D+1)}_{\rho\sigma}
\nonumber\\
&& +{\cal L}_{Nn} P^{\mu\nu}
\ea
that is, only on the constraint surface and only when the (vacuum)
equations of motion
hold, can the Hamiltonian flow of $P^{\mu\nu}$ with respect to $H(N)$ be
interpreted as the action of a diffeomorphism in the direction
perpendicular to $\Sigma_t$. Now, using again the definition of curvature
as the commutator of covariant derivatives it is not difficult to check
that
\ba \label{2.61}
G_{\mu\nu} n^\mu n^\nu &=& \frac{s H}{2\sqrt{\det(q)}}
\nonumber\\
G_{\mu\nu} n^\mu q^\nu_\rho &=& -\frac{s H_\rho}{2\sqrt{\det(q)}}
\ea
so that the constraint equations actually are equivalent to $D+1$ of the
Einstein equations. Since (\ref{2.60}) contains besides $H$ all the spatial
projections of $G_{\mu\nu}$ we see that our interpretation of
$\{H(N),P^{\mu\nu}\}$ holds only on shell, $G_{\mu\nu}=0$.\\
This finishes our geometric analysis of the Hamiltonian flow of the
constraints which shows that the symmetry group of spacetime
diffeomorphisms Diff$(M)$ of Einstein's action is faithfully implemented in
the canonical framework, although in a not very manifest way
(more precisely, it is only the subset
of those symmetries \cite{15} generated by the Lie algebra of that
symmetry group). The importance of this result cannot be stressed
enough: It is often said that every $(D+1)-$ diffeomorphism invariant
quantity should be a Dirac observable since Diff$(M)$ is the
symmetry of the Einstein-Hilbert action. But this would mean that any
higher derivative theory (containing arbitrary scalars built from
polynomials of the curvature tensor) would also have the same Dirac
observables, meaning that to be an observable would be theory
independent. The catch is that $(D+1)$dimensional diffeomorphism
invariance is not only a kinematical statement but involves the
theory dependent dynamics. The fact that the motions generated
by the constraints can be interpreted as spacetime diffeomorphisms
only on (the theory dependent) shell spells this out in the precise way.

What do these considerations tell us ? The Hamiltonian of general relativity
is not a true Hamiltonian but a linear combination of constraints. Rather
than generating time translations it generates spacetime diffeomorphisms.
Since the parameters of these diffeomorphisms, $N,N^a$ are completely
arbitrary unspecified functions, the corresponding motions on the phase space
have to be interpreted as {\it gauge transformations}. This is
quite similar to the gauge motions generated by the Gauss constraint
in Maxwell theory \cite{3}. The basic
variables of the theory, $q_{ab},P^{ab}$ are not observables of the theory
because they are not gauge invariant. Let us count the number of
kinematical and dynamical (true) degrees of freedom : The basic variables are
both symmetric tensors of rank two and thus have $D(D+1)/2$ independent
components per spatial point. There are $D+1$ independent constraints
so that $D+1$ of these phase space variables can be eliminated. $D+1$
of the remaining degrees of freedom can be gauged away by a gauge
transformation leaving us with
$D(D+1)-2(D+1)=(D-2)(D+1)$ phase space degrees of freedom
or $(D-2)(D+1)/2$ configuration space degrees of freedom per spatial
point. For $D=3$ we thus recover the two graviton degrees of freedom.

The further classical analysis of this system could now proceed as
follows :\\
1) One determines a complete set of gauge invariant observables on the
constraint surface $\overline{{\cal M}}$ and computes the induced symplectic
structure
$\overline{\Omega}$ on the so reduced symplectic manifold $\hat{{\cal M}}$.
Equivalently, one obtains the full set of solutions to the equations of
motion, the set of Cauchy data are then the searched for observables.
This programme of ``symplectic reduction" could never be completed due to
the complicated appearance of the Hamiltonian constraint.
In fact, until today one does not know any observable for full general
relativity (with exception of the generators of the Poincar\'e group at
spatial infinity in the case that $(\sigma,q_{ab})$ is asymptotically
flat \cite{19}).\\
2) One fixes a gauge and solves the constraints. Years of research in
the field of solving the Cauchy problem for general relativity reveal that
such a procedure works at most locally, that is, there do not exist, in
general, global gauge conditions. This is reminiscent of the Gribov
problem in non-Abelian Yang-Mills theories.

In summary, general relativity can be cast into Hamiltonian form, however,
its equations of motion are complicated non-linear partial differential
equations of second order and very difficult to solve. Nevertheless,
the Cauchy problem is well-posed and the classical theory is consistent
up to the point where singularities (e.g. black holes) appear \cite{14}. This
is one instance where it is expected that the classical theory is unable to
describe the system appropriately any longer and that the more exact
theory of quantum gravity must take over in order to remove the
singularity. This is expected to be quite in analogy to the case of the
hydrogenium atom whose
stability was a miracle to classical electrodynamics but was easily
explained by
quantum physics. Of course, the quantum theory of gravity is expected to
be even harder to handle mathematically than the classical theory,
however, as a zeroth step an existence proof would already be a triumph.
Notice that up to date a similar existence proof for, say, QCD is lacking
as well \cite{8}.

\subsection{The Programme of Canonical Quantization}
\label{s2.2}

In this section we briefly summarize which steps the method of canonical
quantization consists of. We will not go too much into details, the
interested reader is referred to \cite{20,II,22,47f1}. We then consider the
application of this programme to canonical general relativity in the ADM
formulation and point
out the immediate problems that one is confronted with and which prevented
one to make progress in this field for so long.

\subsubsection{The General Programme}
\label{s2.2.1}

Let be given an (infinite dimensional) constrained symplectic manifold
$({\cal M},\Omega)$ modelled on a Banach space $E$ with strong symplectic
structure $\Omega$ and first class constraint functionals $C_I(N^I)$
(in case of second class constraints one should replace $\Omega$ by the
corresponding Dirac bracket \cite{3}; there could also be an additional
true Hamiltonian which is not constrained to vanish but which is supposed
to be gauge invariant). Here $I$ takes
values in some finite index set and $C_I(N^I)$ is an appropriate pairing
as in the previous section between the constraint density $C_I(x)$,
$x$ a point in the $D$-dimensional manifold of the Hamiltonian
framework, and its
corresponding Lagrange multiplier $N^I$. Unless otherwise specified
no summation over repeated indices $I$ is assumed.\\
The quantization algorithm for this system consists of the following.
\begin{itemize}
\item[I)] {\it Polarization}\\
The phase space can be coordinatized in many ways by what are called
``elementary variables", that is, global coordinates such that
all functions on $\cal M$ can be expressed in terms of them.
One set of elementary variables
may be more convenient than another in the sense that the equations of
motion or the constraint functions $C_I$ look more or less
complicated in terms of them.

Also, one has to split the set of elementary
variables into ``configuration $q_a(x)$ and momentum variables $P^a(x)$"
(here $x$ is a coordinate of the $D$-dimensional time slice and $a$
takes values in a finite set). This means that,
roughly, wave functions should only depend on half of the number of
elementary variables. In the theory of geometric quantization this
``splitting" is called a polarization of the symplectic manifold
\cite{18}. Among the possible choices of elementary variables those are
preferred that come in canonically conjugate pairs $(P^a(x),q_a(x))$,
that is, global
Darboux coordinates in terms of which the symplectic structure looks
as simple as possible. This is important as the quantization of the
elementary variables requires that their commutator algebra mirrors their
Poisson algebra, see below. In general, the set of elementary variables
should
form a subalgebra of the Poisson algebra on $\cal M$ and should be closed
under complex conjugation.

Further complications may arise in case that the phase space does not
admit an independent set of global coordinates. In this case it may be
necessary to work with an overcomplete set of variables and to impose
their relations among each other as conditions on states on the Hilbert
space. Example :\\
Suppose we want to coordinatize the cotangent bundle over
the sphere $S^2$. The sphere cannot be covered by a single coordinate
patch, but we can introduce Cartesian coordinates on $\Rl^3$ and impose
the condition $(\hat{x}^1)^2+(\hat{x}^2)^2+(\hat{x}^3)^2-1=0$ on states
depending on $\Rl^3$.

Finally, in the infinite-dimensional context, it is important how to smear
the elementary variables : for instance, the relation
$\{P(f),P(g)\}=0$ where $P(f)=\int d^Dx f_a(x)P^a(x)$ and $f_a$ is a
smooth co-vector field leads one to conclude the distributional
relation $\{P^a(x),P^b(y)\}=0$. However, this is meaningless without
specifying the space $\cal S$ to which the smearing fields $f_a$ belong.
For instance, we
could also write $\{P^a(x),P^b(y)\}=J^{ab}(x)\delta(x,y)$ where
$J^{ab}(x)$ is a nonsingular anti-symmetric tensor field supported in a
set of $d^Dx$ measure zero without affecting $\{P(f),P(g)\}=0$.
If, on the other hand, one would use distributional $f^a$ then
$\{P(f),P(g)\}$ may not vanish. Of course, this
last point is not independent of the choice of the model space $E$
mentioned above since some model spaces allow distributional smearing
fields while others do not.
\item[II)] {\it Quantum Configuration Space}\\
As experience shows, while the (restriction to the configuration space
$\cal C$ of the) phase space $\cal M$ is typically some space of
smooth fields, complete in a suitable norm, the states of the quantum
theory will depend on a more general, distributional quantum configuration
space $\overline{\cal C}$. For instance, the canonical quantization
of a free, massive, real scalar field \cite{8} comes with a Hilbert space
which is an $L_2$ space with respect to a Gaussian measure for the
corresponding covariance. As is well known, Gaussian measures are
supported on a space $\overline{{\cal C}}$ of tempered distributions
on $\Rl^n$ and the space $\cal C$ is contained in a measurable set $\cal
N$ which has measure zero.

Thus, one has to decide what the quantum configuration space
$\overline{{\cal C}}$ should be. In a sense, this is determined to a large
extent by the space $\cal S$ of the smearing fields $F^a$ : if $q(F)$ is
supposed to be a meaningful random variable for a measure (the
quantum field
$\overline{{\cal C}}\ni\; q\; : \; {\cal S}\to {\cal R}({\cal B});\;F\to
F(q)$ is a map from the smearing fields into the random variables of some
measureable space $\cal B$ and is called a
generalized stochastic process in the language of
probability theory and constructive quantum field theory \cite{8}) then
the object $q$ typically lies in the topological dual
${\cal S}'$ of $\cal S$ in view of the Bochner-Minlos theory \cite{8,25}.
\item[III)] {\it Kinematical Measures}\\
One now has to equip $\overline{{\cal C}}$ with the structure
of a Hilbert space $\cal H$. This will be naturally an $L_2$ space for a
suitable measure $\mu_0$ on $\overline{{\cal C}}$. Certainly, $\cal H$
is not yet the physical Hilbert space as one still has to impose the
constraints, however, one has to start from such a ``kinematical"
Hilbert space $\cal H$ in order to quantize the constraints (the name
``kinematical" stems from the fact that it does not know about the
constraints yet which capture the {\it dynamics} of the system).
The minimal requirements on a measure $\mu_0$ are as follows :\\
A)\\
It should not only be a cylindrical measure but must be
$\sigma-$additive, in other words, one must be able to integrate functions
of an infinite number of degrees of freedom.\\
B)\\
The Hilbert space ${\cal H}:=L_2(\overline{{\cal C}},d\mu_0)$ must be
an irreducible representation of the canonical commutation relations.
More precisely, if $F(\hat{q})$ and $\hat{P}(f)$ are the representations
of $F(q)$ and $P(f)$ as linear operators on $\cal H$ with common dense
domain $\cal D$ which they leave invariant, then we must get, for instance,
$[\hat{P}(f),F(\hat{q})]=i\hbar\{P(f),F(q)\}^\wedge$. Notice that this
condition is well-defined, first because by assumption the Poisson
bracket can be expressed in terms of elementary variables again and,
secondly the commutator makes sense because by assumption the elementary
operators have domain and range $\cal D$. Typically, $F(\hat{q})$ will
act by multiplication, $(F(\hat{q})\Psi)[q']=F(q')\Psi[q']$ where the
prime is to indicate that $q'$ is distributional rather than smooth
while $\hat{P}(f)$ will act as some kind of derivative operator.\\
Irreducibility of this representation means that the basic operators have
a dense range when acting on a cyclic vector. In a reducible representation
one tends to have too many degrees of freedom since every
irreducible subspace already represents a quantization of the
corresponding classical system.\\
\\
Finally, in case that we have to work with an overcomplete system of
elementary variables we must require that the quantizations of their
relations among each other are identically satisfied on $\cal H$.\\
C)\\
The Hilbert space must implement the classical complex conjugation
relations among the elementary variables as adjointness relations on the
corresponding operators. More precisely, by assumption the Poisson
subalgebra of the elementary variables is closed under complex
conjugation, therefore, e.g., a relation of the kind
$\overline{F(q)}=F'_F(q)+P(f'_F)$ will hold for certain smearing fields
$F'_F,f'_F$ depending on $F$. We then require that
$F(\hat{q})^\dagger=F'_F(\hat{q})+\hat{P}(f'_F)$ holds. Here the dagger
denotes the adjoint with respect to $\mu_0$.
Notice that this condition means also that
the domains of the elementary operators and their adjoints coincide.

In summary, we have a representation of the classical $^*$ subalgebra of
the Poisson algebra, given by the elementary variables, on the Hilbert
space.

One can slightly relax these requirements as follows :\\
Given the classical configuration manifold $\cal C$ modelled on a Banach
space we are naturally equipped also with
the space of smooth functions ${\cal F}({\cal C})$ and of smooth vector fields
${\cal V}({\cal C})$ on it. Let us consider elements $(a,b)$ of the product
space ${\cal F}\times{\cal V}$ and let us equip it with a Lie algebra structure
given by $[(a,b),(a',b'))]:=(b[a']-b'[a],[b,b'])$ where
${\cal V}\times {\cal F}\mapsto {\cal F};\;(b,a)\mapsto
b[a]$ denotes the natural action of vector fields on functions and
${\cal V}\times {\cal V}\mapsto {\cal V};\;(b,b')\mapsto
[b,b']$ denotes the natural action of vector fields on themselves by
the Lie bracket, that is, $([b,b'])[a]:=b[b'[a]]-b'[b[a]]$. It is easy to see
that the set of elementary kinematical variables $F(q), P(f)$ can be
identified with points in the set ${\cal F}\times {\cal V}$ by
$F(q)\mapsto (Q_F,0),\;P(f)\mapsto (0,P_f)$ where $Q_F(q):=F(q),\;
(P_f[a])(q):=\{P(f),a\}(q)$. The advantage is now that while the $F(q),P(f)$
may not (be known to) form a closed subalgebra of the Poisson algebra
on $\cal M$,
the set ${\cal F}\times {\cal V}$ always forms a closed Lie algebra. In other
words, it may happen that $\{P(f),P(f')\}$ cannot be (obviously)
written as a function
of the $P(f),F(q)$ again but it is always true that $[P_f,P_{f'}]$ is an element
of $\cal V$ again. It is easy to see that $\{P(f),P(f')\}\mapsto
[P_f,P_{f'}]$ if and only if the Jacobi identity $\{\{P(f),P(f'\},a\}+
\mbox{cyclic}=0$
holds for all $a$ which, of course, requires the knowledge of $\{P(f),P(f')\}$.
Now, the quantization map is given by $(Q_F,P_f)\mapsto
(F(\hat{q}),\hat{P}(f))$
and the requrement on the measure is that this be a Lie algebra homorphism.
We will take this more general approach also in our case.
\item[IV)] {\it Constraint Operators}\\
By assumption we can write the classical constraint functions $C_I(N^I)$
as certain functions $C_I(N^I)=c_I(N^I,\{F(q)\},\{P(f)\})$ of
the elementary variables where the curly brackets denote dependence on an in
general
infinite collection of variables. A naive quantization procedure would
be to define its quantization as
$\hat{C}_I(N^I)=c_I(N^I,\{F(\hat{q})\},\{\hat{P}(f)\})$. This will in general
not work, at least not straightforwardly, for several reasons :\\
A)\\
As is well-known, the quantization of a phase space function is not
unique, to a given candidate we can add arbitrary $\hbar$ corrections
and still the classical limit of the corrected operator will be the
original function. This is called the {\it factor ordering ambiguity}.\\
B)\\
While such corrections in quantum mechanics are relatively harmless,
in quantum field theory they tend to be desasterous, a simple example
is quantum Maxwell theory where the straightforward quantization of the
Hamiltonian gives a divergent nowhere defined operator. It is only
after factor ordering that one obtains a densely defined operator.
This is what is called a {\it factor ordering singularity}. \\
C)\\
More seriously, in general the singularities of an operator are of an
even worse kind and cannot be simply removed by a judicious choice of
factor ordering. One has to introduce a regularization of the operator
and subtract its divergent piece as one removes the regulator again.
This is called the {\it renormalization} of the operator. The end result
must be a densely defined operator on $\cal H$.\\
D)\\
If $C_I(N^I)$ is classically a real-valued function then one would
like to implement $C_I(N^I)$ as a self-adjoint operator on $\cal H$,
the reason being that this would guarantee that its spectrum (and therefore
its measurement values) is contained in the set of real numbers. While
this is
certainly a necessary requirement if $C_I(N^I)$ was a true Hamiltonian
(i.e. not a constraint), in the case of a constraint this condition can be
relaxed as long as the value $0$ is contained in its spectrum because this
is what we are interested in. On the other hand, a self-adjoint constraint
operator is sometimes of advantage when it comes to actually solving the
constraints \cite{22,47f1}.
\item[V)] {\it Imposing the Constraints}\\
We would now like to solve the constraints in the quantum theory. A first
guess of how to do that is by saying that a state $\psi\in {\cal H}$ is
physical provided that $\hat{C}_I(N^I)\psi=0$. The study of the simple
example of a particle moving in $\Rl^2$ with the constraint $C=p_2$
reveals that this does not work in general : in the momentum
representation ${\cal H}=L_2(\overline{{\cal C}}:=\Rl^2,d\mu_0:=d^2p)$ the
physical state condition becomes $p_2\psi(p_1,p_2)=0$ with the general
solution $\psi_f(p_1,p_2)=\delta(p_2)f(p_1)$ for some function $f$. The
problem is that
$\psi_f$ is not an element of $\cal H$. This is a frequent problem of an
operator with continuous spectrum : such an operator does not in general
have eigenfunctions in the ordinary sense. However, it has so-called
``generalized eigenfunctions" of which $\psi_f$ is an example
\cite{26}.\\
The way to solve the constraint is as follows (see \cite{26} for
details and section \ref{si}) : One takes a convex topological vector space
$\cal D$ which is
dense in ${\cal H}$ in the topology of $\cal H$ and which serves as
a common domain for constraint operators and elementary operators.
It is then true that $\cal H$ is contained in the space ${\cal D}^\ast$ of
all linear functionals on $\cal D$ (i.e. ${\cal D}^\ast$ is
the algebraic dual of $\cal D$). We thus have
${\cal D}\subset {\cal H}\subset {\cal D}^\ast$ (in case that the
topology of $\cal D$
is nuclear and we take the topological dual
${\cal D}'$ instead of the algebraic dual,
this triple of spaces is
called a Gel'fand triple). We now say that an element $\Psi$ of ${\cal
D}^\ast$ is a solution of the constraints iff
\be \label{2.62}
\Psi(\hat{C}_I(N^I)\psi)=0
\ee
for all $I, N^I\in{\cal S},\psi \in {\cal D}$. \\
In the example above, we could take for ${\cal D}$ the space of
functions of rapid decrease on $\Rl^2$ and then ${\cal D}'$ as the space of
tempered distributions on $\Rl^2$.\\
The set of solutions in ${\cal D}^\ast$ is called ${\cal D}^\ast_{phys}$.
Notice
that ${\cal D}^\ast_{phys}$ does not carry a natural Hilbert space structure
yet. In the example it is of course natural to take
$<\Psi_f,\Psi_g>_{phys}:=\int dp_1 \bar{f} g$.
\item[VI)] {\it Quantum Anomalies}\\
Even if we finally managed to produce a densely defined, possibly
self-adjoint operator, with a non-trivial kernel in the above sense we
might encounter a quantum anomaly of the
following kind :\\
Recall that by assumption the constraint algebra is first class. This
means that there exist so-called structure maps $f_{IJ}\;^K$ from
${\cal S}^2$ into the functions on $\cal M$ such that
$\{C_I(N^I),C_J(N^J)\}=\sum_K C_K(f_{IJ}\;^K(N^I,N^J))$. The quantum
version of this condition is
\be \label{2.63}
[\hat{C}_I(N^I),\hat{C}_J(N^J)]\psi=
\sum_K [C_K(f_{IJ}\;^K(N^I,N^J))]^\wedge\psi
\ee
for all $\psi\in {\cal D}$.
There are two potential problems with (\ref{2.63}) :\\
First of all, it does not make sense to take a commutator unless the range
of the first operator is contained in the domain of the second. Therefore,
we must require that all operators $\hat{C}_I(N^I)$ leave ${\cal D}$
invariant.\\
Secondly, notice
that especially since, as it is the case in general relativity (see
\ref{2.38}),
$f_{IJ}\;^K$ depends in general on the phase space coordinates, we are
not guaranteed that the right hand side of (\ref{2.63}) can actually be
written in the form
$\sum_K \hat{C}_K(\hat{f}_{IJ}\;^K(N^I,N^J))\psi$ with the $\hat{C}_K$
{\it ordered to the left}. If that is not the case then the following
inconsistency arises : Let $\Psi\in {\cal D}^\ast_{phys}$ and let us evaluate
$\Psi$ on (\ref{2.63}). Then we find that
\be \label{2.64}
0=\sum_K \Psi([C_K(f_{IJ}\;^K(N^I,N^J))]^\wedge\psi)
\ee
for all $\psi\in{\cal D},I,J,N^I,N^J$. Thus, not only does every member
of ${\cal D}^\ast_{phys}$ satisfy the constraints (\ref{2.62}) but also the
additional constraints (\ref{2.64}) which are absent in the classical
theory. Since (\ref{2.64}) will in general be new constraints, algebraically
independent from the original ones, the number of physical degrees of
freedom in the classical and the quantum theory would differ from each
other.\\
In summary, we must make sure that (\ref{2.64}) is automatically satisfied
once (\ref{2.62}) holds which puts additional restrictions on the
freedom to order the constraint operators if at all possible.
\item[VII)] {\it Physical Scalar Product}\\
Suppose that we managed to produce densely defined, anomaly-free
constraint operators and the space of solutions ${\cal D}^\ast_{phys}$. How
can we
arrive at a Hilbert space ${\cal H}_{phys}$ with respect to which which
the solutions are square integrable ? In the most general case not much
is known about a rigorous solution to this problem but an idea is
provided by the following formal ansatz in fortunate cases : \\
Suppose that the constraint operators are all self-adjoint and mutually
commuting, then we can
formally define the functional $\delta$-distribution
\be \label{2.65}
\delta[\hat{C}]:=\lim_{\epsilon\to
0} \prod_{I,\alpha\in A_\epsilon}\delta(\hat{C}_I(\chi_\alpha))
\ee
through the spectral theorem. \\
The $\chi_\alpha$ are the characteristic functions of
mutually non-overlapping regions $B_\alpha$ in $\sigma$ of coordinate
volume $\epsilon^D$ and
$\cup_{\alpha\in A_\epsilon} B_\alpha=\sigma$. Given an element
$f\in{\cal D}$ we define an element of ${\cal D}^\ast_{phys}$ and a physical
inner product between such elements by
\be \label{2.66}
\Psi_f:=\delta[\hat{C}]\cdot f\mbox{ and }
<\Psi_f,\Psi_g>_{phys}:=<\delta[\hat{C}]\cdot f,g>
\ee
where $<.,.>$ is the inner product of $\cal H$. The fact that the
constraints are Abelian reveals that the inner product
(\ref{2.66}) is formally Hermitean, non-negative and sesquilinear. The
completion of ${\cal D}^\ast_{phys}$ (possibly after factoring by a subspace
of null vectors) with respect to $<.,.>_{phys}$ defines the physical
Hilbert space ${\cal H}_{phys}$.\\
For a
successful application of these
ideas in the finite dimensional context see \cite{22,47f1} and in the
infinite-dimensional and even non-Abelian context see \cite{II,27}.
\item[VIII)] {\it Observables}\\
By definition, a quantum observable is a self-adjoint operator
on ${\cal H}_{phys}$. It is actually not difficult to construct such
abstractly defined observables once ${\cal H}_{phys}$ is known, however,
the real problem is to find observables which are quantizations of
classical observables, the latter being gauge invariant functions on the
constraint surface of the phase space. Obviously, this is a hard problem
if not even the classical observables are known as it is the case with
pure general relativity. The situation improves if one couples matter
\cite{28}.\\
In any case, the physics of the system lies in studying the spectra of
the observables. This will in general be a hard problem as well and
approximation methods have to be used. General relativity poses a further
problem : since there is no true Hamiltonian, the physical Hilbert space
is a space ``without dynamics". This is the problem of time. There are
literally hundreds of publications on this issue without clear
conclusions and nothing will be said about it in this article and the
author will not even try to give references. However,
the author agrees with Rovelli (see \cite{29} and references therein) that
the evolution of one physical
quantity in a theory without background metric and thus no background
time can only be studied relative to another one. Therefore, it is possible
to assign to one of the degrees of freedom, say $\hat{O}_1$, the role of a
clock variable
and one may ask the question : ``What is the expectation value of
$\hat{O}_2$ in the state $\Psi$ when $\hat{O}_1$
has the expectation value $t$ in the state $\Psi$ ?
\end{itemize}

This is the outline of the general programme. We will now estimate
how far one can get with this programme in application to canonical
general relativity as displayed in the previous subsection.

\subsubsection{Application to General Relativity in the ADM Formulation}
\label{s2.2.2}

Let us go through the steps of the programme one by one and see what the
immediate problems are :
\begin{itemize}
\item[I)] {\it Polarization}\\
Let us assume, as it is usually done throughout the literature, that we
choose as elementary variables
the ones of the previous section, $F(q):=\int_\sigma d^3x q_{ab} F^{ab},
P(f):=\int_\sigma d^3x P^{ab} f_{ab}$. All fields are smooth and
are symmetric tensors with the appropriate density weight. We choose
the polarization that the configuration variables be the $F(q)$.
\item[II)] {\it Quantum Configuration Space}\\
Experience from scalar quantum field theory motivates to have $q_{ab}$
take values in the space of tempered distributions on $\sigma$.
\item[III)] {\it Kinematical Measures}\\
Notice that ${\overline{\cal C}}$ is an infinite dimensional noncompact
space. Therefore \cite{25}
we cannot take $\mu_0$ to be an infinite product Lebesgue measure but must
take some sort of probability measure, for instance a Gaussian measure
with ``white noise" covariance
$C_{ab,cd}(x,y)=\frac{1}{\sqrt{\det(q^0)(x)}}q^0_{ac}(x)
q^0_{bd}(x) \delta(x,y)$ where $q^0$ is any fixed positive definite
background metric on $\sigma$. In other words, the characteristic
functional \cite{8} of this measure is given by
\be \label{21}
\chi(f)=\int_{\overline{{\cal C}}} d\mu(q) e^{iF(q)}
=\exp(-\frac{1}{2}\int d^3x \frac{1}{\sqrt{\det(q^0)}}F^{ab} F^{cd}
q^0_{ac}q^0_{bd}) \;.
\ee
From the general theory \cite{8} we know that this measure is supported on
$\overline{{\cal C}}$ and that finite linear combinations of states
of the form $(\psi_F)(q):=\exp(i F(q))$ are dense in ${\cal H}$.

It is obvious that this measure fails to be invariant under
three-dimensional diffeomorphisms which turns out to be a major obstacle
in solving the diffeomorphism constraint since, for instance, the
natural representation of the spatial diffeomorphism group
Diff$(\sigma)$ on $\cal H$,
densely defined by $\hat{U}(\varphi)\psi_F=\psi_{\varphi^*F}$, is not
unitary and therefore cannot be generated by a self-adjoint constraint
operator
($\varphi\in$Diff$(\sigma)$ and $\varphi^*$ is the pull-back action).\\
In fact, the author is not aware of any work where a diffeomorphism
invariant measure was rigorously defined for the stochastic process
corresponding to metric quantum fields.

If we let $F(\hat{q})$ act by multiplication and define
\be \label{22}
\hat{P}(f)\psi_F
:=\hbar[ F(f)+\frac{i}{2}\int d^3x \frac{f_{ab}}{\sqrt{\det(q_0)}}
q_0^{ac}q_0^{bd}q_{cd}]\psi_F
\ee
then the canonical commutation relations and the adjointness relations are,
at least formally, indeed satisfied.
\item[IV)] {\it Constraint Operators}\\
So far we did not encounter any particular problems. However, now we will
encounter a major roadblock : Looking at the algebraic structure
of (\ref{2.30}) we see that the classical constraint
functions depend non-polynomially, not even analytically on the metric
$q_{ab}$ (recall that the curvature scalar depends also on the inverse
metric tensor $q^{ab})$.
This, first of all, seems to rule out completely the
polarization for which the $\hat{P}(f)$ are diagonal since then the
$F(\hat{q})$ would become derivative operators.
More seriously, since the $\hat{P}^{ab}(x),\hat{q}_{ab}(x)$ are operator
valued distributions which are multiplied at the same point in (\ref{2.30}),
a simple replacement of variables by operators is hopelessly divergent
and completely meaningless since it is not clear how a distribution in
the denominator can be defined.

The only chance is that one can suitably regularize
expressions (\ref{2.30}) by defining them as limits of functions
of smeared field variables, the limit corresponding to vanishing smearing
volume. However, nobody succeeded up to date to accomplish such a
regularization and renormalization procedure for the quantum operator
corresponding to the Hamiltonian constraint (\ref{2.30}) which is also
famously called the {\it Wheeler-DeWitt Operator}.

Since the subsequent steps of the quantization programme depend on this
one which we could not solve, not much can be said about the remaining
steps.
\item[V)] {\it Imposing the Constraints}\\
Of course, one could try to find formal solutions to the Wheeler-DeWitt
constraint equation which can be seen as the
{\it Quantum-Einstein-Equations}. Not even one solution could be found in
the full
theory (although solutions could be found in certain finite-dimensional
truncations of the theory). Notice that not even the constant state
$\psi(q)=1$ is a solution.
\item[VI)] {\it Quantum Anomalies}\\
Given that one could not even define the Wheeler-DeWitt constraint operator
it seems to be a hopeless enterprise to find an ordering for which
it is free of anomalies or even self-adjoint.
\item[VII)] {\it Physical Scalar Product}\\
DeWitt has defined in his famous three works \cite{2} a formal
inner product which is at least invariant under three-dimensional
diffeomorphisms, however, to the best knowledge of the author nobody could
ever give a rigorous meaning to the construction.
\item[VIII)] {\it Observables}\\
This step has been completely out of reach since one started the analysis.
\end{itemize}

Summarizing, the programme of canonical quantization applied the way
as just displayed was unsuccessful for decades. Thus, most researchers
in the field gave up and turned to different approaches. It should be kept
in mind, however, that the programme is not a rigid algorithm but requires
to make choices at various stages which are not dictated by mathematical
consistency but depend on one's intuition. Already in the very first step
one is asked to make a choice about the elementary variables and the
polarization of the phase space. Until the mid 80's people worked only
with those ADM variables displayed above since they are so natural.
On the other hand, given the complicated structure of (\ref{2.30}) which
was a roadblock for such a long time, it seems mandatory to look for better
suited canonical variables which, preferrably, render the constraints
at least polynomial. This is precisely the achievement of Ashtekar
\cite{5}.

\subsection{The New Canonical Variables of Ashtekar for General Relativity}
\label{s2.3}

In this subsection we focus on the classical aspects of the so-called
``new variables' '.
The history of the the classical aspects of the new variables is
approximately twenty
years old and we wish to give a brief account of the developments
(the history of the quantum aspects will be given in section \ref{s2.4}):
\begin{itemize}
\item {\it 1981-82}\\
The starting point was a series of papers due to Sen \cite{30} who
generalized the covariant derivative $\nabla_\mu$ of the previous section
for $s=-1$ to $Sl(2,\Cl)$ spinors of left (right) handed helicity
resulting in an
(anti) self-Hodge-dual connection which is therefore complex-valued. An
exhaustive treatment on spinors and spinor calculus can be found in
\cite{32l}.
\item {\it1986-87}\\
Sen
was motivated in part by a spinorial proof of the positivity of energy theorem
of general relativity \cite{115,116}. But it was only Ashtekar
\cite{5,10g,20} who realized that modulo a slight modification of his
connection,
Sen had stumbled  on a new canonical formulation of general relativity
in terms of the (spatial
projection of) this connection, which turns out to be a generalization
of $D_\mu$ to this class of spinors, and a conjugate electrical field kind of
variable, such that the initial value constraints of general
relativity (\ref{2.30}) can be written in {\it polynomial form} if one rescales
$H$ by $H\mapsto\tilde{H}=\sqrt{\det(q)} H$ (which looks like a harmless
modification at first sight). In fact  $\tilde{H}$ is only of fourth
order in the
canonical coordinates, not worse than non-Abelian Yang-Mills theory and thus
a major roadblock on the way towards quantization seemed to be removed.
Ashtekar also noted the usefulness of the connection for $s=+1$ in which
case it is actually real-valued \cite{35}.
\item {\it 1987-88}\\
Ashtekar's proofs were in a Hamiltonian context. Samuel as well as
Jacobson and Smolin dicovered independently that there exists in fact
a Lagrangean formulation of the theory by considering only the
(anti) selfdual part of the curvature of Palatini gravity \cite{30a}. Jacobson
also considered the coupling of fermionic matter \cite{30b} and an extension to
supergravity \cite{30c}. Coupling to standard model matter was considered
by Ashtekar et. al. in \cite{30d}.
All of these developments still used a spinorial
language which, although not mandatory, is of course quite natural if one wants
to treat spinorial matter.

A purely tensorial approach to the new variables was given by Goldberg
\cite{30e} in terms of  triads and by Henneax et. al. in terms of tetrads
\cite{30f}.
\item {\it 1989-92}\\
While the Palatini formulation of general relativity uses a connection and
a tetrad field as independent variables, Capovilla, Dell and Jacobson
realized that there is a classically equivalent action which depends only
on a connection and a scalar field, moreover, they were able to solve
both initial value constraints of general relativity {\it algebraically} for
a huge (but not the complete) class of field configurations. Unfortunately,
there is a third constraint besides the diffeomorphism and Hamiltonian
constraint in this new formulation of general relativity, the so-called Gauss
constraint, which is not automatically satisfied by this so-called
``CDJ-Ansatz' '.

This line of thought was further developed by Bengtsson and Peldan \cite{30g}
culminating in the discovery that in the presence of a cosmological constant
the just mentioned scalar field can be eliminated by a field equation, resulting
in a pure connection Lagrangean for general relativity (but not a polynomial
one).
\item {\it 1994-96}\\
As mentioned above, for Lorentzian (Euclidean) signature one considered
complex (real) valued connection variables. Meanwhile it turned out that
it is very hard to implement the reality conditions for the complex valued
case as adjointness conditions on the measure in the quantum theory while
for the real valued case it is relatively easy. This motivated Barbero
\cite{36}
to consider real valued connections also for Lorentzian signature.
Barbero discovered that one can give a Hamiltonian formulation even
{\it for all complex values} of a parameter considered earlier by
Immirzi \cite{34}
for either choice of signature. However, in order to keep polynomiality of the
Hamiltonian constraint when using real valued connections one has to multiply
it by an even higher power of $\det(q)$. Moreover, the constraint becomes
algebraically much more complicated.

This caveat is removed by a so-called ``phase space Wick rotation' '
intoduced in \cite{IV,71} and later considered also in \cite{72} where
one can work with real connections {\it while keeping the algebraic form of the
constraint simple}. This line of development was motivated by a seminal
paper due to Hall \cite{70} who constructed a unitary transform from a Hilbert
space of square integrable functions on a compact gauge group
to a Hilbert space of square integrable, holomorphic
functions on the complexification of that gauge group and this transform
was generalized by Ashtekar, Lewandowski, Marolf, Mour\~ao and Thiemann
to gauge theories for compact gauge groups \cite{72a}.
Mena-Marugan clarified the relation between this phase space
Wick rotation and the usual one (analytic continuation in the time parameter)
\cite{73}.

The last development in this respect is the result of \cite{37} which
states that
polynomiality of the constraint operator is not only unimportant in order
to give a rigorous meaning to it in quantum theory, it is in fact {\it
desastrous}.
The important condition is that the constraint be a scalar of
{\it density of weight one}. This forbids the rescaling from $H$, which is
already a density of weight one, by any non-trivial power of $\det(q)$.
It is only in that case that the quantization of the operator can be done in
a {\it background independent way without picking up UV divergences on
the kinematical Hilbert space}.
For this reason, real connection variables are currently favoured as far
as quantum theory is concerned. In retrospect, what is really important
is that one bases the quantum theory on connections and canonically
conjugate electric fields (which is dual in a metric independent way
to a two-form). The reason is that $n$-forms can be naturally integrated
over $n$-dimensional submanifolds of $\sigma$ without requiring a
background structure, this is not possible for the metric variables of
the ADM formulation and has forbidden progress for such a long time.
We will come back to this point in the next section.
\item {\it 1996-2000}\\
So far a Lagrangean action principle had been given only for the following
values of signature $s$ and Immirzi parameter $\beta$, namely Lorentzian
general relativity $s=-1,\beta=\pm i$ and Euclidean general relativity
$s=+1,\beta=\pm 1$. For arbitrary complex $\beta$ and either signature
a Lagrangean formulation was discovered by Holst and Barros e S\'a
\cite{30i}. Roughly speaking the action is given by a modification of
the Palatini action
\be \label{2.67}
S=\int_M \mbox{tr}(F\wedge [\ast-\beta^{-1}](e\wedge e))
\ee
(it results for $\beta=\infty$) where $\ast$ denotes the Hodge dual,
$F=F(\omega)$ is the curvature of some connection $\omega$ which
is considered as an independent field next to the tetrad $e$. This action
should be considered in analogy with the $\theta$ theta angle modification of
bosonic QCD
\be \label{2.68}
S=\int_M \mbox{tr}(F\wedge [\ast+\theta]F)
\ee
In the gravitational case the $\beta$ term drops out by an equation of motion,
in the QCD case the variation of the $\theta$ term is exact and also drops
out of the equations of motion. This holds for the classical theory, but
it is well
known that in the quantum theory the actions with different values of $\theta$
are not unitarily equivalent. A similar result holds for general
relativity \cite{34}.

Recently Samuel \cite{30k} criticized the use of real connection variables for
Lorentzian gravity because of the following reason: The Hamiltonian analysis
of the action
(\ref{2.67}) leads, unless $\beta=\pm i$ for $s=-1$, to constraints of second
class which one has to solve by imposing a gauge condition. It eliminates
the boost part of the original $SO(1,3)$ Gauss constraint and one is left with
an $SO(3)$ Gauss constraint (which also appears in in the case
$\beta=\pm i$). That gauge condition fixes the direction of an
internal $SO(1,3)$ vector which is automatically preserved by the remaining
$SO(3)$ subgroup and by the evolution derived from the associated Dirac
bracket, so that everything is consistent. Now while for $\beta=\pm i,s=-1$
the spatial connection is simply the pull-back of the (anti)self-dual part of the
four-dimensional spin connection to the spatial slice, for real $\beta$ its
spacetime interpretation is veiled due to the appearance of the second
class constaints and the gauge fixing.

Samuel now asks the following
question: For any value of $\beta$ it can be shown that every
$SO(3)$ gauge invariant function of the spatial connection and the triad
can be expressed in terms of the (pull-back to the spatial slice of the)
spacetime fields $q_{\mu\nu},K_{\mu\nu}$. In the previous section we have
shown that the Hamiltonian evolution of these fields under the Hamiltonian
constraint coincides, on the constraint surface, with their infinitesimal
transformation under a timelike diffeomorphism. Is it then true that
the induced Hailtonian transformation of $SO(3)$ gauge invariant functions of
the connection (such as traces of its holonomy around a loop in a spatial
slice) coincides with that of (the pull-back to the spatial slice of ) a spacetime
connection? He finds that this is the case if and only if $\beta=\pm i$.
The simple algebraic reason is that only for an (anti)self-dual connection
$A^{IJ},\;I,J=0,1,2,3$ the components $A^{0j}$ are already determined
by $A^j=\frac{1}{2}\epsilon_{jkl} A^{kl}$ so that the pull-back to the spatial
slice of $A^j$ determines the pull-back of an $SO(1,3)$ connection with its
full spacetime interpretation only then.

It should be stressed, however,
that Samuel's criticism is purely aesthetical in nature, for interpretational
reasons it is certainly convenient to have a spacetime interpretation of the
spatial connection but it is by no means mandatory, one just has to bear
in mind that the connection does not have the naive transformation behaviour
under Hamiltonian evolution on the constraint surface. In fact, to date
a satisfactory quantum theory has been constructed only for $\beta$ real
(which in turn does not mean that it is impossible to do for $\beta=\pm i$).
In fact, as we will show in this subsection, at the classical
level {\it all complex values of the Immirzi parameter lead to Hamiltonian
formulations completely equivalent to the ADM formulation}.
\end{itemize}
This concludes our historical digression and we come now to the actual
derivation of the new variable formulation. We decided for the extended
phase space approach using triads as this makes the contact and equivalence
with the ADM formulation most transparent and quickest and avoids the
introduction of additional $SL(2,\Cl)$ spinor calculus which would blow up our
exposition unnecessarily.
Also we do this for either signature and any complex value of the Immirzi
parameter. What is no longer arbitrary is the dimension of $\sigma$ : We
will be forced to work with $D=3$ as will become clear in the course of the
derivation.\\
The construction consists of two steps : First an extension of the
ADM phase space and secondly a canonical transformation on the extended
phase space.\\
\\
{\it\bf Extension of the ADM phase space}\\
\\
We would like to consider the phase
space described in section \ref{s2.1} as the symplectic
reduction of a larger symplectic manifold with coisotropic constraint
surface \cite{18}. One defines a so-called co-D-bein field $e_a^i$
on $\sigma$ where the indices $i,j,k,..$ take values $1,2,..,D$.
The D-metric is expressed in terms of $e_a^i$ as
\be \label{2.69}
q_{ab}:=\delta_{jk} e^j_a e^k_b \;.
\ee
Notice that this relation is invariant
under local $SO(D)$ rotations $e_a^i\to O^i_j e_a^j$ and we therefore can
view $e_a^i$, for $D=3$, as an $su(2)$-valued one-form (recall that the adjoint
representation of $SU(2)$ on its Lie algebra is isomorphic with the
defining representation of $SO(3)$ on $\Rl^3$ under the isomorphism
$\Rl^3\to su(2);\; v^i\to v^i\tau_i$ where $\tau_i$ is a basis of
$su(2)$ (also called ``soldering forms" \cite{32}). This observation
makes it already obvious that we have to get rid of the $D(D-1)/2$ rotational
degrees of freedom sitting in $e_a^i$ but not in $q_{ab}$. Since the
Cartan-Killing metric of $so(D)$ is just the Euclidean one we will
in the sequel drop the $\delta_{ij}$ and also do not need to care about
index positions.

Next we introduce yet another, independent one form $K_a^i$ on $\sigma$
which for $D=3$ we also consider as $su(2)$ valued and
from which the extrinsic curvature is derived as
\be \label{2.70}
-2s K_{ab}:=\mbox{sgn}(\det((e_a^i)))K_{(a}^i e^i_{b)} \;.
\ee
We see immediately that $K_a^i$ cannot be an arbitrary $D\times D$
matrix but must satisfy the constraint
\be \label{2.71}
G_{ab}:=K_{[a}^j e_{b]}^j=0
\ee
since $K_{ab}$ was a symmetric tensor field. With the help of the quantity
\be \label{2.72}
E^a_j:=\frac{1}{(D-1)!}\epsilon^{a a_1..a_{D-1}}\epsilon_{j j_1..j_{D-1}}
e_{a_1}^{j_1}.. e_{a_{D-1}}^{j_{D-1}}
\ee
one can equivalently write (\ref{2.72}) in the form
\be \label{2.73}
G_{jk}:=K_{a [j}E^a_{k]}=0
\ee
Consider now the following functions on the extended phase space
\be \label{2.74}
q_{ab}:=E_a^j E_b^j
|\det((E^c_l))|^{2/(D-1)},\;
P^{ab}:= |\det((E^c_l))|^{-2/(D-1)} E^a_k E^d_k K_{[d}^j \delta_{c]}^b
E^c_j
\ee
where $E_a ^j$ is the inverse of $E^a_j$. It is easy to see that when
$G_{jk}=0$, the functions (\ref{2.74}) precisely reduce to the ADM
coordinates. Inserting (\ref{2.74}) into (\ref{2.30}) we can also
write the diffeomorphism and Hamiltonian constraint as functions on the
extended phase space which one can check to be explicitly given by
\ba \label{2.75}
H_a:=-D_b[K_a^j E^b_j-\delta_a^b K_c^j E^c_j] \nonumber\\
H:=-\frac{s}{4\sqrt{\det(q)}}(K_a^l K_b^j-K_a^j K_b^l)E^a_j E^b_l
-\sqrt{\det(q)} R
\ea
where $\sqrt{\det(q)}:=|\det((E^a_j))|^{1/(D-1)}$ and $q^{ab}=
E^a_j E^b_j/\det(q)$ by which $R=R(q)$ is considered as a function of
$E^a_j$. Notice that, using (\ref{2.70}), (\ref{2.72}),
expressions (\ref{2.75}) indeed reduce to (\ref{2.30}) up to
terms proportional to $G_{jk}$.

Let us equip the extended phase
space coordinatized by $(K_a^i,E^a_i)$ with the symplectic
structure (formally, that is without smearing) defined by
\be \label{2.76}
\{E^a_j(x),E^b_k(y)\}=\{K_a^j(x),K_b^k(y)\}=0,\;
\{E^a_i(x),K_b^j(y)\}=\kappa\delta^a_b\delta_i^j\delta(x,y)
\ee
We claim now that the symplectic
reduction with respect to the constraint $G_{jk}$ of the constrained
Hamiltonian system subject to the constraints (\ref{2.73}), (\ref{2.74})
results precisely in the ADM phase space of section \ref{s2.1} together
with the original diffeomorphism and Hamiltonian constraint.

To prove this statement we first of all define the smeared ``rotation
constraints"
\be \label{2.77}
G(\Lambda):=\int_\sigma d^Dx \Lambda^{jk} K_{aj} E^a_k
\ee
where $\Lambda^T=-\Lambda$ is an arbitrary antisymmetric matrix, that is,
an $so(D)$ valued scalar on $\sigma$. They satisfy the Poisson algebra,
using (\ref{2.76})
\be \label{2.78}
\{G(\Lambda),G(\Lambda')\}=G([\Lambda,\Lambda'])
\ee
in other words, $G(\Lambda)$ generates infinitesimal $SO(D)$ rotations as
expected.
Since the functions (\ref{2.74}) are manifestly $SO(D)$ invariant by
inspection they Poisson commute with $G(\Lambda)$, that is,
they comprise a complete set of rotational invariant Dirac observables with
respect to $G(\Lambda)$ for any $\Lambda$. As the
constraints
defined in (\ref{2.75}) are in turn functions of these, $G(\Lambda)$ also
Poisson commutes with the constraints (\ref{2.75}) whence the total
system of constraints
consisting of (\ref{2.77}), (\ref{2.75}) is of first class.

Finally we must check that Poisson brackets among
the $q_{ab}, P^{cd}$, considered as the functions (\ref{2.74}) on the
extended phase space with symplectic structure (\ref{2.76}), is equal
to the Poisson brackets of the ADM phase space (\ref{2.31}, at least when
$G_{jk}=0$. Since $q_{ab}$ is a function of $E^a_j$ only it is clear that
$\{q_{ab}(x),q_{cd}(y)\}=0$. Next we have
\ba \label{2.79}
\kappa \{P^{ab}(x),q_{cd}(y)\} &=&
(\frac{1}{2}[q^{a(e} q^{bf)}-q^{ab} q^{ef}]E^j_f)(x)
\{K_e^j(x),(|\det(E)|^{2/(D-1)} E^k_c E^k_d)(y)\}
\nonumber\\
&=&
(\frac{1}{2}[q^{a(e} q^{bf)}-q^{ab} q^{ef}]E^j_f)(x)
[\frac{2}{D-1} q_{cd}(x) \frac{\{K_e^j(x),|\det(E)|(y)\}}{|\det(E)|(x)}
\nonumber\\
&& +2(\det(q) E^k_{(c}(x)\{K_e^j(x),E^k_{d)}(y)\}]
\nonumber\\
&=&
([q^{a(e} q^{bf)}-q^{ab} q^{ef}]
[-\frac{1}{D-1} q_{cd} q_{ef}+q_{e(c} q_{d)f}])(x)\delta(x,y)
\nonumber\\
&=& \delta^a_{(c} \delta^b_{d)}\delta(x,y)
\ea
where we used $\delta E^{-1}=-E^{-1}\delta E E^{-1},\;
[\delta |\det(E)|]/|\det(E)|=[\delta \det(E)]/\det(E)=E_a^j \delta E^a_j$.
The final Poisson bracket is the most difficult one. By carefully
inserting the definitions, making use of the relations $E^a_j=\det(e) e^a_j,
E_a^j=e_a^j\det(e), e^a_j=q^{ab} e_b^j$ at various steps one finds after
two pages of simple but tedious algebraic manipulations that
\be \label{2.80}
\{P^{ab}(x),P^{cd}(y)\}=-\frac{\det(e)}{8}
[q^{bc} G^{ad}+q^{bd} G^{ac}+q^{ac} G^{bd}+q^{ad} G^{bc}])(x) \delta(x,y)
\ee
where $G^{ab}=q^{ac} q^{bd} G_{cd}$ and so (\ref{2.80}) vanishes only at
$G_{ab}=0$.

Let us summarize : The functions (\ref{2.74}) and (\ref{2.75}) reduce at
$G_{jk}=0$ to the corresponding functions on the ADM phase space, moreover,
their Poisson brackets among each other reduce at $G_{jk}=0$ to those of
the ADM phase space. Thus, as far as rotationally invariant observables are
concerned, the only ones we are interested in, both the ADM system and
the extended one are completely equivalent and we can as well work with
the latter.
This can be compactly described by saying that the symplectic reduction
with respect to $G_{jk}$ of
the constrained Hamiltonian system described by the action
\be \label{2.81}
S:=\frac{1}{\kappa}\int_\Rl dt \int_\sigma d^Dx (\dot{K}_a ^j E^a_j
-[-\Lambda^{jk} G_{jk}+N^a H_a+ N H])
\ee
is given by the system described by the ADM action of section (\ref{s2.1}).
Notice that, in accordance with what we said before, there is no claim
that the Hamiltonian flow of $K_a^j, E^a_j$ with respect to $H_a,H$ is a
spacetime diffeomorphism. However, since the Hamiltonian flow of $H,H_a$
on the constraint surface $G_{jk}=0$ is the same as on the ADM phase
space for the gauge invariant observables $q_{ab}, P^{ab}$,
a representation of Diff$(M)$ is still given on the constraint surface
of $G_{jk}=0$.\\
\\
{\it\bf Canonical Transformation on the Extended Phase Space}\\
\\
Up to now we could work with arbitrary $D\ge 2$, however, what follows works
only for $D=3$. First we introduce the notion of the {\it spin connection}
which is defined as an extension of the spatial covariant derivative $D_a$
from tensors to generalized tensors with $so(D)$ indices. One defines
\be \label{2.82}
D_a u_{b..} v_j:= (D_a u_b)_{..}v_j+..+u_{b..} (D_a v_j) \mbox{ where }
D_a v_j:=\partial_a v_j+\Gamma_{ajk}v^k
\ee
extends by linearity and requires that $D_a$ is compatible with $e_a^j$,
that is
\be \label{2.83}
D_a e_b^j=0\; \Rightarrow\; \Gamma_{ajk}=-e^b_k[\partial_a e_b^j-
\Gamma^c_{ab} e_c^j]
\ee
Obviously $\Gamma_a$ takes values in $so(D)$, that is, (\ref{2.83}) defines
an antisymmetric matrix.

Our aim is now to write the constraint $G_{jk}$ in such a form that it becomes
the Gauss constraint of an $SO(D)$ gauge theory, that is,
we would like to write it in the form $G_{jk}=(\partial_a E^a+[A_a,E^a])_{jk}$
for some $so(D)$ connection $A$. It is here where $D=3$ is singled out :
What we have is an object of the form $E^a_j$ which transforms in the defining
representation of $SO(D)$ while $E^a_{jk}$ transforms in the adjoint
representation of $SO(D)$. It is only for $D=3$ that these two are equivalent.
Thus from now on we take $D=3$.

The canonical transformation that we have in mind consists of two parts :
1) A constant Weyl (rescaling) transformation and 2) an affine
transformation. \\ \\
{\it Constant Weyl Transformation}\\
\\
Observe that for any finite complex number
$\beta\not=0$, called the {\it Immirzi parameter},
the following rescaling $(K_a^j,E^a_j)\mapsto
(^{(\beta)}K_a^j:=\beta K_a^j,^{(\beta)}E^a_j:=E^a_j/\beta)$
is a canonical transformation (the Poisson brackets (\ref{2.76}) are
obviously invariant under this map). We will use the notation
$K=K^{(1)},E=K^{(1)}$. In particular, for the rotational
constraint (which we write in $D=3$ in the equivalent form)
\be \label{2.84}
G_j=\epsilon_{jkl} K_a^k E^a_l=\epsilon_{jkl}
(^{(\beta)}K_a^k)(^{(\beta)}E^a_l)
\ee
is invariant under this rescaling transformation. We will consider the other
two constraints (\ref{2.75}) in a moment.\\
\\
{\it Affine Transformation}\\
\\
We notice from (\ref{2.83})
that $D_a E^b_j=0$. In particular, we have
\be \label{2.85}
D_a E^a_j=[D_a E^a]_j+\Gamma_{aj}\;^k E^a_k=\partial_a E^a_j+\epsilon_{jkl}
\Gamma_a^k E^a_l=0
\ee
where the square bracket in the first identity means that $D$ acts only
on tensorial indices which is why we could replace $D$ by $\partial$ as
$E^a_j$ is an $su(2)$ valued vector density of weight one. We also used the
isomorphism between antisymmetric tensors of second rank and vectors in
Euclidean space to define $\Gamma_a=:\Gamma_a^l T_l$ where
$(T_l)_{jk}=\epsilon_{jlk}$ are the generators of $so(3)$ in the defining -- or,
equivalently, of $su(2)$ in the adjoint representation if the structure
constants are chosen to be $\epsilon_{ijk}$. Next we explicitly solve
the spin connection in terms of $E^a_j$ from (\ref{2.83}) by using the
explicit formula for $\Gamma^a_{bc}$ and find
\ba \label{2.86}
\Gamma_a^i &=&\frac{1}{2}\epsilon^{ijk}e^b_k[e_{a,b}^j-e_{b,a}^j+e^c_j e_a^l
e^l_{c,b}]
\\
&=&\frac{1}{2}\epsilon^{ijk}E^b_k[E_{a,b}^j-E_{b,a}^j+E^c_j E_a^l E^l_{c,b}]
+\frac{1}{4}\epsilon^{ijk}E^b_k[
2E_a^j\frac{(\det(E))_{,b}}{\det(E)}-E_b^j\frac{(\det(E))_{,a}}{\det(E)}]
\nonumber
\ea
where in the second line we used that $\det(E)=[\det(e)]^2$ in $D=3$.
Notice that the second line in (\ref{2.86}) explicitly shows that
$\Gamma_a^j$ is a homogenous rational function of degree zero
of $E^a_j$ and its derivatives. Therefore we arrive at the important
conclusion that
\be \label{2.87}
(^{(\beta)}\Gamma_a^j):=
\Gamma_a^j(^{(\beta)} E)=\Gamma_a^j=\Gamma_a^j(^{(1)} E)
\ee
is itself invariant under the rescaling transformation. This is obviously
also true for the Chritoffel connection $\Gamma^a_{bc}$ since it is a
homogenous rational function of degree zero in $q_{ab}$ and its derivatives
and $q_{ab}=\det(E) E_a^j E_b^j\mapsto (^{(\beta)}q_{ab})=\beta
(^{(1)}q_{ab})$. Thus the derivative $D_a$ is, in fact, independent of
$\beta$ and we therefore have in particular $D_a (^{(\beta)} E^a_j)=0$.
We can then write the rotational constraint in the form
\be \label{2.88}
G_j=0+\epsilon_{jkl} (^{(\beta)} K_a^k)(^{(\beta)} E^a_l)
=\partial_a (^{(\beta)} E^a_j)+\epsilon_{jkl}
[\Gamma_a^j+(^{(\beta)} K_a^k)](^{(\beta)} E^a_l)
=:\;^{(\beta)}D_a \;^{(\beta)}E^a_j
\ee
This equation suggests to introduce the new connection
\be \label{2.89}
(^{(\beta)}A_a^j):=\Gamma_a^j+(^{(\beta)} K_a^j)
\ee
This connection could be called the Sen -- Ashtekar -- Immirzi --
Barbero connection (names in historical order) for the historical
reasons mentioned in the beginning of this section. More precisely
the Sen connection arises for $\beta=\pm i, G_j=0$, the Ashtekar
connection for $\beta=\pm i$, the Immirzi connection for complex $\beta$
and the Barbero connection for real $\beta$. For simplicity we will
refer to it as the {\it new} connection which now replaces the
spin-connection $\Gamma_a^j$ and gives rise to a new
derivative $^{(\beta)}D_a$ acting on generalized tensors as the extension
by linearity of the basic rules $^{(\beta)} D_a v_j:=
\partial_a v_j+\epsilon_{jkl} (^{(\beta)}A_a^k) v_l$ and
$^{(\beta)} D_a u_b:=D_a u_b$. Notice that (\ref{2.89}) has {\it precisely}
the structure of a Gauss law constraint for an  $SU(2)$ gauge theory although
$^{(\beta)}A$ qualifies as the pull-back to $\sigma$ by local sections of
a connection on an $SU(2)$ fibre bundle over $\sigma$ only when $\beta$ is
real. Henceforth we will call $G_j$ the {\it Gauss constraint}.\\
Given the complicated structure of (\ref{2.86}) it is quite surprising that
the variables $(^{(\beta)}A,^{(\beta)}E)$ form a canonically conjugate
pair, that is
\be \label{2.90}
\{^{(\beta)}A_a^j(x),^{(\beta)}A_b^k(y)\}
=\{^{(\beta)}E^a_j(x),^{(\beta)}E^b_k(y)\}=0,\;
\{^{(\beta)}E^a_j(x),^{(\beta)}A_b^j(y)\}=
\kappa\delta^a_b\delta_j^k\delta(x,y)
\ee
This is the key feature for why these variables are at all useful in
quantum theory : If we would not have such a simple bracket structure
classically then it would be very hard to find Hilbert space representations
that turn these Poisson bracket relations into canonical commutation
relations.\\
To prove (\ref{2.90}) by means of (\ref{2.76}) (which is invariant under
replacing $K,E$ by $^{(\beta)}K,^{(\beta)}E)$
we notice that the only non-trivial relation is the first one
since $\{E^a_j(x),\Gamma_b^k(y)\}=0$. That relation is explicitly given
as
\be \label{2.91}
\beta[\{\Gamma_a^j(x),K_b^k(y)\}-\{\Gamma_b^k(y),K_a^j(x)\}]
=\beta\kappa
[\frac{\delta\Gamma_a^j(x)}{\delta E^b_k(y)}
-\frac{\delta\Gamma_b^k(y)}{\delta E^a_j(x)}]=0
\ee
which is just the integrability condition for $\Gamma_a^j$ to have a generating
potential $F$. A promising candidate for $F$ is given by the functional
\be \label{2.92}
F=\int_\sigma d^3x E^a_j(x) \Gamma^a_j(x)
\ee
since if (\ref{2.91}) holds we have
\ba \label{2.93}
\frac{\delta F}{\delta E^a_j(x)}-\Gamma_a^j(x)
&=&\int d^3y E^b_k(y) \frac{\delta\Gamma_b^k(y)}{\delta E^a_j(x)}
=\int d^3y E^b_k(y) \frac{\delta\Gamma_a^j(x)}{\delta E^b_k(y)}
\nonumber\\
=\frac{1}{\kappa}\{\Gamma_a^j(x),\int d^3 y K_b^k(y) E^b_k(y)\}=0
\ea
because the function $\int d^3 y K_b^k(y) E^b_k(y)$ is the canonical
generator of {\it constant} scale transformations under which $\Gamma_a^j$
is invariant as already remarked above. To show that $F$ is indeed a
potential for $\Gamma_a^j$ we demonstrate (\ref{2.93}) in the form
$\int d^3x E^a_j(x) \delta \Gamma_a^j(x)=0$. Starting from (\ref{2.86})
we have (using $\delta e_a^j e^b_j=\delta e_b^j e^b_k=0$ repeatedly)
\ba \label{2.94}
e^a_i \delta\Gamma_a^i &=&
\frac{1}{2}\epsilon^{ijk}
\det(e) e^a_i \delta (e^b_k[e_{a,b}^j-e_{b,a}^j+e^c_j e_a^l e^l_{c,b}])
\nonumber\\
&=& \frac{1}{2}\epsilon^{ijk}\det(e)
[e^a_i \delta (e^b_k(e_{a,b}^j-e_{b,a}^j))+\delta(e^b_k e^c_j e ^i_{c,b})
-(\delta e^a_i ) e^c_j e_a^l e^b_k e^l_{c,b}]
\nonumber\\
&=& \frac{1}{2}\epsilon^{ijk}\det(e)
[e^a_i \delta (e^b_k(e_{a,b}^j-e_{b,a}^j))+\delta(e^b_k e^a_j e ^i_{a,b})
+(\delta e_a^l) e^a_i e^c_j e^b_k e^l_{c,b}]
\nonumber\\
&=& \frac{1}{2}\epsilon^{ijk}\det(e)
[\delta (e^a_i e^b_k(e_{a,b}^j-e_{b,a}^j)+e^b_k e^a_j e ^i_{a,b})
-(\delta e^a_i) e^b_k(e_{a,b}^j-e_{b,a}^j)
+(\delta e_a^l) e^a_i e^c_j e^b_k e^l_{c,b}]
\nonumber\\
&=& \frac{1}{2}\epsilon^{ijk}\det(e)
[\delta (e^b_k (e^a_j e ^i_{a,b}+e^a_i e_{a,b}^j) -e^a_i e^b_k e_{b,a}^j))
+(\delta e^b_k) e^a_i e_{b,a}^j +(\delta e^a_i) e^b_k e_{b,a}^j
\nonumber\\
&& +(\delta e_a^l) e^a_i e^c_j e^b_k e^l_{c,b}]
\nonumber\\
&=& -\frac{1}{2}\epsilon^{abc}
[e_c^j\delta e_{b,a}^j-(\delta e_a^j) e^j_{c,b}]
\nonumber\\
&=& -\frac{1}{2}\epsilon^{abc}
\partial_a [(\delta e_b^j)e_c^j]
\ea
From the first to the second line we pulled $e^a_i$ into the variation of the
the third term of $\delta \Gamma^a_i$ resulting in a correction proportional
to $\delta e_a^i$, in the next line we relabelled the summation index
$c$ into $a$ in the third term
and traded the variation of $e^a_i$ for that of $e_a^l$ in the fourth
term, in the
next line we pulled again $e^a_i$  inside a variation resulting in altogether
six terms, in the next line we collected the total variation terms and
reordered them and in the fourth term we relabelled the summation indices
$a,b$ into $b,a$ and $i,k$ into $k,i$ resulting in a minus sign from the
$\epsilon^{ijk}$, in the next line we realized that the first two terms are
symmetric in $i,j$ which thus drop out due to the $\epsilon^{ijk}$ and
that the $e^a_i$ and $e^b_k$ variation pieces of the third term cancel
against the fourth and fifth term, in the next line we made use of the relations
$\det(e)\epsilon^{ijk} e^b_j e^c_k=\epsilon^{abc} e_a^i,
\det(e)\epsilon^{ijk} e^a_i e^b_j e^c_k=\epsilon^{abc}$ and relabelled $j$ for
$l$ and in the last line finally we relabelled $a$ for $b$ in the second term
resulting in a minus sign and allows us to write the whole thing as a
derivative. It follows that
\be \label{2.95}
\int_\sigma d^3x E^a_j \delta \Gamma_a^j
=-\frac{1}{2}\int_\sigma d^3x \partial_a(\epsilon^{abc}\delta e_b^j e_c^j)
=\frac{1}{2}\int_{\partial\sigma} dS_a \epsilon^{abc} e_b^j \delta e_c^j
\ee
which vanishes since $\partial \sigma=\emptyset$. If $\sigma$ has a boundary
such as spatial infinity then the boundary conditions such as imposing
$e_a^j$ to be an even function on the asymptotic sphere under Cartesian
coordinate reflection guarantee vanishing of (\ref{2.95}) as well, see
\cite{19,30k}.

It remains to write the constraints (\ref{2.75}) in terms of the variables
$^{(\beta)}A,^{(\beta)}E$. To that end we introduce the curvatures
\ba \label{2.96}
R_{ab}^j:=2\partial_{[a}\Gamma_{]a}^j+\epsilon_{jkl} \Gamma_a^k \Gamma_b^l
\nonumber\\
^{(\beta)}F_{ab}^j:=2\partial_{[a}\;^{(\beta)}A_{b]}^j+\epsilon_{jkl}
^{(\beta)}A_a^k\;^{(\beta)}A_b^l
\ea
whose relation with the covariant derivatives is given by
$[D_a,D_b]v_j=R_{abjl} v^l=\epsilon_{jkl} R_{ab}^k v^l$ and
$[^{(\beta)}D_a,^{(\beta)}D_b]v_j=^{(\beta)}F_{abjl} v^l=
\epsilon_{jkl}\;^{(\beta)}F_{ab}^k v^l$. Let us expand $^{(\beta)}F$ in
terms of $\Gamma$ and $^{(\beta)}K$
\be \label{2.97}
^{(\beta)}F_{ab}^j=R_{ab}^j+2\beta D_{[a} K_{b]}^j
+\beta^2 \epsilon_{jkl} K_a^j K_b^k
\ee
Contracting with $^{(\beta)}E$ yields
\be \label{2.98}
^{(\beta)}F_{ab}^j\;^{(\beta)}E^b_j=
\frac{R_{ab}^j E^b_j}{\beta}+2 D_{[a} (K_{b]}^j E^b_j)
+\beta K_a^j G_j
\ee
where we have used the Gauss constraint in the form (\ref{2.84}). We claim
that the first term on the right hand side of (\ref{2.98}) vanishes identically.
To see this we first derive from (\ref{2.83}) due to
torsion freeness of the Christoffel connection in the language of forms
the {\it algebraic Bianchi identity}
\ba \label{2.99}
&& dx^a\wedge dx^b D_a e_b^j=d e^j+\Gamma^j_k\wedge e^k=0
\nonumber\\
&\Rightarrow& 0=-d^2 e^j=d\Gamma^j_k\wedge e^k-\Gamma^j_l\wedge de^l
=[d\Gamma^j_k+\Gamma^j_l\wedge \Gamma^l_k]\wedge e^k
=\Omega^j_k\wedge e^k
\ea
Now $\Omega^j_k=\Omega^i (T_i)_{jk}=:(\Omega)_{jk}$ and we see that
$$
\Omega=d\Gamma+\Gamma\wedge\Gamma=
d\Gamma^i\; T_i+\frac{1}{2}[T_j,T_k]\Gamma^j\wedge\Gamma^k=
\frac{1}{2}dx^a\wedge dx^b R_{ab}^i T_i
$$
Thus the Bianchi identity can be rewritten in the form
\ba \label{2.100}
&& \epsilon_{ijk}\epsilon^{efc}R_{ef}^j e_c^k=0\Rightarrow
\nonumber\\
&& \frac{1}{2}\epsilon_{ijk}\epsilon^{efc}R_{ef}^j e_c^k e_a^i
=\frac{1}{2}E^b_j \epsilon_{cab}\epsilon^{efc}R_{ae}^j
\nonumber\\
&=& R_{ab}^j E^b_j=0
\ea
as claimed. Now we compare with the first line of (\ref{2.75}) and thus arrive
at the conclusion
\be \label{2.101}
^{(\beta)}F_{ab}^j\;^{(\beta)}E^b_j=H_a+\;^{(\beta)}K_a^j G_j
\ee
Next we contract (\ref{2.102}) with $\epsilon_{jkl}\; ^{(\beta)}
E^a_k\;^{(\beta)} E^b_l$ and find
\ba \label{2.102}
&& ^{(\beta)}F_{ab}^j \epsilon_{jkl}\; ^{(\beta)} E^a_k\;^{(\beta)} E^b_l
\nonumber\\
&=&\det(q) \frac{R_{abkl} e^a_k e^b_l}{\beta^2}
-2\frac{E^a_j D_a G_j}{\beta}
+ (K_a^j E^a_j)^2 -(K_b^j E^a_j )(K_a^k E^b_k )
\ea
Expanding $v_j=e^a_j v_a,\;v_a=e_a^j v_i$, using $D_a e_b^j=0$ and comparing
$[D_a,D_b]v_j$ with $[D_a,D_b]v_c$ for any $v_j$ we find
$R_{abij}=R_{abcd} e^c_i e^d_j$ and so (\ref{2.102}) can be rewritten as
\ba \label{2.103}
&& ^{(\beta)}F_{ab}^j \epsilon_{jkl}\; ^{(\beta)} E^a_k\;^{(\beta)} E^b_l
\nonumber\\
&=&-\det(q) \frac{R}{\beta^2} -2\;^{(\beta)}E^a_j D_a G_j
+ (K_a^j E^a_j)^2 -(K_b^j E^a_j )(K_a^k E^b_k )
\ea
and comparing with the second line of (\ref{2.75}) we conclude
\ba \label{2.104}
&& ^{(\beta)}F_{ab}^j \epsilon_{jkl}\; ^{(\beta)} E^a_k\;^{(\beta)} E^b_l
+2\;^{(\beta)}E^a_j D_a G_j
\nonumber\\
&=&\sqrt{\det(q)}[-\sqrt{\det(q)}\frac{R}{\beta^2}
-\frac{(K_b^j E^a_j )(K_a^k E^b_k )-(K_a^j E^a_j)^2}{\sqrt{\det(q)}}]
\nonumber\\
&=&\frac{\sqrt{\det(q)}}{\beta^2}[
-\sqrt{\det(q)}R-
\beta^2\frac{(K_b^j E^a_j )(K_a^k E^b_k )-(K_a^j E^a_j)^2}{\sqrt{\det(q)}}]
\nonumber\\
&=&\frac{\sqrt{\det(q)}}{\beta^2}[
H+(\frac{s}{4}-\beta^2)
\frac{(K_b^j E^a_j )(K_a^k E^b_k )-(K_a^j E^a_j)^2}{\sqrt{\det(q)}}]
\nonumber\\
&=&4s\sqrt{\det(q)}[
-\frac{s}{4\sqrt{\det(q)}}[(K_b^j E^a_j )(K_a^k E^b_k )-(K_a^j E^a_j)^2]
-\frac{s}{4\beta^2} \sqrt{\det(q)} R]
\nonumber\\
&=& 4s\sqrt{\det(q)}[H-(1+\frac{s}{4\beta^2}) \sqrt{\det(q)} R]
\ea
We see that the left hand side of  (\ref{2.104}) is proportional to $H$
if and only
if $\beta=\pm \sqrt{s}/2$, that is, imaginary (real) for Lorentzian (Euclidean)
signature. We prefer, for reasons that become obvious only in a later section,
to solve (\ref{2.104}) for $H$ as follows
\ba \label{2.105}
H&=& \frac{\beta^2}{\sqrt{\det(^{(\beta)}q\beta)}}
[^{(\beta)}F_{ab}^j \epsilon_{jkl}\; ^{(\beta)} E^a_k\;^{(\beta)} E^b_l
+2\;^{(\beta)}E^a_j D_a G_j]
\nonumber\\
&& +(\beta^2-\frac{s}{4})
\frac{(^{(\beta)}K_b^j\; ^{(\beta)}E^a_j )(^{(\beta)}K_a^j\; ^{(\beta)}E^a_j )
-(^{(\beta)}K_c^j\; ^{(\beta)}E^c_j )^2}{\sqrt{\det(^{(\beta)}q\beta)}}
\ea
In formula (\ref{2.105}) we wrote everything in terms of
$^{(\beta)}A,^{(\beta)}E$
if we understand $^{(\beta)}K=^{(\beta)}A-\Gamma$ and we used
$^{(\beta)}q_{ab}=\beta^{-1} q_{ab}=^{(\beta)}E_a^j\; ^{(\beta)}E_b^j
\det(^{(\beta)}E)$.

We notice that both (\ref{2.101}) and (\ref{2.105}) still involve the Gauss
constraint. Since the transformation $K_a^j\mapsto ^{(\beta)}A_a^j,
E^a_j\mapsto ^{(\beta)}A_a^j$ is a canonical one, the Poisson brackets
among the set of first class constraints given by $G_j, H_a, H$ are unchanged.
Let us write
symbolically $H_a=H_a'+f_a^j G_j,H=H'+f^j G_j$ where $H_a',H'$ are the
pieces of $H_a,H$ respectively not proportional to the Gauss constraint.
Since $G_j$ generates a subalgebra of the constraint algebra it follows that
the modified system of constraints given by $G_j,H_a',H'$ not only defines the
same constraint surface of the phase space but also gives a first class system
again, of course, with somewhat modified algebra which however coincides with
the Dirac algebra on the submanifold $G_j=0$ of the phase space. In other
words, it is
completely equivalent to work with the set of constraints $G_j,H'_a,H'$
which we write once more, dropping the prime, as
\ba \label{2.106}
G_j&=&\;^{(\beta)}D_a \;^{(\beta)}E^a_j=\partial_a
\;^{(\beta)}E^a_j+\epsilon_{jkl} \;^{(\beta)}A_a^j \;^{(\beta)}E^a_j
\nonumber\\
H_a&=&^{(\beta)}F_{ab}^j\;^{(\beta)}E^b_j
\nonumber\\
H &=&
[\beta^2\;^{(\beta)}F_{ab}^j +(\beta^2-\frac{s}{4})
\epsilon_{jmn}\; ^{(\beta)} K_a^m\;^{(\beta)} K_b^n]
\frac{\epsilon_{jkl}\; ^{(\beta)} E^a_k\;^{(\beta)}
E^b_l}{\sqrt{\det(^{(\beta)}q\beta)}} \ea
For easier comparison with the literature we also write (\ref{2.106}) in terms
of $^{(\beta)}A_a^j, K_a^j, E^a_j$ which gives
\ba \label{2.107}
G_j&=&(^{(\beta)}D_a E^a_j)/\beta=(\partial_a \;{(\beta)}E^a_j+\epsilon_{jkl}
\;{(\beta)}A_a^j E^a_j)/\beta
\nonumber\\
H_a&=&(^{(\beta)}F_{ab}^j E^b_j)/\beta
\nonumber\\
H &=&
[^{(\beta)}F_{ab}^j +(\beta^2-\frac{s}{4})
\epsilon_{jmn} K_a^m K_b^n]
\frac{\epsilon_{jkl} E^a_k E^b_l}{\sqrt{\det(q)}}
\ea
At this point we should say that our conventions differ slightly from those
in the literature : There one writes the constraint in terms
of $\tilde{K}_a ^j:=K_a^j/2$ and one defines
$^{(\beta)}\tilde{K}:=\beta\tilde{K}=\;^{(\beta)}K/2=\;^{(\beta/2)}K$ and
$^{(\beta)}\tilde{A}:=
\Gamma+\beta\tilde{K}=\Gamma+\beta/2 K=^{(\beta/2)}A$
at the price of $2\;^{(\beta)}E$ being conjugate to $^{(\beta)}\tilde{A}$
instead of $^{(\beta)}E$ being conjugate to $^{(\beta)}A$. Thus
$^{(\beta)}A=\;^{(2\beta)}\tilde{A}=\;^{(\tilde{\beta})}\tilde{A}$
with $\tilde{\beta}=2\beta$. When writing
$H$ in terms of these quantities we find
\be \label{2.108}
H=[^{(\tilde{\beta})}F_{ab}^j +(\tilde{\beta}^2-s)
\epsilon_{jmn} \tilde{K}_a^m \tilde{K}_b^n]
\frac{\epsilon_{jkl} E^a_k E^b_l}{\sqrt{\det(q)}}
\ee
where now $\tilde{\beta}^2=s$ is the preferred value. \\
Summarizing, we have rewritten the Einstein Hilbert action in the following
equivalent form
\be \label{2.108a}
S=\frac{1}{\kappa}\int_{\Rl}dt\int_\sigma d^3x(^{(\beta)}\dot{A}_a^i
\;^{(\beta)}E^a_i-
[\Lambda^i G_i+N^a V_a+N H])
\ee
where the appearing constraints are the ones given by either of
(\ref{2.108}), (\ref{2.107}) or (\ref{2.106}).\\
\\
Several remarks are in order :
\begin{itemize}
\item {\it Four-dimensional Interpretation}\\
Let us try to give a four-dimensional meaning to $^{(\beta)} A$. To that
end we must complete the 3-bein $e^a_i$ to a 4-bein $e^\mu_\alpha$ where
$\mu$ is a spacetime tensor index and $\alpha=0,1,2,3$ an index for
the defining representation of the Lorentz (Euclidean) group for $s=-1 (+1)$.
By definition $g_{\mu\nu} e^\mu_\alpha e^\nu_\beta=\eta_{\alpha\beta}$
is the flat Minkowski (Euclidean) metric. Thus $e^\mu_0,e^\mu_i$ are
orthogonal vectors and we thus choose $e^\mu_0=n^\mu$ and in the
ADM frame with $\mu=t,a$ we choose $(e^\mu_i)_{\mu=a}=e^a_i$.
Using the defining properties of a tetrad basis and the explicit form of
$n^\mu,g_{\mu\nu}$ in the ADM frame derived earlier, above choices are
sufficient to fix the tetrad components completely to be
$e^t_0=1/N,\; e^a_0=-N^a/N,\; e^t_i=0,\;e^a_i$. Inversion gives
(notice that $e_\mu^0=s e_{\mu 0}=s g_{\mu\nu} e^\nu_0=s g_{\mu\nu} n^\mu
=s n_\mu$) $e_t^0=N,\;e_a^0=0,\;e_t^i=N^a e_a^i, e_a^i$. Finally we have
for $q^\mu_\nu=\delta^\mu_\nu-s n^\mu n_\nu=\delta^\mu_\nu-e^\mu_0
 e_\nu^0$ in the ADM frame
$q^t_t=0,\;q^t_a=0,\;q^a_t=N^a, q^a_b=\delta^a_b$. Thus we obtain, modulo
$G_j=0$
\ba \label{2.110}
K_a^j&=& -2s e^b_j K_{ab}=-2s e^b_j q^\mu_a q^\nu_b \nabla_\mu n_\nu
=-2 e^b_j  (\nabla_a e_b)^0=2 e^b_j  (\omega_a)^0\;_\alpha e_b^\alpha
\nonumber\\
&=& 2 e^b_j  (\omega_a)^0\;_k e_b^k
=2   (\omega_a)^0\;_j
\ea
where in the second identity the bracket denotes that $\nabla$ only acts on
the tensorial index and in the third we used the definition of the four
dimensional
spin connection $\nabla_\mu e_nu^\alpha=(\nabla_\mu e_\nu)^\alpha
+(\omega_\mu)^\alpha_\beta e_nu^\beta=0$. On the other hand we have
\be \label{2.111}
(\Gamma_a)^j\;_k e_b^k=-(D_a e_b)^j=-q^\mu_a q^\nu_b (\nabla_\mu e_\nu)^j
=-(\nabla_a e_b)^j=(\omega_a)^j\;_k e_b^k
\ee
whence $\omega_{ajk}=\Gamma_{ajk}$. It follows that
\be \label{2.112}
^{(\beta)}A_{ajk}=\omega_{ajk}-2\beta s\omega_{a0l}\epsilon_{jkl}
\ee
The Hodge dual of an antisymmetric tensor $T_{\alpha\beta}$ is defined
by $\ast T_{\alpha\beta}=\frac{1}{2}\epsilon_{\alpha\beta\gamma\delta}
\eta^{\gamma\gamma'} \eta^{\delta\delta'} T_{\gamma'\delta'}$. Since
$\epsilon_{0ijk}=\epsilon_{ijk}$ we can write (\ref{2.112}) in the form
\be \label{2.113}
^{(\beta)}A_{ajk}=\omega_{ajk}-2\beta \ast\omega_{ajk}
\ee
Now an antisymmetric tensor is called (anti)self-dual provided that
$\ast T_{\alpha\beta}=\pm \sqrt{s} T$ with $\sqrt{s}:=i^{[1-s]/2}$
and the (anti)self-dual piece
of any $T_{\alpha\beta}$ is defined by $T^\pm=\frac{1}{2}[T\pm \ast T/\sqrt{s}]$
since $\ast\circ\ast=s$ id. An (anti)self-dual tensor therefore has only three
linearly independent components. This case happens for (\ref{2.113})
provided that either $s=1,\beta=\mp 1/2$ or $s=-1,\beta=\pm i/2$ and {\it in this
case the new connection is just (twice) the (anti)self-dual piece of the
pull-back
to $\sigma$ of the four-dimensional spin-connection}. In all other cases
(\ref{2.113}) is only half of the information needed in order to build a
four-dimensional connection and therefore we do not know how it transforms
under internal boosts. This is, from this perspective, the reason why one
has to gauge fix the boost symmetry of the action (\ref{2.67}) (by the time
gauge $e_\mu^\alpha n^\mu=\delta^\alpha_0$) in order to remove the then
present second
class constraints and to arrive at the present formulation. Obviously, this
is no obstacle, first, since there {\it does exist} a four-dimensional
interpretation even in that case as we just showed and more explicitly
from (\ref{2.76}) and, secondly, since
we are not interested in the transformation properties under spacetime
diffeomorphisms and internal Lorentz transformations
of non-gauge-invariant objects anyway, although from an aesthetic point of
view it would be desirable to have such an interpretation.
\item {\it Reality Conditions}\\
When $\beta$ is real valued $^{(\beta)}A,\;^{(\beta)}E$ are both real valued and
can directly be interpreted as the canonical pair for the phase space of an
$SU(2)$ Yang-Mills theory. If $\beta$ is complex then these variables are
complex valued. However, they cannot be arbitrary complex functions on
$\sigma$ but are subject to the following reality condtions
\be \label{2.114}
^{(\beta)} E/\beta =\overline{^{(\beta)} E/\beta},\;
[^{(\beta)} A-\Gamma]/\beta =\overline{[^{(\beta)} A-\Gamma]/\beta}
\ee
where $\Gamma=\Gamma(^{(\beta)})$ is a non-polynomial, not even analytic
function. These reality conditions guarantee that there is no doubling of the
number of degrees of freedom and one can check explicitly that they are
preserved by the Hamiltonian flow of the constraints provided that
$\Lambda^j$, the Lagrange multiplier of the Gauss constraint, is real valued.
Thus, only $SU(2)$ gauge transformations are allowed but not general
$SL(2,\Co)$ transformations. The reality conditions are difficult to implement
in the quantum theory directly as already mentioned above.
\item {\it Simplification of the Hamiltonian Constraint}\\
The original motivation to introduce the new variables was that for the
quantization of general relativity it seemed mandatory to
simplify the algebraic sructure of the Hamiltonian constraint which for
$s=-1$ requires $\beta=\pm i/2$ since then the constraint becomes
polynomial after multiplying by a factor proportional to $\sqrt{\det(q)}$.
On the other hand, then the reality conditions become non-polynomial.
Finally, if one wants polynomial reality conditions then
one must have $\beta$ real and then the Hamiltonian constraint is still
complicated. Thus it becomes questionable what has been gained. The
answer is the following : For any choice of $\beta$ one can actually
make both the Hamiltonian constraint {\it and} the reality conditions
polynomial by multiplying by a sufficiently high power of $\det(q)$.
But the real question is whether the associated classical functions
will become well-defined operator-valued distributions in quantum theory
while keeping background independence. As we will see in later sections,
the Hilbert space that we will choose does not support any quantum versions
of these functions rescaled by powers of $\det(q)$ and there are abstract
arguments that suggest that this is a representation independent statement.
The requirement seems to be that the Hamiltonian constraint is a scalar
density of weight one and thus we must keep the factor of $1/\sqrt{\det(q)}$
in (\ref{2.107}) whatever the choice of $\beta$ and therefore the
motivation for polynomiality is lost completely. The motivation to
have a connection formulation rather than a metric formulation is then
that {\it that one can go much farther in the background independent
quantization programme provided that $\beta$ is real}. For instance,
a connection formualtion enables us to employ the powerful arsenal of
techniques that have been developed for the canonical quantization of
Yang-Mills theories, specifically Wilson loop techniques.
\item {\it Choice of Fibre Bundle}\\
In the whole exposition so far we have assumed that we have a trivial
principal $SU(2)$ bundle over $\sigma$ (see e.g. \cite{33a} for a
good textbook on fibre bundle theory and section \ref{sa})
so that we can work with a globally defined
connection potential and globally defined electric field
$^{(\beta)}A,\;^{(\beta)}E$ respectively. What about different bundle
choices ?

Following the notation of appanedix \ref{sa} our situation is that
we are dealing with a principal $SU(2)$ bundle over
$\sigma$ with pull-backs $^{(\beta)}A_I$ by local sections of a connection
and local sections $^{(\beta)}E_I$ of an associated (under the adjoint
representation) vector bundle of two forms and would like to know whether
these bundles are trivial. Since the latter is
built out of the 3-beins we can equivalently look also at the frame bundle
of orthonormal frames in order to decide for triviality. Triviality of
the frame bundle is equivalent to to the triviality of its associated
principal bundle and in turn to $\sigma$ being parallelizable.
But this is automatically the
case for any compact, orientable three manifold provided that $G=SU(2)$,
see \cite{33} paragraph 12, exercise 12-B.
More generally, in order to prove that a principal fibre bundle is trivial
one has to show that the cocycle $h_{IJ}$ of transition functions
between charts of an atlas of $\sigma$ is a coboundary, that is,
its (non-Abelian) \v{C}ech cohomolgy class is trivial. In \cite{33}
one uses a different method, obstruction theory, where triviality
can be reduced to the vanishing of the coefficients
(taking values in the homotopy groups of $G$) of certain cohomology
groups of $\sigma$ related to Stiefel-Whitney classes.

So far we did not make the assumption that $\sigma$ is compact or orientable.
If $\sigma$ is not compact but orientable then one usually requires that
there is a compact subset $B$ of $\sigma$ such that $\sigma-B$ has the
topology of the complement of a ball in $\Rl^3$. Then the result holds in
$B$ and trivially in $\sigma-B$ and thus all over $\sigma$. Thus, compactness
is not essential. If $\sigma$ is not orientable then a smooth nowhere
singular frame cannot exist and the above quaoted result does not hold, there are no
smooth 3-bein fields in this case. In that case we allow non-smooth
3-bein fields, that is, we allow that $\det(e)$ has finite jumps
between $\pm |\det(e)|$ on subsets of $\sigma$ of Lebesgue measure zero
(two surfaces) due to change of sign of one of the three forms $e^j$.
This requires that one works with a fixed trivialization at the
gauge variant level classically. At the gauge invariant level the dependence
on that trivialization disappears, so there is no problem. More specifically,
the constraints
$H,H_a$ as well as the symplectic structure are gauge invariant while
$G_j$ is gauge covariant so that we have independence of the choice of
trivialization again on the constraint surface $G_j$ as expected, we get
equivalence with the ADM formulation.. As we
will see, the choice of the bundle will become completely irrelevant
anyway in the quantum theory.
\item {\it Orientation}\\
So far we did not need to impose any restriction on the orientation of
the $e_a^j$. However, from $E^a_j=e^a_j\det(e)$ we easily obtain in $D=3$ that
$\det(E)=[\det(e)]^2=\det(q)>0$. Thus, classically the $E^a_j$ are not
arbitrary Lie algebra valued vector densities but rather are subject to
the {\it anholonomic} constraint
\be \label{2.115}
\det(E)>0
\ee
One can remove this constraint by multiplying the basic variables by
$\mbox{sgn}(\det(e))$ : $E^a_j:=\sqrt{\det(q)} e^a_j, K_a^j=-2s K_{ab} e^b_j$
(modulo $G_j=0$) so that in fact $\det(E)=\det(q)\mbox{sgn}(\det(e))$
but then the result (\ref{2.94}) fails to hold (the symplectic structure
remains, surprisingly, unchanged), one would get instead
$$
\int d^3x E^a_j\delta\Gamma_a^j
=-\frac{1}{2}\int \mbox{sgn}(\det(e)) \epsilon^{abc} \partial_a(\delta e_b^j
e_c^j) =\frac{1}{4}\int d^3x \partial_a[\mbox{sgn}(\det(e))] \epsilon^{abc}
\delta q_{bc}
$$
which is ill-defined since $0=\epsilon^{abc}\delta q_{bc}$ is
multiplied by the distribution $\partial_a[\mbox{sgn}(\det(e))]$
unless one makes further assumptions classically such as that this
distributional one form has support on a set of measure zero
(motivated by the fact that $q_{ab}$ is smooth.
\end{itemize}
In view of these considerations we will from now on only
consider positive $\beta$ unless otherwise specified.

\subsection{Functional Analytic Description of Classical Connection Dynamics}
\label{s2.4}

In this final subsection of the classical part of this review
we recall some of the elements of the usual infinite
dimensional symplectic geometry that underlies gauge theories.
It turns out to be rather difficult to consistently restrict the space
of classical fields on a given differential manifold in such a way that
the classical action remains functionally differentiable, usually
critically depending on the boundary conditions that one imposes, while
keeping ``enough" solutions of the field equations. Usually the simplest
solutions, those with a high degree of symmetry, are at the verge of lying
outside of the space of fields that the variational principle was based
on. Fortunately, these issues will be not too important for us as the
space of quantum fields tends to be even much larger and generically is of
a distributional kind without leading to any problems. Those issues will
however be of some interest again when we discuss the calssical limit.
We can therefore be brief here and will just sketch some of the main ideas.
The interested reader is referred to the exhaustive treatment in
\cite{32b}.\\

Let $G$ be a compact gauge group, $\sigma$ a $D-$dimensional manifold which
admits a principal $G-$bundle with connection over $\sigma$.
Let us denote the pull-back to $\sigma$ of the connection by
local sections by $A_a^i$
where $a,b,c,..=1,..,D$ denote tensorial indices and $i,j,k,..=1,..,
\dim(G)$ denote indices for the Lie algebra of $G$. We will denote the set
of all smooth connections by $\a$ and endow it with a globally
defined metric topology of the Sobolev kind
\be \label{2.116}
d_\rho[A,A']:=\sqrt{-\frac{1}{N}\int_\sigma d^Dx \sqrt{\det(\rho)(x)}
\mbox{tr}([A_a-A'_a](x)[A_b-A'_b](x))\rho^{ab}(x)}
\ee
where $\mbox{tr}(\tau_i\tau_j)=-N\delta_{ij}$ is our choice of
normalization for the generators of a Lie algebra $Lie(G)$ of rank $N$ and
our conventions are such
that $[\tau_i,\tau_j]=2f_{ij}\;^k\tau_k$ define the structure constants
of $Lie(G)$. Here $\rho_{ab}$ is a fiducial metric on $\sigma$ of
everywhere Euclidean
signature. In what follows we assume that either $D\not=2$ (
for $D=2$, (\ref{2.116}) depends only on the conformal structure
of $\rho$ and cannot guarantee convergence for arbitrary fall-off
conditions on the connections) or that $D=2$ and the fields $A$
are Lebesgue integrable.

Let $F^a_j$ be a Lie algebra valued vector density test field of weight one
and let $f_a^j$ be a Lie algebra valued covector test field. Let,
as before $A_a^j$ be a the pull-back of a connection to $\sigma$ and consider
a vector bundle of electric fields, that is, of Lie algebra valued
vector densities of weight one whose bundle projection to $\sigma$ we denote
by $E^a_i$. We consider the smeared quantities
\be \label{2.117}
F(A):=\int_\sigma d^Dx F^a_i A_a^i\mbox{ and }
E(f):=\int_\sigma d^Dx E^a_i f_a^i
\ee
While both are diffeomorphism covariant it is only the latter which is
gauge covariant, one reason to consider the singular smearing through
holonomies discussed below.
The choice of the space of pairs of test fields $(F,f)\in{\cal S}$
depends on the boundary conditions on
the space of connections and electric fields which in turn depends on the
topology of $\sigma$ and will not be specified in what follows.

We now want to select a subset $\cal M$ of the set
of all pairs of smooth functions $(A,E)$ on $\sigma$ such that (\ref{2.117})
is well defined (finite) for any $(F,f)\in {\cal S}$
and endow it with a manifold structure and a symplectic
structure, that is, we wish to turn it into an infinite dimensional
symplectic manifold.

We define a topology on $\cal M$ through the metric:
\ba \label{2.118}
&& d_{\rho,\sigma}[(A,E),(A',E')]
\\
&:=&\sqrt{-\frac{1}{N}\int_\sigma d^Dx
[\sqrt{\det(\rho)} \rho^{ab} \mbox{tr}([A_a-A'_a][A_b-A'_b])+
\frac{\sigma_{ab} \mbox{tr}([E^a-E^{a\prime}][E^b-E^{b\prime}])}
{\sqrt{\det(\sigma)}}]} \nonumber
\ea
where $\rho_{ab},\sigma_{ab}$ are again fiducial metrics on $\sigma$ of
everywhere Euclidean signature. Their fall-off behaviour has to be suited
to the boundary conditions of the fields $A,E$ at spatial infinity.
Notice that the metric (\ref{2.118}) is gauge invariant (and thus globally
defined, i.e. is independent of the choice of local section) and
diffeomorphism covariant and that
$d_{\rho,\sigma}[(A,E),(A',E')]=
d_\rho[A,A']+d_\sigma[E,E']$ (recall (\ref{2.1})).

Now, while the space of electric fields in Yang-Mills theory is a vector
space, the
space of connections is only an affine space. However, as we have also
applications in general relativity with asymptotically Minkowskian
boundary conditions in mind, also the space of electric fields
will in general not be a vector space.
Thus, in order to induce a norm
from (\ref{2.118}) we proceed as follows: Consider an atlas of $\cal M$
consisting only of $M$ itself and
choose a fiducial background connection and electric field $A^{(0)},
E^{(0)}$ (for instance $A^{(0)}=0$). We define the global chart
\be \label{2.119}
\varphi\; :\; {\cal M}\mapsto {\cal E};\; (A,E)\mapsto (A-A^{(0)},E-E^{(0)})
\ee
of $\cal M$ onto the vector space of pairs $(A-A^{(0)},E-E^{(0)})$. Obviously,
$\varphi$ is a bijection. We topologize $\cal E$ in the norm
\be \label{2.120}
||(A-A^{(0)},E-E^{(0)})||_{\rho\sigma}:=
\sqrt{d_{\rho\sigma}[(A,E),(A^{(0)},E^{(0)})]}
\ee
The norm (\ref{2.120}) is of course no longer gauge and diffeomorphism
covariant since the fields $A^{(0)},E^{(0)}$ do not transform, they are
background fields. We need it, however, only in order to encode the
fall-off behaviour of the fields which are independent of gauge -- and
diffeomorphism covariance.

Notice that the metric induced by this norm coincides with (\ref{2.118}).
In the terminology of weighted Sobolev spaces the completion of $\cal E$ in
the norm (\ref{2.120}) is called the Sobolev space
$H^2_{0,\rho}\times H^2_{0,\sigma^{-1}}$, see e.g. \cite{31}. We will call
the completed space $\cal E$ again and its image under
$\varphi^{-1}$, $\cal M$ again (the dependence of $\varphi$ on
$(A^{(0)},E^{(0)})$ will be suppressed). Thus, $\cal E$ is a
normed, complete
vector space, that is, a Banach space, in fact it is even a Hilbert space.
Moreover, we have modelled $\cal M$ on the Banach space $\cal E$, that is,
$\cal M$ acquires the structure of a (so far only topological) Banach
manifold.
However, since $\cal M$ can be covered by a single chart and the identity map
on $\cal E$ is certainly $C^\infty$, $M$ is actually a smooth manifold.
The advantage of modelling $\cal M$ on a Banach manifold is that one can
take over almost all the pleasant properties from the finite dimensional
case to the infinite dimensional one (in particular, the inverse function
theorem).

Next we study differential geometry on $\cal M$ with the standard techniques
of calculus on infinite dimensional manifolds (see e.g. \cite{32}).
We will not repeat all the technicalities of the definitions involved,
the interested reader is referred to the literature quoted.
\begin{itemize}
\item[i)]
A function $f:\; {\cal M}\mapsto \Cl$ on $\cal M$ is said to be differentiable
at $m$
if $g:=f\circ\varphi^{-1}:\;{\cal E}\mapsto \Cl$ is differentiable
at $u=\varphi(m)$,
that is, there exist {\it bounded} linear operators $Dg_u, Rg_u:\;\cal
E\mapsto\Cl$ such that
\be \label{2.122}
g(u+v)-g(u)=(Dg_u)\cdot v+(Rg_u)\cdot v \mbox{ where }
\lim_{||v||\to 0} \frac{|(Rg_u)\cdot v|}{||v||}=0\;.
\ee
$Df_m:=Dg_u$ is called the functional derivative of $f$ at $m$ (notice
that we identify, as usual, the tangent space of $\cal M$ at $m$ with
$\cal E$). The definition extends in an obvious way to the case where
$\Cl$ is replaced by another Banach manifold. The equivalence class
of functions differentiable at $m$ is called the germ $G(m)$ at $m$.
Here two functions are said to be equivalent provided they coincide in
a neighbourhood containing $m$.
\item[ii)] In general, a tangent vector $v_m$ at $m\in {\cal M}$ is an
equivalence class of triples
$(U,\varphi,v_m)$ where $(U,\varphi)$ is a chart of the atlas of $\cal M$
containing $m$ and $v_m\in {\cal E}$.
Two triples are said to be equivalent provided that
$v'_m=D(\varphi'\circ\varphi^{-1})_{\varphi(m)}\cdot v_m$. In our case
we have only one chart and equivalence becomes trivial. Tangent vectors
at $m$ can be considererd as derivatives on the germ $G(m)$ by defining
\be \label{2.123}
v_m(f):=(Df_m)\cdot v_m=(D(f\circ\varphi^{-1})_{\varphi(m)})\cdot v_m
\ee
Notice that the definition depends only on the equivalence class and not
on the representative. The set of vectors tangent at $m$ defines the tangent
space $T_m({\cal M})$ of ${\cal M}$ at $m$.
\item[iii)] The cotangent space $T'_m({\cal M})$ is the topological dual
of $T_m({\cal M})$, that is, the set of {\it continuous} linear functionals
on
$T_m({\cal M})$. It is obviously isomorphic with ${\cal E}'$, the topological
dual of $\cal E$. Since our model space $\cal E$ is reflexive (it is a
Hilbert space) we can naturally identify tangent and cotangent space
(by the Riesz lemma) which also makes the definition of contravariant
tensors less ambiguous. We will, however, not need them for what follows.
Similarly, one defines the space of
$p-$covariant tensors at $m\in {\cal M}$ as the space of {\it continuous}
$p-$linear forms on the $p-$fold tensor product of $T_m({\cal M})$.
\item[iv)] So far the fact that $\cal E$ is a Banach manifold was not very
crucial. But
while the tangent bundle $T({\cal M})=\cup_{m\in {\cal M}} T_m({\cal M})$ carries a
natural manifold structure modelled on ${\cal E}\times {\cal E}$
for a general Fr\'echet space (or even locally convex space) $\cal E$
the cotangent bundle $T'(M)=\cup_{m\in {\cal M}} T'_m({\cal M})$
carries a manifold structure only when $\cal E$ is a Banach space as one
needs the inverse function theorem to show that each chart is not only a
differentiable bijection but that also its inverse is differentiable.
In our case again there is no problem. We define differentiable vector
fields and $p-$covariant tensor fields as cross sections of the
corresponding fibre bundles.
\item[v)] A differential form of degree $p$ on ${\cal M}$ or $p-$form is a
cross section
of the fibre bundle of completely skew continuous $p-$linear forms.
Exterior product, pull-back, exterior differential, interior product
with vector fields and Lie derivatives are defined as in the finite
dimensional case.
\end{itemize}
\begin{Definition} \label{def2.11}
Let $\cal M$ be a differentiable manifold modelled on a Banach space $\cal E$.
A weak respectively strong symplectic structure $\Omega$ on $M$ is a closed
2-form such that for all $m\in {\cal M}$ the map
\be \label{2.124}
\Omega_m:\; T_m({\cal M})\mapsto T'_m({\cal M});\; v_m\to \Omega(v_m,.)
\ee
is an injection respectively a bijection.
\end{Definition}
Strong symplectic structures are more useful because weak symplectic
structures do not allow us to define Hamiltonian vector fields
through the definition $DL+i_{\chi_L}\Omega=0$ for differentiable
$L$ on $M$ and Poisson brackets through
$\{f,g\}:=\Omega(\chi_f,\chi_g)$, see e.g. \cite{32a} for details.

Thus we define finally a strong symplectic structure for our case by
\be \label{2.125}
\Omega((f,F),(f',F')):=\int_\Sigma d^Dx [F^a_i f^{i\prime}_a
-F^{a\prime}_i f_a^i](x)
\ee
for any $(f,F),(f',F')\in {\cal E}$.
To see that $\Omega$ is a strong symplectic structure
we observe first that the integral kernel of $\Omega$ is constant so that
$\Omega$ is clearly exact, so, in particular, closed. Next, let
$\theta\in {\cal E}'\equiv {\cal E}$. To show that $\Omega$ is a bijection
it suffices
to show that it is a surjection (injectivity follows trivially from
linearity). We must find $(f,F)\in {\cal E}$ so that $\theta(.)=
\Omega((f,F),.)$ for any one-form $\theta$ . Now by the Riesz lemma there exists
$(f_\theta,F_\theta)\in {\cal E}$ such that $\theta(.)=
<(f_\theta,F_\theta),.>$ where $<.,.>$ is the inner product induced
by (\ref{3.4}). Comparing (\ref{2.118}) and (\ref{2.125}) we see that we have
achieved our goal provided that the functions
\be \label{2.129}
F^a_i:=\rho^{ab}\sqrt{\det(\rho)}f_{b\theta}^i,\;
f_a^i:=-\frac{\sigma_{ab}}{\sqrt{\det(\rho)}}F^b_{i\theta}
\ee
are elements of $\cal E$. Inserting the definitions we see that this will
be the case provided that the functions
$\rho^{cd}\sigma_{ca}\sigma_{db}/\sqrt{\det(\rho)}$ and
$\det(\rho)\sigma_{cd}\rho^{ca}\rho^{db}/\sqrt{\det(\sigma)}$
respectively fall off at least as
$\sigma_{ab}/\sqrt{\det(\sigma)}$ and
$\rho^{ab}\sqrt{\det(\rho)}$ respectively. In physical
applications these metrics are usually chosen to be of the form
$1+O(1/r)$ where $r$ is an asymptotical radius function so that these
conditions are certainly satisfied.
Therefore, $(f,F)\in {\cal E}$ and our small lemma is established.

Let us compute the Hamiltonian vector field of a function $L$ on our $\cal M$.
By definition for all $(f,F)\in {\cal E}$ we have at $m=(A,E)$
\be \label{2.130}
DL_m\cdot(f,F)=\int_\Sigma d^Dx [(DL_m)^a_i f_a^i+(DL_m)_a^i F^a_i]
=-\int_\Sigma d^Dx [(\chi_{Lm})^a_i f_a^i-(\chi_{Lm})_a^i F^a_i ]
\ee
thus $(\chi_L)^a_i=-(DL)^a_i$ and $(\chi_L)_a^i=(DL)_a^i$. Obviously,
this defines a bounded operator on $\cal E$ if and only if $L$ is
differentiable. Finally, the Poisson bracket is given by
\be \label{2.131}
\{L,L'\}_m=\Omega(\chi_L,\chi_{L'})=\int_\Sigma d^Dx
[(DL_m)_a^i (DL'_m)^a_i-(DL_m)^a_i (DL'_m)_a^i]
\ee
It is easy to see that $\Omega$ has the symplectic potential $\Theta$,
a one-form on $\cal M$, defined by
\be \label{2.132}
\Theta_m((f,F))=\int_\Sigma d^Dx E^a_i f_a^i
\ee
since
$$
D\Theta_m((f,F),(f',F'))
:=(D(\Theta_m)\cdot(f,F))\cdot (f',F')
-(D(\Theta_m)\cdot(f',F'))\cdot (f,F)
$$
and $DE^a_i(x)_m\cdot (f,F)=F^a_i(x)$ as follows from the definition.

Coming back to the choice of $\cal S$, it will in general be a subspace
of $\cal E$ so that (\ref{3.1}) still converges. We can now compute the
Poisson brackets between the functions $F(A),E(f)$ on $M$ and find
\be \label{2.133}
\{E(f),E(f')\}=\{F(A),F'(A)\}=0,\;\{E(f),A(F)\}=F(f)
\ee
Remark :\\
In physicists' notation one often writes $(DL_m)_a^i(x):=
\frac{\delta L}{\delta A_a^i(x)}$ etc. and one writes the symplectic
structure as
$\Omega=\int d^Dx \;DE^a_i(x)\wedge DA_a^i(x)$.

\newpage

\section{Mathematical Foundations of Modern Canonical Quantum General
Relativity}
\label{s3}

In the previous section we have derived a canonical connection formulation
of classical general relativity. We have emphasized the importance of
$n$-form fields for a background independent quantization of the theory.
In this section we will see that the insistence on background independent
methods results in a Hilbert space that is drastically different from the
usual Fock space employed in perturbative quantum field theory.\\

We begin by sketching the history of the subject:\\
In the previous section we have shown that for $\beta=\pm i,s=-1$
the Hamiltonian constraint greatly simplifies, up to a factor of
$1/\sqrt{\det(q)}$ it becomes a fourth order polynomial in $^\Cl A_a^j,E^a_j$.
In order to find solutions to the quantum constraint one chose a
{\it holomorphic} connection
representation, that is, wave functions are functionals of $^\Cl A$ but not
of $\overline{^\Cl A}$, the connection itself becomes a multiplication
operator while the electric field becomes a functional differential
operator. In formulae for the choice $\beta=-i$,
\be \label{3.1}
(^\Cl \hat{A}_a^j(x)\psi)[^\Cl A]=\;^\Cl A_a^j(x)\psi[^\Cl A] \mbox{ and }
(\hat{E}^a_j(x)\psi)[^\Cl A]=
\ell_p^2 \frac{\delta\psi[^\Cl A]}{\delta ^\Cl A_a^j(x)}
\ee
(notice that $i E/\kappa$ is conjugate to $^\Cl A$, $\ell_p^2=\hbar\kappa$
is the Planck area). With this definition, which is only formal at this
point since one does not know what the functional derivative means without
specifying the function space to which the $^\Cl A$ belong, the canonical
commutation relations
\be \label{3.2}
[^\Cl \hat{A}_a^j(x),^\Cl \hat{A}_b^k(y)]=
[\hat{E}^a_j(x),\hat{E}^b_k(y)]=0,\;
[\hat{E}_a^j(x),^\Cl \hat{A}_b^k(y)]=
\ell_p^2 \delta^a_b \delta_j^k \delta(x,y)
\ee
are formally satisfied. However, the adjointness relations
\be \label{3.3}
(\hat{E}_a^j(x))^\dagger=\hat{E}_a^j(x),\;
^\Cl \hat{A}_a^j(x)+(^\Cl \hat{A}_a^j(x))^\dagger=2\hat{\Gamma}_a^j(x)
\ee
could not be checked because no scalar product was defined with respect to
which  (\ref{3.3}) should hold. Besides simpler mathematical problems
such as domains of definitions of the operator valued distributions
(\ref{3.1}), equation (\ref{3.3}) looks desastrous in view of the
explicit formula (\ref{2.86}) for the spin connection where operator valued
distributions would appear multiplied not only at the same point but also
in the denominator which would be extremely difficult to define if possible
at all and could prevent one from defining a positive definite scalar
product with respect to which the adjointness conditions should hold.

The implementation of the adjointness relations (which one can make
polynomial by multiplying $^\Cl A$ by a sufficiently high power of
the operator corresponding to $\det(q)$) continues to be the major obstacle
with the complex connection formulation even today which, is why
the real connection formulation is favoured at the moment. However,
in these pioneering years at the end of the 90's nobody thought about
using real connections since the simplification of the Hamiltonian constraint
seemed to be the most important property to preserve which is why researchers
postponed the solution of the adjointness relations and the definition
of an inner product to a later stage and focussed first on other problems.
There was no concrete proposal at that time how to do that but the
fact that the complex connection $^\Cl A=\Gamma-i K$ is reminecent of the
harmonic oscillator variable $z=x-ip$ made it plausible that one could
possibly make use of the
technology known from geometric quantization concerning complex
K\"ahler polarizations \cite{18} and the relevant Bargmann-Segal
transformation theory. A concrete proposal in terms of the phase space
Wick rotation transformation mentioned earlier appeared only later
in \cite{IV}
but until today these ideas have not been mathematically rigorously
implemented.

To speak with the critics of the connection variable
approach\footnote{Quotation from a comment given by Karel Kucha\v{r}
to the author just before his talk at the meeting ``Quantum Gravity in the
Southern Cone", Punta del Este, Uruguay, 1996.}
``The new variable people have
a credit card that is called ``Adjointness Relations". Whenever they
meet a problem that they cannot solve, they charge the credit card. But one
day they must pay the price for their charges and I wonder what will happen
then".

If one would always first worry about the potential problems
and not sometimes close one's eyes and push foreward anyway, then progress
would never have been made in physics. There was a mutitude of results
that one could obtain by formal manipulations even in absence of an inner
product. The most important observation at that time, in the opinion of the
author, is the discovery of the importance of the use of holonomy variables
also known as Wilson loop functions. We will drop the superscripts
$\beta, \Cl$ in what follows.

Already in the early 80's Gambini et al \cite{39} pointed out the
usefulness of
Wilson-loop functions for the canonical quantization of Yang-Mills theory.
Given a directed loop (closed path) $\alpha$ in $\sigma$ and a
G-connection $A$ for
some gauge group $G$ one can consider the holonomy $h_\alpha(A)$ of $A$
along $\alpha$. The holonomy of a connection is abstractly defined via
principal fibre bundle theory, but physicists prefer the formula
$h_\alpha(A):={\cal P}\exp(\oint_\alpha A)$ where $\cal P$ stands for
path-ordering the power expansion of the exponential in such a way that
the connection variables are ordered from left to right
with the parameter along the loop on which they depend increasing.
We will give a precise definition later on.
The connection can be taken in any representation of $G$ but we will be
mostly concerned with $G=SU(2)$ and will choose the fundamental
representation (in case of $G=SL(2,\Cl)$ one chooses one of its two
fundamental representations). The Wilson loop functions are then given by
\be \label{35}
T_\alpha(A):=\mbox{tr}(h_\alpha(A))
\ee
where tr denotes the corresponding trace. The importance of such Wilson
loop functions is that, at least for compact groups, one knows that they
capture the full gauge invariant information about the connection
\cite{40}. For the case at hand, $SL(2,\Cl)$, an independent proof exists
\cite{41}.

After the introduction of the new variables which display
general relativity as a special kind of Yang-Mills theory, Jacobson,
Rovelli and Smolin independently rediscovered and applied Gambini's ideas
to canonical quantum gravity
\cite{42}. Since the connection representation was holomorphic, one needed
only one of the fundamental represenations of $SL(2,\Cl)$ (and not its
complex conjugate).

The author does not want to go very much into details
about the rich amount of formal and exact results that were obtained by
working with
these loop variables before 1992 but just list the most important ones.
An excellent review of these issues is contained in the book by Gambini and
Pullin \cite{44} which has become the standard introductory reference
on the loop representation.
\begin{itemize}
\item[1)] {\it Formal Solutions to the Hamiltonian Constraint in the
Connection Representation}\\
By ordering the operators $\hat{E}$ to the right in the
quantization of the rescaled density weight two operator corresponding to
$\tilde{H}$, one can show \cite{42} that formally
$\hat{\tilde{H}}T_\alpha=0$ for every non-intersecting smooth loop $\alpha$
(see also \cite{43} for an extension to more complicated loops).
The formal character of this argument is due to the fact that this is a
regulated calculation where in the limit as the regulator is removed one
multiplies zero by infinity. An important role plays the notion of a
so-called ``area-derivative".
\item[2)] {\it Loop Transform and Knot Invariants}\\
Since the diffeomorphism constraint maps a Wilson loop function to a
Wilson loop function for a diffeomorphic loop one immediately sees
that knot invariants should play an important role. Let $\mu$ be a
diffeomorphism invariant measure on some space of connections, $\alpha$
a loop and $\psi$ any state. One can then define a loop transformed state
by $\psi'(\alpha):=\int d\mu(A)\overline{T_\alpha(A)}\psi(A)$.
The state $\Psi=1$ is
annihilated by the diffeomorphism constraint if we define the action of an
operator
$\hat{O}'$ in this loop representation by
$(\hat{O}'\psi')(\alpha):=\int d\mu(A)
\overline{(\hat{O}T_\alpha)(A)}\psi(A)$ where $\hat{O}$ is its action in
the connection representation. Likewise one sees, at least formally, that
if $\alpha$ is a smooth non-self-intersecting loop then
$\psi'(\alpha)$ is annihilated by the Hamiltonian constraint. Of course,
again this is rather formal because a suitable diffeomorphism
invariant measure $\mu$ was not known to exist.
\item[3)] {\it Chern-Simons Theory}\\
If one considers, in particular, the loop transform
with respect to the formal measure given by Lebesgue measure times
the exponential of $i/\lambda$ times the Chern-Simons action where
$\lambda$ is the cosmological constant then one can
argue to obtain particular knot invariants related to the Jones-polynomial
\cite{44,45} the coefficients of which seem to be formal solutions to
the Hamiltonian constraint in the loop representation with a cosmological
term: Since
the exponential of the Chern-Simons action is also a formal solution to
the Hamiltonian constraint with a cosmological term in the connection
representation \cite{46} with momenta ordered to the left one obtains
solutions to the Hamiltonian constraint (provided a certain formal
integration by parts formula holds) which correspond to
arbitrary, possibly intersecting, loops.
\item[4)] {\it Commutators}\\
Also commutators of constraints were studied formally in the loop
representation reproducing the Poisson algebra up to quantities which
become singular as the regulator is removed (see \cite{44}). These singular
coefficients
will later be seen to come from the fact that $\tilde{H}$ is a density of
weight
two rather than one. Such singularities must be removed but this could be
done for $\tilde{H}$ only by breaking diffeomorphism invariance
which is unacceptable in
quantum gravity. We will come back to this point later.
\item[5)] {\it Model Systems}\\
One could confirm the validity of the connection representation
in exactly solvable model systems such as the familiar mini -- and
midisuperspace models based on Killing -- or dimensional reduction
for which the reality conditions
can be addressed and solved quantum mechanically \cite{38}.
\end{itemize}
These developments in the years 1987-92 confirmed that using Wilson
loop functions was something extremely powerful and a rigorous quantization
of the theory should be based on them. Unfortunately, all the
nice results obtained so far in the full theory, especially concerning
the dynamics as, e.g.,
the existence of solutions to the constraints, were only formal because
there was no Hilbert
space available which would enable one to say in which topology certain
limits might exist or not.

The time had come to invoke rigorous functional analysis into the
approach. Unfortunately, this was not possible so far for quantum theories
of connections for non-compact gauge groups such as $SL(2,\Cl)$ but only for
arbitrary compact gauge groups. The motivation behind pushing these
developments anyway at that time had been, again, that by using
Bargmann-Segal
transformation theory one would be able to transfer the results obtained
to the physically interesting case. Luckily, due to the results of
\cite{37} one could avoid this aditional step and make the results
of this section directly available for Lorentzian quantum gravity,
although in the real connection formulation rather than the complex one.
\begin{itemize}
\item[i)] {\it 1992: Quantum Configuration Space}\\
The first functional analytic ideas appeared in the seminal paper
by Ashtekar and Isham \cite{47} in which they
constructed a quantum configuration space of distributional connections
$\ab$ by using abstract Gel'fand -- Naimark -- Segal (GNS) theory for
Abelean $C^\ast$ algebras, see appendix \ref{se}. In quantum field theory it is generic
that the measure underlying the scalar product of the theory is supported on
a distributional extension of the classical configuration space and therefore
it was natural to look for something similar, although in a background
independent context. Rendall \cite{47a} was able to show that the classical
configuration space of smooth connections $\a$ is topologically densely
embedded into $\ab$.
\item[ii)] {\it 1993 -- 1994: Measure Theory, Projective Techniques}\\
Ashtekar and Lewandowski \cite{47b} then succeeded
in providing $\ab$ with a $\sigma$-algebra of measurable subsets
of $\ab$ and giving a cylindrical defintion of a measure $\mu_0$ which is
invariant
under $G$ gauge transformations and invariant under the spatial
diffeomorphisms of Diff($\sigma$). In \cite{47c}
Marolf and Mour\~ao established that this cylindrically defined measure
has a unique $\sigma$-additive extension to the just mentioned
$\sigma-$algebra. Moreover, they proved that,
expectedly, $\a$ is contained in a measurable subset of $\ab$ of measure
zero and introduced projective techniques into the framework.
In \cite{47d} Ashtekar and Lewandowski developed the projective techniques
further and used them in \cite{47e} to set up integral and differential
calculus of on $\ab$.

Also Baez \cite{47k} had constructed different spatially diffeomorphism
invariant measures
on $\ab$, however, they are not faithful (do not induce positive definite
scalar products).
\item[iii)] {\it 1994: Complex Connections and Heat Kernel Measures}\\
The Segal-Bargmann representation in ordinary quantum mechanics on the
phase space $\Rl^2$ is a representation in which wave functions are
holomorphic, square integrable (with respect to the Liouville measure)
functions of the complex variable $z=q-ip\in\Cl$.
One can obtain this representation by heat kernel evolution followed by
analytic continuation from the usual position space representation.
In \cite{70} Hall generalized this unitary, so-called Segal-Bargmann
transformation,
to phase spaces which are cotangent bundles over arbitrary compact
gauge groups based on the observation that a natural Laplace operator
(generator of the heat kernel evolution) exists on such groups. The role
of $\Cl$ is then replaced by the complexification $G^\Cl$ of $G$.
Since it
turns out that the Hilbert space of functions on $\ab$ labelled by a
piecewise analytical loop reduces to $SU(2)^N$ for some finite natural number
$N$ one can just apply Hall's construction to quantum gravity which would
seem to map us from the real connection representation to the complex one.
This was done in \cite{72a} by Ashtekar, Lewandowski, Marolf, Mour\~ao
and Thiemann. The question remained wether the so obtained inner product
incorporates the correct adjointness -- and canonical commutation relations
among the complexified holonomies.
In \cite{IV,71} this was shown not to be the case but at the same time
a proposal was made for how to modify the transform in such a way that the
correct adjointness -- and canonical commutation relations are guaranteed
to hold. This so-called Wick rotation transformation is a special case
of an even more general method, the so-called complexifier method, which
consists in replacing the Laplacian by a more general operator (the
complexifier) and can be utilized, as in the case of quantum gravity,
to keep the algebraic structure of an operator simple while at the same
time trivializing the adjointness conditions on the inner product.
Unfortunately, the Wick rotation generator for quantum gravity is very
complicated which is why there is no rigorous proof to date for the
existence and the unitarity of the proposed transform.
\item[iv)] {\it 1995: Hilbert Space, Adjointness Relations and Canonical
Commutation Relations}\\
In \cite{II} Ashtekar, Lewandowski, Marolf, Mour\~ao
and Thiemann could show that the Hilbert space ${\cal H}_0=L_2(\ab,d\mu_0)$
in fact solves the adjointness -- and canonical commutation relations
for any canonical quantum field theory of connections that is based on
a comapct gauge group provided one represents the connection as a
multiplication operator and the electric field as a functional derivative
operator, in fact, with these actions of the operators the measure
$\mu_0$ on $\ab$ is almost uniquely
selected. The results of \cite{II} demonstrated that the Hilbert space
${\cal H}_0$ provides in fact a physically correct, kinematical representation
for such
theories. Kinematical here means that the elements of this Hilbert space
carry a representation of the constraint operators but are not
annihilated by them, that is, they are not physical (or dynamical) states.
These authors were also able to provide, in the same paper,
the complete set of solutions of the spatial, analytic
diffeomorphism constraint
(labelled by singular (intersecting) knot classes)
plus a physical (with respect to the spatial diffeomorphism constraint)
inner product using
group averaging methods \cite{47f1} (Gel'fand triple techniques)
and thus, as a side result, showed
that the Husain-Kucha\v{r} \cite{67} model is a completely integrable,
diffeomorphism invariant quantum field theory. Group averaging methods
provide a sytematic framework of how to go from the kinematical Hilbert
space to the physical one.
\item[v)] {\it 1995: Loop -- and Connection Representation:
Spin Network Functions}\\
Quite independently,
Rovelli and Smolin as well as as Gambini and Pullin et al had
pushed another representation of the
canonical commutation relations, the so-called loop representation already
mentioned above for which states of the Hilbert space are to be thought
of as functionals of loops rather than connections.
Since the Wilson loop functionals (polynomials of
traces of holonomies) are not linearly independent, they are subject to the
so-called Mandelstam identities, it was mandatory to first find a set of
linearly independent functions. Using older ideas due to Penrose
\cite{47i} Rovelli and Smolin \cite{47h} were able to write down
such loop functionals, later called spin-network functions, that are
labelled by a smooth $SU(2)$ connection. They then introduced an inner
product between these functions by simply {\it defining} them to be
orthonormal. Baez \cite{47j} then
proved that, using that spin-network functions (considered as functionals
of connections labelled by loops) can in fact be extended
to $\ab$, the spin-network functions are indeed orthonormal with respect
to ${\cal H}_0$, moreover, they form a basis, the two Hilbert spaces
defined by Ashtekar and Lewandowski on the one hand and Rovelli and Smolin
are indeed unitarily equivalent. In \cite{47j1} Thiemann proved a
Plancherel theorem, saying that,
expectedly, the loop representation and the connection representation are
like mutual, non-Abelean Fourier transforms (called the loop transform as
mentioned above) of each other where the role of
the kernel of the transform is played by the spin-network functions as
one would intuitively expect because they are labelled by both loops and
connections.
\item[vi)] {\it 1996 -- 1998: Analytical Versus Smooth and Piecewise Linear
Loops}\\
In all these developments it was crucial, for reasons that will be
explained below, that $\sigma$ is an analytic
manifold and that the loops were piecewise analytic. Baez and Sawin
\cite{47l} were able
to transfer much of the structure to the case that the loops are only
piecewise smooth and intersect in a controlled way (a so-called web) and
some of their results were strengthened by
Lewandowski and Thiemann \cite{47m}. In \cite{47n} Zapata introduced the
concept of piecewise linear loops. The motivation for these modifications
was that the analytical category is rather unnatural from a physical
viewpoint although it is a great technical simplification. For instance,
in the smooth category there is no spin network basis any longer. Both
in the analytic and smooth category the Hilbert space is non-separable
after moding out by analytic or smooth diffeomorphisms respectively while
in the piecewise linear category one ends up with a separable Hilbert space.
The motivation for the piecewise linear category is, however,
unclear from a classical
viewpoint (for instance the classical action is not invariant under
piecewise linear diffeomorphisms). In \cite{47m1} arguments were
given, that support the fact that
the (mutually orthogonal, unitarily equivalent) Hilbert spaces labelled by
the continuous moduli that still
appear in the diffeomorphism invariant analytic and smooth category are
superselected. If one fixes the moduli, the Hilbert space becomes separable.
\item[vii)] {\it 1994 -- 2001: Relation with Constructive Quantum (Gauge)
Field Theory}\\
One may wonder whether the techniques associated with $\ab$ can be applied
to ordinary Yang-Mills theory on a background metric. The rigorous
quantization of Yang-Mills theory on Minkowski space is still one of the
major challenges of theoretical and mathematical physics \cite{47n1}. There
is a vast literature on this subject \cite{47n2} and the
most advanced results in this respect are undobtedly due to Balaban et al
which are so difficult to understand ``... that they lie beyond the limits
of human communicational abilities..." \cite{47n3}. Technically the problem
has been formulated in the context of constructive (Euclidean) quantum field
theory
\cite{8} which is geared to scalar fields propagating on Minkowski space.
In \cite{47n5} a proposal for a generalization of the key axioms of the
framework, the so-called Osterwalder-Schrader axioms \cite{47n6}, has been
given by Ashtekar, Lewandowski, Marolf, Mour\~ao and Thiemann. These were
then successfully applied in \cite{47n7} by the same authors to the
completely solvable Yang-Mills theory in two dimensions by making explicit
use of $\ab,\mu_0$ and spin-network techniques which so far had not been
done before although the literature on Yang-Mills theory in two
dimensions is rather vast \cite{47n8}. These results have been refined
by Fleischhack \cite{47n9}. It became clear these axioms
apply only to background independent gauge field theories which is why
it works in two dimensions only (in two dimensions Yang Mills theory
is not background independent but almost: it is invariant under area
preserving diffeomorphisms which turns out to be sufficient for the
constructions to work out). This motivated Ashtekar, Marolf, Mour\~ao
and Thiemann to generalize the Osterwalder-Schrader framework to general
diffeomorphism invariant quantum field theories \cite{47n10}. Surprisingly
the key theorem of the whole approach, the Osterwalder-Schrader
reconstruction theorem that allows to obtain the Hilbert space of
the canonical quantum field theory from the Euclidean one, can be
straightforwardly adapted to the more general context.

One of the Osterwalder-Schrader axioms is the uniqueness of the vacuum
which is stated in terms of the ergodicity property of the underlying
measure with respect to the time translation subgroup of the Euclidean
group (see e.g. \cite{47n11}) which in turn has consequences for the
support properties of the measure. In \cite{47n12} Mour\~ao, Velhinho and
Thiemann analyzed these issues for $\mu_0$ and found ergodicity with
respect to any infinite, discrete subgroup of the diffeomorphism group
which implied a refinement of the support properties established in
\cite{47c}.
\item[viii)] {\it 1999 -- 2001: Categories and Groupoids, Hyphs and
Gauge Orbit Structure of $\ab$}\\
Following an earlier idea due to Baez \cite{47o} Velhinho \cite{47p} gave
a nice categorical and
purely algebraic chracterization of $\ab$ and all the structure that comes
with it without using $C^\ast$ techniques. The technical simplifications
that are involved rest on the concept of a groupoid of piecewise
analytic paths in $\sigma$ rather than (base-pointed) loops.

In \cite{47o1} Fleischhack, motivated by his results in \cite{47n9},
discussed a new notion of ``loop independence" which has the advantage
of being independent of the differentiability category of the graphs
under consideration and in particular includes the analytical and
smooth category. The new
type of collections of loops are called hyphs. A
hyph is a finite collection of piecewise $C^r$ paths together with an
ordering $\alpha\mapsto
p_\alpha$ of its paths $p_\alpha$ where $\alpha$ belongs to some linearly
ordered index set such that $p_\alpha$ is independent of all the
paths $\{p_\beta;\beta<\alpha\}$. Here a path $p$ is said to be
independent of another path $p'$ if there exists a free point $x$ on $p$
(which may be one of its boundary points), that is, there is a
segment of $p$ incident at $x$ which does not overlap with a segment of $p'$
(although $p,p'$ may intersect in $x$). That is, path independence is
based on the germ of a path. In contrast to graphs or webs (collections
of piecewise analytical or smooth paths), a hyph requires an ordering.
Nevertheless one can get as far with hyphs as with webs but not as far as
with graphs.

Fleischhack also investigated the issue of Gribov copies in $\ab$
\cite{47o2} with respect to
$SU(2)$ gauge transformations. It should be noted that fortunately Gribov
copies are no problem in our context: The measure is a probability
measure and the gauge group therfore has finite volume. Integrals over
gauge invariant functions are therefore well-defined.
\item[ix)] {\it 2000 -- 2001: Infinite Tensor Product Extension}\\
Finally, the Hilbert space ${\cal H}_0$ is sufficient for the
applications of quantum general relativity only if $\sigma$ is compact.
In the non-compact case an extension from compactly supported to
non-compactly supported, piecewise analytic paths becomes necessary.
Thiemann and Winkler
\cite{47q} discovered that the framework of the Infinite Tensor Product
of Hilbert spaces, developed by von Neumann more than 60 years ago, is
ideally suited to deal with this problem. In contrast to ${\cal H}_0$
the extended Hilbert space ${\cal H}^\otimes$ is no $L_2$ space any
longer.
\end{itemize}
We notice that all these developments still use a concrete manifold $\sigma$
and that the loops or paths are embedded into it. However, in order to
describe topology change within quantum gravity it would be desirable
to formulate a Hilbert space using non-embedded (algebraic) graphs
\cite{48}. The state of the abstract Hilbert space itself should tell us
into wich $\sigma$'s the algebraic graph on which it is based can be
embedded. For some ideas into that direction in connection with
semiclassical issues see \cite{49}.   \\

This concludes our historical overview over the development of the subject.
In the next section we try to give a modern introduction into the key
structural theorems by combining most of the above cited literature
which means that we will depart from the historical chronology.

We would like to stress at this point from the outset that
the Hilbert space that we will construct in the course of the section
is {\it just one from infinitely many inequivalent
(kinematical) representations
of the abstract algebra of operators}. On the other hand, as we will show
in section \ref{s4}, it has many extremely natural and physically
appealing properties and is at the moment the one that is most studied.
However, one should not forget that there are many other
(kinematical) representations
which is a freedom that we may need to use exploit in later stages of the
development of the theory. For some examples see section \ref{s8}.

\subsection{The Space of Distributional Connections
for Diffeomorphism Invariant Quantum Gauge Theories}
\label{s3.1}

In this section we will follow closely Velhinho \cite{47p}.
For simplicity we stick to the analytic category. For generalization to the
other categories discussed above, please refer to to the literature cited
there. So in what follows, $\sigma$ is an analytic, connected and orientable
$D-$dimensional manifold which is locally compact (every point has a
compact neighbourhood, automatic if $\sigma$ is finite dimensional) and
paracompact (the countable union of compact sets). Generalization to
non-connected and non-orientable $\sigma$ is straightforward.

\subsubsection{The Label Set: Piecewise Analytic Paths}
\label{s3.1.1}

In all that follows we work with connection potentials, thus we assume that
a fixed trivialization of the principal $G-$bundle has been chosen
(upon passage to the gauge invariant sector nothing will depend on that
choice any more).
\begin{Definition} \label{def3.1}  ~~~~~~~~\\
By $\cal C$ we denote the set of continuous, oriented,
piecewise analytic, parameterized, compactly supported curves embedded into
$\sigma$. That is, an element $c\in {\cal C}$ is given as a map
\be \label{3.1a}
c:\;[0,1]\to\sigma;\;t\mapsto c(t)
\ee
such that there is a finite natural number $n$ and a partition
$[0,1]=[t_0=0,t_1]\cup[t_1,t_2]\cup..\cup[t_{n-1},t_n=1]$ and such that
a) $c$ is continuous at $t_k,\;k=1,..,n-1$, b) real analytic in
$[t_{k-1},t_k],\;k=1,..,n-1$ and c) $c((t_{k-1},t_k),\;k=1,..,n-1$ is an
embedded one-dimensional submanifold of $\sigma$. Moreover, there is a
compact subset of $\sigma$ containing $c$.
\end{Definition}
Recall that a differentiable map $\phi:\;M_1\to M_2$ between finite
dimensional manifolds
$M_1,M_2$ is called an immersion when $\phi$ has everywhere rank $\dim(M_1)$.
An immersion need not be injective but when it is, it is called an embedding.
For an embedding, the map $\phi:\;M_1\to \phi(M_1)$ is a bijection and
the manifold structure induced by $\phi$ on $\phi(M_1)$ is given by the
atlas $\{\phi(U_I),\varphi_I\circ\phi^{-1}\}$ where $\{U_I,\varphi_I\}$
is an atlas of $M_1$. This differentiable structure need not be equivalent
to the submanifold structure of $\phi(M_1)$ which is given by the atlas
$\{V_J\cap\phi(M_1),\phi_J\}$ where $\{V_J,\phi_J\}$ is an atlas of $M_2$.
When both differential structures are equivalent (diffeomorphic in the
chosen differentiability category, say $C^r,\;r\in\Nl\cup\{\infty\}\cup
\{\omega\}$ where $\infty,\omega$ denotes smooth and analytic respectively)
the embedding is called regular. The above definition allows a curve to have
self-intersections and self-overlappings so that it is only an immersion,
but on the open intervals $(t_{k-1},t_k)$ a curve $c$ is a regular embedding,
in particular, it does not come arbitrarily close to itself.
\begin{Definition} \label{def3.2}  ~~~~~~~~~~~~\\
i) The beginning point, final point and range of a curve $c\in {\cal C}$
is defined, respectively, by
\be \label{3.2a}
b(c):=c(0),\;f(c):=c(1),\;r(c):=c([0,1])
\ee
ii) Composition $\circ:\;{\cal C}\times{\cal C}\to {\cal C}$ of composable
curves $c_1,c_2\in {\cal C}$ (those with
$f(c_1)=b(c_2)$) and inversion $^{-1}:\;{\cal C}\to {\cal C}$ of
$c\in {\cal C}$ are defined
by
\be \label{3.3a}
(c_1\circ c_2)(t)\left\{:= \begin{array}{cc}
c_1(2t) & t\in[0,\frac{1}{2}] \\
c_2(2t-1) & t\in[\frac{1}{2},1]
\end{array} \right., \;c^{-1}(t):=c(1-t)
\ee
\end{Definition}
Notice that the operations (\ref{3.3}) do not equip $\cal C$ with the
structure of a group for several reasons: First of all, not every two
curves can be composed. Secondly, composition is notassociative because
$(c_1\circ c_2)\circ c_3,\;c_1\circ (c_2)\circ c_3)$ differ by a
reparametrization. Finally, the retraced curve $c\circ c^{-1}$ is not
really just given by $b(c)$ so that $c^{-1}$ is not the inverse of $c$
and anyway there is no natural ``identity" curve in $\cal C$.
\begin{Definition} \label{def3.3} ~~~~~~~~~\\
Two curves $c,c'\in{\cal C}$ are said to be equivalent, $c\sim c'$
if and only if\\
1) $b(c)=b(c'),\;f(c)=f(c')$ (identical boundaries) and \\
2) $c'$ is identical with $c$ up to a combination of a finite number of
retracings and a reparameterization.
\end{Definition}
It is easy to see that $\sim$ defines an equivalence relation on $\cal C$
(reflexive: $c\sim c$, symmetric: $c\sim c'\Rightarrow c'\sim c$,
transitive: $c\sim c',\;c'\sim c^\dprime \Rightarrow c\sim c^\dprime$).
The equivalence class of $c\in {\cal C}$ is denoted by $p_c$ and the
set of equivalence classes is denoted by $\cal P$. In order to
distinguish the equivalence classes from their representative curves
we wil refer to them as paths. As always, the dependence of $\cal P$ on
$\sigma$ will not be explicitly displayed.
The second condition means that
$c'=c_1'\circ \tilde{c}_1'\circ (\tilde{c}_1')^{-1}\circ..\circ
c_{n-1}'\circ \tilde{c}_{n-1}'\circ (\tilde{c}_{n-1}')^{-1}\circ c'_n$
for some finite natural number $n$ and curves $c'_k,\tilde{c}'_l,\;
k=1,..,n,\;l=1,..,n-1$ and that there exists a diffeomorphism
$f:\;[0,1]\to[0,1]$ such that $c\circ f=c_1'\circ..\circ c'_n$.

Definition \ref{def3.3} has the following fibre bundle theoretic
origin (see e.g. \cite{50} and section \ref{sa}):\\
Recall that a connection $\omega$ on a principal $G$ bundle $P$ maybe
defined in terms of local connection potentials $A_I(x)$ over the chart
$U_I$ of an atlas $\{U_I,\varphi_I\}$ of $\sigma$ which are the pull-backs
to $\sigma$ by local sections $s_I^\phi(x):=\phi_I(x,1_G)$ of $\omega$
where $\phi_I:\;U_I\times G\to \pi^{-1}(U_I)$ denotes the system of
local trivializations of $P$ adapted to the $U_I$ and $\pi$ is the
projection of $P$. The holonomy $h_{cI}:=h_{cI}(1)$ of $A_I$ along a curve
in the domain
of a chart $U_I$ is uniquely defined by the differential equation
\be \label{3.4}
\dot{h}_{cI}(t)=h_{cI}(t)A_{Ia}(c(t))\dot{c}^a(t);\;h_{cI}(0)=1_G
\ee
and one may check that under a gauge transformation
\be \label{3.5}
A_I(x)\mapsto A_J(x)=-dh_{JI}(x) h_{JI}(x)^{-1}+\mbox{ad}{_{h_JI}(x)}(A_I(x))
\ee
the holonomy transforms as
\be \label{3.6}
h_{cI}\mapsto h_{cJ}=h_{JI}(b(c))h_{cI}h_{JI}(f(c))^{-1}
\ee
Denote by $\a_P$ the space of smooth connections (abusing the notation by
identifying the collection of potentials with the connection itself)
over $\sigma$ (the dependence on the bundle is explicitly displayed) and
in what follows we will write $h_c(A)$ for the holonomy of $A$ along $c$
understood as an element of $G$ which is possible once a trivialization
has been fixed. We will denote by $A^g:=-dg g^{-1}+\mbox{ad}_g(A)$ a
gauge transformed connection and have
\be \label{3.6a}
h_c^g(A):=h_c(A^g)=g(b(c))h_c(A)g(f(c))^{-1}
\ee
Besides these transformation properties, the holonomy has the following
important algebraic properties:\\
1) $h_{c_1\circ c_2}(A)=h_{c_1}(A) h_{c_2}(A)$,\\
2) $h_{c^{-1}}(A)=h_c(A)^{-1}$\\
as may be easily checked by using the differential equation (\ref{3.4}).
Furthermore, one can verify that the differential equation (\ref{3.4}) is
invariant under reparametrizations of $c$. These three properties
guarantee that $h_c(A)$ does not depend on $c\in{\cal C}$ but only
on the equivalence class $p_c\in{\cal C}$.

One might therefore also have given the following definition
of equivalence of curves:
\begin{Definition}  \label{def3.4} ~~~~~~~~~\\
Two curves $c,c'\in{\cal C}$ are said to be equivalent, $c\sim c'$
if and only if\\
1) $b(c)=b(c'),\;f(c)=f(c')$ (identical boundaries) and \\
2) $h_c(A)=h_{c'}(A)$ for all $A\in \a$.
\end{Definition}
In fact, defintions \ref{def3.4}) and \ref{def3.3} are equivalent
if $G$ is compact and non-Abelean \cite{47m} since then every group element
can be written as a commutator, that is, in the form
$h=h_1 h_2 h_1^{-1} h_2^{-1}$ so that curves of the form
$c_1\circ c_2\circ c_1^{-1}$ is not equivalent with $c_2$. In the Abelean
case, definition \ref{def3.4}
is stronger than definition \ref{def3.3}. In what follows we will work with
definition \ref{def3.3}.

Property 1) of definition \ref{3.4} implies that the functions $b,f$
can be extended to ${\cal C}$ by $b(p_c):=b(c),
f(p_c)=:f(c)$, the right hand sides are independent of the
representative. However, the function $r$ can be extended only
special elements which we will call edges.
\begin{Definition} \label{def3.5}   ~~~~~~~~~~\\
An edge $e\in{\cal P}$ is an equivalence class of a curve $c_e\in{\cal C}$
which is analytic in all of $[0,1]$. In this case $r(e):=r(c_e)$.
\end{Definition}
For an entire analytic curve we may find an equivalent one which is
not entire analytic and contains a retracing. However, we do not allow such
representatives in the definition of $r(e)$.

It may be checked that $p_{c_1}\circ p_{c_2}:=p_{c_1\circ c_2}$
and $p_c^{-1}:=p_{c^{-1}}$ are well-defined.
The advantage of dealing with paths $\cal P$ rather than curves is
that we now have almost a group structure since composition becomes
associative and the path $p_c\circ p_c^{-1}=b(p_c)$ is
trivial (stays at its beginning point). However, we still do not have
a natural identity element in $\cal P$ and not all of its elements
can be composed. The natural structure behind this is that of a
{\it groupoid}. Let us recall the slightly more definition of a category.
\begin{Definition} \label{def3.6} ~~~~~~~~~\\
i)\\
A category $\cal K$ is a class (in general, more general than a set), the
members of which
are called objects $x,y,z,..$, together with a collection $M({\cal K})$ of
sets hom$(x,y)$ for each ordered pair of objects $(x,y)$, the members
of which are called morphisms. Between the sets of morphisms
there is defined a composition operation
\be \label{3.7}
\circ:\;\mbox{hom}(x,y)\times \mbox{hom}(y,z)\to \mbox{hom}(x,z);
(f,g)\mapsto f\circ g
\ee
which satisfies the two following rules:\\
a) Associativity: $f\circ (g\circ h)=(f\circ g)\circ h$ for all
$f\in \mbox{hom}(w,x),g\in \mbox{hom}(x,y),h\in \mbox{hom}(y,z)$,\\
b) Identities: For every $x\in {\cal K}$ there exists a unique element
id$_x\in \mbox{hom}(x,x)$ such that for all $y\in {\cal K}$ we have
$\mbox{id}_x\circ f=f$ for all $f\in \mbox{hom}(y,x)$ and
$f\circ\mbox{id}_x=f$ for all $f\in \mbox{hom}(x,y)$.\\
ii)\\
A subcategory ${\cal K}'\subset{\cal K}$ is a category which contains
a subclass of the class of objects in ${\cal K}$ and for each pair of
objects $(x,y)$ in ${\cal K}'$ we have for the set of morphisms
$\mbox{hom}'(x,y)\subset\mbox{hom}(x,y)$.\\
iii)\\
A morphism $f\in \mbox{hom}(x,y)$ is called an isomorphism provided
there exists $g\in \mbox{hom}(y,x)$ such that
$f\circ g=\mbox{id}_y,\;g\circ f=\mbox{id}_x$.\\
iv)\\
If ${\cal K}_1,\;{\cal K}_2$ are categories with collections of
sets of morphisms $M({\cal K}_1),\;M({\cal K}_2)$ respectively, then
a map $F:\;[{\cal K}_1,M({\cal K}_1)]\to[{\cal K}_2,M({\cal K}_2)]$
is called a covariant [contravariant] functor, also denoted by
$F_\ast\;[F^\ast]$, provided that the algebraic
structures are preserved, that is\\
1) $f\in\mbox{hom}(x,y)\;\Rightarrow\; F(f)\in \mbox{hom}(F(x),F(y))
\;[\mbox{hom}(F(y),F(x))]$ \\
2) $F(f\circ g)=F(f)\circ F(g) \;[F(g)\circ F(f)]$\\
3) $F(\mbox{id}_x)=\mbox{id}_{F(x)}$.\\
v)\\
A category in which every morphism is an isomorphism is called a groupoid.
\end{Definition}
This definition obviously applies to our situation with the following
identifications:\\
Category: $\sigma$.\\
Objects: points $x\in\sigma$.\\
Morphisms: paths between points $hom(x,y):=\{p\in {\cal P};
b(p)=x,\;f(p)=y\}$. Obviously, every morphism is an isomorphism. \\
Collection of sets of morphisms : all paths $M(\sigma)={\cal P}$\\
Composition: composition of paths $p_{c_1}\circ p_{c_2}=
p_{c_1\circ c_2}$\\
Identities: $\mbox{id}_x=p\circ p^{-1}$ for any
$p\in {\cal P}$ with $b(p)=x$.\\
We wil call this category $\sigma$ the category of points and paths and
denote it synonymously by $\cal P$ as well.\\
Subcategories: $\gamma\subset{\cal P}$ consisting of a subset of $\sigma$
as the set of objects and for each two such objects $x,y$ a subset
$\mbox{hom}'(x,y)\subset\mbox{hom}(x,y)$.

It is clear that every path is a composition of edges, however, $\cal P$
is not freely generated by edges (free of algebraic relations among edges)
because the composition $e\circ e'$ of two edges $e,e'$ defined
as the equivalence class of entire analytic curves $c_e,c_{e'}$ which are
analytic continuations of each other defines a new edge $e^\dprime$ again.
Notice that hom$(x,y)\not=\emptyset$ for any $x,y\in\sigma$ because we
have assumed that $\sigma$ is connected, one says that $\cal P$ is connected.
Moreover, hom$(x,x)$ is actually a group with the identity element id$_x$
being given by the trivial path in the equivalence class of the curve
$c(t)=x,\;t\in[0,1]$. The groups hom$(x,x)$ are all isomorphic: Fix
an arbitrary path $p_{xy}\in\mbox{hom}(x,y)$, then hom$(x,x)=p_{xy}\circ
\mbox{hom}(y,y)\circ p_{xy}^{-1}$.
\begin{Definition} \label{def3.7}   ~~~~~~~~~~\\
Fix once and for all $x_0\in\sigma$. Then ${\cal Q}:=\mbox{hom}(x_0,x_0)$ is
called the hoop group in the literature.
\end{Definition}
The name ``hoop" is an acronym for ``holonomical equivalence class of a
loop based at $x_0$". We use the word hoop to distinguish a hoop
(a closed path) from its representative loop (a closed curve).
\begin{Lemma} \label{la3.1} ~~~~~~~~\\
Fix once and for all a system of paths $p_x\in\mbox{hom}(x_0,x)$ with
$p_{x_0}=\mbox{id}_{x_0}$. Then for any $p\in {\cal P}$ there is a unique
$\alpha\in {\cal Q}$ such that
\be \label{3.8}
p=p_{b(p)}^{-1}\circ\alpha\circ  p_{f(p)}
\ee
\end{Lemma}
The proof consists in solving equation (\ref{3.8}) for $\alpha$.
\begin{Lemma} \label{la3.2} ~~~~~~~~~\\
Denote, for any subgroupoid
$l\subset{\cal P}$ containing $x_0$ as an object, by
$\mbox{hom}_l(x_0,x_0)$ the subgroup of $\cal Q$ consisting of hoops
within $\gamma$.

Let ${\cal Q}'$ be any subgroup of $\cal Q$ and let $X\subset\sigma$ be any
subset
containing $x_0$. Then $l:=\{p_x^{-1}\circ\alpha\circ p_y;\;x,y\in
X,\;\alpha\in {\cal Q}'\}$ is a connected subgroupoid of $\cal Q$
($p_x$ the above fixed path system) and
${\cal Q}'=\mbox{hom}_l(x_0,x_0)$.
\end{Lemma}
Proof of Lemma \ref{la3.2}:\\
i) $l$ is a connected subgroupoid:\\
Given $p\in l$ there exist $x,y\in X,\;\alpha\in {\cal Q}'$ such that
$p=p_x^{-1}\circ\alpha\circ p_y$. Thus
$p^{-1}=p_y^{-1}\circ\alpha^{-1}\circ p_x\in l$ since ${\cal Q}'$ is
a subgroup. Also given $p'=p_y^{-1}\circ\beta\circ p_z\in l$ we have
$p\circ p'=p_x^{-1}\circ\alpha\circ\beta\circ p_z\in l$ since
${\cal Q}'$ is a subgroup. $l$ is trivially connected since by
construction every $x\in X$ is connected to $x_0\in X$ through the path
$p_x^{-1}\circ\alpha\circ p_y$ with $y=x_0,\alpha=\mbox{id}_{x_0}$.\\
ii) \\
We have
\be \label{3.9}
\mbox{hom}_l(x_0,x_0)=\{p\in{\cal Q};\;p\in l\}=
\{p_{x_0}^{-1}\circ\alpha\circ p_{x_0};\;\alpha\in{\cal Q}'\}={\cal Q}'
\ee
since $p_{x_0}=\mbox{id}_{x_0}$.\\
$\Box$

\subsubsection{The Topology: Tychonov Topology}
\label{s3.1.2}

We have noticed above that for an element $A\in\a$ its holonomy
$h_c(A)$ (understood as taking values in $G$, subject to a fixed
trivialization)
depends only on $p_c$. To express this we will use the notation
\be \label{3.10}
A(p_c):=h_c(A)
\ee
It follows then that
\be \label{3.11}
A(p\circ p')=A(p)A(p'),\;A(p^{-1})=A(p)^{-1}
\ee
in other words, every $A\in \a_P$ defines a {\it groupoid morphism}.
\begin{Definition} \label{def3.8}  ~~~~~~~\\
Hom$({\cal P},G)$ is the set of all (algebraic, no continuity assumptions)
groupoid morphisms from the set of paths in $\sigma$ into the gauge group.
\end{Definition}
What we have just shown is that $\a$ can be
understood as a subset of $\mbox{Hom}({\cal P},G)$ via the injection
$H:\;\a_P\to\mbox{Hom}({\cal P},G);\;A\mapsto H_A$ where
$H_A(p):=A(p)$. That $H$ is an injection ($H_A=H_{A'}$ implies $A=A'$)
is the content of Giles' theorem \cite{40}  and can easily be understood
from the fact that for a smooth connection $A\in \a$ we have for
short curves $c_\epsilon:\;[0,1]\to \sigma;\;c_\epsilon(t)=c(\epsilon
t),\;0<\epsilon<1$ an expansion of the form
$h_{c_\epsilon}(A)=1_G+\epsilon \dot{c}^a(0)A_a(c(0))+o(\epsilon^2)$
so that $(\frac{d}{d\epsilon})_{\epsilon=0} h_{c_\epsilon}(A)=
\dot{c}^a(0)A_a(c(0))$, that is, by varying the curve $c$ we can recover
$A$ from its holonomy.

We now show that $\a$ is certainly not all of
Hom$({\cal P},G)$, i.e. $H$ is not a surjection, suggesting that
Hom$({\cal P},G)$ is a natural distributional extension of $\a$:\\
First of all, as we have said before, unless $\sigma$ is three dimensional
and $G=SU(2)$ the bundle $P$ is not necessarily trivial and the classical
spaces $\a$ are all different for different bundles. However, the space
Hom$({\cal P},G)$
depends only on $\sigma$ and not on any $P$ which means that it contains
all possible classical spaces $\a$ at once and thus is much larger.
Beyond this union of all the $\a$ it contains distributional elements,
for instance the following: Let $f:\;S^2\to G$ be any map, $x\in\sigma$
any point. Given a path
$p$ choose a representant $c_p$. The curve $c_p$ can pass through $x$ only
a finite number of times, say $N$ times, due to piecewise analyticity (see
below). At the k-th passage denote by $n^\pm_k$ the direction of
$\dot{c}_p(t)$ at $x$ when it enters (leaves) $x$. Then define
$H(p):=[f(-n^-_1)^{-1}f(n^+_1)]..[[f(-n^-_N)^{-1}f(n^+_N)]$ (for $N=0$
defined to be
$1_G$). Notice that a retracing through $x$ does not affect this formula
because in that case $n^+_k=-n^-_k$ and since we are taking only the
direction of a tangent, also reparameterizations do not affect it. It follows
that it depends only on paths rather than curves. It is easy to check
that this defines an element of Hom$({cal P},G)$. It is not of the form
$H_A,\;A\in\a_P$ because $H$ has support only at $x$, it is distributional.
More examples of distributional elements can be found in \cite{47b}.

Having motivated the space Hom$({\cal P},G)$ as a distributional
extension of $\a_P$, the challenge is now to equip this so far only
algebraically defined space with a topology. The reason is that, being
distributional, it is a natural candidate for the support of a quantum
field theory measure as we have stressed before but measure
theory becomes most powerful in the context of topology. In order to define
such a topology, projective
techniques \cite{25} suggest themselves. We begin quite general.
\begin{Definition} \label{def3.9} ~~~~~~~~~\\
i)\\
Let $\cal L$ be some abstract label (index) set. A partial order $\prec$
on $\cal L$ is a relation, i.e. a subset of ${\cal L}\times {\cal L}$,
which is reflexive ($l\prec l$), symmetric
($l\prec l',\;l'\prec l\Rightarrow l=l'$) and transitive
($l\prec l',\;l'\prec l^\dprime\Rightarrow l\prec l^\dprime$).
Not all pairs of elements of $\cal L$ need to be in relation and if they
are, $\cal L$ is said to be linearly ordered.\\
ii)\\
A partially ordered set $\cal L$ is said to be directed if for any
$l,l'\in {\cal L}$ there exists $l^\dprime\in {\cal L}$ such that
$l,l'\prec l^\dprime$.\\
iii)\\
Let $\cal L$ be a partially ordered, directed index set. A projective
family $(X_l,p_{l'l})_{l\prec l'\in {\cal L}}$ consists of sets $X_l$
labelled by $\cal L$ together with surjective projections
\be \label{3.12}
p_{l' l}:\;X_{l'}\to X_l \;\forall\;l\prec l'
\ee
satisfying the consistency condition
\be \label{3.13}
p_{l' l}\circ p_{l^\dprime l'}=p_{l^\dprime l}\;\forall\;l\prec l'\prec
l^\dprime
\ee
iv)\\
The projective limit $\overline{X}$ of a projective family $(X_l,p_{l' l})$
is the subset of the direct product
$X_\infty:=\prod_{l\in {\cal L}} X_l$ defined by
\be \label{3.14}
\overline{X}:=\{(x_l)_{l\in {\cal L}};\; p_{l' l}(x_{l'})=x_l\;\forall\;
l\prec l'\}
\ee
\end{Definition}
The idea to use this definition for our goal to equip Hom$({\cal P},G)$
with a topology is the following: We will readily see that
Hom$({\cal P},G)$ can be displayed as a projective limit. The compactness
of the Hausdorff space $G$ will be responsible for the fact that every
$X_l$ is compact and Hausdorff.
Now on a direct product space (independent of the cardinality of the
index set) in which each factor is compact and Hausdorff one can naturally
define a topology, the so-called Tychonov topology, such that $X_\infty$ is
compact again. If we manage to show that $\overline{X}$ is closed in
$X_\infty$ then $\overline{X}$ will be compact and Hausdorff as well in
the subspace topology (see, e.g. \cite{51}). However, for compact Hausdorff
spaces powerful measure theoretic theorems hold which will enable us to
equip Hom$({\cal P},G)$ with the structure of a $\sigma-$algebra and
to develop measure theory thereon.

In order to apply definition \ref{def3.9} then to our situation, we must
decide on the label set $\cal L$ and the projective family.
\begin{Definition} \label{def3.10}
i)\\
A finite set of edges $\{e_1,..,e_n\}$ is said
to be independent provided
that the $e_k$ intersect each other at most in the points $b(e_k),f(b_k)$.
ii)\\
A finite set of edges $\{e_1,..,e_n\}$ is said to be
algebraically independent provided none of the $e_k$ is a finite
composition of the $e_1,..,e_{k-1},e_{k+1},..,e_n$ and their inverses.\\
iii)\\
An independent set of edges $\{e_1,..,e_n\}$ defines an oriented
graph $\gamma$ by $\gamma:=\cup_{k=1}^n r(e_k)$ where $r(e_k)\subset\gamma$
carries the arrow induced by $e_k$ ($e\cup e':=p_{c_e\cup c_{e'}}$).
From $\gamma$ we can recover
its set of edges $E(\gamma)=\{e_1,..,e_n\}$ as the maximal analytic
segments of $\gamma$ together with their orientations as well as
set of vertices of
$\gamma$ as $V(\gamma)=\{b(e),f(e);\;e\in E(\gamma)\}$.
Denote by $\Gamma^\omega_0$ the set all of all graphs.\\
iv)\\
Given a graph $\gamma$ we denote by $l(\gamma)\subset{\cal P}$ the
subgroupoid generated by $\gamma$ with $V(\gamma)$ as the set of objects
and with the
$e\in E(\gamma)$ together with their inverses and finite compositions
as the set of homomorphisms.
\end{Definition}
Notice that independence of sets of edges implies algebraic independence
but not vice versa (consider independent $e_1,e_2$ with $f(e_1)=b(e_2)$
and define $e_1'=e_2,\;e'_2=e_1\circ e_2$. Then $e'_1,e'_2$ is
algebraically independent but not independent) and that $l(\gamma)$ is
freely generated by the $e\in E(\gamma)$ due to their algebraic
independence. Also, $l(\gamma)$
does not depend on the orientation of the graph since $e_1,..,e_n$ and
$e_1^{s_1},..,e_n^{s_n},\;s_k=\pm 1$ generate the same subgroupoid.
The labels $\omega,0$ in $\Gamma^\omega_0$ stand for ``analytic" and
``of compact support" respectively for obvious reasons.

The following theorem finally explains why it was important to stick with
the analytic, compact category.
\begin{Theorem} \label{th3.1}   ~~~~\\
Let $\cal L$ be the set of all tame subgroupoids $l(\gamma)$ of $\cal P$,
that is, those determined by graphs $\gamma\in\Gamma^\omega_0$. Then the relation
$l\prec l'$ iff $l$ is a subgroupoid of $l'$ equips $\cal L$ with
the structure of a partially ordered and directed set.
\end{Theorem}
Proof of Theorem \ref{th3.1}:\\
Since $l$ is a subgroupoid of $l'$ iff all objects of $l$ are objects of
$l'$ and all morphisms of $l$ are morphisms of $l'$ it is clear that
$\prec$ defines a partial order. To see that $\cal L$ is directed
consider any two graphs $\gamma,\gamma'\in\Gamma^\omega_0$ and consider
$\gamma^\dprime:=\gamma\cup\gamma'$. We claim that $\gamma^\dprime$
has a finite number of edges again, that is, it is an element of
$\Gamma^\omega_0$. For this to be the case it is obviously sufficient
to show that any two edges $e,e'\in{\cal P}$ can only have a finite
number of isolated intersections or they are analytic extensions of
each other. Clearly they are analytic extensions of each other if
$e\cap e'$ is a common finite segment. Suppose then that $e\cap e'$
is an infinite discrete set of points. We may choose parameterizations
of their representatives $c,c'$ such that each of its
component functions $f(t)^a:=e'(t)^a-e(t)^a$ vanishes in at least a
countably infinite number of points $t_m,\;m=1,2,..$. We now show
that for any function $f(t)$ which is real analytic in $[0,1]$ this
implies $f=0$. Since $[0,1]$ is compact there is an accumulation
point $t_0\in [0,1]$ of the $t_m$ (here the compact support of the $c\in
{\cal C}$ comes into play) and we may assume without loss of generality that
$t_m$ converges to $t_0$ and is strictly monotonous. Since $f$ is
analytic we can
write the absolutely convergent Taylor series $f(t)=\sum_{n=0}^\infty f_n
(t-t_0)^n$ (here analyticity comes into play). We show $f_n=0$ by induction
over $n=0,1,..$. The induction start $f_0=f(t_0)=\lim_{m\to\infty}
f(t_m)=\lim_{m\to\infty} 0=0$ is clear.
Suppose we have shown already that $f_0=..=f_n=0$. Then $f(t)=f_{n+1}
(t-t_0)^{n+1}+r_{n+1}(t)(t-t_0)^{n+2}$ where $r_{n+1}(t)$ is
uniformly bounded in $[0,1]$.
Thus $0=f(t_m)/(t_m-t_0)^{n+1}=f_{n+1}+r_{n+1}(t_m) (t_m-t_0)$ for all $m$,
hence $f_{n+1}=\lim_{m\to \infty}[f_{n+1}+r_{n+1}(t_m) (t_m-t_0)]=0$.\\
$\Box$\\
Notice that the subgroupoids $l\in{\cal L}$ also conversely define
a graph up to orientation through its edge generators.

Now that we have a partially ordered and directed index set $\cal L$
we must specify a projective family.
\begin{Definition} \label{def3.11}   ~~~~~~~~\\
For any $l\in {\cal L}$ define $X_l:=\mbox{Hom}(l,G)$ the set of all
homomorphisms from the subgroupoid $l$ to $G$.
\end{Definition}
Notice that for $l=l(\gamma)$ any $x_l\in X_l$ is completely determined
by the group elements $x_l(e),\;e\in E(\gamma)$ so that we have a bijection
\be \label{3.15}
\rho_\gamma:\;X_l\to G^{|E(\gamma)|};\;x_l\mapsto (x_l(e))_{e\in E(\gamma)}
\ee
Since $G^n$ for any finite $n$ is a compact Hausdorff space (here compactness
of $G$ comes into play) in its natural manifold topology we can equip
$X_l$ with a compact Hausdorff topology through the identification
(\ref{3.15}). This topology is independent of the choice of edge generators
of $l$ since any map $(e_1,..,e_n)\mapsto
(e_{\pi(1)}^{s_1},..,e_{\pi(n)}^{s_n})$ for any element $\pi\in S_n$ of
the permutation group of $n$ elements induces a homeomorphism
(topological isomorphism) $G^n\to G^n$.

Next we must define the projections.
\begin{Definition} \label{def3.12}  ~~~~~~~~~\\
For $l\prec l'$ define a projection by
\be \label{3.16}
p_{l' l}:\;X_{l'}\to X_l;\;x_{l'}\mapsto (x_{l'})_{Ýl}
\ee
restriction of the homomorphism $x_{l'}$ defined on the groupoid $l'$
to its subgroupoid $l\prec l'$.
\end{Definition}
It is clear that the projection (\ref{3.16}) satisfies the consistency
condition (\ref{3.13}) since for $l\prec l^\dprime$
we have $(x_{l^\dprime})_{Ýl}=((x_{l^\dprime})_{Ýl'})_l$ for any
intermediate $l\prec l'\prec l^\dprime$. Surjectivity is less obvious.
\begin{Lemma} \label{la3.3} ~~~~~~~~~~\\
The projections $p_{l' l},\;l\prec l'$ are surjective, moreover, they are
continuous.
\end{Lemma}
Proof of Lemma \ref{la3.3}:\\
Let $l=l(\gamma)\prec l'=l(\gamma')$ be given. Since $l$ is a subgroupoid
of $l'$ we may decompose any generator $e\in E(\gamma)$ in the form
\be \label{3.17}
e=\circ_{e'\in E(\gamma')} (e')^{s_{e e'}}
\ee
where $s_{ee'}\in \{\pm 1,0\}$. Notice that $|s_{ee'}|>2$ is not allowed
and that any $e'$ appears at most once in (\ref{3.17})
because $e$ is an edge (cannot overlap itself). \\
Surjectivity:\\
We must show that for any $x_l\in X_l$ there exists an $x_{l'}\in X_{l'}$
such that $p_{l' l}(x_{l'})=x_l$. Since $x_l$ is completely determined
by $h_e:=x_l(e)\in G,\;e\in E(\gamma)$ and
$x_{l'}$ is completely determined
by $h'_{e'}:=x_{l'}(e')\in G,\;e'\in E(\gamma')$ and since $h_e$ could be
any value in $G$, what we have to show is that there exist group elements
$h'_{e'}\in G,\;e'\in E(\gamma')$ such that for any group elements
$h_{e}\in G,\;e\in E(\gamma)$ we have
\be \label{3.18}
h_e=\circ_{e'\in E(\gamma')} h_{e'}^{s_{e,e'}}
\ee
However,
since the $e\in E(\gamma)$ are disjoint up to their boundaries
we have $s_{e e'} s_{e\tilde{e}'}=0$
for any $e'\not=\tilde{e}'$ in $E(\gamma')$ so that we may specify
some $e'(e)\in E(\gamma')$ for any $e\in E(\gamma)$ and the $e'(e)$ are
disjoint up to their boundaries. Since also the $h'_{e'}$ can independently
take any value we may choose $h'_{e'(e)}=h_e,\;h'_{e'}=1_G$ for $e'\not\in
\{e'(e)\}_{e\in E(\gamma)}$.\\
Continuity:\\
Under the identification (\ref{3.15}) the projections are given as maps
\be \label{3.19}
p_{l' l}:\;G^{ÝE(\gamma')Ý}\to G^{ÝE(\gamma)Ý};\;(h'_{e'})_{e'\in E(\gamma')}
\mapsto (\prod_{e'\in E(\gamma')}(h'_{e'})^{s_{ee'}})_{e\in E(\gamma)}
\ee
By definition, a net $(h_k^\alpha)_{k=1}^n$ converges in $G^n$ to
$(h_k)_{k=1}^n$ if an only if every net
$\lim_\alpha(h_k^\alpha)=h_k,\;k=1,..,n$ individually converges
(i.e., $Ý(h_k^\alpha)_{AB}-(h_k)_{AB}Ý\to 0$ for all matrix
elements $AB$). Suppose then that $(h^{\prime\alpha}_{e'})_{e'\in E(\gamma')}$
converges to $(h'_{e'})_{e'\in E(\gamma')}$. By definition, in a Lie group
inversion and finite multiplication are continuous operations. Therefore
$(\prod_{e'\in E(\gamma')}
(h^{\prime\alpha}_{e'})^{s_{ee'}})_{e\in E(\gamma)}$
converges to
$(\prod_{e'\in E(\gamma')}(h'_{e'})^{s_{ee'}})_{e\in E(\gamma)}$
(as one can check also explicitly).\\
$\Box$\\
We can now form the projective limit $\overline{X}$ of the $X_l$.
In order to equip it with a topology we start by providing the direct
product $X_\infty$ with a topology. The natural topology on the direct
product is the Tychonov topology.
\begin{Definition} \label{def3.13}   ~~~~~~~~~~~\\
The Tychonov topology on the direct product $X_\infty=\prod_{l\in{\cal L}}
X_l$
of topological spaces $X_l$ is the weakest topology such that all the
projections
\be \label{3.20}
p_l:\;X_\infty\to X_l;\;(x_{l'})_{l'\in {\cal L}}\mapsto x_l
\ee
are continuous, that is, a net $x^\alpha=(x^\alpha_l)_{l\in {\cal L}}$
converges to $x=(x_l)_{l\in {\cal L}}$ iff $x^\alpha_l\to x_l$
for every $l\in {\cal L}$ pointwise (not necessarily uniformly) in
$\cal L$.
\end{Definition}
We then have the following non-trivial result.
\begin{Theorem}[Tychonov] \label{th3.2} ~~~~~~~~~~~\\
Let $\cal L$ be an index set of arbitrary cardinality and suppose
that for each $l\in {\cal L}$ a compact topological space $X_l$ is given.
Then the direct product space $X_\infty=\prod_{l\in {\cal L}} X_l$
is a compact toplogical space in the Tychonov topology.
\end{Theorem}
An elegant proof of this theorem in terms of universal nets is given in
section \ref{sb} where also other relevant results from general topology
inluding proofs can be found.

Since $\overline{X}\subset X_\infty$ we may equip it with the subspace
topology, that is, the open sets of $\overline{X}$ are the sets
$U\cap \overline{X}$ where $U\subset X_\infty$ is any open set in $X_\infty$.
\begin{Lemma} \label{la3.4}  ~~~~~\\
The projective limit $\overline{X}$ is a closed subset of $X_\infty$.
\end{Lemma}
Proof of Lemma \ref{la3.4}:\\
Let $(x^\alpha):=((x^\alpha_l)_{l\in {\cal L}})$ be a convergent net in
$X_\infty$ such that $x^\alpha:=(x^\alpha_l)_{l\in{\cal L}}\in \overline{X}$
for any $\alpha$. We must show that the limit point $x=(x_l)_{l\in {\cal L}}$
lies in $\overline{X}$. By lemma \ref{la3.3}, the
projections $p_{l'l}:\;X_{l'}\to X_l$ are continuous, therefore
\be \label{3.21}
p_{l'l}(x_{l'})=\lim_\alpha p_{l' l}(x^\alpha_{l'})
=\lim_\alpha x^\alpha_l=x_l
\ee
where the second equality follows from $x^\alpha\in\overline{X}$.
Thus, the point $x\in X_\infty$ qualifies as a point in $\overline{X}$.\\
$\Box$\\
Since closed subspaces of compact spaces are compact in the subspace topology
(see section \ref{sb}) we conclude that $\overline{X}$ is compact in the
subspace topology induced by $X_\infty$.
\begin{Lemma} \label{la3.5} ~~~~~~~~~~\\
Both $X_\infty,\overline{X}$ are Hausdorff spaces.
\end{Lemma}
Proof of Lemma \ref{la3.5}:\\
By assumtion, $G$ is a Hausdorff topological group. Thus $G^n$ for any
finite $n$ is a Hausdorff topological group as well and since $X_l$
is topologically identified with some $G^n$ via \ref{3.15} we see that
$X_l$ is a topological Hausdorff space for any $l\in {\cal L}$.
Let now $x\not=x'$ be points in $X_\infty$. Thus, there is at least one
$l_0\in {\cal L}$ such that $x_{l_0}\not=x'_{l_0}$. Since $X_{l_0}$
is Hausdorff we find disjoint open neighbourhoods $U_{l_0},U'_{l_0}\subset
X_{l_0}$ of $x_{l_0},x'_{l_0}$ respectively. Let
$U:=p_{l_0}^{-1}(U_{l_0}),U':=p_{l_0}^{-1}(U'_{l_0})$. Since the
topology of $X_\infty$ is generated by the continuous functions
$p_l:\;X_\infty\to X_l$ from the topology of the $X_l$, it follows that
$U,U'$ are open in $X_\infty$. Moreover, $U,U'$ are obviously neighbourhoods
of $x,x'$ respectively since $p_l(U)=X_l=p_l(U')$ for any $l\not=l_0$.
Finally, $U\cap U'=\emptyset$ since $p_{l_0}(U\cap U')=U_{l_0}\cap U'_{l_0}
=\emptyset$ so that $U,U'$ are disjoint open neighbourhoods of $x\not=x'$
and thus $X_\infty$ is Hausdorff.

Finally, to see that $\overline{X}$ is Hausdorff, let $x\not=x'$ be
points in $\overline{X}$, then we find respective disjoint open
neighbourhoods $U,U'$ in $X_\infty$ whence $U\cap \overline{X},
U'\cap \overline{X}$ are disjoint open neighbourhoods in $\overline{X}$
by definition of the subspace topology.\\
$\Box$\\
Let us collect these results in the following theorem.
\begin{Theorem} \label{th3.3}  ~~~~~~~~~\\
The projective limit $\overline{X}$ of the spaces
$X_l=\mbox{Hom}(l,G),\;l\in {\cal L}$
where ${\cal L}$ denotes the set of all tame
subgroupoids of ${\cal P}$
is a compact Hausdorff space in the induced Tychonov topology whenever $G$
is a compact Hausdorff topological group.
\end{Theorem}
The purpose of our efforts was to equip Hom$({\cal P},G)$ with a topology.
Theorem \ref{th3.2} now enables us to do this provided we manage to
identify Hom$({\cal P},G)$ with the projective limit $\overline{X}$
via a suitable bijection. Now an elementary exercise is that any point
of Hom$({\cal P},G)$ defines a point in $\overline{X}$ if we define
$x_l:=H_{|l}$ since the projections $p_{l' l}$ encode the algebraic
relations that are induced by asking that $H$ be a homomorphism. That
this map is actually a bijection is the content of the following theorem.
\begin{Theorem} \label{th3.4} ~~~~~~~~~~~\\
The map
\be \label{3.22}
\Phi:\;\mbox{Hom}({\cal P},G)\to\overline{X};\;H\mapsto (H_{|l})_{l\in
{\cal L}}
\ee
is a bijection.
\end{Theorem}
Proof of theorem \ref{th3.3}:\\
Injectivity:\\
Suppose that $\Phi(H)=\Phi(H')$, in other words, $H_{|l}=H'_{|l}$ for any
$l\in {\cal L}$. Thus, if $l=l(\gamma)$ we have $H(e)=H'(e)$ for any
$e\in E(\gamma)$. Since $l$ is arbitrary we find $H(p)=H'(p)$ for any
$p\in {\cal P}$, that is, $H=H'$.\\
Surjectivity:\\
Suppose we are given some $x=(x_l)_{l\in {\cal L}}\in\overline{X}$. We must
find $H_x\in\mbox{Hom}({\cal P},G)$ such that $\Phi(H_x)=x$. Let
$p\in{\cal P}$ be any path, then we can always find a graph $\gamma_p$ such
that $p\in l:=l(\gamma_p)$. We may then define
\be \label{3.23}
H_x(p):=x_{l(\gamma_p)}(p)
\ee
Of course, the map $p\mapsto \gamma_p$ is one to many and therefore the
definition (\ref{3.23}) seems to be ill-defined. We now show that this is
not the case, i.e., (\ref{3.23}) does not depend on the choice of
$\gamma_p$. Thus, let $\gamma'_p$ be any other graph such that $p\in
l':=l(\gamma'_p)$. Since ${\cal L}$ is directed we find $l^\dprime$ with
$l,l'\prec l^\dprime$. But then by the definition of a point $x$ in the
projective limit
\be \label{3.24}
x_l(p)=[p_{l^\dprime l}(x_{l^\dprime})](p)=(x_{l^\dprime})_{|l}(p)
\equiv x_{l^\dprime}(p)\equiv
(x_{l^\dprime})_{|l'}(p)
=[p_{l^\dprime l}(x_{l^\dprime})](p)=x_{l'}(p)
\ee
It remains to check that $H_x$ is indeed a homomorphism. We have
for any $p,p',p\circ p'\in l$ with $f(p)=b(p')$
\be \label{3.25}
H_x(p^{-1})=x_l(p^{-1})=(x_l(p))^{-1}=H_x(p)^{-1} \mbox{ and }
H_x(p\circ p')=x_l(p\circ p')=x_l(p) x_l(p')=H_x(p) H_x(p')
\ee
since $x_l\in \mbox{Hom}(l,G)$.\\
$\Box$
\begin{Definition} \label{def3.14}
The space $\ab:=\mbox{Hom}({\cal P},G)$ of homomorphisms from the set of
piecewise analytical paths into the compact Hausdorff topological group
$G$, identified set-theoretically and
topologically via (\ref{3.22}) with the projective limit $\overline{X}$
of the spaces $X_l=\mbox{Hom}(l,G)$, where $l\in{\cal L}$ runs through
the tame subgroupoids of $\cal P$, is called the space of distributional
connections over $\sigma$. In the induced Tychonov topology inherited
from $\overline{X}$ it is a compact Hausdorff space.
\end{Definition}
Once again it is obvious that the space of distributions $\ab$ does not
carry any sign anymore of the bundle $P$, it depends only on the base
manifold $\sigma$ via the set of embedded paths $\cal P$.

\subsubsection{Gauge Invariance: Distributional Gauge Transformations}
\label{s3.1.3}

The space $\ab$ contains connections (from now on considered as
morphisms ${\cal P}\to G$) which are nowhere continuous as we will see
later on and these turn out to be measure-theoretically much more important
than the smooth ones contained in $\a$. Therefore it is motivated
to generalize also the space of smooth gauge transformations
$\g:=C^\infty(\sigma,G)$ to the space of {\it all} functions
\be \label{3.26}
\gb:=\mbox{Fun}(\sigma,G)
\ee
with no restrictions (e.g. continuity). It is clear that $g\in \g$ may be
thought of as the net $(g(x))_{x\in\sigma}$ and thus $\g$ is just the
continuous infinite direct product $\g=\prod_{x\in \sigma} G$.

The transformation property of $\a$ under $\g$ (\ref{3.6a}) can be understood
as an action $\lambda:\;\g\times \a\to \a;\;(g,A)\to A^g:=\lambda_g(A):=
\lambda(g,A)$ where $A^g(p):=g(b(p))A(p)g(f(p))^{-1}$ for any $p\in {\cal P}$
which we may simply lift to $\ab,\gb$ as
\ba \label{3.27}
\lambda &: &\;\gb\times \ab\to \ab;\;(g,A)\to A^g:=\lambda_g(A):=
\lambda(g,A) \mbox{ where } \nonumber\\
A^g(p)&:=&g(b(p))A(p)g(f(p))^{-1}\;\forall \;
p\in {\cal P}
\ea
Notice that this is really an action, i.e. $A^g$ really is an
element of $\ab=\mbox{Hom}({\cal P},G)$, that is, it satisfies the
homomorphism property
\ba \label{3.28}
A^g(p^{-1}) &=& g(b(p^{-1}))A(p^{-1})g(f(p^{-1}))^{-1}
= g(f(p))A(p)^{-1}g(b(p))^{-1}=(A^g(p))^{_1}
\nonumber\\
A^g(p) A^g(p')&=&
[g(b(p))A(p)g(f(p))^{-1}][g(b(p'))A(p')g(f(p'))^{-1}]
=g(b(p))A(p)A(p')g(f(p'))^{-1}
\nonumber\\
&=&
g(b(p\circ p'))A(p\circ p')g(f(p\circ p'))^{-1}
=A^g(p\circ p')
\ea
because $f(p)=b(p'),b(p)=b(p\circ p'),f(p')=f(p\circ p')$. The action
(\ref{3.27}) is also continuous on $\ab$, that is, for any $g\in\gb$
the map $\lambda_g:\;\ab\to\ab$ is continuous. To see this, let
$(A^\alpha)$ be a net in $\ab$ converging to $A\in \ab$. Then
$\lim_\alpha \lambda_g(A_\alpha))=\lambda_g(A)$ if and only if
$\lim_\alpha p_l(\lambda_g(A_\alpha)))=p_l(\lambda_g(A))$ for any
$l\in {\cal L}$. Identifying $\ab_{|l}$ with some $G^n$ via (\ref{3.15})
and using the bijection (\ref{3.22}) we have for any $p\in l$
\ba \label{3.29}
[p_l(\lambda_g(A^\alpha))](l)&=&[(\lambda_g(A^\alpha))_{|l}](p)=
[\lambda_g(A^\alpha)](p)=g(b(p))A^\alpha(p)g(f(p))^{-1}
\nonumber\\
&=& g(b(p))[p_l(A^\alpha)](p)g(f(p))^{-1}
\ea
Since group multiplication and inversion are continuous in $G^n$ we
easily get
$\lim_\alpha [p_l(\lambda_g(A^\alpha))](l)=[p_l(\lambda_g(A))](l)$
for any $p\in l$, that is,
$\lim_\alpha p_l(\lambda_g(A^\alpha))=p_l(\lambda_g(A))$, thus $\lambda_g$
is continuous for any $g\in\gb$.

Since $\ab$ is a compact
Hausdorff space and $\lambda$ is a continuous group action on $\ab$ it then
follows immediately
from abstract results (see section \ref{sb}) that the quotient space
\be \label{3.30}
\abgb:=\{[A];\;A\in\ab\}\mbox{ where } [A]:=\{A^g;\;g\in \gb\}
\ee
is a compact Hausdorff space in the quotient topology. The quotient topology
on the quotient $\ab/\gb$ is defined as follows: The open sets in
$\ab/\gb$ are precisely those whose preimages under the quotient map
\be \label{3.31}
[]:\;\ab\to\ab/\gb;\;A\mapsto [A]
\ee
are open in $\ab$, that is, the quotient topology is generated by asking
that the quotient map be continuous.

Now as $\gb$ is a continuous direct product of the compact Hausdorff spaces
$G$ it is a compact Hausdorff space in the Tychonov topology by the
theorems proved in section \ref{s3.1.2}. More explicitly, the projective
construction of $\gb$ proceeds as follows: Given $l\in {\cal L}$ with
$l=l(\gamma)$ we define $\gb_l:=\prod_{v\in V(\gamma)} G$ and extend
the surjective projection $p_l:\;\ab\to \ab_l;\;A\mapsto A_{|l}$ to
$p_l:\;\gb\to \gb_l;\;g\mapsto g_{|l}$ and for $l\prec l'$ the
surjective projection
$p_{l'l}:\;\ab_{l'}\to \ab_l;\;A_{l'}\mapsto (A_{l'})_{|l}$ to
$p_{l'l}:\;\gb_{l'}\to \gb_l;\;g_{l'}\mapsto (g_{l'})_{|l}$.
These projections are obviously surjective again because $\gb$ is
actually a direct product of copies of $G$, one for every $x\in \sigma$.

Notice that the projective limit $\gb=\{(g_l)_{l\in {\cal L}};\;
p_{l' l}(g_{l'})=g_l\}$ is a group since
$p_{l' l}(g_{l'} g'_{l'})=g_l g'_l=p_{l' l}(g_{l'}) p_{l' l}(g'_{l'})$
and
$p_{l'l}((g^{-1})_{l'})=(g^{-1})_l=g_l^{-1}=p_{l' l}(g_{l'})^{-1}$
so that actually the $p_{l' l}$ are surjective group homomorphisms.
Since the $\gb_l$ are compact Hausdorff topological groups it follows
that $\gb$ is also a compact Hausdorff topological group.

Summarizing: $\ab/\gb$ is the quotient of two projective limits both of which
are compact Hausdorff spaces.

On the other hand observe that for $l\prec l'$ we have
\be \label{3.32}
p_{l'l}(\lambda_{g_{l'}}(A_{l'}))=\lambda_{g_l}(A_l)
\ee
for any $A\in\ab,g\in\gb$, one says the group action $\lambda$ is
{\it equivariant}. Consider then the quotients
\be \label{3.33}
[\ab_l]_l:=\ab_l/\gb_l:=\{[A_l]_l;\;A_l\in \ab_l\} \mbox{ where }
[]_l:\;\ab_l\to
\ab_l/\gb_l;\;A_l\mapsto [A_l]_l:=\{\lambda_{g_l}(A_l);\;g_l\in \gb_l\}
\ee
Due to the equivariance property for $l\prec l'$
\be \label{3.34}
p_{l' l}([A_{l'}]_{l'})
=\{p_{l' l}(\lambda_{g_{l'}}(A_{l'});\;g_{l'}\in \gb_{l'}\}
=\{\lambda_{g_l}(A_l);\;g_l\in \gb_l\}=[A_l]_l
\ee
since the projections $p_{l'l}:\;\gb_{l'}\to \gb_l$ are surjective.
Now $\ab_l$ is a compact Hausdorff space and $\lambda$ a continuous
group action og $\gb_l$ thereon, thus $[\ab_l]_l$ is a compact Hausdorff
space in the quotient topology induced by $[]_l$. By the results proved
in section \ref{s3.1.2} we find that the projective limit of these
quotients, denoted by $\agb$, is again a compact Hausdorff space in
the induced Tychonov topology.

We therefore have two compact Hausdorff spaces associated with gauge
invariance, on the one hand
the quotient of projective limits $\abgb$ and on the other hand the
projective limit of the quotients $\agb$. The question arises what the
relation between the spaces $\abgb,\agb$ is. In what follows we will show
by purely algebraic and topological methods (without using $C^\ast$
algebra techniques) that they are homeomorphic.

We begin by giving a characterization of $\agb$ similar to the
characterization of $\ab$ as Hom$({\cal P},G)$.
\begin{Definition} \label{def3.15}  ~~~~~~~~~~~~\\
Let, as in definition \ref{def3.7} a point $x_0\in\sigma$ be fixed once
and for all and denote by ${\cal Q}:=\mbox{hom}(x_0,x_0)$ the hoop group
of $\sigma$.\\
i)\\
A finite set $\{\alpha_1,..,\alpha_n\}$ of hoops is said to be independent
if any $\alpha_k$ contains an edge that is traversed precisely once and
that is intersected by any $\alpha_l,\;l\not=k$ in at most a finite
number of points.
ii)\\
An independent set of hoops $\{\alpha_1,..,\alpha_n\}$ defines an unoriented,
closed graph $\check{\gamma}$ by $\check{\gamma}:=\cup_{k=1}^n r(\alpha_k)$
($\alpha\cup \alpha':=p_{c_\alpha\cup c_{\alpha'}}$) up to $x_0$.
Here closed up to $x_0$ means that
every vertex is at least bivalent except, possibly for the vertex $x_0$.
From an oriented graph $\gamma$ we can recover
one set $H(\gamma)=\{\beta_1,..,\beta_n\}$ of independent hoops generating
the
fundamental group $\pi_1(\gamma)$ of $\gamma$ (although not a canonical one
whence possibly
$\{\alpha_k\}\not=\{\beta_k\}$ but the number $n$ is identical for both
sets) as well as the set of vertices of
$\gamma$ as $V(\gamma)=\{b(e),f(e);\;e\in E(\gamma)\}$.
We fix once and for all generators of $\pi_1(\gamma)$ for every
oriented graph $\gamma$. \\
iii)\\
Given a graph $\gamma$ we denote by $s(\gamma)\subset{\cal Q}$ the
(so-called tame) subgroup generated by the generators of $\pi_1(\gamma)$,
that is, $s(\gamma)=\pi_1(\gamma)$.
\end{Definition}
We now have an analogue of theorem \ref{th3.1}
\begin{Theorem} \label{th3.5}   ~~~~\\
Let $\cal S$ be the set all tame subgroups $s(\gamma)$ of $\cal Q$, that is,
those
freely generated by graphs $\gamma\in\Gamma^\omega_0$. Then the relation
$s\prec s'$ iff $s$ is a subgroup of $s'$ equips $\cal Q$ with
the structure of a partially ordered and directed set.
\end{Theorem}
Let now $Y_s:=\mbox{Hom}(s,G)$. As with $X_l=\mbox{Hom}(l,G)$ we can
identify $Y_s$ with some $G^n$ displaying it as a compact Hausdorff space.
Likewise we have surjective projections for $s\prec s'$ given by the
restriction map,
$p_{s' s}:\;Y_{s'}\to Y_s;\;x_{s'}\mapsto (x_{s'})_{|s}$ which satisfy the
consistency condition $p_{s' s}\circ p_{s^\dprime s'}=p_{s^\dprime s}$
for any $s\prec s'\prec s^\dprime$. We therefore can form the direct product
$Y_\infty=\prod_{s\in {\cal S}} Y_s$ and its projective limit subset
\be \label{3.35}
\overline{Y}=\{y=(y_s)_{s\in {\cal S}};\;p_{s' s}(y{s'})=y_s\;\forall\;
s\prec s'\}
\ee
which in the Tychonov topology induced from $Y_\infty$ is a compact
Hausdorff space. Repeating step by step the proof of theorem
\ref{th3.3}) we find that the map
\be \label{3.36}
\Phi:\;\mbox{Hom}({\cal Q},G)\to \overline{Y};\;H\mapsto
(H_{|s})_{s\in{\cal S}}
\ee
is a bijection so that we can identify Hom$({\cal Q},G)$ with $\overline{Y}$
and equip it with the topology of $\overline{Y}$ (open sets of
Hom$({\cal Q},G)$ are the sets $\Phi^{-1}(U)$ where $U$ is open in
$\overline{Y}$). This topology is the weakest one so that all the projections
$p_s:\;\overline{Y}\to Y_s;\;y\mapsto y_s$ are continuous.

The action $\lambda$ of $\gb$ on $\ab=\overline{X}$ reduces on
$\overline{Y}$ to
\be \label{3.37}
\lambda:\;\gb\times\overline{Y}\to\overline{Y};\;(g,y)\mapsto
\lambda(g,y)=\lambda_g(y)=\mbox{Ad}_g(y);\;
[\mbox{Ad}_g(y)]_s=\mbox{Ad}_{g(x_0)}(y_s)
\ee
where for $\alpha\in s$ we have $[\mbox{Ad}_{g(x_0)}(y_s)](\alpha)=
\mbox{Ad}_{g(x_0)}(y(\alpha))$ and $\mbox{Ad}:\;G\times G\to G;\;
(g,h)\mapsto ghg^{-1}$ is the adjoint action of $G$ on itself.
In other words, $(\lambda_{\gb}){|\overline{Y}}=\mbox{Ad}_G$ where
$G$ can be identified with the restriction of $\gb$ to $x_0$. Clearly
$\mbox{Ad}$ acts continuously on $\overline{Y}$.

Consider then the quotient space $\mbox{Hom}({\cal Q},G)/G$ (notice that we
mod out by $G$ and not $\gb$ !) which by the results obtained in the
previous section is a compact Hausdorff space in the quotient topology.
Now the action Ad on $\overline{Y}$ is completely independent of the
label $s$, that is
\be \label{3.38}
\mbox{Ad}_g\circ p_{s' s}=p_{s' s}\circ \mbox{Ad}_g
\ee
so that the points in $\overline{Y}/G$ are given by the equivalence classes
\be \label{3.38a}
(y):=\{\mbox{Ad}_g(y);\;g\in G\}=
\{(\mbox{Ad}_g(y_s))_{s\in {\cal S}};\;g\in G\}=
((y_s)_s)_{s\in {\cal S}}
\ee
where $()_s:\;Y_s\to (Y_s)_s\;y_s\mapsto (y_s)_s=
\{\mbox{Ad}_g(y_s);\;g\in G\}$ denotes the quotient map in $Y_s$.
It follows that Hom$({\cal Q},G)/G$ is the projective limit of the
$(Y_s)_s$. On the other hand, consider the quotients $[X_l]_l$ discussed
above. If $l'=l(\gamma')$ and $\gamma'$ is not a closed graph then
by the action of $\gb$ on $X_{l'}$ we get $[X_{l'}]_{l'}=[X_l]_l$
where $l=l(\gamma)$ and $\gamma$ is the closed graph obtained from $\gamma'$
by deleting its open edges (monovalent vertices). Next, if
$x_0\not\in\gamma$
then we add a path to $\gamma$ connecting any of its points
to $x_0$ without intersecting $\gamma$ otherwise and obtain a third
graph $\gamma^\dprime$ where again $[X_{l^\dprime}]_{l^\dprime}=[X_l]_l$
with $l^\dprime=l(\gamma^dprime)$ due to quotienting by the action of the
gauge group. But now $\gamma^\dprime$ is a closed graph up to $x_0$.
Thus we see that the projective limit of the
$[X_l]_l,\; l\in {\cal L}$ and of the
$[Y_s]_l,\; s\in {\cal S}$ coincides, in other words we have the identity
\be \label{3.39}
\agb=\mbox{Hom}({\cal Q},G)/G
\ee

Our proof of the existence of a homeomorphism between $\abgb$ and
$\agb$ will be based on the identity (\ref{3.39}) and the fact that
$\ab=\mbox{Hom}({\cal P},G)$. We will break this proof into several lemmas.\\
\\
Fix once and for all a system of edges
\be \label{3.40}
{\cal E}:=\{e_x\in\mbox{Hom}(x_0,x);\;x\in\sigma\}
\ee
where $e_{x_0}$ is the trivial hoop based at $x_0$.
Let $\gb_{x_0}:=\{g\in \gb;\;g(x_0)=1_G\}$ be the subset of all gauge
transformtions that are the identity at $x_0$ and
consider the following map
\ba \label{3.41}
f_{{\cal E}}&:& \mbox{Hom}({\cal P},G)\to
\mbox{Hom}({\cal Q},G)\times\gb_{x_0};
\;A\mapsto (B,h) \mbox{ where} \nonumber\\
B(\alpha)&:& =A(\alpha)\;\forall\;\alpha\in{\cal Q} \mbox{ and }
h(x):=A(e_x)\;\forall x\in \sigma
\ea
Clearly $g(x_0)=A(e_{x_0})=1_G$. From the known action $\lambda$ of $\gb$
on $\ab$ we induce the following action of $\gb$ on
$\mbox{Hom}({\cal Q},G)\times\gb_{x_0}$
\ba \label{3.42a}
\lambda' &:& \gb\times(\mbox{Hom}({\cal Q},G)\times\gb_{x_0})\to
(\mbox{Hom}({\cal Q},G)\times\gb_{x_0});\;(g,(B,h))\mapsto (B^g,h^g)
=\lambda'_g(B,h)  \nonumber\\
&& \mbox{ where}
B^g(\alpha)=\mbox{Ad}_{g(x_0)}(B(\alpha));\forall\;\alpha\in{\cal Q}
\mbox{ and }
h^g(x)=g(x_0)h(x) g(x)^{-1} \;\forall\; x\in\sigma
\ea
The action (\ref{3.42a}) evidently splits into a
$G-$action by Ad on Hom$({\cal Q},G)$ (with $G\equiv \gb_{|x_0}$)
as already observed above and
a $\gb-$action on $\gb_{x_0}$ (indeed $h^g(x_0)=1_G)$).\\
\begin{Theorem} \label{th3.6}  ~~~~~~~~~~~\\
For any choice of $\cal E$ the map $f_{{\cal E}}$ in (\ref{3.41}) is a
homeomorphism which is $\lambda-$equivariant, that is,
\be \label{3.43}
f_{{\cal E}}\circ\lambda=\lambda'\circ f_{{\cal E}}
\ee
\end{Theorem}
Proof of Theorem \ref{th3.6}:\\
Bijection:\\
The idea is to construct explicitly the inverse $f_{{\cal E}}^{-1}$.
The ansatz is of course, that given any $p\in{\cal P}$ we can construct
a hoop based at $x_0$ by using $\cal E$, namely
$\alpha_p:=e_{b(p)}\circ p\circ e_{f(p)}^{-1}$, which we can use in order to
evaluate a given $B\in\mbox{Hom}({\cal Q},G)$. Since we want that
$A^g(p)=g(b(p))A(p)g(f(p))^{-1}$ we see that given $h\in \gb_{x_0}$
the only possibility is
\ba \label{3.44}
f_{{\cal E}}^{-1} &:& \mbox{Hom}({\cal Q},G)\times\gb_{x_0}\to
\mbox{Hom}({\cal P},G);\;(B,h)\mapsto A \mbox{ where} \nonumber\\
A(p)&:=& h(b(p))^{-1} B(e_{b(p)}\circ p\circ e_{f(p)}^{-1})h(f(p))
\ea
One can verify explicitly that this is the inverse of (\ref{3.41}). \\
Equivariance:\\
Trivial by construction.\\
Continuity:\\
By definition of the topology on the spaces
$\mbox{Hom}({\cal P},G),\;\mbox{Hom}({\cal Q},G),\;\gb$ respectively,
a corresponding net $(A^\alpha),(B^\alpha),(g^\alpha)$ converges
to $A,B,g$ iff the nets
$(A^\alpha_l)=(p_l(A^\alpha)),(B^\alpha_s)=(p_s(B^\alpha)),
(g^\alpha_x)=(p_x(g^\alpha))$ converge to
$A_l=p_l(A),B_s=p_s(B),g_x=p_x(g)$ where $g_x=g(x)$ for all
$l\in {\cal L},s\in {\cal S},x\in\sigma$.

Continuity of $f_{{\cal E}}$ then means that
$(p_s\times p_x)\circ f_{{\cal E}}$ is continuous for all $s\in{\cal S},
x\in\sigma$ while continuity of $f_{{\cal E}}^{-1}$ means that
$p_l\circ f_{{\cal E}}^{-1}$ is continuous for all $l\in{\cal L}$.
Recalling the map (\ref{3.15}) it is easy to see that
\be \label{3.45}
p_x\circ f_{{\cal E}}=\rho_{e_x}\circ p_{l(e_x)}
\ee
and since the $\rho_\gamma$ are by definition continuous we easily get
continuity of $p_x\circ f_{{\cal E}}$ as the composition of two continuous
maps.

To establish the continuity of
$p_s\circ f_{{\cal E}},p_l\circ f_{{\cal E}}^{-1}$ requires more work.
\begin{Lemma} \label{la3.6}   ~~~~~~~~~~\\
i)\\
For all $s\in {\cal S}$ there exists a connected subgroupoid
$l\in {\cal L}$ such that $s$ is a subgroup of $l$, i.e. $s\prec l$
($s\in {\cal L}$ in particular). The projection
\be \label{3.47}
p_{ls}:\;X_l\to Y_s;\;x_l\mapsto (x_l)_{|s}
\ee
is continuous and satisfies $p_s\circ f_{{\cal E}}=p_{ls}\circ p_l$
for any choice of $\cal E$. \\
ii)\\
For any $l\in {\cal L}$ there exists $s\in {\cal S}$ and a conncted
subgroupoid $l'\in {\cal L}$ such that with $l=l(\gamma),l'=l(\gamma')$ we
have $V(\gamma')=V(\gamma)\cup\{x_0\}$, moreover $l\prec l'$ and
hom$_{l'}(x_0,x_0)=s$. Let $\gb_{x_0}(l'):=\mbox{Fun}(V(\gamma'),G)\cap
\gb_{x_0}$ and let $\pi_{l'}:\;\gb_{x_0}\to \gb_{x_0}(l')$ be the
restriction map. The projection $p_{l' l}:\;X_{l'}\to X_l$ induces a
continuous map
$p_{s l}:\;Y_s \times \gb_{x_0}(l')\to X_l $ which satisfies
\be \label{3.48}
p_l\circ f_{{\cal E}(l)}^{-1}=p_{sl}\circ(p_s\times \pi_l)
\ee
for an appropriate choice ${\cal E}(l)$ of $\cal E$.\\
iii)\\
For any two choices ${\cal E},{\cal E}'$ the map
\be \label{3.48a}
f_{{\cal E}}\circ f_{{\cal E}'}^{-1}:\;
\overline{Y}\times \gb_{x_0}\to \overline{Y}\times \gb_{x_0}
\ee
is a homeomorphism.
\end{Lemma}
Proof of Lemma \ref{la3.6}:\\
i)\\
Let $s\in{\cal S}$ be freely generated by the independent hoops
$\alpha_1,..,\alpha_m$,
let $\check{\gamma}$ be the unoriented graph they determine and
choose some orientation for it. Then every $\alpha_k$ is a finite
composition of the edges $e_1,..,e_n\in E(\gamma)$
demonstrating that $s$ is a
subgroup of $l=l(\gamma)$ consisting of hoops based at $x_0\in V(\gamma)$.
We have bijections $\rho_{\alpha_1,..,\alpha_m}:\;Y_s\to G^m$ and
$\rho_{e_1,..,e_n}:\;Y_s\to G^n$ as in (\ref{3.15}) which can be used
to define the projection $p_{ls}:\;X_l\to Y_s$. In particular we get
$X_s=Y_s$ so that $p_{ls}$ is continuous. It follows that
$p_s\circ f_{{\cal E}}(A)=A_s=p_{ls}(A_l)=(p_{ls}\circ p_l)(A)$
so that $p_s\circ f_{{\cal E}}$ is continuous.\\
ii)\\
Let $l\in {\cal L}$ be freely generated by independent edges $e_1,..,e_n$
and let $\gamma$ be the oriented graph they determine. If $x_0\in V(\gamma)$
invert the orientation of $e_k$ if necessary in order to achieve that
$f(e_k)\not=x_0$ for any $k=1,..,n$. For every vertex $v\in V(\gamma)$ not
yet
connected to $x_0$ through one of the edges $e_1,..,e_n$ add another edge
$e_v$ connecting $x_0$ with $v$ to the set $\{e_1,..,e_n\}$ so that the
extended set remains independent. The extended set $\{e_1,..,e_{n'}\}$
determines an oriented graph $\gamma'$ with $x_0\in V(\gamma')$ and
every vertex of $\gamma'$ is conncted to $x_0$ through at least one edge.
Given $v\in V(\gamma')$ choose one edge $e^l_v\in\mbox{hom}(x_0,v)$ from
$e_1,.,e_{n'}$ with the convention that $e^l_{x_0}$ be the trivial hoop.
Define ${\cal E}'(l):=\{e^l_v;\; v\in V(\gamma)\cup\{x_0\}\}$ and let
$\{e'_1,..,e'_m\}:=\{e_1,..,e_{n'}\}-{\cal E}'(l)$. The hoops based at $x_0$
given by $\alpha_k:=e^l_{b(e'_k)}\circ e'_k\circ (e^l_{f(e'_k)})^{-1},
\;k=1,..,m$ are
independent due to the segments $e'_k$ traversed precisely once and which
are intersected by the other $\alpha_l$ in only a finite number of points
(namely the end points). Let $s$ be the subgroup of $\cal Q$ generated by
the $\alpha_k$ and let $l'\in {\cal L}$ be the subgroupoid generated
by the $(e^l_x)^{-1}\circ\alpha_k\circ e^l_y,\;x,y\in V(\gamma)\cup \{x_0\},
k=1,..,m$ (we know that it is a connected subgroupoid with
hom$_{l'}(x_0,x_0)=s$ from lemma \ref{la3.2}). We claim $l\prec l'$.
To see this, consider the original set of edges $\{e_1,..,e_n\}$.
Each $e_k,\;k=1,..,n$ is either one of the
$e^l_v,\;v\in V(\gamma)\cup\{x_0\}$ or one of the $e'_j,\;j=1,..,m$.
In the first case we have
$e_k=e^l_v=e_{x_0}^{-1}\circ e_{x_0}\circ e^l_v\in l'$
where $e_{x_0}$ is the trivial hoop. In the latter case by definition
$e_k=e'_j=(e^l_{b(e'_j)})^{-1}\circ \alpha_j\circ e^l(f(e'_j))\in l'$.

Consider now the bijection
\be \label{3.49}
f^{l'}_{{\cal E}'(l)}:\;X_{l'}\to Y_s\times \gb_{x_0}(l')
\ee
defined exactly as in (\ref{3.41}) but restricted to $X_{l'}$ so that
only the system of edges ${\cal E}'(l)$ is needed in order to define
it. We can define now
\be \label{3.50}
p_{s l}:=p_{l' l}\circ (f^{l'}_{{\cal E}'(l)})^{-1}:\;Y_s\times
\gb_{x_0}(l')\to X_l
\ee
which is trivially continuous again because both $X_l$ and $Y_s\times
\gb_{x_0}(l')$ are identified with powers of $G$.

Let finally ${\cal E}(l)$ be any system of paths $e_x\in\mbox{hom}(x_0,x)$
that contains ${\cal E}'(l)$. Then for any $B\in\mbox{Hom}({\cal Q},G),
g\in \gb_{x_0},p\in l$ we have
\ba \label{3.51}
&& [(p_l\circ f_{{\cal E}(l)}^{-1})(B,g)](p)=
[f_{{\cal E}(l)}^{-1}(B,g)](p)
=g(b(p))^{-1} B(e^l_{b(p)}\circ p \circ (e^l_{f(p)})^{-1}) g(f(p))
\nonumber\\
&=&
(\pi_l \circ g)(b(p))^{-1}
(p_s\circ B)(e^l_{b(p)}\circ p \circ (e^l_{f(p)})^{-1})
(\pi_l\circ g)(f(p))
\nonumber\\
&=&
[(f^{l'}_{{\cal E}'(l)})^{-1})(p_s\circ B,\pi_l\circ g)](p)
=(p_{l'l}\circ (f^{l'}_{{\cal E}'(l)})^{-1})(p_s\circ B,\pi_l\circ g)](p)
\nonumber\\
&=& [p_{sl}\circ (p_s\times \pi_l)(B,g)](p)
\ea
where in the second line we exploited that $b(p),f(p)\in V(\gamma)$ and that
$e^l_{b(p)}\circ p \circ (e^l_{f(p)})^{-1}\in s$, in the third we observed
that only the subset ${\cal E}'(l)\subset {\cal E}(l)$ is being used
and that $p\in l\prec l'$ and finally we used (\ref{3.50}).
Thus, $p_l\circ f_{{\cal E}(l)}^{-1}=p_{sl}\circ (p_s\times \pi_l)$
is a composition of continuous maps and therefore continuous.\\
iii)\\
Let ${\cal E}=\{e_x,\;x\in\sigma\},{\cal E}'=\{e'_x,\;x\in\sigma\}$
and $\alpha\in {\cal Q},x\in \sigma$, then
\ba \label{3.52}
&&
[f_{{\cal E}}\circ f_{{\cal E}'}^{-1}(B,g)](\alpha,x)
=([f_{{\cal E}'}^{-1}(B,g)](\alpha),[f_{{\cal E}'}^{-1}(B,g)](e_x))
\nonumber\\
&=&
(g(b(\alpha))^{-1}B(e'_{b(\alpha)}\circ\alpha\circ(e'_{f(\alpha)})^{-1})
g(f(\alpha)),
(g(b(e_x))^{-1}B(e'_{b(e_x)}\circ e_x \circ(e'_{f(e_x)})^{-1})
g(f(e_x)))
\nonumber\\
&=&(B(\alpha),B(e_x\circ (e'_{x})^{-1})g(x))
\ea
where in the last step we noticed that $f(e_x)=x,\;
b(\alpha)=f(\alpha)=b(e_x)=x_0,\;
g(x_0)=1_G$ because $g\in \gb_{x_0}$ and that $e'_{x_0}$ is the trivial
hoop based at $x_0$. It follows that the map (\ref{3.48a}) is given
by $(B,g)\mapsto (B',g')$ with $B'=B,g'(.)=B(e_x\circ (e'_x)^{-1})g(.)$.
The inverse map is given similarly by
$(B,g)\mapsto (B',g')$ with $B'=B,g'(.)=B(e'_x\circ (e_x)^{-1})g(.)$
so that it will be sufficient to demonstrate continuity of the former.

To show that $f_{{\cal E}}\circ f_{{\cal E}'}^{-1}$ is continuous requires
to show that $(p_s\times p_x)\circ f_{{\cal E}}\circ f_{{\cal E}'}^{-1}$
is continuous for all $s\in{\cal S},x\in \sigma$. Now obviously
$p_s\circ f_{{\cal E}}\circ f_{{\cal E}'}^{-1}=p_s$ is continuous
by definition. Next $[p_x\circ f_{{\cal E}}\circ f_{{\cal E}'}^{-1}(B,g)](x)
=B(e_x\circ (e'_x)^{-1}) g(x)$. Define the restriction map
\be \label{3.53}
f^{{\cal E},{\cal E}'}_x:=p_{e_x\circ (e'_{x})^{-1}}\times p_x:\;
\overline{Y}\times \gb_{x_0}\to Y_{e_x\circ (e'_{x})^{-1}}\times
(\gb_{x_0})_{|x}
\ee
and denote by $m:\;G\times G \to G;\;(g_1,g_2)\to g_1 g_2$ multiplication
in $G$. Then
\be \label{3.54}
p_x\circ f_{{\cal E}}\circ f_{{\cal E}'}^{-1}=
m\circ(p_{e_x\circ (e'_{x})^{-1}}\times p_x )
\ee
is a composition of continuous maps and therefore continuous.
Hence, $f_{{\cal E}}\circ f_{{\cal E}'}^{-1}$ is a homeomorphism.\\
$\Box$\\

We can now complete the proof of continuity of both $f_{{\cal E}}$ and
$f_{{\cal E}}^{-1}$ for a given, fixed $\cal E$.
We showed already that $p_x\circ f_{{\cal E}}$ is continuous for all
$x\in\sigma$ and by lemma \ref{la3.6}i) we have that
$p_s\circ f_{{\cal E}}$ is continuous for all $s\in {\cal S}$, hence
$f_{{\cal E}}$ is continuous. Next
\be \label{3.55}
p_l\circ f_{{\cal E}}^{-1}=
[p_l\circ f_{{\cal E}(l)}^{-1}]\circ[f_{{\cal E}(l)}\circ f_{{\cal E}}^{-1}]
\ee
is a composition of two continuous functions since the function in the
first bracket is continuous by lemma \ref{la3.6}ii)
and the second by lemma \ref{la3.6}iii), thus $f_{{\cal E}}^{-1}$ is
continuous.\\
$\Box$\\
\begin{Theorem} \label{th3.7}  ~~~~~~~~~~~\\
The spaces $\abgb=\mbox{Hom}({\cal P},G)/\gb$ and
$\agb=\mbox{Hom}({\cal Q},G)/G$ are homeomorphic.
\end{Theorem}
Proof of Theorem \ref{th3.7}:\\
By theorem \ref{th3.6} we know that\\
1)  Hom$({\cal P},G)$ and Hom$({\cal Q},G)\times \gb_{x_0}$ are
homeomorphic and \\
2) $\gb$ acts equivariantly on both spaces via $\lambda,\lambda'$
respectively.\\
We now use the abstract result that if a group acts (not necessarily
continuously) equivariantly on
two homeomorphic spaces then the corresponding spaces continue to be
homeomorphic in their respective quotient topologies (see section \ref{sb}).
We therefore know that $\mbox{Hom}({\cal P},G)/\gb$ and
$(\mbox{Hom}({\cal Q},G)\times \gb_{x_0})/\gb$ are homeomorphic. But
$\gb$ is a direct product space, that is, $\gb=\gb_{x_0}\times G$ whence
$(\mbox{Hom}({\cal Q},G)\times \gb_{x_0})/\gb=\mbox{Hom}({\cal Q},G)/G$.
More explicitly, recalling the action of $\lambda'$ in (\ref{3.42}) and
writing $g\in\gb$ as
$g=(g_1,g_0)\in \gb_{x_0}\times G$ where $g(x)=g_1(x)$ for $x\not=x_0$ and
$g(x_0)=g_0$ we see that $B^g(\alpha)=\mbox{Ad}_{g_0}(B(\alpha))$ and
$h^g(x)=g_0 h(x)g(x)^{-1}$ which gives $h^g(x_0)=h(x_0)=1_G$ and
$h^g(x)=g_0 h(x) g_1(x)^{-1}$ for $x\not=x_0$. It follows that, given
$h\in \gb_{x_0}$, for any
choice of $g_0$ we can gauge $h^g(x)=1_G$ for all $x\in\sigma$ by choosing
$g_1(x)=g_0 h(x)$. The remaining gauge freedom expressed in $g_0$ then
only acts by Ad on Hom$({\cal Q},G)$.\\
$\Box$

\subsection{The $C^\ast$ Algebraic Viewpoint}
\label{s3.2}

In the previous sections we have defined the quantum configuration spaces
of (gauge equivalence casses of) distributional connections $\ab\;(\abgb)$
as Hom$({\cal P},G)$ (Hom$({\cal P},G)/\gb$) and equipped them with
the Tychonov topology through projective techniques. We could be satisfied
with this because we know that these spaces are compact Hausdorff spaces
and this is a sufficiently powerful result in order to develop measure
theory on them as we will se below.

However, the result that we want to establish in this section, namely that
both spaces can be seen as the Gel'fand spectra of certain $C^\ast$ algebras,
has the advantage to make the connection with so-called cylindrical
functions on these spaces explicit which then helps to construct (a priori only
cylindrically defined) measures on them. Moreover, it has a wider
range of applicability in the sense that it does not make use of the concrete
label sets used in the previous section. It therefore establishes a
concrete link with constructive quantum gauge field theories.
A brief introduction to Gel'fand -- Naimark -- Segal theory can be found
in section \ref{se}. We will follow closely Ashtekar and Lewandowski
\cite{47e}\\
\\
We begin again quite generally and suppose that we are given a partially
ordered and directed index set $\cal L$ which label compact Hausdorff
spaces $X_l$ and that we have surjective and continuous projections
$p_{l' l}: X_{l'}\to X_l$ for $l\prec l'$ satisfying the consistency
condition $p_{l'l}\circ p_{l^dprime l'}=p_{l^\dprime l}$ for
$l\prec l'\prec l^\dprime$. Let $X_\infty,\overline{X}$ be the corresponding
direct product and projective limit respectively with Tychonov topology
with respect to which we know that they are Hausdorff and compact from the
previous sections.
\begin{Definition} \label{def3.16} ~~~~~~~~\\
i)\\
Let $C(X_l)$ be the continuous, complex valued functions on $X_l$ and
consider their union
\be \label{3.56}
\mbox{Cyl}'(\overline{X}):=\cup_{l\in {\cal L}} C(X_l)
\ee
Given $f,f'\in \mbox{Cyl}(\overline{X})$ we find $l,l'\in {\cal L}$ such
that $f\in C(X_l),f'\in C(X_{l'})$ and we say that
$f,f'$ are equivalent,
denoted $f\sim f'$ provided that
\be \label{3.57}
p_{l^dprime l}^\ast f =p_{l^\dprime l'} f'\;\forall\; l,l'\prec l^\dprime
\ee
(pull-back maps)
ii)\\
The space of cylindrical functions on the projective limit $\overline{X}$
is defined to be the space of equivalence classes
\be \label{3.58}
\mbox{Cyl}(\overline{X}):=\mbox{Cyl}'(\overline{X})/\sim
\ee
We will denote the equivalence class of $f\in \mbox{Cyl}'(\overline{X})$
by $[f]_\sim$.
\end{Definition}
Notice that we are actually abusing the notation here since an element
$f\in \mbox{Cyl}(\overline{X})$ is not a function on $\overline{X}$
but an equivalence class of functions on the $X_l$. We will justify this
later by showing that $\mbox{Cyl}(\overline{X})$ can be identified
with $C(\overline{X})$, the continuous functions on $\overline{X}$.

Condition (\ref{3.57}) seems to be very hard to check but it is sufficient
to find just one single $l^\dprime$ such that (\ref{3.57}). For suppose
that $f_{l_1}\in C(X_{l_1}),f_{l_2}\in C(X_{l_2})$ are given and that we find
some $l_1,l_2\prec l_3$ such that
$p_{l_3 l_1}^\ast f_{l_1}=p_{l_3 l_2}^\ast f_{l_2}$ Now let any
$l_1,l_2\prec l_4$ be given. Since $\cal L$ is directed we find
$l_1,l_2,l_3,l_4\prec l_5$ and due to the consistency condition among the
projections we have
\be \label{3.59}
\mbox{i) }p_{l_4 l_1}\circ p_{l_5 l_4}=p_{l_5 l_1}=
p_{l_3 l_1}\circ p_{l_5 l_3} \mbox{ and }
\mbox{ii) }p_{l_4 l_2}\circ p_{l_5 l_4}=p_{l_5 l_2}=
p_{l_3 l_2}\circ p_{l_5 l_3}
\ee
whence
\be \label{3.60}
p_{l_5 l_4}^\ast p_{l_4 l_1}^\ast f_{l_1} =_{\mbox{\tiny i)}}\;
p_{l_5 l_3}^\ast p_{l_3 l_1}^\ast f_{l_1}=
p_{l_5 l_3}^\ast p_{l_3 l_2}^\ast f_{l_2}=_{\mbox{\tiny ii)}}\;
p_{l_5 l_4}^\ast p_{l_4 l_2}^\ast f_{l_2}
\ee
where in the middle equality we have used (\ref{3.57}) for $l^\dprime=l_3$.
We conclude that
$p_{l_5 l_4}^\ast [p_{l_4 l_1}^\ast f_{l_1}-p_{l_4 l_2}^\ast f_{l_2}]=0$.
Now for any $f_{l_4}\in C(X_{l_4})$ the condition
$f_{l_4}(p_{l_5 l_4}(x_{l_5}))=0$ for all $x_{l_5}\in X_{l_5}$ means that
$f_{l_4}=0$ because $p_{l_5 l_4}:\;X_{l_5} \to X_{l_4}$ is surjective.
\begin{Lemma} \label{la3.7} ~~~~~~~~~~~~\\
Given $f,f'\in \mbox{Cyl}(\overline{X})$ there exists a common label
$l\in {\cal L}$ and $f_l,f'_l\in C(X_l)$ such that
$f=[f_l]_\sim,f'=[f'_l]_\sim$.
\end{Lemma}
Proof of Lemma \ref{la3.7}:\\
By definition we find $l_1,l_2\in {\cal L}$ and representatives
$f_{l1}\in C(X_{l1}),f_{l2}\in C(X_{l2})$ such that
$f=[f_{l_1}]_\sim,f'=[f_{l_2}]_\sim$. Choose any $l_1,l_2\prec l$
then $f_l:=p_{l l_1}^\ast f_{l1}\sim f_{l_1}$ (choose $l^\dprime=l$
in (\ref{3.57} and use $p_{ll}=\mbox{id}_{X_l}$) and
$f'_l:=p_{l l_2}^\ast f_{l2}\sim f_{l_2}$. Thus
$f=[f_l]_\sim,f'=[f'_l]_\sim$. \\
$\Box$\\
\begin{Lemma} \ref{la3.8}  ~~~~~~~~~~~~~~~\\
i)\\
Let $f,f'\in \mbox{Cyl}(\overline{X})$ then the following operations are
well defined (independent of the representatives)
\be \label{3.61}
f+f':=[f_l+f'_l]_\sim,\; f f':=[f_l f'_l]_\sim,\;zf:=[z f_l]_\sim,\;
\bar{f}:=[\bar{f}_l]_\sim
\ee
where $l,f_l,f_l'$ are as in lemma \ref{la3.7}, $z\in \Cl$ and
$\bar{f}_l$ denotes complex conjugation.\\
ii)\\
$\mbox{Cyl}(\overline{X})$ contains the constant functions.\\
iii)\\
The sup -- norm for $f=[f_l]_\sim$
\be \label{3.62}
||f||:=\sup_{x_l\in X_l} |f_l(x_l)|
\ee
is well-defined.
\end{Lemma}
Proof of Lemma \ref{la3.8}:\\
i)\\
We consider only pointwise multiplication, the other cases are similar.
Let $l,f_l,f'_l$ and $l',f_{l'},f'_{l'}$ as in lemma \ref{la3.7}. We find
$l,l'\prec l^\dprime$ and have
$p_{l^\dprime l}^\ast f_l=p_{l^\dprime l'}^\ast f_{l'}$ and
$p_{l^\dprime l}^\ast f'_l=p_{l^\dprime l'}^\ast f'_{l'}$. Thus
\be \label{3.63}
p_{l^\dprime l}^\ast(f_l f'_l)
=p_{l^\dprime l}^\ast(f_l)p_{l^\dprime l}^\ast(f'_l)
=p_{l^\dprime l'}^\ast(f_{l'})p_{l^\dprime l'}^\ast(f'_{l'})
=p_{l^\dprime l'}^\ast(f_{l'} f'_{l'})
\ee
so $f_l f'_l\sim f_{l'} f'_{l'}$.\\
ii)\\
The function $f_l^z:\; X_l\to \Cl;\;x_l\to z$ for any $z\in \Cl$ certainly
is an element of $C(X_l)$ and for any $l,l'\prec l^\dprime$ we have
$z=(p_{l^\dprime l}^\ast f_l^z)(x_{l^\dprime})
=(p_{l^\dprime l'}^\ast f_{l'}^z)(x_{l^\dprime})$ for all $x_{l^\dprime}\in
X_{l^\dprime}$ so $f^z:=[f_l^z]_\sim$ is well-defined.\\
iii)\\
If $f=[f_l]_\sim=[f_{l'}]_\sim$ is given, choose any $l,l'\prec l^\dprime$
so that we know that $p_{l^\dprime l}^\ast f_l=p_{l^\dprime l'}^\ast f_{l'}$.
Then from the surjectivity of $p_{l^\dprime l},p_{l^\dprime l'}$ we have
\be \label{3.64}
\sup_{x_l\in X_l} |f_l(x_l)|
=\sup_{x_{l^\dprime}\in X_{l^\dprime}}
|(p_{l^\dprime l}^\ast f_l)(x_{l^\dprime})|
=\sup_{x_{l^\dprime}\in X_{l^\dprime}}
|(p_{l^\dprime l'}^\ast f_{l'})(x_{l^\dprime})|
=\sup_{x_{l'}\in X_{l'}} |f_{l'}(x_{l'})|
\ee
$\Box$\\
Lemma \ref{la3.8}i) tells us that $\mbox{Cyl}(\overline{X})$ is an Abelean,
$^\ast-$algebra defined by the pointwise operations \ref{3.61}.
Lemma \ref{la3.8}ii) tells us that $\mbox{Cyl}(\overline{X})$ is also
unital, the unit being given by the constant function
$1=[1_l]_\sim,\;1_l(x_l)=1$. Finally,
lemma \ref{la3.8}iii) tells us that $\mbox{Cyl}(\overline{X})$ is a normed
space and that the norm is correctly normalized, that is, $||1||=1$.
Notice that here the compactness of the $X_l$ comes in since
the norm (\ref{3.62}) certainly does not make sense any longer on
$C(X_l)$ for non-compact $X_l$. If $X_l$ is at least locally compact we
can replace the $C(X_l)$ by $C_0(X_l)$, the continuous complex valued
functions of compact support and still would get an Abelean $^\ast$
algebra with norm although no longer a unital one. One can always embed
an algebra isometrically into a larger algebra with identity (even preserving
the $C^\ast$ property, see below) but this does not solve all problems
in $C^\ast-$algebra theory. Fortunately, we have not to deal with
these complications in what follows.

Recall that a norm induces a metric on a linear space via $d(f,f'):=
||f-f'||$ and that a metric space is said to be complete whenever all
its Cauchy sequences converge. Any incomplete metric space can be
uniquely (up to isometry) embedded into a complete metric space by
extending it by its non-converging Cauchy sequences (see e.g. \cite{47n11}).
We can then complete $\mbox{Cyl}(\overline{X})$ in the norm $||.||$
in this sense and obtain an Abelean, unital Banach $^\ast-$algebra
$\overline{\mbox{Cyl}(\overline{X})}$. But we notice that not only
the submultiplicativity of the norm ($||f f'||\le ||f||\;||f'||$) holds
but in fact the $C^\ast$ property $||f \bar{f}||=||f||^2$.
Thus $\overline{\mbox{Cyl}(\overline{X})}$ is in
fact an unital, Abelean $C^\ast-$algebra. This observation
suggests to apply Gel'fand-Naimark-Segal theory to which an elementary
introduction can be found in section \ref{se}.

Denote by $\Delta(\mbox{Cyl}(\overline{X}))$ the spectrum of
$\mbox{Cyl}(\overline{X})$, that is, the set of {\it all} (algebraic,
i.e. not necessarily continuous) homomorphism from $\mbox{Cyl}(\overline{X})$
into the complex numbers and denote the Gel'fand isometric isomorphism by
\be \label{3.65}
\bigvee:\;\overline{\mbox{Cyl}(\overline{X})}\to
C(\Delta(\overline{\mbox{Cyl}(\overline{X})}));\;f\mapsto\check{f}
\mbox{ where } \check{f}(\chi):=\chi(f)
\ee
where the space of continuous functions on the spectrum
is equipped with the sup-norm. The spectrum is automatically a compact
Hausdorff space in the Gel'fand topology, the weakest topology in which all
the $\check{f},\;f\in \mbox{Cyl}(\overline{X})$ are continuous.

Notice the similarity between the spaces
$\overline{\mbox{Cyl}(\overline{X})}$ and
$C(\Delta(\overline{\mbox{Cyl}(\overline{X})}))$: both are spaces of
continuous functions over compact Hausdorff spaces and on both spaces the
norm is the sup-norm. This suggests that there is a homeomorphism between
the projective limit space $\overline{X}$
and the spectrum $\mbox{Hom}(\overline{\mbox{Cyl}(\overline{X})},\Cl)$.
This is what we are going to prove in what follows.

Consider the map
\be \label{3.66}
{\cal X}:\;\overline{X}\to \Delta(\overline{\mbox{Cyl}(\overline{X})});\;
x=(x_l)_{l\in{\cal L}}\mapsto {\cal X}(x) \mbox{ where }
[{\cal X}(x)](f):=f_l(p_l(x))\mbox{ for } f=[f_l]_\sim
\ee
Notice that (\ref{3.66}) is well-defined since $f=p_l^\ast f_l=
p_{l'}^\ast f_{l'}$ for any $f_l\sim f_{l'}$ which follows from
\be \label{3.67}
p_l^\ast f_l(x)=f_l(x_l)=(p_{l^\dprime l}^\ast f_l)(x_{l^\dprime})
=(p_{l^\dprime l'}^\ast f_{l'})(x_{l^\dprime})=f_{l'}(x_{l'})
=p_{l'}^\ast f_{l'}(x)
\ee
for any $x\in \overline{X},\;l,l'\prec l^\dprime$. Notice also that
(\ref{3.66}) a priori defines
${\cal X}(x)$ only on $\mbox{Cyl}(\overline{X})$ and not on the completion
$\overline{\mbox{Cyl}(\overline{X})}$. We now show that every ${\cal X}(x)$
is actually
continuous: Let $(f^\alpha)$ be a net converging in
$\mbox{Cyl}(\overline{X})$ to $f$, that is, $\lim_\alpha ||f_\alpha-f||=0$.
Then ($f^\alpha=[f^\alpha_{l_\alpha}]_\sim,\;f=[f_l]_\sim,\;
l,l_\alpha\prec l_{\alpha,l}$)
\ba \label{3.68}
|[{\cal X}(x)](f^\alpha)-[{\cal X}(x)](f)|&=&
|(p_{l_\alpha}^\ast f^\alpha_{l_\alpha}- p_l^\ast f_l)(x)|
=|(p_{l_{\alpha,l} l_\alpha}^\ast f^\alpha_{l_\alpha}
-p_{l_{\alpha,l} }^\ast f_l)(x_{l_{\alpha,l}})|
\\
&=&
|f^\alpha_{l_{\alpha,l}}-f_{l_{\alpha,l}})(x_{l_{\alpha,l}})|
\le \sup_{x_{l_{\alpha,l}}\in X_{l_{\alpha,l}}}
|(f^\alpha_{l_{\alpha,l}}-f_{l_{\alpha,l}})(x_{l_{\alpha,li}})|
=||f^\alpha-f||\nonumber
\ea
hence $\lim_\alpha [{\cal X}(x)](f^\alpha)=[{\cal X}(x)](f)$ so
${\cal X}(x)$ is
continuous. It follows that ${\cal X}(x)$ is a continuous linear
(and therefore
bounded) map from the normed linear space $\mbox{Cyl}(\overline{X})$ to the
complete, normed linear space $\Cl$. Hence, by the bounded linear
transformation theorem \cite{47n11} each ${\cal X}(x)$ can be uniquely
extended to a bounded linear transformation (with the same bound) from the
completion $\overline{\mbox{Cyl}(\overline{X})}$ of
$\mbox{Cyl}(\overline{X})$ to $\Cl$ by taking the limit of the evaluation
on convergent series in $\overline{\mbox{Cyl}(\overline{X})}$ which are
only Cauchy in $\mbox{Cyl}(\overline{X})$. We will denote the extension
of ${\cal X}(x)$ to $\overline{\mbox{Cyl}(\overline{X})}$ by
${\cal X}(x)$ again and it is then easy to check that this extended map
${\cal X}$ is an element of
$\Delta(\overline{\mbox{Cyl}(\overline{X})})$
(a homomorphism), e.g. if $f_n\to f, f'_n\to f'$ then
\be \label{3.69}
[{\cal X}(x)](ff'):=\lim_{n\to\infty}[{\cal X}(x)](f_n f'_n)=
\lim_{n\to\infty}([{\cal X}(x)](f_n))\;([{\cal X}(x)](f'_n))=
([{\cal X}(x)](f))\;([{\cal X}(x)](f'))
\ee
The map ${\cal X}$ in (\ref{3.66}) is to be understood in this extended
sense.
\begin{Theorem}  \label{th3.8}  ~~~~~~~~~~~\\
The map ${\cal X}$ in \ref{3.66} is a homeomorphism.
\end{Theorem}
Proof of Theorem \ref{th3.8}:\\
Injectivity:\\
Suppose ${\cal X}(x)={\cal X}(x')$, then in particular
$[{\cal X}(x)](f)=[{\cal X}(x')](f)$
for any $f\in\mbox{Cyl}(\overline{X})$. Hence $f_l(x_l)=f_l(x'_l)$ for any
$f_l\in C(X_l),\;l\in {\cal L}$. Since $X_l$ is a compact Hausdorff space,
$C(X_l)$ separates the points of $X_l$ by the Stone-Weierstrass theorem
\cite{47n11}, hence $x_l=x'_l$ for all $l\in{\cal L}$.
It follows that $x=x'$.\\
Surjectivity:\\
Let $\chi\in \mbox{Hom}(\overline{\mbox{Cyl}(\overline{X})},\Cl)$ be given.
We must construct $x^\chi\in\overline{X}$ such that ${\cal X}(x^\chi)=\chi$.
In particular for any $f=[f_l]_\sim\in \mbox{Cyl}(\overline{X})$ we have
$f_l(x_l^\chi)=\chi([f_l]_\sim)$. Given $l\in {\cal L}$ the character
$\chi$ defines an element $\chi_l\in \mbox{Hom}(C(X_l),\Cl)$ via
$\chi_l(f_l):=\chi([f_l]_\sim)$ for all $f_l\in C(X_l)$. Since
$X_l$ is a compact Hausdorff space, it is the spectrum of the Abelean,
unital $C^\ast-$algebra $C(X_l)$, hence $X_l=\mbox{Hom}(C(X_l),\Cl)$
(see section \ref{se}). It follows that there exists $x^\chi_l\in X_l$
such that $\chi_l(f_l)=f_l(x^\chi_l)$ for all $f_l\in C(X_l)$.
We define $x^\chi:=(x^\chi_l)_{l\in{\cal L}}$ and must check that it defines
an element of the projective limit.

Let $l\prec l'$ and $f=[f_l]_\sim$. Then $f_l\sim f_{l'}:=p_{l' l}^\ast f_l$
(choose $l^\dprime =l'$ and use $p_{l' l'}=\mbox{id}_{X_{l'}}$) and
therefore
\be \label{3.70}
f_l(x^\chi_l)=\chi_l(f_l)=\chi([f_l]_\sim)=\chi([f_{l'}]_\sim)=
\chi_{l'}(f_{l'})=f_{l'}(x^\chi_{l'})=f_l(p_{l' l}(x^\chi_{l'}))
\ee
for any $f_l\in C(X_l),\;l\in {\cal L}$. Since $C(X_l)$ separates the points
of $X_l$ we conclude $x^\chi_l=p_{l' l}(x^\chi_{l'})$ for any
$l\prec l'$, hence $x^\chi\in \overline{X}$.\\
Continuity:\\
We have established that $\cal X$ is a bijection. We must show that both
${\cal X}, {\cal X}^{-1}$ are continuous.

The topology on $\Delta(\overline{\mbox{Cyl}(\overline{X})})$ is the
weakest topology such that the Gel'fand transforms $\check{f},\;f\in
\overline{\mbox{Cyl}(\overline{X})})$ are continuous while the topology on
$\overline{X}$ is the weakest topology such that all the projections
$p_l$ are continuous, or equivalently that al the $p_l^\ast f_l,\;
f_l\in C(X_l)$ are continuous.\\
Continuity of $\cal X$:\\
Let $(x^\alpha)$ be a net in $\overline{X}$ converging to $x$, that
is, every net $(x^\alpha_l)$ converges to $x_l$. Let
first $f=[f_l]_\sim\in\mbox{Cyl}(\overline{X})$. Then
\be \label{3.71}
\lim_\alpha [{\cal X}(x^\alpha)](f)=\lim_\alpha (p_l^\ast f_l)(x^\alpha)
=(p_l^\ast f_l)(x)=[{\cal X}(x)](f)
\ee
for any $f\in \mbox{Cyl}(\overline{X})$.
Now given $\epsilon>0$ for general $f\in \overline{\mbox{Cyl}(\overline{X})}$
we find $f_\epsilon\in \mbox{Cyl}(\overline{X})$ such that
$||f-f_\epsilon||<\epsilon/3$ because $\mbox{Cyl}(\overline{X})$ is dense in
$\overline{\mbox{Cyl}(\overline{X})}$. Also, by (\ref{3.71}), we find
$\alpha(\epsilon)$ such that
$|[{\cal X}(x^\alpha)(f_\epsilon)-[{\cal X}(x)](f_\epsilon)|\le
\epsilon/3$ for any $\alpha(\epsilon)\prec\alpha$. Finally, since
${\cal X}(x^\alpha),{\cal X}(x)$ are characters they are bounded
(by one) linear functionals on $\overline{\mbox{Cyl}(\overline{X})}$ as we have
shown above (continuity of the ${\cal X}(x)$). It follows that
\ba \label{3.72}
|[{\cal X}(x^\alpha)](f)-[{\cal X}(x)](f)|
&\le&
|[{\cal X}(x^\alpha)](f-f_\epsilon)|
+|[{\cal X}(x)](f-f_\epsilon)|
+|[{\cal X}(x^\alpha)](f_\epsilon)-[{\cal X}(x)](f_\epsilon)|
\nonumber\\
& \le& 2||f-f_\epsilon||+\epsilon/3\le \epsilon
\ea
for all $\alpha(\epsilon)\prec \alpha$. Thus
\be \label{3.73}
\lim_\alpha \check{f}({\cal X}(x^\alpha))=\check{f}({\cal X}(f))
\ee
for all $f\in\overline{\mbox{Cyl}(\overline{X})}$, hence
${\cal X}(x^\alpha)\to {\cal X}(x)$ in the Gel'fand topology.\\
${\cal X}^{-1}$:\\
Let $(\chi^\alpha)$ be a net in $\Delta(\overline{\mbox{Cyl}(\overline{X})})$
converging to $\chi$, so
$\chi^\alpha(f)\to \chi(f)$ for any $f\in\overline{\mbox{Cyl}(\overline{X})}$
and so in particular for  $f=[f_l]_\sim\in\mbox{Cyl}(\overline{X})$.
Therefore
\be \label{3.74}
\chi^\alpha(f)=\chi^\alpha(p_l^\ast f_l)=(p_l^\ast f_l)(x^{\chi_\alpha})
=(p_l^\ast f_l)({\cal X}^{-1}(\chi_\alpha))
\to (p_l^\ast f_l)({\cal X}^{-1}(\chi))=\chi(f)
\ee
for all $f_l\in C(X_l),l\in{\cal L}$. Hence
${\cal X}^{-1}(\chi_\alpha)\to {\cal X}^{-1}(\chi)$ in the Tychonov
topology.\\
$\Box$\\
\begin{Corollary} \label{col3.1}  ~~~~~~~~~~\\
The closure of the space of cylindrical functions
$\overline{\mbox{Cyl}(\overline{X})}$
may be identified with the space of continuous functions $C(\overline{X})$
on the sprojective limit $\overline{X}$.
\end{Corollary}
This follows from the fact that via theorem \ref{th3.8} we may identify
$\overline{X}$ set-theoretically and topologically with the spectrum
$\Delta(\overline{\mbox{Cyl}(\overline{X})})$ and the fact that the
Gel'fand transform between $\overline{\mbox{Cyl}(\overline{X})}$ and
$C(\Delta(\overline{\mbox{Cyl}(\overline{X})}))$ is an (isometric)
isomorphism. This justifies in retrospect the notation Cyl$(\overline{X})$
although cylindrical functions are not functions on $\overline{X}$ but
rather equivalence classes of functions on the $X_l$ under $\sim$.\\
\\
Next we give an abstract and independent $C^\ast-$algebraic proof for the
fact that the spaces $\overline{X}/G$ and $\overline{X/G}$
are homeomorphic whenever a topological group $G$ acts continuously and
equivariantly on the projective limit $\overline{X}$, that is, we reprove
theorem \ref{th3.7}.

Suppose then that for each $l\in{\cal L}$ we have a group action
\be \label{3.75}
\lambda^l:\;G\times X_l\to X_l;\;(g,x_l)\mapsto \lambda^l_g(x_l)
\ee
where $\lambda^l_g$ is a continuous map on $X_l$ which is equivariant
with repspect to the projective structure, that is,
\be \label{3.76}
p_{l' l}\circ \lambda^{l'}=\lambda^l\circ p_{l' l}\; \forall l\prec l'
\ee
Due to continuity of the group action and since $X_l$ is Hausdorff and
compact, the quotient space $X_l/G$ is again
compact and Hausdorff in the quotient topology (see section \ref{sb}) and
due to equivariance the net of equivalence classes $([x_l]_l)_{l\in
{\cal L}}$ is a projective net again (with respect to the same
projections $p_{l' l}$) so that we can form the
projective limit $\overline{X/G}$ of the $X_l/G$ which then is a compact
Hausdorff space again. Here $[.]_l:\;X_l\to X_l/G$ denotes the individual
quotient maps with respect to the $\lambda^l$.

On the other hand we may directly define an action of $G$ on $\overline{X}$
itself by
\be \label{3.77}
\lambda:\;\overline{X}\times G\to \overline{X};\;x=(x_l)_{l\in {\cal L}}
\mapsto \lambda_g(x):=(\lambda^l_g(x_l))_{l\in {\cal L}}
\ee
Since $\overline{X}$ is compact and Hausdorff and $\lambda_g$ is a continuous
map on $\overline{X}$ (since it is continuous iff all the $\lambda^l_g$
are continuous) it follows that the quotient space $\overline{X}/G$
is again a compact Hausdorff space.

We now want to know what the relation between $\overline{X}/G$ and
$\overline{X/G}$ is. Let $[.]:\;\overline{X}\to\overline{X}/G$ be the
quotient map with respect to $\lambda$. We then may define a map
\be \label{3.78}
\Phi:\;\overline{X}/G\to \overline{X/G};\;[x]=[(x_l)_{l\in {\cal L}}]\mapsto
([x_l]_l)_{l\in {\cal L}}
\ee
as follows: we have
\be \label{3.79}
[x]=\{\lambda_g(x);\;g\in G\}:=\{(\lambda^l_g(x_l))_{l\in {\cal L}}\;g\in G\}
\ee
Now take an arbitrary representative in $[x]$, say $\lambda_{g_0}(x)$ for
some $g_0\in G$ and compute its class in $\overline{X/G}$, that is,
\be \label{3.80}
\Phi([x]):=([p_l(\lambda_{g_0}(x))]_l)_{l\in {\cal L}}=
(\{\lambda^l_g(\lambda^l_{g_0}(x_l));\;g\in G\})_{l\in {\cal L}}
=(\{\lambda^l_g(x_l);\;g\in G\})_{l\in {\cal L}}
\ee
which shows that $\Phi$ is well-defined, that is, independent of the
choice of $g_0$.
\begin{Theorem} \label{th3.9}  ~~~~~~~~~~~\\
The map $\Phi$ defined in (\ref{3.78}) is a homeomorphism.
\end{Theorem}
Proof of Theorem \ref{th3.9}:\\
The strategy of the proof is to 1) first show that the pull-back map
\be \label{3.81}
\Phi^\ast:\;C(\overline{X/G})\to C(\overline{X}/G)
\ee
is a bijection and then 2) to show that for any compact Hausdorff spaces
$A, B$ such that $\Phi^\ast:\; A\to B$ is a bijection it follows that
$\Phi$ is a homeomorphism.\\
Step 1)\\
Let $f\in C(\overline{X/G})$ be given. Via corollary
\ref{col3.1} we may think of $f$ as an element of
$\overline{\mbox{Cyl}(\overline{X/G})}$ and elements of
$\mbox{Cyl}(\overline{X/G})$ lie dense in that space. Now any
$f\in \mbox{Cyl}(\overline{X/G})$ is given by $f=[f_l]_\sim$ where
$f_l$ is a $\lambda^l$ invariant function on $X_l$. Then
\be \label{3.82}
f_l([x_l]_l)=p_l^\ast f_l(\Phi([x]))
\ee
Thus the functions on $\mbox{Cyl}(\overline{X/G})$ are obtained as
$p_l^\ast f_l$ for some $l\in {\cal L}$ where $f_l$ is $\lambda^l$ invariant
and then $\Phi^\ast p_l^\ast f_l$ is a $\lambda-$invariant function on
$\overline{X}$. But such functions are precisely those that lie dense in
$C(\overline{X}/G)$ because a function $f\in C(\overline{X}/G)$ is
simply a $\lambda-$invariant function in $C(\overline{X})$, that is,
via corollary \ref{col3.1} a $\lambda-$invariant function in
$\overline{\mbox{Cyl}(\overline{X})}$ in which the $\lambda-$invariant
functions in $\mbox{Cyl}(\overline{X})$ lie dense and the latter
are of the form $p_l^\ast f_l$ for some $l\in {\cal L}$ and
$\lambda-$invariant.

To see that
$\Phi^\ast$ is injective on $\mbox{Cyl}(\overline{X/G})$ suppose that
$\Phi^\ast p_l^\ast f_l=\Phi^\ast p_{l'}^\ast f'_{l'}$ for some $l, l'$.
Then trivially
$p_l^\ast f_l(x)=p_{l'}^\ast f'_{l'}(x)$ for all $x\in \overline{X}$.
Let $l,l'\prec l^\dprime$ then
\be \label{3.82a}
p_l^\ast f_l(x)=f_l(x_l)=p_{l^\dprime l}^\ast f_l(x_{l^\dprime})
=p_{l'}^\ast f'_{l'}(x)=f_{l'}(x_{l'})=p_{l^\dprime l'}^\ast
f_{l'}(x_{l^\dprime})\;\forall\; x_{l^\dprime}\in X_{l^\dprime}
\ee
which shows that $f_l\sim f'_{l'}$, hence $[f_l]_\sim=[f'_{l'}]_\sim$
define the same element of $\mbox{Cyl}(\overline{X/G})$.

To see that $\Phi^\ast$ is a surjection we notice that it maps the dense set
of functions in $\overline{\mbox{Cyl}(\overline{X/G})}$ of the form
$p_l^\ast f_l$ ($f_l$ being $\lambda^l-$invariant) into the dense set
of functions in $\overline{\mbox{Cyl}(\overline{X}/G)}$ of the form
$\Phi^\ast p_l^\ast f_l$ that are $\lambda-$invariant. If we can show that
$\Phi^\ast:\;\mbox{Cyl}(\overline{X/G)}\to
\overline{\mbox{Cyl}(\overline{X}/G)}$ is continuous then it can be
uniquely extended as a continuous map to the completion
$\Phi^\ast:\;\overline{\mbox{Cyl}(\overline{X/G)}}\to
\overline{\mbox{Cyl}(\overline{X}/G)}$ by the bounded linear transformation
theorem and it will be a surjection since any
$f\in\overline{\mbox{Cyl}(\overline{X}/G)}$ can be approximated
arbitrarily well by elements in $\mbox{Cyl}(\overline{X}/G)$ which
we know to lie in the image of $\Phi^\ast$ already. To prove that
$\Phi^\ast$ is continuous (bounded), we show that it is actually an isometry
and therefore has unity bound.
\ba \label{3.83}
||\Phi^\ast p_l^\ast f_l||_{\mbox{Cyl}(\overline{X}/G)}
&=& \sup_{[x]\in\overline{X}/G}|f_l(p_l(\Phi([x])))|
\nonumber\\
&=& \sup_{([x_{l'}]_{l'})_{l'\in {\cal L}}\in\overline{X/G}}
|f_l(p_l(([(x_{l'}]_{l'})_{l'\in {\cal L}})))|
=||p_l^\ast f_l||_{\overline{\mbox{Cyl}(\overline{X/G})}}
\ea
Step 2)\\
Let $\Phi:\;A\to B$ be a map between compact Hausdorff spaces such that
$\Phi^\ast:\;C(B)\to C(A)$ is a bijection. \\
Injectivity:\\
Suppose $\Phi(a)=\Phi(a')$. Then for any $F\in C(B)$ we have
$(\Phi^\ast F)(a)=(\Phi^\ast F)(a')$. Since $\Phi^\ast$ is a surjection
and $C(A)$ separates the points of $A$ it follows that $a=a'$.\\
Surjectivity:\\
Since $A,B$ are the Gel'fand spectra
$\mbox{Hom}(C(A),\Cl),\mbox{Hom}(C(B),\Cl)$ of $C(A),C(B)$ respectively
and $\Phi^\ast$ is a bijection we obtain a corresponding
bijection between $A,B$ (since the spectrum can be constructed
algebraically from the algebras) via
\be \label{3.84}
\Phi_\ast:\;A=\Delta(C(A))\to B=\Delta(C(B));a\mapsto a\circ\Phi^\ast
\ee
where
\be \label{3.85}
f(a)\equiv a(f)=a(\Phi^\ast F)
=(a\circ\Phi^\ast)(F)=F(\Phi(a))=(\Phi(a))(F)
\ee
for any $f=\Phi^\ast F\in C(A),\;F\in C(B)$. It follows that any $b\in B$
can be written in the form $b=\Phi(a)$ for some $a\in A$.\\
Continuity:\\
We know that both $\Phi^{-1},(\Phi^\ast)^{-1}$ exist. Then
$(\Phi^\ast)^{-1}=(\Phi^{-1})^\ast$ since
\ba \label{3.86}
f(a)
&=&[(\Phi^\ast\circ (\Phi^\ast)^{-1})f](a)=[(\Phi^\ast)^{-1} f](\Phi(a))
\nonumber\\
&=& f((\Phi^{-1}\circ \Phi)(a))=[(\Phi^{-1})^\ast f](\Phi(a))
\ea
for any $f\in C(A),a\in A$. Let now $(a^\alpha)$ be a net in $A$ converging
to $a$. This is equivalent with $\lim_\alpha f(a^\alpha)=f(a)$ for all
$f\in C(A)$ which in turn implies
$\lim_\alpha F(\Phi(a^\alpha))=F(\Phi(a))$
for all $F\in C(B)$ since any $f$ can be written as $\Phi^\ast F$ which
then is equvalent with the convergence of the net $\Phi(a^\alpha)$ to
$\Phi(a)$ in $B$. The proof for $\Phi^{-1}$ is anlogous.\\
$\Box$

\subsection{Regular Borel Measures on the Projective Limit: The Uniform
Measure}
\label{s.3.3}

In this section we describe a simple mechanism, based on the Riesz
representation theorem, of how to construct
$\sigma-$additive measures on the projective limit $\overline{X}$ starting
from a so-called self-consistent family of (so-called cylindrical) measures
$\mu_l$ on the various $X_l$. See section \ref{sf} for some useful
measure theoretic terminology and the references cited there for further
reading.\\
\\
Our spaces $X_l$ are compact Hausdorff spaces and in particular
topological spaces and are therefore naturally equipped with the
$\sigma-$algebra ${\cal B}_l$ of Borel sets (the smallest $\sigma-$algebra
containing
all open (equivalently closed) subsets of $X_l$). Let $\mu_l$ be a positive,
regular, Borel, probability measure on $X_l$, that is, a positive
semi-definite, $\sigma-$additive function on ${\cal B}_l$
with $\mu_l(X_l)=1$ and regularity means that the measure of every
measurable set can be approximated arbitrarily well by open and compact sets
(hence closed since $X_l$ is compact Hausdorff) respectively.
Since the measure is Borel, the continuous functions $C(X_l)$ are
automatically measurable.
\begin{Definition}  \label{def3.17}   ~~~~~~~~~~~~\\
A family of measures $(\mu_l)_{l\in {\cal L}}$ on the projections
$X_l$ of a projective family $(X_l,p_{l l'})_{l\prec l'\in {\cal L}}$
where the $p_{l' l}:\; X_{l'}\to X_l$ are continuous and surjective
projections is said
to be consistent provided that
\be \label{3.87}
(p_{l' l})_\ast \mu_{l'}:=\mu_{l'}\circ p_{l' l}^{-1}=\mu_l
\ee
for any $l\prec l'$. The measure $(p_{l' l})_\ast \mu_{l'}$ on $X_l$
is called the push-forward of the measure $\mu_{l'}$.
\end{Definition}
The meaning of condition (\ref{3.87}) is the following: Let
${\cal B}_l\ni U_l\subset X_l$ be measurable. Since $p_{l' l}$ is
continuous the pre-images of open sets in $X_l$ are open in $X_{l'}$
and therefore measurable, hence $p_{l' l}$ is measurable. Since
$U_l$ is generated from countable unions and intersections of open sets
it follows that $p_{l' l}^{-1}(U_l)$ is measurable. Then we require that
\be \label{3.88}
\mu_{l'}(p_{l' l}^{-1}(U_l))=\mu_l(U_l)
\ee
for any measurable $U_l$. We can rewrite condition (\ref{3.88}) in the
form
\be \label{3.89}
\int_{X_{l'}} d\mu_{l'}(x_{l'}) \chi_{p_{l' l}^{-1}(U_l)}(x_{l'})
=\int_{X_l} d\mu_l(x_l) \chi_{U_l}(x_l)
\ee
where $\chi_S$ denotes the characteristic function of a set $S$.
Here it is strongly motivated to have surjective projections $p_{l' l}$
as otherwise $p^{-1}_{l' l}(X_l)$ is a proper subset of $X_{l'}$
so that $1=\mu_l(X_l)=\mu_{l'}(p_{l' l}^{-1}(X_l)$ could give a
contradiction with the $\mu_l$ being probability measures if
$X_{l'}-p_{l' l}^{-1}(X_l)$ is not a set of measure zero with respect
to $\mu_{l'}$.

Condition (\ref{3.89}) extends linearly to linear combinations of
characteristic functions, so-called simple functions (see section
\ref{sf}) and the (Lebesgue) integral of any measurable function
is defined in terms of simple functions (see section \ref{sf}).
Therefore we may equivalently write (\ref{3.87}) as
\be \label{3.90}
\int_{X_{l'}} d\mu_{l'}(x_{l'}) [p_{l' l}^\ast f_l](x_{l'})
=\int_{X_l} d\mu_l(x_l) f_l(x_l)
\ee
for any $l\prec l'$ and any $f_l\in C(X_l)$ since every measurable function
can be approximated by simple functions and measurable simple functions
can be approximated by continuous functions (which are automatically
measurable). In the form (\ref{3.90}) the consistency condition means
that integrating out the degrees of
freedom in $X_{l'}$ on which $p_{l' l}^\ast f_l$ does not depend, we end up
with with the same integral as if we had integrated over $X_l$ only.

To summarize:\\
Let $f=[f_l]_\sim\in\mbox{Cyl}(\overline{X})$ with $f_l\in C(X_l)$.
Then (\ref{3.90}) ensures that the linear functional
\be \label{3.91}
\Lambda:\;\mbox{Cyl}(\overline{X})\to \Cl;\;
f=[f_l]_\sim\mapsto \Lambda(f):=\int_{X_l} d\mu_l(x_l) f_l(x_l)
\ee
is well defined, i.e independent of the representative
$f_l\sim p_{l' l}^\ast f_l$ of $f$. Moreover, it is a positive linear
functional (integrals of positive functions are positive) because the
$\mu_l$ are positive measures. Since
$\mbox{Cyl}(\overline{X})\subset\overline{\mbox{Cyl}(\overline{X})}$
is a subset of a unital $C^\ast-$algebra, $\Lambda$ is automatically
continuous (see the end of section \ref{sf}) and therefore extends uniquely
and continuously to the completion $\overline{\mbox{Cyl}(\overline{X})}$
by the bounded linear transformation theorem.
Now in sections \ref{s3.1.2}, \ref{s3.1.3} we showed that the Gel'fand
isomorphism applied to $\overline{\mbox{Cyl}(\overline{X})}$ leads to an
(isometric) isomorphism of $\overline{\mbox{Cyl}(\overline{X})}$ with
$C(\overline{X})$ given by
\be \label{3.92}
\bigvee:\;\mbox{Cyl}(\overline{X})\to C(\overline{X});\;
f=[f_l]_\sim\mapsto p_l^\ast f_l
\ee
(and extended to $\overline{\mbox{Cyl}(\overline{X})}$ using that
$\mbox{Cyl}(\overline{X})$ is dense). It follows that we may
consider (\ref{3.90}) as a positive linear functional on
$C(\overline{X})$. Since $\overline{X}$ is a compact Hausdorff space we
are in position to apply the Riesz representation theorem.
\begin{Theorem}  \label{th3.10}   ~~~~~~~~~~~~~\\
Let $(X_l,p_{l' l})_{l\prec l'\in {\cal L}}$ be a compact Hausdorff
projective family with continuous and surjective projections
$p_{l' l}:\; X_{l'}\to X_l$,
projective limit $\overline{X}$ and projections $p_l:\;\overline{X}\to
X_l$.\\
i)\\
If $\mu$ is a regular Borel probability measure on $\overline{X}$ then
$(\mu_l:=\mu\circ p_l^{-1})_{l\in {\cal L}}$ defines a consistent family of
regular Borel probability measures on $X_l$.\\
ii)\\
If $(\mu_l)_{l\in {\cal L}}$ defines a consistent family of regular
Borel probability measures on $X_l$ then there exists a unique, regular
Borel probability measure $\mu$ on $\overline{X}$ such that
$\mu\circ p_l^{-1}=\mu_l$.\\
iii)\\
The measure $\mu$ is faithful if and only if every $\mu_l$ is faithful.
\end{Theorem}
Proof of Theorem \ref{th3.10}:\\
i)\\
Define the following positive lineal functional on $C(X_l)$:\\
\be \label{3.93}
\Lambda_l:\; C(X_l)\to\Cl;\;f_l\mapsto \int_{\overline{X}} d\mu(x)
(p_l^\ast f_l)(x)
\ee
which satisfies $\Lambda_l(1)=1$.
Since $X_l$ is a compact Hausdorff space, by the Riez representation theorem
there exists a unique, positive, regular Borel probability measure $\mu_l$
on $X_l$ that represents $\Lambda_l$, that is
\be \label{3.94}
\Lambda_l(f_l)=\int_{X_l} d\mu_l(x_l) f_l(x_l)
\ee
Since $p_{l' l}\circ p_{l'}=p_l$, the consistency condition (\ref{3.90})
is obviously met.\\
ii)\\
As was shown above, the positive linear functional on $C(\overline{X})$
\be \label{3.95}
\Lambda:\;C(\overline{X})\to \Cl;\;f=p_l^\ast f_l\equiv[f_l]_\sim\mapsto
\int_{X_l} d\mu_l(x_l) f_l(x_l)
\ee
is well-defined due to the consistency condition and satisfies
$\Lambda(1)=1$. Since $\overline{X}$
is a compact Hausdorff space the Riesz representation theorem guarantees
the existence of a unique, positive, regular Borel probability measure
$\mu$ on $\overline{X}$ representing $\Lambda$, that is
\be \label{3.96}
\Lambda(f)=\int_{\overline{X}} d\mu(x) f(x)
\ee
iii)\\
Consider $f\in C(\overline{X})$ of the form $f=p_l^\ast f_l$ for some
$l\in {\cal L},\;f_l\in C(X_l)$.
Functions of the form $p_l^\ast f_l$ lie dense in $C(\overline{X})$.
Now $f=p_l^\ast f_l$
is non-negative iff $f_l$ is non-negative because $p_l$ is a surjection.
It follows that we can restrict attention to all non-negative functions
of the form $f=p_l^\ast f_l$ for arbitrary $f_l\in C(X_l),\;l\in {\cal L}$
as far as faithfulness is concerned. Let $\Lambda_\mu,\Lambda_{\mu_l}$
be the positive linear functionals determined by $\mu,\mu_l$ respectively.
Then:\\
$\mu$ faithful $\Leftrightarrow\;\Lambda_\mu(p_l^\ast f_l)=
\Lambda_{\mu_l}(f_l)=0$ for any
non-negative $f_l\in C(X_l)$ and any $l\in {\cal L}$ implies $f=0$
$\Leftrightarrow$ For any $l\in {\cal L}$ and any
non-negative $f_l\in C(X_l)$ the condition $\Lambda_{\mu_l}(f_l)$ implies
$f_l=0$ $\Leftrightarrow$ all $\mu_l$ are faithful.\\
$\Box$\\
\\
We now define a natural measure on the spectrum of interest namely
$\ab$, the so-called uniform measure. To do this we must specify
the space of cylindrical functions. Given a subgroupoid $l\in {\cal L}$
with $l=l(\gamma)$ we think of an element $x_l\in X_l$ as a collection of
group elements $\{x_l(e)\}_{e\in E(\gamma)}=\rho_l(x_l)$ and $X_l$ can be
identified
with $G^{|E(\gamma)|}$ (see \ref{3.15}). Thus, an element $f_l\in C(X_l)$
is simply given by
\be \label{3.96a}
f_l(x_l)=F_l(\{x_l(e)\}_{e\in E(\gamma)})=(\rho_l^\ast F_l)(x_l)
\ee
where $F_l$ is a continuous complex valued function on $G^{|E(\gamma)|}$.
For $l\prec l'$ with $l=l(\gamma),l'=l(\gamma')$ we define
$\rho_{l' l}:\;G^{|E(\gamma')|}\to G^{|E(\gamma)|}$ by
$\rho_l\circ p_{l' l}=\rho_{l' l}\circ \rho_{l'}$ (recall that
$\rho_l$ is a bijection.
\begin{Definition} \label{def3.18}  ~~~~~~~~~~\\
Let $\cal L$ be the set of all tame subgroupoids of the set of piecewise
analytic paths $\cal P$ in $\sigma$ and $X_l=\mbox{Hom}(l,G)$ identified
with $G^{|E(\gamma)|}$ if $l=l(\gamma)$ via (\ref{3.15}). Then we define
for any $f\in C(X_l)$
\be \label{3.96b}
\mu_{0l}(f_l)=\int_{X_l} d\mu_{0l}(x_l) \rho_l^\ast F_l(x_l)
:=\int_{G^{|E(\gamma)|}}
[\prod_{e\in E(\gamma)} d\mu_H(h_e)]\; F_l(\{h_e\}_{e\in E(\gamma)})
\ee
where $\mu_H$ is the Haar probability measure on $G$ which thanks to the
comapctness of $G$ is invariant under left -- and right translations and
under inversions.
\end{Definition}
\begin{Lemma} \label{la3.8} ~~~~~~~\\
The linear functionals $\mu_l$ in (\ref{3.96b}) are positive and consistently
defined.
\end{Lemma}
Proof of Lemma \ref{la3.8}:\\
That $\mu_l$ defines a positive linear functional follows from the
explicit formula (\ref{3.96a}) in terms of the positive Haar measure
on $G^n$. That $(\mu_{0l})_{l\in {\cal L}}$ defines a consistent family
follows from the observation that if $l\prec l'$ with $l=l(\gamma),
l'=l(\gamma')$ then we can reach $l$ from $l'$ by a finite combination
of the following three steps:\\
a) $e_0\in E(\gamma')$ but $e_0\cap \gamma\subset \{b(e_0),f(e_0)\}$
(deletion of an edge).\\
b) $e_0\in E(\gamma')$ but $e_0^{-1}\in E(\gamma)$ (inversion of an edge).\\
c) $e_1,e_2\in E(\gamma')$ but $e_0=e_1\circ e_2\in E(\gamma)$
(composition of edges)\\
It therefore suffices to establish consistency with respect to all of these
elementary steps.

In general we have
\be \label{3.96c}
p_{l' l}^\ast f_l
=p_{l' l}^\ast \rho_l^\ast F_l
=\rho_{l'}^\ast \rho_{l' l}^\ast F_l
\ee
whence
\be \label{3.96d}
\mu_{0l'}(p_{l' l}^\ast f_l)
=\mu_{0l'}(\rho_{l'}^\ast [\rho_{l' l}^\ast F_l])
=\int_{G^{|E(\gamma')|}}
[\prod_{e\in E(\gamma')} d\mu_H(h_e)]\;
[\rho_{l' l}^\ast F_l](\{h_e\})_{e\in E(\gamma')})
\ee
In what follows we will interchange freely orders of integration
and break the integral over $G^n$ in integrals over $G^m,G^{n-m}$.
This is allowed by Fubini's theorem since the integrand, being bounded, is
absolutely integrable in any order.\\
a)\\
We have $\rho_{l' l}(\{h_e\}_{e\in E(\gamma')})=\{h_e\}_{e\in E(\gamma)}$
thus
\be \label{3.96e}
\mu_{0l'}(p_{l' l}^\ast f_l)=
\{\int_{G^{|E(\gamma)|}} [\prod_{e\in E(\gamma)} d\mu_H(h_e)]
F_l(\{h_e\}_{e\in E(\gamma)})\}\{\int_G d\mu_H(h_{e_0}) \;1\}
=\mu_{0l}(f_l)
\ee
since $\mu_H$ is a probability measure.\\
b)\\
We have $\rho_{l' l}(\{h_e\}_{e\in E(\gamma')})=
\{\{h_e\}_{e\in E(\gamma)-\{e_0\}},h_{e_0}^{-1}\}$ thus
\ba \label{3.96f}
\mu_{0l'}(p_{l' l}^\ast f_l) &=&
\int_{G^{|E(\gamma)|-1}} [\prod_{e\in E(\gamma)-\{e_0\}} d\mu_H(h_e)]
\int_G d\mu_H(h_{e_0})
F_l(\{h_e\}_{e\in E(\gamma)-\{e_0\}},h_{e_0^{-1}})
\nonumber\\
&=&
\int_{G^{|E(\gamma)-1|}} [\prod_{e\in E(\gamma)-\{e_0\}} d\mu_H(h_e)]
\int_G d\mu_H(h_{e_0}^{-1})
F_l(\{h_e\}_{e\in E(\gamma)-\{e_0\}},h_{e_0}^{-1})
\nonumber\\
&=&
\int_{G^{|E(\gamma)|}} [\prod_{e\in E(\gamma)} d\mu_H(h_e)]
F_l(\{h_e\}_{e\in E(\gamma)})=\mu_{0l}(f_l)
\ea
since the Jacobian of the Haar measure with respect to the inversion map
on $G$ equals unity and where we have defined a new integration variable
$h_{e_0^{-1}}:=h_{e_0}^{-1}$.\\
c)\\
We have $\rho_{l' l}(\{h_e\}_{e\in E(\gamma')})=
\{\{h_e\}_{e\in E(\gamma)-\{e_0\}},h_{e_1} h_{e_2}\}$ thus
\ba \label{3.9gf}
&& \mu_{0l'}(p_{l' l}^\ast f_l) \nonumber\\
&=&
\int_{G^{|E(\gamma)|-1}} [\prod_{e\in E(\gamma)-\{e_0\}} d\mu_H(h_e)]
\int_{G^2} d\mu_H(h_{e_1})d\mu_H(h_{e_2})
F_l(\{h_e\}_{e\in E(\gamma)-\{e_0\}},h_{e_1} h_{e_2})
\nonumber\\
&=&
\int_{G^{|E(\gamma)|-1}} [\prod_{e\in E(\gamma)-\{e_0\}} d\mu_H(h_e)]
\int_G d\mu_H(h_{e_1})\int_G d\mu_H(h_{e_1}^{-1}h_{e_1\circ e_2})
\times \nonumber\\
&& \times F_l(\{h_e\}_{e\in E(\gamma)-\{e_0\}},h_{e_1\circ e_2})
\nonumber\\
&=&
\int_{G^{|E(\gamma)|-1}} [\prod_{e\in E(\gamma)-\{e_0\}} d\mu_H(h_e)]
\int_G d\mu_H(h_{e_1\circ e_2})
F_l(\{h_e\}_{e\in E(\gamma)-\{e_0\}},h_{e_1\circ e_2})
\times \nonumber\\
&& \times [\int_G d\mu_H(h_{e_1}) 1]
\nonumber\\
&=&
\int_{G^{|E(\gamma)|}} [\prod_{e\in E(\gamma)} d\mu_H(h_e)]
F_l(\{h_e\}_{e\in E(\gamma)})=\mu_{0l}(f_l)
\ea
since the Jacobian of the Haar measure with respect to the left or right
translation map
on $G$ equals unity and where we have defined a new integration variable by
$h_{e_1\circ e_2}:=h_{e_1}h_{e_2}$.\\
$\Box$\\
It follows from theorem \ref{th3.10} that the family $(\mu_{0l})$ defines
a regular Borel probability measure on $\overline{X}$.

We now can equip the quantum configuration space $\ab$ with a Hilbert
space structure.
\begin{Definition} \label{def3.18a}  ~~~~~~~~~~~\\
The Hilbert space ${\cal H}^0$ is defined as the space of square integrable
functions over $\ab$ with respect to the uniform measure $\mu_0$, that is
\be \label{3.95a}
{\cal H}_0:=L_2(\ab,d\mu_0)
\ee
\end{Definition}
Notice that since we have identified cylindrical functions over
$\agb$ with gauge invariant, cylindrical functions over $\ab$ the measure
$\mu_0$ can also be defined as a measure on $\agb$: Simply restrict
the $\mu_{0l}$ to the invariant elements which still defines a
positive linear functional on $C([X_l]_l)$ and then use the
Riez representation theorem. It is easy to check that the obtained measure
coincides with the restriction of $\mu_0$ to $\agb$ with $\sigma-$algebra
given by the sets $U\cap \agb$ where $U$ is measurable in $\ab$. We
will denote the restricted and unrestricted measure by the same symbol
$\mu_0$.

At this point the physical significance of the Hilbert space is unclear
because we did not show that it supports a representation of the
canonical commutation relations and adjointness relations. We will
demonstrate this to be the case in section \ref{s4}.

\subsection{Functional Calculus on a Projective Limit}
\label{s3.3a}

This section rests on the simple but powerful obeservation that
in the case of interest the projections $p_{l'l}$ are not only continuous
and surjective but also analytic. This can be seen by using the
bijection (\ref{3.15}) between $X_l$ and $G^n$ for some $n$ and using
the standard differentiable structure on $G^n$. \\
\\
{\it Functions}\\
We have seen
that we can identify $C(\overline{X})$ with the (completion of the) space
of) cylindrical functions $f=[f_l]/\sim=p_l^\ast f_l,\;f_l\in C(X_l)$.
This suggests to proceed analogously with the other differentiability
categories. Let $n\in\{0,1,2,..\}\cup\{\infty\}\cup \{\omega\}$ then
we define
\be \label{3.95.1}
\mbox{Cyl}^n(\overline{X}):=(\bigcup_{l\in {\cal L}} C^n(X_l))/\sim
\ee
That is, a typical element $f=[f_l]_\sim\in \mbox{Cyl}^n(\overline{X})$ can
be thought of as an equivalence class of elements of the form
$f_l\in C^n(X_l)$ where $f_l\sim f_{l'}$ iff there exists
$l,l'\prec l^\dprime$ such that
$p_{l^\dprime l}^\ast f_l=p_{l^\dprime l'}^\ast f_{l'}$. As in the previous
section, the existence of one such $l^\dprime$ implies that this equation
holds for all $l,l'\prec l^\dprime$. Notice that $f_l\in C^n(X_l)$ implies
$p_{l' l}^\ast f_l\in C^n(X_l)$ due to the analyticity of the projections,
this is where their analyticity becomes important. \\
\\
{\it Differential Forms}\\
In fact, since the Grassman algebra of differential forms on $X_l$ is
generated by finite linear combinations of monomials of the form
$f_l^{(0)}\; df^{1)}_l\wedge..\wedge df^{(p)}_l$ with
$0\le p\le \dim(X_l),\;f^{(0)}_l\in C^n(X_l),\;
f^{(k)}_l\in C^{(n+1)}(X_l),\;k=1,..,p$ we can define the space of
cylindrical $p-$forms and the cylindrical Grassman algebra by
\be \label{3.95.2}
\bigwedge^n(\overline{X})=(\bigcup_{l\in {\cal L}} \bigwedge^n(X_l))/\sim
\ee
because the pull-back commutes with the exterior derivative, that is,
$p_{l^\dprime l}^\ast f_l=p_{l^\dprime l'} f_{l'}$ implies
$p_{l^\dprime l}^\ast df_l=d(p_{l^\dprime l'} f_{l'})$, in other words,
the exterior derivative is a well-defined operation on the Grassmann
algebra. Notice that if $\omega=[\omega_l]_\sim\in \bigwedge(\overline{X})$
and $\omega_l$ has degree $p$ then also $p_{l' l}^\ast \omega_l$
has degree $p$, hence the degree of forms on $\overline{X}$ is
well-defined. \\
\\
{\it Volume Forms}\\
The case
of volume forms is slightly different because a volume form
on an orientable $X_l$ is a nowehere vanishing differential form of degree
$\dim(X_l)$ so that the degree varies with
the label $l$. However, volume forms on $\overline{X}$ (even in the
non-orientable case) are nothing else
than cylindrically defined measures satisfying the consistency
condition $\mu_{l'}\circ p_{l' l}=\mu_l$ for all $l\prec l'$. If they
are probability measures we can extend them to $\sigma-$additive measures
on $\overline{X}$ using the Riesz-Markow theorem as in the previous
section.  \\
\\
{\it Vector Fields}\\
Differentiable
vector fields $V^n(X_l)$ on $X_l$ are conveniently
introduced algebraically
on $X_l$ as derivatives, that is, they are
linear functionals $Y_l;\;C^{n+1}(X_l)\to C^{n}(X_l)$ annihilating
constants and satisfying the Leibniz rule. We want to proceed
similarly with respect to $\overline{X}$ and the first impulse
would be to define
$$
V^n(\overline{X})=(\bigcup_{l\in {\cal L}} V^n(X_l))/\sim
$$
where the equivalence relation is given through the push-forward map.
The push-forward is defined by
\be \label{3.95.3}
(p_{l' l})_\ast:\;V^n(X_{l'})\to V^n(X_{l'});
p_{l'l}^\ast([(p_{l' l})_\ast Y_{l'}](f_l)):=Y_{l'}(p_{l' l}^\ast f_l)
\ee
and we could try to define $Y_l\sim Y_{l'}$ iff for any
$l^\dprime\prec l,l'$ we have
$(p_{l' l^\dprime})_\ast Y_{l'}=(p_{l l^\dprime})_\ast Y_l$. The problem
with this definition is that the push-forward moves us ``down" in the
directed label set $\cal L$ instead of ``up" as is the case with the
pull-back so that it is not guaranteed that, given $l,l'$ there exists any
$l^\dprime$ at all that satisfies $l,l'\prec l^\dprime$ whence the
consistency condition might be empty. This forces us to adopt
a different strategy, namely to define $V^n(\overline{X})$ as
projective nets $(Y_l)_{l_0\prec l\in {\cal L}}$ with the consistency
condition
\be \label{3.95.4}
(p_{l' l})_\ast Y_{l'}=Y_l\Leftrightarrow
p_{l' l}^\ast [Y_l(f_l)]=Y_{l'}(p_{l' l}^\ast f_l)\;\forall\;
f_l\in C^n(X_l)\;l_0\prec l\prec l'
\ee
The necessity to restrict attention to $l_0\prec l$ is that it may not
be possible or necessary to define $Y_l$ for all $l\in {\cal L}$ or to
have (\ref{3.95.4}) satisfied.
This question never came up of course for the pull-back. Notice
that (\ref{3.95.4}) means that if $f_{l'}=p_{l' l}^\ast f_l$ then
$Y_{l'}(f_{l'})=p_{l' l}^\ast Y_l(f_l)$ for $l_0\prec l\prec l'$, that is
consistently defined vector fields map cylindrical fucntions to
cylindrical functions.

It is clear that for $f=[f_l]_\sim=p_l^\ast f_l$ with $l_0\prec l$
the formula
\be \label{3.95.5}
Y(p_l^\ast f_l):=p_l^\ast Y_l(f_l)=:p_l^\ast [(p_l)_\ast Y](f_l)
\ee
is well-defined for suppose that $f_l\sim f_l'$ with $l_0\prec l'$ then
we find $l_0\prec l, l'\prec l^\dprime$ such that
$p_{l^\dprime l}^\ast f_l=p_{l^\dprime l'}^\ast f_{l'}$
whence, using $p_{l^\dprime l}\circ p_{l^\dprime}=p_l,
p_{l^\dprime l'}\circ p_{l^\dprime}=p_{l'}$
\be \label{3.95.6}
p_{l'}^\ast Y_{l'}(f_{l'})
=p_{l^\dprime}^\ast p_{l^\dprime l'}^\ast Y_{l'}(f_{l'})
=p_{l^\dprime}^\ast Y_{l^\dprime}(p_{l^\dprime l'}^\ast f_{l'})
=p_{l^\dprime}^\ast Y_{l^\dprime}(p_{l^\dprime l}^\ast f_{l})
=p_l^\ast Y_l(f_l)
\ee
{\it Lie Brackets}\\
Suppose that $Y=(Y_l)_{l_0\prec l\in {\cal L}},
Y'=(Y'_l)_{l'_0\prec l\in {\cal L}}\in V^n(\overline{X})$ are consistently
defined vector fields. We certainly find $l_0,l_0'\prec l_0^\dprime$
and claim that $[Y,Y']:=([Y_l,Y_l])_{l_0^\dprime l\in {\cal L}}\in
V^{n-1}(\overline{X})$ is again consistently defined. To see this,
consider $l_0^\prec l\prec l'$ then for any $f_l\in C^n(X_l)$ we have
due to $l_0\prec l$ and $l_0'\prec l$
\be \label{3.95.7}
p_{l' l}^\ast([Y_l,Y'_l](f_l))=
Y_{l'}[p_{l' l}^\ast (Y'_l(f_l))]-Y'_{l'}[p_{l' l}^\ast (Y_l(f_l))]
=[Y_{l'},Y'_{l'}](p_{l' l}^\ast f_l)
\ee
{\it Vector Field Divergences}\\
Recall that the Lie derivative of an element $\omega_l\in \bigwedge^n(X_l)$
with respect to a vector field $Y_l\in V^n(X_l)$ is defined by
$L_{Y_l}\omega_l=[i_{Y_l} d+d i_{Y_l}]\omega_l$ where
$$
i_{Y_l} f^{(0)}_l\;df^{(1)}_l\wedge ..\wedge df^{(p)}_l
=f^{(0)}_l\sum_{k=1}^p (-1)^{k+1} Y_l(f^{(k)}_l)
\;df^{(1)}_l\wedge ..df^{(k-1)}_l \wedge df^{(k+1)}_l\wedge ..\wedge
df^{(p)}_l
$$
denotes contraction of forms with vector fields, annihilating zero
forms. Let now $\mu_l$ be a volume form on $X_l$. Since $X_l$ is finite
dimesional, all smooth volume forms are absolutely continuous with respect to
each other and there exists a well-defined function, called the
divergence of $Y_l$ with respect to $\mu_l$, uniquely defined by
\be \label{3.95.8}
L_{Y_l}\mu_l=:[\mbox{div}_{\mu_l} Y_l] \mu_l
\ee
We say that a vector field $Y=(Y_l)_{l_0\prec l\in {\cal L}}$ is
compatible with a volume form $\mu=(\mu_l)_{l\in {\cal L}}$ provided
that the family of divergences defines a cylindrical function, that is
\be \label{3.95.9}
p_{l' l}^\ast [\mbox{div}_{\mu_l} Y_l]=\mbox{div}_{\mu_{l'}} Y_{l'}
\; \forall l_0\prec l\prec l'
\ee
Hence there exists a well defined cylindrical function
$\mbox{div}_\mu Y:=[\mbox{div}_{\mu_l} Y_l]_\sim$, called the divergence
of $Y$ with respect to $\mu$.
\begin{Lemma}  \label{la3.9}
Let $\mu$ be a smooth volume form,
$Y,Y'$ $\mu-$compatible vector fields
and $f,f'\in \mbox{Cyl}^1(\overline{X})$ cylindrical functions
on $\overline{X}$.\\
i)\\
If $\partial X_l=\emptyset$ has no boundary then
\be \label{3.95.10}
\int_{\overline{X}} \mu\; f\;Y(f')=-
\int_{\overline{X}} \mu\; (Y(f)+f\;[\mbox{div}_\mu Y])f'
\ee
ii)\\
The Lie bracket $[Y,Y']$ is again $\mu-$compatible and
\be \label{3.95.11}
\mbox{div}_\mu [Y,Y']=Y(\mbox{div}_\mu Y')-Y'(\mbox{div}_\mu Y)
\ee
\end{Lemma}
Proof of Lemma \ref{la3.9}:\\
i)\\
We find $l_0,l_0'\prec l$ such that $f=p_l^\ast f_l,f'=p_l^\ast f'_l$.
Then
\ba \label{3.95.12}
\mu(f Y(f'))&=& \mu([p_l^\ast f_l] [p_l^\ast Y_l(f'_l)])
=\mu_l(f_l L_{Y_l}[f'_l])=\int_{X_l}
\{L_{Y_l}[\mu_l f_l f'_l]-(L_{Y_l}[\mu_l f_l]) f'_l\}
\nonumber\\
&=& \int_{X_l}
\{d\;i_{Y_l}[\mu_l f_l f'_l]
-\mu_l (Y_l(f_l)+f_l\;[\mbox{div}_{\mu_l} Y_l]) f'_l\}
\nonumber\\
&=&-\mu((Y(f)+f\;[\mbox{div}_\mu Y])f')
\ea
where in the third line we have applied Stokes' theorem and that the Lie
derivative satisfies the Leibniz rule.\\
ii)\\
We find $l_0,l_0'\prec l_0^\dprime$ so that
$([Y_l,Y'_l])_{l_0^\dprime\prec l\in {\cal L}}$ is consistently defined
as shown above. From the fact that the Lie derivative is an isomorphism
between the Lie algebra of vector fields and the derivatives
respectively on $C^n(X_l)$, $L_{[Y_l,Y'_l]}=[L_{Y_l},L_{Y'_l}]$,
and the fact that Lie derivation and exterior derivation commute,
$[d, L_{Y_l}]=0$, we have
\ba \label{3.95.13}
(\mbox{div}_{\mu_l}[Y_l,Y'_l])\mu_l&=&
[L_{Y_l},L_{Y'_l}] (\mu_l)
=L_{Y_l} ([\mbox{div}_{\mu_l} Y'_l] \mu_l)
-L_{Y'_l} ([\mbox{div}_{\mu_l} Y_l] \mu_l)
\nonumber\\&=&
[Y_l(\mbox{div}_{\mu_l} Y'_l)
-Y'_l(\mbox{div}_{\mu_l} Y_l)] \mu_l
\ea
It follows from the consistency of the $Y_l$ and the compatibility with
the $\mu_l$ that for $l\prec l'$
\be \label{3.95.14}
p_{l l'}^\ast Y_l(\mbox{div}_{\mu_l} Y'_l)
=Y_{l'}(p_{l l'}^\ast(\mbox{div}_{\mu_l} Y'_l))
=Y_{l'}(\mbox{div}_{\mu_{l'}} Y'_{l'})
\ee
$\Box$\\
{\it Momentum Operators}\\
Let $Y$ be a vector field compatible with $\sigma-$additive measure
(volume form) $\mu$ such that it is together with its divergence
$\mbox{div}_\mu Y$ is real valued. We consider the Hilbert space
${\cal H}_\mu:=L_2(\overline{X},\mu)$ and define the momentum operator
\be \label{3.95.15}
P(Y):=i(Y+\frac{1}{2} (\mbox{div}_\mu Y) 1_{{\cal H}_\mu})
\ee
with dense domain $D(P(Y))=\mbox{Cyl}^1(\overline{X})$. From
(\ref{3.95.10}) we conclude that for $f,f'\in D(P(Y))$
\be \label{3.95.16}
<f,P(Y) f'>_\mu=\mu(\overline{f} P(Y)f)=\mu(\overline{P(Y)f} f)
=<P(Y)f, f'>_\mu
\ee
from which we see that
$$
D(P(Y))\subset D(P(Y)^\dagger):=\{f\in {\cal H}_\mu;\;
\sup_{||f'||>0} |<f,P(Y) f'>|/||f'||<\infty
\mbox{ and } D(P(Y))^\dagger_{|D(P(Y))}=P(Y)
$$
whence $P(Y)$ is a symmetric unbounded operator.

Finally we notice that if $Y,Y'$ are both $\mu-$compatible
then
\be \label{3.95.16a}
[P(Y),P(Y')]=iP([Y,Y'])
\ee
by a straightforward computation using lemma \ref{la3.9}.\\
\\
Remark:\\
That div$_{\mu_l} Y_l$ is a cylindrical function is a sufficient
criterion for $P(Y)$ to be well defined, but it is too strong a
requirement
because it means that for given $l$ on any other $l\prec l'$
the function $div_{\mu_{l'}} Y_{l'}\equiv p_{l' l}^\ast (div_{\mu_l} Y_l)$
does not depend
on the additional degrees of freedom contained in $X_{l'}$. That is, if
not some special graphs are too be distinguished then div$_\mu Y=const.$
is the only possibility.
So compatibility between $\mu$ and $Y$ is only sufficient but has not
been shown to be necessary in order to define interesting momentum
operators. It would be important to replace the compatibility criterion
by a weaker one.\\
\\
{\it General Operators}\\
More generally we have the following abstract situation:\\
a)\\
We have a partially ordered and directed index set ${\cal L}$,
a family of Hilbert spaces ${\cal H}_l:={\cal H}_{\mu_l}:=
L_2(X_l,d\mu_l)$
and isometric monomorphisms (linear injections)
\be \label{3.95.17}
\hat{U}_{l l'}:\;{\cal H}_l\to {\cal H}_{l'}
\ee
for every $l\prec l'$ which in our special case is given by
$\hat{U}_l f_l:=p_{l'l}^\ast f_l$.
The isometric monomorphisms satisfy the compatibility condition
\be \label{3.95.18}
\hat{U}_{l' l^\dprime}\hat{U}_{l l'}=\hat{U}_{l l^\dprime}
\ee
for any $l\prec l'\prec l^\dprime$ due to
$p_{l' l}\circ p_{l^\dprime l'}=p_{l^\dprime l}$. A system
$({\cal H}_l,\hat{U}_{l l'})_{l\prec l'\in {\cal L}}$ of this sort
is called a directed system of Hilbert spaces. A Hilbert space
${\cal H}$ is called the inductive limit of a directed system of
Hilbert spaces provided that there exist isometric monomorphisms
\be \label{3.95.18a}
\hat{U}_l:\;{\cal H}_l\to {\cal H}
\ee
for any $l\in {\cal L}$ such that the compatibility condition
\be \label{3.95.19}
\hat{U}_{l'}\hat{U}_{l l'}=\hat{U}_l
\ee
holds. In our case, obviously $\hat{U}_l f_l:=p_l^\ast f_l$ provides
these monomorphisms so that we have displayed ${\cal H}_\mu$ as the
inductive limit of the ${\cal H}_{\mu_l}$.

Likewise we have a family of operators $\hat{O}_l=P(Y_l)$ with
dense domain $D(\hat{O}_l)=C^1(X_l)$ in ${\cal H}_l$ which are
defined for a cofinal subset ${\cal L}(\hat{O})=\{l\in {\cal L};\;
l_0\prec l\}$ (that is, for any $l\in {\cal L}$ there exists
$l\prec l'\in {\cal L}(\hat{O})$) of $\cal L$. These families of
domains and operators satisfy the following compatibility conditions:
\be \label{3.95.20}
\hat{U}_{l l'} D(\hat{O}_l)\subset D(\hat{O}_{l'})
\ee
for any $l\prec l'\in {\cal L}(\hat{O})$ since
$p_{l'l}^\ast C^1(X_l)\subset C^1(X_{l'})$ (the pull-back of functions
is $C^1$ with respect to the $X_l$ arguments but $C^\omega$ with respect
to the remaining arguments in $X_{l'})$. Furthermore
\be \label{3.95.21}
\hat{U}_{l l'}\hat{O}_l=\hat{O}_{l'}\hat{U}_{l l'}
\ee
for any $l\prec l'\in {\cal L}(\hat{O})$ since
$p_{l' l}^\ast(Y_l(f_l)+[\mbox{div}_{\mu_l} Y_l] f_l/2)=
(Y_{l'}(p_{l' l}^\ast f_l)+[\mbox{div}_{\mu_{l'}} Y_{l'}] p_{l' l}^\ast f_l/2)$
due to consistency and compatibility. A structure of this kind is called
a directed system of operators. An operator $\hat{O}$ with dense
domain $D(\hat{O})$ is called
the inductive limit of a directed system of operators provided the
above defined isometric isomorphisms interact with domains and operators in
the expected way, that is,
\be \label{3.95.22}
\hat{U}_{l} D(\hat{O}_l)\subset D(\hat{O})
\ee
and
\be \label{3.95.23}
\hat{U}_l\hat{O}_l=\hat{O}\hat{U}_l
\ee
In our case this is by definition statisfied since $p_l^\ast C^1(X_l)
\subset \mbox{Cyl}^1(\overline{X})$ and
$p_l^\ast(Y_l(f_l)+[\mbox{div}_{\mu_l} Y_l] f_l/2)\equiv
(Y(p_l^\ast f_l)+[\mbox{div}_\mu Y] p_l^\ast f_l/2)$.

It turns out that directed systems of Hilbert spaces and operators
always have an inductive limit which is unique up to unitary equivalence.
\begin{Lemma} \label{la3.10}   ~~~~~~~~~~~~~~~~\\
i)\\
Given directed systems of Hilbert spaces
$({\cal H}_l,\hat{U}_{ll'})_{l\prec l'\in{\cal L}}$ and operators
$(\hat{O}_l,D(\hat{O}_l),\hat{U}_{ll'})_{l\prec l'\in{\cal L}(\hat{O})}$
with a cofinal index set ${\cal L}(\hat{O})$, there is a, up to unitary
equivalence, unique inductive
limit Hilbert space $({\cal H},\hat{U}_l)_{l\in{\cal L}}$ as well as
a unique inductive limit operator
$(\hat{O},D(\hat{O}),\hat{U}_l)_{l\in {\cal L}(\hat{O})}$ densely defined
on the inductive limit Hilbert space.\\
ii)\\
If the $\hat{O}_l$ are essentially self-adjoint with core $D(\hat{O}_l)$
then $\hat{O}$ is essentially self-adjoint with core $D(\hat{O})$.\\
iii)\\
If the $\hat{O}_l$ are essentially self-adjoint then
$(\hat{O}'_l,D(\hat{O}'_l),\hat{U}_{ll'})_{l\prec l'\in{\cal L}(\hat{O})}$
is a directed system of operators where $O'_l$ denotes the self-adjoint
extension of $\hat{O}_l$.
\end{Lemma}
Proof of Lemma \ref{la3.10}:\\
i)\\
In the case of bounded operators, that is $D(\hat{O}_l)={\cal H}_l$,
part i) is standard in operator theory, see e.g. vol. 2 of
the first reference of \cite{53}
for more details
and an extension of the theorem to directed systems of $C^\ast-$algebras
and von Neumann algebras which have a unique inductive limit up to
algebra isomorphisms. \\
\\
We consider the vector space $V$ of equivalence classes of nets
$f=(f_l)_{l_0\prec l \in {\cal L}(f)}$ for some cofinal
${\cal L}(f)\subset{\cal L}$
with $f_l\in {\cal H}_l$
satisfying $\hat{U}_{ll'} f_l=f_{l'}$ for any $l_0\prec l\prec l'$
and where $f\sim f'$ are equivalent if $f_l=f'_l$ for all
$l\in {\cal L}(f)\cap {\cal L}(f')$. Let us write $[f]_\sim$ for the
equivalence class of $f$.
We define
\be \label{3.95.24}
\hat{U}_l:\;{\cal H}_l\to V;\; f_l\mapsto
[(\hat{U}_{ll'} f_l)_{l\prec l'\in {\cal L}}]_\sim
\ee
Due to isometry of the $\hat{U}_{ll'}$ the norm on $V$ given by
$||[f]_\sim||:=||f_l||_l$ is independent of the choice of
$l\in {\cal L}(f)$,
in particular, $\hat{U}_l$ becomes an isometry. We have for $l\prec l'$
$$
\hat{U}_{l'}\hat{U}_{ll'} f_l=
[(\hat{U}_{l'l^\dprime}\hat{U}_{ll'}f_l)_{l'\prec l^\dprime}]_\sim
=[(\hat{U}_{l l^\dprime} f_l)_{l'\prec l^\dprime}]_\sim
=[(\hat{U}_{l l^\dprime} f_l)_{l\prec l^\dprime}]_\sim
=\hat{U}_l f_l
$$
Finally we consider the subspace of $V$ given by the span of elements
of the form $\hat{U}_l f_l$ with $f_l\in {\cal H}_l$ and complete it
to arrive at the Hilbert space
\be \label{3.95.25}
{\cal H}:=\overline{\bigcup_l \hat{U}_l {\cal H}_l}
\ee
to which $\hat{U}_l$ can be extended uniquely as an isometric monomorphism
by continuity.
To see the uniqueness one observes that given another inductive
limit $({\cal H}',\hat{V}_l)$ we may define
$W_l:=\hat{V}_l \hat{U}_l^{-1}:\;\hat{U}_l {\cal H}_l\to
\hat{V}_l {\cal H}_l$ which one checks to be an isometry. Also
for $l\prec l'$ we have
$W_{l'}\hat{U}_l=\hat{W}_{l'}\hat{U}_{l'}\hat{U}_{ll'}=
\hat{V}_{l'}\hat{U}_{ll'}=V_l=\hat{W}_l \hat{U}_l$, in other words,
$W_{l'}$ is an extension of $W_l$ for $l\prec l'$. This means that
we have a densely defined isometry
$\hat{W}:\;\bigcup \hat{U}_l{\cal H}_l\to \bigcup \hat{V}_l{\cal H}_l$
defined by $\hat{W}_{|\hat{U}_l{\cal H}_l}=W_l$ which extends by continuity
uniquely to an isometry between the two Hilbert spaces.

Next, define an operator on the dense subspace of ${\cal H}$ given by
$D(\hat{O}):=\bigcup_{l\in{\cal L}(\hat{O})}\hat{U}_l D(\hat{O}_l)$
\be \label{3.95.26}
\hat{O} [(f_l)_{l\in {\cal L}(\hat{O})}]_\sim:=
[(\hat{O}_l f_l)_{l\in {\cal L}(\hat{O})}]_\sim
\ee
Since ${\cal L}(\hat{O})\cap \{l'\in {\cal L};\;l\prec l'\}
=\{l'\in {\cal L}(\hat{O});\;l\prec l'\}$ is cofinal we have
\ba \label{3.95.27}
\hat{O}\hat{U}_l f_l & =& \hat{O}
[(\hat{U}_{ll'} f_l)_{l\prec l'\in {\cal L}}]_\sim
=\hat{O}[(\hat{U}_{ll'} f_l)_{l\prec l'\in {\cal L}(\hat{O})}]_\sim
=[(\hat{O}_{l'}\hat{U}_{ll'} f_l)_{l\prec l'\in {\cal L}(\hat{O})}]_\sim
\nonumber\\
& = &
[(\hat{U}_{ll'} \hat{O}_l f_l)_{l\prec l'\in {\cal L}(\hat{O})}]_\sim
=[(\hat{U}_{ll'} \hat{O}_l f_l)_{l\prec l'\in {\cal L}}]_\sim
\nonumber\\
& = & \hat{U}_l\hat{O}_l f_l
\ea
ii)\\
By the basic criterion of essential self-adjointness we know that
$(\hat{O}_l\pm i\cdot 1_{{\cal H}_l})D(\hat{O}_l)$ is dense in ${\cal H}_l$.
It follows that
\ba \label{3.95.28}
(\hat{O}\pm i\cdot 1_{{\cal H}})D(\hat{O})
&=& \bigcup_{l\in {\cal L}(\hat{O})}
(\hat{O}\pm i\cdot 1_{{\cal H}})\hat{U}_l D(\hat{O}_l)
\nonumber\\
&=& \bigcup_{l\in {\cal L}(\hat{O})}
\hat{U}_l (\hat{O_l}\pm i\cdot 1_{{\cal H}_l})D(\hat{O}_l)
\ea
hence $(\hat{O}\pm i\cdot 1_{{\cal H}})D(\hat{O})$ is dense in
${\cal H}$ so that $\hat{O}$ is essentially self-adjoint by the
basic criterion of essential self-adjointness.\\
iii)\\
Recall that the self-adjoint extension $\hat{O}'_l$ of an essentially
self-adjoint
operator $\hat{O}_l$ with core $D(\hat{O}_l)$ is unique and given by
its closure, that is, the set $D(\hat{O}'_l)$ given by those
$f_l\in {\cal H}_l$ such that
$(f_l,\hat{O}_l f_l)\in \overline{\Gamma}_{\hat{O}_l}$, the closure
in ${\cal H}_l\times{\cal H}_l$ of the graph
$\Gamma_{\hat{O}_l}=\{(f_l,\hat{O}_l f_l);\;f_l\in D(\hat{O}_l)\}$
of $\hat{O}_l$ with respect to the norm $||(f_l,f_l')||^2=
||f_l||^2+||f'_l||^2$.

To see that $\hat{U}_{ll'} D(\hat{O'}_l)\subset D(\hat{O}'_{l'})$
we notice that $\hat{U}_{ll'} D(\hat{O}_l)\subset D(\hat{O}_{l'})$.
Hence, the closure $D(\hat{O}'_{l'})$ of $D(\hat{O}_{l'})$ will contain
the closure of $\hat{U}_{ll'} D(\hat{O}_l)$ which coincides with
$\hat{U}_{ll'} D(\hat{O}'_l)$ because $\hat{U}_{ll'}$ is bounded.

To see that $\hat{U}_{ll'}\hat{O}'_l=\hat{O}'_{l'}\hat{U}_{ll'}$
holds on $D(\hat{O}'_l)$ we notice
that $\hat{U}_{ll'}\hat{O}_l=\hat{O}_{l'}\hat{U}_{ll'}$ holds on
$D(\hat{O}_l)$. Since $\hat{O'}_l,\hat{O}'_{l'}$ are just the extensions
of $\hat{O}_l,\hat{O}_{l'}$ from $D(\hat{O}_l),D(\hat{O}_{l'})$ to
$D(\hat{O}'_l),D(\hat{O}'_{l'})$ and since $\hat{U}_{ll'}D(\hat{O}'_l)
\subset D(\hat{O}'_{l'})$ the claim follows.\\
$\Box$\\
\\
{\it Passage to the Quotient Space}\\
Finally we consider the case of interest, namely the quotient space
$\agb$ projective limit. The significance of the result $\abgb=\agb$
is that we can identify cylindrical functions on $\agb$ simply with
$\gb-$invariant functions on $\ab$. More precisly, if
$\lambda:\;\gb\times \ab\to \ab;\; A\mapsto \lambda_g(A)$ is the
$\gb-$action and $f\in \mbox{Cyl}^n(\ab)$ is $\gb-$invariant then we
may define $\tilde{f}\in\mbox{Cyl}^n(\agb)$ by
$\tilde{f}([A]):=f(A)=f(\lambda_g(A))$ for all $g\in \gb$ where
$[.]:\;\ab\to\abgb\equiv \agb$ denotes the quotient map.
Thus we define zero forms on $\agb$ as zero forms on $\ab$ which
satisfy $f=\lambda_g^\ast f$ for any $g\in \gb$. Notice that this
is possible for any differentiability category because the
$\gb-$action is evidently not only continuous but even analytic !

Since pull-backs commute with exterior derivation we can likewise
define the Grassman algebra $\bigwedge(\agb)$ as the subalgebra of
$\bigwedge(\ab)$ given by the $\gb-$invariant differential forms,
that is, those that satisfy $\lambda_g^\ast\omega=\omega$ for
all $g\in \gb$ (if $f$ is $\gb-$invariant, so is $df$ because
$\lambda_g^\ast df=d\lambda_g^\ast f=df$).

Next, volume forms on $\agb$ are just $\gb-$invariant volume forms
on $\ab$, that is $(\lambda_g)_\ast\mu=
\mu\circ \lambda_g^{-1}=\mu\circ \lambda_{g^{-1}}=\mu$ for all
$g\in \gb$. Given any volume form $\mu$ on $\ab$
we may derive a measure $\overline{\mu}$ on
$\agb$ by $\overline{\mu}(f):=\mu(f)$ for all
$\gb-$invariant functions $f$ on $\ab$. If we denote the
Haar probability measure on $\gb\equiv \prod_{x\in\sigma} G$ by $\mu_H$
then from $\mu(f)=\mu(\lambda_g^\ast f)=[(\lambda_g)_\ast\mu](f)$
for all $\gb-$invariant measurable functions we find
\be \label{3.95.29}
\overline{\mu}([A])=\int_{\gb} \mu_H(g) \;[(\lambda_g)_\ast\mu](A)
\ee

Finally, we define vector fields on $\agb$ as $\gb-$invariant vector
fields $\ab$, that is, those satisfying $(\lambda_g)_\ast Y=Y$
for all $g\in \gb$, more precisely, if $Y=(Y_l)_{l_0\prec l}$ then
\be \label{3.95.30}
(\lambda^l_g)^\ast ([(\lambda^l_g)_\ast Y_l](f_l)):=
Y_l[(\lambda^l_g)^\ast f_l]=(\lambda^l_g)^\ast (Y_l(f_l))
\ee
for any $f_l\in C^n(\ab)$ and $l_0\prec l$.

\subsection{Density and Support Properties of $\a,\ag$ with Respect to
$\ab,\agb$}
\label{s3.4}

In this section we will see that $\a$ lies topologically dense, but
measure theoretically thin in $\ab$ (similar results apply to $\ag$
with respect to $\agb=\abgb$) with respect to the uniform measure
$\mu_0$. More precisely, there is a dense embedding
(injective inclusion) $\a\to \ab$ but $\a$ is embedded into a measurable
subset of $\ab$ of measure zero. The latter result demonstrates
that the measure is concentrated on non-smooth (distributional)
connections so that $\ab$ is indeed much larger than $\a$.

We have seen in section \ref{s3.1.2} that every element $A\in \a$ defines
an element of $\mbox{Hom}({\cal P},G)$ and that this space can be identified
with the projective limit $\overline{X}\equiv \ab$. Now via the
$C^\ast-$algebraic framework
we know that $\overline{\mbox{Cyl}(\overline{X})}$ can be identified with
$C(\overline{X})$ and the latter space of functions separates the
points of $\overline{X}$ by the Stone-Weierstrass theorem since it is
Hausdorff and compact. The question is whether the smaller set of
functions $\mbox{Cyl}(\overline{X})$ separates the smaller set of points
$\a$. This is almost obvious and we will do it for $G=SU(N)$, other compact
groups can be treated similarly:\\
Let $A\not=A'$ be given
then there exists a point $x\in \sigma$ such that
$A(x)\not=A'(x)$. Take $D=\dim(\sigma)$ edges $e_{x,\alpha}\in {\cal P}$
with
$b(e_{x,\alpha})=x$ and linearly independent tangents
$\dot{e}_{x,\alpha}(0)$ at $x$. Consider the cylindrical function
\be \label{3.96g}
F^\epsilon_x:\;\a\to \Cl;\;A\mapsto \frac{1}{\epsilon^2} \sum_{\alpha,j}
[\mbox{tr}(\tau_j A(e^\epsilon_{x,\alpha}))]^2
\ee
where $\tau_j$ is a basis of Lie$(G)$ with normalization
$\mbox{tr}(\tau_j \tau_k)=-N\delta_{jk}$ and
$e^\epsilon_{x,\alpha}(t)=e_{x,\alpha}(\epsilon t)$. Using smoothness
of $A$ it easy to see that (\ref{3.96}) can be expanded in a convergent
Taylor series with respect to $\epsilon$ with zeroth order component
$\sum_{j, e_\alpha} |A_a^j(x)\dot{e}^a_{x,\alpha}(0)|^2$ whence
$F^\epsilon_x\in \mbox{Cyl}(\overline{X})$ separates our given
$A\not=A'$. The proof for $\a$ replaced by $\ag$ is similar and was
given by Giles \cite{40} and will not be repeated here. In that proof
it is important that $G$ is compact.

We thus have the following abstract situation: A collection
${\cal C}=\mbox{Cyl}(\overline{X})$ of bounded complex valued functions
on a set
$X=\a$ including the constants which separate the points of $X$. The set $X$
maybe equipped with its own topology (e.g. the Sobolov topology that we
defined in section \ref{s2}) but this will be irrelevant for the
following result which is an abstract property of Abelean unital
$C^\ast-$algebras.
\begin{Theorem}  \label{th3.11}   ~~~~~~~~~~~~~~~\\
Let $\cal C$ be a collection of real-valued, bounded functions on a set $X$
which contain the constants and separate the points of $X$.
Let $\overline{{\cal C}}$ be the Abelean, unital $C^\ast-$ algebra
generated from ${\cal C}$ by pointwise addition, multiplication, scalar
multiplication and complex conjugation, completed in the sup-norm.
Then the image of $X$ under its natural embedding into the Gel'fand spectrum
$\overline{X}$ of $\overline{{\cal C}}$ is dense with respect to the
Gel'fand topology on the spectrum.
\end{Theorem}
Proof of Theorem \ref{th3.11}:\\
Consider the following map
\be \label{3.97}
J:\;X\to \overline{X};\;x\mapsto J_x \mbox{ where } J_x(f):=f(x)
\;\forall\; f\in \overline{{\cal C}}
\ee
This is an injection since $J_x=J_{x'}$ implies in particular
$f(x)=f(x')$ for all $f\in {\cal C}$, thus $x=x'$
since ${\cal C}$ separates the
points of $X$ by assumption, hence $J$ provides an embedding.

Let $\overline{J(X)}$ be the closure of $J(X)$ in the Gel'fand topology
on $\overline{X}$ of pointwise convergence on $\overline{{\cal C}}$.
Suppose that $\overline{X}-\overline{J(X)}\not=\emptyset$ and take any
$\chi\in\overline{X}-\overline{J(X)}$. Since $\overline{X}$ is a compact
Hausdorff space we find $a\in C(\overline{X})$ such that
$1=a(\chi)\not=a(J_x)=0$ for any $x\in X$ by Urysohn's lemma.
(In Hausdorff spaces one point sets are closed, hence $\{\chi \}$ and
$\overline{J(X)}$ are disjoint closed sets and finally compact Hausdorff
spaces are normal spaces).

Since the Gel'fand map $\bigvee:\;\overline{{\cal C}}\to C(\overline{X})$
is an isometric isomorphism we find $f\in \overline{{\cal C}}$ such
that $\check{f}=a$. Hence $0=a(J_x)=\check{f}(J_x)=J_x(f)=f(x)$
for all $x\in X$, hence $f=0$, thus $a\equiv 0$ contradicting
$a(\chi)=1$. Therfore $\chi$ in fact does not exist whence
$\overline{X}=\overline{J(X)}$.\\
$\Box$\\
Of course in our case $\overline{{\cal C}}=\overline{\mbox{Cyl}(\ab)}$
and $\overline{X}=\ab$.\\
\\
Our next result is actually much stronger than merely showing that
$\a$ is contained in a measurable subset of $\ab$ of $\mu_0-$measure
zero.

Let $e$ be an edge and if $e(t)$ is a representative curve then
consider the family of segments $e_s$ with $e_s(t):=e(st),\;s\in [0,1]$.
Consider the map
\be \label{3.98}
h^e:\;\ab\to \mbox{Fun}([0,1],G);\;A\mapsto h^e_A \mbox{ where }
h^e_A(s):=A(e_s)
\ee
The set Fun$([0,1],G)$ of all functions from the interval $[0,1]$ into
$G$ (no continuity assumptions) can be thought of as the uncountable
direct product $G^{[0,1]}:=\prod_{s\in [0,1]} G$ via the bijection
$E:\;\mbox{Fun}([0,1],G)\to G^{[0,1]};\;h\to (h_s:=h(s))_{s\in [0,1]}$.
The latter space can be equipped with the Tychonov topology
generated by the
open sets on $G^{[0,1]}$ which are generated from the sets
$P_s^{-1}(U_s)=[\prod_{s'\not= s} G]\times U_s$ (where
$U_s\subset G$ is open in $G$) by finite intersections and arbitrary
unions. Here $P_s:\; G^{[0,1]}\to G$ is the natural projection.
Now the pre-image of such sets under $h^e$ is given by
\ba \label{3.99}
&& (h^e)^{-1}(P_s^{-1}(U_s))=\{A\in \ab;\; h^e_A\in p_s^{-1}(U_s)\}
\nonumber\\
&=& \{A\in \ab;\; h^e_A(s)\in U_s,\; h^e_A(s')\in G\mbox{ for } s'\not=s\}
\nonumber\\
&=&\{A\in \ab;\; A(e_s)\in U_s\}=p_{e_s}^{-1}(U_s)
\ea
where $p_{e_s}:\;\ab\to \mbox{Hom}(e_s,G)$ is the natural projection
in $\ab$. Since $\ab$ is equipped with the Tychonov topology, the maps
$p_{e_s}$ are continuous and since $\ab$ is equipped with the Borel
$\sigma$ algebra, continuous functions (pre-images of open sets are
open) are automatically measurable (pre-images of open sets are measurable.
Hence we have shown that $h^e$ is a measurable map.

Let $f$ be a function on $G^{[0,1]}$, that is, a complex valued function
$h\mapsto f(\{h_s\}_{s\in [0,1]})$. We have an associated map of the
form (\ref{3.15}), that is,
$\rho_{l^e}:\; X_{l^e}\to G^{[0,1]};\;
A_{l^e} \mapsto (A_{l^e}(e_s)=h^e_A(s))_{s\in [0,1]}$ where $l^e$ is the
subgroupoid generated by the algebraically independent edges $e_s$.
Thus $h^e=\rho_{l^e}\circ p_{l^e}$.
The push-forward of the uniform measure
$\nu:=h^e_\ast \mu_0=\mu_0\circ (h^e)^{-1}$ is then the measure on
$G^{[0,1]}$ given by
\ba \label{3.100}
\int_{G^[0,1]} d\nu(h) f(h)
&=&\mu_0((h^e)^\ast f)=\mu_{0l^e} (\rho_{l^e}^\ast f)=
\int_{G^{[0,1]}} \prod_{s\in [0,1]} d\mu_H(h_{e_s})
f(\{h_{e_s}\}_{s\in [0,1]})
\nonumber\\
&\equiv& \int_{G^{[0,1]}} \prod_{s\in [0,1]} d\mu_H(h_s)
f(\{h_s\}_{s\in [0,1]})
\ea
\begin{Theorem} \label{th3.12}  ~~~~~~~~~~~~~~\\
The measure $\mu_0$ is supported on the subset $D_e$ of $\ab$ defined
as the set of those $A\in \ab$ such that $h^e_A$ is nowhere continuous
on $[0,1]$.
\end{Theorem}
Proof of Theorem \ref{th3.12}:\\
Trivially
\ba \label{3.101}
D_e &=& \{A\in\ab;\;h^e_A \mbox{ nowhere continuous in }[0,1] \}
\\
&=& (h^e)^{-1}(\{h\in G^{[0,1]};\;s\mapsto h_s
\mbox{ nowhere continuous in }[0,1] \}=:(h^e)^{-1}(D)
\nonumber
\ea
If we can show that $D$ contains a measurable set of $\nu-$measure one or
that $G^{[0,1]}-D$ is contained in a measurable set $D'$ of $\nu-$measure
zero then we have shown that $D_e$ contains a measurable set
$D_e'=(h^e)^{-1}(G^{[0,1]}-D')$ of measure
one because $\mu_0(D_e)=[\mu_0\circ (h^e)^{-1}](G^{[0,1]}-D')=
\nu(G^{[0,1]}-D')=1$ and because $h^e$ is measurable (since
$G^{[0,1]}$ is equipped with the Borel $\sigma-$algebra). In other
words, $D_e$ will be a support for $\mu_0$.

Let us then show that
$G^{[0,1]}-D=\{h\in G^{[0,1]};\; \exists s_0\in [0,1]\ni\;h
\mbox{ continuous at }s_0\}$ is contained in a measurable set of
$\nu-$measure zero. Let $h_0\in G^{[0,1]}-D$, then we find
$s_0\in [0,1]$ such that $h_0$ is continuous at $s_0$. Fix any
$0<r<1$ and consider an open cover of $G$ by sets $U$ with Haar
measure $\mu_H(U)=r$. Since $G$ is compact, we find a finite subcover,
say $U_1,..,U_N$. Now there is $k_0\in \{1,..,N\}$ such that
$h_0(s_0)\in U_{k_0}$. By definition of continuity at a point we find an
open interval $I\subset [0,1]$ such that $h(I)\subset U_{k_0}$.
This motivates to consider the subsets
$S_k:=\{h\in G^{[0,1]};\;\exists I\subset [0,1] \mbox{ open }\ni
h(I)\subset U_k\}\subset G^{[0,1]}$ and obviously $h_0\in S_{k_0}$.
Our aim is to show that these sets are contained in measure zero sets.

Let $B(q,1/m):=\{s\in [0,1];\;|s-q|<1/m\}$ with $q\in \Ql,\;m\in \Nl$.
It is easy to show that these sets are a countable basis for the topology for
$[0,1]$ (every open set can be obtained by arbitrary unions and finite
intersections). Hence any open interval is given as a countable union
of these open balls, i.e. $I=\bigcup_{B(q,m)\subset I} B(q,m)$.
Since $h(I\cup J)=h(I)\cup h(J)$ we have
\ba \label{3.102}
S_k &=&\{h\in G^{[0,1]};\;\exists I\subset [0,1]\ni
\bigcup_{B(q,m)\subset I} h(B(q,m))\subset U_k\}=
\bigcup_{(q,m)\in (\Ql\times \Nl)_k} S_{k,q,m}
\nonumber\\
S_{k,q,m}&:=&\{h\in G^{[0,1]};\; h(B(q,m))\subset U_k\}
\ea
where $(\Ql\times\Nl)_k$ are defined to be the subsets of rational and
natural numbers $(q,m)$ respectively such that $S_{U_k,q,m}\not=\emptyset$.
(We could also remove that restriction).

We now show that $S_{k,q,m}$ is contained in a measure zero set.
Let $(s_n)$ be a sequence of points in $B(k,q,m)$. Then
$S_{k,q,m}\subset\{h\in G^{[0,1]};\;h(s_n)\in U_k \;\forall s_n\}
=\cap_n \{h\in G^{[0,1]};\;h(s_n)\in U_k\}$. Now the sets
$\{h\in G^{[0,1]};\;h(s_n)\in U_k\}=P_s^{-1}(U_k)$ are measurable because
$P_s$ is continuous and $U_k$ is open, hence so is
$\cap_n \{h\in g^{[0,1]};\;h(s_n)\in U_k\}$. But
\be \label{3.103}
\nu(\cap_n \{h\in G^{[0,1]};\;h(s_n)\in U_k\})=
\nu([\prod_{s\not=s_n} G]\times[\prod_n U_k])=\prod_n \mu_H(U_k)=\prod_n r
=0
\ee
since $r<1$. Hence $S_{k,q,m}$ is contained in a measure zero subset
and since $\nu$ is $\sigma-$additive also $S_k$ is since (\ref{3.102})
is a countable union.

Finally, any $h_0\in G^{[0,1]}-D$ is contained in one of the $S_k$, thus
$G^{[0,1]}-D\subset \bigcup_{k=1}^N S_k$ is contained in a measurable
subset of measure zero.\\
$\Box$

\subsection{Spin -- Network Functions, Loop Representation,
Gauge -- and Diffeomorphism Invariance of $\mu_0$ and Ergodicity}
\label{s3.5}

In order to study the ergodicity properties of $\mu_0$ we need to
introduce an important concept, the so-called spin-network basis.
We will distinguish between gauge variant and gauge invariant
spin-network states. For representation theory on compact Lie groups,
the Peter\&Weyl theorem and Haar measures
the reader is referred to \cite{57a}.
\begin{Definition} \label{def3.19}   ~~~~~~~~~~~~~~\\
Fix once and for all a representative from each equivalence class
of irreducible representations of the compact Lie group $G$ and
denote the collection of these representatives by $\Pi$.
Let $l=l(\gamma)$ be given. Associate with every edge $e\in E(\gamma)$
a non-trivial, irreducible representation $\pi_e\in \Pi$
which we assemble in a vector $\vec{\pi}=(\pi_e)_{e\in E(\gamma)}$.\\
i)\\
The gauge variant spin-network functions are given by
\be \label{3.109}
T_{\gamma,\vec{\pi},\vec{m},\vec{n}}:\;\ab\to \Cl;\;
A\mapsto \prod_{e\in E(\gamma)} \sqrt{d_{\pi_e}} [\pi_e(A(e))]_{m_e n_e}
\ee
where $d_\pi$ denotes the dimension of $\pi$ and
$\vec{m}=\{m_e\}_{e\in E(\gamma)},\vec{n}=\{n_e\}_{e\in E(\gamma)}$
with $m_e,n_e=1,..,d_{\pi_e}$ label the matrix elements of the
representation.\\
ii)\\
Given a vertex
$v\in V(\gamma)$ consider the subsets of edges given by
$E_v^b(\gamma):=\{e\in E(\gamma);\;b(e)=v\}$
and $E_v^f(\gamma):=\{e\in E(\gamma);\;f(e)=v\}$.
For each $v\in V(\gamma)$, consider the tensor product representation
\be \label{3.110}
(\otimes_{e\in E^b_v(\gamma)} \pi_e)\otimes
(\otimes_{e\in E^f_v(\gamma)} \pi_e^c)
\ee
where $h\mapsto \pi^c(h):=\pi(h^{-1})^T$ denotes the representation
contragredient to $\pi$ ($(.)^T$ denotes matrix transposition).
Since $G$ is compact, every representation is completely reducible
and decomposes into an orthogonal sum of irreducible representations
(not necessarily mutually inequivalent).
Let ${\cal I}_v(\vec{\pi},\pi'_v)$ be the set of all
representations that appear in that decomposition of (\ref{3.109})
and which are equivalent to $\pi'_v\in \Pi$ with $\pi_t\in \Pi$ a
representative of the trivial representation.
An element
$I_v\in{\cal I}_v(\vec{\pi},\pi'_v)$ is called an intertwiner and we assemble
a given choice of intertwiners into a vector $\vec{I}=(I_v)_{v\in
V(\gamma)}$. By construction,
we can project the representation (\ref{3.110}) into the representation
$I_v\in {\cal I}_v(\vec{\pi},\pi')$ by contracting (\ref{3.110})
with a corresponding intertwiner. Since the function
\be \label{3.111}
A\mapsto (\otimes_{e\in E^b_v(\gamma)} \pi_e(A(e)))\otimes
(\otimes_{e\in E^f_v(\gamma)} \pi_e(A(e)))
\ee
transforms in the representation (\ref{3.110}) under gauge transformations
at $v$ it therefore transforms in the representation $I_v$ at $v$
when contracted with the intertwiner $I_v\in {\cal I}_v(\vec{\pi},\pi'_v)$.
We now take the function
\be \label{3.112}
A\mapsto \otimes_{e\in E(\gamma)} \pi_e(A(e))
\ee
and for each vertex $v$ consider the subproduct (\ref{3.110}) and then
contract with an appropriate intertwiner $I_v$. The result
is a cylindrical function on $\ab$ over $l=l(\gamma)$ which we denote
by $T_{\gamma,\vec{\pi},\vec{I}}(A)$ and which transforms in the
representation $I_v$ at $v$. If we vary the $\pi'_v,I_v$ then
the set of
functions $T_{\gamma,\vec{\pi},\vec{I}}$ span the same vector space as the
space of functions $T_{\gamma,\vec{\pi},\vec{m},\vec{n}}$. In particular,
we may take these functions to be normalized with respect to ${\cal H}_0$.

The gauge invariant spin network functions result when we restrict
the $\pi'_v$ to be trivial, that is, to equal $\pi_t$ with the convention
that $T_{\gamma,\vec{\pi},\vec{I}}$ vanishes if
${\cal I}_v(\vec{\pi},\pi^t)=\emptyset$ for any $v\in V(\gamma)$.
Since these functions are gauge invariant, we may consider them as
functions $T_{\gamma,\vec{\pi},\vec{I}}:\;\agb\to \Cl$.
\end{Definition}
Since spin-network functions are fundamental for what follows, let us
discuss an example to make the definition clear:\\
\\
{\bf Example}\\
Consider the case of ultimate interest $G=SU(2)$ whose irreducible
representations are labelled by non-negative, half-integral {\it spin}
quantum
numbers (from which the name ``spin-network" originates). Consider a graph
$\gamma$ consisting of
$N$ edges $e_I,\;I=1,..,N$ and two vertices $v_1,v_2$ such that
$b(e_I)=v_1,f(e_I)=v_2$ for $1\le I\le M$ and
$b(e_I)=v_2,f(e_I)=v_1$ for $M+1\le I\le N$ for some $1<M<N$.
%
%

Associate with $e_I$ an irreducible representation of $SU(2)$ labelled
by the spin quantum number $j_I$. Under gauge transformations the
function
\be \label{3.113}
A\mapsto \prod_{I=1}^N \pi_{j_I}(A(e_I))_{m_I n_I}
=[\otimes_{I=1}^N \pi_{j_I}(A(e_I))]_{m_1,..,m_N;n_1,..,n_N}
\ee
with $m_I,n_I=1,..,2j_I+1=d_{\pi_{j_I}}$ is mapped into
\ba \label{3.114}
A &\mapsto&
[\otimes_{I=1}^N
\pi_{j_I}(g(b(e_I))A(e_I)g(f(e_I))^{-1})]_{m_1,..,m_N;n_1,..,n_N}
\\
&=&
[\{\otimes_{I=1}^M \pi_{J_I}(g(v_1))\pi_{j_I}(A(e_I))\pi_{j_I}(g(v_2))^{-1}\}
\otimes\nonumber\\
&& \otimes \{\otimes_{I=M+1}^N \pi_{J_I}(g(v_2))\pi_{j_I}(A(e_I))
\pi_{j_I}(g(v_1))^{-1}\}]_{m_1,..,m_N;n_1,..,n_N}
\nonumber\\
&=&
[\{\otimes_{I=1}^M \pi_{J_I}(g(v_1))\}\otimes
\{\otimes_{I=M+1}^N \pi_{J_I}(g(v_1))^{-1}\}
]_{m_1,..,m_M,l_{M+1},..,l_N;k_1,..,k_M,n_{M+1},..,n_N}
\times \nonumber\\
&&\times [\otimes_{I=1}^N \pi_{j_I}(A(e_I))]_{k_1,..,k_N;l_1,..,l_N}
\times \nonumber\\
&& \times
[\{\otimes_{I=1}^M \pi_{J_I}(g(v_2))^{-1}\}\otimes
\{\otimes_{I=M+1}^N
\pi_{J_I}(g(v_2))\}]_{l_1,..,l_M,m_{M+1},..,m_N;n_1,..,n_M,k_{M+1},..,k_N}
\nonumber\\
&=&
[\{\otimes_{I=1}^M \pi_{J_I}(g(v_1))\}\otimes
\{\otimes_{I=M+1}^N \pi_{J_I}^c(g(v_1))\}
]_{m_1,..,m_M,n_{M+1},..,n_N;k_1,..,k_M,l_{M+1},..,l_N}
\times\nonumber\\
&& \times
[\otimes_{I=1}^N \pi_{j_I}(A(e_I))]_{k_1,..,k_N;l_1,..,l_N}
\times\nonumber\\
&& \times
[\{\otimes_{I=1}^M \pi_{J_I}^c(g(v_2))\}\otimes
\{\otimes_{I=M+1}^N \pi_{J_I}(g(v_2))\}
]_{n_1,..,n_M,m_{M+1},..,m_N;l_1,..,l_M,k_{M+1},..,k_N}
\nonumber
\ea
from which we see how the contragredient representation enters the stage.
Now $SU(2)$ is special in the sense that a representation and its
contragredient one are equivalent which follows from
$g^c=\tau_2 g \tau_2^{-1}$ for any $g\in SU(2)$ where $-\tau_2=i\sigma_2$ is
the spinor metric. We are thus lead to consider tensor products of the form
\ba \label{3.115}
&&
[\{\otimes_{I=1}^M \pi_{J_I}(g)\}\otimes\{\otimes_{I=M+1}^N \pi_{j_I}^c(g)\}]
=[\{\otimes_{I=1}^M \pi_{J_I}(1)\}\otimes\{\otimes_{I=M+1}^N
\pi_{j_I}(\tau_2)\}] \cdot
\nonumber\\
&\cdot&
[\otimes_{I=1}^N \pi_{J_I}(g)]
\cdot
[\{\otimes_{I=1}^M \pi_{J_I}(1)\}\otimes\{\otimes_{I=M+1}^N
\pi_{j_I}(\tau_2)^{-1}\}
\ea
In order to decompose the tensor product representation
$j_1\otimes j_2\otimes..\otimes j_N$ into irreducibles we must agree
on a {\it recoupling scheme}, that is, we must decide on a bracketing
of this tensor product. We choose
$(..((j_1\otimes j_2)\otimes j_3)\otimes..)\otimes j_N$ and apply the
Clebsch-Gordan theorem $j_1\otimes j_2=j_1+j_2 \oplus j_1+j_2-1\oplus
..\oplus |j_1-j_2|$ starting from the inner most bracket and working our
way outwards. For instance in the case $N=3$ we have
\ba \label{3.116}
&& (j_1\otimes j_2)\otimes j_3=
(j_1+j_2 \oplus ..\oplus |j_1-j_2|)\otimes j_3
\\
&=& (j_1+j_2+j_3\oplus..\oplus |j_1+j_2-j_3|)\oplus
(j_1+j_2-1+j_3\oplus..\oplus |j_1+j_2-1-j_3|)\oplus..
(|j_1-j_2|+j_3\oplus..\oplus ||j_1-j_2|-j_3|)
\nonumber
\ea
Notice that all appearing representations, even those that appear with
multiplicity higher than one (and are therefore mutually equivalent), are
realized on mutually orthogonal subspaces of the $(2j_1+1)(2j_2+1)(2j_3+1)$
dimensional representation space of $j_1\otimes j_2\otimes j_3$.
We see that in this case $N=3$
${\cal I}_{v_{1,2}}(\vec{j},j=0)$ is empty unless
$j_3\in\{j_1+j_2,..,|j_1-j_2|\}$ and if that is the case there is only
one trivial representation contained in (\ref{3.116}) no matter how
large $j_1,j_2$ are. If $N>3$ this is no longer true, the space of
spin-network states on graphs with at least one vertex of valence larger
than three is generically more than one-dimensional for given values of the
$j_I$.

We see that the theory of spin-network states for $SU(2)$ is largely
governed by the representation theory of $SU(2)$ and Clebsh-Gordan
coefficients which give the precise numerical coefficients
in the orthogonal sums (\ref{3.116}). In particular,
changing of recoupling schemes gives
rise to the complicated $3Nj$ symbols which can be decomposed into $6j$
symbols. It is this complicated recoupling theory that determines the
spectrum of the volume operator, see \cite{58} for an introduction using
the terminology of the present review.

Once we have then isolated all possible trivial representations in the
decomposition (\ref{3.115}) we insert one of them back into (\ref{3.114})
in place of (\ref{3.115}) and have found a suitable, gauge invariant
intertwiner. This
we do for all possible (mutually orthogonal) intertwiners and vertices
and have then found all possible gauge invariant states over the given
$\gamma$ given the $j_I$. This concludes our example.\\
\\
The importance of spin-network functions is that they provide a basis
for ${\cal H}^0$.
\begin{Theorem} \label{th3.13} ~~~~~~~~~~~\\
i)\\
The gauge variant spin-network states provide an orthonormal basis
for the Hilbert space $L_2(\ab,d\mu_0)$.\\
ii)\\
The gauge invariant spin-network states provide an orthonormal basis
for the Hilbert space $L_2(\agb,d\mu_0)$.
\end{Theorem}
Proof of Theorem \ref{th3.13}:\\
i)\\
The inner product on $L_2(\ab,d\mu_0)$ is defined by
\be \label{3.117}
<f,f'>_{L_2(\ab,d\mu_0)}:=\Lambda_{\mu_0}(\overline{f}f')
\ee
where $\Lambda_{\mu_0}$ is the positive linear functional
on $C(\ab)$ determined by $\mu_0$ via the Riesz representation theorem.
The cylinder functions of the form $p_l^\ast f_l,\;f_l\in C(X_l)$ are
dense in $C(\ab)$ (in the sup-norm) and since $\ab$ is a (locally) compact
Hausdorff space
and $\mu_0$ comes from a positive linear functional on the space of
continuous functions on $\ab$ (of compact support), these functions are
dense in $L_2(\ab,d\mu_0)$ (in the $L_2$ norm
$||f||_2=<f,f>^{1/2}$, see e.g. \cite{57}). It follows that
$L_2(\ab,d\mu_0)$ is the completion of $\mbox{Cyl}(\ab)$ in the $L_2$
norm. Now
\be \label{3.118}
\mbox{Cyl}(\ab)=\bigcup_{l\in {\cal L}} p_l^\ast C(X_l)
\ee
and since by the same remark $C(X_l)$ is dense in $L_2(X_l,d\mu_{0l})$
it follows that
\be \label{3.118a}
L_2(\ab,d\mu_0)=\overline{\bigcup_{l\in {\cal L}} p_l^\ast
L_2(X_l,d\mu_{0l})}
\ee
Now by definition $(\rho_l)_\ast \mu_{0l}=\otimes_{e\in E(\gamma)} \mu_H$
for $l=l(\gamma)$ so that $L_2(X_l,d\mu_{0l})$ is isometric isomorphic with
$L_2(G^{|E(\gamma)|},\otimes^{|E(\gamma)|}d\mu_H)$ which in turn is
isometric isomorphic with $\otimes_{e\in E(\gamma)} L_2(G,d\mu_H)$ since
$\otimes^{|E(\gamma)|}\mu_H$ is a finite product of measures.
By the Peter\&Weyl theorem the matrix
element functions
\be \label{3.119}
\pi_{mn}:\;G\to \Cl;\;h\mapsto \sqrt{d_\pi} \pi_{mn}(h),\pi\in
\Pi,\;m,n=1,..,d_\pi
\ee
form
an orthonormal basis of $L_2(G,d\mu_H)$ for any compact gauge group $G$,
that is,
\be \label{3.120}
<\pi_{mn},\pi'_{m'n'}>:=\int_G d\mu_H(h) \overline{\pi_{mn}(h)}
\pi'_{m'n'}(h)=\frac{\delta_{\pi\pi'}\delta_{mm'}\delta_{nn'}}{d_\pi}
\ee
This shows that functions of the form (\ref{3.109}) span
$L'_2(X_l,d\mu_{0l}):\cong \otimes^{|E(\gamma)|}
L_2'(G,d\mu_H)$ where
$L'_2(G,d\mu_H)$ is the closed linear span of the functions $\pi_{mn}$
with $\pi\not=\pi_t$ (only non-trivial representations allowed).

It remains to prove 1) that $p_l^\ast L'_2(X_l,d\mu_{0l})\perp
p_{l'}^\ast L'_2(X_{l'},d\mu_{0l'})$ unless $l=l'$ and 2) that
$L_2(X_l,d\mu_{0l})=\overline{\oplus_{l'\prec l} L'_2(X_{l'},d\mu_{0l'})}$.
where completion is with respect to $L_2(X_l,d\mu_{0l})$.\\ To see
the former, notice that if $l=l(\gamma)\not=l'=l(\gamma')$ there is
$l,l'\prec l^\dprime:=l(\gamma\cup\gamma')$. Since $\gamma\not=\gamma'$
are piecewise analytic, there must be an edge $e\in E(\gamma)$ which
contains a segment $s\subset e$ which is disjoint from $\gamma'$
(reverse the roles of $\gamma,\gamma'$ if necessary) and this segment
is certainly contained in $\gamma\cup \gamma'$. Let
$f_l\in L'_2(X_l,d\mu_{0l}), f_{l'}\in L'_2(X_{'l},d\mu_{0l'})$
then
\be \label{3.121}
<p_l^\ast f_l,p_{l'}^\ast f_{l'}>=
\mu_{0l^\dprime}
(\overline{p_{l^\dprime l}^\ast f_l} p_{l^\dprime l'}^\ast f_{l'})=0
\ee
since $p_{l^\dprime l}^\ast f_l, p_{l^\dprime l'}^\ast f_{l'}$ are
(Cauchy sequences of) functions of the form (\ref{3.109}) over
$\gamma\cup\gamma'$ where the dendence on $s$ of the former function
is through a non-trivial representation and of the latter through a
trivial representation, so the claim follows from formula (\ref{3.120}).\\
To see the former, observe that
$L_2(G,d\mu_H)=\overline{L_2'(G,d\mu_H)\oplus\mbox{span}(\{1\})}$
and that a function cylindrical over $\gamma$ which depends on $e\in
E(\gamma)$ through the trivial representation is cylindrical over
$\gamma-e$ as well.

Summarizing, if we define ${\cal H}^0_l:=p_l^\ast
L_2(X_l,d\mu_{0l}),\;{\cal H}^{0l}:=p_l^\ast L'_2(X_l,d\mu_{0l})$
then
\be \label{3.122}
{\cal H}^0=\overline{\bigcup_{l\in {\cal L}} p_l^\ast {\cal H}^0_l}
=\overline{\oplus_{l\in {\cal L}} p_l^\ast {\cal H}^{0l}}
\ee
ii)\\
The assertion follows easily from i) and the fact that $L_2(\agb,d\mu_0)$
is simply the restriction of $L_2(\ab,d\mu_0)$ to the gauge invariant
subspace:
That subspace is the closed linear span of gauge invariant
spin-network states by i) and the specific choice that we have made in
definition \ref{def3.19} shows that they form an orthonormal system
since we have chosen them to be normalized and the intertwiners to be
projections onto mutually orthogonal subspaces of a tensor product
representation space of $G$.\\
More specifically, the inner product
between two spin network functions
$T_{\gamma,\vec{\pi},\vec{I}},T_{\gamma',\vec{\pi}',\vec{I}'}$
is nonvanishing only if $\gamma=\gamma'$ and
$\vec{\pi}=\vec{\pi}'$. In that case, consider $v\in V(\gamma)$
and assume w.l.g. that all edges $e_1,..,e_N$ incident at $v$ are outgoing.
An intertwiner $I_v\in {\cal I}_v(\vec{\pi},\pi_t)$ can be thought of as
a vector $I_v^{n_1,..,n_N}:=(I_v)_{m_1^0,..,m_N^0;n_1,..,n_N}$ in the
representation space of the representation $\otimes_{I=1}^N \pi_I$
where $m^0_I$ are some matrix elements that we fix once and for all.
Since $I_v$ is a trivial representation and in particular represents
$1_G=(1_G)^T$ we have
$(I_v)_{m_1^0,..,m_N^0;n_1,..,n_N}=
I_v^{n_1,..,n_N}:=(I_v)_{n_1,..,n_N;m_1^0,..,m_N^0}$, moreover the
intertwiners are real valued because the functions
$\pi_{mn}(h)$ depend analytically on $h$ and $1_G$ is real valued. Now the
spin-network state restricted to its dependence on $e_1,..,e_N$ is of the
form
\be \label{3.123}
I_v^{n_1,..,n_N} [\otimes_{I=1}^N \pi_I(A(e_I))]_{n_1,..,n_N;k_1,..,k_N}
\ee
It follows from (\ref{3.120}) that the inner product between
$T_{\gamma,\vec{\pi},\vec{I}},T_{\gamma,\vec{pi},\vec{I}'}$
will be proportional to
\be \label{3.124}
I_v^{n_1,..,n_N} (I')_v^{n_1,..,n_N}
=[(I_v) (I'_v)]_{m_1^0,..,m_N^0;m_1^{\prime 0},..,m_N^{\prime 0}}
\propto \delta_{I_v I'_v}
\ee
(if $I_v=I'_v$ then $m_I^0=m_I^{0\prime}$ by construction)
since the $I_v$ are representations on mutually orthogonal subspaces.\\
$\Box$\\
We remark that the spin-network basis is not countable because the
set of graphs in $\sigma$ is not countable, whence ${\cal H}^0$ is not
separable. We will see that this is even the case after moding out by
spatial diffeomorphisms although one can argue that after moding
out by diffeomorphisms the remaining space is an orthogonal, uncountably
infinite sum of superselected, mutually isomorphic, separable
Hilbert spaces \cite{47m1}.
\begin{Definition}  \label{def3.20}  ~~~~~~~~~~~~\\
The gauge variant spin-network representation is a vector space
$\tilde{{\cal H}}^0$
of complex valued functions
\be \label{3.125}
\psi:\;{\cal S}\to \Cl;\;s\mapsto \psi(s)
\ee
where $\cal S$ is the set of quadruples $(\gamma,\vec{\pi},\vec{m},\vec{n})$
which label a spin-network state. Likewise, the loop representation
is the gauge invariant spin-network representation defined analogously.
This vector space is equipped with the scalar product
\be \label{3.126}
<\psi,\psi'>_{\tilde{{\cal H}}^0}:=\sum_{s\in {\cal
S}}\overline{\psi(s)}\psi'(s) \ee
between square summable functions.
\end{Definition}
Clearly the uncountably infinite sum (\ref{3.126}) converges if and only
if $\psi(s)=0$ except for countably many $s\in{\cal S}$. The next
corollary shows that the connection representation that we have been
dealing with so far and the spin-network representation are in a precise
sense Fourier transforms of each other where the role of the kernel
of the transform is played by the spin-network functions.
\begin{Corollary} \label{col3.2}
The spin-network (or loop) transform
\be \label{3.127}
T:\;{\cal H}^0\to \tilde{{\cal H}}^0;\;f\mapsto \tilde{f}(s):=<T_s,f>_{{\cal
H}^0} \ee
is a unitary transformation between Hilbert spaces with inverse
\be \label{3.128}
(T^{-1}\psi)(A):=\sum_{s\in {\cal S}} \psi(s) T_s(A)
\ee
\end{Corollary}
Proof of Corollary \ref{col3.2}:\\
If $f\in {\cal H}^0$ then
\be \label{3.129}
f=\sum_{s\in {\cal S}} <T_s,f> T_s
\ee since the
$T_s$ form an orthonormal basis (Bessel's inequality is saturated).
Since the $T_s$ form an orthonormal system we conclude
that $||f||^2=\sum_s |<T_s,f>|^2$ converges, meaning in particular
that $<T_s,f>=0$ except for finitely many $s\in {\cal S}$.
It follows that $||T f||^2:=\sum_s |\tilde{f}(s)|^2=||f||^2$ which shows
that $T$ is a partial isometry. Comparing (\ref{3.28}) and (\ref{3.29}) we
see that $T^{-1}\tilde{f}=f$ is indeed the inverse of $T$. Finally again by
the orthogonality of the $T_s$ we have
$||T^{-1}\psi||^2=\sum_s |\psi(s)|^2=||\psi||^2$ so that $T^{-1}$
a partial isometry as well. Since $T$ is a bijection, $T$ is actually
an isometry. Notice that $\tilde{T}_s(s')=\delta_{s,s'}$.\\
$\Box$\\
Whenever it is convenient we may therefore think of states either in the loop
or the connection representation. In this review we will work entirely in
the connection representation.\\

In the previous section we have investigated the topological and measure
theoretical relation between $\a$ and $\ab$. In this section
we will investigate the action of the gauge and diffeomorphism group
on $\ab$.
The uniform measure has two important further properties: it is invariant
under both the gauge group $\gb$ and the Diffeomorphism group
Diff$^\omega(\sigma)$ (analytic diffeomorphisms). To see this, recall the
action of $\gb$ on $\ab$
defined through its action on the subspaces $X_l$ by
$x_l\mapsto \lambda_g(x_l)$ with $[\lambda_g(x_l)](p)=
g(b(p))x_l(p)g(f(p))^{-1}$ for any $p\in l$. This action has the feature
to leave the $X_l$ invariant for any $l\in {\cal L}$ and therefore
lifts to $\overline{X}$ as $x\mapsto \lambda_g(x)$ with
$[\lambda_g(x)](p)=g(b(p))x(p)g(f(p))^{-1}$ for any $p\in {\cal L}$.
Likewise we have an action of Diff$^\omega(\sigma)$ on $\overline{X}$ defined
by
\be \label{3.104}
\delta^l:\;\mbox{Diff}^\omega(\sigma)\times X_l\to X_{\varphi^{-1}(l)};\;
(\varphi,x_l)\mapsto \delta^l_\varphi(x_l)
=x_{\varphi^{-1}(l)}
\ee
where $\varphi^{-1}=l(\varphi^{-1}(\gamma))$ if $l=l(\gamma)$. This action
does not preserve the various $X_l$. The action on all of
$\overline{X}$ is then evidently defined by
\be \label{3.105}
\delta:\;\mbox{Diff}^\omega(\sigma)\times \overline{X}\to \overline{X};\;
(\varphi,x=(x_l)_{l\in {\cal L}})\mapsto \delta_\varphi(x)
=(\delta^l_\varphi(x_l))_{l\in{\cal L}}
\ee
Clearly $\delta_\varphi(x)$ is still an element of the projective
limit since it just permutes the various $x_l$ among each other.
Moreover, $l\prec l'$ iff $\varphi^{-1}(l)\prec \varphi^{-1}(l')$
so the diffeomorphisms preserve the partial order on the label set.
Therefore
\be \label{3.106}
p_{\varphi^{-1}(l')\varphi^{-1}(l)}(\delta^{l'}_\varphi(x_{l'})
=x_{\varphi^{-1}(l)}=\delta^l_\varphi(p_{l' l}(x_{l'})
\ee
for any $l\prec l'$, so we have equivariance
\be \label{3.106a}
p_{\varphi^{-1}(l')\varphi^{-1}(l)}\circ \delta^{l'}_\varphi
=\delta^l_\varphi\circ p_{l' l}
\ee
It is now easy to see that for the push-forward measures we have
$(\lambda_g)_\ast \mu_0=\mu_0,(\delta_\varphi)_\ast \mu_0=\mu_0$.
For any $f=p_l^\ast f_l\in C(\overline{X}),\;f_l=\rho_l^\ast F_l\in C(X_l),\;
F_l\in C(G^{|E(\gamma)|}),\;l=l(\gamma)\in {\cal L}$ we have
\ba \label{3.107}
\mu_0(\lambda_g^\ast f)&=&\mu_0(p_l^\ast (\lambda^l_g)^\ast f_l)
=\mu_{0l}((\lambda^l_g)^\ast f_l)
\nonumber\\
&=&\int_{G^{|E(\gamma)|}} [\prod_{e\in E(\gamma)} d\mu_H(h_e)]
F_l(\{g(b(e))h_e g(f(e))^{-1}\}_{e\in E(\gamma)})
\nonumber\\
&=&\int_{G^{|E(\gamma)|}} [\prod_{e\in E(\gamma)}
d\mu_H(g(b(e))^{-1}h_e g(f(e)))]
F_l(\{h_e\}_{e\in E(\gamma)})
\nonumber\\
&=&\int_{G^{|E(\gamma)|}} [\prod_{e\in E(\gamma)} d\mu_H(h_e)]
F_l(\{h_e\}_{e\in E(\gamma)})=\mu_0(f)
\ea
where we have made a change of integration variables
$h_e\to g(b(e))h_e g(f(e))^{-1}$ and used that the associated Jacobian
equals unity for the Haar measure (translation invariance). Next
\ba \label{3.108}
\mu_0(\delta_\varphi^\ast f)&=&\mu_0(p_{\varphi^{-1}(l)}^\ast
(\delta^l_g)^\ast f_l)
=\mu_{0\varphi^{-1}(l)}((\delta^l_\varphi)^\ast f_l)
\nonumber\\
&=&\int_{G^{|E(\varphi^{-1}(\gamma))|}}
[\prod_{e\in E(\varphi^{-1}(\gamma))} d\mu_H(h_e)]
F_l(\{h_e\}_{e\in E(\varphi^{-1}(\gamma))})
\nonumber\\
&=&\int_{G^{|E(\gamma)|}} [\prod_{e\in E(\gamma)} d\mu_H(h_e)]
F_l(\{h_e\}_{e\in E(\gamma)})=\mu_0(f)
\ea
where we have written
$\{h_e\}_{e\in E(\varphi^{-1}(\gamma))}=
\{h_{\varphi^{-1}(e)}\}_{e\in E(\gamma)}$ and have
performed a simple relabelling $h_{\varphi^{-1}(e)}\to h_e$.
It is important to notice that in contrast to other measures on some
space of connections the ``volume of the gauge group is finite":
The space $C(\agb)$ is a subspace of $C(\ab)$ and we may integrate
them with the measure $\mu_0$ which is the same as integrating them with
the restricted measure. We do not have to fix a gauge and never have to
deal with the problem of Gribov copies.

One may ask now why one does not repeat with the diffeomorphism group
what has been done with the gauge group: Passing from analytic
diffeomorphisms Diff$^\omega(\sigma)$ to distributional ones
$\overline{\mbox{Diff}(\sigma)}$ and passing to the quotient space
$(\agb)/\overline{\mbox{Diff}(\sigma)}$. There are two problems: First,
in the case of $\gb$ there was a natural candidate for the extension
$\g\to \gb$ but this is not the case for diffeomorphisms because
distributional diffeomorphisms will not lie in any differentiability
category any more and therefore are not diffeomorphisms in the strict
sense. Secondly, as we will now show, even the analytic
diffeomorphisms act ergodically on the measure space which means that
there are no non-trivial invariant functions.
Thus, one either has to proceed differently
(e.g. downsizing rather than extending the diffeomorphism group),
change the measure or solve the diffeomorphism constraint differently.
We will select the third option in section \ref{s4}. It should be
pointed out, however, that the last word of how to deal with
diffeomorphism invariance has not been spoken yet. In a sense, it is
one of the {\it key questions} for the following reason: The concept
of a smooth spacetime should not have any meaning in a quantum theory
of the gravitational field where probing distances beyond the Planck
length must result in black hole creation which then evaporate in Planck
time, that is, spacetime should be fundamentally discrete. But clearly
smooth diffeomorphisms have no room in such a discrete quantum spacetime.
The fundamental symmetry is probably something else, maybe a combinatorial
one, that looks like a diffeomorphism group at large scales.
Also, if one wants to allow for topology change in quantum gravity then
talking about the diffeomorphism group for a fixed $\sigma$ does not make
much sense. We see that there is a tension between classical diffeomorphism
invariance and the discrete structure of quantum spacetime which
in our opinion has not been satisfactorily resolved yet and which we consider
as one of the most important conceptual problems left open so far.

Let us then move on to establish ergodicity:\\
The above discussion reveals that as far as $\gb$ and Diff$(\sigma)$ are
concerned we have the following abstract situation (see section
\ref{sf}): We have a measure space with a measure preserving group action
of both groups (so that the pull-back maps
$\lambda_g^\ast,\delta_\varphi^\ast$ provide unitary actions on the
Hilbert space) and the question is whether that action is ergodic.
That is certainly not the case with respect to $\gb$ since the subspace
of gauge invariant functions is by far not the span of the constant
functions as we have shown.
\begin{Theorem} \label{th3.14}  ~~~~~~\\
The group Diff$^\omega_0(\sigma)$ of analytic diffeomorphisms on an analytic
manifold $\sigma$ connected to the identity acts ergodically on the
measure space $\ab$ with respect to the Borel measure $\mu_0$.
\end{Theorem}
Proof of Theorem \ref{th3.14}:\\
The diffeomorphism group acts unitarily on ${\cal H}^0$ via
\be \label{3.130}
[\hat{U}(\varphi)f](A)=f(\delta_\varphi(A))
\ee
which means for spin-network states that $\hat{U}(\varphi) T_s=
T_{\varphi^{-1}(s)}$ where
\be \label{3.131}
\varphi^{-1}(s)=(\varphi^{-1}(\gamma),\;
\{\pi_{\varphi^{-1}(e)}=\pi_e\}_{e\in E(\gamma)},\;
\{m_{\varphi^{-1}(e)}=m_e\}_{e\in E(\gamma)},\;
\{n_{\varphi^{-1}(e)}=n_e\}_{e\in E(\gamma)}
\ee
for $s=(\gamma,\vec{\pi},\vec{m},\vec{n})$. Let now
$f=\sum_{s\in {\cal S}} c_s\; T_s\in {\cal H}^0$ be given with $c_s=0$
except for countably many. Suppose that $\hat{U}(\varphi)f=f$ $\mu_0-$a.e.
for any $\varphi\in\mbox{Diff}^\omega_0(\sigma)$. Since
${\cal S}$ is left invariant by diffeomorphisms, this means
that
\be \label{3.132}
\sum_s c_s T_{\varphi^{-1}(s)}=\sum_s c_{\varphi(s)} T_s=\sum_s c_s T_s
\ee
for all $\varphi$. Since the $T_s$ are mutually orthogonal we conclude that
$c_s=c_{\varphi(s)}$ for all $\varphi\in\mbox{Diff}^\omega_0(\sigma)$.
Now for any $s\not=s_0=(\emptyset,\vec{0},\vec{0},\vec{0})$ the orbit
$[s]=\{\varphi(s);\;\varphi\in \mbox{Diff}^\omega_0(\sigma)\}$
contains infinitely many different elements (take any vector field
that does not vanish in an open set which contains the graph determined by
$s$ and consider the one parameter subgroup of diffeomorphisms determined
by its integral curve -- this is where we can make the restriction to
the identity component). Therefore
$c_s=$const. for infinitely many $s$. Since $f$ is normalizable, this is
only possible if const.$=0$, hence $f=c_{s_0} T_{s_0}$ is constant
$\mu_0-$a.e. and therefore $\delta$ ergodic.\\
$\Box$\\
We see that the theorem would still hold if we would replace
$\mbox{Diff}^\omega_0(\sigma)$ by any infinite subgroup $D$ with respect
to which each orbit $[s],\;s\not=s_0$ is infinite. An example would be
the case $\sigma=\Rl^D$ and $D$ a discrete subgroup of the translation
group given by integer multiples of translations by a fixed non-zero vector.

The theorem shows that the only vectors in ${\cal H}^0$ invariant under
diffeomorphisms are the constant functions, hence we cannot just pass
to that trivial subspace in order to solve the diffeomorphism constraint.
The solution to the problem lies in passing to a larger space of functions,
distributions over a subspace of ${\cal H}^0$ in which one can solve the
constraint. The proof of the theorem shows already how that distributional
space must look like: it must allow for uncountably infinite linear
combinations of the form $\sum_s c_s T_s$ where $c_s$ is a generalized
knot invariant (i.e. $c_s=c_{\varphi{s}}$ for any $\varphi$, generalized
because $\gamma(s)$ has in general self-intersections and is not a
regular knot). This brings us to the next section.

\newpage

\section{Quantum Kinematics}
\label{s4}

This section is concerned with the following issues: In the previous
section we have introduced a distributional extension $\ab$ of the
space of smooth connections $\a$ which we choose as our quantum
configuration space. We equipped it with a topology and the natural
Borel $\sigma-$algebra that comes with it and have defined a natural
measure $\mu_0$ thereon. The measure is natural because it is invariant
under both gauge transformations and spatial diffeomorphisms. However,
in order to be physically meaningful we must show that the corresponding
$L_2$ Hilbert space implements
the correct adjointness relations and canonical commutation relations.
This will be our first task. Next we must solve the quantum constraints
by the methods of refined algebraic quantization (RAQ) to which we will
give a brief introduction in section \ref{si}.
In order to do this we must define the constraints
as closed, densely defined
operators on the Hilbert space ${\cal H}^0$ and look for solutions
in the algebraic dual of a certain subspace thereof.
Since the solutions to the constraints are not elements of ${\cal H}^0$
as we already saw at the end of the previous section, one must
define a new inner product on the space of solutions.
We will do this in this section restricted to the kinematical
constraints, that is, the Gauss and Diffeomorphism constraint.
The inner product on the space of solutions of the Diffeomorphism
constraint that we will derive by using RAQ methods is, however,
only of mathematical interest because it is not possible to
solve the Diffeomorphism and Hamiltonian constraint in two separate
steps, the (dual) Hamiltonian constraint does not leave the space of
diffeomorphism invariant distributions invariant.

\subsection{Canonical Commutation -- and Adjointness Relations}
\label{s4.1}

In this section as well as the two following ones it will not be
important that $G=SU(2)$ or that $\sigma$ is threedimensional,
hence we will leave the discussion at the level of general compact,
connected gauge groups $G$ and $D-$dimensional analytic manifolds.

\subsubsection{Classical Lie Algebra of Functions and Vector Fields on
$\a$: Electric Fluxes}
\label{s4.1.1}

In order to convince ourselves that ${\cal H}^0=L_2(\ab,d\mu_0)$
implements the correct adjointness and canonical commutation relations
we must first decide on an appropriate set of classical functions
that separate the points of the classical phase space $\cal M$.
For the configuration space we have already seen several times
that the holonomy functions $p\mapsto A(p):=h_p(A)$ with
$p\in {\cal P}$ separate the
points of the space of classical connections $\a$. We now have to
look for appropriate momentum space functions.

Let $S$ be an analytic, orientable, connected, embedded $(D-1)$-dimensional
submanifold of $\sigma$ (a surface) which we choose to be open.
Since $E^a_j$ is a vector density of weight one, the function
$(\ast E_j)_{a_1.. a_{D-1}}:=E^c_j\epsilon_{ca_1..a_{D-1}}$ is a $(D-1)$-form
which we may integrate
in a background independent way over $S$, that is,
\be \label{4.1}
E_j(S):=\int_S  \ast E_j
\ee
These functions, which we will refer to as {\it electric flux} variables,
certainly separate the space $\cal E$ of smooth electric
fields on $\sigma$:
To see this consider a surface of the form $S:\;(-1/2,1/2)^{D-1}\to
\sigma;\;(u_1,..,u_{D-1})\mapsto S(u_1,..,u_{D-1})$ with analytic functions
$S(u_1,..,u_{D-1})$ and let
$S_\epsilon(u_1,..,u_{D-1}):=S(\epsilon u_1,..,\epsilon u_{D-1})$. Then
(\ref{4.1}) becomes
\ba \label{4.2}
E_j(S_\epsilon) &=& \int_{(-\epsilon/2,\epsilon/2)^{D-1}} du_1.. du_{D-1}
\epsilon_{a a_1..a_{D-1}}
(\partial S^{a_1}/\partial u_1)(u_1,..,u_{D-1})..
\\
&& ..(\partial S^{a_{D-1}}/\partial u_{D-1})(u_1,..,u_{D-1})
E^a_j(S(u_1,..,u_{D-1}))
\nonumber\\
&=& \epsilon^{D-1}
\epsilon_{a a_1..a_{D-1}}
(\partial S^{a_1}/\partial u_1)(0,..,0)..
(\partial S^{a_{D-1}}/\partial u_{D-1})(0,..,0)
E^a_j(S(0,..,0)) +O(\epsilon^D)
\nonumber
\ea
where we have written the lowest order term in the Taylor expansion
in the second line. It follows that
\be \label{4.3}
\lim_{\epsilon\to 0} \frac{E_j(S_\epsilon)}{\epsilon^{D-1}}
=\epsilon_{a a_1..a_{D-1}}
(\partial S^{a_1}/\partial u_1)(0,..,0)..
(\partial S^{a_{D-1}}/\partial u_{D-1})(0,..,0)
E^a_j(S(0,..,0)) +O(\epsilon^D)
\ee
and by varying $S$ we may recover every component of $E^a_j(x)$ at
$x=S(0,..,0)$.

The proposal then is to start classically from the set functions
$A(p),E(S)$. Notice that in contrast to the holonomy
functions the functions $E_j(S)$ do not have a simple behaviour
under gauge transformations. This is not troublesome at the level
of gauge-variant functions but will become a problem when passing
to the gauge invariant sector. Fortunately, as we will see, one can
construct gauge invariant functions from the $E_j(S)$ by certain
limiting procedures and the amazing fact is that the corresponding
operators continue to be well-defined in quantum theory. Thus,
we will use the functions $E_j(S)$ directly only in an intermediate step
in order to verify that the reality conditions and canonical brackets
are correctly implemented.

\subsubsection{Regularization of the Magnetic and Electric Flux
Poisson Algebra}
\label{s4.1.2}

The reality conditions are simply that $A(p)$ is $G-$valued and that
$E(S)=E_j(S)\tau_j$ is Lie$(G)-$valued, i.e., $E_j(S)$ is real valued.
The Poisson brackets among $A(p),E(S)$ are, however, a priori ill-defined
because the Poisson brackets that we derived in section \ref{s2} required
that the fields $A,E$ be smeared in $D$ directions by smooth functions
while the functions $A(p),E(S)$ represent only one -- and $(D-1)-$
dimensional smearings only. Therfore it is not possible to simply
compute their Poisson brackets: The aim to have a background independent
formulation of the quantum theory forces us to consider such singular
smearings and prevents us from using the Poisson brackets on $\cal M$
directly.
The strategy will therfore be to regularize the functions $A(p), E(S)$
in order to arrive at a three-dimensional smearing, then to compute the
Poisson brackets of the regulated functions and finally we will
remove the regulator and hope to arrive at a well-defined symplectic
structure for the $A(p),E(S)$.

The simplest way to do this is to define a {\it tube} $T^\epsilon_p$
with central path $p$ to be a smooth function of the form
\be \label{4.4}
T^{\epsilon t}_p:\; \Rl^{D-1}\times [0,1]\to \sigma;\;
T^{\epsilon t}_p(s_1,.,s_{D-1},t'):=
\delta^\epsilon(t'-t)\delta^\epsilon(s_1,..,s_{D-1})
p_{s_1,..,s_{D-1}}(t')
\ee
where $p_{s_1,..,s_{D-1}}$ is a smooth assignment of mutually
non-intersecting paths diffeomorphic
to $p:=p_{0,..,0}$ (a congruence) and $\delta^\epsilon$ is a smooth
regularization of the
$\delta-$distribution in $\Rl^{D-1}$ and $\Rl$ respectively. We then define
(recall formula (\ref{a.13a}) for the holonomy)
\be \label{4.5}
h^\epsilon_p(A):={\cal P}
e^{\int_{\Rl^{D-1}} d^{D-1}s \;f^\epsilon(s_1,..,s_{D-1})\int_0^1 dt
\int_{p_{s_1,..,s_{D-1}}} \delta^\epsilon_t A}
\ee
where path ordering is with respect the $t$ parameter. We obviously
have $\lim_{\epsilon\to 0} h_{T^\epsilon_p}=h_p$ pointwise in $\a$ for
any choice of $\delta^\epsilon$.
Likewise we define
a {\it disk} $D^\epsilon_S$
with central surface $S$ to be a smooth function of the form
\be \label{4.6}
D^\epsilon_S:\; \Rl\times U\to \sigma;\;
D^\epsilon_p(s;u_1,.,u_{D-1}):=\delta^\epsilon(s)
S_s(u_1,..,u_D)
\ee
where $S_s$ is a smooth assignment of mutually non-intersecting surfaces
diffeomorphic
to $S:=S_0$ (a congruence). Here $U$ denotes the subset of
$\Rl^{D-1}$ in the pre-image of $S$. We then define
\be \label{4.7}
E^\epsilon(S):=\int_\Rl ds\; \delta^\epsilon(s) E(S_s)
\ee
We obviously
have $\lim_{\epsilon\to 0} E(D^\epsilon_S)=E(S)$ pointwise in $\cal E$.

Next recall that the Poisson bracket algebra among the functions
$F(A)=\int d^Dx A_a^j F^a_j$, $E(f)=\int d^Dx E^a_j f_a^j$ of section
\ref{s2} is isomorphic with a subalgebra of the Lie algebra
$C^\infty(\a)\times V^\infty(\a)$ of smooth functions and vector fields
(derivatives on functions) on $\a$ respectively . This Lie algebra is
defined by
\be \label{4.8}
[(\phi,\nu),(\phi',\nu')]:=(\nu(\phi')-\nu'(\phi),[\nu,\nu'])
\ee
where $\nu(\phi)$ denotes the action of the vector field $\nu$ on the
function
$\phi$ and $[\nu,\nu']$ denotes the Lie bracket of vector fields. The
subalgebra of $C^\infty(\a)\times V^\infty(\a)$ which is isomorphic
to the Poisson subalgebra generated by the functions $F(A),E(f)$
is given by the elements $(F(A),E(f))\mapsto (\phi_F,\beta\kappa\nu_f)$ with
algebra \be \label{4.9}
[(\phi_F,\nu_f),(\phi_{F'},\nu_{f'})]:=(F'(f)-F(f'),0)
\ee
and if one would like to quantize the system based on the real-valued
functions and vector fields $\phi_F,\nu_f$ respectively,
then one would ask to promote them to self-adjoint operators
with commutator algebra isomorphic with (\ref{4.9}).

We are interested in quantizing the system based on another subalgebra
of $C^\infty(\a)\times V^\infty(\a)$  $(A(p),E(S)\mapsto
(\phi_p,\beta\kappa\nu_S)$
which we now must derive using the above regularization.
Let
\ba \label{4.10}
F^{\epsilon k t}_p(x)^a_j &:=& \delta_j^k
\int_{\Rl^{D-1}} d^{D-1}s \;\delta^\epsilon(s_1,..,s_{D-1})
\int_0^1 dt' \delta^\epsilon(t'-t)
\dot{p}^a_{s_1,..,s_{D-1}}(t') \delta(x,p_{s_1,..,s_{D-1}}(t'))
\nonumber\\
f^{\epsilon k}_S(x)_a^j &:=& \delta_j^k
\int_{\Rl} ds\;\delta^\epsilon(s)\int_U d^{D-1}u
\epsilon_{a a_1..a_{D-1}}
\frac{\partial S^{a_1}_s(u_1,..,u_{D-1})}{\partial u_1}..
\frac{\partial S^{a_{D-1}}_s(u_1,..,u_{D-1})}{\partial u_{D-1}}
\times \nonumber\\
&& \times \delta(x,S_s(u_1,..,u_{D-1}))
\ea
then we trivially have
\ba \label{4.11}
h^\epsilon_p(A)&=&{\cal P} e^{\int_0^1 F^{\epsilon j t}_p(A)\tau_j/2}
\nonumber\\
E^\epsilon_j(S)=E(f^{\epsilon j}_S)
\ea
Notice that
the smearing functions (\ref{4.10}) are not quite smooth due to
the sharp cut-off at the boundary of the family of paths and surfaces
respectively but this does not cause any trouble, the smeared
functions are still differentiable because the functional derivatives
(\ref{4.10}) define a bounded linear functional on $\cal M$ (see section
\ref{s2}).

Formula (\ref{4.11}) enables us to map our regulated holonomy and surface
variables
into the Lie algebra $C^\infty(\a)\times V^\infty(\a)$ via
\be \label{4.12}
h^\epsilon_p(A)\mapsto \phi^\epsilon_p:=
{\cal P}e^{\int_0^1 dt \phi_{F^{\epsilon j t}_p}\tau_j/2}
\mbox{ and }
E^\epsilon_j(S)\mapsto \nu^\epsilon_{S j}:=\nu_{f^{\epsilon j}_S}
\ee
compute their algebra and then take the limit $\epsilon\to 0$ where
we may use the known action of $\nu_f$ on $\phi_F$.

Now the following issue arises:\\
By (\ref{4.9}) the
vector fields $\nu_{f^{\epsilon j}_S}$ are Abelean at finite
$\epsilon$. On the other hand, we will compute a vector
field $\nu_S$ by $\nu_S(\phi_p):=\lim_{\epsilon\to 0}
\nu^\epsilon_{S j}(\phi^\epsilon_p)$. But taking the limit $\epsilon\to 0$
and computing Lie brackets of vector fields might not commute, as we will
see, it does not, the algebra of the $\nu_S$ will be non-Abelean,
specifically, $[\nu_S,\nu_{S'}]$ is generically non-vanishing
if $S\cap S'\not=\emptyset$.
This is no cause of trouble because we will take the resulting limit Lie
algebra as a starting point for quantization. It is here where it was
important to have started with the Lie algebra of functions and vector
fields, the commutator $[\nu_S,\nu_{S'}]$ is no longer of the form
$\nu_{S^\dprime}$ and thus does not come from some $E(S^\dprime)$, hence,
if we would have based quantization on a Poisson algebra of functions
we would get into trouble as it would not be a closed Poisson algebra of
functions any longer. However, the Lie bracket algebra of vector fields
is always closed and hence our elementary classical algebra that quantization
will be based on will be the smallest closed
subalgebra of $C^\infty(\a)\times V^\infty(\a)$ generated by the
$\phi_p,\nu_S$. Of course, only vector fields of the form $\nu_S$
will have a classical interpretation as some $E(S)$ which is good enough
in order to take the classical limit.

Let us then actually compute $\phi_p,\nu_S$:\\ The calculation is quite
lengthy and involves expanding out carefully the path ordered exponential
in (\ref{4.12}) and using the known action $\nu_f(\phi_F)=F(f)=\int d^Dx
F^a_j(x) f_a^j(x)$.
We find
\ba \label{4.13}
\nu^{\epsilon'}_{Sj}(\phi^\epsilon_p)
&=&\sum_{n=1}^\infty \int_0^1 dt_n\int_0^{t_n}
dt_{n-1}..\int_0^{t_2} dt_1 \sum_{k=1}^n \times\nonumber\\
&\times&
(\phi_{F^{\epsilon j_1 t_1}_p}\tau_{j_1}/2)..
(\phi_{F^{\epsilon j_{k-1}t_{k-1}}_p}\tau_{j_{k-1}}/2)
[\nu_{f^{\epsilon' j}_S}(\phi_{F^{\epsilon j_kt_k}_p})\tau_{j_k}/2]
\times \nonumber\\
&\times&
(\phi_{F^{\epsilon j_{k+1} t_{k+1}}_p}\tau_{j_{k+1}}/2)..
(\phi_{F^{\epsilon j_n t_n}_p}\tau_{j_n}/2)
\ea
Using
\ba \label{4.14}
\nu_{f^{\epsilon' j}_S}(\phi_{F^{\epsilon k t}_p})
&=& \delta_{jk} \int_{\Rl^{D-1}} d^{D-1}s \delta^\epsilon(s_1,..,s_{D-1})
\int_\Rl ds \delta^{\epsilon'}(s)
\int_0^1 dt' \delta^\epsilon(t'-t) \int_U d^{D-1}u  \times\nonumber\\
&\times &
\dot{p}^a_{s_1,..,s_{D-1}}(t')
\epsilon_{a a_1..a_{D-1}}
\frac{\partial S^{a_1}_s(u_1,..,u_{D-1})}{\partial u_1} ..
\frac{\partial S^{a_{D-1}}_s(u_1,..,u_{D-1})}{\partial u_{D-1}}
\times \nonumber\\
&\times&
\delta(S_s(u_1,..,u_{D-1}),p_{s_1,..,s_{D-1}}(t'))
\ea
we can now take {\it first} the limit $\epsilon\to 0$ and
{\it then} $\epsilon'\to 0$ (the reason for doing this will become
transparent below).
The
result is
\ba \label{4.15}
\nu^{\epsilon'}_{Sj}(\phi_p)
&:=& \sum_{n=1}^\infty \int_0^1 dt_n\int_0^{t_n}
dt_{n-1}..\int_0^{t_2} dt_1 \sum_{k=1}^n \times\nonumber\\
&\times&
A(t_1)..A(t_{k-1})
[\lim_{\epsilon \to 0}\nu_{f^{\epsilon' j}_S}(\phi_{F^{\epsilon j_k t_k}_p})
\tau_{j_k}/2] A(t_{k+1}) ..A(t_n)
\ea
where the limit in the square bracket is given by the distribution
\ba \label{4.16}
&& \delta_{j j_k} \int_\Rl ds \delta^{\epsilon'}(s)
\int_U d^{D-1}u
\dot{p}^a(t_k)
\epsilon_{a a_1..a_{D-1}}
\frac{\partial S_s^{a_1}(u_1,..,u_{D-1})}{\partial u_1}..
\frac{\partial S_s^{a_{D-1}}(u_1,..,u_{D-1})}{\partial u_{D-1}}
\times \nonumber\\
&& \times \delta(S_s(u_1,..,u_{D-1}),p(t_k))
\ea
Luckily, there is an additional $t_k$ integral involved in (\ref{4.15})
so that the end result will be non-distributional. Let $t\mapsto F(t)$
be any (integrable) function and consider the integral
\ba \label{4.17}
&& \int_\Rl ds \delta^{\epsilon'}(s)
\int_U d^{D-1}u\int_0^{t_{k+1}} dt
F(t) \dot{p}^a(t)
\epsilon_{a a_1..a_{D-1}}
\frac{\partial S_s^{a_1}(u_1,..,u_{D-1})}{\partial u_1}..
\frac{\partial S_s^{a_{D-1}}(u_1,..,u_{D-1})}{\partial u_{D-1}}
\times \nonumber\\
&& \times \delta(S_s(u_1,..,u_{D-1}),p(t))
\ea
Notice first of all that the derivative $\dot{p}$ is well-defined since
$p$ is piecewise analytic. Next, we can subdivide $p$ into analytic
segments $e$ (edges) of the following four types:\\
{\it up}\\
$e$ intersects $S$ in one of its endpoints only, i.e.
$q:=S\cap e=b(e)$ or $S\cap e=f(e)$ (but not both, subdivide an
edge into two halves if necessary). Let
$T_q(S)$ be the $D-1$
dimensional subspace of the tangent space $T_q(\sigma)$ at $q$ spanned by
the vectors $\partial S/\partial u_k(u_1,..,u_{D-1})_{S(u)=q}$
tangential to $S$ at $q$ carrying the orientation induced from
$S$, that is,
\be \label{4.18}
n_a(q):=
\epsilon_{a a_1..a_{D-1}}
(\frac{\partial S^{a_1}(u_1,..,u_{D-1})}{\partial u_1}..
\frac{\partial S^{a_{D-1}}(u_1,..,u_{D-1})}{\partial u_{D-1}})_{S(u)=q}
\ee
is the outward normal direction.
Consider all derivatives $(d^n e/dt^n)(t)_{e(t)=q})$ and
take
the first one, $(d^{n_q} e/dt^{n_q})(t)_{e(t)=q}$ which, considered as a
tangential vector, does not lie in $T_q(S)$. Then we require that
\be \label{4.19}
(-1)^{(n_q-1) \theta(q,e)}(d^{n_q} e^a/dt^{n_q})(t)_{e(t)=q}n_a(q)>0
\ee
where $\theta(q,e)=0$ if $q=b(e)$ and $\theta(q,e)=1$ if $q=f(e)$.
If the orientation of $e$ induced from $p$ is such that it is outgoing
from $q$, that is $q=b(e)$ then there is a neighbourhood $U$ of $q$ such
that $U\cap r(e)$ lies ``above" $S$. If $e$ is ingoing, that is $q=f(e)$,
then $q=b(e^{-1})$ so that $e^{-1}$ is outgoing and we have
$e^{(n)}(1)=(-1)^n (e^{-1})^{(n)}(0)$ where $e^{-1}(t)=e(1-t)$, so
(\ref{4.19}) makes sure that $U\cap r(e^{-1})$ lies ``below" $S$.
Since $r(e)=r(e^{-1})$ also $U\cap r(e)$ lies below $S$ in this case.
One could summarize this by saying that the
{\it up} case corresponds to edges whose orientation points ``upwards" the
normal direction of $S$. \\
{\it down}\\
The same as in the ``up" case but now
\be \label{4.20}
(-1)^{(n_q-1) \theta(q,e)}(d^{n_q} e^a/dt^{n_q})(t)_{e(t)=q}n_a(q)<0
\ee
Now  $U\cap r(e)$ lies ``below" $S$ if outgoing and ``above" if ingoing,
so the orientation of $e$ is such that it points ``downwards" the
normal direction of $S$.\\
{\it inside}\\
The segment lies entirely inside $S$, that is $S\cap e=e$, so that
for each $q\in e$ and any $n=1,2,..$ we have
\be \label{4.21}
(d^n e^a/dt^n)(t)_{e(t)=q} n_a(q)=0
\ee
{\it outside}\\
The segment $e$ does not intersect $S$ at all, that is, $e\cap
S=\emptyset$.\\
\\
First of all we notice that due to piecewise analyticity of $p$,
$p$ is a finite composition of entire analytic segments and for each
of them the
number of edges of the {\it up} and {\it down} type must be finite.
Namely, otherwise we could draw an analytic, inextendable curve $c$ within
$S$ ($c$ is then analytic because it lies in the analytic surface $S$)
through this infinite number of isolated intersection points which means
that actually $S\cap e\subset c$ since $e$ is analytic, that is, $e$ has no
isolated intersection points at all which is a contradiction.
On the other hand, if $e$ contains a segment of the {\it inside} type
then $e$ cannot have a segment of the {\it up} or {\it down} type
because of analyticity, that is, $e=e_1\circ (e\cap U) \circ e_2$ where
$e_1\cap U=e_2\cap U=\emptyset$ are of the {\it outside} type and $e\cap U$
is an (open, since $U$ is open) segment of the {\it inside} type.

We conclude that $p$ is a composition
$p=e_1\circ e_2\circ ..\circ e_N$ where each $e_k$ is an
analytic edge of a definite type. Let
$e_l=p([t'_{l-1},t'_l])$ for some $t'_l,\;l=0,..,N,\;0=t'_0<t'_1<..<t'_N=1$
and define $0\le l(t)\le N-1$ to be the largest number such that
$t'_{l(t)}\le t$. Then (\ref{4.17}) becomes
\ba \label{4.17a}
&& \sum_{l=1}^{l(t_{k+1})}\int_\Rl ds \delta^{\epsilon'}(s)
\int_U d^{D-1}u\int_{t'_{l-1}}^{t'_l} dt
F(t) \dot{p}^a(t)
\epsilon_{a a_1..a_{D-1}}
\frac{\partial S_s^{a_1}(u_1,..,u_{D-1})}{\partial u_1}..
\nonumber\\
&..&
\frac{\partial S_s^{a_{D-1}}(u_1,..,u_{D-1})}{\partial u_{D-1}}
\delta(S_s(u_1,..,u_{D-1}),p(t))
\nonumber\\
&&+ \int_\Rl ds \delta^{\epsilon'}(s)
\int_U d^{D-1}u\int_{t'_{l(t_{k+1})}}^{t_{k+1}} dt
F(t) \dot{p}^a(t)
\epsilon_{a a_1..a_{D-1}}
\frac{\partial S_s^{a_1}(u_1,..,u_{D-1})}{\partial u_1}..
\nonumber\\
&..&
\frac{\partial S_s^{a_{D-1}}(u_1,..,u_{D-1})}{\partial u_{D-1}}
\delta(S_s(u_1,..,u_{D-1}),p(t))
\nonumber\\
&=&
\sum_{l=1}^{l(t_{k-1})}\int_\Rl ds \delta^{\epsilon'}(s)
\int_U d^{D-1}u\int_{0}^1 dt
F(\tilde{t}_l(t)) \dot{e_l}^a(t)
\epsilon_{a a_1..a_{D-1}}
\frac{\partial S_s^{a_1}(u_1,..,u_{D-1})}{\partial u_1}..
\nonumber\\
&..&
\frac{\partial S_s^{a_{D-1}}(u_1,..,u_{D-1})}{\partial u_{D-1}}
\delta(S_s(u_1,..,u_{D-1}),p(t))
\nonumber\\
&&+ \int_\Rl ds \delta^{\epsilon'}(s)
\int_U d^{D-1}u\int_{t'_{l(t_{k+1})}}^{\delta_{l(t_{K+1})}(t_{K+1})} dt
F(\tilde{t}_l(t)) \dot{e}_{l(t_{k+1})+1}^a(t)
\epsilon_{a a_1..a_{D-1}}
\frac{\partial S_s^{a_1}(u_1,..,u_{D-1})}{\partial u_1}..
\nonumber\\
&..&
\frac{\partial S_s^{a_{D-1}}(u_1,..,u_{D-1})}{\partial u_{D-1}}
\delta(S_s(u_1,..,u_{D-1}),p(t))
\ea
where in the second step we have used reparameterization invariance
of (\ref{4.17}) and a reparameterization of
$e_l$ given by $[t'_{l-1},t'_l]=\tilde{t}_l([0,1])$ for $l\le l(t_{k-1})$
and $[t'_l,t_{k+1}]=\tilde{t}_l([0,\delta_{l(t_{k+1})}(t_{k+1})])$
for $l=l(t_{k+1})+1$ where $\delta_l(t)=0,1/2,1$ if
$t=t'_{l-1},t'_{l-1}<t<t'_l,t=t'_l$.

Consider then for $t'=0,1/2,1$ the integral
\ba \label{4.22}
&& \int_\Rl ds \delta^{\epsilon'}(s)
\int_U d^{D-1}u\int_0^{t'} dt
F(t) \dot{e}^a(t)
\epsilon_{a a_1..a_{D-1}}
\frac{\partial S_s^{a_1}(u_1,..,u_{D-1})}{\partial u_1}..
\nonumber\\
&..&
\frac{\partial S_s^{a_{D-1}}(u_1,..,u_{D-1})}{\partial u_{D-1}}
\delta(S_s(u_1,..,u_{D-1}),e(t))
\ea
where $e\in \{e_1,..,e_N\}$ and $t'\in \{0,1/2,1\}$. We compute the
value of (\ref{4.22}) for each possible type of $e$ separately.
Obviously, (\ref{4.22}) vanishes if $t'=0$ so we only consider
$t'=1/2,1$.
\\
Case {\it outside}:\\
This case is trivial, we let $\epsilon'\to 0$ in (\ref{4.22}) and see that
the value of the integral becomes arbitrarily small and vanishes in
the limit because the integrand has then support on an empty set.\\
Case {\it inside}\\
Since $s\mapsto S_s$ is a congruence it is clear that
$\delta(S_s(u_1,..,u_{D-1}),e(t))$ has support at $s=0$ and the unique
solution $u_1(t),..,u_{D-1}(t)$ (which are interior points of $U$ since
$S$ is open) of the equation $S(u)=e(t)$. Thus (\ref{4.22}) becomes
\be \label{4.23}
\delta^{\epsilon'}(0) \int_0^{t'} dt
F(t)
\frac{\dot{e}^a(t)
\epsilon_{a a_1..a_{D-1}}
[\frac{\partial S^{a_1}}{\partial u_1}..
\frac{\partial S^{a_{D-1}}}{\partial u_{D-1}}]_{u(t)}}
{|\det(\partial S_s(u)/\partial(s,u_1,..,u_{D-1}))_{s=0,u=u(t)}|}
\ee
which vanishes at finite $\epsilon'$ since the denominator is finite
while the numerator vanishes by definition of an {\it inside} edge.
Since (\ref{4.23}) vanishes at finite $\epsilon'$ its limit $\epsilon'\to
0$ vanishes as well. Expression (\ref{4.23}) is the precise reason for
why we have not synchronized the limits $\epsilon\to 0,\epsilon'\to 0$
as otherwise we would have obtained an ill-defined result of the
form $0\cdot\infty$.\\
Case {\it up}\\
Suppose first that $q=S\cap e=b(e)$. Expanding $e(t)$ around $t=0$ yields
$e(t)=q+\frac{t^{n_q}}{n!}(d^{n_q} e/dt^{n_q})(0)+O(t^{n_q+1})$. Likewise
let $u(q)$ be the unique solution of $S(u)=q$ and let us expand
$S(u)=q+\sum_{k=1}^{D-1} (\partial S/\partial u_k)(u(q))[u_k-u(q)_k]
+O((u-u(q))^2)$. Let us introduce new coordinates
$v_k=u_k-u(q)_k,\;k=1,..,D-1;\;v_D=t^n/(n !)$ and the matrix
$(M_q)^a_k$ with $a,k=1,..,D$ and entries
$M^a_k= (\partial S^a/\partial u_k)(u(q))$ for $k<D$ and
$M^a_D=-(d^{n_q} e^a/dt^{n_q})(0)$. Letting $\epsilon'\to 0$, (\ref{4.22})
becomes
\ba \label{4.24}
&&
\int_U d^{D-1}u\int_0^{t'} dt
F(t) [(d^n e^a/dt^{n_q})(0)t^{n_q-1}/((n_q-1)!)+O(t^{n_q})]
\epsilon_{a a_1..a_{D-1}}
\times\nonumber\\
&\times& [\frac{\partial S^{a_1}}{\partial u_1}(u(q))+O(v_1)]..
[\frac{\partial S^{a_{D-1}}}{\partial u_{D-1}}(u(q))+O(v_{D-1})]
\delta(M(q)\cdot v+O(v))
\nonumber\\&=&
\int_{(U-u(q))\times[0,(t')^{n_q}/((n_q)!)]}
d^Dv F([(n_q !)v_D]^{1/n_q}) (-\det(M_q)+O(v^{1/n_q}))
\delta(M(q)\cdot v+O(v))
\nonumber\\
&=&
-\frac{\det(M_q)}{|\det(M_q)|}\frac{F(0)}{2}=\frac{F(0)}{2}
\ea
where in the last step we noticed that
$-\det(M_q)=n_a(q)(e^{(n_q)})^a(0)>0$ by definition of the {\it up}
type. The factor of $1/2$ is due to the fact that the support $v_D=0$
of the $\delta$ distribution
is at the boundary of the $v_D$ integral (while $v_k=0$ is in the interior
of the $v_k$ integral for $k<D$ since $S$ is open), in other words,
$\int_0^1 dt \delta(t) F(t)=F(0)/2$. One may ask whether we could not have
chosen a different prescription for the value of the integral, say
$\int_0^1 dt\delta(t)F(t)=sF(0)=F(0)-\int_{-1}^0 dt \delta(t) F(t)$ for some
$1<s<1$. However, it is only for the value $s=1/2$ that the area operator,
to be derived below, is invariant under switch of the orientation
of the surface that it measures as it physically must be, see below. The reason
is that now $\epsilon(S,e)=2s$ for the up type while $\epsilon(S,e)=-2(1-s)$
for the down type. Under switch of orientation of $S$ this becomes
$\epsilon(S,e)=2(1-s)$ for the up type while $\epsilon(S,e)=-2s$
for the up type.

Suppose now $q=S\cap e=f(e)$. Then, taking the limit $\epsilon'\to 0$ we see
that (\ref{4.22}) vanishes if $t'=1/2$ while by a similar calculation
it takes the value $F(1)/2$ if $t'=1$ (switch to
$v_D=(1-t)^{n_q}/(n_q !)$ instead noticing that
$e(t)=q+(-1)^{n_q} e^{(n_q)}(1) v_D+..,\;
\dot{e}(t)=(-1)^{n_q-1} |\dot{v}_D|$).

We conclude that the value of the integral (\ref{4.22}) is given by
$[F(0)\theta(t'-1/4)\delta_{S\cap e,b(e)}+F(1)\theta(t'-1)
\delta_{S\cap e,f(e)}]/2$
where $\theta(t')=1$ for $t'\ge 0$ and $\theta(t')=0$ for $t'<0$
denotes the step function.
\\
Case {\it down}\\
The calculation is completely analogous with the result that
(\ref{4.22}) is given by
$-[F(0)\theta(t'-1/4)\delta_{S\cap e,b(e)}+F(1)\theta(t'-1)
\delta_{S\cap e,f(e)}]/2$. \\
\\
We can summarize the analysis by defining $\epsilon(e,S)$ to be
$+1,-1,0$ whenever $e$ has type {\it up, down or in(out)side} respectively
whence the value of (\ref{4.22}) is given by
\be \label{4.25}
\epsilon(e,S)
[F(0)\theta(t'-1/4)\delta_{S\cap e,b(e)}+F(1)\theta(t'-1)
\delta_{S\cap e,f(e)}]/2
\ee
Inserting (\ref{4.25}) into (\ref{4.15}) we obtain
\ba \label{4.26}
&& \nu_{Sj}(\phi_p):= \lim_{\epsilon'\to 0} \nu^{\epsilon'}_{Sj}(\phi_p)
\nonumber\\
&=&\frac{1}{2}\sum_{n=1}^\infty \int_0^1 dt_n\int_0^{t_n}
dt_{n-1}..\int_0^{t_2} dt_1 \sum_{k=1}^n
A(t_1)..A(t_{k-1})  \times\nonumber\\
&\times&
\{\sum_{l=1}^{l(t_{k+1})} \epsilon(e_l,S)
[\delta(t_k-\tilde{t}_l(0))\delta_{S\cap
e_l,b(e_l)}+\delta(t_k-\tilde{t}_l(1))
\delta_{S\cap e_l,f(e_l)}]
\nonumber\\
&&+\epsilon(e_{l(t_{k+1})+1},S)
[\delta(t_k-\tilde{t}_{l(t_{k+1})+1}(0))
\theta(\delta_{l(t_{k+1})}(t_{k+1})-1/4)
\delta_{S\cap e_{l(t_{k+1})+1},b(e_{l(t_{k+1})+1})}
\nonumber\\
&&+
\delta(t_k-\tilde{t}_{l(t_{k+1})+1}(1))
\theta(\delta_{l(t_{k+1})}(t_{k+1})-1)
\delta_{S\cap e_{l(t_{k+1})+1},f(e_{l(t_{k+1})+1})}]\}
\tau_j/2\times\nonumber\\
A(t_{k+1}) ..A(t_n)
\nonumber\\
&=& \frac{1}{2}\sum_{n=1}^\infty \int_0^1 dt_n\int_0^{t_n}
dt_{n-1}..\int_0^{t_{k+2}} dt_{k+1}
\sum_{k=1}^n
\times\nonumber\\
&\times&
\{\sum_{l=1}^{l(t_{k+1})} \epsilon(e_l,S)
[\delta_{S\cap e_l,b(e_l)}\int_0^{\tilde{t}_l(0)} dt_{k-1}
+\delta_{S\cap e_l,f(e_l)}\int_0^{\tilde{t}_l(1)} dt_{k-1}]
\nonumber\\
&&+\epsilon(e_{l(t_{k+1})+1},S)
[\theta(\delta_{l(t_{k+1})}(t_{k+1})-1/4)
\delta_{S\cap e_{l(t_{k+1})+1},b(e_{l(t_{k+1})+1})}
\int_0^{\tilde{t}_{l(t_{k+1})+1}(0)} dt_{k-1}
\nonumber\\
&& +
\theta(\delta_{l(t_{k+1})}(t_{k+1})-1)
\delta_{S\cap e_{l(t_{k+1})+1},f(e_{l(t_{k+1})+1})}
\int_0^{\tilde{t}_{l(t_{k+1})+1}(1)} dt_{k-1}
]\}
\times\nonumber\\
&&\times
\int_0^{t_{k-1}} dt_{k-2}.. \int_0^{t_2} dt_1
A(t_1)..A(t_{k-1})\tau_j/2 A(t_{k+1}) ..A(t_n)
\ea
Now, using the algebraic properties of the holonomy
$h_{p\circ p'}=h_p h_{p'},\;h_{p^-1}=h_p^{-1}$, working out the consequences
in terms of path ordred exponentials,
one can convince oneself after tedious algebra that
\ba \label{4.27}
\nu_{Sj}(\phi_p)
&=& \sum_{l=1}^N \frac{\epsilon(e_l,S)}{2}
h_{e_1}(A)..h_{e_{l-1}}(A)
\nonumber\times \\
&=&
[\delta_{e_l\cap S,b(e_l)}\frac{\tau_j}{2} h_{e_l}(A)+
\delta_{e_l\cap S,f(e_l)}h_{e_l}(A)\frac{\tau_j}{2}]
h_{e_{l+1}}(A)..h_{e_N}(A)
\ea
This is our end result. Notice that the details of the regularization of
the delta-distribtions did not play any role. It was seemingly
important that we smeared  via congruences of curves and surfaces
as compared to more general smearings, however, any ``reasonable" smearing
admits a foliation via curves and surfaces respectively. Thus, the result
(\ref{4.27}) is fairly general.

\subsubsection{Invariant Vector Fields on $G^n$}
\label{s4.1.3}

The amazing feature of expression (\ref{4.27}) is that it is again a product
of a finite number of holonomies, the harvest of having started from
a manifestly background independent formulation. If we would
have started from a function of $E$ which is smeared in all
$D$ directions then this would be no longer true, (\ref{4.27}) would be
replaced by a more complicated expression in which an additional
integral over the extra dimension would appear.

The fact that (\ref{4.27}) is again a product of holonomies enables us
to generalize the action of $\nu_{j S}$ to arbitrary cylindrical
functions, restricted to smooth connections.
Let $f\in \mbox{Cyl}^1(\ab)$, then we find
a subgroupoid $l=l(\gamma)\in {\cal L}$ and $f_l\in C^1(X_l)$ such that
$f=p_l^\ast f_l=[f_l]_\sim$ and a complex valued function $F_l$ on
$G^{|E(\gamma)|}$ such that $f(A)=f_l(p_l(A))=F_l(\rho_l(p_l(A)))$ with
$\rho_l(A_l)=\{A_l(e)\}_{e\in E(\gamma)}=\{A(e)\}_{e\in E(\gamma)}$.
We may choose $\gamma$ in such a way that it is {\it adapted} to a given
surface $S$, that is, each edge of $\gamma$ has a definite type
with respect to $S$. This will make the following computation simpler.
Notice that every graph can be chosen to be adapted by subdividing edges
appropriately. Let us now restrict $f$ to $\a$ then
\be \label{4.28}
[\nu_{Sj}(f)](A)=\frac{1}{2}\sum_{e\in E(\gamma)} \epsilon(e,S)
[\delta_{e\cap S,b(e)}\frac{\tau_j}{2} A(e)
+\delta_{e\cap S,f(e)}A(e)\frac{\tau_j}{2}]_{AB}
\frac{\partial F_l}{\partial A(e)_{AB}}(\{A(e')\}_{e'\in E(\gamma)})
\ee
Evidently, (\ref{4.28}) leaves $C^\infty(X_l)$ restricted to $\a$ invariant
{\it which is why we can extend it to all of $\ab$}!

More precisely:\\
Define the so-called right -- and left invariant vector fields on $G$
by
\ba \label{4.29}
(R_j f)(h) &:=& (\frac{d}{dt})_{t=0} f(e^{t\tau_j}h)
=:(\frac{d}{dt})_{t=0} [L^\ast_{e^{t\tau_j}}f](h)
\nonumber\\
(L_j f)(h) &:=& (\frac{d}{dt})_{t=0} f(h e^{t\tau_j})
=:(\frac{d}{dt})_{t=0} [R^\ast_{e^{t\tau_j}}f](h)
\ea
where $R_h(h')=h' h,\;L_h(h')=hh'$ denotes the right and left action
of $G$ on itself. The right (left) invariance of $R_j$ ($L_j$),
that is, $(R_h)_\ast R_j=R_j$ ($(L_h)_\ast L_j=L_j$), follows  immediately
from the commutativity of left and right translations
$L_h R_{h'}=R_{h'} L_h$. Notice, however, that the right invariant field
generates left translations and vice versa. Then we can write (\ref{4.28})
in the compact form
\be \label{4.30}
Y^j_l(S)[f_l]=\frac{1}{4} \sum_{e\in E(\gamma)} \epsilon(e,S)
[\delta_{e\cap S,b(e)} R^j_e+\delta_{e\cap S,f(e)} L^j_e] f_l
\ee
where $R^j_e$ is $R^j$ on the copy of $G$ labelled by $e$ and where from
now on we just identify $X_l$ with $G^{|E(\gamma)|}$ via $\rho_l$.
Expression (\ref{4.30}) obviously does not require us to restrict
$f=p_l^\ast f_l$ to $\a$ any more. Notice that while $Y^j_l(S)$, just as
$E_j(S)$ does not have a simple transformation behaviour under gauge
transformations, $R^j_e,L^j_e$ in fact do
\ba \label{4.30a}
[(\lambda^e_g)^\ast([(\lambda^e_g)_\ast R^j_e](f_e))](h_e)
&=&[R^j_e((\lambda^e_g)^ast f_e)](h_e)
=(\frac{d}{dt})_{t=0}f_e(g(b(e))e^{t\tau_j}h_e g(f(e))^{-1})
\nonumber\\
&=&(\frac{d}{dt})_{t=0}
f_e(e^{t\mbox{ad}_{g(b(e))}(\tau_j)}g(b(e))h_e g(f(e))^{-1})
\nonumber\\
&=& [(\lambda^e_g)^\ast (R^{\mbox{ad}_{g(b(e))}(\tau_j)} f_e)](h_e)
\ea
so that $(\lambda^e_g)_\ast R^j_e=[\mbox{ad}_{g(b(e))}(\tau_j)]_k R^k_e$
where $\mbox{ad}_{g(b(e))}(\tau_j)=:[\mbox{ad}_{g(b(e))}(\tau_j)]_k \tau_k$.
Similarly $(\lambda^e_g)_\ast L^j_e=[\mbox{ad}_{g(f(e))}(\tau_j)]_k L^k_e$.
This shows once more that $R^j_e$ ($L^j_e$) is right (left) invariant.

We thus have found a family of vector fields $Y^j_l(S)$ whenever
$l$ is adapted to $S$. If $l=l(\gamma)$ is not adapted then we can
produce an adapted one $l_S=l(\gamma')$ e.g. by choosing
$r(\gamma)=r(\gamma')$ and by subdividing edges of $\gamma$
into those with definite type with respect to $S$ and
where the edges of $\gamma'$ carry the orientation induced by the
edges of $\gamma$. Since $p_{l_S l}^\ast f_l\sim f_l$ we then simply
{\it define}
\be \label{4.31}
p_{l_S l}^\ast(Y^j_l(S)(f_l)):=Y^j_{l_S}(p_{l_S l}^\ast f_l)
\ee
We must check that (\ref{4.31}) does not depend on the choice
of an adapted subgroupoid. Hence, let $l_S'$ be another adapted subgroupoid
then we find $l_S,l_S'\prec l_S^\dprime$ which is still adapted
(take for instance the union of the corresponding graphs and subdivide
edges as necessary). Since (\ref{4.31}) is supposed to be a
cylindrical function and
$p_{L_s l}\circ p_{l_S^\dprime l_S}=p_{L_s l}\circ p_{l_S^\dprime l_S}$
we must show that
\be \label{4.32}
p_{l^\dprime_S l_S}^\ast Y^j_{l_S}(S)(p_{l_S l}^\ast f_l)
=p_{l^\dprime_S l'_S}^\ast Y^j_{l'_S}(S)(p_{l'_S l}^\ast f_l)
\ee
As usual, if (\ref{4.32}) holds for one such adapted $l_S^\dprime$ then it
holds for all. To see that (\ref{4.32}) holds,
it will be sufficient to show
that for any adapted subgroupoids $l_S\prec l_S^\dprime$ we have
\be \label{4.33}
p_{l^\dprime_S l_S}^\ast Y^j_{l_S}(S)(f_{l_S})
=Y^j_{l_S^\dprime}(S)(p_{l_S^\dprime l_S}^\ast f_{l_S})
\ee
from which then (\ref{4.32}) will follow due to
$p_{l_s l}\circ p_{l_S^\dprime l_S}=p_{l'_s l}\circ p_{l_S^\dprime l'_S}$.
We again need to check three cases:\\
a)\\
$e\in E(\gamma_S^\dprime)$ but $e\not\in E(\gamma_S)$, then
(\ref{4.33}) holds because $p_{l_S^\dprime l_S}^\ast f_{l_S}$
does not depend on $A(e)$ so that the additional terms proportional to
$R^j_e,L^j_e$ in (\ref{4.30}) drop out.\\
b) \\
$e\in E(\gamma_S^\dprime)$ but $e^{-1}\in E(\gamma_S)$.
We observe that with $\tilde{f}(h)=f(h^{-1})$
\be \label{4.34}
(R^j \tilde{f})(h)=(\frac{d}{dt})_{t=0} f((e^{t \tau_j}h)^{-1})
=(\frac{d}{dt})_{t=0} f(h^{-1}e^{-t \tau_j})=-(L^j f)(h^{-1})
\ee
But then
\ba \label{4.35}
&&\epsilon(e,S)
[\delta_{e\cap S,b(e)} R^j_e+\delta_{e\cap S,f(e)} L^j_e] \tilde{f}_e
\nonumber\\
&=& -\epsilon(e,S)
[\delta_{e\cap S,b(e)} L^j_{e^{-1}}+
\delta_{e\cap S,f(e)} R^j_{e^{-1}}] f_{e^{-1}}
\nonumber\\
&=& \epsilon(e^{-1},S)
[\delta_{e\cap S,b(e)} L^j_{e^{-1}}+
\delta_{e\cap S,f(e)} R^j_{e^{-1}}] f_{e^{-1}}
\nonumber\\
&=& \epsilon(e^{-1},S)
[\delta_{e^{-1}\cap S,f(e^{-1})} L^j_{e^{-1}}+
\delta_{e^{-1}\cap S,b(e^{-1})} R^j_{e^{-1}}] f_{e^{-1}}
\ea
where $\tilde{f}_e=p_{e e^{-1}}^\ast f_e$ in obvious notation.
Hence (\ref{4.33}) is satisfied.\\
c)\\
$e_1,e_2\in E(\gamma_S^\dprime)$ but $e=e_1\circ e_2\in E(\gamma_S)$.
If $e\cap S=b(e)$ then $e_1\cap S=b(e_1)$ and $\epsilon(e,S)=\epsilon(e_1,S)$
while
$e_2\cap S=\emptyset$ and $\epsilon(e_2,S)=0$ (recall that
$\epsilon(e,S)\not=0$ implies that $e,S$ intersect in only one point).
Similarly,
if $e\cap S=f(e)$ then $e_2\cap S=f(e_2)$ and $\epsilon(e,S)=\epsilon(e_2,S)$
while
$e_1\cap S=\emptyset$ and $\epsilon(e_1,S)=0$.
Let $f_1(h_1):=f_2(h_2)=f(h_1 h_2)$ then due to left and right invariance
\be \label{4.36}
(R^j f_1)(h_1)=(R^j f)(h_1 h_2) \mbox{ and }
(L^j f_2)(h_2)=(L^j f)(h_1 h_2)
\ee
hence
\ba \label{4.37}
&&\sum_{I=1,2}\epsilon(e_I,S)
[\delta_{e_I\cap S,b(e_I)} R^j_{e_I}+\delta_{e_I\cap S,f(e_I)} L^j_{e_I}]
p_{e_1,e_2),e_1\circ e_2}^\ast f_e
\nonumber\\
&=&
\left\{ \begin{array}{cc}
\epsilon(e_1,S) R^j_{e_1} p_{e_1,e_2),e_1\circ e_2}^\ast f_e &
\mbox{ if } e\cap S=b(e) \\
\epsilon(e_2,S) L^j_{e_2} p_{e_1,e_2),e_1\circ e_2}^\ast f_e &
\mbox{ if } e\cap S=f(e)
\end{array} \right.
\nonumber\\
&=&
\left\{ \begin{array}{cc}
\epsilon(e,S) R^j_e f_e &
\mbox{ if } e\cap S=b(e) \\
\epsilon(e,S) L^j_e f_e &
\mbox{ if } e\cap S=f(e)
\end{array} \right.
\nonumber\\
&=& \epsilon(e,S)
[\delta_{e\cap S,b(e)} R^j_e+\delta_{e\cap S,f(e)} L^j_e]
f_e
\ea
as claimed.\\
\\
Hence our family of vector fields $(Y^j_l(S))_{l\in {\cal L}}$ is now
defined for all possible $l\in {\cal L}$, in the language of the
previous section we have the cofinal set $l_0:=l(\emptyset)\prec {\cal L}$.
Let us check that it is a consistent family, that is
\be \label{4.38}
p_{l' l}^\ast(Y^j_l(S)(f_l))=Y^j_{l'}(p_{l' l}^\ast f_l)
\ee
for all $l\prec l'$ which are not necessarily adapted. Given $l\prec l'$ we
find always an adapted subgroupoid $l,l'\prec l_S$.
Now by the just established independence on the adapted graph we
may equivalently show that
\be \label{4.39}
p_{l_s l'}^\ast p_{l' l}^\ast(Y^j_l(S)(f_l))=
p_{l_s l'}^\ast Y^j_{l'}(p_{l' l}^\ast f_l)
\ee
Now since
$p_{l_s l'}^\ast p_{l' l}^\ast=p_{l_S l}^\ast$
the left hand side equals
$p_{l_S l}^\ast(Y^j_l(S)(f_l)\equiv Y_{l_S}(p_{l_S l}^\ast f_l)$
by definition of $Y_l$ on arbitrary, not necessarily adapted graphs
and the right hand side equals because of the same reason
$Y^j_{l_S}(p_{l_s l'}^\ast p_{l' l}^\ast f_l)=Y_{l_S}(p_{l_S l}^\ast f_l)$.\\
\\
We thus have established that the family of vector fields
$(Y^j_l(S))_{l\in {\cal L}}$ is a consistent family and defines
a vector field $Y^j(S)$ on $\ab$. Notice moreover that $Y^j_l(S)$ is
real valued: From (\ref{4.30}) this will follow if $R_j,L_j$ are real
valued. Now we have embedded $G$ into a unitary group which means
that $\bar{h}^T=h^{-1}$, in particular $\bar{\tau}_j^T=-\tau_j$.
Hence
\ba \label{4.39a}
\overline{R^j_h} &=&\overline{(\tau_j h)_{AB}}\partial/\partial\bar{h}_{AB}
=-(h^{-1}\tau_j )_{BA}\partial/\partial\bar{h}^{-1}_{BA}
\nonumber\\&=&-(h^{-1}\tau_j )_{AB}(\partial h_{CD}/\partial\bar{h}^{-1}_{AB})
\partial/\partial h_{CD}
=(h^{-1}\tau_j )_{AB} h_{CA} h_{BD} \partial/\partial h_{CD}
=R^j_h
\ea
where use was made of $\delta h^{-1}=h^{-1} h h^{-1}$ and the fact that
the symbol $\partial/\partial h_{AB}$ acts as if all components of
$h_{AB}$ were independent by definition of $R_j (f)=(\tau_j h)_{AB}
\partial f/\partial h_{AB}$.

Next we consider its family of divergences with respect to the
uniform measure $\mu_0$. Now the projection $\mu_{0l}$ is simply
the Haar measure on $G^{|E(\gamma)|}$. Since the Haar measure is right
and left invariant, i.e. $(L_h)_\ast \mu_H=\mu_H=(R_h)_\ast \mu_H$
we have $\mbox{div}_{\mu_H} R_j=\mbox{div}_{\mu_H} L_j=0$ as the following
calculation shows:
\be \label{4.40}
-\int_G \mu_h [\mbox{div}_{\mu_H} R_j]f=+\int_G \mu_H R_j(f)=
(\frac{d}{dt})_{t=0}\int_G \mu_H L_{e^{t\tau j}}^\ast f
=(\frac{d}{dt})_{t=0}\int_G (L_{e^{t\tau j}})_\ast\mu_H  f
=0
\ee
It follows that $\mbox{div}_{\mu_{0l}} Y^j_l(S)=0$ so that
$Y^j(S)$ is automatically $\mu_0$ compatible (and the divergence
is real valued).

\subsubsection{Essential Self-Adjointness of Electric Flux Momentum
Operators}
\label{s4.1.4}

Since $Y^j_S$ is a consistently defined smooth vector field on $\ab$
which is $\mu_0-$compatible, all the results
from section \ref{s3.3a} with respect to the definition of
corresponding momentum operators apply and the
remaining question is whether the
family of symmetric operators $P^j_l(S):=i Y^j_l(S)$ with dense
domain $D(P^j_l(S))=C^1(X_l)$ is an essentially
self-adjoint family.

Looking at (\ref{4.30}), essential self-adjointness of $P^j_l(S)$
on $L_2(X_l,d\mu_{0l})$ will
follow if we can show that $i R_j, iL_j$ are essentially self-adjoint
on $L_2(G,d\mu_H)$ with core $C^1(G)$. That they are symmetric operators
we know already. Now we we invoke the Peter\&Weyl theorem that tells
us that
\be \label{4.41}
L_2(G,d\mu_H)=\overline{\oplus_{\pi\in \Pi} L_2(G,d\mu_H)_{|\pi}}
\ee
where $\Pi$ is a collection of representatives of irreducible representations
of $G$, one for each equivalence class, and  $L_2(G,d\mu_H)_{|\pi}$ is the
closed subspace of $L_2(G,d\mu_H)$
spanned by the matrix element functions $h\mapsto \pi_{mn}(h)$. The
observation is now that $R_j,L_j$ leave each $L_2(G,d\mu_H)_{|\pi}$
separately invariant. For instance
\be \label{4.42}
(R_j\pi_{mn})(h)=(\frac{d\pi_{mm'}(e^{t\tau_j})}{dt})_{t=0}\pi_{m'n}(h)
\ee
It follows that $iR_j,iL_j$ are symmetric operators on the finite dimensional
Hilbert space $L_2(G,d\mu_H)_{|\pi}$ of dimension $\dim(\pi)^2$ and therefore
are self-adjoint. Since the matrix element functions are smooth,
by the basic criterion of essential self-adjointness
it follows that $(i (R_j)_{|\pi}\pm i\cdot 1_{\pi})C^\infty(G)_{|\pi}$
is dense in $L_2(G,d\mu_H)_{|\pi}$, hence so is
$(i (R_j)_{|\pi}\pm i\cdot 1_{\pi})C^1(G)_{|\pi}$.
Correspondingly,
\be \label{4.43}
(i R_j\pm i\cdot 1)C^\infty(G)=
\oplus_{\pi \in \Pi}
(i (R_j)_{|\pi}\pm i\cdot 1_\pi) C^\infty(G)_{|\pi}
\ee
is dense in $L_2(G,d\mu_H)$ and thus $iR_j$ is essentially self-adjoint.
The proof for $i L_j$ is the same.

\subsubsection{Selection of the Uniform Measure by Adjointness Conditions}
\label{s4.1.5}

We are now in the position to establish that ${\cal H}^0$ is
a physically relevant Hilbert space. In fact, we ask the more general
question whether this Hilbert space is in some sense naturally selected
by just imposing the canonical commutation relations and the adjointness
conditions: \\
\\
The classical system is a Lie subalgebra of $C^\infty(\a)\times
V^\infty(\a)$ generated by pairs of the form $\phi_p,\nu_{jS}$
associated with $A(p)$ and $E_j(S)$ where $\phi_p$ is $G$ valued and
$\nu_{jS}$ is real valued. We are therefore asked to find a representation
of the Lie subagebra generated by the $\phi_p,\nu_{js}$ by operators
$\hat{A}(p),\hat{E}_j(S)$ on a Hilbert space ${\cal H}^0$ such
that
\be \label{4.43a}
(\hat{A}(p)_{AB})^\dagger=(\hat{A}(p)^{-1})_{BA}
\mbox{ and }
\hat{E}_j(S)^\dagger=\hat{E}_j(S)
\ee
(remember that $G$ is w.l.g. a subgroup of some $U(N)$). More precisely,
$\hat{A}(p)_{AB}$
is supposed to be a bounded operator (no domain questions therefore)
taking values in $G$ and $\hat{E}_j(S)$ should be self-adjoint. Moreover,
we must represent the bracket relations, that is
the classical Lie algebra relations
\be \label{4.44}
[((\phi_p)_{AB},\beta\kappa\nu_{jS}),((\phi_{p'})_{A'B'},\beta
\kappa\nu_{j'S'})]
=\beta\kappa(\nu_{jS}((\phi_{p'})_{A' B'})-\nu_{j'S'}((\phi_p)_{AB}),
\beta\kappa [\nu_{j S},\nu_{j' S'}])
\ee
must be promoted to canonical commutation relations
\ba \label{4.45}
&& [((\hat{\phi}_p)_{AB},\beta\kappa
\hat{\nu}_{jS}),((\hat{\phi}_{p'})_{A'B'}, \beta\kappa \hat{\nu}_{j'S'})]
\nonumber\\
&=&i\beta \ell_p^2((\nu_{jS}((\phi_{p'})_{A'
B'}))^\wedge-(\nu_{j'S'}((\phi_p)_{AB})))^\wedge,
\beta\kappa ([\nu_{j S},\nu_{j' S'}])^\wedge)
\ea
where the Planck area $\ell_p^2=\hbar\kappa$ has naturally come into
play. Here $\hat{A}(p):=\phi_p,\;\beta\kappa\hat{\nu}_{j S}=:\hat{E}_j(S)$.
It is clear that (\ref{4.45}) is trivially satisfied if we represent
an element $(\phi,\nu)\in C^\infty(\a)\times V^\infty(\a)$
on a Hilbert space of the form $L_2(\ab,d\mu)$ with some distributional
extension $\ab$ of $\a$ and some measure $\mu$ thereon by
\be \label{4.64}
(\hat{\phi}\psi)(A):=\phi(A)\psi(A)\mbox{ and }
(\hat{\nu}\psi)(A):=i\hbar[\nu[\psi](A)+(\phi_\nu \psi)(A)]
\ee
where $\phi_\nu$ is a linear function of $\nu$,
provided that the triple $\phi,\nu,\phi_\nu$ can be extended from the smooth
space $\a$ to
the distributional space $\ab$. This will be true for our choice
of $\ab,\phi_p,\nu_{jS}$ provided $\phi_{\nu_{jS}}$ can be chosen
appropriately. So the canonical commutation relations are
formally (since we did not discuss domain questions yet) satisfied then.

Now we come to the adjointness relations. Since $(\hat{\phi}_p)_{AB}$ is
just a $G-$valued multiplication operator, the adjointness relation
for $G$ is trivially satisfied. Now $\nu_{jS}$ is real valued and
in order to get $\hat{\nu}_{jS}$ symmetric to start with one should
choose $\phi_{\nu_{jS}}=\frac{1}{2}\mbox{div}_\mu \nu_{jS}$.
Let now any measure $\mu$ be given and denote its push-forward to $X_l$
by $\mu_l$. Since $X_l$ is finite dimensional, provided that $\mu_l$
is {\it absolutely
continuous} with respect to $\mu_{0l}$, there exists a
mon-negative function $\rho_l$ on $X_l$ such that $\mu_l=\rho_l\mu_{0l}$.
Since $\nu_{jS}$ leaves $C^\infty(X_l)$ invariant and has a push-forward
$(\nu_{jS})_l$ to $C^\infty(X_l)$ given by (\ref{4.30}) which is
a linear combination of left and right invariant vector fields,
it follows that $\mbox{div}_{\mu_l}(\nu_{jS})_l=(\nu_{jS})_l[\ln(\rho_l)]$.
We therefore
see that {\it the uniform measure $\mu\equiv \mu_0$ is uniquely picked}
once we require $\mbox{div}_\mu \nu_{jS}=0$ and that $\mu$ is a probability
measure which is regular with respect to $\mu_0$. In other words, if we had
not constructed $\mu_0$ before,
guided by diffeomorphism -- and gauge invariance, we would have found this
measure anyway now if we use the very natural condition of divergence
freeness. Besides, it is not known whether there exists any other choice
for $\mu$ such that the $\nu_{jS}$ are $\mu-$compatible.

Together with the former results we therefore arrive at the following
classification theorem.
\begin{Theorem}  \label{th4.1}   ~~~~~~~~~~\\
i)\\
Suppose that we want to base quantization on the classical Lie algebra
$\mbox{Cyl}^\infty(\a)\times V^\infty(\a)$. Then the Lie subalgebra
generated by holonomy functions and the vector fields corresponding to
electric fluxes is well-defined and can be extended to a Lie subalgebra of
$\mbox{Cyl}^\infty(\ab)\times V^\infty(\ab)$.\\
ii)\\
The electric flux vector fields arise from a consistent family of vector
fields which are real-valued and compatible with the uniform measure,
the corresponding divergence (which in fact vanishes) being real-valued.
The corresponding momentum operator is essentially self-adjoint with
respect to ${\cal H}^0=L_2(\ab,d\mu_0)$ with core $\mbox{Cyl}^1(\ab)$.\\
iii)\\
The uniform measure $\mu_0$ is uniquely selected among all
regular, Borel probability measures $\mu$ on $\ab$ regular with respect
to it by imposing that 1)
the adjointness -- and canonical commutation relations are implemented
on $L_2(\ab,d\mu)$ and that 2) the electric flux vector fields are
divergence free.
\end{Theorem}

\subsection{Implementation of the Gauss Constraint}
\label{s4.2}

We do not really need to implement the Gauss constraint since we can
directly work with gauge invariant functions (that is, one solves the
constraint classically and quantizes only the phase space reduced with
respect to the Gauss constraint). However, we will nevertheless
show how to get to gauge invariant functions starting from gauge variant
ones by using the technique of refined algebraic quantization
outlined in section \ref{si}.

\subsubsection{Derivation of the Gauss Constraint Operator}
\label{s4.2.1}

We proceed similarly as in the case of the electric flux operator
and start from the classical expression
\be \label{4.65}
G(\Lambda):=-\int d^Dx [D_a \Lambda^j] E^a_j\equiv-E(D\Lambda)
\ee
where $D_a \Lambda^j=\partial_a \Lambda^j+f^j\;_{kl} A_a^k \Lambda^l$
is the covariant derivative of the smearing field $\Lambda^j$.
Notice that (\ref{4.65}) is almost an electric field smeared in $D$
dimensions except that the smearing field $D\Lambda$ depends on
the configuration space. Nevertheless the vector field on $\a$ corresponding
to it is given by $-\kappa\beta\nu_{D\Lambda}$. Next we apply it to
$\mbox{Cyl}(\a)$ by first computing its action on the special functions
$\phi_p$ and then use the chain rule. In order to compute its action
on $\phi_p$ we must regulate it as in the previous section and
then define $\nu_{D\Lambda}(\phi_p):=\lim_{\epsilon\to 0}
\nu_{D\Lambda}(\phi^\epsilon)$. Finally we hope that the end result
is again a cylindrical function which we then may extend to $\ab$
and thus derive a cylindrical family of hopefully consistent vector
fields on $\ab$.

We will not write all the steps, the details are precisely as in the previous
section just that the additional limit $\epsilon'\to 0$ is missing.
For the same reason a split of $p$ into edges of different type is not
necessary because $E$ is smeared in $D$ directions.
One finds
\be \label{4.66}
\nu_{D\Lambda}(\phi_p)=\beta\kappa\int_0^1 dt \dot{p}^a(t)
(D_a\Lambda^j)(p(t)) h_{p([0,t])}(A)\frac{\tau_j}{2}h_{p([t,1])}(A)
\ee
Let us use the notation $\Lambda=\Lambda^j \tau_j$ and
$A(p(t))=\dot{p}^a(t) A_a^j(p(t))\tau_j/2$. Using
$[\tau_j,\tau_k]=2f_{jk}\;^l\tau_l$ we can then recast (\ref{4.66})
into the form
\be \label{4.67}
\nu_{D\Lambda}(\phi_p)=\frac{\beta\kappa}{2}\int_0^1 dt
h_{p([0,t])}(A)
\{\frac{d}{dt}\Lambda(p(t))+[A(p(t)),\Lambda(p(t))]\}
h_{p([t,1])}(A)
\ee
Now we invoke the parallel transport equation for the holonomy
\be \label{4.68}
\frac{d}{dt} h_{p([0,t])}(A)=h_{p([0,t])}(A)
\ee
and use $h_{p([t,1])}(A)=h_{p([0,t])}(A)^{-1} h_p(A)$, then it is
easy to see that (\ref{4.67}) becomes
\be \label{4.69}
\nu_{D\Lambda}(\phi_p)=\frac{\beta\kappa}{2}\int_0^1 dt
\frac{d}{dt} \{h_{p([0,t])}(A)\Lambda(p(t))h_{(p([t,1])}(A)\}
=\frac{\beta\kappa}{2}[-\Lambda(b(p))h_p(A)+h_p(A)\Lambda(f(p))]
\ee
where we have performed an integration by parts in the last step.
So indeed we {\it are} lucky: (\ref{4.69}) is a cylindrical function again.
Let us write $\nu_\Lambda:=-\nu_{D\Lambda}$ then for any
$f_l\in C^\infty(X_l)$ for any subgroupoid $l=l(\gamma)$ we have
\ba \label{4.70}
[\nu_\Lambda(f_l)](A)&=&\frac{\beta\kappa}{2}\sum_{e\in E(\gamma)}
[\Lambda(b(e))A(e)-A(e)\Lambda(f(e))]_{AB}
(\partial f_l/\partial A(e)_{AB})(A)
\nonumber\\
&=&
\frac{\beta\kappa}{2}\sum_{e\in E(\gamma)}
([\Lambda_j(b(e))R^j_e-\Lambda_j(f(e)) L^j_e] f_l)(A)
\ea
Finally we write this as a sum over vertices in the compact form
\be \label{4.71}
G_l(\Lambda)[f_l]:=\nu_\Lambda(f_l)
=\frac{\beta\kappa}{2}\sum_{v\in V(\gamma)}\Lambda_j(v)
[\sum_{e\in E(\gamma);\;v=b(e)}R^j_e
-\sum_{e\in E(\gamma);\;v=f(e)}L^j_e] f_l
\ee
Hence we have successfully derived a family of vector fields
$G_l(\Lambda)\in V^\infty(X_l)$ for any $l\in {\cal L}$. No adaption of
the graph was necessary this time. Since $\Lambda_j$ is real valued
for compact $G$, it follows from our previous analysis that $G_l(\Lambda)$
is real valued. Using the steps a), b) and c) of section \ref{s4.1.3}
one quickly verifies that it is a consistent family and that it is trivially
$\mu_0-$compatible because it is divergence-free since it is a linear
combination of left -- and right invariant vector fields. For the same
reason, the associated momentum operator
\be \label{4.72}
\hat{G}_l(\Lambda)[f_l]=
=\frac{i\beta \ell_p^2}{2}\sum_{v\in V(\gamma)}\Lambda_j(v)
[\sum_{e\in E(\gamma);\;v=b(e)}R^j_e
-\sum_{e\in E(\gamma);\;v=f(e)}L^j_e] f_l
\ee
is essentially self-adjoint with dense domain $C^1(\ab)$.

\subsubsection{Complete Solution of the Gauss Constraint}
\label{s4.2.2}

Using the Lie algebra of the left -- and right invariant vector fields
on $X_l$ given by
\be \label{4.73}
[R^j_e,R^k_{e'}]=-2 \delta_{ee'}f^{jk}\;_l R^l,\;\;
[L^j,L^k]=2 \delta_{ee'}f^{jk}\;_l L^l,\;\;
[R^j,L^k]=0
\ee
(e.g. $([R^j,R^k]f)(h)=(\frac{\partial^2}{\partial s\partial s'})_{s=s'=0}
f([e^{s'\tau_k},e^{s\tau_j}]h)$)
we find
\ba \label{4.74}
[G_l(\Lambda),G_l(\Lambda')]
&=&
(\frac{\beta\kappa}{2})^2\sum_{e\in E(\gamma)}
\{\Lambda_j(b(e))\Lambda'_k(b(e))
[R^j_e,R^k_e]+\Lambda_j(f(e))\Lambda_k(f(e)) [L^j_e,L^k_e]\}
\nonumber \\ &=&-\beta\kappa G([\Lambda,\Lambda'])
\ea
where we have defined $\Lambda(x):=\Lambda_j(x)\tau_j/2$. We see that
the Lie algebra of the $G_l(\Lambda)$ represents the Lie algebra
Lie$(G)$ for each $l\in {\cal L}$ seprately and also represents the
classical Poisson brackets among the Gauss constraints, see section
\ref{s2}. This is already a strong hint that the condition
$\hat{G}(\Lambda)=0$ for all smooth
$\Lambda_j$ really means imposing gauge invariance.

Let us see that this is indeed the case.
According to the programme of RAQ
we must choose a dense subspace of ${\cal H}^0$ which we choose to
be ${\cal D}:=\mbox{Cyl}^\infty(\ab)$. Let $f=[f_l]_\sim$ be a
smooth cylindrical function, that is, $f_l\in C^\infty(X_l)$, then
$\hat{G}(\Lambda)f= p_l^ast(\hat{G}_l(\Lambda)f_l)$. We are looking
for an algebraic distribution $L\in {\cal D}^\ast$ such that
\be \label{4.75}
L(p_l^\ast\hat{G}_l(\Lambda)f_l)=0
\ee
for all $\Lambda_j,\;l\in {\cal L},\;f_l\in C^\infty(X_l)$. Since,
given $l$ the smooth function
$\Lambda$ is still arbitrary, we may restrict its support to one of the
vertices of $\gamma$ with $l=l(\gamma)$ and see that (\ref{4.75})
is completely equivalent with
\be \label{4.76}
L(p_l^\ast[\sum_{e\in E(\gamma);\;v=b(e)}R^j_e
-\sum_{e\in E(\gamma);\;v=f(e)}L^j_e] f_l)=0
\ee
for any $v\in V(\gamma),\;l\in {\cal L},\;f_l\in C^\infty(X_l)$.

We now use the fact that any function in ${\cal D}=C^\infty(\ab)$ is a
finite linear combination of spin-network functions $T_s$. Therefore,
an element $L\in {\cal D}^\ast$ is completely specified by the
complex values $L(T_s)$ with no growth condition on these complex
numbers (an algebraic distribution is well-defined if it is defined
pointwise in $\cal D$). We conclude that any element $L\in{\cal D}^\ast$
can be written in the form
\be \label{4.77}
L=\sum_{s\in {\cal S}} L_s <T_s,.>
\ee
where $<.,.>$ denotes the inner product on $L_2(\ab,d\mu_0)$
and ${\cal S}$ denotes the set of all spin-network labels.
Now, first of all (\ref{4.76})
is therefore completely equivalent with
\be \label{4.79}
L(p_{l(\gamma(s))}^\ast[\sum_{e\in E(\gamma(s));\;v=b(e)}R^j_e
-\sum_{e\in E(\gamma(s));\;v=f(e)}L^j_e] T_s)=0
\ee
for any $v\in V(\gamma(s)),\;s\in {\cal S}$ where $\gamma(s)$ is the
graph that underlies $s$. Since the opertor involved in (\ref{4.79})
leaves $\gamma(s),\vec{\pi}(s)$ invariant and spin-network functions
are mutually orthogonal we find that
\be \label{4.80}
\sum_{s'\in {\cal S},\;\gamma(s')=\gamma(s);\vec{\pi}(s')=\vec{\pi}(s)}
L_{s'}
<T_{s'},[\sum_{e\in E(\gamma(s));\;v=b(e)}R^j_e
-\sum_{e\in E(\gamma(s));\;v=f(e)}L^j_e] T_s>=0
\ee
for any $v\in V(\gamma(s)),\;s\in {\cal S}$.
Effectively the sum over $s'$ is now reduced over all $\vec{m},\vec{n}$
with $m_e,n_e=1,..,d_{\pi_e}$ for any $e\in E(\gamma(s))$ and is
therefore finite. From this it follows already that the most general
solution $L$ is an arbitrary linear combination of solutions of the
form $<\psi,.>$ where $\psi$ is actually normalizable.

Consider now an infinitesimal gauge transformation
$g_t(x)=e^{t\Lambda_j(x)\tau_j}$ for some function
$\Lambda_j(x)$ with $t\to 0$.
Since $\gb\cong G^\sigma$ we may arrange that $g=1$ at all
vertices of $\gamma(s)$ except for $v$. Our spin network
function is of the form
\be \label{4.81}
T_s=[\prod_{e\in E(\gamma(s));\;b(e)=v} f_e(h_e)] \;[\prod_{e\in
E(\gamma(s));\;f(e)=v} f_e(h_e)] F_s
\ee
where $F_s$ is a cylindrical function that does not depend on the edges
incident at $v$. Then under an infinitesimal gauge transformation
the spin-network function changes as
\ba\ \label{4.82}
&& (\frac{d}{dt})_{t=0} \lambda_{g_t}^\ast T_s
=(\frac{d}{dt})_{t=0}
[\prod_{e\in E(\gamma(s));\;b(e)=v} f_e(g_t(v)h_e)] \;[\prod_{e\in
E(\gamma(s));\;f(e)=v} f_e(h_e g_t(v)^{-1})]\; F_s
\nonumber\\
&=&(\frac{d}{dt})_{t=0}
[\circ_{e\in E(\gamma(s));\;b(e)=v} (L^e_{g_t(v)})^\ast]\circ
[\circ_{e\in E(\gamma(s));\;f(e)=v} (R^e_{g_t(v)^{-1}})^\ast]\; T_s
\nonumber\\
&=&\Lambda_j(v)
[\sum_{e\in E(\gamma(s));\;b(e)=v} R^j_e-\sum_{e\in E(\gamma(s));\;f(e)=v}
L^j_e]\; T_s
\nonumber\\
&=& G_{l(\gamma(s))}(\Lambda)[T_s]
\ea
which proves that $G_l(\Lambda)$ is the infinitesimal generator of
$\lambda^l_{e^{t\Lambda}}$. It is therefore clear that the general
solution $L$ is a linear combination of solutions of the form
$<\psi,.>$ where $\psi\in {\cal H}^0$ is gauge invariant. Strictly
speaking, $\psi$ has to be invariant under infinitesimal gauge
transformations only but since $G$ is connected there is no difference
with requiring it to be invariant under all gauge transformations
(the exponential map between Lie algebra and group is surjective since there
is only one component, that of the identity).

We could therefore also have equivalently required that
\be \label{4.83}
L(\lambda_g^\ast f)=L(f)
\ee
for all $g\in\gb$ and all $f\in {\cal D}:=C^\infty(\ab)$.
In passing we recall that we have defined in the previous section a
unitary representation of $\gb$ on ${\cal H}^0$ defined densely
on $C(\ab)$ by $\hat{U}(g)f:=\lambda_g^\ast$. Let $t\mapsto g_t$
be a continuous one-parameter subgroup of $\gb$, meaning that
$\lim_{t\to 0}g_t(x)=g_0(x)\equiv 1_G$ for any $x\in \sigma$, meaning that
$t\mapsto g_{tx}:=g_t(x)$ is a continuous one parameter subgroup of
$G$ for any $x\in \sigma$ (if $g_t$ is continuous at $t=0$ then also
at every $s$ since $\lim_{t\to s}g_t=\lim_{t\to 0}g_t g_s=g_s$ since group
multiplication is continuous). We claim that the one parameter subgroup
of unitary operators $\hat{U}(t):=\hat{U}(g_t)$ is strongly continuous,
that is, $\lim_{t\to 0}||\hat{U}(t)\psi-\psi||=0$ for any $\psi\in
{\cal H}^0$. Since any $\hat{U}(t)$ is bounded and $C^\infty(\ab)$ is dense
in ${\cal H}^0$ it will be sufficient to show that strong continuity
holds when restricted to ${\cal D}$. Also, strong continuity follows already
from weak continuity (i.e. $<\psi,\hat{U}(t)\psi'>\to <\psi,\psi'>$
for any $\psi,\psi'\in {\cal H}^0$) since
$||\hat{U}(t)\psi-\psi||^2=2(||\psi||^2-\Re(<\psi,\hat{U}(t)\psi>)$.
Since $\cal D$ is spanned by finite linear combinations of
mutually orthonormal spin network functions (they are in fact smooth), it
will then be sufficient to
show that $<T_s,\hat{U}(t) T_{s'}>\to <T_s,T_{s'}>=\delta_{ss'}$.
If $s=(\gamma,\vec{\pi},\vec{m},\vec{n})\;,s'=(\gamma',\vec{\pi}',\vec{m}',
\vec{n}')$ then a short computation, using that $\lambda_g$ leaves
$\gamma(s),\vec{\pi}(s)$ invariant, shows that
\be \label{4.84}
<T_s,\hat{U}(t) T_{s'}>=\delta_{\gamma,\gamma'}\delta_{\vec{\pi},\vec{\pi}'}
\prod_{e\in E(\gamma)}[\pi_e(g_t(b(e)))_{m'_e m_e}\;
\pi_e(g_t(f(e))^{-1})_{n_e n'_e}
\ee
and since the matrix element functions are smooth, the claim follows.
We conclude therefore from Stone's theorem that for
$g_t(x)=\exp(t\Lambda(x))$ the operator $\hat{G}(\Lambda)$ is the
self-adjoint generator of $\hat{U}(t)$.

Finally we display the corresponding rigging map. Since $\gb$ is a group,
the obvious ansatz is
\be \label{4.85}
\eta(f):=<\int_{\gb} \mu_H(g) \lambda_g^\ast f,.>
\ee
which, since $\lambda_g^\ast$ preserves $C(C_l)$, is actually a map
${\cal D}\to {\cal D}$. Since $\mu_0$ is a probability measure
we could therefore immediately take the inner product on ${\cal H}^0$
for the solutions $\eta(f)$. But let us see where the rigging map
proposal takes us. By definition
\ba \label{4.86}
<\eta(f),\eta(f')>_\eta &:=&\eta(f')[f]=
\int_{\gb} \mu_H(g) <\lambda_g^\ast f,f'>
\nonumber\\
&=& \int_{\gb} \mu_H(g)\int_{\gb} \mu_H(g') <\lambda_g^\ast f,\lambda_{g'}f'>
=<\eta(f)^\dagger,\eta(f')^\dagger>
\ea
where in the second equality we have observed that $<\lambda_g^\ast f,f'>$
is invariant under gauge transformations of $f'$ and
$\eta(f)^\dagger:=<.,\int_{\gb} \mu_H(g) \lambda_g^\ast f>$.
So, indeed the gauge invariant inner product is just the restricted gauge
variant
inner product. Finally, for any gauge invariant observable we trivially
have $\hat{O}'\eta(f)=\eta(\hat{O} f)$.

\subsection{Implementation of the Diffeomorphism Constraint}
\label{s4.3}

Again we could just start from the fact that we have a reasonable unitary
representation of the diffeomorphism group already defined in section
\ref{s3} but we wish to make the connection to the classical diffeomorphism
constraint more clear in order to show that the representation defined
really comes from the classical constraint. We will work at the
gauge variant level in this section for convenience, however, we could
immediately work at the gauge invariant level and all formulae in this
section go through with obvious modifications. The reason for this
is that the Gauss constraint not only forms a subalgebra in the
full constraint algebra but actually an ideal, that is, since the
Diffeomorphism and Hamiltonian constraint are actually gauge invariant,
the corresponding operators leave the space of gauge invariant
cylindrical functions invariant. Hence one can solve the Gauss
constraint independently before or after solving the other two
constraints.

\subsubsection{Derivation of the Diffeomorphism Constraint Operator}
\label{s4.3.1}

The representation $\hat{U}(\varphi)$ of Diff$(\sigma)$ was densely defined
on spin network functions as
\ba \label{4.87}
&& \hat{U}(\varphi)T_s:= T_{\varphi\cdot s} \mbox{ where }
\varphi\cdot s:=(\varphi\cdot e:=\varphi^{-1}(e),
\\
&&
(\varphi\cdot\vec{\pi}(s))_{\varphi^{-1}(e)}:=\pi_e,
(\varphi\cdot\vec{m}(s))_{\varphi^{-1}(e)}:=m_e,
(\varphi\cdot\vec{n}(s))_{\varphi^{-1}(e)}:=n_e)_{e\in E(\gamma(s))}
\nonumber
\ea
Let $u$ be an analytic vector field on $\sigma$ and consider the
one parameter subgroup $t\to \varphi^u_t$ of Diff$^\omega(\sigma)$
(analytic diffeomorphisms) determined by the integral curves of
$u$, that is, solutions to the differential equation
$\dot{c}_{u,x}(t)=u(c(t)),\;c_{u,x}(0)=x$ with
$\phi^u_t(x):=c_{u,x}(t)$.
The classical diffeomorphism constraint is given by
\be \label{4.88}
V_a=H_a-A_a^j G_j=2(\partial_{[a} A^j_{b]})E^b_j-A_a^j\partial_b E^b_j
\ee
Smearing it with $u$ gives
\be \label{4.89}
V(u)=\int d^3x ({\cal L}_u A^j)_a(x) E^a_j(x)=E({\cal L}_u A)
\ee
where ${\cal L}$ denotes the Lie derivative.
Since the constraint is again linear in momenta we can associate with
it a vector field $\beta\kappa\nu_{{\cal L}_u A}$ on $\a$ which again depends
on $A$ as well. Proceeding similarly as with the Gauss constraint
we find for its action on holonomies of smooth connections
\be \label{4.90}
\nu_{{\cal L}_u A}\phi_p=\int_0^1 ds h_{p([0,s])}(A)({\cal L}_u A)(p(s))
h_{p([s,1])}(A)
\ee
We claim that (\ref{4.90}) equals
\be \label{4.91}
(\frac{d}{dt})_{t=0} h_p((\varphi^u_t)^\ast A)
\ee
To see this, one uses the expansion $(\varphi^u_t)^\ast A=A+t({\cal L}_u A)
+O(t^2)$ and the fact that with $p=p_1\circ ..\circ p_N$ we have
$h_p=h_{p_1}..h_{p_N}$ with $p_k=p([t_{k-1},t_k]),\;0=t_0<t_1<..<t_N=1,\;
t_k-t_{k-1}=1/N$. Denote $\delta h_{p_k}:=h_{p_k}(A+\delta A)-h_{p_k}(A)$
Hence
\ba \label{4.92}
&& h_p(A+\delta A)-h_p(A)=\sum_{n=1}^N\sum_{1\le k_1<..<k_n\le N}
(h_{p_1\circ..\circ p_{k_1-1}}(A)[\delta h_{p_{k_1}}])
(h_{p_{k_1+1}\circ..\circ p_{k_2-1}}(A)[\delta h_{p_{k_2}}])..
\nonumber\\ &&..(h_{p_{k_{n-1}+1}\circ..\circ p_{k_n-1}}(A)[\delta h_{p_{k_n}}])
(h_{p_{k_n+1}\circ..\circ p_N}(A))
\ea
which holds at each finite $N$. Now using the formula
$h_{p_k}(A)={\cal P}\exp(A(p_k))$ where
$A(p_k)=\int_{p_k} A^j\tau_j/2$ we obtain
\be \label{4.93}
\delta h_{p_k}={\cal P} \{e^{[A+\delta A](p_k)}-e^{A(p_k)}\}
\ee
so that $\delta h_{p_k}$ is at least linear in $\delta A$ and therefore
in $t$ for $\delta A=(\varphi^u_t)^\ast A-A$.
Thus, dividing
(\ref{4.92}) by $t$ and taking the limit $t\to 0$ we find
\be \label{4.94}
(\frac{d}{dt})_{t=0} h_p((\varphi^u_t)^\ast A)
=\sum_{k=1}^N h_{p_1\circ..\circ p_{k-1}}(A)
[(\frac{d}{dt})_{t=0} h_{p_k}((\varphi^u_t)^\ast A)]
h_{p_{k+1}\circ..\circ p_N}
\ee
Finally we have $h_{p_k}(A+\delta A)-h_{p_k}(A)=\delta A(p_k)+O(1/N^2)$
so that in the limit $t\to 0$ indeed (\ref{4.94}) turns into
(\ref{4.90}).

Unfortunately, (\ref{4.90}) is no longer a cylindrical function and therefore
we cannot construct a consistent family of cylindrically defined vector
fields on $\ab$, in other words, (\ref{4.90}) cannot be extended to
$\ab$. Of course for each $s$ the functions $h_{p([0,s])}(A)=A(p([0,s])$
can directly be extended to $\ab$, however, ${\cal L}_u A$ makes only sense
for smooth $A$. Moreover, we recall from section \ref{s3} that the measure
$\mu_0$ is supported on connections $A$ such that for any $p\in {\cal P}$
the function $s\mapsto A([0,s])$ is nowhere continuous and therefore
unlikely to be mesurable with respect to $ds$. Thus, we are not able
to define an operator that correspeonds to the infinitesimal
diffeomorphism constraint.

The way out is the observation that {\it
the action of finite diffeomorphisms} can be extended to $\ab$.
In fact, the identity $\nu_{{\cal L}_u}h_p(A)=
(\frac{d}{dt})_{t=0} h_p((\varphi^u_t)^\ast A)$ suggests to consider
the exponentiation of the vector field $\nu_{{\cal L}_u A}$ which
then gives the action $h_p(A)\mapsto h_p((\varphi^u_t)^\ast A)$.
Since classically we can always recover the infinitesimal action
from the exponentiated one, we do not lose any information.
Moreover, we may consider general finite diffeomorphisms $\varphi$
which unlike the $\varphi^u_t$ are not necessarily connected to the identity.
Now, by the duality between $p-chains$ and $p-forms$ we have
for smooth $A$
\be \label{4.95}
h_p(\varphi^\ast A)
={\cal P} e^{\int_p \varphi^\ast A}
={\cal P} e^{\int_{\varphi^{-1}(p)}A}
=h_{\varphi^{-1}(p)}(A)
\ee
which is the reason for taking the inverse diffeomorphism in (\ref{4.87}).
In the form (\ref{4.95}), it is clear that the finite action of
Diff$^\omega(\sigma)$ on $\a$ can be extended to $\ab$ when considering
it as a map between homomorphisms. Hence
\be \label{4.96}
\delta:\;\mbox{Diff}^\omega(\sigma)\times \ab\to \ab;\;\;
(\varphi,A)\mapsto \delta_\varphi(A) \mbox{ where }
[\delta_\varphi(A)](p):=A(\varphi^{-1}(p))
\ee
This furnishes the derivation of the action (\ref{4.96}) already defined
in section \ref{s3}  from the classical diffeomorphism constraint.
Notice that by bconstruction {\it the diffeomorphism quantum constraint
algebra is free of anomalies}
\be \label{4.96a}
\hat{U}(\varphi)\hat{U}(\varphi')
\hat{U}(\varphi^{-1})
\hat{U}((\varphi')^{-1})=\hat{U}(\varphi\circ\varphi'\circ\varphi^{-1}
\circ (\varphi')^{-1})
\ee

\subsubsection{General Solution of the Diffeomorphism Constraint}
\label{s4.3.2}

We have seen that we can define a unitary representation of
Diff$^\omega(\sigma)$
on ${\cal H}^0$ by (\ref{4.87}) and that it is impossible
to construct an action of the Lie algebra of Diff$^\omega(\sigma)$ on $\ab$.
We will now see that this has a counterpart for the representation
$\hat{U}(\varphi)$: If there would be a quantum operator $\hat{V}(u)$
which generates infinitesimal diffeomorphisms, then it would be the
self-adjoint generator of the one parameter subgroup
$t\mapsto \hat{U}(\varphi^u_t)$, that is, we would have
$\hat{U}(\varphi^u_t)=e^{it\hat{V}(u)}$. However, that generator exists
only if the one parameter group is stronly continuous. We will now
show that it is not strongly continuous. To see this, take any
non-zero vector field and find an open subset $U\subset \sigma$ in which
it is non-vanishing. We find a non-trivial graph $\gamma$ contained in $U$
and an infinite decreasing sequence $(t_n)$ with limit $0$ such that
the graphs $\varphi^{-1}(\gamma)$ are mutually different. Take
any spin network state $T_s$ with $\gamma(s)=\gamma$. Since spin-network
states over different graphs are orthogonal we have
$||\hat{U}(\varphi^u_0)T_s-T_s||^2=2$, thus proving our claim.
This small computation demonstrates once again how distributional
$\ab$ in fact is: Once a path just differs infinitesimally from a second
one, they are algebraically independent and a distributional homomorphism
is able to assign to them completely independent values, there is no
continuity at all. This behaviour is drastically different from that
of Gaussian measures and is deeply rooted in the background independence
of our formalism: The covariance of a Gaussian measure depends on a
background metric which is able to tell us {\it how far apart two points
are}. However, in a diffeomorphism invariant theory there is no distinguished
background metric, in contrast, there are diffeomorphisms which, with respect
to {\it any} background metric, can take the
two points as far apart or as close together as we desire, the
positions of the two points are not gauge invariant.

The absence of an infinitesimal generator of diffeomorphisms is not
necessarily bad because we can still impose diffeomorphism invariance
via finite diffeomorphisms, in fact finite diffeomorphisms are even better
suited to constructing a rigging map as we will see. However, it should
be kept in mind that the passage from the connected component
Diff$^\omega_0(\sigma)$ to all of Diff$^\omega(\sigma)$ is a non-trivial
step which
is not forced on us by the formalism. Since the so-called mapping class
group Diff$^\omega(\sigma)/$Diff$^\omega_0(\sigma)$ is huge and not very
well understood (see e.g. \cite{59}),
to take all of Diff$^\omega(\sigma)$ is at least the most practical option
then.
Furthermore, one should stress once more that while analytic diffeomorphisms
are not too bad (every smooth paracompact manifold admits a real analytic
differentiable structure
which is unique up to smooth diffeomorphisms, see e.g \cite{60})
they are at least rather unnatural because the classical action has smooth
diffeomorphisms as its symmetry group and also because an analytic
diffeomorphism is determined already by its restriction to an arbitrarily
small open subset $U$ of $\sigma$. In particular, an analytic diffeomorphism
cannot be the identity in $U$ and non-trivial elsewhere.
Hopefully these fine details
will no longer
be important in the final picture of the theory in which diffeomorphisms
of any differentiability category should have at most a semiclassical
meaning anyway.

Let us then go ahead and solve the finite diffeomorphism
constraint. That is, by the methods of RAQ we are looking for
algebraic distributions $L\in {\cal D}^\ast$ with ${\cal
D}=C^\infty(\ab)$ such that \be \label{4.97}
L(\hat{U}(\varphi)f)=L(f)\;\forall \varphi\in
\mbox{Diff}^\omega(\sigma),\; f\in {\cal D} \ee Here we have
explicitly written out the invariance condition in terms of
analytic diffeomorphisms. Since the span of spin network functions
is dense in ${\cal D}$, (\ref{4.97}) is equivalent with
\be
\label{4.98}
L(\hat{U}(\varphi)T_s)=L(T_s)\;\forall \varphi\in
\mbox{Diff}^\omega(\sigma),\; s\in {\cal S}
\ee
%
In order to solve (\ref{4.97}) recall from section \ref{s4.2.2}
that every element of ${\cal D}^\ast$ can be written in the form
$L=\sum_s L_s <T_s,.>$ where $L_s$ are some complex numbers. Then
(\ref{4.98}) becomes a very simple condition on the coefficients
$L_s$ given by
\be \label{4.99} L_{\varphi\cdot s}=L_s\;\forall \varphi\in
\mbox{Diff}^\omega(\sigma),\; s\in{\cal S}
\ee
Equation (\ref{4.99}) suggests to introduce the orbit $[s]$ of
$s$ given by
\be \label{4.100}
[s]=\{\varphi\cdot s;\;\varphi\in \mbox{Diff}^\omega(\sigma)\}
\ee
and therefore (\ref{4.99}) means that $s\mapsto L_s$ is constant
on every orbit. Obviously, ${\cal S}$ is the disjoint union of orbits
which motivates to introduce the space of orbits $\cal N$ whose elements
we denote by $\nu$. Introducing the elementary distributions
$L_\nu:=\sum_{s\in\nu} <T_s,.>$ we may write the general solution
of the diffeomorphism constraint as
\be \label{4.101}
L=\sum_{\nu\in {\cal N}} c_\nu L_\nu
\ee
for some complex coefficients $c_\nu$ which depend only on the orbit
but not on the representative. Notice that $L_\nu(T_s)=\chi_\nu(s)$
where $\chi$ denotes the characteristic function.

We still do not have a rigging map but the structure of the
solution space suggests to define
\be \label{4.102}
\eta(T_s):=\eta_{[s]} L_{[s]}
\ee
for some complex numbers $\eta_\nu$ for each $\nu\in{\cal N}$ and
to extend (\ref{4.102}) by linearity to all of ${\cal D}$, that is,
one writes a given $f\in {\cal D}$ in the form $f=\sum_s f_s T_s$
with complex numbers $f_s=0$ except for finitely many $s$ and
then defines $\eta(f)=\sum_s f_s \eta(T_s)$. This way the
map $\eta$ is tied to the spin network basis. The crucial
question is now whether the coefficients can be chosen in such a way that
$\eta$ satisfies all requirements to be a rigging map.

Notice that $\eta$ is almost an integral over the diffeomorphism group:
One could have considered instead of $\eta$ the following transformation
\be \label{4.103}
T_s\mapsto \sum_{\varphi\in \mbox{Diff}^\omega(\sigma)}
<\hat{U}(\varphi)T_s,.>
\ee
and the right hand side is certainly diffeomorphism invariant.
The measure that is being used here is a {\it counting
measure} which is trivially translation invariant.

Unfortunately (\ref{4.103}) does not even define an element of
${\cal D}^\ast$
because there are uncountably infinitely many
analytic diffeomorphisms
which leave $\gamma(s)$ invariant. To see this, notice that
if $e\in E(\gamma)$ is an analytic curve (edge) which is left
invariant by $\varphi$ then $\varphi$ must leave invariant also
the maximal analytic extension $\tilde{e}$ of $e$ in $\sigma$
due to analyticity (By definition, an analytic edge is given by
$D$ functions $t\mapsto e^a(t)$ so
$\varphi_{|e}:\;t\mapsto \varphi(e(t))$ is again analytic
which is why there must be an analytic reparameterization
$t_{\varphi}$ such that $\varphi(e(t))=e(t_\varphi(t))$
if $\varphi$ leaves $e$ invariant. But then the range of the
maximal analytic
extension of $e\circ \tilde{t}_\varphi$ coincides with $\tilde{e}$
and equals $\varphi(\tilde{e})$). There is an analytic
function $F_{\tilde{e}}$ which vanishes precisely on $\tilde{e}$
because the condition $F_{\tilde{e}}(\tilde{e}(t))=0$ for all $t$ is
equivalent
to the condition that an analytic function on $\sigma$ should vanish
only on the $x_1-$axis in local coordinates, which is easily satisfied
by the analytic function $x_2^2+..+x_D^2$ for instance. Hence,
the function $F_{\tilde{\gamma}}:=\prod_{e\in E(\gamma)} F_{\tilde{e}}$
vanishes precisely on $\tilde{\gamma}$. Thus, if we choose
a constant vector field $u$ on $\sigma$ (which is trivially analytic)
then
$u_{\tilde{\gamma}}:=F_{\tilde{\gamma}}^2 e^{-F_{\tilde{\gamma}}^2} u$
vanishes precisely on $\tilde{\gamma}$ and is analytic. It follows
that its integral curves define a diffeomorphism which is trivial
precisely on $\tilde{\gamma}$. Since there are uncountably many
$u,F_{\tilde{\gamma}}$ the claim follows. As a consequence,
(\ref{4.103}) contains uncountably many times the same functional
$<T_s,.>$ so that its value on $T_s$ diverges.

In a sense then, $\eta$ is a group averaging map in which these
{\it trivial action} diffeomorphisms have been factored out.
Now while one can find a subgroup
Diff$^\omega_{[s]}(\sigma)$ of Diff$^\omega(\sigma)$
such that
\be \label{4.104}
\eta(T_s)=\eta_{[s]}
\sum_{\varphi\in \mbox{Diff}^\omega_{[s]}(\sigma)}
<\hat{U}(\varphi)T_s,.>
\ee
(just choose precisely one diffeomorphism that maps
$s$ to a given $s'\in [s]$),
unfortunately these subgroups depend on $[s]$ so that
one cannot view (\ref{4.104}) as a regularized
rigging map.

Let us see whether we can choose the coefficients
$\eta_{[s]}$ in such a way that the rigging inner product
is well-defined. By definition
\be \label{4.105}
<\eta(T_s),\eta(T_{s'})>_\eta:=\eta(T_s')[T_s]=
\eta_{[s']}\chi_{[s']}(s)
\ee
Thus, positivity requires that $\eta_{[s]}>0$. Imposing hermiticity
then requires that
\be \label{4.106}
\eta_{[s']}\chi_{[s']}(s)
=\overline{<\eta(T_{s'}),\eta(T_s)>}=\overline{\eta(T_s)[T_{s'}]}=
\eta_{[s]}\chi_{[s]}(s')
\ee
Now both the right and left hand side are non vanishing if and only
if $[s]=[s']$ so that (\ref{4.106}) is correct with
no extra condition on the $\eta_{[s]}$.

Finally we come to the issue of diffeomorphism invariant
observables. We call an operator
$\hat{O}$ a strong observable if $\hat{U}(\varphi)\hat{O}
\hat{U}(\varphi)^{-1}=\hat{O}$. We call it a weak observable
if $\hat{O}'$ leaves the solution space invariant, in other words
\ba \label{4.107}
L(\hat{U}(\varphi)f)&=&L(f) \;\;\forall\;\varphi\in
\mbox{Diff}^\omega(\sigma)
\\
&\Rightarrow&
[\hat{O}'L](\hat{U}(\varphi)f)
=L(\hat{O}^\dagger\hat{U}(\varphi)f)
=L(\hat{U}(\varphi)^{-1}\hat{O}^\dagger\hat{U}(\varphi)f)
={\hat{O}'L}(f)
\nonumber
\ea
We now show that restricting attention to strong observables
would lead to superselection sectors. Namely, suppose that
$\hat{O}$ is a densely defined, closed, strongly diffeomorphism
invariant operator and consider any two spin-network functions
$T_s,T_{s'}$ with $\tilde{\gamma}(s)\not=\tilde{\gamma}(s')$
where $\tilde{\gamma}$ denotes the maximal analytic extension of
$\gamma$.
Then by the above construction we an at least countably infinite
number of analytic diffeomorphisms $\varphi_n$ with
$\varphi_n(\gamma(s))=\gamma(s)$ but such that the
$\varphi_n(\gamma(s'))$ are mutually different. Hence for any
$n$
\be \label{4.108}
<T_{s'},\hat{O} T_{s}>
=<T_{s'},\hat{U}(\varphi_n)^{-1}\hat{O}\hat{U}(\varphi_n)T_{s}>
=<\hat{U}(\varphi_n)T_{s'},\hat{O}T_{s}>
\ee
Since the states $\hat{U}(\varphi_n)T_{s'}$ are mutually orthogonal
and since
\be \label{4.109}
||\hat{O} T_s||^2=\sum_{s^\dprime\in {\cal S}} |<T_{s^\dprime},\hat{O}
T_s>|^2\ge \sum_{n=1}^\infty |<\hat{U}(\varphi_n)T_{s'},\hat{O}T_{s}>|
=|<T_{s'},\hat{O} T_{s}>|^2\sum_{n=1}^\infty\; 1
\ee
we conclude that $<T_{s'},\hat{O} T_{s}>=0$. In other words,
strongly diffeomorphism invariant, closed and densely defined
operators cannot have matrix elements between spin network
states defined over graphs with different maximal analytic
extensions so that the Hilbert space would split into mutually
orthogonal superselection sectors. If $\sigma$ is compact, the
total spatial volume would be an operator of that kind, it
actually preserves the graph on which it acts. More generally,
operators which are built entirely from electric field operators
will have this property. However, classically the theory
contains many diffeomorphism invariant functions which
are not built entirely from electric fields but depend on the
curvature of the connection (for instance the Hamiltonian
constraint) and hence, as operators, do not leave the
graph on which they act invariant (see the next section).
Thus, it is not enough to consider only strongly invariant
operators which is why no superselection takes place
\cite{27}.

Next we show that there exists a choice for the
$\eta_{[s]}$ such that $\hat{O}'\eta(f)=\eta(\hat{O} f)$
at least for strongly invariant operators
which then by the general theory of section
\ref{si} implies that the reality conditions
$(\hat{O}')^\star=(\hat{O}^\dagger)'$ are satisfied.
To choose the $\eta_\nu$ appropriately we must
discuss the so-called symmetry group $P_{[s]}$ of
$[s]$, defined as follows: Let $p$ be a permutation
on the set $E(\gamma(s))$ and define
$p\cdot s:=(p(e),\pi_{p(e)}:=\pi_e,m_{p(e)}:=m_e,
n_{p(e)}:=n_e)_{e\in E(\gamma)}$ (in the gauge
invariant case a similar action is defined).
Then
$P_{[s]}$ is the subgroup of the permutation group
consisting of those permutations such
that for each $p\in P_{[s]}$ there exists an analytic
diffeomorphism $\varphi_p$ such that
$\varphi_p\cdot s=p\cdot s$. It is clear that
this definition is independent
of the choice of the representative $s'\in [s]$.
For instance,
if $\gamma$ is the figure eight loop (with
intersection) and $e,e'$ are its two edges
then $P_{[s]}$ has two generators for
$s=(\gamma,\pi_e=\pi_{e'},m_e=m_{e'},n_e=n_{e'})$
while there would be none if e.g. $\pi_e\not=\pi_{e'}$.
This demonstrates that the orbit generating groups
Diff$^\omega_{[s]}(\sigma)$ can have different sizes
for $[s],[s']$ even if $\gamma(s),\gamma(s')$ are
diffeomorphic.

Now we have just seen that a strong
observable has matrix elements at most between
$T_s,T_{s'}$ where $\tilde{\gamma}(s)=\tilde{\gamma}(s')$.
The point is then the following: Let
$[\gamma]=\{\varphi(\gamma);\;
\varphi\in \mbox{Diff}^\omega(\sigma)\}$ be the
orbit of a graph and let us consider
a smaller subgroup Diff$^\omega_{[\gamma(s)]}(\sigma)$
of Diff$^\omega(\sigma)$ contained in
Diff$^\omega_{[s]}(\sigma)$ and consisting of diffeomorphisms
which map $\gamma(s)$ into precisely one of its orbit elements.
Now, due to analyticity we can in fact choose
Diff$^\omega_{[\gamma]}(\sigma)=$Diff$^\omega_{[\tilde{\gamma}]}(\sigma)$:
If $\varphi\not=\varphi'\in \mbox{Diff}^\omega_{[\gamma]}(\sigma)$
then $\varphi(\gamma)\not=\varphi'(\gamma)$ so certainly
$\varphi(\tilde{\gamma})\not=\varphi'(\tilde{\gamma})$.
Conversely if
If $\varphi\not=\varphi'\in \mbox{Diff}^\omega_{[\tilde{\gamma}]}(\sigma)$
then $\varphi^{-1}\circ\varphi'(\tilde{s})\not=\tilde{s}$
for at least a segment $\tilde{s}$ of some edge of $\tilde{\gamma}$.
However, $\tilde{s}$ belongs to the analytic extension
of some edge $e$ of $\gamma$. Suppose that
$\varphi^{-1}\circ\varphi'(e)=e$. This is a contradiction
because we have seen above that then
$\varphi^{-1}\circ\varphi'$ preserves the whole analytic
extension of $e$.

We conclude that the orbit size of $[s]$ is
$|P_{[\tilde{\gamma}(s)]}|/|P_{[s]}|$ times the orbit size
of $[\tilde{\gamma}(s)]$ where
$P_{[\gamma]}$ is defined similarly as $P_{[s]}$
just that now $\varphi_p(\gamma)=p(\gamma)$
is required. (Again, if $\varphi_p$ is a symmetry
of $\gamma$ then it is a symmetry of $\tilde{\gamma}$
by analyticity). We can therefore write
\be \label{4.110}
\eta(T_s)=\frac{\eta_{[s]}}{|P_{[s]}|}\;\;
\sum_{\varphi\in \mbox{Diff}^\omega_{[\tilde{\gamma}(s)]}(\sigma),\;
p\in P_{[\tilde{\gamma}(s)]}} <\hat{U}(\varphi)\hat{U}(\varphi_p)T_s,.>
\ee
Let now $\hat{O}$ be a strong observable
then
\ba \label{4.111}
<\eta(f),\hat{O}' \eta(T_s)>_\eta
&=& [\hat{O}' \eta(T_s)](f)=[\eta(T_s)](\hat{O}^\dagger f)
\nonumber\\
&=& \frac{\eta_{[s]}}{|P_{[s]}|}
\sum_{\varphi\in \mbox{Diff}^\omega_{[\tilde{\gamma}(s)]}(\sigma),\;
p\in P_{[\tilde{\gamma}(s)]}}
<\hat{U}(\varphi)\hat{U}(\varphi_p) T_s,\hat{O}^dagger f>
\nonumber\\
&=& \frac{\eta_{[s]}}{|P_{[s]}|}
\sum_{\varphi\in \mbox{Diff}^\omega_{[\tilde{\gamma}(s)]}(\sigma),\;
p\in P_{[\tilde{\gamma}(s)]}}
<\hat{U}(\varphi)\hat{U}(\varphi_p) \hat{O}T_s,f>
\nonumber\\
&=&
<f, \eta(\hat{O}T_s)>_\eta
\ea
where in the last step we have used that $\hat{O} T_s$
is a countable linear combination of spin-network states
$T_{s'}$ with $\tilde{\gamma}(s)=\tilde{\gamma}(s')$.

Hence there are in fact no additional
conditions on $\eta_{[s]}$ as far as strong observables
are concerned. If even the rigging map ansatz is
general enough with respect to the weak observables
is a completely different issue and not known at the
moment. However, whether or not there is a rigging
map with respect to the diffeomorphism constraint
is of marginal interest anyway for the following reason:
Remember that the classical constraint algebra between
the Hamiltonian constraint $H(N)$ and Diffeomorphism constraint
$\vec{H}(\vec{N})$ respectively has the structure
\ba \label{4.112}
&& \{\vec{H}(\vec{N}),\vec{H}(\vec{N}')\}\propto
\vec{H}([\vec{N},\vec{N}']),
\\
&& \{\vec{H}(\vec{N}),H(N)\}\propto H(\vec{N}[N]),\;
\{H(N),H(N')\}\propto \vec{H}(q^{-1}(N d N'-N' dN)
\nonumber
\ea
Thus, the Poisson Lie algebra of diffeomorphism constraints
is actually a subalgebra (the first identity) of the full constraint
algebra but it is not an ideal (the second identity).
It is therefore not possible to solve the full constraint
algebra in two steps by first solving the diffeomorphism
constraint and then solving the Hamiltonian constraint in
a second step: As (\ref{4.112}) shows, the dual Hamiltonian
constraint operator cannot leave the space of diffeomorphism
invariant distributions invariant and it is therefore
meaningless to try to construct an inner product that
solves only the diffeomorphism constraint. Rather, one has
to construct the space of solutions of all constraints
first before one can tackle the issue of the physical inner
product.


\newpage

\section{Kinematical Geometrical Operators}
\label{s5}

In this section we will describe the so-called kinematical geometrical
operators of Canonical Quantum Relativity. These are gauge invariant
operators which measure the length, area and volume
respectively
of coordinate curves, surfaces and volumes for $D=3$.
The area and volume operators were
first considered by Rovelli and Smolin
in the loop representation \cite{61}. In \cite{61a}
Loll divovered that the volume operator vanishes on gauge
invariant states with at most trivalent vertices and used
area and volume operators in her lattice theoretic
framework \cite{63}. Ashtekar
and Lewandowski \cite{64} used the connection representation
defined in
previous sections and could derive the full spectrum of the
area operator while their volume operator differs from that
of Rovelli and Smolin on graphs with vertices of valence
higher than three which can be seen as the result of using
different diffeomorphism classes of regularizations.
In \cite{65} de Pietri and Rovelli computed the matrix
elements of the RS volume operator in the loop representation
and de Pietri created a computer code for the
actual case by case evaluation of the eigenvalues.
In \cite{58} the connection representation was used
in order to obtain the complete set of matrix elements of the
AL volume operator. Area and volume operator could be quantized
using only the known quantizations of the electric flux
of section \ref{s4.1} but the construction of the length operator
\cite{66a} required a new quantization technique which was actually first
employed for the Hamiltonian constraint, see section \ref{s6}.
To the same category of operators also belong the
ADM energy surface integral \cite{X}, angle operators \cite{67b}
and similar other operators that test components of the three metric
tensor \cite{67c}.

In $D$ dimensions we have
analogous objects corresponding to $d-$dimensional submanifolds of
$\sigma$ with $1\le d\le D$. To get an idea of the constructions involved
it will be sufficient here to describe the simplest operator, the
so-called area operator which we construct
in $D$ dimensions and which measures the area of an open
$D-1$ dimensional submanifold of $\sigma$.
A common feature of all these operators is that they are essentially
self-adjoint, positive semi-definite unbounded operators with
pure point (discrete) spectrum which has a length, area, volume ...
gap respectively of the order of the Planck length, area, volume etc.
(that is, zero is not an accumulation point of the spectrum).

We call these operators kinematical because they do not (weakly)
commute with the Diffeomorphism or Hamiltonian constraint operator.
One may therefore ask what their physical significance should be.
As a partial answer we will sketch a proof that if the curves, surfaces
and regions are not coordinate manifolds but are invariantly defined
through matter, then they not only weakly commute with the Diffeomorphism
constraint but also their spectrum remains unaffected. There is no such
argument with respect to the Hamiltonian constraint however.
We will follow the treatment in \cite{64}.

\subsection{Derivation of the Area Operator}
\label{s5.1}

Let $S$ be an oriented, embedded, open, compactly supported,
analytical surface and let
$X:\;U\to S$ be the associated embedding where $U$ is an open submanifold
of $\Rl^{D-1}$. The area functional Ar$[S]$ of the $D-$metric tensor
$q_{ab}$ is the volume of $X^{-1}(S)$ in the induced $(D-1)-$metric
\be \label{5.1}
\mbox{Ar}[S]:=\int_U d^{D-1}u \sqrt{\det([X^\ast q](u))}
\ee
which coincides with the Nambu-Goto action for the bosonic Euclidean
$(D-1)-$brane propagating in a $D-$dimensional target spacetime
$(\sigma,q_{ab})$. Using the covector densities
\be \label{5.2}
n_a(u):=\epsilon_{aa_1..a_{D-1}} \prod_{k=1}^{D-1}
\frac{\partial X^{a_k}}{\partial u_k}(u)
\ee
familiar from section \ref{s4.1}
it is easy to see that we can write (\ref{5.1} in the form
\be \label{5.3}
\mbox{Ar}[S]:=\int_U d^{D-1}u \sqrt{n_a(u) n_b(u) E^a_j(X(u)) E^b_j(X(u))}
\ee
Let now $U=\bigcup_{n=1}^N U'_n$ be a partition of $U$ by closed sets $U'_n$
with open interior $U_n$ and let $\cal U$ be the collection of these
open sets. Then the Riemann integral (\ref{5.3})
is the limit as $N\to \infty$ of the Riemann sum
\be \label{5.4}
\mbox{Ar}_{{\cal U}}[S]:=\sum_{U\in {\cal U}} \sqrt{E_j(S_U) E_j(S_U)}
\ee
where $S_U=X(U)$ and $E_j(S)$ is the electric flux function of
section \ref{s4.1}. The strategy for quantizing
(\ref{5.4}) will be to use the known quantization of $E_j(S)$,
to plug it into (\ref{5.4}), to apply it to cylindrical functions
and to hope that in the limit $N\to\infty$ we obtain a
consistently defined family of positive semi-definite operators.
Notice that the square root involved makes sense because its
argument will be a sum of squares of (essentially) self-adjoint
operators which has non-negative real spectrum and we may
therefore define the square root by the spectral resolution of
the operator.

Let then $l=l(\gamma)$ be any subgroupoid and $f_l\in C^2(X_l)$.
Using the results of section \ref{s4.1} we obtain for any
surface $S$
\be \label{5.5}
\hat{E}_j(S)\hat{E}_j(S)p_l^\ast f_l
=-p_{l_S}^\ast \frac{\ell_p^4\beta^2}{16}
\{\sum_{e\in E(\gamma_S)}\epsilon(e,S)
[\delta_{e\cap S=b(e)} R^j_e+\delta_{e\cap S=f(e)} L^j_e]\}^2
p_{l_S l}^\ast f_l
\ee
where $l_S=l(\gamma(S))$ is any adapted subgroupoid $l\prec l_S$.

When we now plug (\ref{5.5}) into (\ref{5.4}) we can exploit
the following fact: Since (\ref{5.4}) classically approaches
(\ref{5.3}) for {\it any} uniform refinement of the partition
$\overline{{\cal U}}$, for given $l$ and adapted $l_S$
we can refine in such a way that for all $e\in E(\gamma)$ with
$\epsilon(e,S)\not=0$ ($e$ is of the up or down type with respect
to $S$) we have always that $e\cap S$ is an interior
point of some $U\in {\cal U}$. Notice that then
$\epsilon(e,S)=\epsilon(e,S_U)$ and $e\cap S=e\cap S_U$. If on the other
hand $\epsilon(e,S)=0$ but $S\cap e\not=\emptyset$ ($e$ is of the
inside type with respect to $S$) then for those $U$ with
$U\cap e\not=\emptyset$ we also have $\epsilon(e,S_U)=0$. Clearly, if
$e\cap S=\emptyset$ then $e\cap U=\emptyset$ for all $U\in {\cal U}$
so again $\epsilon(e,S)=\epsilon(e,S_U)$. We conclude that
under such refinements the subgroupoid $l_S$ stays adapted for
all $S_U$. Let us denote an adapted partition and their refinements
by ${\cal U}_l$. Then
\ba \label{5.6}
\widehat{\mbox{Ar}}_{{\cal U}_l}[S] p_l^\ast f_l
&=&
\frac{\ell_p^2\beta}{4} p_{l_S}^\ast \sum_{U\in {\cal U}}
\times\nonumber\\
&&\times
\sqrt{-\{\sum_{e\in E(\gamma_S)}\epsilon(e,S_U)
[\delta_{e\cap S_U=b(e)} R^j_e+\delta_{e\cap S_U=f(e)} L^j_e]\}^2}
p_{l_S l}^\ast f_l
\ea
Let us introduce the set of
isolated intersection points between $\gamma$ and $S$
\be \label{5.6a}
P_l(S):=\{e\cap S;\;\epsilon(e,S)\not=0,\;e\in E(\gamma_S)\}
\ee
which is independent of the choice of $\gamma_S$ of course.
After sufficient refinement, every $S_U$ will contain at most
one point which is the common intersection point of edges of the
up or down type respectively.
Let then for each $x\in P_l(S)$ the surface that contains $x$ be
denoted by $S_{U_x}$. From our previous discussion we know that
then $\epsilon(e,S)=\epsilon(e,S_{U_x})$ for any $e\in E(\gamma_S)$
with $x\in \partial e$.
It follows that (\ref{5.6}) simplifies after sufficient refinement to
\ba \label{5.6b}
\widehat{\mbox{Ar}}_{{\cal U}_l}[S] p_l^\ast f_l
&=&
\frac{\ell_p^2\beta}{4} p_{l_S}^\ast
\sum_{x\in P_l(S)}
\times\nonumber\\
&&\times
\sqrt{-\{\sum_{e\in E(\gamma_S),x\in\partial e}\epsilon(e,S)
[\delta_{x=b(e)} R^j_e+\delta_{x=b(e)} L^j_e]\}^2}
p_{l_S l}^\ast f_l
\ea
Now the right hand side no longer depends on the degree of the
adapted refinement and hence the limit becomes trivial
\ba \label{5.7}
\widehat{\mbox{Ar}}_{|l}[S] p_l^\ast f_l
&=&
\frac{\ell_p\beta}{4} p_{l_S}^\ast
\sum_{x\in P_l(S)}
\times\nonumber\\
&&\times
\sqrt{-\{\sum_{e\in E(\gamma_S),x\in\partial e}\epsilon(e,S)
[\delta_{x=b(e)} R^j_e+\delta_{x=f(e)} L^j_e]\}^2}
p_{l_S l}^\ast f_l
\ea
Thus, we have managed to derive a family of operators
$\widehat{\mbox{Ar}}_l[S]$ with dense domain $\mbox{Cyl}^2(\ab)$.
The independence of (\ref{5.7}) of the adapted graph follows
from that of the $\hat{E}_j(S)$. Here we have encountered for
the first time a common theme throughout the formalism: A state
(or graph) dependent regularization. One must make sure therefore
that the resulting family of operators is consistent.

\subsection{Properties of the Area Operator}
\label{s5.2}

The following properties go through with minor modifications
also for the length and volume operators.
\begin{itemize}
\item[1)] {\it Consistency}\\
We must show that for any $l\prec l'$ holds that
a) $\hat{U}_{ll'} C^2(X_l)\subset C^2(X_{l'})$ and that
$\hat{U}_{ll'} \widehat{\mbox{Ar}}_l[S]=
\widehat{\mbox{Ar}}_{l'}[S]\hat{U}_{ll'}$ where
$\hat{U}_{ll'}f_l=p_{l' l}^\ast f_l$. Since
the $p_{ll'}^\ast$ are analytic, a) is trivially satisfied.
To verify b) we notice that (\ref{5.7}) can be written
as
\be \label{5.8}
\widehat{\mbox{Ar}}_{|l}[S]=
\hat{U}_{l_S}
\widehat{\mbox{Ar}}_{l_S}[S]\hat{U}_{ll_S}
\ee
where $\widehat{\mbox{Ar}}_{l_S}[S]$ is simply the midlle
operator in (\ref{5.7}) between the two pull-backs for the
case that $l$ is already adapted.
First we must check that (\ref{5.8}) is independent of the adapted
subgroupoid $l\prec l_S$. Let $l\prec l'_S$ be another subgroupoid
and take a third adapted subgroupoid  with $l_S,l'_S\prec l_S^\dprime$.
If we can show that for any adapted subgroupoids with
$l_S\prec l_S^\dprime$ we have
\be \label{5.9}
\widehat{\mbox{Ar}}_{l_S^\dprime}[S]\hat{U}_{l_S l_S^\dprime}
=\hat{U}_{l_S l_S^\dprime}\widehat{\mbox{Ar}}_{l_S}[S]
\ee
then we will be done. To verify (\ref{5.9}) we must make
a case by case analysis as in section \ref{s4.1.3} for the
electric flux operator. But since (\ref{5.7}) is essentially
the sum of square roots of the sum of squares of electric
flux operators, the analysis is completely analogous and
will not be repeated here.

Finally, let $l\prec l'$. We find an adapted subgroupoid
$l,l'\prec l_S$. Then
\ba \label{5.10}
\widehat{\mbox{Ar}}_{|l'}[S]\hat{U}_{ll'}=
=\hat{U}_{l_S}
\widehat{\mbox{Ar}}_{l_S}[S]\hat{U}_{l'l_S}\hat{U}_{ll'}
=\hat{U}_{l_S}
\widehat{\mbox{Ar}}_{l_S}[S]\hat{U}_{ll_S}
=\widehat{\mbox{Ar}}_{|l}[S]
\ea
which is equivalent with consistency.\\
That the operator exists at all is like a small miracle:
Not only did we multiply two functional derivatives
$\hat{E}^a_j(x)$ at the same point, even worse, we took
the square of it. Yet it is a densely defined, positive
semi-definite operator without that we encounter any
need for renormalization after taking the regulator
(here the fineness of the partition) away.
The reason for the existence of the
operator is the {\it pay -- off for having constructed a
manifestly background independent representation}.
We will see more examples of this ``miracle" in the sequel.
\item[2)] {\it Essential Self-Adjointness}\\
To see that the area operator is symmetric, let
$f_l\in C^2(X_l),f_{l'}\in C^2(X_{l'})$. Then
we find an adapted subgroupoid $l,l'\prec l_S$ whence
\ba \label{5.11}
<p_l^\ast f_l,\widehat{\mbox{Ar}}[S]p_{l'}^\ast f_{l'}>
&=&
<p_{l_S l}^\ast f_l,
\widehat{\mbox{Ar}}_{l_S}[S]p_{l_S l'}^\ast f_{l'}>_{L_2(X_{l_S}),d\mu_{0l_S})}
\nonumber\\
&=&
<\widehat{\mbox{Ar}}_{l_S}[S] p_{l_S l}^\ast f_l,
p_{l_S l'}^\ast f_{l'}>_{L_2(X_{l_S}),d\mu_{0l_S})}
\nonumber\\
&=& <\widehat{\mbox{Ar}}[S] p_l^\ast f_l,p_{l'}^\ast f_{l'}>
\ea
where in the second step we used that
$\widehat{\mbox{Ar}}_{l_S}[S]$ is symmetric on $L_2(X_{l_S}),d\mu_{0l_S})$
with $C^2(X_{l_S})$ as dense domain.

Thus, the area operator is certainly a symmetric, positive semi-definite
operator. Therefore we know that it possesses at least one
self-adjoint extension, the so-called Friedrich's extension.
However, we can show that $\widehat{\mbox{Ar}}[S]$ is even
essentially self-adjoint. The proof is quite similar to
proving essential self-adjointness for the electric flux operator:
Let ${\cal H}^0_{\gamma,\vec{\pi}}$ be the finite dimensional
Hilbert subspace of ${\cal H}^0$ given by the closed linear
span of spin-network functions over $\gamma$ where all
edges are labelled with the same irreducible representations
given by $\vec{\pi}$. Then the Hilbert space maybe written as
\be \label{5.12}
{\cal H}^0=
\overline{\oplus_{\gamma\in \Gamma^\omega_0,\vec{\pi}}
{\cal H}^0_{\gamma,\vec{\pi}}}
\ee
Given a surface $S$ we can without loss of generality
restrict the sum over graphs to adapted ones because
for $r(\gamma)=r(\gamma_S)$ we have
${\cal H}^0_{\gamma,\vec{\pi}}\subset {\cal H}^0_{\gamma_S,\vec{\pi}'}$
for the choice $\pi'_{e'}=\pi_e$ with
$E(\gamma_S)\ni e'\subset e\in E(\gamma)$.
Since then $\widehat{\mbox{Ar}}[S]$ preserves each
${\cal H}^0_{\gamma,\vec{\pi}}$ its restriction is a symmetric
operator on a finite dimensional Hilbert space, therefore it is
self-adjoint. It follows that
$\widehat{\mbox{Ar}}_{|\gamma,\vec{\pi}}[S]\pm i\cdot 1_{\gamma,\vec{\pi}}$
has dense range on
${\cal H}^0_{\gamma,\vec{\pi}}=
C^\infty(X_{l(\gamma)})_{\vec{\pi}}\subset
C^2(X_{l(\gamma)})_{\vec{\pi}}$.
Therefore
\ba \label{5.13}
&& [\widehat{\mbox{Ar}}[S]\pm i\cdot 1_{{\cal H}^0}] C^2(\ab)
=\oplus_{\gamma,\vec{\pi}}[\widehat{\mbox{Ar}}_{|\gamma,\vec{\pi}}[S]
\pm i\cdot 1_{\gamma,\vec{\pi}}] C^2(X_{l(\gamma)})_{\vec{\pi}}
\nonumber\\
&& \supset
\oplus_{\gamma,\vec{\pi}}[\widehat{\mbox{Ar}}_{|\gamma,\vec{\pi}}[S]
\pm i\cdot 1_{\gamma,\vec{\pi}}] {\cal H}^0_{\gamma,\vec{\pi}}
=\oplus_{\gamma,\vec{\pi}}{\cal H}^0_{\gamma,\vec{\pi}}
\ea
is dense in ${\cal H}^0$.
\item[3)] {\it Spectral Properties}
\begin{itemize}
\item[i)] {\it Discreteness}\\
Since $\widehat{\mbox{Ar}}[S]$ leaves the ${\cal H}^0_{\gamma,\vec{\pi}}$
invariant it is simply a self-adjoint matrix there with non-negative
eigenvalues. Since
$$
{\cal H}^0_{\gamma}=\overline{\oplus_{\vec{\pi}}
{\cal H}^0_{\gamma,\vec{\pi}}}
$$
and the set of $\vec{\pi}$ is countable it follows that
${\cal H}^0_\gamma$ has a countable basis of eigenvectors
for $\widehat{\mbox{Ar}}[S]$ so that
the spectrum is pure point (discrete), i.e. it does not
have a continuous part. Now, as we vary $\gamma$ we get
a non-separable Hilbert space, however, the spectrum
of $\widehat{\mbox{Ar}}[S]$
depends only a) on the number of intersection points with edges of the
up and down type, b) on their respective number per such intersection
point and c) on the irreducible representations they carry
and not on any other intersection characteristics. These possibilities
are countable whence the entire spectrum is pure point and each
eigenvalue comes with an uncountably infinite multiplicity.\\
\item[ii)] {\it Complete Spectrum}\\
It is even possible to compute the complete spectrum directly
and to prove the discreteness from an explicit formula.
Such a closed formula is unfortunately not available for
the volume and length operator while badly needed for
purposes in particular connected with quantum dynamics as we will see
in the next section.

From the explicit formula (\ref{5.7}) it is clear that
we may compute the eigenvalues for each intersection
point $x$ of $S$ with edges of $\gamma_S$ of the up
or down type separately. Since the operator is independent
of the choice of adapted graph, we may assume that
all edges $e\in E(\gamma_S)$ have outgoing orientation,
that is, $x=b(e)$ for each edge incident at $x$.
Then (\ref{5.7}) reduces to
\be \label{5.14}
\widehat{\mbox{Ar}}_{l_S}[S]
= \frac{\ell_p^2\beta}{4}
\sum_{x\in P_l(S)}
\sqrt{-\{\sum_{e\in E(\gamma_S),x\in\partial e}\epsilon(e,S) R^j_e\}^2}
\ee
Let $E_{x,\star}=(\gamma_S)=\{e\in E(\gamma_S);\;x=b(e);\;e=\star\mbox{type}\}$
where $\star=$u,d,i for ``up, down, inside" respectively and let
$R^j_{x,\star}=\sum_{e\in E_{x,\star}(\gamma_S)} R^j_e$.
Then we have
\ba \label{5.15}
&&\{\sum_{e\in E(\gamma_S),x\in\partial e}\epsilon(e,S) R^j_e\}^2
=[R^j_{x,u}-R^j_{x,d}]^2
\\
&=& (R^j_{x,u})^2+(R^j_{x,d})^2-2R^j_u R^j_d
=2(R^j_{x,u})^2+2(R^j_{x,d})^2-(R^j_u +R^j_d)^2
\nonumber
\ea
where we have used that $[R^j_{x,u},R^k_{x,d}]=0$
(independent degrees of freedom).
We check that
$[R^j_{x,\star},R^k_{x,\star}]=-2f_{jk}\;^l R^j_{x,\star}$
so that also
$[R^j_{x,u+d},R^k_{x,u+d}]=-2f_{jk}\;^l R^j_{x,u+d}$
with $R^j_{x,u+d}=R^j_u+R^j_v$.
From this follows that
$[R^k_\star,(R^j_u)^2]=[R^k_\star,(R^j_d)^2]=0$ so that
$\Delta_u=(R^j_{x,u})^2/4,\Delta_d=(R^j_{x,d})^2/4,
\Delta_{u+d}=(R^j_{x,u+d})^2/4$ are mutually commuting
operators and each of $R^j_{x,u},R^j_{x,d},R^j_{x,u+d}$
satisfies the Lie algebra of right invariant vector fields.
Thus their respective spectrum is given by the eigenvalues
$-\lambda_\pi<0$ of the Laplacian $4\Delta=(R^j)^2=(L^j)^2$
on $G$ in irreducible
representations $\pi$ for which all matrix element functions
$\pi_{mn}$ are simultaneous eigenfunctions with the same
eigenvalue. It follows that
\be \label{5.16}
\mbox{Spec}(\widehat{\mbox{Ar}}[S])=\{\frac{\ell_p^2\beta}{2} \sum_{n=1}^N
\sqrt{2\lambda_{\pi^1_n}+2\lambda_{\pi^1_n}-\lambda_{\pi^{12}_n}};\;
N\in\Nl,\;\pi^1_n,\pi^2_n,\pi^{12}_n\in \Pi;\;\pi^{12}_n\in
\pi^1_n\otimes \pi^2_n\}
\ee
where the last condition means that $\pi^{12}_n$ is an irreducible
representation that appears in the decomposition into irreducibles
of the tensor product representation $\pi^1_n\otimes \pi^2_n$.
In case that we are looking only at gauge invariant states we actually
have $R^j_{x,u+v}=-R^j_{x,i}$. The spectrum (\ref{5.16})
is manifestly discrete by inspection.
It is bounded from below by zero and is unbounded
from above and depends explicitly on the Immirzi parameter.\\
\item[iii)] {\it Area Gap}\\
Let us discuss the spectrum more closely for $G=SU(2)$. Then
per intersection point we have eigenvalues of the form
\be \label{5.17}
\lambda=
\frac{\ell_p^2 \beta}{2}\sqrt{2j_1(j_1+1)+2j_2(j_2+1)-j_{12}(j_{12}+1)}
\ee
where $|j_1-j_2|\le j_{12}\le j_1+j_2$ by recoupling theory.
Recoupling theory \cite{69}, that is, coupling of $N$ angular momenta
also tells us how to build the corresponding eigenfunctions through
an appropriate recoupling scheme.
The lowest positive eigenvalue is given by the minimum
of (\ref{5.17}). At given $j_1,j_2$ the minimum is given at $j_{12}=j_1+j_2$
which gives
\be \label{5.18}
\frac{\ell_p^2 \beta}{2}\sqrt{(j_1-j_2)^2+j_1+j_2}=
\frac{\ell_p^2 \beta}{2}\sqrt{(j_2-(j_1-1/2))^2+2j_1-1/4}
\ee
Since (\ref{5.18}) vanishes at $j_1=j_2=0$ at least one
of them must be greater than zero, say $j_1$. Then
(\ref{5.18}) is minimized at $j_2=j_1-1/2\ge 0$ and proportional
to $\sqrt{2j_1-1/4}$ which takes its minimum at $j_1=1/2$.
Thus we arrive at the {\it area gap}
\be \label{5.19}
\lambda_0=\frac{\sqrt{3}\ell_p^2 \beta}{4}
\ee
\item[iv)] {\it Main Series}\\
It is sometimes claimed \cite{70a} that the regularization of
the area operator is incorrect and that a different regularization
gives eigenvalues proportional to $\sqrt{j(j+1)}$ rather than
(\ref{5.17}). If that would be the case then this would be of some
significance for black hole physics as we will see in section
\ref{s8.2}. However, first of all regularizations in quantum field
theory are never unique and may lead to different answers, the only
important thing is that all of them give the same classical limit.
Secondly, even if the regularization performed in \cite{70a} is
more aesthetic to some authors it is incomplete: In \cite{70a}
one looks only at the so-called main series which results if
we choose $j_1=j_2=j,\;j_{12}=0$ and then just gives
$$
\ell_p^2 \beta\sqrt{j(j+1)}
$$
(plus a quantum correction $j(j+1)\mapsto j(+1/2)^2$ due to the different
regularization which results in integral quantum numbers).
However, the complete spectrum (\ref{5.17}) is much richer, the
side series have physical significance for the black hole spectrum
as we will see and lead to a {\it correspondence principle}, that
is, at large quantum numbers the spectrum approaches a continuum.
To see this notice that at large eigenvalue $\lambda$ changes as
\be \label{5.20}
\frac{\delta\lambda}{\lambda}
\approx \frac{2(2j_1+1)\delta j_1+2(2j_2+1)\delta j_2-(2j_{12}+1)\delta j_{12}}
{2[(j_1+1)j_1+(j_2+1)j_2-(j_{12}+1)j_{12}]}
\ee
Suppose we choose $j_1=j_2=j\gg 1$. Then $0\le j_{12}\le 2j$ and we may
choose $j_{12}=0,\delta j_{12}=1/2,\delta j_1=\delta j_2=0$
(notice that such a transition
is ignored if we do not discuss the side series). Then (\ref{5.20})
can be written
\be \label{5.21}
\delta\lambda\approx -\frac{(\lambda_0)^2}{\lambda}
\ee
which becomes arbitrarily small at large $j$. The subsequent
eigenvalues have been calculated numerically in \cite{71a} displaying
a rapid transition to the continuum.
\item[v)] {\it Sensitivity to Topology}\\
The eigenvalues (\ref{5.17}) do detect some topological properties of
$\sigma$ as well. For instance, in the gauge invariant sector the spectrum
depends on whether $\partial S=\emptyset$ or not. Moreover, for
$\partial S=\emptyset$ the spectrum depends on whether $S$ divides
$\sigma$ into two disjoint ergions or not.
\end{itemize}
\end{itemize}

\subsection{Diffeomorphism Invariant Volume Operator}
\label{s5.3}

We now sketch how to make the geometrical operators at least
a weak observable with respect to spatial diffeomorphisms.
This is easiest for the volume functional.

Let $R$ be a coordinate region, i.e.
a $D-$dimensional submanifold of $\sigma$ then
the volume functional is defined by
\be \label{5.30}
\mbox{Vol}[R]
:=\int_R d^D x \sqrt{\det(q)}
=\int_\sigma d^D x \chi_R \sqrt{\det(q)}
\ee
where $\chi_R$ denotes the characteristic function
of the set $R$. Suppose now that we couple gravity
to matter (which is possible, see section \ref{s7})
and that $\rho$ is a positive definite scalar density of any
weight of the matter (and gravitational) degrees of freedom.
Here by positive definite we mean that $\rho(x)=0$ if and only
if the matter field vanishes at $x$. For instance, if
we have an electromagnetic field scalar field $\phi$ we could use
the electromagnetic field energy density
$$
\rho=\frac{q_{ab}}{2\sqrt{\det(q)}}[E^a E^b+B^a B^b]
$$
Consider now the intrinsically defined region
\be \label{5.31}
R_\rho:=\{x\in \sigma;\;\rho(x)>0\}
\ee
Then
\be \label{5.32}
\mbox{Vol}[R_\rho]
=\int_\sigma d^D x \tilde{\theta}(\rho) \sqrt{\det(q)}
\ee
where $\tilde{\theta}$ is the modified step function
with $\tilde{\theta}(x)=1$ if $x>0$ and
$\tilde{\theta}(x)=0$ otherwise.
We claim that (\ref{5.32}) is in fact diffeomorphism invariant.
To see this, it is sufficient to show that $F_\rho(x):=
\tilde{\theta}(\rho(x))$ is a scalar of density weight zero.
Let $\rho$ be of density weight $n$, then under a diffeomorphism
$$
F_\rho(x)\mapsto
\tilde{\theta}(|\det(\partial\varphi(x)/\partial x)|^n\rho(\varphi(x)))
=\tilde{\theta}(\rho(\varphi(x)))
=(\varphi^\ast F_\rho)(x)
$$
since $\tilde{\theta}(cx)=\tilde{\theta}(x)$ for any $c>0$.

The use of matter is not really essential, we could also
have used a gravitational degree of freedom say
$\rho=\sqrt{\det(q)}R^2$ where $R$ is the curvature scalar.
The point is now that for scalar densities of weight one
we can actually define $\hat{\rho}$ as an operator valued
distribution (see section \ref{s6}) {\it if and only if $\rho$ has
density weight one}. Let $\cal U$ be a partion of $\sigma$.
If it is fine enough and $\rho(x)>0$ then also
$\rho[U]:=\int_U d^D x \rho(x)>0$ for $x\in U\in {\cal U}$, therefore
(\ref{5.32}) is approximated by
\be \label{5.33}
\mbox{Vol}_{{\cal U}}[R_\rho]
=\sum_{U\in {\cal U}} \tilde{\theta}(\rho[U])
\mbox{Vol}[U]
\ee
Now $\rho[U]$ can be turned into a densely defined positive definite
operator and thus $\tilde{\theta}(\hat{\rho}[U])$
can be defined by the spectral theorem. Moreover, since
$\tilde{\theta}(x)^2=\tilde{\theta}(x)$ we can order
(\ref{5.33}) symmetrically and define
\be \label{5.34}
\widehat{\mbox{Vol}}_{{\cal U}}[\rho]
=\sum_{U\in {\cal U}} \tilde{\theta}(\hat{\rho[U]})
\widehat{\mbox{Vol}}[U] \tilde{\theta}(\hat{\rho[U]})
\ee
where for an adapted subgroupoid $l_U=l(\gamma_U)$
\be \label{5.35}
\widehat{\mbox{Vol}}_{l_U}[U]=\frac{\beta^{3/2} \ell_p^3}{4}
\sum_{v\in V(\gamma)}
\sqrt{|\frac{1}{3!}
\sum_{e,e',e^\dprime\in E(\gamma_U);\;v=b(e)=b(e')=b(e^\dprime)}
\epsilon(e,e'e^\dprime) f_{jkl} R^j_e R^k_{e'} R^l_{e^\dprime}|}
\ee
is the volume operator for coordinate regions. The adaption
consists in orienting each edge to be outgoing from each
vertex (for a given graph, subdivide each edge into
two halves if necessary to get an adapted graph), the
sum is over unordered triples of edges and
$$
\epsilon(e,e',e^\dprime)=\mbox{sgn}(\det(\dot{e}(0),\dot{e}'(0),\dot{e}^\dprime(0)))
$$
The action on unadapted subgroupoids is defined similarly
as for the area operator.

One now has to refine the partition and show that the
final operator $\widehat{\mbox{Vol}}[\rho]$, if it exists,
is consistently defined. Since the spectrum of
$\tilde{\theta}(\hat{\rho[U]})$ is given by $\{0,1\}$, the spectra
of that final operator and the coordinate volume operator should
coincide and in that sense the discreteness of the spectrum
is carried over to the diffeomorphism invariant context.
Of course there remain technical issues, for instance
$\widehat{\mbox{Vol}}[U], \tilde{\theta}(\hat{\rho[U]})$ do not commute
and cannot be diagonalized simultaneously, the existence of the
limit is unclear etc. The details will appear elsewhere \cite{74}.

What this sketch shows are three points:\\
1) Kinematical Operators have a chance to become
full Dirac observables by defining their coordinate
regions invariantly through matter (for invariance under
the Hamiltonian evolution, this requires them to be
smeared over time intervals as well). Actually,
this is physically the way that one defines regions !\\
2) The discreteness of the spectrum then has a chance
to be an invariant property of the physical observables.\\
3) If true, then something amazing has happened:\\
We started out with an analytic manifold $\sigma$ and smooth
area functions. Yet, their spectra are entirely discrete,
hinting at a discrete Planck scale physics, quantum geometry
is distributional rather than smooth. Hopefully, the analytic
structure that we needed at the classical level everywhere
can be lifted to a purely combinatorial structure in the
final picture of the theory, as it happened for $2+1$ gravity,
see the fourth reference in \cite{38}.

\newpage

\cleardoublepage

\part{Current Research}
\label{p2}


\section{Quantum Dynamics}
\label{s6}

We now come to the ``Holy Grail" of Canonical Quantum General
Relativity, the definition of the Hamiltonian constraint.
We will see that although one can, surprisingly, densely define a
closed constraint operator at all, there is much less control
on the correctness of the proposed operator than for the
Gauss -- and Diffeomorphism constraint operators.
Also actually solving the proposed operator is not only
techniclly much more difficult but also conceptually:
For instance, while RAQ
gives some guidelines for how to do that and although it actually works
(with some limitations) if we restrict ourselves to the spatial
Diffeomorphism constraint as we have seen above, the
definition of the physical inner product for all constraints, even
for the already mentioned proposal, is an open problem so far.
The reason is that
the concept of a rigging map is currently out of control if the
constraints do not form a Lie algebra as is the case for quantum gravity.
Summarizing, the implementation of the correct quantum dynamics is not
yet completed and one of the most active research directions at the moment.

While the situation with the proposed operator is certainly
not completely satisfactory at the moment, in order to appreciate
nevertheless its existence one should keep in mind that the situation with
canonical quantum general relativity had {\it come to a
sort of crisis in 1996}:\\
\\
There were rigorous as well as formal results derived.

On the rigorous
side one had constructed a $^\ast$representation of
the classical Poisson algebra for a suitable elementary set
of ``string-surface" variables, that is, a kinematical Hilbert space
realized as an $L_2$ space with respect to a diffeomorphism invariant
measure on a suitable quantum configuration space. Unfortunately, these
results were not immediately useful for quantum gravity because the
gravitational connection for the way the theory was defined at that
time was complex valued rather than real valued and the kinematical Hilbert
space defined above depends crucially on the fact that the connection is
real-valued. It was considered impossible to quantize the density one
valued unrescaled Hamiltonian constraint $H$ in real variables
because it is not polynomial.\\
{\it This was the first big problem: The reality structure of Ashtekar's new
variables had not been addressed yet, not even at the kinematical level}.

On the formal side there were proposals for the quantization of the
constraints and even for their kernel, however, the way they were defined was
lacking diffeomorphism
covariance, they included singular parameters and although they were
meant for the complex Ashtekar connection, since the complex theory was
not equipped with any Hilbert space it was unclear in which topology
certain limits were performed and what the singular nature of the
quantum field operators (and their products) was.
For Euclidean gravity $\tilde{H}$ becomes actually polynomial
in real variables but then one could show with the existing
kinematical framework that the constraint operator was ill-defined
in the given representation.\\
{\it This was the second big problem:
There existed no rigorous quantization of the constraints, especially
not of the Wheeler-DeWitt constraint, all proposals were singular}.

\newpage

{\bf It seemed that one had a rigorous kinematical framework
at one's disposal which was unphysical if one insisted in using complex
variables (which was considerded mandatory) and which even in the
unphysical representation did not support the Hamiltonian constraint
operator !}\\
\\
There was some hope in terms of the Wick rotation proposal which we are
going to sketch below which should keep the constraint polynomial and
solve the reality conditions at the same time, however, that construction
could be called at best formal and, moreover, the polynomial constraint
for the real variables would suffer from the same singularities as the
one for the complex variables.

The operator that we will describe below in principle kills both problems in 
one stroke:\\
The crucial point was to realize that it is impossible to quantize
the density weight two Hamiltonian constraint $\tilde{H}$ without
breaking background independence. Could one then quantize the original
density one valued Hamiltonian constraint $H$ ? Since $H$ is non-polynomial
even in complex variables, the desire to have a complex connection
formulation turned out to be of marginal interest. We then could show
that with a new regularization technique, $\hat{H}$ can be turned into
a well-defined operator using real-valued variables and using the
established rigorous kinematical framework which now had become
physically relevant. As a side result,
also the generator of the Wick transform can be defined in
principle using the same technique which could be a starting
point for introducing the aesthetically more satisfactory
complex variables into the framework again.\\
\\
These considerations should be sufficient to indicate
that the proposed operator, which we will describe in this section,
is merely a first rigorous ansatz for the final operator
but it is at least a promising hint that the kinematical framework that was 
developed {\it can support the Hamiltonian constraint operator}. It is
arguably the most precisely defined ansatz that exists so far and hopefully
it is a good starting point for improvements,
generalizations and more drastic modifications (if necessary).

We will follow the only and exhaustive treatment in \cite{IV,37,V,VI,47m1}.\\
\\
Remark:\\
Recently, a second approach towards solving the Hamiltonian
constraint has been proposed \cite{91,92} which is constructed on
(almost) diffeomorphism invariant distributions which are based
on Vasiliev invariants. What is exciting about this is that one
can define something like an area derivative \cite{44}
in this space and therefore the arc attachment which we
will deal with exhaustively in what follows becomes much less
ambiguous. In this review we will not describe this rather recent formalism
because at the moment it falls outside of a Hilbert space context.
Hopefully we can return to this in a future edition of this review when
the theory has evolved more.

\subsection{The Wick Transform Proposal}
\label{s6.1}

The Bargmann-Segal Transform for quantum gravity discussed in
\cite{72a} gives a rigorous construction of quantum kinematics
on a space of complexified, distributional connections
by means of key results obtained by Hall \cite{70}. Since
the transform depended on a background structure, it was clear
that the associated scalar product did not implement the correct
reality conditions. To fix this was the purpose of \cite{IV}
where a general theory was developed of how to trivialize
reality conditions while keeping the algebraic structure
of a functional as simple as when complex variables are being used.
The same idea proves very useful in order to obtain a very general
class of coherent states as we will see in section \ref{s8}.
Moreover, as a side result, we
were able to improve the coherent state transform as defined by Hall
in the following sense :\\
Notice that the prescription given by Hall turns out to establish
indeed a unitary transformation but that it was ``pulled out of the hat",
that is, it was guessed by an analogy consideration with the transform
on $\Rl^n$ and turned out to work. It would be much more satisfactory to
have a derivation of the transform $\hat{U}_t$ and the measure $\nu_t$
on the complexified configuration space
from first principles, that is, one should be able to compute them just
from the knowledge of the two polarizations of the phase space.
We will first describe the general scheme in formal terms and then apply
it to quantum gravity.

\subsubsection{The General Scheme}
\label{s6.1.1}

Consider an arbitrary phase space $\cal M$, finite or infinite, with local
real canonical coordinates $(p,q)$ where $q$ is a configuration variable
and $p$ its conjugate momentum (we suppress all discrete and continuous
indices in this subsection). Furthermore, we have a Hamiltonian
(constraint) $H'(p,q)$ which unfortunately looks rather complicated in the
variables $p,q$ (the reason for the prime will become evident in a moment).
Suppose that, however, we are able to perform a
{\it canonical} transformation on $\cal M$ which leads to the complex
canonical pair $(p_\Co,q_\Co)$ such that the Hamiltonian becomes
algebraically simple
(e.g.) a polynomial $H_\Co$ in terms of $p_\Co,q_\Co$. That is, we have a
complex symplectomorphism $(p_\Co,q_\Co):=W^{-1}(p,q)$ such that
$H_\Co=H'\circ W$ is algebraically simple. Notice that we are not
complexifying the phase space, we just happen to find it convenient
to coordinatize it by complex valued coordinates. The reality conditions
on $p_\Co,q_\Co$ are encoded in the map $W$.

We now wish to quantize the system. We choose two Hilbert spaces,
the first one, $\cal H$, for which the $q$'s become a maximal set
of mutually commuting, diagonal operators and a second
one, ${\cal H}_\Co$, for which the $q_\Co$'s
become a maximal set of mutually commuting, diagonal operators.
According to the canonical commutation relations we represent
$\hat{p},\hat{q}$ on $\psi\in {\cal H}$ by
$(\hat{p}\psi)(x)=i\hbar\partial\psi(x)/\partial x$
and $(\hat{q}\psi)(x)=x\psi(x)$. Likewise, we represent
$\hat{p}_\Co,\hat{q}_\Co$ on $\psi_\Co\in {\cal H}_\Co$ by
$(\hat{p}_\Co\psi_\Co)(z)=i\hbar\partial\psi_\Co(z)/\partial z$
and $(\hat{q}_\Co\psi)(x)=z\psi_\Co(z)$.
The fact that $p,q$ are real-valued force us to set
${\cal H}:=L_2({\cal C},d\mu_0)$ where $\cal C$ is the quantum
configuration space and $\mu_0$ is the uniform (translation
invariant) measure on $\cal C$ in order that $\hat{p}$ be self-adjoint.

In order to see what the Hilbert space ${\cal H}_\Co$ should be, we also
represent the operators $\hat{p}_\Co,\hat{q}_\Co$ on $\cal H$ by choosing
a particular ordering of the function $W^{-1}$ and substituting $p,q$
by $\hat{p},\hat{q}$. In order to avoid confusion, we will write
them as $(\hat{p}',\hat{q}'):=W^{-1}(\hat{p},\hat{q})$ where the prime
means that the operators are defined on $\cal H$ but are also
quantizations of the classical functions $p_\Co,q_\Co$. Now, the point is
that the operators $\hat{p}',\hat{q}'$, possibly up to $\hbar$
corrections, automatically satisfy the correct adjointness relations
on $\cal H$ declining from the reality conditions on $p_\Co,q_\Co$.
This follows simply by expanding the function $W^{-1}$ in terms of
$\hat{p},\hat{q}$, computing the adjoint and defining the result to be the
quantization of $\bar{p}_\Co,\bar{q}_\Co$ on ${\cal H}$ which equals any valid
quantization prescription up to $\hbar$ corrections. Thus, if we could find
a unitary operator $\hat{U}\; :\;{\cal H}\to {\cal H}_\Co$ such that
\be \label{6.1}
\hat{p}_\Co=\hat{U}\hat{p}'\hat{U}^{-1} \mbox{ and }
\hat{q}_\Co=\hat{U}\hat{q}'\hat{U}^{-1}
\ee
then we have automatically implemented the reality conditions on
${\cal H}_\Co$ as well because by unitarity
\be \label{6.2}
(\hat{p}_\Co)^\dagger=\hat{U}(\hat{p}')^\dagger\hat{U}^{-1} \mbox{ and }
(\hat{q}_\Co)^\dagger=\hat{U}(\hat{q}')^\dagger\hat{U}^{-1}
\ee
where the $\dagger$ operations in (\ref{6.2}) on the left and right hand
side respectively are to be understood in terms of ${\cal H}_\Co$ and
$\cal H$ respectively. In other words, the adjoint of the operator
on ${\cal H}_\Co$ is the image of the correct adjoint of the operator
on $\cal H$.

To see what $\hat{U}$ must be, let $\hat{K}\; : \; {\cal H}\cap
\mbox{Ana}({\cal C})
\to {\cal H}_\Co$ be the operator of analytical extension of real
analytical elements of $\cal H$ and likewise
$\hat{K}^{-1}$ the operator that restricts the elements of
${\cal H}_\Co$ (all of which are holomorphic) to real values. We then have
the identities
\be \label{6.3}
\hat{p}_\Co=\hat{K}\hat{p}\hat{K}^{-1} \mbox{ and }
\hat{q}_\Co=\hat{K}\hat{q}\hat{K}^{-1} \;.
\ee
We now exploit that $W^{-1}$ was supposed to be a canonical
transformation (an automorphism of the phase space that preserves the
symplectic structure but not the reality structure). Let $C$ be its
infinitesimal generator, called the {\it complexifier}, that is,
for any function $f$ on $\cal M$,
\be \label{6.4}
f(p_\Co,q_\Co):=f_\Co(p,q):=((W^{-1})^\ast f)(p,q)=\sum_{n=0}^\infty
\frac{i^n}{n !}\{C,f\}_{(n)}
\ee
where the multiple Poisson bracket is inductively defined by
$\{C,f\}_{(0)}=f$ and $\{C,f\}_{(n+1)}=\{C,\{C,f\}_{(n)}\}$.
Using the substitution rule that Poisson brackets become commutators times
$1/(i\hbar)$ we can quantize (\ref{6.4}) by
\be \label{6.5}
\hat{f}':=f_\Co(\hat{p},\hat{q}):=\sum_{n=0}^\infty
\frac{1}{\hbar^n n !}[\hat{C},\hat{f}]_{(n)}
=(\hat{W}_t)^{-1}\hat{f}\hat{W}_t
\ee
where we have defined the generalized ``heat kernel" operator
\be \label{6.6}
\hat{W}_t:=e^{-t\hat{C}}
\ee
and $t=1/\hbar$. That is, the generator $C$ motivates a natural ordering
of $W^{-1}(p,q)$.

Substituting (\ref{6.6}) into (\ref{6.4}) we find
\be \label{6.7}
\hat{p}_\Co=\hat{U}_t\hat{p}'\hat{U}_t^{-1} \mbox{ and }
\hat{q}_\Co=\hat{U}_t\hat{q}'\hat{U}_t^{-1}
\ee
where we have defined the generalized coherent state or {\it Wick rotation
transform}
\be \label{6.8}
\hat{U}_t:=\hat{K}\hat{W}_t
\ee
with $t=1/\hbar$. The reason for the names we chose will become obvious
in the next subsection.

It follows that if $\hat{C},\hat{W}_t$ exist on real analytic functions and
if we can then extend $\hat{U}_t$ to a unitary operator from $\cal H$ to
${\cal H}_\Co:=L_2({\cal C}_\Co,d\nu_t)\cap\mbox{Hol}({\cal C}_\Co)$
where ${\cal C}_\Co$ denotes the complexification of $\cal C$ then
we have completed the programme.

Moreover, as a bonus we would have
{\it simplified the spectral analysis} of the operator that corresponds
to the quantization of $H'$ : \\
First of all we define an unphysical Hamiltonian (constraint) operator
$\hat{H}$ on $\cal H$ simply by choosing a suitable ordering of the
function
\be \label{6.9}
H(p,q):=H_\Co(p_\Co,q_\Co)_{|p_\Co\to p,q_\Co\to q}=(K^{-1}\cdot H_\Co)(p,q)
\ee
and substituting $p,q$ by the operators $\hat{p},\hat{q}$. Thus we obtain
an operator $\hat{H}_\Co$ on ${\cal H}_\Co$ by
$\hat{H}_\Co:=\hat{K}\hat{H}\hat{K}^{-1}$. It follows that if we
{\it define}
the quantization of the physical Hamiltonian (constraint) $H'$ on $\cal H$
by $\hat{H}':=\hat{W}_t^{-1}\hat{H}\hat{W}_t$ then in fact
$\hat{H}_\Co=\hat{U}_t\hat{H}'\hat{U}_t^{-1}$ and since $\hat{U}_t$ is
unitary the spectra of $\hat{H}'$ on $\cal H$ and of
$\hat{H}_\Co$ on ${\cal H}_\Co$ {\it coincide}. But since $\hat{H}_\Co$
is an algebraically simple function of the elementary operators $\hat{p}_\Co,
\hat{q}_\Co$ it follows that one has drastically simplified the spectral
analysis of the complicated operator $\hat{H}'$ ! Finally, given
a (generalized) eigenstate $\psi_\Co$ of $\hat{H}_\Co$, we obtain
a (generalized) eigenstate $\psi:=\hat{U}_t^{-1}\psi_\Co$ of
$\hat{H}'$ by the {\it inverse} of the coherent state transform.

The crucial question then is whether we can actually make $\hat{U}_t$
unitary. In \cite{IV} we derived the following formula for the unitarity
implementing measure $\nu_t$ on ${\cal C}_\Co$ :
\ba \label{6.9a}
d\nu_t(z,\bar{z})&:=&\nu_t(z,\bar{z})
d\mu_0^\Co(z)\otimes d\bar{\mu}_0^\Co(\bar{z}) \nonumber\\
\nu_t(z,\bar{z})&:=&
(\hat{K}[\overline{[\hat{W}_t]^\dagger}]\hat{K}^{-1})^{-1}
(\overline{(\hat{K}[\overline{[\hat{W}_t]^\dagger}]\hat{K}^{-1})})^{-1}
\delta(z,\bar{z}) \;.
\ea
The adjoint operation is meant in the sense of $\cal H$, $\hat{K}$
means analytical extension as before and the bar
means complex conjugation of the expression of the operator (i.e. any
appearance of multiplication or differentiation by $z$ is
replaced with multiplication or differentiation by $\bar{z}$ and vice
versa, and, of course, also numerical coefficients are complex conjugated).
Here $\mu_0^\Co$ and $\bar{\mu}_0^\Co$ are just the analytic and
anti-analytic extensions of the measure $\mu_0$ on $\cal C$ (they are
just complex conjugates of each other thanks to the positivity of
$\mu_0$) and
the distribution in the second line of (\ref{6.9}) is defined by
\be \label{6.10}
\int_{{\cal C}_\Co} d\mu_0^\Co(z)d\bar{\mu}_0^\Co(\bar{z})
f(z,\bar{z})\delta(z,\bar{z})=\int_{{\cal C}} d\mu_0(x) f(x,x)
\ee
for any smooth function $f$ on the complexified configuration space of
rapid decrease with respect to $\mu_0$.\\
Whenever (\ref{6.9}) exists (it is straightforward to check
that (\ref{6.9}) does the job formally), the extension of $\hat{U}_t$
to an unitary
operator (isometric, densely defined and surjective) in the sense above
can be expected \cite{IV}. A concrete proof is model-dependent.

In summary, we have solved {\it two problems in one stroke} :\\
We have implemented the correct adjointness relations and we have
simplified the Hamiltonian (constraint) operator.

A couple of remarks are in order :
\begin{itemize}
\item
The method does not require that $\hat{C}$ is self-adjoint, positive,
bounded or at least normal. All that is important is that $\hat{W}_t$
exists on real analytic functions in the sense of Nelson's analytic
vector theorem.
\item
It reproduces the cases of the harmonic oscillator and the case considered
by Hall \cite{70}. But it also explains {\it why} it works the way it works,
namely it answers the question of how to identify analytic continuation with
a given complex polarization of the phase space as is obvious from
$\hat{K}=\hat{U}_t\hat{W}_t^{-1}$.
The computation of $\nu_t$ with our metod via (\ref{6.9}), (\ref{6.10})
is considerably simpler. The harmonic oscillator corresponds to the
complexifier $C=\frac{1}{2}p^2$.
\item
On might wonder why one should compute $\nu_t$ at all and bother with
${\cal H}_\Co$ \cite{72} ? Could one not just forget about the analytic
continuation and work only on $\cal H$ simply by studying the spectral
analysis of the unphysical operator $\hat{H}$ and defining the physical
operator by $\hat{H}':=\hat{W}_t^{-1}\hat{H}\hat{W}_t$ ? The problem
is that, while it is true that restrictions to real arguments
of (generalized) eigenvectors of $\hat{H}_\Co$ are
{\it formal} eigenvectors of $\hat{H}$, these are typically not (generalized)
eigenvectors in the sense of the topology of $\cal H$. Intuitively, what
happens is that the measure $\nu_t$ provides for the necessary much
stronger fall-off in order to turn the analytic extension of the badly
behaved formal eigenvectors $\hat{W}_t^{-1} \psi$ of $\hat{H}'$ into
well-defined (generalized) eigenvectors $\hat{K}\psi$ of $\hat{H}_\Co$. \\
One can see this also from another point of view : by unitarity, whenever
$\hat{H}_\Co$ is self-adjoint, so is $\hat{H}'$ but in general $\hat{H}$
is not. Thus, one would not expect the spectra of $\hat{H},\hat{H}'$
to coincide. See the appendix of \cite{IV} for a
discussion of this point.
\item
There are also other applications of this transform, for example in
Yang-Mills theory it can be used to turn the Hamiltonian from a fourth
order polynomial into a polynomial of order three only !
\end{itemize}

This completes the outline of the general framework. We will now turn to
the interesting case of quantum gravity.

\subsubsection{Wick Transform for Quantum Gravity}
\label{s6.1.2}

As Barbero \cite{36} correctly pointed out, all the machinery that is
associated with the quantum configuration space $\ab$ and the
uniform measure $\mu_0$ is actually also available
for Lorentzian quantum general relativity if one chooses the Immirzi
parameter $\beta$ to be real. However,
the Hamiltonian constraint then does not simplify at all as
compared to the ADM expression and so the virtue of the
new variables would be lost. The coherent state transform as derived below
in principle combines both advantages, namely a well-defined calculus on
$\overline{{\cal A}}$ and a simple Wheeler-DeWitt constraint.

Let us then apply the framework of the previous subsection. The phase space
of Lorentzian general relativity
can be given a real polarization through
the canonical pair $(A_a^j:=\Gamma_a^j+K_a^j,E^a_j/\kappa)$
(the case considered by Barbero with $\beta=1$) and a complex
polarization through
canonical pair $((^\Co A_a^i):=\Gamma_a^j-i K_a^j, (^\Co E^a_j):=i E^a_j/\kappa)$
(the case considered by Ashtekar). The rescaled Hamiltonian constraint
looks very simple in the complex variables, namely
\be \label{6.11}
\tilde{H}_\Co(A_\Co,E_\Co)=\epsilon_{ijk}(^\Co F_{ab}^i)
(^\Co E^a_j) (^\Co E^b_k)
\ee
but if we write $A_\Co,E_\Co$ in terms of $A,E$ then the resulting
Hamiltonian $\tilde{H}'(A,E)$ becomes extremely complicated. Let us compute
the map $W$. We first of all see that we can go from $(A,E)$ to
$(A_\Co,E_\Co)$ in a sequence of three canonical transformations given by
$$
(A=\Gamma+ K,E/\kappa)\to(K,E/\kappa)\to(-iK,iE/\kappa)\to(A_\Co=\Gamma-i K,E_\Co=iE/\kappa)\;.
$$
That the first and third step are indeed canonical transformations was
already shown in section \ref{s2.3}. The second step is a {\it phase space
Wick
rotation}. Since $(K,E)$ is a canonical pair it is trivial to see that
we have
\be \label{6.12}
-iK=\sum_{n=0}^\infty \frac{i^n}{n !}\{C,K\}_{(n)} \mbox{ and }
iE=\sum_{n=0}^\infty \frac{i^n}{n !}\{C,E\}_{(n)}
\ee
where the {\it complexifier} or generator of the Wick transform
is given by
\be \label{6.13}
C=-\frac{\pi}{2\kappa}\int_\sigma d^3x K_a^i E^a_i
\ee
which is easily seen to be the integrated densitized trace of the
extrinsic curvature. $C$ generates infinitesimal constant scale
transformations. It now seems that we need to compute the
generator of the transform that adds and subtracts the
spin-connection $\Gamma$.
However, we have seen in section \ref{s2} that the spin-connection in
three dimensions is a homogeneous polynomial of degree zero in $E$ and its
derivatives and since a {\it constant} scale factor is unaffected by derivatives
we have $\{\Gamma,C\}=0$. Thus in fact we have
\be \label{6.14}
A_\Co=\sum_{n=0}^\infty \frac{i^n}{n !}\{C,A\}_{(n)} \mbox{ and }
E_\Co=\sum_{n=0}^\infty \frac{i^n}{n !}\{C,E\}_{(n)} \;.
\ee
The task left is to define the operator $\hat{C}$ and to compute the
corresponding measure $\nu_t$. This seems to be a very hard problem
because $K_a^i=A_a^i-\Gamma_a^i$ and $\Gamma_a^i$ is just a very complicated
function to quantize. Nevertheless it can be done as we will see in the
next section.

We conclude this section with a few remarks :\\
1) The Wick transform is a phase space Wick
rotation and has {\it nothing to do with analytical continuation in the time
parameter $t$} ! Mena Marug\'an \cite{73}
has given a formal relation with the usual Wick rotation corresponding to
an analytical continuation of time together with a complex conformal
rescaling of the four-dimensional metric.\\
2) As we will see in the next section, one {\it can} construct a well-defined
operator $\hat{C}$, whether its exponential makes any sense
though is an open question. But we will derive an even stronger result : one
can really dispense with the complex variables altogether because
one can give meaning to the {\it unrescaled, original} Hamiltonian
constraint $H'=\tilde{H}'/\sqrt{\det(q)}$
in terms of the real variables $(A,E)$. Although the complexifier
$C$ is then not used any more for the purpose of a Wick rotation, it still
plays a crucial role in the quantization scheme displayed there.
That it comes out rigorously quantized of that scheme is more a side
result than a premise. The corresponding operator $\hat{H}$ which we
construct directly on the Hilbert space ${\cal H}^0$
is surprisingly not terribly complicated. Still, it maybe important to
construct a Wick transform one day because 1) it could simplify the
construction of rigorous solutions and since 2) a coherent state
transform
always has a close connection with semi-classical physics which is
important for the interpretation and the classical limit of the theory.\\
3) Not surprisingly, the unphysical Hamiltonian $\tilde{H}(A,E)
:=\tilde{H}_\Co(A_\Co:=A,E_\Co:=E)$ can be recognized as the
Hamiltonian constraint that one obtains from the Hamiltonian
formulation of Riemannian general relativity (i.e. ordinary general
relativity just that one considers four-metrics of Euclidean signature).\\
4) The Wick transform derived in \cite{IV} is the first
honest proposal for a
solution of the reality conditions for the {\it complex} connection
variables. For a different proposal geared to a Minkowski
space background, see \cite{75}.

From now on we remove the prime in $H'$ again and will only work
with physical, unrescaled functions of real variables.

\subsection{Derivation of the Hamiltonian Constraint Operator}
\label{s6.2}

In view of the previous section, a crucial question that remained
was whether one could make the Wick transform to work, that is, whether
one could realize its generator as a self-adjoint operator on ${\cal H}^0$
to begin with. In the
course of efforts towards this aim, a new perspective came to
the foreground which
enables one to get rid of the difficult complex variables
altogether and to work entirely with the real ones.
This then also made the existence of the
Wick transform a question of marginal interest. In retrospect, it is now
clear that in any case one could never have succeeded working with an operator
corresponding to $\tilde{H}$ even if one could make
the Wick transform work : The reason for this is so simple that it is
surprising that it was not pointed out long before. It has nothing to do
with the use of complex valued variables but rather with the fact
that $\tilde{H}$, regardless of whether written in terms of real or complex
variables, is a scalar with a density weight of {\it two} rather than
{\it one}. Let us clarify this point from the outset:\\
\\
When one quantizes an integrated scalar density $s(x)$ of weight $k$ then
one replaces
the canonical variables by multiplication operators and functional
derivatives respectively. When one applies the local operator $\hat{s}(x)$
to a
state $\psi$, which is, in particular, a scalar of density weight zero,
the various
multiplication operators and functional derivatives produce a new state
which is roughly of the form $D(x)\psi'(x)$ where $\psi'(x)$ is a
well-defined
scalar with density weight zero and $D(x)$ is a distribution. However,
the operator remembers the density weight of its classical counterpart
$s$ and therefore the density weight $k$ must be encoded in the
distribution $D(x)$. The only distribution of density weight different
from zero that can appear is the delta-distribution (and derivatives
thereof). We conclude that $D(x)$ is proportional to $\delta(x)^k$
(and derivatives thereof) which is meaningless unless $k=1$.

Why does one not see this problem in ordinary quantum field theory as for
instance the Maxwell Hamiltonian is a density of weight two as well ?
The answer is that one actually {\it does} see this problem : the divergence
that appears can be cured in this case by normal ordering, one subtracts
an infinite constant from the Hamiltonian. Such a procedure is possible
in free quantum field theory on a fixed background but in background-free
general relativity this cannot be done : the infinite constant
contributes to the vacuum energy and cannot be discarded. Also a
regularization and renormalization does not work : Consider for instance
a point splitting regularization. That is, one measures distances by
a background metric. If one subtracts the divergence and removes the
regulator, the result is necessarily a background dependent operator
destroying diffeomorphism covariance.

Actually, this problem was noted by many working on formal solutions to
the Hamiltonian constraint (see, e.g., \cite{42,43,44,76,77,78} and
references therein) but its underlying reason in terms of density weights
had not been spelled out clearly.

In order to solve the problem even multiplicative renormalizations were
considered, that is, one multiplies the operator by a regulator which
vanishes in the limit. While this removes the background dependence one now
has a quantum operator whose classical limit is zero.\\
Another suggestion was to take the square root of the Hamiltonian
constraint $\tilde{H}$ since this reduces the density weight to one
and to quantize this square root (see \cite{79}, in particular in connection
with matter coupling \cite{80}). However,
since $\tilde{H}$ is famously indefinite it is unclear how to define the
square root of an infinite number of non-self-adjoint, non-positive and
non-commuting operators, moreover, classically the square
root of a constraint has an ill-defined Hamiltonian vector field and
therefore does not generate gauge transformations. \\
A brute force method finally to remove the singularities is to go to a
lattice formulation but
the problem must undoubtedly reappear when one takes the continuum limit
(see, e.g., \cite{81} and references therein).

For those reasons, {\it the factor $1/\sqrt{\det(q)}$ in
$H$ as compared to $\tilde{H}$ is, in fact, needed and one cannot work
with the rescaled
constraint}. Since $H$ in either real or complex connection variables is as
non-polynomial
as in the ADM variables, it seems at first that the whole virtue of
introducing connection variables is lost (even if the Wick transform
could be made to work since the non-polynomial pre-factor does not get
removed by it).

However, this is by far not the case, the advantage of connection variables
is twofold, there are very powerful kinematical and dynamical reasons
for using them:\\
The kinematical reason is it has been possible to give a rigorous,
background independent mathematical formulation (using real connections)
only using form fields (here one forms and $(D-1)-$ forms). This has
not been achieved using ADM metric variables so far. Only connections
provide us with the powerful calculus on $\ab$.\\
The dynamical reason is that, as we are about to show,
one can actually give rigorous meaning to $H$ as a quantum operator
on ${\cal H}^0$ {\it despite its non-polynomial nature} !
By means of a novel quantization technique the
non-polynomial prefactor is absorbed into a commutator between
well-defined operators.
Since a commutator is essentially a derivation one can intuitively understand
that this operation will express a denominator in terms of a numerator
which has a better chance to be well-defined as an operator.

This technique then removed the two major roadblocks that plagued
Canonical Quantum  General Relativity until 1996 in one stroke :\\
\\
{\bf First, it showed that $\hat{H}$ can be made well-defined in terms of
real connections and therefore, secondly, the full machinery of
$L_2(\overline{{\cal A}},d\mu_0)$ could be accessed.}
\\
Even more is true : the new technique turns out to be so general that
it applies to any kind of field theory for which a Hamiltonian formulation
exists \cite{V,VI,47m1,VIII,IX,X,XI}. The series of these papers
is entitled ``Quantum
Spin Dynamics (QSD)" for the following reason : the Hamiltonian constraint
$\hat{H}$ acting on a spin-network state creates and annihilates the spin
quantum numbers with which the edges of the underlying graph are coloured.
On the other hand, the ADM energy surface Hamiltonian operator \cite{X} is
essentially
diagonal on spin-network states where its eigenvalue is also determined
by the spin-quantum numbers. Thus, we may interpret the spin-network
representation as the {\bf non-linear Fock representation of quantum
general relativity}, the spin quanta playing the role of the occupation
numbers of momentum excitations of the usual Fock states of, say, Maxwell
theory. The excitations
of the gravitational quantum field are string-like, labelled
by the edges of a graph, and the degree of freedom corresponding
to an edge can be excited only according to half-integral spin quantum
numbers.\\
\\
The rest of this section is devoted to a hopefully pedagogical
explanation of the main idea on which \cite{V} is based.
(see also \cite{37,83} for an even less technical introduction).

Usually, the Hamiltonian constraint is written in terms of the
real connection variables as follows \cite{36,81} (we set $\beta=1/2$
in this section, the generalization to arbitrary positive values is
trivial, and drop the label $\beta$ from all formulas)
\be \label{6.15}
H=\frac{1}{\kappa\sqrt{\det(q)}}\mbox{tr}([F_{ab}-R_{ab}][E^a,E^b])
\ee
(we have a trace and a commutator for the Lie algebra valued quantities
and kept explicitly a factor of $1/\kappa$ coming from an overall
factor of $1/\kappa$ in front of the action).
The reason for this clear : since $A,E$ are the elementary variables
one better avoids the appearance of $K_a^i=A_a^i-\Gamma_a^i$. We, however,
will work paradoxically with the following identical formula (up to
an overall numerical factor)
\be \label{6.16}
H=\frac{2}{\kappa\sqrt{\det(q)}}\mbox{tr}([K_a,K_b][E^a,E^b])-H_E
\ee
where
\be \label{6.17}
H_E=\frac{1}{\kappa\sqrt{\det(q)}}\mbox{tr}(F_{ab}[E^a,E^b])
\ee
is called the {\it Euclidean Hamiltonian constraint}, that is, the
(unrescaled) unphysical Hamiltonian constraint that one would employ into
the Wick rotation transform as alluded to in section \ref{s6.1}. Its
natural appearance here is not a coincidence as we will see. The reason
for doing this will become clear in a moment. Notice that we have
correctly introduced the overall factor $1/\kappa$ in front of the
action into $H_E,H$ which will get the dimensionalities right.

Consider the following two quantities,\\
(i) The volume of an open region $R$ of $\sigma$ :
\be \label{6.18}
V(R):=\int_R d^3x \sqrt{|\det(q)|} \mbox{ and}
\ee
(ii) the integrated densitized trace of the extrinsic curvature
\be \label{6.19}
K:=\int_\sigma d^3x K_a^i E^a_i
\ee
the latter of which is nothing else than the generator of the Wick transform
up to a factor of $-\pi/(2\kappa)$. Notice that in (\ref{6.19}) we
have taken absolute values under the square root. However,
$\det((q_{ab}))=[\det((e_a^i))]^2$ is anyway positive so that we could
drop the absolute value at the classical level. At the quantum level,
however, it will be important to keep it. On the other hand,
if we define $E^a_j=\det(e_b^k) e^a_j$ then
$E^a_i$ satisfies the following anholonomic constraint
\be \label{6.19a}
\det((E^a_i))=\det((q_{ab}))\ge 0
\ee
as pointed out before. Strictly speaking, one could argue to have to
impose (\ref{6.19}) on quantum states later on.
On the other hand, quantum theory is an extension of
the classical theory anyway and (\ref{6.19}) could be required to hold
on {\it semiclassical states} only in the sense of expectation values. One
could also work instead
with $E^a_j=|\det(e_b^k)| e^a_j$ in which case (\ref{6.19}) would no
longer hold, however, then one has to absorb a factor of
$\mbox{sgn}(\det(e_b^k))$ into the lapse function in the following
formulas so that $N$ is allowed to take both signs.
Then one might want to
argue that $N\mbox{sgn}(\det(e_b^k))\ge 0$ should hold which is peculiar
since it would mean to quantize the lapse, which is in contradiction with
the whole formalism and therefore must be dropped as well in the quantum
theory. Whether one strategy
is preferred over the other is not yet clear. What is clear,
however, is that the condition (\ref{6.19}) in the strong form (i.e.
that the equality sign is excluded) is not preserved under
the quantum evolution: The right hand side of (\ref{6.19}) becomes in
quantum theory, roughly, the square of the volume operator. Now while
the volume eigenvalues are non-negative, the value zero is attained on
trivalent vertices and these are among the types of vertices created by
the Hamiltonian constraint. Therefore one cannot quantize $1/\sqrt{\det(q)}$
by replacing it by the inverse volume operator. We choose here the first
alternative and simply drop (\ref{6.19}) while working with
$E^a_j=\det(e) e^a_j$.

The following two classical identities are {\it key} for all that follows :
\be \label{6.20}
(\frac{[E^a,E^b]_i}{\sqrt{\det(q)}})(x)=\epsilon^{abc}
(\mbox{sgn}(\det(e))e_c^i)(x)
=2\epsilon^{abc}\frac{\delta V(R)}{\delta E^a_i(x)}=
2\epsilon^{abc}\{V(R),A_a^i(x)\}/\kappa
\ee
for any region $R$ such that $x\in R$ and
\be \label{6.21}
K_a^i(x)=\frac{\delta K}{\delta E^a_i(x)}=\{K,A_a^i(x)\}/\kappa
\ee
where (\ref{6.21}) relies on $\{\Gamma_a^i,K\}=0$ as already pointed out in
section \ref{s6.1}. In the sequel we will use the notation $R_x$ for
any open neighbourhood of $x\in\sigma$.\\
Using these key identities the reader can quickly convince himself that
\ba
\label{6.22}
(H-H_E)(x)&=&
-8\epsilon^{abc}\mbox{tr}(\{A_a,K\}\{A_b,K\}\{A_c,V(R_x)\})/\kappa^4
\\
\label{6.23}
H_E(x)&=&-2\epsilon^{abc}\mbox{tr}(F_{ab}\{A_c,V(R_x)\})/\kappa^2
\ea
or, in integrated form, $H(N)=\int_\sigma d^3x N(x) H(x)$ etc. for some
lapse function $N$ and any smooth neighbourhood-valued function
$R\; : x\mapsto R_x$
\ba
\label{6.24}
(H-H_E)(N)&=&
-8\int_\sigma N\mbox{tr}(\{A,K\}\wedge\{A,K\}\wedge\{A,V(R)\})/\kappa^4
\\
\label{6.25}
H_E(N)&=&-2\int_\sigma N\mbox{tr}(F\wedge\{A,V(R)\})/\kappa \;.
\ea
What we have achieved in (\ref{6.22}), (\ref{6.23}) or
(\ref{6.24}), (\ref{6.25}) is to remove the problematic
$1/\sqrt{\det(q)}$ from the denominator by means of Poisson brackets.

The reader will now ask what the advantage of all this is. The idea
behind these formulas is the following :\\
What we want to quantize is $H(N)$ on ${\cal H}^0$ and since ${\cal H}^0$ is
defined in terms of generalized holonomy variables $A(e)$ we first
need to write (\ref{6.24}), (\ref{6.25}) in terms of holonomies. This
can be done by introducing a triangulation $T(\epsilon)$ of $\sigma$
by tetrahedra which fill all of $\sigma$ and intersect each other
only in lower dimensional submanifolds of $\sigma$. The
small parameter $\epsilon$ is to indicate
how fine the triangulation is, the limit $\epsilon\to 0$ corresponding
tetrahedra of vanishing volume (the number of tetrahedra grows in this
limit as to always fill out $\sigma$; we will not be specific here
about what $\epsilon$ actually is, the interested reader is referred to
\cite{V}). So let $e_I(\Delta)$ denote
three edges of an analytic tetrahedron $\Delta\in T(\epsilon)$ and let
$v(\Delta)$
be their common intersection point with outgoing orientation (the
quantities $\Delta, e_I(\Delta), v(\Delta)$, of course, also depend on
$\epsilon$ but we do not display this in order not to clutter the
formulae with too many symbols). The matrix consisting of the tangents of
the edges $e_1(\Delta),e_2(\Delta),e_3(\Delta)$ at $v(\Delta)$ (in that
sequence) has non-negative determinant which induces an orientation of
$\Delta$. Furthermore, let $a_{IJ}(\Delta)$ be the arc on the boundary of
$\Delta$ connecting the endpoints of $e_I(\Delta),e_J(\Delta)$
such that the loop
$\alpha_{IJ}(\Delta)=e_I(\Delta)\circ a_{IJ}(\Delta)\circ e_J(\Delta)^{-1}$
has positive orientation in the induced orientation of the boundary
for $(I,J)=(1,2),(2,3),(3,1)$ and negative in the remaining cases.
One can then see that in the limit as $\epsilon\to 0$ the quantities
\ba
\label{6.26}
(H^\epsilon-H^\epsilon_E)(N)&=&\frac{8}{3\kappa^4}
\sum_{\Delta\in T(\epsilon)} \epsilon^{IJK}
N(v(\Delta))\times\\
&\times& \mbox{tr}(h_{e_I(\Delta)}\{h_{e_I(\Delta)}^{-1},K\}
h_{e_J(\Delta)}\{h_{e_J(\Delta)}^{-1},K\}
h_{e_K(\Delta)}\{h_{e_K(\Delta)}^{-1},V(R_{v(\Delta)})\}) \nonumber\\
\label{6.27}
H^\epsilon_E(N)&=&\frac{2}{3\kappa^2}
\sum_{\Delta\in T(\epsilon)}N(v(\Delta))\epsilon^{IJK}
\mbox{tr}(h_{\alpha_{IJ}(\Delta)}
h_{e_K(\Delta)}\{h_{e_K(\Delta)}^{-1},V(R_{v(\Delta)})\})
\nonumber\\
&&
\ea
converge to (\ref{6.24}), (\ref{6.25}) respectively pointwise on $\cal
M$ {\it for any choice of triangulation} !
This independence of the limit, for the classical theory, from the choice
of the family of triangulations enables us to choose the triangulations
state-dependent just as for the area operator, see below.

Suppose now that we can turn $V(R)$ and $K$ into well-defined operators on
$\cal H$, densely defined on cylindrical functions. Then, according to the
rule that upon quantization
one should replace Poisson brackets by commutators times $1/(i\hbar)$
(\ref{6.26}), (\ref{6.27}) {\it would become densely defined regulated
operators
on ${\cal H}^0$ without any divergences for a specific choice of factor
ordering} ! We will discuss the issue of what happens upon removal of the
regulator $\epsilon$ in a moment.

Is it then true that $\hat{V}(R)$ and $\hat{K}$ exist ?
We have seen in section \ref{s5} that the answer is affirmative for the
case of the volume operator. We use the version
of the volume operator that was constructed in \cite{64}
as compared to the one in \cite{61} because it turns out that only the
operator \cite{64} gives a densely defined
Hamiltonian constraint operator in the regularization scheme that we
advertize here, it is important that the volume vanishes on planar
vertices (that is, the tangent space at the vertex spanned there by the
tangents of the edges incident at it is at most two-dimensional) .

Recall from section \ref{s5} that the volume operator of \cite{64} acts
on a function cylindrical over a graph $\gamma$ as follows :
\be \label{6.28}
\hat{V}(R)f_\gamma:=\frac{\ell_p^3}{4}\sum_{v\in V(\gamma)\cap R}
\sqrt{|\frac{i}{3!} \sum_{e\cap e'\cap \tilde{e}=v}
\epsilon(e,e',\tilde{e})
\epsilon_{ijk} R^i_e R^j_{e'} R^k_{\tilde{e}}|} \;f_\gamma
\ee
where the sum is over the set $V(\gamma)$ of all vertices $v$ of the graph
$\gamma$ that lie in $R$ and
over all unordered triples of edges that start at $v$ (we can take the
orientation of each edge incident at $v$ to be outgoing by suitably
splitting an edge into two halves if necessary). The function
$\epsilon(e,e',\tilde{e})$ takes the values $+1,-1,0$ if the tangents of
the three edges at $v$ (in that sequence) form a matrix of positive,
negative or vanishing determinant and the right invariant vector fields
$R^i_e$ were defined in section \ref{s4}. The absolute value $|\hat{B}|$
of the operator $\hat{B}$ indicates that one
is supposed to take the square root of the operator $\hat{B}^\dagger\hat{B}$.
The dense domain of this operator are the thrice differentiable
cylindrical functions. Notice that planar vertices of arbitrary valence
do not contribute. Surprisingly, also arbitrary tri-valent vertices do
not contribute \cite{61a} if the corresponding state is gauge-invariant.
(Proof: We have $-(R^j_1+R^j_2)=R^j_3$ due to gauge invariance where
$R^j_I=R^j_{e_I},\;I=1,2,3$. Substituting this into
$\epsilon_{jkl}R^j_1 R^j_2 R^j_3$ and using
$[R^j_I,R^k_J]=-2\delta_{IJ}\epsilon_{jkl} R^l_I$ completes the proof).

Thus, it seems that one can make sense out of a regulated operator
corresponding to
(\ref{6.26}) for each $N$, in particular for $N=1$. Now recall the
classical identity that the integrated densitized trace of the extrinsic
curvature is the ``time derivative" of the total volume
\be \label{6.29}
K=\{H^E(1),V(\sigma)\}=\{H(1),V(\sigma)\}\;.
\ee
where $N=1$ is the constant lapse equal to unity.
This formula makes sense even if $\sigma$ is not compact (see \cite{V}
for the details). Notice that
(\ref{6.29}) holds for either signature (i.e. it does not matter which
Hamiltonian constraint of the Hamiltonian formulation
of general relativity we use, the one corresponding to four metrics of
Euclidean or
Lorentzian signature). But if we then replace again Poisson brackets
by commutators times $1/(i\hbar)$ and define
\be \label{6.30}
\hat{K}^\epsilon:=-\frac{i}{\hbar}[\hat{H}_E^\epsilon(1),\hat{V}(\sigma)]
\ee
using the already defined quantities
$\hat{H}^\epsilon(1),\hat{V}(\sigma)$ it seems that we can also define
a regulated operator corresponding to (\ref{6.27}) !

This concludes the explanation of the main idea. The next subsection
comments on the concrete implementation of this idea.

\subsection{Mathematical Definition of the Hamiltonian Constraint
Operator}
\label{s6.3}

Obviously, central questions regarding the concrete implementation of
the technique are :\\
I) What are the allowed, physically relevant choices for a family of
triangulations $T(\epsilon)$ ?\\
II)
How should one treat the limit $\epsilon\to 0$
for the operator $\hat{H}^\epsilon(N)$ ? That is, should one keep $\epsilon$
finite and just refine $\gamma\to \sigma$ or
is there an operator topology such
that this limit can be given a meaning ? Secondly, does the
refined or limit operator
remember something about the choice of the family $T(\epsilon)$
or is there some notion of universality ? \\
III) What is the commutator algebra of these (limits of) operators, is it
free of anomalies ? \\
We will address these issues separately.

\subsubsection{Concrete Implementation}
\label{s6.3.1}

A natural choice for a triangulation turns out to
be the following (we simplify the presentation drastically, the details
can be found in \cite{V}):\\
Given a graph $\gamma$ one constructs a triangulation $T(\gamma,\epsilon)$
of $\sigma$ {\it adapted} to $\gamma$ which satisfies the following basic
requirements :
\begin{itemize}
\item[a)] The graph $\gamma$ is embedded in $T(\gamma,\epsilon)$ for all
$\epsilon>0$.
\item[b)] The valence of each vertex $v$ of $\gamma$, viewed as a vertex of
the
infinite graph $T(\epsilon,\gamma)$, remains constant and is equal to the
valence of $v$, viewed as a vertex of $\gamma$, for each $\epsilon>0$.
\item[c)] Choose a system of analytic arcs $a^\epsilon_{\gamma,v,e, e'}$,
one for each pair of edges $e,e'$ of
$\gamma$ incident at a vertex $v$ of $\gamma$, which do not intersect
$\gamma$ except in its endpoints where they intersect transversally. These
endpoints are interior points
of $e,e'$ and are those vertices of $T(\epsilon,\gamma)$ contained
in $e,e'$ closest to $v$ for each $\epsilon>0$ (i.e. no others are in
between). For each
$\epsilon,\epsilon'>0$ the arcs $a^\epsilon_{\gamma,v,e e'},
a^{\epsilon'}_{\gamma,v,e,e'}$ are diffeomorphic with respect to
analytic diffeomorphisms. The segments of $e,e'$ incident at $v$ with
outgoing orientation that are determined by the endpoints of the
arc $a^\epsilon_{\gamma,v,e,e'}$ will be denoted by
$s^\epsilon_{\gamma,v,e},s^\epsilon_{\gamma,v,e'}$ respectively.
Finally, if $\varphi$ is an analytic diffeomorphism then
$s^\epsilon_{\varphi(\gamma),\varphi(v),\varphi(e)},
a^\epsilon_{\varphi(\gamma),\varphi(v),\varphi(e),\varphi(e')}$ and
$\varphi(s^\epsilon_{\gamma,v,e}),\varphi(a^\epsilon_{\gamma,v,e,e'})$
are analytically diffeomorphic.
\item[d)] Choose a system of mutually disjoint neighbourhoods
$U^\epsilon_{\gamma,v}$,
one for each vertex $v$ of $\gamma$, and require that for each $\epsilon>0$
the $a^\epsilon_{\gamma,v,e,e'}$ are contained in $U^\epsilon_{\gamma,v}$.
These neighbourhoods are nested in the sense that
$U^\epsilon_{\gamma,v}\subset U^{\epsilon'}_{\gamma,v}$ if
$\epsilon<\epsilon'$ and $\lim_{\epsilon\to 0}U^\epsilon_{\gamma,v}=\{v\}$.
\item[e)] Triangulate $U^\epsilon_{\gamma,v}$ by tetrahedra
$\Delta(\gamma,v,e,e',\tilde{e})$,
one for each ordered triple of distinct edges $e,e',\tilde{e}$ incident at
$v$, bounded by the segments
$s^\epsilon_{\gamma,v,e},s^\epsilon_{\gamma,v,e'},
s^\epsilon_{\gamma,v,\tilde{e}}$ and the arcs
$a^\epsilon_{\gamma,v,e,e'},a^\epsilon_{\gamma,v,e',\tilde{e}},
a^\epsilon_{\gamma,v,\tilde{e},e}$
from which loops $\alpha^\epsilon(\gamma;v;e,e')$ etc. are built
and triangulate the rest of $\sigma$ arbitrarily. The ordered triple
$e,e',\tilde{e}$ is such that their tangents at $v$, in this sequence, form a
matrix of positive determinant.
\end{itemize}
Requirement a) prevents the action of the Hamiltonian constraint operator
from being trivial. Requirement b) guarantees that the regulated operator
$\hat{H}^\epsilon(N)$ is densely defined for each $\epsilon$. Requirements
c), d) and e) specify the triangulation in the neighbourhood of each
vertex of $\gamma$ and leave it unspecified outside of them.
The more
detailed prescription of \cite{V} shows that triangulations satisfying
all of these requirements always exist and can also deal with degenerate
situations, e.g., how to construct a tetrahedron for a planar vertex.
More specifically, what we have done in \cite{V} is to fix the routing
of the analytical arcs through the ``forest" of the already present
edges in such a way that it is invariant under smooth diffeomorphisms
that leave $\gamma$ invariant and the arcs analytic. The use of
smooth diffeomorphisms here is not in contradiction to having only an
analytic manifold as we use them only in order to choose the routing,
they do not play any other role e.g. in imposing the diffeomorphism
constraint. In particular, since we do not need that arcs corresponding
to different pairs of edges are analytically diffeomorphic, there
is no contradiction. Here we are more general than in \cite{V}
in that we just use the {\it axiom of choice}. That is, we only use that
a choice function
\be \label{6.31}
a^\epsilon:\;\Gamma^\omega_0\to \Gamma^\omega_0\;
\gamma\mapsto \{a^\epsilon_{\gamma,v,e,e'}\}_{v\in V(\gamma);\;
e,e'\in E(\gamma);\;v\in \partial e\cap\partial e'}
\ee
subject to requirements a) -- e) always exists.

The reason
for why those tetrahedra lying outside the neighbourhoods of the vertices
described above are irrelevant rest crucially on the choice of ordering
(\ref{6.27}) with $[\hat{h}_s^{-1},\hat{V}]$ on the most right and on
our choice of the volume operator \cite{64}: If $f$ is a cylindrical
function over $\gamma$ and $s$ has support outside the neighbourhood of
any vertex of $\gamma$, then $V(\gamma\cup s)-V(\gamma)$ consists of planar
at most four-valent vertices only so that
$[\hat{h}_s^{-1},\hat{V}]f=0$.
Notice, however, that \cite{61} does not vanish on planar vertices
and so $[\hat{h}_s^{-1},\hat{V}]f$ would not vanish even on trivalent
vertices in $V(\gamma\cup s)-V(\gamma)$ because it is not gauge invariant.
In other words, in the limit of small $\epsilon$ the operator would
map us out of the space of cylindrical functions. Therefore the
Hamiltonian constraint operator inherits
from the volume operator a basic property : It annihilates all states
cylindrical with respect to graphs with only co-planar
vertices as
can be understood from the fact that the volume operator enters the
construction of both $\hat{H}_E^\epsilon(N), \hat{H}^\epsilon(N)$. In other
words, the dynamics ``happens only at the vertices of a graph".

Notice that a)-e) are natural
extensions to arbitrary graphs of what one does in lattice gauge theory
\cite{84} with one exception : what we will get is not an operator
$\hat{H}^\epsilon(N)$ to begin with, but actually a family of operators
$\hat{H}^\epsilon_\gamma(N)$, one for each graph $\gamma$. This happened
because we adapted the triangulation to the graph of the state on which
the operator acts. One must then worry that this does not define a linear
operator any more, that is, that it is not cylindrically consistently
defined. Here we circumvent that problem as follows: We do not define
the operator on functions cylindrical over graphs but cylindrical over
{\it coloured graphs}, that is, we define it on spin-network functions.
Since every smooth cylindrical function, the domain for the operator
that we will choose, is a finite linear combination of spin-network
functions this defines the operator uniquely as a linear operator.
Any operator automatically becomes consistent if one defines it
on a basis, the consistency condition simply drops out.

Moreover, the regulated operator $\hat{H}^\epsilon(N)$ is
by construction
background independently  defined for each $\epsilon$ but not symmetric
which, as described in section \ref{si}, is not a necessary requirement
for a constraint operator and even argued to be better not the case
\cite{85} in order for the constraint algebra to be non-anomalous
for open constraint algebras.

Finally, we point out that beyond the freedom of a choice
function (\ref{6.31}) even requirements a)-e) can be generalized
and even the regularization itself can be generalized. For instance in
\cite{86} one uses instead of $\mbox{tr}(\tau_j h_\alpha)$ the function
\be \label{6.32a}
\frac{\sum_{k=1}^N \mbox{tr}(\tau_j h_\alpha^{n_k})}{\sum_{k=1}^N n_k}
\ee
for any choice of integers $n_k$ such that the denominator is non-vanishing
which again gives the correct continuum limit since all the functions
(\ref{6.31}) are identical in the leading order that we need.
Hence, there is vast room for generalizations. Which choice
is ``more physical" than another, whether they all are
equivalent or whether all of them are unphysical can only be decided
in the investigation of the classical limit.\\
\\
Let us then display the action of the Hamiltonian constraint on a
spin-network function $f_\gamma$ cylindrical with respect to a graph
$\gamma$. It is given by
\ba
\hat{H}^\epsilon_E(N) f_\gamma&=&\frac{16}{3i\kappa\ell_p^2}
\sum_{v\in V(\gamma)} \frac{N(v)}{E(v)}\sum_{v(\Delta)=v}\epsilon^{IJK}
\mbox{tr}(h_{\alpha_{IJ}(\Delta)}
h_{e_K(\Delta)}[h_{e_K(\Delta)}^{-1},\hat{V}(U_\epsilon(v))])\;f_\gamma
\nonumber\\
&& \label{6.32}\\
\label{6.33}
(\hat{H}^\epsilon-\hat{H}^\epsilon_E)(N)f_\gamma &=&
\frac{64}{3\kappa(i\ell_p^2)^3}
\sum_{v\in V(\gamma)}\frac{N(v)}{E(v)}\sum_{v(\Delta)=v}\epsilon^{IJK}
\times\\
&\times&
\mbox{tr}(h_{e_I(\Delta)}[h_{e_I(\Delta)}^{-1},\hat{K}^\epsilon]
h_{e_J(\Delta)}[h_{e_J(\Delta)}^{-1},\hat{K}^\epsilon]
h_{e_K(\Delta)}[h_{e_K(\Delta)}^{-1},\hat{V}(U_\epsilon(v))]) \;f_\gamma
\nonumber
\ea
where $\hat{K}_\epsilon$ is defined by (\ref{6.30}).
The first sum is over all the vertices of a graph and
the second sum over all ordered tetrahedra of the triangulation
$T(\epsilon,\gamma)$
that saturate the vertex (the remaining tetrahedra drop out). The symbols
$e_I(\Delta)$ etc. mean the same as in (\ref{6.26}), (\ref{6.27}) just that
now the tetrahedra in question are the particular ones as specified in
a)-e) above. Here the numerical factors
$E(v)=\left( \begin{array}{c} n(v)\\ 3 \end{array} \right)$, where $n(v)$
is the valence of the vertex $v$, come about as follows:\\
\\
Given a triple of edges $(e,e',e^\dprime)$
incident at $v$ with outgoing orientation consider the tetrahedron
$\Delta^\epsilon(\gamma,v,e,e',e^\dprime)$ bounded by the three
segments $s^\epsilon_{\gamma,v,e}\subset e,
s^\epsilon_{\gamma,v,e'}\subset e',
s^\epsilon_{\gamma,v,e^\dprime}\subset e^\dprime$
incident at $v$ and the three arcs
$a^\epsilon_{\gamma,v,e,e'},a^\epsilon_{\gamma,v,e',e^\dprime},
a^\epsilon_{\gamma,v,e^\dprime,e}$. We now define the ``mirror
images"
\ba \label{6.34}
s^\epsilon_{\gamma,v,\bar{p}}(t)&:=&2v-s^\epsilon_{\gamma,v,p}(t)
\nonumber\\
a^\epsilon_{\gamma,v,\bar{p},\bar{p}'}(t)
&:=&2v-a^\epsilon_{\gamma,v,p,p'}(t)
\nonumber\\
a^\epsilon_{\gamma,v,\bar{p},p'}(t)
&:=&a^\epsilon_{\gamma,v,\bar{p},\bar{p}'}(t)
-2t[v-s^\epsilon_{\gamma,v,p'}(1)]
\nonumber\\
a^\epsilon_{\gamma,v,p,\bar{p}'}(t)
&:=&a^\epsilon_{\gamma,v,p,p'}(t)
+2t[v-s^\epsilon_{\gamma,v,p'}(1)]
\ea
where $p\not= p'\in \{e,e',e^\dprime\}$ and we have chosen any
parameterization of segments and arcs. Using the data (\ref{6.34})
we build seven more ``virtual" tetrahedra bounded by these quantities
so that we obtain altogether eight tetrahedra that saturate $v$ and
triangulate a neighbourhood $U^\epsilon_{\gamma,v,e,e',\tilde{e}}$
of $v$. Let $U^\epsilon_{\gamma,v}$ be the union of these neigbourhoods
as we vary the ordered triple of edges of $\gamma$ incident at $v$.
The $U^\epsilon_{\gamma,v},\;v\in V(\gamma)$ were
chosen to be mutually disjoint in point d) above.
Let now
\ba \label{6.35}
\bar{U}^\epsilon_{\gamma,v,e,e',e^\dprime}
&:=&U^\epsilon_{\gamma,v}-U^\epsilon_{\gamma,v,e,e',e^\dprime}
\nonumber\\
\bar{U}^\epsilon_{\gamma}&:=&\sigma-\bigcup_{v\in V(\gamma)}
U^\epsilon_{\gamma,v}
\ea
then we may write any classical integral (symbolically) as
\ba \label{6.36}
\int_\sigma&=& \int_{\bar{U}^\epsilon_\gamma}+\sum_{v\in V(\gamma)}
\int_{U^\epsilon_{\gamma,v}}
\nonumber\\
&=&\int_{\bar{U}^\epsilon_\gamma}+\sum_{v\in V(\gamma)}
\frac{1}{E(v)} \sum_{v=b(e)\cap b(e')\cap b(e^\dprime)}
[\int_{U^\epsilon_{\gamma,v,e,e',e^\dprime}}+
\int_{\bar{U}^\epsilon_{\gamma,v,e,e',e^\dprime}}]
\nonumber\\
&\approx&\int_{\bar{U}^\epsilon_\gamma}+\sum_{v\in V(\gamma)}
\frac{1}{E(v)} [\sum_{v=b(e)\cap b(e')\cap b(e^\dprime)}
8\int_{\Delta^\epsilon_{\gamma,v,e,e',e^\dprime}}+
\int_{\bar{U}^\epsilon_{\gamma,v,e,e',e^\dprime}}]
\ea
where in the last step we have noticed that classically the integral
over $U^\epsilon_{\gamma,v,e,e',e^\dprime}$ converges to eight times
the integral over $\Delta^\epsilon_{\gamma,v,e,e',e^\dprime}$.
Now when triangulating the regions of the integrals over
$\bar{U}_{v,e,e',e^\dprime}$ in
(\ref{6.36}), regulation and quantization gives operators that vanish
on $f_\gamma$ because the corresponding regions do not contain a
non-planar vertex of $\gamma$.\\
\\
Notice that (\ref{6.32}) and (\ref{6.33}) are {\it finite}
for each $\epsilon>0$, that is, densely defined without that any
renormalization is necessary and with range in the smooth
cylindrical functions again. Furthermore, the adjoints of the
expressions (\ref{6.32}) and (\ref{6.33}) are densely defined
on smooth cylindrical functions again so that we get in fact
a consistently and densely defined family of closed operators
on ${\cal H}^0$.

Let us check
the dimensionalities: The volume operator in (\ref{6.32}) is
given by $\ell_p^3$ times a dimension free operator, hence (\ref{6.32})
is given by $\ell_p/\kappa=m_p$ times a dimension free operator.
Hence the correct dimension of Planck mass $m_p=\sqrt{\hbar/\kappa}$
has popped out. Therefore, by inspection, (\ref{6.30}) has dimension
of $\ell_p^3 m_p/\hbar=\ell_p^2$ which is correct since
$K(x)=\sqrt{\det(q)}(x)K_{ab}(x)q^{ab}(x)$ dimension cm$^{-1}$
so that $K=\int d^3x K(x)$ has dimension cm$^2$. Finally therefore
(\ref{6.33}) has the correct dimension of
$(\ell_p^2)^2\ell_p^3/(\kappa \ell_p^6=m_p$ again.

\subsubsection{Operator Limits}
\label{s6.3.2}

Basically there are two, technically equivalent viewpoints towards
treating the limit $\epsilon\to 0$.
\begin{itemize}
\item[A)] {\it Effective Operator Viewpoint}\\
The more radical proposal is {\it to drop the parameter $\epsilon$}
from all formulas. That is, take a choice function $a$ once and for all.
One gets a densely defined family of closed operators. One may object
that on a given graph $\gamma$ one does not get a quantization of the full
expressions (\ref{6.26}), (\ref{6.27}), however, that is only because
the graph $\gamma$ does not fill all of $\sigma$. In other words,
the continuum limit of infinitely fine triangulation of the
Riemann sum expressions (\ref{6.26}), (\ref{6.27}) in the classical
theory is nothing else than taking the graphs, on which the operator
is probed, finer and finer. This is a new viewpoint not previously
reported in the literature and could be called the {\it effective
operator viewpoint} because on fine but not infinitely fine graphs
the classical limit of the operator will only {\it approximate}
the exact classical expression in the same way as (\ref{6.26})
and (\ref{6.27}) only approximate
(\ref{6.24}) and (\ref{6.25}). However, it may be that this {\it is the
fundamental theory} and classical physics is just an approximation to it.
This way the UV regulator
$\epsilon$ corresponding to the continuum limit is trivially
removed and our family of operators is really defined on ${\cal H}^0$.
Whether the operator $\hat{H}$ that we then obtain has the correct classical
limit cannot decided at this stage but is again subject to a rigorous
semiclassical analysis which requires new input, see section
\ref{s8}.
\item[B)] {\it Limit Operator Viewpoint}\\
The challenge is to find an operator topology in which
the one-parameter family of operators $\hat{H}^\epsilon$ converges.\\
The operators (\ref{6.32}) and (\ref{6.33}) are easily seen to be unbounded
(already the volume operator has this property). Thus, a convergence in
the uniform or strong operator topology is excluded. Next, one may try
the weak operator topology (matrix elements converge pointwise) but
with respect to this topology the limit would be the zero operator
(it is too weak) : for instance, a matrix element between two
spin-network states is non-zero for at most one value
of $\epsilon$. Finally, we try the weak$^\ast$ topology, that is,
we must check whether $\Psi(\hat{H}^\epsilon(N)f)$ converges for
each $\Psi\in{\cal D}',f\in{\cal D}$ where ${\cal D}=C^\infty(\ab)$
with its natural nuclear topology is a dense domain and ${\cal D}'$ is its
topological dual. It turns out that this topology is a
little bit too strong, however, convergence holds with respect to a
topology which we might call {\it Uniform Rovelli-Smolin Topology} (URST)
in appreciation of the fact that Rovelli and Smolin first pointed out in
\cite{79} that, if instead
of ${\cal D}'$ we consider the space ${\cal D}^\ast$ of {\it diffeomorphism
invariant algebraic distributions}
on $\cal D$, then objects of the form $\Psi(\hat{H}^\epsilon(N)f)$ do not
depend at all on the position or shape of the arcs
$a^\epsilon_{\gamma,v,e,e'}$
alluded to above. In their original work \cite{79} Rovelli and Smolin did not
spell out this property in the context of ${\cal H}^0$ and also they did not
have a well-defined constraint operator but their observation applies
to a huge class of operators, their only feature being an analog of
property c) above. This is how we proceeded in \cite{V,VI,47m1}.

Therefore, since all the triangulations
$T(\gamma,\epsilon)$
restricted to each of the neighbourhoods $U^\epsilon_{\gamma,v}$ are
diffeomorphic by property c) above, the numbers
$\Psi(\hat{H}^\epsilon(N)f)$ are actually {\it already independent of
$\epsilon$} ! Therefore we have the striking result that with respect
to the URST
\be \label{6.38}
\hat{H}(N):=\lim_{\epsilon\to 0} \hat{H}^\epsilon(N)=\hat{H}^{\epsilon_0}(N)
\ee
where $\epsilon_0$ is an arbitrary but fixed positive number. Notice that
we require that for each $\delta>0$ there exists an $\epsilon'(\delta)>0$
such that for each $f\in{\cal D},\Psi\in{\cal D}^\ast_{Diff}$
$$
|\Psi(\hat{H}^\epsilon(N)f)-\Psi(\hat{H}^{\epsilon_0}(N)f)|<\delta
$$
for all $\epsilon<\epsilon'(\delta)$ where $\epsilon'(\delta)$ depends
only on $\delta$
but not on $f,\Psi$. In other words, we have convergence {\it uniform} in
${\cal D}\times {\cal D}^\ast_{Diff}$ rather than pointwise. This will be
important in what follows.

Notice that therefore the convergence
in the URST is very similar to the effective operator viewpoint
in the sense that it gives a topology in which it is allowed to
drop the label $\epsilon$ from the choice function altogether.

In particular we stress that {\it in contrast to the viewpoint
taken in \cite{87,88} we still have the operator defined on ${\cal H}^0$
and not on the dual subspace ${\cal D}^\ast_{Diff}\subset {\cal D}^\ast$
or an extension thereof}, precisely in the same sense as the limit
of a family of operators which converges in the
weak $^\ast$ topology on ${\cal D}$ is still considered an operator on
${\cal D}$ and not a dual operator on ${\cal D}'$. In fact, the
dual of $\hat{H}(N)$ cannot be
defined on ${\cal D}^\ast_{Diff}$ because that space is not left
invariant by $\hat{H}(N)'$ as we pointed out frequently which is why
the authors of \cite{87,88} have to take an extension to the so-called
``vertex smooth" distributions
${\cal D}^\ast_{Diff}\subset {\cal D}^\ast_\star\subset {\cal D}^\ast$
which is genuinely bigger than ${\cal D}^\ast_{Diff}$ and therefore
unphysical. Our viewpoint is completely different: {\it We do not
want to define $\hat{H}'(N)$ at all, we just use ${\cal D}^\ast_{Diff}$
as a means to define a topology} !

On the other hand, the physical reason for why testing
convergence of the operator only on ${\cal D}^\ast_{Diff}$ rather than on
a bigger space is precisely because we are eventually going to look
for the space of solutions to all constraints which in turn must be a
subspace ${\cal D}^\ast_{phys}$ of ${\cal D}^\ast_{Diff}$, so in a sense
we do not need stronger convergence. Notice that ${\cal D}^\ast_{phys}$
{\it is} left invariant by the dual action of $\hat{H}(N)$
(namely it is mapped to zero).

Again, whether the continuum operator thereby obtained has the
correct classical limit must be decided in an additional step.
\end{itemize}
Which viewpoint one takes is a matter of taste, technically they are
completely equivalent. The limit operator viewpoint has the advantage
that it shows that many choice functions are going to be physically
equivalent and thus decreases (but does not remove) the degree
of redundancy. In what follows we will therefore drop the label
$\epsilon$.\\
\\
The limit (\ref{6.38}) certainly only depends on the diffeomorphism invariant
characteristics of the particular triangulation $T(\gamma,\epsilon)$ that
we chose. For instance, the limit would be different if we would use
arcs that intersect the graph tangentially or which are smooth rather
than analytical. Other than that, there is no residual ``memory" of
the triangulation.

\subsubsection{Commutator Algebra}
\label{s6.3.3}

We now come to question III) whether the commutator between two
Hamiltonian constraints and between Hamiltonian and diffeomorphism
constraints exists and is free of anomalies.
\begin{itemize}
\item[1)] {\it Hamiltonian and Diffeomorphism Constraint}\\
Recall that the infinitesimal generator of diffeomorphisms is
ill-defined so that we must check the commutator algebra in terms of
finite diffeomorphisms. The classical infinitesimal
relation $\{\vec{H}(u),H(N)\}=H(u[N])$ can be exponentiated
and gives
$$
e^{t{\cal L}_{\chi_{\vec{H}(u)}}}\cdot H(N)=H((\varphi^u_t)^\ast N)
$$
where $\chi_{\vec{H}(u)}$ denotes the Hamiltonian vector field
of $\vec{H}(u)$ on the classical continuum phase space $\cal M$
and $\varphi^u_t$ the one parameter family of diffeomorphisms
generated by the integral curves of the vector field $u$.
It tells us that $H(x)$ is a scalar density of weight one.
Therefore we expect to have in quantum theory the relation
\be \label{6.39}
\hat{U}(\varphi)\hat{H}(N)\hat{U}(\varphi)^{-1}=\hat{H}(\varphi^\ast N),
\ee
To check whether (\ref{6.39}) is satisfied, we notice that for a spin network
function $f_\gamma$ we have by the definition of the action
of the diffeomorphism group $\hat{U}(\varphi)$ on ${\cal H}^0$
on the one hand
\ba \label{6.40}
\hat{U}(\varphi)\hat{H}(N)f_\gamma
&=&\hat{U}(\varphi)
\sum_{v\in V(\gamma)} N(v) \hat{H}_{v,a(\gamma)} f_\gamma
\nonumber\\
&=&\sum_{v\in V(\gamma)} N(v)
\hat{H}_{\varphi^{-1}(v),\varphi^{-1}(a(\gamma))} f_{\varphi^{-1}(\gamma)}
\nonumber\\
&=&\sum_{v\in V(\gamma)} (\varphi^\ast N)(\varphi^{-1}(v))
\hat{H}_{\varphi^{-1}(v),\varphi^{-1}(a(\gamma))} f_{\varphi^{-1}(\gamma)}
\nonumber\\
&=&
[\hat{U}(\varphi)\hat{H}(N)\hat{U}(\varphi)^{-1}]
\hat{U}(\varphi)f_\gamma
\nonumber\\ &=&
[\hat{U}(\varphi)\hat{H}(N)\hat{U}(\varphi)^{-1}]
f_{\varphi^{-1}(\gamma)}
\ea
and on the other hand
\ba \label{6.41}
\hat{H}(\varphi^\ast N)f_{\varphi^{-1}(\gamma)}
&=&\sum_{v\in V(\varphi^{-1}(\gamma))} (\varphi^\ast N)(v)
\hat{H}_{v,a(\varphi^{-1}(\gamma))} f_{\varphi^{-1}(\gamma)}
\nonumber\\
&=&\sum_{v\in V(\gamma)} (\varphi^\ast N)(\varphi^{-1}(v))
\hat{H}_{\varphi^{-1}(v),a(\varphi^{-1}(\gamma))} f_{\varphi^{-1}(\gamma)}
\ea
Here $\hat{H}_{v,a(\gamma)}$ is the operator coefficient of $N(v)$
in (\ref{6.32}), (\ref{6.33}) which depends on the graph $a(\gamma)$
assigned to $\gamma$ through the choice function $a$, that is,
the segments $s_{\gamma,v,e}$ and arcs $a_{\gamma,v,e,e'}$.
Comparing (\ref{6.40}) and (\ref{6.41}) we get equality provided
that
\be \label{6.42}
\varphi\circ a=a\circ\varphi \;\forall \varphi\in \mbox{Diff}^\omega(\sigma)
\ee
This seems to burden us with the proof that such a choice function
really exists and in fact we do not have a proof although it would be very
nice to have one since it would decrease the possible number of choice
functions.
However, we can avoid
this by the observation that our choice function was constructed in such
a way that the assignments $a(\gamma)$ and $a(\varphi(\gamma))$ are
analytically diffeomorphic. In other words we always find an
analytical diffeomorphism $\varphi'_{\varphi^{-1}(\gamma)}$
which preserves $\varphi^{-1}(\gamma)$ such that
\be \label{6.43}
[\hat{U}(\varphi)\hat{H}(N)\hat{U}(\varphi)^{-1}]f_{\varphi^{-1}(\gamma)}
[\hat{U}(\varphi'_{\varphi^{-1}(\gamma)})
\hat{H}(\varphi^\ast
N)\hat{U}(\varphi'_{\varphi^{-1}(\gamma)})^{-1}]f_{\varphi^{-1}(\gamma)}
\ee
for any $\gamma$ and any $f_{\varphi^{-1}(\gamma)}$. Thus, while
(\ref{6.39}) is violated, it is violated in an allowed way because
the ``anomaly" is a constraint operator again. Put differently,
the ``anomaly" is not seen in the URST so that (\ref{6.39}) is an
{\it exact operator identity} in the URST.

In that sense then, $\hat{H}(N)$ is a diffeomorphism
covariant, densely defined, closed  operator on ${\cal H}^0$.
\item[2)] {\it Hamiltonian and Hamiltonian Constraint}\\
There are three
important properties of the operator $\hat{H}(N)$ that follow from our
class of choice functions (properties a)-e)) :\\
A) First of all,
we observe that $\hat{H}(N)$ has dense domain and range consisting of
smooth (in the sense of $\cal D$) cylindrical functions. Therefore it makes
sense to multiply operators and in particular to compute commutators.\\
B) Secondly, it annihilates planar vertices.\\
C) Thirdly, {\it for no other choice of triangulation proposed so
far other than the one we proposed in
\cite{V} and only when using the volume operator of \cite{64} rather than
the one of \cite{61}} is it true that in fact any finite product of operators
$\hat{H}^{\epsilon_1}(N_1)..\hat{H}^{\epsilon_n}(N_n)$ is independent of
the parameters $\epsilon_1,..,\epsilon_n$ in the URST.

The second and third properties do not hold for a more general class of
operators
considered in the papers \cite{87,88} so that there is no convergence
in the URST not even of the operators themselves, not to speak of
their commutators. Since certainly none (of the duals) of these
operators leaves the space ${\cal D}^\ast_{Diff}$ invariant, in order
to compute commutators these authors suggest
to introduce the larger, unphysical  space ${\cal D}^\ast_\star$ already
mentioned on which one can compute limits
$\hat{H}'(N)=\lim_{\epsilon\to 0}(\hat{H}^\epsilon(N))'$ {\it pointwise}
in ${\cal D}^\ast_\star\times {\cal D}$ of their duals
and products of these limits.

Let again $f_\gamma$ be a spin-network function over some graph $\gamma$.
Then we compute
\ba \label{6.44}
&&[\hat{H}(N),\hat{H}(N')]f_\gamma
=
\sum_{v\in V(\gamma)}
[N'(v)\hat{H}(N)-N(v)\hat{H}(N')]\hat{H}_{a(\gamma)_{|v}}f_\gamma \\
&=&
\sum_{v\in V(\gamma)}\sum_{v'\in V(\gamma)\cup a(\gamma)_{|v})}
[N'(v)N(v')-N(v)N'(v')]
\hat{H}_{a(\gamma\cup a(\gamma)_{|v})_{|v'}}\hat{H}_{a(\gamma)_{|v}}f_\gamma
\nonumber
\ea
where for clarity we have written $\hat{H}_{a(\gamma)_{|v}}\equiv
\hat{H}_{v,a(\gamma)}$ in order to indicate that
$\hat{H}_{v,a(\gamma)}$ does not depend on all of $a(\gamma)$ but only
on its restriction to the arcs and segments around $v$.
We are abusing somewhat the notation in the second step because
one should really expand $\hat{H}_{a(\gamma)_{|v}}f_\gamma$ into spin network
functions over $\gamma\cup a(\gamma)_{|v}$ and then apply the second operator
to that expansion into spin-network functions. In particular,
$\hat{H}_{a(\gamma)_{|v}}f_\gamma$
it really is a finite linear combination of terms where each of them depends
only on $\gamma\cup a(\gamma)_{|v,e,e'}$ for some edges $e,e'$
incident at $v$ and each of those should be expanded into spin-network
functions. We will not write this
explicitly because it is just a book keeping exercise
and does not change anything
in the final argument. So either one writes out all the details or one just
assumes for the sake of the argument that
$\hat{H}_{a(\gamma)_{|v}}f_\gamma$ is a spin network function over
$\gamma\cup a(\gamma)_{|v}$. Everything we say is more or less obvious
for the Euclidean Hamiltonian constraint but a careful analysis shows that
it extends to the Lorentzian one as well.

Let us now analyze (\ref{6.44}).
The right hand side surely vanishes for $v'=v$. We notice that any
vertex $v'\in V(\gamma\cup a(\gamma)_{|v})-V(\gamma)$ is planar and
since $\hat{H}_{v',a(\gamma\cup a(\gamma))}$ has an operator
of the form $[h_s^{-1},\hat{V}]$ to the outmost right hand side
where $s$ is a segment, incident at $v'$, of an edge incident at
$v'$, it follows that none of these vertices contributes. Here
it was again crucial that we used the operator \cite{64} rather
than the operator \cite{61} ! Thus (\ref{6.44}) reduces to
\ba \label{6.45}
[\hat{H}(N),\hat{H}(N')]f_\gamma
&=&
\sum_{v\not= v'\in V(\gamma)} [N'(v)N(v')-N(v)N'(v')]
\hat{H}_{a(\gamma\cup a(\gamma)_{|v}))_{|v'}}\hat{H}_{a(\gamma)_{|v}}f_\gamma
\nonumber\\ &=&
\frac{1}{2}\sum_{v\not= v'\in V(\gamma)} [N'(v)N(v')-N(v)N'(v')]
\times \nonumber\\
&& \times
[\hat{H}_{a(\gamma\cup a(\gamma)_{|v}))_{|v'}}\hat{H}_{a(\gamma)_{|v}}
-\hat{H}_{a(\gamma\cup a(\gamma)_{|v'}))_{|v}}\hat{H}_{a(\gamma)_{|v'}}]f_\gamma
\ea
where in the second step we used the antisymmetry of the expression
$[N'(v)N(v')-N(v)N'(v')]$ in $v,v'$. Now the crucial point is that
for $v\not=v'\in V(\gamma)$ the prescription of how to attach the arcs first
around $v$ and then around $v'$ as compared to the opposite may not be the same
because our prescription depends explicitly on the graph to which we apply
it, however, they are certainly analytically diffeomorphic.

Thus, there exist analytical diffeomorphisms $\varphi_{\gamma,v,v'}$
preserving $\gamma\cup a(\gamma)_{|v}$ such that
\be \label{6.46}
\hat{H}_{a(\gamma\cup a(\gamma)_{|v}))_{|v'}}\hat{H}_{a(\gamma)_{|v}}f_\gamma
=\hat{U}(\varphi_{\gamma,v,v'})
\hat{H}_{a(\gamma)_{|v'}}\hat{H}_{a(\gamma)_{|v}}f_\gamma
\ee
for any $v\not=v'\in V(\gamma)$. It follows that
\ba \label{6.47}
[\hat{H}(N),\hat{H}(N')]f_\gamma
&=&
\frac{1}{2}\sum_{v\not= v'\in V(\gamma)} [N'(v)N(v')-N(v)N'(v')]
\times \nonumber\\
&& \times [\hat{U}(\varphi_{\gamma,v,v'})-\hat{U}(\varphi_{\gamma,v',v})]
\hat{H}_{a(\gamma)_{|v'}}\hat{H}_{a(\gamma)_{|v}}f_\gamma
\ea
where we have used $[\hat{H}_{a(\gamma)_{|v'}},\hat{H}_{a(\gamma)_{|v}}]=0$
for $v\not=v'$ since the derivative operators involved act on disjoint
sets of edges.

Expression (\ref{6.47}) is to be compared with the classical formula
$\{H(N),H(N'\}=\vec{H}(q^{-1}[(dN) N'-(dN') N])$. The fact that we get a
difference between finite diffeomorphism constraint operators
looks promising at first because for next neighbour vertices $v,v'$
this could be interpreted as a substitute for the operator $\hat{H}_a$
which somehow had to be written in terms of finite diffeomorphism anyway
because we know that the infinitesimal generator dos not exist. Unfortunately
there are also contributions from pairs $v,v'$ which are far apart.
This we could avoid by specifying the choice function more closely in the
sense that the arcs $a_{\gamma,v,e, e'}$ should, for a given vertex $v$, not depend
on all of $\gamma$ but only on $\gamma_v\subset \gamma$, the subset of $\gamma$
consisting of all edges incident at $v$. But still (\ref{6.47}) does
not, at least not obviously,
resemble the classical calculation too closely because there it is crucial
that $\{H(x),H(x')\}\not=0$ as $x\to x'$
while $[\hat{H}_{a(\gamma)_{|v'}},\hat{H}_{a(\gamma)_{|v}}]=0$
for any $v\not=v'$.

Certainly then for $\Psi\in {\cal D}^\ast_{Diff}, f\in {\cal D}$
we have in the URST
\be \label{6.48}
\Psi([\hat{H}(N),\hat{H}(N')]f):=\lim_{\epsilon\to 0}\lim_{\epsilon'\to 0}
\Psi([\hat{H}^{\epsilon}(N),\hat{H}^{\epsilon'}(N')]f) =0
\ee
where the limit is again {\it uniform} in both $\Psi,f$. But this would not be
surprising even if the right hand side would be a manifset quantization
of the right and side (with $\vec{H}$ replaced by
$\hat{U}(\varphi)-1_{{\cal H}^0}$ and ordered to the outmost left).
In other words, in the URST we do not see the difference between any
operators which are, like (\ref{6.47}), of the form of a difference between
two finite diffeomorphism operators ordered to the outmost left times
any other operators. Here it proves useful to take the effective operator
point of view which in fact {\it can} detect those differences.
Again it requires more work, that is, semiclassical analysis,
in order to decide whether the classical limit
of the right hand side of (\ref{6.47}) has anything to do with
$\vec{H}(q^{-1}([(dN) N'-(dN') N])$. It is worthwhile, however,
to point out that (\ref{6.47}) proves the {\it absence of a
strong anomaly}. In other words, if (\ref{6.48}) would not hold,
then the quantization that we have proposed would be {\it mathematically
inconsistent}. What is possible though is that (\ref{6.47}) could represent
a {\it weak anomaly} in the sense that the quantum dynamics that $\hat{H}(N)$
generates is {\it physically inconsistent}, that is, the classical dynamics
is not reproduced in the classical limit. This is precisely what has to be
analyzed in the future and, if true, to find out how to cure the problem.
\end{itemize}
To summarize: The constraint algebra of the Hamiltonian constraints among
each other is mathematically consistent but possibly has a physical anomaly.

Three remarks are in order:\\
i)\\
In \cite{87,88} the authors prove a statement similar to
(\ref{6.48}) on their space ${\cal D}^\ast_\star$. The algebra of their
dual constraint operators becomes Abelean for a large
class of operators which even classically do not need to be proportional
to a diffeomorphism constraint. They then argue that the quantization
method proposed here cannot be correct because it either implies a
physical anomaly or, even worse, that the (dual of the) quantum metric
operator $\hat{q}^{ab}$ vanishes identically.\\
\\
We disagree with this conclusion for two reasons:\\
1)\\
Their limit dual operators are defined by
\be \label{6.49}
[\hat{H}'(N)\Psi](f):=\lim_{\epsilon\to 0} \Psi(\hat{H}^\epsilon(N)f)
\ee
where convergence is only pointwise, that is, for any
$\delta>0,\;\Psi\in {\cal D}^\ast_\star,f\in {\cal D}$
there exists $\epsilon(\delta,\Psi,f)$ such that
\be \label{6.49a}
|[\hat{H}'(N)\Psi](f)-\Psi(\hat{H}^\epsilon(N)f)|<\delta
\ee
for any $\epsilon<\epsilon(\delta,\Psi,f)$. Thus, while they have
blown up ${\cal D}^\ast_{Diff}$ to ${\cal D}^\ast_\star$, their
convergence is weaker when restricted to ${\cal D}^\ast_{Diff}$
so that it is not easy to compare the two operator topologies (notice that
we can also define a dual operator vial (\ref{6.49}) restricted
to ${\cal D}^\ast_{Diff}$ considered as a subspace of ${\cal D}^\ast$,
this subspace is just not left invariant so that we cannot compute
commutators of duals). However it is clear that the subspace
${\cal D}^\ast_\star$ is a sufficiently small extension of
${\cal D}^\ast_{Diff}$ in order to make sure that a much wider class
of operators converges in their topology than the class that we have
mind for our topology since our topology roughly  requires that
$\Psi(\hat{H}^\epsilon(N)f)$ is already independent of $\epsilon$
while their topology only requires that the $\epsilon$ dependence
rests in the smearing functions $N$ which are required to be smooth
at vertices.\\
\\
Therefore our first conclusion is that it is not surprising
that in their topology more operators converge.\\
\\
Next, let us turn to commutators. In our topology, what is required
is that the expression $\Psi([\hat{H}^\epsilon(N),\hat{H}^{\epsilon'}(N')]f)$
{\it just equals zero independently of how large the graph is on which $f$
depends} because we have identified $\hat{H}^\epsilon(N)$ with the
continuum operator. In their topology what happens is that unless
the operator $\hat{H}(N)$ has also the properties B), C) besides
A) then one gets for the commutator an expression of the form
(\ref{6.45}) on which one acts with an element
$\Psi\in {\cal D}^\ast_\star$ the result of which is that one gets
\ba \label{6.50}
([\hat{H}(N')',\hat{H}(N)']\Psi)(f_\gamma)
&=& \lim_{\epsilon\to 0}\lim_{\epsilon'\to 0}\sum_{v\in V(\gamma)}
\sum_{v'\in V(\gamma\cup a^\epsilon(\gamma)_{|v})-V(\gamma)}
[N'(v)N(v')-N(v)N'(v')] \times\nonumber\\
&&\times
\Psi[\hat{H}_{a^{\epsilon'}(\gamma\cup a^\epsilon(\gamma)_{|v}))_{|v'}}
\hat{H}_{a^\epsilon(\gamma)_{|v}}f_\gamma]
\ea
For the same reason as for $\Psi\in {\cal D}^\ast_{Diff}$ each evaluation
of $\Psi$ that appears on the right hand side is already independent
of $\epsilon,\epsilon'$ for any $\Psi\in{\cal D}^\ast_\star$ by definition
of that space. Therefore the only $\epsilon,\epsilon'$ dependence rests
in the function $[N'(v)N(v')-N(v)N'(v')]$. Now, while each of the
roughly $|V(\gamma)|$ $\Psi-$evaluations is nonvanishing, since
we take the limit pointwise and the $N,N'$ are smooth, the limit vanishes.
If we would not have taken pointwise convergence, then for each finite
$\epsilon,\epsilon'$ we can find $f_\gamma,\Psi$ such that the right hand side
of (\ref{6.50}) takes an arbitrarily large value. The reason for why this
happens is that since one of the conditions B), C) does not hold, now
the vertices $V(\gamma\cup a^\epsilon(\gamma)_{|v})-V(\gamma)$
in fact do contribute.

We conclude that their topology is too weak in order to detect even a
mathematical anomaly, not to mention a physical anomaly, and suffices
even less to select the physically relevant operators. The details of this
calculation will appear in \cite{89}.\\
2)\\
Finally, coming to their second conclusion, we will explicitly display
in the next subsection a quantization of $\vec{H}(q^{-1}[(dN) N'-(dN') N])$.
Now in their topology, the dual of that operator again
annihilates ${\cal D}^\ast_\star$ but this is again only because
one takes only pointwise rather than uniform limits. If one tests
this operator on a finite graph then, again because there are finitely
many contributions each of which evidently proportional to a term
of the form $[N'(v)N(v')-N(v)N'(v')]$, the limit must vanish pointwise,
however, uniformly it blows up. In particular, this does not show that
$\hat{q}^{ab}$ is the zero operator.\\
ii)\\
In \cite{90} we find the that claim the action of the Hamiltonian constraint
is too local in order to allow for interesting critical points
in the renormalization flow of the theory
and that therefore the Hamiltonian constraint must be changed drastically
if possible at all.

Four comments are appropriate:\\
First of all the claim
is not even technically true, how non-local the operator $\hat{H}(N)$
is depends on our choice function $a$ which builds a new graph
around any vertex of a given graph $\gamma$ and the details of that new
graph around $v$ may depend on an {\it arbitrarily large neighbourhood
of $v$} (where a neighbourhood of degree $n$ can be background independently
defined as the set of edges that one can trace within $\gamma$
if one performs a closed loop with endpoints $v$ using at most $n$ edges).
Secondly, as we have said right at the beginning: We are here just
exploring the first naive definition of a Hamiltonian constraint,
not even the author of \cite{V} believes that the operator proposed
gives the final answer. Thirdly, it is unclear what role a renormalization
group should play in a diffeomorphism invariant theory, after all
renomalization group analysis has much to do with scale transformations
(integrating out momentum degrees of freedom above a certain scale)
which are difficult to deal with in absence of a background metric.
Finally, suppose that we would manage to write down a
physically correct Hamiltonian
operator of the type of $\hat{H}(N)$. We could order it symmetrically
and presumably find a self-adjoint extension. It would then be possible
to diagonalize it and in the associated ``eigenbasis" the operator
would act in an ultralocal way ! Thus any non-local operator can be
made ultralocal in an appropriate basis. A good example is given by the
Laplace
operator in $\Rl^n$ which is non-local in position space but ultralocal
in momentum space. Of course the momentum eigenfunctions are not eigenfunctions
but rather distributions and we must take an uncountably infinite linear
combination of them (rather, an integral against a sufficiently nice
function, that is, a Fourier transform)
in order to obtain an $L_2$ function on which the Laplacian looks
rather non-local. Thus, non-locality is hidden in infinite linear
combinations  which is the reason for why
we are working with ${\cal D}^\ast$ rather than with ${\cal D}$.\\
iii)\\
The proposal for a Hamiltonian constraint whose dual action is
restricted to distributions based on Vasiliev invariants \cite{91,92}
also has an Abelean dual algebra. Presumably this also will no longer
be the case after strengthening the topology but this must wait
until the space of Vasiliev distributions has been turned into
a Hilbert space.

\subsubsection{The Quantum Dirac Algebra}
\label{s6.3.4}

Recall from section \ref{s6.3.3} that in the URST the commutator of two
Hamiltonian constraints vanishes : The non-zero
operator on ${\cal H}^0$ given by $[\hat{H}(N),\hat{H}(N')]$,
is indistinguishable from the zero-operator in the URST.
We would like to know whether there exists an operator
corresponding to $\vec{H}(q^{-1}[(dN) N'-(dN') N])$
and if it is also indistinguishable from the zero operator
in the URST. If that would be true, then we could equate the
two operators in the URST. Notice that this is still not
satisfactory because one cannot test the correctness of the
algebraic form of an operator on its kernel, but it is still
an important consistency check whether an operator corresponding
to $\vec{H}(q^{-1}[(dN) N'-(dN') N])$ exists at all. More explicitly,
we wish to study whether we can quantize
\be \label{6.56}
O(N,N'):=\int d^3x (N N'_{,a}-N_{,a} N')q^{ab} V_b
\ee
In \cite{47m1} we answer this question affirmatively, that is, we manage
to quantize a regulated operator $\hat{O}^\epsilon(N,N')$ corresponding to
(\ref{6.56}) and
prove that it converges in the URST to an operator $\hat{O}(N,N')$.
We will not derive the operator but merely give its final expression.
However, let us point out once more that while $H_a$ and $q^{ab}$ are
known not to have well-defined quantizations because the infinitesimal
generator of diffeomorphisms does not exist (section \ref{s5.3}) and since
$q^{ab}$ has the wrong density weight (section \ref{s6.3.1}) the
combination $\omega_a q^{ab}V_b$ is a scalar density of weight one and
therefore has a chance to result in a well-defined operator for any
co-vector field $\omega_a$ such as
$\omega_a=N N'_{,a}-N_{,a} N'$.\\

Let $\gamma$ be a graph, $V(\gamma)$ its set of vertices, $v\in V(\gamma)$ a
vertex of $\gamma$, introduce
the triangulation $T(\gamma)$ of section \ref{s6.3} adapted to
$\gamma$, let $\Delta$ be a tetrahedron of that triangulation such
that $v(\Delta)=v$, let $\chi_{\epsilon,v}(x)$ be the
smoothened out characteristic
function of the neighbourhood $U(v)$ (using, for instance, a
partition of unity) and finally let
$s_I(\Delta)$ be the endpoint of the edge $e_I(\Delta)$ of $\Delta$
incident at $v$. We define a vector field on $\sigma$ of compact support
by
\be \label{6.58}
\xi^a_{\epsilon,v,\Delta,I}(x):=\chi_{U(v)}(x)
\frac{s_I^a(\Delta)-v^a}{\epsilon}
\ee
where $\epsilon^3$ is the coordinate volume of $U(v)$
and for any vector field $\xi$ on $\sigma$ let $\varphi^\xi_t$ be the
one-parameter group of diffeomorphisms that it generates. Let us also
introduce the short-hand notation $\hat{V}(v):=
\hat{V}(U(v))$. It was shown in \cite{47m1} that
there is a classical object $O_\gamma(N,N')$ which uses the
triangulation $T(\gamma)$ and whose limit, as $\gamma\to \sigma$,
in the topology of the phase space coincides with (\ref{6.56}).
The quantizations of these objects define
densely defined operators $\hat{O}(N,N')$ with consistent
cylindrical projections $\hat{O}_\gamma(N,N')$ given by their
action on functions $f_\gamma$ cylindrical over a graph $\gamma$. The
explicit form of these projections is given by
\ba \label{6.59}
\hat{O}(N,N')f_\gamma &=&
-i\frac{16\epsilon_{ijk}\epsilon_{ilm}}{\hbar\ell_p^2}
\sum_{v\in V(\gamma)}\;\sum_{v(\Delta)=v(\Delta')=v}
[\hat{U}(\varphi_\epsilon^{\xi_{\epsilon,v,\Delta',R}}-\mbox{id}_{{\cal H}}]
\times\\
&\times& \epsilon^{RST}\epsilon^{NPQ}
[N(v) N'(s_N(\Delta))-N(s_N(\Delta)) N'(v)]\times\nonumber\\
&\times& \mbox{tr}(\tau_j h_{e_P(\Delta)}[h_{e_P(\Delta)}^{-1},
\sqrt{\hat{V}(v)}])
\mbox{tr}(\tau_k h_{e_Q(\Delta)}[h_{e_Q(\Delta)}^{-1},
\sqrt{\hat{V}(v)}])
\times\nonumber\\
&\times& \mbox{tr}(\tau_l h_{e_S(\Delta')}[h_{e_S(\Delta')}^{-1},
\sqrt{\hat{V}(v)}])
\mbox{tr}(\tau_m h_{e_T(\Delta')}[h_{e_T(\Delta')}^{-1},
\sqrt{\hat{V}(v)}])f_\gamma
\nonumber
\ea
Basically, what happened in the quantization step was that one had to
introduce a point splitting which is why one has a double sum over
tetrahedra and again factors of $1/\sqrt{\det(q)}$ got absorbed into
Poisson brackets which then were replaced by commutators. Notice that
in (\ref{6.59}) the square root of the volume operator appears.\\
The fact that the combination
$[\hat{U}(\varphi_\epsilon^{\xi_{\epsilon,v,\Delta',R}}-
\mbox{id}_{{\cal H}}]$ stands {\it to the left} shows that
$\Psi(\hat{O}(M,N')f)=0$ uniformly
in $\Psi\in {\cal D}^\ast_{Diff}$ and $f\in {\cal D}$ for any $N,N'$.

\subsection{The Kernel of the Wheeler-DeWitt Constraint Operator}
\label{s6.4}

In \cite{VI} it was investigated to what extent one can
solve the {\it Quantum Einstein Equations} for $\Psi\in{\cal D}^\ast_{Diff}$
\be \label{6.50a}
\Psi(\hat{H}(N)f)=0
\ee
for all $N\in C^\infty(\sigma),f\in{\cal D}$. This section is devoted to an
outline of an explicit construction of the complete and rigorous kernel of
the proposed operator $\hat{H}(N)$. While $\hat{H}(N)$ will certainly have
to be modified in the future, hopefully the methods that we display here
will prove useful for other candidates of $\hat{H}(N)$.
Notice that these solutions were really the first honest
solutions to the Wheeler-DeWitt constraint in full four dimensional
quantum general relativity in terms of connections that have appeared in the
literature because the result of calculations performed in \cite{42,43,44}
was of the type zero times infinity. Also, they were the first ones that
have non-zero volume and which do not need non-zero cosmological
constant.

We first want to give an intuitive
picture of the way that the Hamiltonian constraint acts on cylindrical
functions.
When looking at (\ref{6.32}) and (\ref{6.33}) one realizes the following :\\
The Euclidean Hamiltonian Constraint operator, when acting on, say, a
spin-network state $T$ over a graph $\gamma$, looks at each non-planar
vertex
$v$ of $\gamma$ and for each such vertex considers each triple of
distinct edges
$e,e',\tilde{e}$ incident at it. For each such triple, the constraint
operator contains three terms labelled by the three possible pairs of
edges that one can form from $\{e,e',\tilde{e}\}$. Let us look at one of
them, say (neglecting numerical factors)
\be \label{6.51}
\mbox{tr}([h_{\alpha(v;e,e')}-h_{\alpha(v;e,e')^{-1}}]h_{\tilde{s}}
[h_{\tilde{s}}^{-1},\hat{V}(U(v))]) T\;.
\ee
The notation is as follows : $s,s',\tilde{s}$ are the segments of
$e,e',\tilde{e}$
incident at $v$ that end in the endpoints of the three arcs
$a(v;,e,e')$ etc., $\alpha(v;e,e')$ is the loop
$s\circ a(v;e,e')\circ (s')^{-1}$ and $U(v)$ is any system
of mutually disjoint neighbourhoods, one for ech vertex $v$.
For notational simplicity we have dropped the graph label.
Let $j,j',\tilde{j}$ be the spins of the edges $e,e',\tilde{e}$ in
$T$. First of all it is easy to see that the piece
$h_{\tilde{s}} [h_{\tilde{s}}^{-1},\hat{V}(U_{\epsilon_0}(v))] $
is invariant under a gauge transformation at the endpoint $\tilde{p}$
of $\tilde{s}$.
Therefore the state (\ref{6.51}) is also invariant at $\tilde{p}$ and since
$\tilde{p}$ is a two-valent vertex this is only possible if the segments
$\tilde{s}$ and $\tilde{e}-\tilde{s}$ of $\tilde{e}$ carry the same spin
in the decomposition of (\ref{6.51}) into spin-network states $T'$. But
since $\tilde{e}-\tilde{s}$ carries still spin $\tilde{j}$ (no holonomy
along $\tilde{e}-\tilde{s}$ appears in (\ref{6.51})) we conclude that
the spin of $\tilde{e}$ is unchanged in $T'$ as compared to $T$.

However, the same is not true for $e,e'$ : The piece
$[h_{\alpha(v;e,e')}-h_{\alpha(v;e,e')^{-1}}]$ is a multiplication operator
and raises the spin of $a(v;,e,e')$ from zero to $1/2$ and
(\ref{6.51}) decomposes into, in general, four spin-network states $T'$
where the spins of the segments $s,s'$ are raised or lowered in units of
$1/2$ as compared to $T$, that is, they are $j\pm 1/2,j'\pm 1/2$
respectively while
the spins of the segments $e-s,e'-s'$ remain unchanged, namely $j,j'$.
All this follows from basic Clebsh-Gordan decomposition theory for $SU(2)$.

Next we look at the remaining piece $\hat{H}(N)+\hat{H}_E(N)$ of the
Lorentzian Hamiltonian constraint. Its most important ingredient are the
two factors of the
operator $\hat{K}$ which, up to a numerical factor, equals
$[\hat{V}(\sigma),\hat{H}_E(1)]$. Now as shown in \cite{V}, when
inserting this operator into (\ref{6.33}) what survives in the term
corresponding to the vertex $v$ of the graph is just
$[\hat{V}(U(v)),\hat{H}_E(U(v))]$.
Thus, since the volume operator does not change any spins, the
spin-changing ingredient of the action of the remaining piece of
$\hat{H}(N)$ at $v$ are two successive actions of
$\hat{H}_E(U(v))$ as just outlined. \\

In summary, the Hamiltonian constraint operator has an action similar
to a fourth order polynomial consisting of creation of annihilation
operators. What is being created or annihilated are the spins of edges of
a graph (notice that an edge with spin zero is the same as no edge at all).

Let us now look at this action in more detail. We will restrict attention
only to the Euclidean piece, for the more complicated full action see
\cite{VI}.\\
Notice that the Euclidean constraint operator creates edges of a special
kind, called {\it extraordinary edges}, namely the arcs
$a=a(v;e,e')$. What is special about them is that they end in
planar vertices which are either bi- or tri-valent. If they are tri-valent
then, moreover, the vertex is the intersection of the two analytical edges
$a,e$ where $a$ just ends on an interiour point of $e$. Moreover, let
$e,e'$ be the edges on which $a$ ends. Then the analytical extensions of
$e,e'$ end in at least one point and the two possible earliest of their
intersection points away from $a\cap e,a\cap e'$ are, together with these
analytical extensions, non-planar vertices of $\gamma$. However,
not only are these edges special, also the spin they carry is special, namely
the arc $a$ carries always spin $1/2$. We will continue to call this whole
set of extraordinary structures an extraordinary edge.

The special nature
of these edges allows to classify the full set of labels $\cal S$
of spin-network states, called {\it spin-nets}, as follows. Denote by
${\cal S}_0\subset{\cal S}$, called {\it sources}, the set of spin-nets,
corresponding to graphs with no extraordinary edges at all.

From these sources
one constructs iteratively derived sets ${\cal S}_n(s_0),\;n=0,1,2,..$ for
each source $s_0\in{\cal N}_0$, called spin-nets of level $n$ based on $s_0$.
Put ${\cal S}_0(s_0):=\{s_0\}$ and define ${\cal S}_{n+1}(s_0)$ as
follows : Take each $s\in {\cal S}_n(s_0)$, compute $\hat{H}_E(N) T_s$
for all possible lapse functions $N$, decompose it into spin-network
states and enter the appearing spin-nets into the set ${\cal S}_{n+1}(s_0)$.

In \cite{VI} it is shown that the sets
${\cal S}_n(s_0),{\cal S}_{n'}(s_0')$ are disjoint unless $s_0=s_0'$ and
$n=n'$. It is easy to see that the complement of the set of
sources $\overline{{\cal S}_0}={\cal S}-{\cal S}_0$ coincides with the set
of derived spin-nets of level greater than zero. Moreover, for each
$s\in {\cal N}$ there is a unique integer $n$ and a unique source
$s_0$ such that $s\in {\cal S}_n(s_0)$.

The purpose for doing all this is, of course, that this classification
leads to a simple construction of all rigorous solutions of the
Euclidean Hamiltonian constraint based on the observation that
\be \label{6.52}
\hat{H}_E(N)\cdot\mbox{span}\{T_s\}_{s\in{\cal S}_n(s_0)}
\subset \mbox{span}\{T_s\}_{s\in{\cal S}_{n+1}(s_0)}  \;.
\ee
Since a solution $\Psi$ of (\ref{6.50a}) is a diffeomorphism invariant
distribution in ${\cal D}^\ast_{Diff}$ we define first
$[{\cal S}_n(s_0)]:=\{[s]\}_{s\in {\cal S}_n(s_0)}$ where
$[s]$ is the label for the diffeomorphism invariant distribution
$T_{[s]}$ (recall section \ref{s5.3}).
We can now make an ansatz for a basic solution of the form
\be \label{6.53}
\Psi:=\Psi_{[s_0],\vec{n}}:=\sum_{k=1}^N \;
\sum_{[s]\in [{\cal S}_{n_k}(s_0)]} c_{[s]} T_{[s]}
\ee
with complex coefficients $c_{[s]}$ which are to be determined from the
Quantum Einstein Equations (\ref{6.50a}). Now
from (\ref{6.53}) it is clear that
$\Psi_{[s_0],[\vec{n}]}(\hat{H}_E(N) T_s)$ can be non-vanishing
if and only if $[s]\in [{\cal S}_{n_k-1}(s_0)]$ for some $k=1,..,n$,
say $k=l$. Choose a representant $s\in [s]$ and let $\gamma$ be the graph
underlying $s$ and $V(\gamma)$ its set of vertices. We then
find, writing $\hat{H}_E(N)=\sum_{v\in V(\gamma)} N(v)\hat{H}_E(v)$,
that
\be \label{6.54}
\Psi_{[s_0],\vec{n}}(\hat{H}_E(N)T_s)=
\sum_{[s']\in [{\cal S}_{n_l}(s_0)]} c_{[s']}\sum_{v\in V(\gamma)}
N(v) T_{[s']}(\hat{H}_E(v) T_s)
\ee
should vanish for any choice of lapse function $N(v)$. Since $N(v)$ can
be any smooth function we find the condition that
\be \label{6.55}
\Psi_{[s_0],\vec{n}}(\hat{H}_E(N)T_s)=
\sum_{[s']\in [{\cal S}_{n_l}(s_0)]} c_{[s']} T_{[s']}(\hat{H}_E(v) T_s)=0
\ee
should vanish for each choice of the finite number of vertices
$v\in V(\gamma)$ and for each of the finite number of spin-nets $s\in
{\cal S}_{n_l-1}(s_0)$. This follows from the fact that the numbers
$T_{[s']}(\hat{H}_E(v) T_s)$ are diffeomorphism invariant and therefore
do not actually depend on $v$ itself but only on the diffeomorphism
invariant information that is contained in the graph $\gamma$ together
with the vertex $v$ singled out.

Therefore, (\ref{6.55}) is a finite system of linear equations for the
coefficients $c_{[s']}$. As the cardinality of the sets ${\cal S}_n(s_0)$
exponentially grows with $n$ this system is far from being overdetermined
and we arrive at an infinite number of solutions. The most general
solution will be a linear combination of the elementary solutions
(\ref{6.55}). Qualitatively the same result holds for the Lorentzian
constraint \cite{VI}, however, it is more complicated because
coefficients from different levels get coupled and so one gets solutions
labelled also by the highest level that was used (possibly one has to
allow all levels, that is, the highest level is always infinity).
Nevertheless it is remarkable
how the solution of the Quantum Einstein Equations is reduced to an exercise
in finite-dimensional linear algebra (although the computation of the
coefficients $T_{[s']}(\hat{H}_E(v) T_s)$ is far from easy, see, e.g.,
\cite{93} which, although the authors restrict to tri-valent graphs and
$\hat{H}_E(N)$ only, is already rather involved). On the other hand, it is
expected that physically interesting solutions will actually be infinite
linear combinations of coupled solutions, that is, solutions of infinite
level, an intuition coming from \cite{VIII}.

Notice that the solutions (\ref{6.55}) are bona fide elements of
${\cal D}^\ast_{Diff}$ and therefore give, for the first time, rigorously defined
solutions to the diffeomorphism and the Hamiltonian constraint of full,
four-dimensional Lorentzian Quantum General Relativity in the continuum,
subject to the reservation that we still have to prove that the classical
limit of this theory in fact is general relativity. One should
now organize these solutions into a Hilbert space such that
adjointness and canonical commutation relations of full Dirac observables
are faithfully implemented. Since group averaging does not work
for open algebras, there is no good proposal at this point for how to
do that and is a very important open research problem.

\subsection{Further Related Results}
\label{s6.5}

We list here further results that are directly connected to the issues
that we have touched upon in this section already.

\subsubsection{Generator of the Wick Transform}
\label{s6.5.1}

In principle we could dispense with the rigorous construction of
the Wick transform
since we could work entirely with the operators (\ref{6.32}) and
(\ref{6.33}) rather than with the modified ones described in this
subsection. However, since the availability of a complex connection
representation could be conceptually important in particular
when making contact with the path integral formulation, we will
make a short digression on available first ans\"atze for how
to get there.

Recall that the generator of the Wick transform $\hat{C}$ is given, up to
numerical
factors, by $i[\hat{V}(\sigma),\hat{H}_E(1)]$. One would like to invoke the
spectral theorem in order to define its exponential and it is therefore
motivated to have an at least symmetric operator $\hat{H}_E(1)$. This,
however, is not the case the way $\hat{H}_E(1)$ is defined : take
for example $s\in {\cal S}_0,s'\in{\cal S}_1(s)$ such that
$<T_{s'},\hat{H}_E(1) T_s>\not=0$, but then
$\overline{<T_s,\hat{H}_E(1) T_{s'}>}=0$ by definition of ${\cal S}_0$.
Of course, one can symmetrize the operator by defining the matrix elements
of the symmetric operator to be the half the sum of the matrix elments of
the unsymmetric operator plus the transpose of its complex conjugate.
This operator is also well-defined but in \cite{VI} we did not
succeed in
proving existence of self-adjoint extensions of it. What works is the
following : one {\it marks} the extraordinary edges by taking them to be
smooth but not analytical. This way it becomes possible to tell whether
a given state was obtained by the action of the constraint on a function
cylindrical over a piecewise analytical graph. Then the repeated action
of the constraint adds always the same smooth extraordinary edge
$a(v;e,e')$ to the graph and this turns the operator into a
symmetric one when factor-ordering its expression symmetrically.
The formalism is not disturbed by the fact that one leaves the purely
analytical category of graphs. The so symmetrized operator is
free of mathematical anomalies as well but the structure of its
solutions becomes more
complicated. One can then invoke von Neumann's theorem (that says
that if a densely defined symmetric operator commutes on its domain with a
conjugation operator that preserves its domain then there exist self-adjoint
extensions) to show that $\hat{H}_E(N)$ and in fact also $\hat{C}$ have
self-adjoint extensions. (A conjugation operator
is a bounded, anti-linear operator which squares to the identity). This
method of proof does not work, however, for the constraint $\hat{H}(N)$
because
the Lorentzian operator is a sum consisting of two operators which are
symmetric and have self-adjoint extensions but it is unclear whether they
have extensions to the same domain (the explicit extensions are not even
known although it is likely that all operators in question are essentially
self-adjoint in which case the answer would be given by their closure).

In any case, we could in principle define self-adjoint operators
$\hat{H}_E(N),\hat{C}$
and define the operator $\hat{W}_t:=\exp(-t\hat{C})$ by its
spectral resolution. Then, according to the philosophy of the Wick transform
of section \ref{s6.1} we should analytically continue the operator
$\hat{W}_t^{-1}\hat{H}_E(N)\hat{W}_t$ and define it to be the Lorentzian
Hamiltonian constraint. The problem is that the spectrum of $\hat{C}$
is far from known and it is not even clear, although extremely likely, that
$\hat{H}_E(N)$ and $\hat{C}$ can be extended as self-adjoint operators
to the same domain. A method of proof, as sketched in \cite{VI},
could probably be based on Nelson's analytic vector theorem but the proof
was not completed there.

\subsubsection{Testing the New Regularization Technique by
Models of Quantum Gravity}
\label{s6.5.2}

Presently there are two positive tests for the quantization
procedure that we applied to the Hamiltonian constraint, namely
Euclidean $2+1$ gravity \cite{VIII} and isotropic and homogeneous
BIanchi cosmologies quantized in a non-standard fashion \cite{94}.

The first model is a dimensional reduction of $3+1$ gravity which
one can formulate also as a quantum theory of $SU(2)$ connections and
$su(2)$ electric fluxes with precisely the same algebraic form
of all constraints. Hence, one can introduce the full mathematical
structure of $\ab,\mu_0,{\cal H}^0$ as well as the quantum constraints
$G_j={\cal D}_a E^a_j,\;V_a=F_{ab}^j E^b_j,\;
H_E=F_{ab}^j E^a_k E^b_l\epsilon_{jkl}/\sqrt{\det(q)}$
the only difference with the Lorentzian $3+1$ theory being that
now indices $a,b,c,..=1,2$ have range in one dimension less and that
there is only the Euclidean constraint.

The second model is $3+1$ Lorentzian gravity but instead of
performing the usual Killing reduction one looks for
(distributional) states in the {\it full Hilbert space}
${\cal H}^0$ of the theory which are compatible with the
Killing symmetries of the model.

In both models one then follows step by step the regularization
procedure outlined in sections \ref{s6.2},\ref{s6.3}.
The outcomes are as follows:
\begin{itemize}
\item{Euclidean 2+1 Gravity}\\
The quantization of 2+1 general relativity is an exhaustively studied
problem (see, e.g., \cite{95,96,97,98,99,100,101}, third and fourth
references in \cite{38} and references in all of those).
Several different quantization techniques have been applied
and were shown to give consistent results.
The reader might wonder why 2+1 {\it Euclidean}
quantum gravity should serve as a test model for 3+1 {\it Lorentzian}
quantum gravity. The reason for this is that, as pointed out in
\cite{100,101}, the Hamiltonian formulation of 2+1 gravity via
connections leads to the non-compact gauge group $SU(1,1)$ for three-metrics
of Lorentzian signature while for three-metrics of Euclidean signature
we have the same compact gauge group as in Lorentzian 3+1 gravity, namely
$SU(2)$. Thus, in order to maximally simulate the 3+1 theory, we should
consider Euclidean 2+1 gravity.

However, in order to maximally test the new technique introduced in
sections \ref{s6.2}, \ref{s6.3} and the constraints of the 3+1 theory
one has to develop techniques different from those that people
normally employ in 2+1 gravity which make \cite{VIII} of interest by itself.
In particular, it contains a full fledged derivation of the
$2+1$ volume operator. The reason is the following:\\
Pure $2+1$ gravity on a Riemann surface of some fixed
genus is a topological field theory, that is, there are only
finitely many degrees of freedom. This can be easily seen
from the fact that we have six canonical pairs and six first class
constraints. When the metric $q_{ab}$ is non-degenerate, the Diffeomorphism
and Hamiltonian constraint together are equivalent to the
{\it Flatness Constraint} $C^j:=\epsilon^{ab} F_{ab}^j=0$.
Almost exclusively the theory is quantized using $C_j$ rather than
$V_a,H$, see in particular \cite{96} and reference three and four
of \cite{38}. But of course we must use $V_a,H$ in order to test the
$3+1$ theory appropriately.

The result is that all steps of the quantization programme can be carried
up to and including the construction of the full solution space to
all constraint. The structure of that solution space is as complicated as
in the $3+1$ theory, therefore it is not easy to find a suitable physical
inner product. {\it However, one finds that the full space of solutions to
the flatness constraint is contained as a subspace in the space of
solutions to the Diffeomorphism and Hamiltonian constraint}.
Therefore, the validity of the quantization method is confirmed in
this model. Futhermore,
the curvature operator $F_{ab}$ (which of course becomes substituted by
a holonomy along contractible loop) {\it must be ordered to the most left}.
Thus, ordering is important here, in particular
the Hamiltonian constraint is by far not even symmetric in that ordering.
An inner product on that subspace of the full solution space is then suggested
to be usual product introduced in \cite{96}. The full solution space
of the Diffeomorphism and Hamiltonian constraint is much larger
which could be related to the fact that it contains a huge number of states with
vanishing volume \cite{VIII}, however, this speculation is yet
unconfirmed. What to do with zero
volume states (degenerate three metrics) in $3+1$ connection quantum
gravity has always been a puzzle \cite{95}.
\item{Isotropic, Homogeneous Bianchi Cosmologies}\\
In an outstanding series of papers \cite{94} Bojowald has introduced a
method for embedding the quantum theory of a Killing -- or
dimensionally reduced model of a given field theory of connections
into the quantum theory of the full unreduced theory.
Paraphrasing somewhat the procedure, roughly what happens is
the following:\\
In contrast to the usual mini (midi) superspace quantization
procedure of first reducing the classical theory by the Killing
symmetry and then quantizing the resulting reduced theory, here
one starts with the Hilbert space of the full unreduced theory
and imposes the Killing symmetry on states. Since the Killing
symmetry group is in a sense a subgroup of the diffeomorhism group
it is clear that one gets symmetric (distributional) states by a kind
of group averaging procedure together with a natural group averaging
inner product. One then ``projects" the constraint operators of the full theory,
regularized by the same technique as in section \ref{s6.3}, on the
space of symmetric states. The method is very general but most
is known for the isotropic and homogeneous Bianchi models and
has culminated in a series of papers entitled ``Loop Quantum Cosmology
I -- IV" plus successors thereof \cite{94}. The results are
indeed spectacular:\\
\begin{itemize}
\item[1)] {\it Absence of Initial Singularity}\\
Let us take as a detector for the big bang singularity
in a Friedman -- Robertson -- Walker model the divergence of the
inverse scale factor $1/a$ at $a=0$.
{\it In quantum theory the big bang singularity evaporates !!!} It is simply
not there. The technical reason is as follows:\\
The function $1/a$ is classically a negative power of the
volume functional which we can display as the Poisson bracket between
a positive power of the volume functional and a holonomy by using the
same key identities that were employed in section \ref{s6.2}.
{\bf Now one simply replaces the Poisson bracket by a commutator exactly
as we have done in the full theory and
in that very precise sense, the results by Bojowald confirm the
validity of our quantization technique of \cite{V}}. The crucial fact
is now that the commutator cannot blow up at low volume. Even better,
one can show that the quantization of $1/a$ is a positive semi-definite
operator which is bounded from above ! Thus, in loop quantum cosmology
one of the dreams about quantum gravity, that it cures classical singularities
seems to come true.
\item[2)] {\it Rapid Convergence to the Classical Regime}\\
One might imagine that a quantization of $1/a$ with such bizarre properties
should have a spectrum far off the classcal curve $a\mapsto 1/a$. However,
this is far from true: As one can imagine, (linear combinations of
symmetric subsets of) spin network states diagonalize the inverse scale
factor operator and its spectrum is purely discrete. Already for spins as low
as $j>O(10)$ the spectrum lies exaclty on the classical curve.
\item[3)] {\it Discrete Time Evolution}\\
The scale factor itself makes a good time observable in
the present model. As the solution algorithm for the
Hamiltonian constraint of the full theory, section \ref{s6.4}, already
indicates, the {\it time evolution becomes discrete} and can be solved in
closed form. Amazingly, given initial data one can {\it quantum
evolve through the classical singularity} to negative times.
\item[4)] {\it No Boundary Conditions Necessary}\\
When analyzing the solution space of the quantum evolution
equation (Wheeler -- Dewitt equation) one discovers that
there is a {\it unique} solution with the correct semiclassical
properties. Thus, one does not need to carefully adjust the
boundary condition on the initial state \cite{95}
in order to get
a state today with the desired properties, {\it the theory itself
chooses that state}!
\end{itemize}
\end{itemize}
One should, of course, consider more examples and transfer these results
to the full theory in order to gain confidence into the
quantization method, ultimatively a semiclassical analysis is
unavoidable, however, these two results described are
hopefully promising
enough in order to take the proposal for the regularization
of the Hamiltonian constraint proposed sufficiently serious.

\subsubsection{Quantum Poincar\'e Algebra}
\label{s6.5.3}

In \cite{X} an investigation was started in order to settle the question
whether ${\cal H}^0$ supports the quantization of the ADM energy
surface integral
\be \label{6.70}
E_{ADM}(N)=-\frac{2}{\kappa} \int_{\partial\sigma} dS_a
\frac{N}{\sqrt{\det(q)}}E^a_j\partial_b E^b_j
\ee
for an asymptotically flat spacetime $M$ (here $\partial\sigma$
corresponds to spatial infinity $i^0$ in the Penrose diagramme
describing the conformal completion of $M$). It should be stressed
that (\ref{6.70}) is the value of the graviatational energy
(at unit lapse $N=1$) only
when the constraints are satisfied, otherwise one has to add to
(\ref{6.70}) the Hamiltonian constraint $H(N)$. In particular
one has to use $H_{ADM}(N)=H(N)+E_{ADM}(N)$ in order to compute
the equations of motion. If $N$ is, say of rapid decrease, then
$H_{ADM}(N)=H(N)$ generates gauge transformations (time reparameterizations),
if it is asymptotically constant then it generates symmetries.
There are nine more surface integrals of the type (\ref{6.70})
and together they generate the asymptotic Poincar\'e algebra.
They are the only ten Dirac obeservables known for full, Lorentzian,
asymptotically flat gravity in four dimensions.
For a discussion of these and related issues, see e.g.
\cite{30l} and references therein.

In \cite{X}
we were only able to cover time translations (\ref{6.70}), spatial
translations and spatial rotations. Boosts, which are much harder
to define, were not considered there but there is no principal
problem to do so. We will focus here only on the quantization
of (\ref{6.70}) for reasons of brevity. The method of regularization
and quantization completely parallel those displayed in sections
\ref{s6.2}, \ref{s6.3} and will not be repeated here.
The only new element that goes into the classical regularization
is the exploitation of the fall off conditions on the classical
fields, in particualr that $A=O(1/r^2)$ in an asymptotic radial coordinate.
This enables one to replace, effectively, in (\ref{6.70})
$\partial_b E^b_j$ by
the gauge invariant quantity $G_j={\cal D}_b E^b_j$ in (\ref{6.70}),
that is, the Gauss constraint. At first sight one is tempeted to set
it equal to zero. However, a detailed analysis shows that
for the Gauss constraint to be functionally differentiable,
its Lagrange multiplier must fall off as $1/r^2$ which means that
the Gauss constraint does not need to hold at $\partial\sigma$.
Thus, it would be physically incorrect to require $G_j=0$ at
$\partial\sigma$, in other words, quantum states do not need to be gauge
invariant at $\partial\sigma$ or, put differently, the motions
generated by $G_j$ at $\partial\sigma$ are not gauge transformations but
symmetries.

The final
answer is ($E_{ADM}=E_{ADM}(1)$)
\be \label{6.71}
\hat{E}_{ADM}f_\gamma=-2 m_p
\sum_{v\in V(\gamma)\cap\partial\sigma}
\frac{\ell_p^3}{\hat{V}_v}R^j_v R^j_v f_\gamma
\ee
where $R^j_v=\sum_{f(e)=v} R^j_e,\;\hat{V}_v=\lim_{R_v\to\{v\}}
\hat{V}(R_v)$ and $x\mapsto R_x$ is an
open region valued function with $x\in R_x$. The operator (\ref{6.71})
is defined actually on an extension of ${\cal H}^0$ which allows
for edges that are not compactly supported. Moreover we must require
that 1) for each $v\in\gamma\cap\partial\sigma$ the eigenvalues
of $\hat{V}_v$ are non-vanishing and that 2) $e\cap\partial\sigma$
is a discrete set of points for every $e\in E(\gamma)$. We have
assumed w.l.g. that all edges with $e\cap \partial\sigma\not=\emptyset$
are of the ``up" type with respect to the surface $\partial\sigma$.

Under these assumptions one can show the following:
\begin{itemize}
\item[i)] {\it Positive Semi-Definiteness}\\
(\ref{6.71}) defines a self-consistent family of essentially
self-adjoint, positive semi-definite operators. This is like
a quantum positivity of energy theorem but it rests heavily on the two
assumptions 1) and 2) made above whose physical origin is poorly
understood.
\item[ii)] {\it Fock Space Interpretation}\\
Since the volume
operator is gauge invariant, it follows that it commutes with
the Laplacian $\Delta_v=(R^j_v)^2$ and therefore we can simultaneously
diagonalize these operators. It is clear that the eigenstates are
certain linear combinations of spin-network states and the eigenvalues
are of the form $j_v(j_v+1)/\lambda_v$ (where $\lambda_v$ is a volume
eigenvalue) times $m_p$. Thus we can complete the intuitive picture
that the Hamiltonian constraint gave us: while the constraint changes
the spin quantum numbers, the energy is diagonal in it in very much
the same way as the annihilation and creation operators of quantum
mechanics change the occupation number of an energy eigenstate.
{\it We may thus interpret the spin quantum numbers as occupation
numbers of a non-linear Fock representation}. In quantum field theory
we label Fock states by occupation numbers $n_k$ for momentum modes
$k$. Here we have occupation numbers $j_e$ for ``edge modes" $e$.
\item[iii)] {\it Spectral Properties}\\
The eigenvalues are discrete and unbounded from above but in contrast to the
geometry operators there is no energy gap. Rather there is
an accumulation point at zero because $[\Delta_v,\hat{V}_v]=0$
(we can choose the state to be very close to being gauge invariant but to
have arbitrarily large volume). This is to be expected on physical
grounds because we should be able to detect arbitrarily soft
gravitons at spatial infinity.
\item[iv)] {\it Quantum Dirac Observable and Schr\"odinger equation}\\
(\ref{6.71}) trivially commutes with all constraints (since diffeomorphisms
$\varphi$ and lapses $N$ that generate gauge transformations are trivial
(identity and zero) at
$\partial\sigma$) and therefore represents a true quantum Dirac observable.
In principle we can now solve ``the problem of time" since a
physically meaningful time parameter is selected by the one parameter
unitary groups generated by $\hat{E}_{ADM}$, in other words, we have
a Schr\"odinger equation
\be \label{6.72}
-i\hbar\frac{\partial\Psi}{\partial t}=\hat{E}_{ADM}\Psi
\ee
\end{itemize}
Actually in \cite{X} concepts that go beyond ${\cal H}^0$
were needed and introduced heuristically. They go under the name
``Infinite Tensor Product Extension" and were properly defined only
later in \cite{47q}. They will be discussed briefly in section \ref{s8.2}.

\newpage

\section{Extension to Standard Matter}
\label{s7}

The exposition of section \ref{s6} would be unserious if we would not
be able to extend the framework also to matter, at least to the
matter of the standard model. This is straightforward for gauge
field matter, however for fermionic and Higgs matter one must first
develop a background independent mathematical framework \cite{XI}.
We will discuss the essential steps in the next subsection and
then sketch the quantization of the matter parts of the
total Hamiltonian constraint in the section after that, see \cite{IX}
for details.

We should point out that these representations are geared towards
a background independent formulation. The matter Hamiltonian
operator of the standard model in a background spacetime {\it is not
carried} by these representations. They make sense {\it only if we couple
quantum gravity}. Also, while we did not treat supersymmetric matter
explicitly, the following exposition reveals that it is straightforward to
extend the formalism to Rarita -- Schwinger fields.

\subsection{Kinematical Hilbert Spaces for Diffeomorphism Invariant Theories
of Fermionic and Higgs Fields}
\label{s7.1}

First attempts to couple quantum field theories of fermions to
quantum general relativity
gravity were made in the pioneering work \cite{102}.
However, this paper was still written in
terms of the complex-valued Ashtekar variables for which the kinematical
framework was missing.
Later on \cite{103} appeared in which a
kinematical Hilbert space for diffeomorphism invariant theories for
fermions was proposed which were coupled to arbitrary gauge fields and
real-valued Ashtekar variables using the kinematical framework of
section \ref{s3}. Also, the diffeomorphism
constraint was solved there but not the Hamiltonian constraint. However,
that fermionic Hilbert
space did implement the correct reality conditions for the fermionic
degrees of freedom only for a subset of all kinematical observables. In
\cite{XI} we removed this problem by
introducing new fermionic variables, so-called {\it Grassmann-valued
half-densities} and also extended the framework to Higgs fields.
This section is accordingly subdivided into one section each
for the fermionic and the Higgs sector respectively and in the third section
we collect results and define the most general gauge and diffeomorphism
invariant states of connections, fermions and Higgs fields.

\subsubsection{Fermionic Sector}
\label{s7.1.1}

We will take the fermionic fields to be Grassmann-valued, see \cite{104,105}
for a mathematical introduction into these concepts.
Furthermore, the
Grassmann field $\eta_{A\mu}$ is a scalar with respect to diffeomorphisms
of $\sigma$ which carries two indices, $A,B,C,..=1,2$ and
$\mu,\nu,\rho=1,..,\dim(G)$ corresponding to the fact that it transforms
according to the fundamental representation of $SU(2)$ and the
defining representation of the compact, connected, unimodular gauge group
$G$ of a Yang-Mills gauge
theory to which it may couple. This can be generalized to arbitrary
representations of $SU(2)\times G$ but we refrain from doing that for the
sake of concreteness. Notice that it is no loss of generality to restrict
ourselves to only one helicity of the fermion as we can always perform
a canonical transformation $(i\bar{\sigma}^{A'},\sigma_{A'})\to
(i\epsilon^{A B'}\sigma_{B'},\epsilon_{A B'}\bar{\sigma}^{B'})
=:(i\bar{\eta}^A,\eta_A)$. We will restrict to only one fermionic species
in order not to clutter the formulae.

It turns out that the
real-valued action in Hamiltonian form for any diffeomorphism invariant
theory of fermions is given by
\be  \label{7.1}
S_F=\int_\Rl dt\int_\sigma d^3x(\frac{i}{2}\sqrt{\det(q)}
[\bar{\eta}^{A\mu}\dot{\eta}_{A\mu}
-\dot{\bar{\eta}}^{A\mu}\dot{\eta}_{A\mu}]-[\mbox{more}])
\ee
where ``more" stands for various constraints and possibly a Hamiltonian
and $\det(q)$ is the determinant of the gravitational three-metric
which appears because in four spacetime dimensions one needs a metric
to define a diffeomorphism invariant theory of fermions. Notice that
(\ref{7.1}) is real valued with respect to the usual involution
$(\theta_1..\theta_n)^\ast=\bar{\theta}_n..\bar{\theta}_1$ for Grassmann
variables $\theta_1,..,\theta_n$ since indices $A,\mu$ are raised and
lowered with the Kronecker symbol (the involution is just complex
conjugation with respect to bosonic variables).

The immediate problem with (\ref{7.1}) is that it is not obvious what the
momentum $\pi^{A\mu}$ conjugate to $\eta_{A\mu}$ should be. One
strategy would be to
integrate the second term in (\ref{7.1}) by parts (the corresponding
boundary term being the generator of the associated canonical
transformation) and to conclude
that it is given by $i\sqrt{\det(q)}\bar{\eta}^{A\mu}$. However, there
is a second term from the integration by parts given by
$i\dot{E}^a_i e_a^i\bar{\eta}^{A\mu}\eta_{A\mu}$ which after a further
integration by parts combines with the symplectic potential of the
real-valued Ashtekar variables to the effect that $A_a^i$ is replaced
by $(^\Co A_a^i)=A_a^i- i e_a^i\bar{\eta}^{A\mu}\eta_{A\mu}$ (recall that
$E^a_i$ is the momentum conjugate to $A_a^i$).
This is bad because the connection is now complex-valued and
the techniques from section \ref{s3} do not apply any longer so that we
are in fact
forced to look for another method. The authors of \cite{103} also
noticed this subtlety in the following form: If one assumes that the
connection is still real-valued while
$\pi=i\sqrt{\det(q)}\eta$ is taken as the momentum conjugate to $\eta$
then one discovers the
following contradiction: By assumption we have the classical
Poisson bracket $\{\pi(x),A(y)\}=0$. Taking the involution of this equation
results in $0=-i\eta(x)\{\sqrt{\det(q)}(x),A(y)\}\not=0$. If we, however,
insert instead of $A$ the complex variable $(^\Co A)$ into these
equations then in fact there is no contradiction as was shown in \cite{XI}.

The idea of how preserve the real-valuedness of $A_a^i$ {\it and} to
simplify the reality conditions on the fermions is as
follows : Notice that if we define the Grassmann-valued half-density
\be \label{7.2}
\xi_{A\mu}:=\root4\of{\det(q)}\eta_{A\mu}
\ee
then (\ref{7.1}) in fact equals
\be  \label{7.3}
S_F=\int_\Rl dt\int_\sigma d^3x(\frac{i}{2}
[\bar{\xi}^{A\mu}\dot{\xi}_{A\mu}
-\dot{\bar{\xi}}^{A\mu}\dot{\xi}_{A\mu}]-[\mbox{more}])
\ee
{\it without} picking up a term proportional to $d\det(q)/dt$. Thus
the momentum conjugate to $\xi_{A\mu}$ and the reality conditions
respectively are simply given by
\be \label{7.4}
\pi^{A\mu}=i\bar{\xi}_{A\mu}\mbox{ and }
(\xi)^\ast=-i\pi,\;(\pi)^\ast=-i\xi\;.
\ee
The fact that $\xi,\pi$ are half densities may seem awkward at first
sight but it does not cause any immediate problems. Also, recall that
``half-density-quantization" is a standard procedure in the theory of
geometric quantization of phase spaces with real polarizations \cite{18}.

It is in fact possible to base the quantization on the half-density $\xi$
as a quantum configuration variable as far as the solution to the Gauss
constraint is concerned. Namely, as has been pointed out by many (see,
e.g., \cite{102}) an example for a natural, classical, gauge invariant
observable is given by
\be \label{7.5}
P_e(\xi,A,\underline{A}):= \xi_{A\mu}(e(0))C_1^{A\mu,C\rho}(h_e(A))_{CD}
(\underline{\pi}(\underline{h}_e(\underline{A})))_{\rho\sigma}
C_2^{D\sigma,B\nu} \xi_{B\nu}(e(1))
\ee
where the notation is as follows : By
$(A,h_e,\pi(h_e))$ and
$(\underline{A},\underline{h}_e,\underline{\pi}(\underline{h}_e))$
respectively we denote (connection, holonomy along an edge $e$,
irreducible representation evaluated at the holonomy) of the gravitational
$SU(2)$ and the Yang-Mills gauge group $G$ respectively. The matrices
$C^{A\mu,B\nu}$ are projectors on singlet representations of the
decomposition into irreducibles of tensor product representations that
appear under gauge transformations on both ends of the path
$[0,1]\ni t\to e(t)$ and the irreducible representation $\underline{\pi}$
has to be chosen in such a way that a singlet can occur. For example,
if $G=SU(N)$ then we can choose $\underline{\pi}$ to be the complex
conjugate of the defining representation. In particular, if $G=SU(2)$
as well we can take $\underline{\pi}$ to be the fundamental representation
and $C_1^{A\mu,C\rho}=\epsilon^{AC}\epsilon^{\mu\rho},
C_2^{D\sigma,B\nu}=\delta^{DB}\delta^{\sigma\nu}$. For more general
groups we may have to take more than one spinor field at each end of the path
in order to satisfy gauge invariance.

All this works fine until it comes to diffeomorphism invariance : notice
that the objects (\ref{7.5}) behave strangely under a diffeomorphism
$\varphi$, namely $\varphi\cdot P_e=P_{\varphi(e)}
(J_\varphi(e(0)) J_\varphi(e(1)))^{-1/2}$ where $J_\varphi(x)=
|\det(\partial\varphi(x)/\partial x)|$ is the Jacobian. Since there are
analyticity preserving diffeomorphisms which leave $e$ invariant but such
that, say, $J_\varphi(e(0))$ can take any positive value it follows that
the average of $P_e$ over diffeomorphisms is meaningless. We are
therefore forced to adopt another strategy.

The new idea \cite{XI} is to ``dedensitize" $\xi$
by means of the $\delta$-distribution $\delta(x,y)$ which itself transforms
as a density of weight one in either argument. Let $\theta(x)$ be a
smooth Grassmann-valued scalar (we drop the indices $A\mu$) and we
{\it define} $\xi(x)$ not to be a smooth function but rather a distribution
(already classically). Let $\delta_{x,y}=1$ for $x=y$ and zero
otherwise (a Kronecker $\delta$, not a distribution). Then on the
space of test functions of rapid decrease the distribution
$\sqrt{\delta(x,y)\delta(z,y)}$ is well-defined and equals
$\delta_{x,z}\delta(x,y)$ \cite{XI}. As shown in \cite{XI} the
following
transformations (and corresponding ones for the complex conjugate
variables)
\ba \label{7.6}
\theta(x):=\int_\sigma d^3y \sqrt{\delta(x,y)}\xi(y)\\
\label{7.7}
\xi(x)=\sum_{y\in\sigma} \delta(x,y)\theta(y)
\ea
are canonical transformations between the symplectic structures
defined by the symplectic potentials
$i\int_\sigma d^3x \bar{\xi}(x)\dot{\xi}(x)$ and
$i\sum_{x\in\sigma}\bar{\theta}(x)\dot{\theta}(x)$ respectively. Notice
that (\ref{7.6}) makes sense precisely when $\xi$ is a distributional
half-density and in fact one can show that $\xi=\eta\root4\of{\det(q)}$
will precisely display such a behaviour (at least upon quantization)
since $\sqrt{\det(q)}$ becomes an operator valued distribution
proportional to the $\delta$ distribution (recall the formula
for the volume operator).
The non-trivial anti-Poisson brackets in either case are given by
\be \label{7.8}
\{\xi(x),\bar{\xi}(y)\}_+=-i\delta(x,y) \mbox{ and }
\{\theta(x),\bar{\theta}(y)\}_+=-i\delta_{x,y}\;.
\ee
In summary, we conclude that we can base the quantization of the
fermionic degrees of freedom on $\theta$ as a configuration variable
with conjugate momentum and reality structure given by
\be \label{7.9}
\pi^{A\mu}=i\bar{\theta}_{A\mu}\mbox{ and }
(\theta)^\ast=-i\pi,\;(\pi)^\ast=-i\theta\;.
\ee

We now have to develop integration theory. This will be based, of course,
on the Berezin ``integral" \cite{104,105}. Let ${\cal S}(x)$ be the
superspace
underlying the $2d$ fermionic configuration degrees of freedom
$\theta_{A\mu}(x)$ for any $x\in\Sigma$ where
$d=2\dim(G)$. Of course, all these spaces are just copies of a single
space $\cal S$. This superspace can be turned into a trivial $\sigma$-algebra
${\cal B}(x)$ consisting of ${\cal S}(x)$ and the empty set. On
${\cal B}(x)$ one can define a probability ``measure" $dm_x$ with the
additional property that it is positive on ``holomorphic" functions
(that is, those which depend $\theta(x)$ only and not on $\bar{\theta}(x)$)
in the sense that
$\int_{\cal S} dm_x f(\theta(x))^\ast
f(\theta(x))\ge 0$ where equality holds if and only if $f=0$. This measure
is given by
\be \label{7.10}
dm(\bar{\theta},\theta)=\prod_{A\mu} (1+\bar{\theta}^{A\mu}\theta_{A\mu})
d\bar{\theta}^{A\mu} d\theta_{A\mu}
\ee
and $dm_x=dm(\bar{\theta}(x),\theta(x))$.\\
Let now $\overline{\cal S}:=\times_{x\in\sigma}{\cal S}_x$ be the
fermionic quantum configuration space with $\sigma$-algebra
$\cal B$ given by the {\it direct product} of the ${\cal B}(x)$. The
Kolmogorov-theorem \cite{25} for uncountable direct products of probability
measures ensures that
\be \label{7.11}
d\mu_F(\bar{\theta},\theta):=\otimes_{x\in\sigma} dm_x
\ee
is a rigorously defined probability measure on $\overline{{\cal S}}$.
It can be recovered as the direct product limit (rather than projective
limit) from its finite dimensional joint distributions defined by
cylindrical functions. Here a function $F$ on $\overline{{\cal S}}$
is said to be cylindrical over a finite number of points $x_1,..,x_n$
if it is a function only of the finite number of degrees of freedom
$\theta(x_1),..,\theta(x_n)$ and their complex conjugates, that is,
$F(\theta)=
f_{x_1,..,x_n}(\bar{\theta}_1(x_1),\theta_1(x_1),..,
\bar{\theta}_n(x_n),\theta_n(x_n))$ where $f_{x_1,..,x_n}$
is a function on ${\cal S}^n$. We then have
\be \label{7.12}
\int_{\overline{{\cal S}}} d\mu_F F=\int_{{\cal S}^n}
dm(\bar{\theta}_1,\theta_1)..dm(\bar{\theta}_n,\theta_n)
f_{x_1,..,x_n}
(\bar{\theta}_1,\theta_1,..,\bar{\theta}_n,\theta_n)\;.
\ee
Basic cylindrical functions are the fermionic vertex functions. These are
defined as follows : order the labels $A\mu$ from $1$ to $2d$ and denote
them by $i,j,k,..$ (confusion with the $su(2)$ labels should not arise).
Denote by $I$ an array $1\le i_1<..<i_k\le 2d$ and define $|I|=k$
in this case (confusion with the Lie$(G)$ or spin-network labels should not
arise). Then for each set of distinct points $v_1,..,v_n$ we define
\be \label{7.13}
F_{\vec{v},\vec{I}}=\prod_{l=1}^n F_{v_l,I_{v_l}},\;
F_{v_l,I_{v_l}}=\prod_{j=1}^{|I_{v_l}|}\theta_{i_j(v_l)}(v_l)
\ee

Is this the correct measure, that is, are the adjointness relations
$\hat{\pi}^\dagger=-i\hat{\theta},\;\hat{\theta}^\dagger=-i\hat{\pi}$
and the canonical anti-commutation relations
$[\hat{\theta}_{A\mu}(x),\hat{\pi}^{B\nu}(y)]_+=i\hbar\delta_A^B
\delta_\mu^\nu\delta_{x,y}$ faithfully implemented ? It is sufficient
to check this to be the case on cylindrical subspaces if we represent
$\hat{\theta}(x)$ as a multiplication operator and $\hat{\pi}(x)$
as $i\hbar\partial^l/\partial\theta(x)$ where the superscript stands
for the left {\it ordinary} derivative (not a functional derivative).
In fact, the measure $d\mu_F$ is uniquely selected by these relations
given the representation just as in the case of the theory of
distributional connections $\ab$.
Also, it is trivially diffeomorphism invariant since the
integrals of a function cylindrical over $n$ points and of
its diffeomorphic image coincide.

In summary, the correct kinematical
Fermion Hilbert space is
therefore defined to be ${\cal H}_F:=L_2(\overline{{\cal S}},d\mu_F)$.
It follows immediately from these considerations that the quantum fermion
field {\it at a point} (i.e. totally unsmeared) becomes a densely
defined operator.
This seems astonishing at first sight but it is only a little bit more
surprising than to assume that Wilson loop operators, the quantum
connection being smeared in one direction only, are densely defined.
When quantizing diffeomorphism invariant theories which lack a background
structure one has to give up standard representations and construct new
ones.

\subsubsection{Higgs Sector}
\label{s7.1.2}

It turns out that it is also not possible to combine the well-developed
theory of Gaussian measures for scalar field theories with diffeomorphism
invariance in order to obtain a kinematical framework for diffeomorphism
invariant theories of Higgs fields.
The basic obstacle is that a Gaussian
measure is completely defined by its covariance which, however, depends
on a background structure (see \cite{IX} for a detailed discussion of
this point). We are therefore again led to a new non-standard
representation.

In the following we restrict ourselves to real-valued Higgs-fields $\phi_I$
which transform according to the adjoint representation of $G$. Other cases
can be treated by similar methods. We also allow for scalar fields
(without internal degrees of freedom).\\
Since in the previous subsection we already got used to dealing with
representations for
which the quantum configuration field at a point becomes a well-defined
quantum operator it is perhaps not so awkward anymore to do the same for
the Higgs field. Actually, we are not going to deal with $\phi_I$
itself but with the {\it point-holonomies}, which also play a crucial
role in Bojowald's series \cite{94}
\be \label{7.14}
U_x(\phi):=\exp(\phi_I(x)\tau_I)
\ee
where $\tau_I$ denotes a basis of the Lie algebra Lie$(G)$ of the Yang-Mills
gauge group. The name stems from the fact that under a gauge
transformation $g(x)$ at $x$ we have that
$U(x)\to \mbox{Ad}_{g(x)}(U(x))$ which is precisely the transformation
behaviour of a holonomy $\underline{h}_e$ starting at $x$ in the limit of
vanishing edge length. In the case of a simple scalar field we define
$U_x=e^{i\phi(x)}$. These variables play a role similar to the Wilson
loop variables in lattice gauge theory \cite{93} and it is understood
that any action written in terms of $\phi_I$ should be rewritten in terms of
the $U(x)$ in analogy to the replacement of the Yang-Mills action by the
Wilson action.

This analogy with holonomies suggests a step by step
repetition of the Ashtekar-Isham-Lewandowski framework of section
\ref{s3} \cite{XI} which we are going to
sketch below. Before doing that we must decide on the elementary variables.
With $\phi_I$ being a scalar its conjugate momentum $p^I$ is a scalar density
of weight one. Therefore the integrated quantity
\be \label{7.15}
p^I(B):=\int_B d^3x p^I(x)
\ee
for any open region $B$ in $\sigma$ is diffeomorphism covariantly defined
and the formal Poisson brackets\\
$\{p^I(x),\phi_J(y)\}=\delta^I_J \delta(x,y)$ translate into
\be \label{7.16}
\{p^I(B),U_x\}=\chi_B(x)\frac{1}{2}[\tau_I U_x+U_x\tau_I]
\ee
(in order to see this one must regularize $U_x$ as in \cite{XI} and then
remove the regulator). The other elementary Poisson bracket is
$\{U_x,U_y\}=0$. Actually one has to generalize the Poisson algebra to
the Lie algebra of functions on smooth $\phi_I$'s and vector fields thereon
just as in the case of connections in order to obtain a true Lie algebra
which one can quantize.
Finally, $p^I(B)$ is real-valued and $U_x$ is $G$-valued.\\

The construction of a quantum configuration space $\overline{{\cal U}}$
and a diffeomorphism invariant measure $d\mu_U$ thereon now proceeds just
in analogy with section \ref{s3} :\\
A Higgs vertex function $H_{\vec{v},\vec{\pi},\vec{\mu},\vec{\nu}}$ is just
given by
\be \label{7.17}
H_{\vec{v},\vec{\pi},\vec{\mu},\vec{\nu}}
=\prod_{k=1}^n \sqrt{d_{\pi_k}}(\pi_k(U(v_k)))_{\mu_k \nu_k}
\ee
where $\pi_k$ are chosen from a complete set of irreducible,
inequivalent representations of $G$ and $v_1,..,v_k$ are distinct points
of $\sigma$. \\
Consider the Abelian C$^\ast$ algebra given by finite linear combinations
of Higgs vertex functions and completed in the sup-norm over the set
of smooth Higgs fields $\cal U$. Then $\overline{{\cal U}}$, the quantum
configuration space of distributional Higgs fields, is the spectrum
of that algebra equipped with the weak$^\ast$ topology (Gel'fand topology).

The characterization of the spectrum is as follows : points $\bar{\phi}$
in $\overline{{\cal U}}$ are in one to one correspondence with the set
Fun$(\sigma,G)$ of $g$-valued functions on $\sigma$, the correspondence
being given by $\bar{\phi}\leftrightarrow U_{\bar{\phi}}$ where
$\sqrt{d_{\pi_0}}(U_{\bar{\phi}})_{\mu\nu}(v)=
\bar{\phi}(H_{v,\pi_0,\mu,\nu})$ and $\pi_0$ is the fundamental
representation of $G$.

Again, since the spectrum is a compact Hausdorff space one can define
a regular Borel probability measure $\mu$ on it through positive, normalized,
linear functionals $\Gamma$ on the set of continuous functions $f$ thereon,
the correspondence being given by $\Gamma(f)
=\int_{\overline{{\cal U}}} d\mu f$. We
define the measure $\mu_U$ by
\be \label{7.18}
\Gamma_{\mu_U}(H_{\vec{v},\vec{\pi},\vec{\mu},\vec{\nu}})
=\left\{ \begin{array}{ll}
1 & H_{\vec{v},\vec{\pi},\vec{\mu},\vec{\nu}}=1\\
0 & \mbox{otherwise}
\end{array} \right.
\ee
and one easily sees that this measure is just the Haar measure on
$G^n$ for functions cylindrical over $n$ distinct points. In particular,
the Higgs vertex functions form a complete orthonormal basis by an appeal
to the Peter\&Weyl theorem. The measure $\mu_U$ can be shown \cite{XI} to be
concentrated on nowhere continuous Higgs fields, in particular
$\mu_U({\cal U})=0$.

Finally, $\hat{U}(x)$ is just a multiplication operator on
cylindrical functions and if we replace $p^I$ by $-i\hbar\delta/\delta\phi_I$
then we find for a function $F=f_{\vec{v}}$
cylindrical over $n$ points $\vec{v}$ that
$\hat{p}^I(B)F=-i\hbar \sum_{k=1}^n \chi_B(v_k) X^I_v f_{\vec{v}}$
where $X^I_v=X^I(U(v)),\; X^I(g)=\frac{1}{2}[X^I_R(g)+X^I_L(g)]$
and $X_L,X_R$ are, respectively, left and right invariant vector fields on
$G$. This representation shows that the canonical commutation relations
as well as the adjointness relations are faithfully implemented
and that the appropriate kinematical Higgs field Hilbert space can be
chosen to be ${\cal H}_U:=L_2(\overline{{\cal U}},d\mu_U)$.

\subsubsection{Gauge and Diffeomorphism Invariant Subspace}
\label{s7.1.3}

We now put everything together to arrive at the complete solution to
the Gauss and Diffeomorphism constraint for quantum gravity coupled to
gauge fields, Higgs fields and Fermions.

We begin with the kinematical Hilbert space
\be \label{7.19}
{\cal H}=
L_2(\overline{{\cal A}}_{SU(2)},d\mu_0^{SU(2)})\otimes
L_2(\overline{{\cal A}}_G,d\mu_0^G)\otimes
L_2(\overline{{\cal F}},d\mu_F)\otimes
L_2(\overline{{\cal U}},d\mu_U)
\ee
and now consider its subspace consisting of gauge invariant functions.
A basis of such functions is labelled by a graph $\gamma$, a labelling
of its edges $e$ by spins $j_e$ and colours $c_e$ corresponding to
irreducible representations of $SU(2)$ and $G$ and a labelling of its
vertices $v$ by an array $I_v$, another colour $C_v$ and
two projectors $p_v,q_v$. The array $I_v$ indicates a fermionic
dependence at $v$ by $F_{v,I_v}$ and $C_v$ stands
for an irreducible representation of $G$ evaluated at $U(v)$. Finally,
decompose the tensor product of irreducible representations of
$SU(2)$ given by the fundamental representations corresponding to
$F_{v, I_v}$ and the representations $\pi_{j_e}$ for those edges
$e$ incident at $v$ and project with $p_v$ on a singlet that appears.
Likewise,
decompose the tensor product of irreducible representations of
$G$ given by the fundamental representations corresponding to
$F_{v, I_v}$, the representations
$\underline{\pi}_{c_e}$ for those edges $e$
incident at $v$ and the representation $\underline{\pi}_{C_v}$
and project with $q_v$ on a singlet that appears.

The result is a gauge invariant state
$T_{\gamma,[\vec{j},\vec{I},\vec{p}],[\vec{c},\vec{C},\vec{q}]}$ called
a {\it spin-colour-network state} extending the definition of a purely
gravitational spin-network state. Consider the action $\overline{\cal G}$
of the gauge group $SU(2)\times G$ on all distributional fields. Then the
spin-colour network states contain the space of gauge invariant
functions which is the same as the Hilbert space
\be \label{7.20}
{\cal H}=
L_2([\overline{{\cal A}}_{SU(2)}\times\overline{{\cal A}}_G
\times\overline{{\cal F}}\times\overline{{\cal U}}]/\overline{{\cal G}},
d\mu_0^{SU(2)}\otimes d\mu_0^G\otimes d\mu_F\otimes d\mu_U),
\ee
that is, the $L_2$ space on the moduli space.

To get the solution to the diffeomorphism constraint one considers
the spaces\\
 ${\cal D}_{SU(2)},{\cal D}_G,{\cal D}_F,{\cal D}_U$ of smooth
cylindrical functions (smooth in the sense of the nuclear topology of\\
$SU(2)^n,G^n,{\cal S}^n,G^n$ respectively) and their corresponding
algebraic duals. Then we form the gauge invariant subspaces of
the spaces
\be \label{7.21}
{\cal D}:={\cal D}_{SU(2)}\times{\cal D}_G\times{\cal D}_F\times{\cal D}_U
\mbox{ and }
{\cal D}^\ast:={\cal D}^\ast_{SU(2)}\times{\cal D}^\ast_G\times
{\cal D}^\ast_F\times{\cal D}^\ast_U \;.
\ee
Now the spin-colour-network states span the invariant subspace of $\cal D$
and the diffeomorphism group acts unitarily by
\be \label{7.22}
\hat{U}(\varphi)
T_{\gamma,[\vec{j},\vec{I},\vec{p}],[\vec{c},\vec{C},\vec{q}]}
=T_{\varphi(\gamma),[\vec{j},\vec{I},\vec{p}],[\vec{c},\vec{C},\vec{q}]}
\ee
and similar as in the purely gravitational case we get diffeomorphism
invariant distributions in ${\cal D}^\ast$ by group judiciously averaging
the action (\ref{7.22}).

\subsection{Quantization of Matter Hamiltonian Constraints}
\label{s7.2}

We will restrict ourselves, for the sake of clarity, to only one kind of
matter, namely pure Yang-Mills fields
for a compact, connected
gauge group $G$. See \cite{IX} for the general case.

To anticipate the result, what we find is that certain
ultraviolet divergences, which
appear when we consider Yang-Mills fields propagating on a background
spacetime, {\it disappear} when we let the spacetime metric fluctuate as
well. We do not claim that this proves finiteness of quantum gravity
because, first, we must prove that the quantum theory constructed
has general relativity as its classical limit, secondly,
besides the Hamiltonian constraint we also must show that
quantizations of classical observables
of the theory are finite and, thirdly, we must establish that those
operators remain non-singular upon passing to the physical Hilbert space.
However, these are first promising indications anyway.\\
\\
We will first explain how this works in canonical quantum gravity for
the Yang-Mills field and then describe the general mechanism.\\
The canonical pair coordinatizing the Yang-Mills phase space is given by
$(\underline{E}^a_I,\underline{A}_a^I)$ with symplectic structure
formally given by
\be \label{7.23}
\{\underline{E}^a_I(x),\underline{A}_b^J(y)\}=\delta^a_b\delta_I^J
\delta(x,y)
\ee
where as before $I,J,K,...=1,..,\dim(G)$ denote $L(G)$ indices.
The contribution of the Yang-Mills field to the Hamiltonian constraint
turns out to be
\be \label{7.24}
H_{YM}=\frac{q_{ab}}{2Q^2\sqrt{\det(q)}}[
\underline{E}^a_I\underline{E}^b_I+\underline{B}^a_I\underline{B}^b_I]
\ee
where $Q$ is the Yang-Mills coupling constant, $B^a_I:=\frac{1}{2}
\epsilon^{abc} F_{bc}^I$ the magnetic field of the connection
$\underline{A}_a^I$ and $F_{ab}^I$ its curvature.
The integrated form is given by $H_{YM}(N)=\int_\sigma d^3x N H_{YM}$
where $N$ is the lapse function.

In a background spacetime, say in Minkowski space ($N=1,
q_{ab}=\delta_{ab}$) the integrated
Hamiltonian constraint becomes just the Hamiltonian of the theory.
Let us see what happens if we try to quantize this field theory
propagating in Minkowski space non-perturbatively. It will be enough to
consider Maxwell-Theory. In order not to have to worry about infrared
divergencies, for the sake of the argument, we will add for this paragraph a
mass term $m^2 q^{ab}\sqrt{\det(q)} \underline{A}_a
\underline{A}_b$ to the
Hamiltonian density so that we are actually looking at the Proca field.
We can then define a Hilbert space $L_2({\cal S}',d\mu_G)$ where
$\mu_G$ is some Gaussian measure on the space of tempered distributional
connections. The reality conditions and the canonical commutation
relations are satisfied if we let $\hat{\underline{A}}_a$ act by
multiplication and
$\hat{\underline{E}}^a=-i\hbar Q^2\delta/\delta\underline{A}_a+F^a$
where $F^a$ is a function of the connection chosen in such a way
that
$\hat{\underline{E}}^a$ is formally a self-adjoint operator (for instance
if $\mu_G$ is the white noise measure then $F^a=\lambda\underline{A}_a$
for some constant $\lambda$).

Let us try to compute the ground state of the Hamiltonian. Since its
density is proportional to
$$
-[\frac{\delta}{\delta\underline{A}_a}+\lambda F^a]^2+
\underline{A}_a D \underline{A}_a$$
for a positive, invertible differential operator $D$ one will make the
ansatz
$$
\Psi=\exp(-\frac{1}{2}\int d^3x [\lambda (\underline{A}_a)^2
+\underline{A}_a \sqrt{D} \underline{A}_a])
$$
but
$$
\int d^3x \hat{H}(x)\Psi\propto\Psi\cdot  (\int d^3x)\cdot
[\sqrt{D}_x\delta(x)]_{x=0} $$
diverges.
This divergence is
still rather harmless since it can be removed by factor ordering but in
the interacting case things get worse.\\
What is the origin of the divergence ? As we have already pointed out
earlier in this review, it is rooted in the fact that (\ref{7.24})
becomes a density of weight {\it two} if we take $q_{ab}$ to be
non-dynamical, for instance a constant as in the case of Minkowski space.
This is because $\underline{E}^a$ has density weight one whether or not
we couple gravity. The density
weight shows up, for example, in the evaluation of the two functional
derivatives at one and the same point, giving a meaningless result
since the Lebesgue measure $d^3x$ can only absorb one of the resulting
$\delta$ distributions. On
the other hand, if we treat $q_{ab}$ as a dynamical field then
(\ref{7.24}) adopts a density weight of {\it one} again because
$\sqrt{\det(q)}$ has density weight one and is in the denominator.
As we have seen in section \ref{s6.3}, the fact that
$\sqrt{\det(q)}$ is in the denominator does not need to prevent us from
being able to give rigorous meaning to an operator corresponding to
$H_{YM}(N)$. We have already seen
that gravity regulates itself in this restricted sense.

In the next subsection we explain the quantization of $H_{YM}(N)$ and in
the subsequent one we will explain the general scheme how coupling gravity is
able to regulate certain ultra-violet divergences.

\subsubsection{Quantization of Einstein-Yang-Mills-Theory}
\label{s7.2.1}

We will focus first on the electric part of $H_{YM}(N)$ which we write
in the form
\be \label{7.24a}
H_{YM,el}(N)=\frac{1}{2 Q^2}\int d^3x N
\frac{[e_a^i \underline{E}^a_I]}{\sqrt{\det(q)}}[e_b^i \underline{E}^b_I]
\ee
where $Q$ is the Yang-Mills coupling constant.
Using the same notation as in section \ref{s6} we can also write
this as
\be \label{7.25}
H_{YM,el}(N)=\frac{1}{8\kappa^2 Q^2}\int d^3x N(x)
\frac{[\{A_a^i(x),V(R_x)\} \underline{E}^a_I(x)]}{\sqrt{\det(q)}(x)}
[\{A_b^i(x),V(R_x)\} \underline{E}^b_I(x)]
\ee
Since $\underline{E}^a_I=\frac{1}{2}\epsilon^{abc}\underline{e}_{bc}^I$
is Hodge dual to a two-form $\underline{e}^I$ we can also write this as
\be \label{7.26}
H_{YM,el}(N)=\frac{1}{8\kappa^2 Q^2}\int d^3x N(x)
\frac{[\{A_a^i(x),V(R_x)\} \underline{E}^a_I(x)]}{\sqrt{\det(q)}(x)}
[\{A_i(x),V(R_x)\}\wedge \underline{e}_I(x)]
\ee
which suggests to approximate the integral by a Riemann sum utilizing a
triangulation of $\sigma$ as in section \ref{s6.2}. Using the same
notation as there we get
\ba \label{7.27}
H^\epsilon_{YM,el}(N)&=&\frac{1}{8\kappa^2 Q^2}\sum_{\Delta\in T(\epsilon)}
N(v(\Delta))
\frac{[\{A_a^i(v(\Delta)),V(R_{v(\Delta)})\}
\underline{E}^a_I(v(\Delta))]}{\sqrt{\det(q)}(v(\Delta))}
\times\nonumber\\
&\times& \epsilon^{LMN}
[\mbox{tr}(\tau_i h_{e_L(\Delta)}\{h_{e_L(\Delta)}^{-1},V(R_{v(\Delta)})\})
E_I(S_{MN}(\Delta))]
\ea
where we have used that
$S_{MN}(\Delta)$ is any oriented triangular surface with boundary
$e_M(\Delta)\circ a_{MN}(\Delta)\circ e_N(\Delta)^{-1}$. \\
We now
apply the same trick that we used already in previous sections: Let
$\chi_{\epsilon,x}(y)$ be the characteristic function of a box
$U_\epsilon(x)$ with coordinate volume $\epsilon^3$ and centre $x$.
Then
\ba \label{7.28}
V(U_\epsilon(x))&=&\epsilon^3\sqrt{\det(q)}(x)+o(\epsilon^4) \mbox{ and}\\
\label{7.29}
\int \chi_{\epsilon,x}(y)
\frac{[\{A_i(y),V(R_y)\}\wedge \underline{e}_I(y)]}{\sqrt{V(U_\epsilon(y))}}
&=&\epsilon^3
\frac{[\{A_a^i(x),V(R_x)\} \underline{E}^a_I(x)]}{\sqrt{V(U_\epsilon(x))}}
+o(\epsilon^3)
\ea
which allows us to replace (\ref{7.27}) by
\ba \label{7.30}
H^\epsilon_{YM,el}(N)&=&\frac{1}{2\kappa^2 Q^2}
\sum_{\Delta,\Delta'\in T(\epsilon)}
N(v(\Delta)) \chi_{\epsilon,v(\Delta)}(v(\Delta'))
\epsilon^{LMN}\epsilon^{RST} \times \nonumber\\
&\times&
\frac{\mbox{tr}(\tau_i h_{e_L(\Delta)}
\{h_{e_L(\Delta)}^{-1},V(R_{v(\Delta)})\})
E_I(S_{MN}(\Delta))}{2\sqrt{V(U_\epsilon(v(\Delta)))}}\times\nonumber\\
&\times&
\frac{\mbox{tr}(\tau_i h_{e_R(\Delta')}
\{h_{e_R(\Delta')}^{-1},V(R_{v(\Delta')})\})
E_I(S_{ST}(\Delta'))}{2\sqrt{V(U_\epsilon(v(\Delta')))}}
\ea
Again, the region-valued function $x\to R_x$ is completely arbitrary up
to this point and if we choose $R_x=U_\epsilon(x)$ then we obtain the
final formula
\ba \label{7.31}
H^\epsilon_{YM,el}(N)&=&\frac{1}{2\kappa^2 Q^2}
\sum_{\Delta,\Delta'\in T(\epsilon)}
N(v(\Delta))\chi_{\epsilon,v(\Delta)}(v(\Delta'))
\epsilon^{LMN}\epsilon^{RST} \times \nonumber\\
&\times&
[\mbox{tr}(\tau_i h_{e_L(\Delta)}
\{h_{e_L(\Delta)}^{-1},\sqrt{V(U_\epsilon(v(\Delta)))}\})
E_I(S_{MN}(\Delta))]\times\nonumber\\
&\times&
[\mbox{tr}(\tau_i h_{e_R(\Delta')}
\{h_{e_R(\Delta')}^{-1},\sqrt{V(U_\epsilon(v(\Delta')))}\})
E_I(S_{ST}(\Delta'))]
\ea
in which the $1/\sqrt{\det(q)}$ was removed from the denominator and so
qualifies as the starting point for the quantization. The pointwise limit of
(\ref{7.31}) on the phase space gives back (\ref{7.24})
for {\it any} triangulation.

The theme repeats : in order to arrive at a well-defined result on a
dense set of vectors given by functions cylindrical over graphs $\gamma$
one must adapt the triangulation to the $\gamma$ in question. The
limit of (\ref{7.31}) with respect to the so obtained $T(\epsilon,\gamma)$
gives still back (\ref{7.24}). The only new ingredient of the triangulation
as compared to the one outlined in section \ref{s6.3} is that, at fixed
$\epsilon$, we deform the surfaces $S_{MN}(\Delta)$, controlled by a further
parameter $\delta$, to the effect that $\lim_{\delta\to 0}
S_{MN}(\Delta,\delta)=S_{MN}(\Delta)$ and at finite $\delta$ the edge
$e_L(\Delta),\;\epsilon^{LMN}=1$ is the only one that intersects
$S_{MN}(\Delta,\delta)$ transversally. This can be achieved by detaching
$S_{MN}(\Delta)$ slightly from $v(\Delta)$ and otherwise choosing the
shape of $S_{MN}(\Delta)$ appropriately.
After replacing Poisson brackets by commutators times $1/(i\hbar)$ and the
Yang-Mills
electric field by $-i\hbar Q^2$ times functional derivatives we first get a
family
of  operators $\hat{H}^{\epsilon,\delta}_{YM,el}(N)_\gamma$, the limit
$\delta\to 0$ of which, in the topology of smooth connections, converges to
a family of operators $\hat{H}^\epsilon_{YM,el}(N)_\gamma$ which can be
extended to all of $\overline{{\cal A}}$. One verifies that this family of
operators, for sufficiently small $\epsilon$ depending on $\gamma$
qualifies as the set of cylindrical projections of an
operator $\hat{H}^\epsilon_{YM,el}(N)$ and the limit
$\hat{H}_{YM,el}(N)$ as $\epsilon\to 0$
in the URST exists and is given by
$\hat{H}^{\epsilon_0}_{YM,el}(N)$ for any arbitrary but fixed
$\epsilon_0>0$. We give the final result
\ba \label{7.32}
\hat{H}_{YM,el}(N)f_\gamma&=&-\frac{m_p\alpha_Q}{2\ell_p^3}
\sum_{v\in V(\gamma)}\;\sum_{v(\Delta)=v(\Delta')=v}
\frac{N(v)}{E(v)^2}
\mbox{tr}(\tau_i h_{e_M(\Delta)}
[h_{e_M(\Delta)}^{-1},\sqrt{\hat{V}(U_{\epsilon_0}(v(\Delta)))}])
\times \nonumber\\
&\times&
\mbox{tr}(\tau_i h_{e_N(\Delta')}
[h_{e_M(\Delta')}^{-1},\sqrt{\hat{V}(U_{\epsilon_0}(v(\Delta')))}])
\underline{R}^I_{e_M(\Delta)}\underline{R}^I_{e_N(\Delta')}\; f_\gamma
\ea
where the Planck mass $m_p=\sqrt{\hbar/\kappa}$ and the dimensionless
fine structure constant $\alpha_Q=\hbar Q^2$ have peeled out (in our
notation, $Q^2$ has the dimension of $1/\hbar$) while
the Planck volume $\ell_p^3$ in the denominator makes the rest of
expression dimensionless.
As before, $\underline{R}^I_e=\underline{R}^I(\underline{h}_e)$ and
$\underline{R}^I(\underline{g})$ is the right invariant vector field
on $G$ and $\underline{h}_e$ is the holonomy of $\underline{A}$ along $e$.
Expression (\ref{7.32}) is manifestly gauge invariant and diffeomorphism
covariant.

Notice that, expectedly, (\ref{7.32}) resembles (minus) a Laplacian.
Indeed, one can show \cite{IX} that $\hat{H}_{YM,el}(N=1)$ is an essentially
self-adjoint, positive semi-definite operator on $\cal H$. In particular,
(\ref{7.32}) is densely defined and does not suffer from any singularities,
it is {\it finite} ! This extends to the magnetic part of the Yang-Mills
Hamiltonian whose action on cylindrical functions is given by
\ba \label{7.33}
\hat{H}_{YM,mag}(N)f_\gamma&=&-\frac{m_p}{2\alpha_Q
(12 N^2\ell_p^3}
\sum_{v\in V(\gamma)}\;\sum_{v(\Delta)=v(\Delta')=v}
\frac{N(v)}{E(v)^2}\epsilon^{LMN}\epsilon^{RST} \times \nonumber\\
&\times&
\mbox{tr}(\tau_i h_{e_L(\Delta)}
[h_{e_L(\Delta)}^{-1},\sqrt{\hat{V}(U_\epsilon(v(\Delta)))}])
\times\\
&\times&
\mbox{tr}(\tau_i h_{e_R(\Delta')}
[h_{e_R(\Delta')}^{-1},\sqrt{\hat{V}(U_\epsilon(v(\Delta')))}])
\mbox{tr}(\underline{\tau}_I\underline{h}_{\alpha_{MN}(\Delta)})
\mbox{tr}(\underline{\tau}_I\underline{h}_{\alpha_{ST}(\Delta')})
\; f_\gamma\nonumber
\ea
(we use the normalization $\mbox{tr}(\underline{\tau}_I\underline{\tau}_J)
=-\delta_{IJ}/N$ for the normalization of the generators of
Lie$(G)$). Notice the non-perturbative dependence of (\ref{7.33})
on the fine structure constant.

In summary, the Yang-Mills contribution to the Hamiltonian constraint can be
densely defined on $\cal H$. We can see explicitly the regularizing role
that the gravitational quantum field has played in the quantization
process : the volume operator acts only at vertices of a graph and
therefore also restricts the Yang-Mills Hamiltonian to an action at those
points. Therefore, {\it the volume operator acts as an Infra-Red-Cutoff} !
Next, the divergent factor $1/\epsilon^3$ stemming from the
point-splitting of the two Yang-Mills electric fields was {\it absorbed}
by the volume operator which {\it must happen} in order to preserve
diffeomorphism covariance as the point splitting volume should not be
measured by the coordinate background metric but by the dynamical metric
itself. Therefore, {\it the volume operator also acts as an
Ultra-Violet-Cutoff} ! The volume operator thus plays a key role in the
quantization process which is why a more detailed knowledge about its spectrum
would be highly desirable.

One can verify that the Quantum Dirac algebra
of the complete Hamiltonian constraint $\hat{H}(N)=\hat{H}_{Einstein}(N)
+\hat{H}_{YM}(N)$ closes in a similar fashion as outlined in section
\ref{s6.4}. As shown in \cite{IX}, this extends to the
Fermionic and Higgs sector as well.

That all of this is not coincidence will be the subject of the next
subsection.

\subsubsection{A General Quantization Scheme}
\label{s7.2.2}

Looking at what happened in sections \ref{s6.3} and \ref{s7.2.1}
it seems that one can quantize any Hamiltonian constraint which is a
scalar density
of weight one in such a way that it is densely defined. Indeed, in \cite{IX}
we gave a proof for this which we sketch below (we restrict ourselves
here to non-fermionic matter and to $D=3$ spatial dimensions for the
sake of clarity). It applies to any
field theory in any dimension $D\ge 2$ which is given in Hamiltonian form, that
is, any generally covariant field theory deriving from a Lagrangian (for
theories including higher derivatives as in higher derivative gravity
\cite{106} or as predicted by the effective action of string theory
\cite{11} one can apply the Ostrogradsky method \cite{107} to bring it
into Hamiltonian form).

Suppose then that we are given a scalar density $H(x)$ of weight one.
Without loss of generality we can assume that all the momenta $P$ of the
theory are tensor densities of weight one and act by functional
derivation with respect to
the configuration variables $Q$ which are associated dual tensor densities
of weight zero. By contracting them with triad and co-triad fields we obtain
new canonical variables without tensor indices but with $su(2)$ indices. The
corresponding canonical
transformation is generated by a functional which changes the definition
of the real-valued connection variable $A_a^i$ but preserves its
real-valuedness and thus does not spoil the kinematical Hilbert space of
section \ref{s3}. Spatial covariant derivatives are then with respect to
$A_a^i$.

The general form of this density $H(x)$ is then a sum of homogeneous
polynomials of the form (not displaying internal indices)
\be \label{7.34}
H_{m,n}(x)=[P(x)]^n E^{a_1}(x)..E^{a_m}(x) f_{m,n}[Q]_{a_1..a_m}(x)
\frac{1}{[\sqrt{\det(q)}(x)]^{m+n-1}}
\ee
where $f$ is a local tensor depending only on configuration variables
and their covariant derivatives with respect to $A_a^i$. In order to
quantize (\ref{7.34}) we must point split the momenta $P, E^a$.
Multiply (\ref{7.34}) by $1=[\frac{|\det((e_a^i))|}{\sqrt{\det(q)}}]^k$
where $k=0,1,2,...$ is an integer to be specified later on.
Since up to a numerical constant
$|\det((e_a^i))|$ equals $\epsilon^{abc}\epsilon_{ijk}
\{A_a^i,V(R)\}\{A_b^j,V(R)\}\{A_c^k,V(R)\}$ for some appropriately chosen
region we see that this factor is worth $Dk$ volume functionals in the
numerator and $k$ factors of $\sqrt{\det(q)}$ in the denominator.
We now introduce $m+n+k-1$ point splittings by the point-splitting
functions $\chi_{\epsilon,x}(y)/\epsilon^D$ of the previous subsection
to point split both the momenta and the factors of $|\det((e_a^i))|$.
The factor $1/\epsilon^{D(m+n+k-1)}$ can be absorbed into the
$\sqrt{\det(q)}$'s as before
so that we get a power
of $m+n+k-1$ of volume functionals of the form $V(U_\epsilon(x))$ in the
denominator. Now choose $k$ large enough until $Dk>m+n+k-1$ or
$(D-1)k>m+n-1$. By suitably choosing the arguments in the process of
point-splitting and choosing $R_{.}=U_\epsilon(.)$ we can arrange, as in the
previous subsection,
that the only dependence of (\ref{7.34}) on the volume functional
is through $Dk$ factors of the form
\be \label{7.35}
\frac{\{A_a^i,V(U_\epsilon)\}}{V(U_\epsilon)^{\frac{m+n+k-1}{Dk}}}
=\frac{\{A_a^i,V(U_\epsilon)^{1-\frac{m+n+k-1}{Dk}}\}}
{1-\frac{m+n+k-1}{Dk}}
\ee
so that the volume functional is removed from the denominator. The rest
of the quantization proceeds by choosing a triangulation of $\sigma$
replacing connections by holonomies along its edges, Higgs fields
by point holonomies at vertices, momenta by functional derivatives and
Poisson brackets by commutators. By carefully choosing the factor ordering
(momenta to the right hand side)
one always finds a densely defined operator whose limit (as the regulator
is removed) exists in the URST and whose commutator algebra is non-anomalous.

The proof shows that the density weight of one for $H(x)$ was crucial :
If it would be lower than one then point splitting would result in a
regulated operator whose limit is the zero operator and if it is higher
than one then the limit diverges as said already earlier. Notice that
the final result suffers from factor ordering {\it ambiguities} but {\it
not} from factor ordering {\it singularities}.

\newpage

\section{Semiclassical Analysis}
\label{s8}

Despite the positive mathematical results concerning the quantization of
$H(N)$, there are good reasons to be at least careful about accepting
that it is  physically correct. There are a number of reasons for this:
\begin{itemize}
\item[i)] {\it Commutator Algebra}\\
We have seen that the commutator algebra of the Hamiltonian constraints
among each other on ${\cal H}^0$ does not obviously resemble the
classical Poisson bracket algebra. One possible reaction would be:
``I could not care less\footnote{Comment by Bryce DeWitt on a talk
by the author during the Meeting ``MG IX", Rome, July 2000.} !"
The reason is that all that is physically important is that the constraint
algebra be represented correctly on the physical Hilbert space.
Let us give an example: Suppose we have a classical Poisson bracket
algebra of functions $J_j=\epsilon_{jkl} x_k p_l$ on the phase
space $T^\ast \Rl^3$ given by $\{J_j,J_k\}=\epsilon_{jkl} J_l$
and that we would like to impose the constraints $\hat{J}_j\psi=0$.
Certainly we can quantize $\hat{J}_j=\epsilon_{jkl} \hat{x}_k \hat{p}_l$
and obtain a representation of the Poisson bracket algebra on the
kinematical Hilbert space ${\cal H}^0=L_2(\Rl^3, d^3x)$ in the usual
fashion. Consider now instead the representation
$\hat{J}'_1=\partial/\partial\theta,\; \hat{J}'_2=\partial/\partial\varphi,\;
\hat{J}'_3=0$ where $\theta,\varphi$ are the usual angular coordinates.
This choice is motivated by the fact that the $\hat{J}_k$ are linear
combinations of the $\hat{J}_k'$. Now, although the algebra of the
$\hat{J}'_k$ is Abelean on ${\cal H}^0$ both sets of constraints select
the same space of solutions, namely the wave functions that depend on the
radial coordinate only.

The example shows that one could be lucky, that is, the quantum evolution
of unphysical states may not resemble the classical evolution of
unphysical functions while there is a match for physical states and
physical observables, however, it would be far more convincing in our
case to have a match at the kinematical level as well because {\it we do
not know what the physical observables of the theory are}.
\item[ii)] {\it Non-Perturbative Hilbert Space}\\
The kinematical Hilbert space ${\cal H}^0$ is a non-perturbative one
which is drastically different from the usual perturbative Fock spaces
of quantum field theory on a given background spacetime. Therefore,
all the beautiful coherent state machinery that is available for
Fock spaces and the associated intuition that quantum field theorists
have developed over the last seven decades is completely lost.
That a given operator has the correct classical limit is no longer
``obvious by inspection".
\item[iii)] {\it Tremendous Non-Linearity}\\
Not only is $H(N)$ is no polynomial in the basic variables $A,E$,
it is not even an analytic function of those. The way we defined it
involved the volume functional which itself is not an analytic function
and the replacement of its Poisson brackets by commutators. Next, the basic
operators, due to background independence are not smeared in the standard
way. Finally, we had to invoke a choice function. All these steps are
{\it never performed in standard quantum field theory} since
one is always dealing with polynomials and can drop the choice function
from the beginning because the available background metric fixes
the (point splitting) regularization in a unique way.
\end{itemize}
Thus,
what we need are new tools in order to investigate what the (semi)classical
limit of the theory is and to get control on it, it is in fact the next
logical step in the quantization programme.
In particular, we want to verify that the Hamiltonian constraint operator
really has the classical Hamiltonian
constraint as its classical limit or to find out in the course of the
analysis whether it must be modified and how. It is therefore
necessary to understand in general what one means by the classical
limit of the background independent quantum field theory that we have at
our disposal now.

Roughly, the idea is to construct states with
respect to which the gravitational degrees of freedom
behave almost classical, that is, their fluctuations are
minimal. It is clear that in order to test the correctness
of any constraint operator one has to construct
first of all kinematical semiclassical states since one
cannot test an operator on its kernel. After we have made sure that
the quantization of the Hamiltonian constraint is admissable, we
will pass to physical coherent states.

Three proposals for semiclassical states have appeared in the literature
so far:\\
Historically the first ones are the so-called ``geometric weaves"
\cite{108} which try
to approximate kinematical geometric operators only.
Also ``connection weaves" have been considered \cite{109} (see also
\cite{109a} for a related proposal)
which are geared to approximate kinematical holonomy operators.
Finally, one can get rid of a
certain graph dependence of geometrical weaves
through a clever statistical average \cite{110}
resulting in ``statistical weaves".

The second proposal is based on the construction
of coherent states for full nonlinear, non-Abelean
 Quantum General Relativity
\cite{111,112,113,114} with all the desired properties
like overcompleteness, saturation of the Heisenberg
uncertainty relation, peakedness in {\it phase space}
(thus both connection and electric flux are well approximated),
construction of annihilation and creation operators and
corresponding Ehrenfest theorems. Given such a coherent
state, its excitations can be interpreted as the analogue
of the usual graviton states \cite{114a}. One can combine these
methods with a statistical
average of the kind considered above to elimintae the
graph dependence. The states are naturally cylindrical
projections of distributions in $C^\infty(\ab)^\ast$.

Finally, the third proposal \cite{117} seems to be especially
well suited for the semi-classcal analysis for free Maxwell theory
and linearized gravity. It uses a striking isomorphism between the
the usual Poisson algebra in terms of connections smeared in
$D$ dimensions and unsmeared electric fields on the one hand
and the algebra obtained by one-dimensionally
smeared connections and electric fields smeared in $D$
dimensions on the other hand. Using
this observation, which however does not carry over to the non-Abelean case,
one can carry Fock like coherent states into distributions
over $C^\infty(\ab)$ and drag the Fock inner product
into an inner product on the space of these distributions.
See also \cite{117a} for closely related work.
In \cite{119} it is shown that, for the Abelean case, the
dragged Fock measure and the uniform
measure are mutually singular with respect to each other and
that the dragged Fock measure does not support an electric
field operator smeared in $D-1$ dimensions which are essential to
use in the non-Abelean case. This indicates that
all the nice structure that comes with $U(1)$ does not generalize
to $SU(2)$. Nontheless the formula for these
distributions suggests a transcription to the non-Abelean case \cite{118}
but it remains to be seen whether the non-distributional
cylindrical projections of these distributional
Fock states (called ``shadows" there) have the desired semiclassical
properties.

In what follows we will describe these proposals in some detail,
however, since many details are still in flow we will
restrict to presenting the main ideas without going too much
into the technicalities.

\subsection{Weaves}
\label{s8.1}

\begin{itemize}
\item[a)] {\it Geometrical Weaves}\\
The early geometric weaves (first reference in \cite{108}) were
constructed as follows:\\
Let $q^0_{ab}$ be a background metric. Notice that we are not
introducing some background dependence here, all states still
belong to the background independent Hilbert space ${\cal H}^0$,
we are just looking for states that have low fluctuations around a given
classical three -- metric.  Using that metric, sprinkle
non-intersecting (but possibly linked), circular, smooth loops at random
with mean separation $\epsilon$ and mean radius $\epsilon$ (as
measured by $q^0_{ab}$. The union of these
loops is a graph, more precisely a link $\gamma$ without intersections.
The used random process was, however,
not specified in \cite{108}. Consider the state given by the product
of the traces of the holonomies along those loops.
The reason for choosing non-intersecting loops
was that such a state was formally annihilated by the Hamiltonian constraint.
Consider any
surface $S$. From our discussion in section \ref{s5} it is clear
that this state is an eigenstate of the area operator
$\widehat{\mbox{Ar}}(S)$ with eigenvalue $\ell_p^2\sqrt{3} N(S,q^0,\epsilon)/4$
where $N(S,q^0,\epsilon)$ is the number of intersections of $S$ with
the link $\gamma$. If $q^0$ does not vary too much at the scale $\epsilon$
then this number is roughly given by $\mbox{Ar}_{q^0}(S)/\epsilon^2$. Notice that
all of this was done still in the complex connection representation
and therefore outside of a Hilbert space context. Yet, the eigenvalue
equation $\ell_p^2 \mbox{Ar}_{q^0}(S)/\epsilon^2$ tells us {\it that canonical
quantum gravity seems to have a built in finiteness}: It does not make
sense to take an arbitrarily fine graph $\epsilon\to 0$ since the eigenvalue
would blow up. In order to get the corrrect eigenvalue one must
take $\epsilon\approx \ell_p$, that is, the loops have to be sprinkled at
Planck scale separation. This observation rests crucially on the fact
that there is an area gap.\\
\\
These calculations were done for metrics $q^0$ that are close to being
flat. In the second reference of \cite{108} weaves for Schwarzschild
backgrounds were considered and require an adaption of the sprinkling process
to the local curvature of $q^0$ in order that one obtains reasonable
results.\\
\\
Finally, in the third reference of \cite{108} the link $\gamma$
was generalized to disjoint collections of triples of smooth multi -- loops.
Each triple intersects in one point with linearly independent tangents
there. The motivation for this generalization was that then the volume
operator (which vanishes if there are no intersections) could also be
approximated by the same technique.
\item[b)] {\it Connection Weaves}\\
For an element $h$ of $SU(2)$ we have $\mbox{Tr}(h)\le 2$ where equality
is reached only for $h=1$. Thus $h\mapsto 2-\mbox{tr}(h)$ is a non-negative
function. Let now $\alpha$ be one of the loops
considered in the third reference of \cite{108} and let $A\in \ab$. Then
$A\mapsto e^{-\beta[2-\mbox{tr}(A(\alpha))]}$ is sharply peaked
at those $A\in \ab$ with $A(\alpha)=0$, that is, at a flat connection
(since the $\alpha$ are contractible). Arnsdorf \cite{109} then
considers the product of all those functions which
is concentrated on those distributional connections which, when
restricted to the subgroupoid $l=l(\gamma)$, are flat (this
function is precisely of the form of the exponential of the Wilson
action employed in lattice gauge theory \cite{84}).

Since \cite{109} is written in the context of the Hilbert space
${\cal H}^0$ and since non-compact topologies of $\sigma$ were considered,
in contrast to \cite{108} one had to deal with the case that the graph
$\gamma$ becomes infinite (the number of loops becomes infinite).
Since such a state is not an element of ${\cal H}^0$, Arnsdorf
constructed a positive linear functional on the algebra of local
operators using that formal state and then used the GNS construction
(see section \ref{sg}) in order to obtain a new Hilbert space
in which one can now compute expectation values of various operators.
Expectedly, holonomy operators along paths in $l$ have expectation values
close to their classical value at flat connections while the semi-classical
behaviour of electric flux operators is less clear.
\item[c)] {\it Statistical Weaves}\\
In both the geometric and connection weave construction an arbitrary but
fixed graph $\gamma$ had to be singled out. This is unsatisfactory
because it involves a huge amount of arbitrariness. Which graph should
one take ? Also, unless the graph $\gamma$ is sufficiently random
the expectation values, say of the area operator in a geometric
weave for a flat background metric $q^0$ is not rotationally invariant.

To improve this, Bombelli \cite{110} has employed the Dirichlet -- Voronoi
construction, often used in statistical mechanics \cite{110a}, to the
geometrical weave. Roughly, this works as follows:\\
Given a background metric $q^0$, a {\it compact} hypersurface $\sigma$,
and a density parameter $\lambda$
one can construct a subset $\Gamma(q^0,\lambda)\subset \Gamma^\omega_0$
of piecewise analytic graphs each of which, in $D$ spatial dimensions, is
such that each of its vertices is $(D+1)-$valent.
A member $\gamma_{x_1,..,x_N}\in\Gamma(q^0,\lambda)$ is labelled by
$N\approx[\lambda \mbox{Vol}_{q^0}(\sigma)]$ points $x_k\in\Sigma$ where
$[.]$ denotes the Gauss bracket. The graph $\gamma_{x_1,..,x_N}$
is obtained unambiguously from the set of points $x_1,..,x_N$
and the metric $q^0$ (provided that it is close to being flat)
by employing natural notions like minimal geodesic distances etc.
Next, given a spin label $j$ and an intertwiner $I$ we can construct
a gauge invariant spin-net $s_{x_1,..,x_N}(j,I)$ by colouring each edge
with the same spin and each vertex with the same intertwiner.
From these data one can construct the ``density operator"
\be \label{8.1}
\hat{\rho}(q^0,\lambda,I,j)
:=\int_{\sigma^N} d\mu_{q^0}(x_1)..
d\mu_{q^0}(x_N)\; T_{s_{x_1,..,x_N}(j,I)}
<T_{s_{x_1,..,x_N}(j,I)},.>
\ee
where
\be \label{8.2}
d\mu_{q^0}(x):=\frac{\sqrt{\det(q^0)(x)} d^Dx}{\mbox{Vol}_{q^0}(\sigma)}
\ee
is a probability measure (it is here where compactness of $\sigma$ is
important). The reason for the inverted commas in ``density operator"
is that (\ref{8.1}) actually is the zero operator \cite{110b}.
To see this, notice that for any spin-network state $T_s$ we have
$<T_{s_{x_1,..,x_N}(j,I)},T_s>=\delta_{s_{x_1,..,x_N}(j,I),s}$
which in particular means that $\gamma_{x_1,..,x_N}=\gamma(s)$.
But the set of points satisfying this is certainly thin with respect to
the measure (\ref{8.2}). What happens is that although for any
spin-network state $T_s$ the one-dimensional projector $T_s <T_s.>$
is a trace class operator of unit trace, the trace operation does not
commute with the integration in (\ref{8.1}). However, one can then define
a positive linear functional $\omega_{q^0,\lambda,I,j}$ on the algebra
of linear operators on ${\cal H}^0$ by
\be \label{8.3}
\omega_{q^0,\lambda,j,I}(\hat{O}):=
\int_{\sigma^N} d\mu_{q^0}(x_1)..
d\mu_{q^0}(x_N)\; <T_{s_{x_1,..,x_N}(j,I)},\hat{O}T_{s_{x_1,..,x_N}(j,I)}>
\ee
which would equal $\mbox{Tr}(\hat{\rho}(q^0,\lambda,j,I)\hat{O})$
if integration and trace would commute. Via the GNS construction one can
now define a new representation ${\cal H}^0_{q^0,\lambda,j,I}$ which
now depends on a background structure. The representations ${\cal H}^0$
and ${\cal H}^0_{q^0,\lambda,j,I}$ are certainly not comparable in the sense
that one can embed one space into the other and presumably they are
(unitarily) inequivalent.

What is interesting about (\ref{8.3}) is that for an exactly flat
background the expectation values of, say the area operator, are
{\it Euclidean invariant}. In order to match the expectation values
of $\widehat{\mbox{Ar}}(S)$ with the value $\mbox{Ar}_{q^0}(S)$
one must choose $j$ according to
$[\sqrt{j(j+1)}\ell_p^2\beta\lambda^{2/3}/2]=1$. A similar calculation
for the volume operator presumably fixes the value $I$ for the intertwiner.
\end{itemize}

\subsection{Coherent States}
\label{s8.2}

Especially the statistical weave construction of the previous subsection
looks like a promising starting point for semiclassical analysis.
However, there are several drawbacks with weaves:
\begin{itemize}
\item[i)] {\it Phase Space Approximation}\\
All the weaves discussed above seem to approximate either the connection
or the electric field appropriately although the degree of their
approximation has never been checked (are the fluctuations small ?).
However, what we really need are states which approximate the connection
and the electric field simultaneously with small fluctuations.
\item[ii)] {\it Arbitrariness of Spins and Intertwiners}\\
All weaves proposed somehow seem to arbitrarily single out special
and uniform values for spin and intertwiners. Drawing an anology
with a system of uncoupled harmonic oscillators, it is like trying to
build a semiclassical state by choosing an arbitrary but fixed occupation
number (spin) for each mode (edge). However, we know that the preferred
semiclassical states for the harmonic oscillator are coherent states
which depend on all possible occupation numbers. As we will see,
issue i) and ii) are closely related.
\item[iii)] {\it Arbitrariness of Graphs}\\
Even in the statistical weave construction we select arbitrarily only
a certain subclass of graphs. Again, drawing an anology with the
harmonic oscillator picture, this is like selecting a certain subset
of modes in order to build a semiclassical state. However, then not all
modes can behave semi-classically.
\item[iv)] {\it Missing Construction Principle}\\
The weave states constructed suffer from a missing enveloping construction
principle that would guarantee from the outset that they possess
desired semi-classical properties.
\end{itemize}
The aim of the series of papers \cite{111,112,113,114} was to decrease
this high level of arbitrariness, to look for a systematic construction
principle and to make semiclassical states for
quantum gravity look more similar to the semiclassical states for free
Maxwell theory which are in fact {\it coherent states} and have been
extremely successful, see e.g. \cite{120} and referencese therein.

\subsubsection{Semiclassical States and Coherent States}
\label{s8.2.1}

Recall that quantization is, roughly speaking, an attempt to construct a
$^\ast$ homomorphism
\be \label{8.4}
\bigwedge:\;({\cal M},\{.,.\},{\cal O},\overline{(.)})\to
({\cal H},\frac{[.,.]}{i\hbar},\widehat{{\cal O}},(.)^\dagger)
\ee
from a subalgebra ${\cal O}\subset C^\infty({\cal M})$ of the Poisson
algebra of complex valued functions on the symplectic manifold
$({\cal M},\{.,.\})$ to a subalgebra $\widehat{{\cal O}}\subset
{\cal L}({\cal H})$ of the algebra of linear operators on a
Hilbert space ${\cal H}$ with inner product $<.,.>$ such that
Poisson brackets turn into commutators and complex conjugation into
the adjoint operation. Notice that the map cannot be extended to all
of $C^\infty({\cal M})$ (only up to quantum corrections) unless
one dives into deformation quantization, see e.g. \cite{121} and references
therein, the subalgebra for which it holds is referred to as the algebra
of elementary functions (operators). The algebra ${\cal O}$ should be
sufficiently large in order that more complicated functions can be expressed
in terms of elements of it so that they can be
quantized by choosing a suitable factor ordering.

{\it Dequantization} is the inverse of the map (\ref{8.4}).
A possible way to phrase this more precisely is:
\begin{Definition} \label{def8.1} ~~~~~~~~~~~~~\\
A system of states $\{\psi_m\}_{m\in {\cal M}}\in{\cal H}$ is said
to be semiclassical for an operator subalgebra
$\overline{{\cal O}}\subset {\cal L}({\cal H})$ provided that
for any $\hat{O},\hat{O}'\in \widehat{{\cal O}}$ and any
generic point $m\in {\cal M}$
\begin{itemize}
\item[[1.]] {\it Expectation Value Property}\\
\be \label{8.5}
|\frac{<\psi_m,\hat{O}\psi_m>}{O(m)}-1|\ll 1
\ee
\item[[2.]] {\it Infinitesimal Ehrenfest Property}\\
\be \label{8.6}
|\frac{<\psi_m,[\hat{O},\hat{O}]\psi_m>}{i\hbar\{O,O'\}(m)}-1|\ll 1
\ee
\item[[3.]] {\it Small Fluctuation Property}\\
\be \label{8.7}
|\frac{<\psi_m,\hat{O}^2\psi_m>}{<\psi_m,\hat{O}\psi_m>^2}-1|\ll 1
\ee
The quadruple $({\cal M},\{.,.\},{\cal O},\overline{(.)})$
is then called the classical limit of
$({\cal H},\frac{[.,.]}{i\hbar},\widehat{{\cal O}},(.)^\dagger)$.
\end{itemize} \end{Definition}
Clearly definition \ref{def8.1} makes sense only when none of the
denominators displayed vanish so they will hold at most at generic points
$m$ of the phase space (meaning a subset of ${\cal M}$ whose
complement has Liouville measure comparable to a phase cell) which will be
good enough for all practical
applications. Notice that if [1.] holds for $\hat{O}$ then it holds for
$\hat{O}^\dagger$ automatically. Condition $[1.]$ is for polynomial
operators sometimes required in the stronger form that (\ref{8.5})
should vanish exactly which can always be achieved by suitable
(normal) ordering prescriptions. Condition [2.] ties the commutator to
the Poisson bracket and makes sure that the infinitesimal quantum
dynamics mirrors the infinitesimal classical dynamics. If the error
in [2.] vanishes then we have a finite Ehrenfest property which in
non-linear systems is very hard to achieve. Finally, [3.] controls the
quantum error, the fluctuation of the operator.

Coherent states have further properties which can be phrased roughly as
follows:
\begin{Definition} \label{def8.2} ~~~~~~~~~~~\\
A system of states $\{\psi_m\}_{m\in {\cal M}}\in{\cal H}$ is said
to be coherent for an operator subalgebra
$\hat{{\cal O}}\subset {\cal L}({\cal H})$ provided that
for any $\hat{O},\hat{O}'\in \widehat{{\cal O}}$ and any
generic point $m\in {\cal M}$ in addition to properties
[1.], [2.] and [3.] we have
\begin{itemize}
\item[[4.]] {\it Overcompleteness Property}\\
There is a resoltion of unity
\be \label{8.8}
1_{{\cal H}}=\int_{{\cal M}} d\nu(m) \psi_m <\psi,.>
\ee
for some measure $\nu$ on $\cal M$.
\item[[5.]] {\it Annihilation Operator Property}\\
There exist elementary operators $\hat{g}$ (forming a complete system)
such that
\be \label{8.9}
\hat{g}\psi_m=g(m)\psi_m
\ee
\item[[6.]] {\it Minimal Uncertainty Property}\\
For the self-adjoint operators
$\hat{x}:=(\hat{g}+\hat{g}^\dagger)/2,\;
\hat{y}:=(\hat{g}-\hat{g}^\dagger)/(2i)$
the (unquenched) Heisenberg uncertainty relation is saturated
\be \label{8.10}
<(\hat{x}-<\hat{x}>_m)^2>_m=<(\hat{y}-<\hat{y}>_m)^2>_m=
\frac{1}{2}|<[\hat{x},\hat{y}]>_m|
\ee
\item[[7.]] {\it Peakedness Property}\\
For any $m\in {\cal M}$, the overlap function
\be \label{8.11}
m'\mapsto |<\psi_m,\psi_{m'}>|^2
\ee
is concentrated in a phase cell of Liouville volume
$\frac{1}{2}|<[\hat{p},\hat{h}]>_m|$ if $\hat{p}$
is a momentum operator and $\hat{h}$ a configuration operator.
\end{itemize} \end{Definition}
These four conditions are not completely independent of each other,
in particular, [5.] implies [6.] but altogether [1.] -- [7.]
comprises a fairly complete list of desirable properties for
semiclassical (coherent states).

\subsubsection{Construction Principle: Complexifier Method and Heat
Kernels}
\label{s8.2.2}

Usually one introduces coherent states for the harmonic oscillator
as eigenstates of the annihilation operator in terms of superpositions
of energy eigenstates. This method has the disadvantage that one needs
a preferred Hamiltonian, that is, dynamical input in order to define
suitable annihilation operators. Even if one has a Hamiltonian, the
construction of annihilation operators is no longer straightforward
if we are dealing with a non-linear system. Since we neither have
a Hamiltonian nor a linear system and since for the time being we are
anyway interested in kinematical coherent states, we have to look
for a different constructive strategy.

A hint comes from a different avenue towards the harmonic oscillator
coherent states. Let the Hamiltonian be given by
\be \label{8.12}
H:=\frac{1}{2}[p^2/m+m\omega^2 X^2]
=\omega \bar{z} z \mbox{ where }
z=\frac{\sqrt{m\omega}x-ip/\sqrt{m\omega}}{\sqrt{2}}
\ee
Define the {\it complexifier} function
\be \label{8.13}
C:=\frac{p^2}{2m\omega}
\ee
then it is easy to see that
\be \label{8.14}
z=\sqrt{\frac{m\omega}{2}}\sum_{n=0}^\infty \frac{(-i)^n}{n!}
\{C,x\}_{n}
\ee
(recall that in our terminology $\{p,x\}=1$). Translating this equation
into quantum theory we find
\be \label{8.15}
\hat{z}=\frac{\sqrt{m\omega}}{\sqrt{2}}\sum_{n=0}^\infty \frac{(-i)^n}{n!}
\frac{[\hat{C},\hat{x}]_n}{(i\hbar)^n}
=e^{-t (-\Delta/2)}\frac{\hat{x}\sqrt{m\omega}}{\sqrt{2}}
(e^{-t(-\Delta/2)})^{-1}
\ee
where the {\it classicality parameter}
\be \label{8.16}
t:=\hbar/(m\omega)
\ee
has naturally appeared and
which for this system has dimension cm$^2$. The operator $\hat{z}$
is usually chosen by hand as the annihilation operator. Let us
define that coherent states $\psi_z$ are eigenstates of $\hat{z}$.
Given formula (\ref{8.16}) we can trivially construct them as follows:
Let $\delta_x$ be the $\delta-$distribution, supported at $x$,
with respect to the Hilbert space measure $dx$.
Define $\psi_x:=e^{-t\hat{C}/\hbar^2}\delta_x$. Then formally
\be \label{8.17}
\hat{z}\psi_x=e^{-t\hat{C}/\hbar^2}\frac{\sqrt{m\omega}\hat{x}}{\sqrt{2}}
\delta_x
=\frac{x\sqrt{m\omega}}{\sqrt{2}}\psi_x
\ee
because $\delta_x$ is an eigendistribution of the operator $\hat{x}$.
The crucial point is now that $\psi_x$ is an analytic function of
$x$ as one can see by using the Fourier representation for the
$\delta-$distribution $\delta_x=\int_\Rl dk/(2\pi) e^{ikx}$.
We can therefore analytically extend $\psi_x$ to the complex plane
$x\to x-ip/(m\omega)$ and arrive with the trivial redefinition
$\psi_{x-ip/(m\omega)}\mapsto \psi_z$ at
\be \label{8.18}
\hat{z}\psi_z=z\psi_z
\ee
One can check that the state $\psi_z/||\psi_z||$ coincides with the usual
harmonic oscillator coherent states up to a phase.\\
\\
We see that the harmonic oscillator coherent states can be naturally
put into the language of the Wick rotation transform of section
\ref{s6.1}. This observation, stripping off the particulars of the
harmonic oscillator, admits a generalization that
applies to any symplectic manifold ${\cal M},\{.,.\}$ which is a
cotangent bundle ${\cal M}=T^\ast {\cal C}$ where ${\cal C}$ is the
configuration base space of $\cal M$. The essential steps can be
summarized in the following algorithm (we suppress all indices, discrete
and continuous):
\begin{itemize}
\item[1)] {\it Hilbert Space and $\delta-$Distribution}\\
The Hilbert space is supposed to be an $L_2$ space, that is, there exists
a measure $\mu$ on $\cal C$ such that ${\cal H}=L_2({\cal C},d\mu)$.
With respect to the measure $\mu$ we may define the $\delta-$distribution
supported at $x\in {\cal C}$ by the formula
$\delta_x(f):=\int_{{\cal C}} d\mu(x') \delta_x(x') f(x')=f(x)$
for any $f\in C^\infty_0({\cal C})$ (or any other dense space
of tests functions). Here we have denoted the integral kernel of the
distribution by $\delta_x(x')$.
\item[2)] {\it Complexifier and Heat Kernel Evolution}\\
Find a non-negative function $C$ on $\cal M$ which can be quantized
on ${\cal H}$ as a positive definite, self-adjoint operator.
Moreover, the dimensions of $x,\{C,x\}$ should coincide.
Then
$e^{-\hat{C}/\hbar}$ is a bounded operator and can be defined via the
spectral theorem.
Furthermore, we need that the heat kernel evolution of the
$\delta-$distribution $\psi_x:=e^{-\hat{C}/\hbar}\delta_x$ is a square
integrable function in ${\cal H}$ which at the same time is analytic
in $x$.
\item[3)] {\it Analytic Continuation and Annihilation Operators}\\
Let $\psi_z$ be the analytic continuation of $\psi_x$ and define the
anniliation operator $\hat{z}:=e^{-\hat{C}/\hbar} \hat{x}
(e^{-\hat{C}/\hbar})^{-1}$. Then automatically
$\hat{z}\psi_z=z\psi_z$ is an eigenstate. Notice that the inverse
$(e^{-\hat{C}/\hbar})^{-1}$ is only densely defined on functions of
the form $e^{-\hat{C}/\hbar}f,\;f\in {\cal H}$ which have been
smoothened out by $e^{-\hat{C}/\hbar}$.
\item[4)] {\it Classicality Parameter and Physical Interpretation}\\
The quantity $\hat{C}/\hbar$ is dimensionfree by construction. The
classicality parameter $t$ is defined by $\hat{C}/\hbar=-t\Delta/2$
where $\Delta$ is a negative definite differential operator of order greater
than one in order that, according to the rule $\hat{p}=i\hbar
\partial/\partial x$, the parameter $t$ is proportional to a positive
power of $\hbar$ and therefore small. It is clear that $\hat{z}$ is a
quantization of $z:=\sum_{n=0}^\infty \frac{(-i)^n}{n!}\{C,x\}_n$.
We require further that $C$ has been chosen in such a way that
the functions $z,\bar{z}$ suffice to isolate configuration and
momentum function $x,p$ respectively. Thus we have an invertible map
$m=(x,p)\mapsto (z,\bar{z})$ and can finally define $\psi_m:=\psi_{z(m)}$.
\end{itemize}
Certainly, steps 1) -- 4) are only formal and have to be justified
mathematically in the model at hand. However, given an $L_2$ Hilbert
space over the configuration space, they merely require {\it one
input}: the choice of the complexifier $C$ and which one selects depends
on some physical input.

The complexifier method is extremely natural:
Besides the fact that, as one can show, any coherent states that have
been constructed for linear field theories actually fall into
the catgory of states that have been constructed by the complexifier method,
automatically the following coherent state properties (formally) hold:
The annihilation operator property [5.] trivially holds by construction
and hence the saturation of the unquenched minimal uncertainty
relation, property [6.], as well. Moreover, the expectation value
property [1.] automatically holds for any normal ordered polynomial
of the $\hat{z},\hat{z}^\dagger$. The overercompleteness property
[4.] is equivalent to showing that the coherent state transform
$(\hat{U}f)(z):=[e^{-\hat{C}/\hbar}f(x)]_{|x\to z}$ introduced
in section \ref{s6.1} is unitary and we have given there a formal
recipe that constructs a measure $\nu$ on the complexification ${\cal C}^\Cl$
such that the transform becomes a partial isometry at least (the hard part
is to show that the transform is onto the space of holomorphic
$\nu-$square integrable functions on ${\cal C}^\Cl$).
The peakedness property [7.] is at least rather likely to hold because
what $e^{-\hat{C}/\hbar}$ does to the $\delta-$distribution (which is
{\it sharply peaked}) is to decrease the size of the peak and to increase
its width (of the order $\sqrt{t}$) at least in the configuration
representation. Next, again for polynomials of $\hat{z},\hat{z}^\dagger$
the infinitesimal Ehrenfest property [2.] should follow from the correct
quantization of $\hat{p},\hat{x}$ (less trivial are non-polynomial
functions, which however crucially appear in our applications).
Finally, the small fluctuation property [3.] trivially holds for polynomials
of $\hat{z}$ alone or $\hat{z}^\dagger$ alone and therefore holds for
more general polynomials as well if [2.] holds.

This concludes our motivation for considering complexifier coherent states.

\subsubsection{Coherent States for Canonical Quantum General Relativity}
\label{s8.2.3}

Let us now apply the framework of the previous subsection
to canonical quantum general relativity. We have a (quantum) configuration
space ${\cal C}=\ab$ and a measure $\mu_0$ thereon which together
build the Hilbert space ${\cal H}^0=L_2(\ab,d\mu_0)$. The
$\delta-$distribution on $\ab$ with respect to the measure $\mu_0$ is given
by
\be \label{8.18a}
\delta_A=\sum_{s\in {\cal S}} T_s(A)<T_s,.>
\ee
So in principle, all that remains to do is to find a suitable
complexifier. A natural choice is the volume operator
$\hat{\mbox{Vol}}(\sigma)$ as the complexifier
\cite{111} because it is background independent, gauge invariant, spatially
diffeomorphism invariant, a differential operator of order
$3/2>1$, positive
semi-definite and selfadjoint. In
order that classically $A,\{C,A\}$ have the same dimension
we will choose a parameter $a$ with dimension of length
whose physical significance will become clear only later
and define $C:=2\mbox{Vol}(\sigma)/(\kappa a)$ (again we take $\beta=1$).
Then
\be \label{8.19}
\hat{C}/\hbar=\frac{\ell_p}{a}\frac{\hat{V}(\sigma)}{\ell_p^3}
\ee
is dimensionfree and the classicality parameter is given by $t=\ell_p/a$
which should be much smaller than unity.

It is easy to see that for any path $p\in {\cal P}$ we have
\be \label{8.20}
\sum_{n=0}^\infty \frac{(-i)^n}{n!}\{C,h_p\}(A,E)=h_p(A-ie/a)
\ee
where $e=e_a^j dx^a \tau_j/2$ is the co-triad one-form, that is,
we get the holonomy of an $SL(2,\Cl)$ connection. Therefore, we
know the classical correspondence of the annihilation operators
\be \label{8.21}
\hat{g}_p:=e^{-\hat{C}/\hbar}\hat{h}_p e^{\hat{C}/\hbar}
\ee
The crucial question is now whether $e^{-\hat{C}/\hbar}\delta_A$
is analytic in $A$. In order to compute this quantity we switch
to a new orthonormal basis $s=(\gamma(s),\lambda(s),I(s))$ where for given
$\gamma$ we have diagonalized all spin-network states over $\gamma$
with eigenvalues $\lambda$ of $\hat{V}(\sigma)/\ell_p^3$ and degeneracy
labels $I$. Then
\be \label{8.22}
e^{-\hat{C}/\hbar}\delta_A
=\sum_s e^{-t\lambda(s)} T_s(A) <T_s,.>
\ee
Since the functions $T_s(A)$ are analytic in $A$ we may
define our coherent states to be
\be \label{8.23}
\psi_{A^\Cl}
:=\sum_s e^{-t\lambda(s)} T_s(A^\Cl) <T_s,.>
\ee
where $A^\Cl\in \abc:=\mbox{Hom}({\cal L},SL(2,\Cl))$.
Of course, (\ref{8.23}) can be applied to any suitable
complexifier $C$.

There are several problems with (\ref{8.23}):\\
1)\\
Although
it defines an element of ${\cal D}^\ast$, it is not an element
of ${\cal H}^0$ and in order to use these states in expectation value
calculations one would need to introduce a new inner product
for them just as we have to do for solutions of the Hamiltonian constraint.

There is no obvious choice for such an inner product at all.
One could of course try to normalize (\ref{8.23}) but the
resulting
expressions become hard to control because the eigenbasis of the
volume operator is not known.\\
2)\\
One could consider the ``cut-off" states
\be \label{8.24}
\psi_{\gamma,A^\Cl}:=\sum_{\gamma'\subset\gamma}
\sum_{s;\;\gamma(s)=\gamma'} e^{-t\lambda(s)} T_s(A^\Cl) <T_s,.>
\ee
where the sum is over all subgraphs of $\gamma$, that is,
those that arise by removing edges from $E(\gamma)$, one by one
in all possible ways. (Here a difference arises depending
on whether we work at the gauge variant or gauge invariant
level because in the latter case one can remove edges only
in such a way that $\gamma'$ has no univalent vertices. Also
if $\gamma$ contains three edges $e_1,e_2,e_3$ which meet
in one point and such that $e_1\circ e_2$ is analytic, then
after removing $e_3$ we take the convention that the point
$e_1\cap e_2$ is still a vertex of $\gamma-\{e_3\}$ in the
gauge variant case).
But even for those it is neither clear how to calculate anything
nor is it clear whether (\ref{8.24}) is an element of ${\cal H}^0$
at all because the degeneracy $N_\lambda$ of almost all $\lambda$ on
any given graph could exceed the damping factor $e^{-t\lambda}$. So
again the complicated spectrum of the volume operator makes
(\ref{8.24}) at least highly unpractical.\\
3)\\
If we would use cut-off states to do
semi-classical physics, then they are presumably
inadequate for computing expectation values of
operators with non-vanishing matrix elements between
spin network states over different graphs like the
Hamiltonian -- or Diffeomorphism Constraint.\\
4)\\
The Poisson algebra of the classical functions $h_p(A^\Cl),
\overline{h_p(A^\Cl)}$ does not close and therefore the
commutators between the associated operators should look horrible,
that is, the infinitesimal Ehrenfest property will be difficult to
verify.\\
\\
One way out is too look for a different classical function $C$,
maybe background dependent,
which at least does not have the problems 2), 3). However,
{\it for non-Abelean gauge groups there seems to be no $C$,
polynomial in the electric fields, such that
$\hat{C}$ leaves the space of cylindrical functions invariant and
simultaneously 2) and 3) disappear} \cite{122} !\\
\\
Another option is to construct a family of coherent states
$(\psi_{\gamma,(A,E)})_{\gamma\in \Gamma}$ by hand, that is,
for each $\gamma$ we choose a complexifier $C_\gamma$
and repeat the above procedure restricted to the
Hilbert space ${\cal H}_\gamma$. The function
$C_\gamma$ should cure the problems 2) and 3) just mentioned
but it is no longer required that $C_\gamma$ is the discretization
of some well-defined function $C$ on ${\cal M}$, in particular,
it will be not be the case that the family of operators
$\hat{C}_\gamma$ is consistent (although this can always be cured
by defining them in the spin-network basis). This has been
proposed in \cite{111} and works as follows:
Define
\ba \label{8.25}
\delta^\gamma_A:=\sum_{s\in {\cal S};\;\gamma(s)=\gamma} T_s(A)<T_s,.>
\nonumber\\
\delta_{\gamma,A}:=\sum_{\gamma'\subset\gamma} \delta^{\gamma'}_A
\ea
We evidently have the identitity
\be \label{8.26}
\delta_A=\sum_{\gamma\in \Gamma^\omega_0} \delta_A^\gamma
\ee
so that the second line in (\ref{8.25}) is the ``distribution
cut off at $\gamma$". A simplification arises at the
gauge variant level since then evidently
\be \label{8.27}
\delta_{\gamma,A}=\prod_{e\in E(\gamma)}\delta_{e,A}
\ee
factorizes. Now $\delta_{e,A}=\delta_{A(e)}$ where the latter
distribution is with respect to the Haar measure.
Due to the Peter\&Weyl theorem
\be \label{8.28}
\delta_{h}(h')=\sum_{\pi\in \Pi} d_\pi \chi_\pi(h (h')^{-1})
\ee
which demonstrates that with $\delta^e_A=\delta_{e,a}-1$
we also have
\be \label{8.29}
\delta^\gamma_A=\prod_{e\in E(\gamma)}\delta^e_A
\ee
Let us now specify $C_\gamma$. Given a graph $\gamma$
consider a system of mutually disjoint, open surfaces
$(S_e)_{e\in E(\gamma)}$
where $e\cap S_{e'}=\emptyset$ if $e\not=e'$ and
$x_e:=e\cap S_e$ is an interior point of both $e,S_e$.
Moreover, $S_e$ carries the orientation such that
$e$ is of the ``up" type and the collection $S_e$
is supposed to form a polyhedronal decomposition of
$\sigma$ (add some surfaces that do not intersect
$\gamma$ at all if necessary). Next, choose
a system of non-self-intersecting paths
$\rho_e(x)$ within $S_e$, one for every point $x\in S_e$
with $b(\rho_e(x))=x_e$ and
$f(\rho_e(x))=x$. From these data construct
the functions
\be \label{8.30}
P^e_j(A,E):=-\frac{1}{2a_e^2}\mbox{Tr}(\tau_j \int_{S_e}
\mbox{Ad}_{A(e_{x_e}\circ\rho_e(x))}(\ast E(x))
\ee
where $e_{x_e}$ is the segment of $e$ with
$b(e_{x_e})=b(e),\;f(e_{x_e})=x_e$ and
$\ast E=\epsilon_{abc} E^a_j\tau_j dx^b\wedge dx^c$.
Again the length parameter $a_e$ will receive its
physical meaning only later in concrete physical
applications.\\
\\
The crucial
fact about the system of functions $h_e,P^e$
is that they are gauge covariant, $\lambda_g^\ast P^e=\mbox{Ad}_{g(b(e))}(P^e)$,
in contrast to the $E_j(S)$ of section \ref{4.1}, diffeomorphism
covariant if $a_e=a$ is a constant (all edges, paths, surfaces just get mapped
to diffeomorphic
images) and they form a {\it closed Poisson subalgebra} of
$C^\infty({\cal M})$ given by
\ba \label{8.31}
\{h_e,h_{e'}\} &=& 0
\nonumber\\
\{P^e_j,h_{e'}\} &=& \frac{\kappa}{a_e^2}\delta^e_{e'}\frac{\tau_j}{2}h_e
\nonumber\\
\{P^e_j,P^{e'}_k\} &=& -\delta^{e e'}\frac{\kappa}{a_e^2}
\epsilon_{jkl} P^e_l
\ea
However, this Poisson algebra is isomorphic to the natural
Poisson algebra on ${\cal M}_\gamma:=\prod_{e\in E(\gamma)} T^\ast(SU(2))$
so what we have achieved is construct a map
\be \label{8.32}
\Phi'_\gamma:\;{\cal M}\to {\cal M}_\gamma;\;(A,E)\mapsto
(h_e(A),P^e_j(A,E))_{e\in E(\gamma)}
\ee
which is a {\it partial symplectomorphism}. (Notice that it is neither
one to one nor onto for fixed $\gamma$. Here we are abusing the
notation somewhat because $\Phi'_\gamma$ certainly also depends
on the $S_e,\rho_e(x)$). This fact is going to be
fundamental for all that follows for the following reason:
What we are really going to do is to construct coherent states for the
phase space $M_\gamma:=[T^\ast(SU(2))]^{|E(\gamma)|}$ and since
the Poisson
structures of the phase spaces $\Phi'_\gamma({\cal M})$ and
${\cal M}_\gamma$ coincide we automatically have proved the Ehrenfest
property for $\Phi'_\gamma({\cal M})$. Now, if $\gamma$ gets
sufficiently fine, we can approximate any function on $\cal M$
by functions in $\Phi'_\gamma({\cal M})$ and in that sense we are
constructing {\it approximate coherent states} for $\cal M$.

Next we must construct $C_\gamma$. In analogy to the harmonic
oscillator we choose a function which is quadratic in the momenta
because this will lead to similar Gaussian peakedness properties.
Thus we define
\be \label{8.33}
C_\gamma:=\frac{1}{2\kappa}\sum_{e\in E(\gamma)} a_e^2 (P^e_j)^2
\ee
One may check, that this leads to the complexification
\be \label{8.34}
g_e:=\sum_{n=1}^\infty \frac{(-i)^n}{n!}
\{C_\gamma,h_e\}_n=e^{-iP^e_j\tau_j/2} h_e
\ee
where the Poisson brackets are those of $\cal M$.
Something amazing has happened in (\ref{8.34}): We have stumbled
naturally on the diffeomorphism
\be \label{8.35}
T^\ast(SU(2))\to SL(2,\Cl);\;(h,P)\mapsto e^{-i P^j\tau_j/2}h
\ee
where the inverse of (\ref{8.35}) is given by polar decomposition.
Now, while the complexification of $\Rl$ is given by
$\Cl$, the complexification of a Lie group $G$ with Lie algebra
Lie$(G)$ is given by the image under the exponential map of the
complexification of its Lie algebra (that is, we allow arbitrary
complex coefficients $\theta^j$ of the Lie algebra basis $\tau_j$)
and (\ref{8.34}) tells us precisely how this is induced by the
complexifier. The map (\ref{8.35}) allows us to identify
${\cal M}_\gamma$ with $SL(2,\Cl)^{|E(\gamma)|}$ so that
we have altogether a map
\be \label{8.32a}
\Phi_\gamma:\;{\cal M}\to {\cal M}_\gamma;\;(A,E)\mapsto
m_\gamma(A,E):=(g_e(A,E):=e^{-i P^e_j\tau_j/2} h_e)_{e\in E(\gamma)}
\ee
The Poisson algebra (\ref{8.31}) suggests
on ${\cal H}^0_\gamma$ the quantization
$\hat{P}^e_j=i t_e R^j_e/2$ while $\hat{h}_e$ is a
multiplication operator. Here the classicality parameters
\be \label{8.36}
t_e:=\frac{\ell_p^2}{a_e^2}
\ee
have naturally appeared and it follows that
\be \label{8.36a}
\hat{C}_\gamma/\hbar=-\frac{1}{2}\sum_{e\in E(\gamma)} t_e \Delta_e
\ee
where $\Delta_e=(R^j_e)^2/4$. Our annihilation operators
become
\be \label{8.37}
\hat{g}_e:=e^{-\hat{C}_\gamma/\hbar}\hat{h}_e (e^{-\hat{C}_\gamma/\hbar})^{-1}
=e^{-t_e\tau_j^2/8}e^{-i\hat{P}^e_j\tau_j/2}\hat{h}_e
\ee
which up to a quantum correction is precisely
the quantization of (\ref{8.34}). Then we can
define abstract coherent states for ${\cal H}\gamma$ by
\ba \label{8.38}
\psi_{\gamma,m_\gamma}&:=&[e^{-\hat{C}_\gamma/\hbar}
\delta_{\gamma,h_\gamma}]_{h_\gamma\to m_\gamma}
\nonumber\\
&=& \prod_{e\in E(\gamma)} [e^{t_e\Delta_e/2}\delta_{h_e}]_{h_e\to g_e}
\nonumber\\
\psi^\gamma_{m_\gamma}&:=&[e^{-\hat{C}_\gamma/\hbar}
\delta^\gamma_{h_\gamma}]_{h_\gamma\to m_\gamma}
\nonumber\\
&=& \prod_{e\in E(\gamma)} [e^{t_e\Delta_e/2}\delta_{h_e}-1]_{h_e\to g_e}
\nonumber\\
\psi_g &:=& e^{t\Delta/2}\delta_h]_{h\to g}=\sum_{j=0,1/2,1,3/2,..} (2j+1)
e^{-t j(j+1)/2} \chi_j(g h^{-1})
\ea
and coherent states on ${\cal H}^0$ by
\be \label{8.39}
\psi_{\gamma,m}:=\hat{U}_\gamma \psi_{\gamma,\Phi_\gamma(m)} \mbox{ and }
\psi^\gamma_m:=\hat{U}_\gamma \psi^\gamma_{\Phi_\gamma(m)}
\ee
where $\hat{U}_\gamma:\;{\cal H}^0_\gamma\to {\cal H}^0$ is the usual
isometric monomorphism.

In \cite{112} we have proved peakedness --, expectation value --, small
fluctuation and Ehrenfest properties for the
gauge variant states mathematical states $\psi_{\gamma,m_\gamma}$ and the
algebra of operators ${\cal L}(H_\gamma^0)$. All proofs can be reduced to proving
it for single copy of $SU(2)$. Overcompleteness follows from the
results due to Hall \cite{70}  for the states $\psi_g$ on $L_2(SU(2),d\mu_H)$.
Annihilation operators have been defined above and for those minimal uncertainty
properties follow.

Next, given a system of elements $g_e\in SL(2,\Cl)$, one for each
analytic path $e\in {\cal P}$ we can form the distribution
\be \label{8.40}
\psi_g:=\sum_{\gamma\in\Gamma^\omega_0} <\psi^\gamma_{\overline{g_\gamma}},.>
\psi^\gamma_{g_\gamma}
\ee
where $g_\gamma=\{g_e\}_{e\in E(\gamma)}$. Now as shown in
\cite{72a} it is indeed possible to define an operator $\hat{C}$
through its cylindrical projections $\hat{C}_\gamma$ provided
the system of positive numbers $t_e$ satisfies the two conditions
\be \label{8.41}
t_{e_1}+t_{e_2}=t_{e_1\circ e_2} \mbox{ and } t_e=t_{e^{-1}}
\ee
which implies that the $t_e$ are in this case not constants.
The $t_e$ thus have all the properties of a length function
and we may use the background to be approximated in order to
define it. The distribution (\ref{8.40}) is then precisely
of the type (\ref{8.22}). Of course, by defining $\hat{C}_\gamma$
on spin-network functions rather than cylindrical functions
this can also be achieved if we do not have (\ref{8.41}).
\\
Remark:\\
We could then extend the definition of the operators
$\hat{g}_e$ by
\be \label{8.42}
\hat{g}_e:=e^{-\hat{C}/\hbar}\hat{h}_e (e^{-\hat{C}/\hbar})^{-1}
\ee
from which the properties $\hat{g}_e\hat{g}_{e'}=\hat{g}_{e\circ e'}$
and $\hat{g}_{e^{-1}}=\hat{g}_e^{-1}$ due to similar properties for the
operator $\hat{h}_e$. It follows that if the label $g$ in (\ref{8.40})
is to reproduce all the properties of the operators $\hat{g}_e$ then we
should have $g_e g_{e'}=g_{e\circ e'},\;g_{e^{-1}}=g_e^{-1}$ in other words,
$g$ qualifies as a generalized connection, that is, an element of
Hom$({\cal P},SL(2,\Cl))$. The question is then whether the images
$\Phi_e(m)$ defined in (\ref{8.32}) do have those properties for
all $m\in {\cal M}$ by choosing the $S_e,\rho_e(x)$ appropriately.
Notice that for the volume operator as the complexifier this
property would be trivially satisfied but in our case the answer
is less clear.\\
\\
In any case, the purpose of (\ref{8.40}) is to
demonstrate that with our family of coherent states it
is possible to form a distribution which is graph independent
but unfortunately one does not have an inner product on these objects
available and thus the best thing that one can do at this point is
to take the cut-off states $\psi_{\gamma,m}$ or $\psi^\gamma_m$.
Notice that if for $\gamma'\subset \gamma$ we choose the
$S_e,\rho_e(x);\;e\in E(\gamma')$ to be those that we chose for
$\gamma$ then $\psi_{\gamma,m}=\sum_{\gamma'\subset\gamma} \psi^\gamma_m$.
The coherent state properties that we established hold for the
$\psi_{\gamma,m}$ and to some extent also for the $\psi^\gamma_m$.

We then must deal with the question of how to choose $\gamma$. This
question is analyzed in detail in \cite{114}. One possibility is
to form a density matrix similar to the one we discussed above but
averaging only over a countable number of states (thus not leaving
${\cal H}^0$). Another would be to choose for $\gamma$ a generic
random graph which does not display any direction dependence on large
scales. In any of these scenarios the picture that arises is the
following:\\
Given $\gamma,m$ we can extract from these two data
two scales: The first is a
{\it graph scale} $\epsilon$ given by the average edge length as
measured by $m$. The second is a {\it curvature scale}
$L$ which is determined both by the mean curvature radius
of the four-dimensional metric determined by $m$ and the mean
curvature of the induced metric on the embedded submanifolds
$e,S_e,\rho_e(x)$ (so that even in the case that $m$ are exactly flat
initial data the scale $L$ is not necessarily infinity). We then must decide
which (kinematical) observables should behave maximally semi-classically.
This is a choice that must be made and the choice of $\gamma$ will
depend largely on this physical input. In \cite{114} we chose these
obervables to be electric and magnetic fluxes. When one then tries
to minimize the fluctuations of these obervables the parameters
$\epsilon$ and $a$ (the parameter that appears in $t=\ell_p^2/a^2$, we have
chosen $a_e=const.$ for simplicity) get locked at $a\approx L$ and
$\epsilon=\ell_p^\alpha L^{1-\alpha}$ for some $0<\alpha<1$
which in that case takes the value $\alpha=1/6$. These considerations
suggest the following conclusions:
\begin{itemize}
\item[1)] {\it Three Scales}\\
There are altogether three scales, the microscopic Planck scale $\ell_p$,
the mesoscopic scale $\epsilon$ and the macroscopic scale $L$.
Since $\ell_p\ll L$ we have $\ell_p\ll \epsilon \ll L$ provided that
(as in this case) $\alpha$ is not too close to the values $0,1$.
\item[2)] {\it Geometric Mean}\\
The mesoscopic scale takes a {\it geometric mean} between
the microscopic and macroscopic scales. In particular, it lies
well above the microscopic scale $\ell_p$ in contrast to the
geometric weave states. The reason for this is that not only electric
fluxes had to be well approximated but also magnetic ones:
The weave states are basically spin-network functions which in
turn are very similar to momentum eigenfunctions. Since then
electric fluxes are very sharply peaked, magnetic ones are
not peaked at all due to the Heisenberg uncertainty relation.
This can best be seen by the observation that
$<T_s,(\hat{h}_p)_{AB} T_s>=0$ for any spin-network state
and any $A,B=1,2$ (and therefore
also $\omega_{q^0,\lambda,j,I}(\hat{h}_p)=0$ for the statistical
weave) which is an unacceptable expectation value since
$\hat{h}_p$ should be $SU(2)-$valued. In order to approximate
holonomies {\it one must take an average over large numbers of spins}.
This is precisely what our coherent states do. As a consequence, the
elementary observables, those that are defined at the smallest scale
which still allows semiclassical behaviour, are now defined at scales
not smaller than $\epsilon\gg\ell_p$.
\item[3)] {\it Continuum Limit}\\
Notice that all our states and operators are defined in the continuum,
therefore no continuum limit has to be taken.
Yet, the scale $\epsilon$ could be associated with a measure for
closeness to the continuum in which the graphs with which we probe operators
tend to the continuum. The relation $\epsilon=\ell_p^\alpha L^{1-\alpha}$
reveals that not only one cannot take $\epsilon\to 0$ at finite
$\ell_p$ because fluctuations would blow up, but also that the
{\it ``continuum limit" $\epsilon\to 0$ and the classical limit
$\ell_p\to 0$ get synchronized}.
\end{itemize}
We expect many of those properties to hold generically for any
semiclassical states that one may want to build for canonical
quantum general relativity and that the extensive proofs
in \cite{112} will be useful for a whole class of states of this
kind.\\
\\
Remark:\\
Let us come back once more to the issue of using kinematical rather
than dynamical coherent states. We already said that
the full solution
to the Hamiltonian constraint is not known at the moment and that
even the operator with respect to which we want to compute these
solutions is not under sufficient control. Therefore in a first step
we must use kinematical states in order to make sure that we have the
correct operator. Suppose then that we would have found the correct operator,
then certainly the king's way of doing things would be to work with
dynamical coherent states, but probably
this would be highly impractical because the space of solutions for all
constraints is very complicated (even classically we do not know all the
solutions !). Thus, the poor man's way will be to consider
kinematical coherent states $\psi_m$ where $m$ is a point on the
constraint surface of the full phase space. The virtue of this
is that the expectation value of full Dirac observables is {\it approximately
gauge invariant} since
$$
\delta_N <\psi_m,\hat{O}\psi_m>=
<\psi_m,\frac{[\hat{H}(N),\hat{O}]}{i\hbar}\psi_m>
\approx \{H(N),O\}(m)=0
$$
because $O$ is a Dirac observable. Moreover
$$
<\psi_m,\hat{O}\psi_m>\approx O(m)=O([m])
$$
does not depend on the point $m$ in the gauge orbit $[m]$ for the same
reason. Thus, at least to zeroth order in $\hbar$
the expectation values of full Dirac observables and their infinitesimal
dynamics should coincide whether we use kinematical or dynamical
coherent states. This attitude is similar as in numerical classical
gravity where one cannot just compute the time evolution of a given
initial data set because for practical reasons one can only evolve
approximately. The art is then to gain control on the error of these
computations.

\subsubsection{The Infinite Tensor Product Extension}
\label{s8.2.4}

Quantum field theory on curved spacetimes is best understood if the
spacetime is actually flat Minkowski space on the manifold $M=\Rl^4$.
Thus, when one wants to compute the low energy limit of canonical quantum
general relativity to show that one gets the standard model (plus
corrections) on a background metric one should do this first for the
Minkowski background metric. Any classical metric is macroscopically
non-degenerate. Since the quantum excitations of the
gravitational field are concentrated on the edges of a
graph, in order that, say, the expection values of the volume operator
for any macroscopic region is non-vanishing and changes smoothly as
we vary the region, the graph must fill the
initial value data slice densely enough, the mean separation between
vertices of the graph must be much smaller than the size of the region
(everything is measured by the three metric, determined by the four metric
to be approximated, in this case the Euclidean one). Now $\Rl^4$ is spatially
non-compact and therefore
such a graph must necessarily have an at least {\it countably infinite}
number of edges whose union has {\it non-compact} range.

However, the Hilbert spaces in use for loop quantum gravity have as dense
subspace the space of cylindrical functions labelled either by a piecewise
analytic graph with
a {\it finite} number of edges or by a so-called web, a piecewise smooth
graph determined by the union of a {\it finite} number of smooth
curves that intersect in a controlled way, albeit possibly
a countably infinite number of times. Moreover, in both cases the
edges or curves respectively are contained in {\it compact} subsets of the
initial data hypersurface. These categories of graphs will be denoted by
$\Gamma^\omega_0$ and $\Gamma^\infty_0$ respectively
where $\omega,\infty,0$ stands for analytic, smooth and compactly supported
respectively. Thus, the only way that the current Hilbert
spaces can actually produce states depending on a countably infinite graph
of non-compact range is by choosing elements in the closure of these
spaces, that is, states that are countably infinite linear combinations
of cylindrical functions.

The question is whether it is possible to produce semi-classical states of
this form, that is, $\psi=\sum_n z_n \psi_{\gamma_n}$ where
$\gamma_n$ is either a finite piecewise analytic graph or a web, $z_n$ is
a complex number and we are summing over the intergers. It is easy to see
that this is not the case : Minkowski
space has the Poincar\'e group as its symmetry group and thus we will have
to construct a state which is at least invariant under
(discrete) spatial translations. This forces the $\gamma_n$ to be
translations of $\gamma_0$ and $z_n=z_0$. Moreover, the dependence
of the state on each of the edges has to be the same and therefore
the $\gamma_n$ have to be mutually disjoint. It follows that the norm
of the state is given by
$$
||\psi||^2=|z|^2([\sum_n 1][1-|<1,\psi_{\gamma_0}>|^2]+
[\sum_{n} 1]^2 |<1,\psi_{\gamma_0}>|^2)
$$
where we assumed without loss of generality that $||\psi_{\gamma_0}||=1$
and we used the diffeomorphism invariance of the measure and $1$ is the
normalized constant state. By the Schwartz inequality the first term is
non-negative and convergent only if $\psi_{\gamma_0}=1$ while the second
is non-negative and convergent only if $<1,\psi_{\gamma_0}>=0$. Thus
the norm diverges unless $z=0$.

This caveat is the source of its removal : We notice that the formal state
$\psi:=\prod_n \psi_{\gamma_n}$
really depends on an infinite graph and has unit norm if we formally
compute it by $\lim_{N\to\infty} ||\prod_{n=-N}^N \psi_{\gamma_n}||=1$
using disjointness of the $\gamma_n$. The only problem is that this
state is not any longer in our Hilbert space, it is not the Cauchy
limit of any state in the Hilbert space : Defining $\psi_N:=
\prod_{n=-N}^N \psi_{\gamma_n}$ we find
$|<\psi_N,\psi_M>|=|<1,\psi_{\gamma_0}>|^{2|N-M|}$ so that $\psi_N$ is
not a Cauchy sequence unless $\psi_{\gamma_0}=1$. However, it turns out
that it belongs to the {\it Infinite Tensor Product (ITP) extension} of the
Hilbert space.

To construct this much larger Hilbert space \cite{113} we must first describe
the class of graphs that we want to consider. We will consider graphs
of the category $\Gamma^\omega_\sigma$ where $\sigma$ now stands for
countably infinite. More precisely, an element of
$\Gamma^\omega_\sigma$ is the union of a countably infinite
number of analytic,
mutually disjoint (except possibly for their endpoints) curves called
edges of compact or non-compact range which have no accumulation points of
edges or vertices. In other words, the restriction of the graph
to any compact subset of the hypersurface looks like an element of
$\Gamma^\omega_0$. These are precisely the kinds of graphs that one
would consider in the thermodynamic limit of lattice gauge theories
and are therefore best suited for our semi-classical considerations since
it will be on such graphs that one can write actions, Hamiltonians and the
like.

The construction of the ITP of Hilbert spaces is due to von Neumann
\cite{123} and already more than sixty years old. We will try to outline
briefly some of the notions involved, see \cite{113} for a concise
summary of all definitions and theorems involved.

Let for the time being $I$ be any
index set whose cardinality $|I|=\aleph$ takes values in the set
of non-standard numbers (Cantor's alephs). Suppose that for each
$e\in I$ we have a Hilbert space ${\cal H}_e$ with scalar product
$<.,.>_e$ and norm $||.||_e$.
For complex numbers $z_e$ we say that $\prod_{e\in I} z_e$
{\it converges} to the number $z$
provided that for each positive number $\delta>0$ there
exists a finite set $I_0(\delta)\subset I$ such that for any other finite
$J$ with $I_0(\delta)\subset J\subset I$ it holds that
$|\prod_{e\in J} z_e -z|<\delta$. We say that $\prod_{e\in I} z_e$
is {\it quasi-convergent} if $\prod_{e\in I} |z_e|$ converges.
If $\prod_{e\in I} z_e$ is quasi-convergent but not convergent
we define $\prod_{e\in I} z_e:=0$. Next we say that for
$f_e\in {\cal H}_e$ the ITP $\otimes_f:=\otimes_e f_e$ is a $C_0$ vector
(and $f=(f_e)$ a $C_0$ sequence) if
$||\otimes_f||:=\prod_{e\in I} ||f_e||_e$ converges to
a non-vanishing number. Two $C_0$ sequences $f,f'$ are said to be
strongly resp. weakly equivalent provided that
$$
\sum_e |<f_e,f'_e>_e-1| \mbox{ resp. } \sum_e ||<f_e,f'_e>_e|-1|
$$
converges. The strong and weak equivalence class of $f$ is denoted by
$[f]$ and $(f)$ respectively and the set of strong and weak equivalence
classes by $\cal S$ and $\cal W$ respectively. We define the ITP
Hilbert space ${\cal H}^\otimes:=\otimes_e {\cal H}_e$ to be the closed
linear span of all $C_0$ vectors. Likewise we define
${\cal H}^\otimes_{[f]}$ or ${\cal H}^\otimes_{(f)}$ to be the closed
linear spans of only those $C_0$ vectors which lie in the same strong
or weak equivalence class as $f$. The importance of these notions is that
the determine much of the structutre of ${\cal H}^\otimes$, namely : \\
1) All the ${\cal H}^\otimes_{[f]}$ are isomorphic and mutually orthogonal.\\
2) Every ${\cal H}^\otimes_{(f)}$ is the closed direct sum of all the
${\cal H}^\otimes_{[f']}$ with $[f']\in {\cal S}\cap (f)$.\\
3) The ITP ${\cal H}^\otimes$ is the closed direct sum of all the
${\cal H}^\otimes_{(f)}$ with $(f)\in {\cal W}$.\\
4) Every ${\cal H}^\otimes_{[f]}$ has an explicitly known orthonormal
von Neumann basis.\\
5) If $s,s'$ are two different strong equivalence classes in the same
weak one then there exists a unitary operator on ${\cal H}^\otimes$
that maps ${\cal H}^\otimes_s$ to ${\cal H}^\otimes_{s'}$, otherwise
such an operator does not exist, the two Hilbert spaces are unitarily
inequivalent subspaces of ${\cal H}^\otimes$.\\
Notice that two isomorphic Hilbert spaces can always be mapped into each
other such that scalar products are preserved (just map some orthonormal
bases) but here the question is whether this map can be extended
unitarily to all of
${\cal H}^\otimes$. Intuitively then, strong classes within the same weak
classes describe the same physics, those in different weak classes describe
different physics such as an infinite difference in energy, magnetization,
volume etc. See \cite{124} and references therein for illustrative examples.

Next, given a (bounded) operator
$a_e$ on ${\cal H}_e$ we can extend it in the natural way to
${\cal H}^\otimes$ by defining $\hat{a}_e$ densely on $C_0$ vectors through
$\hat{a}_e \otimes_f=\otimes_{f'}$ with $f'_{e'}=f_{e'}$ for $e'\not=e$
and $f'_e= a_e f_e$. It turns out that the algebra of these extended
operators for a given edge is automatically a von Neumann algebra
\cite{62a,8b,53,55,56}
for ${\cal H}^\otimes$ (a weakly closed subalgebra of the algebra of
bounded operators on a Hilbert space) and we
will call the weak closure of all these
algebras the von Neumann algebra ${\cal R}^\otimes$ of local operators.
This way, adjointness relations and canonical commutation relations
(Weyl algebra) are preserved.

Given these notions, the strong equivalence class Hilbert spaces can be
characterized
further as follows. First of all, for each $s\in {\cal S}$ one can
find a representant $\Omega^s\in s$ such that $||\Omega^s||=1$. Moreover,
one can show
that ${\cal H}^\otimes_s$ is the closed linear span of those $C_0$ vectors
$\otimes_{f'}$ such that $f'_e=\Omega^s_e$ for all but finitely
many $e$. In other words, the strong
equivalence class Hilbert spaces are irreducible subspaces for
${\cal R}^\otimes$, $\Omega^s$ is a cyclic vector for ${\cal H}^\otimes_s$
on which the local operators annihilate and create local excitations
and thus, if $I$ is countable, ${\cal H}^\otimes_s$ is actually separable.
We see that we make naturally contact with Fock space structures,
von Neumann algebras and their factor type classification \cite{8b}
(modular theory) and algebraic quantum field theory \cite{62a}.
The algebra of operators on the ITP which are not local do not have
an immediate interpretation but it is challenging that they
map between different weak equivalence classes and thus change the physics
in a drastic way.

A number of warnings are in order :\\
1) Scalar multiplication is not multi-linear ! That is, if $f$ and
$z\cdot f$ are $C_0$ sequences where $(z\cdot f)_e=z_e f_e$ for some
complex numbers $z_e$ then $\otimes_f=(\prod_e z_e)\;\otimes_f$ is in general
wrong, it is true if and only if $\prod_e z_e$ converges.\\
2) Unrestricted use of the associative law of tensor products is false !
Let us subdivide the index set $I$ into mutually disjoint
index sets $I=\cup_\alpha I_\alpha$ where $\alpha$ runs over some other index
set $A$. One can now form the different ITP ${\cal H}^{\prime\otimes}
=\otimes_\alpha {\cal H}^\otimes_\alpha,\;
{\cal H}^\otimes_\alpha=\otimes_{e\in I_\alpha} {\cal H}_e$.
Unless the index set $A$ is finite, a generic $C_0$ vector of
${\cal H}^{\prime\otimes}$ is orthogonal to all of
${\cal H}^\otimes$. This fact has implications for quantum gravity
which we outline below.

Let us now come back to canonical quantum general relativity. In applying
the above concepts we arrive at the following surprises :
\begin{itemize}
\item[i)] First of all, we
fix an element $\gamma\in \Gamma^\omega_\sigma$
and choose the countably infinite index set $E(\gamma)$, the edge
set of $\gamma$. If $|E(\gamma)|$ is finite then the ITP Hilbert space
${\cal H}^\otimes_\gamma:=\otimes_{e\in E(\gamma)} {\cal H}_e$ is naturally
isomorphic with the subspace ${\cal H}^0_\gamma$ of ${\cal H}^0$
obtained as the closed linear span of cylinder functions
over $\gamma$. However, if $|E(\gamma)|$ is truly infinite then
a generic $C_0$ vector of ${\cal H}^\otimes_\gamma$ is orthogonal
to any possible ${\cal H}^0_{\gamma'}$, $\gamma'\in \Gamma^\omega_0$.
Thus, even if we fix only one $\gamma\in \Gamma^\omega_\sigma$, the total
${\cal H}^0$ is orthogonal to almost every element of ${\cal
H}^\otimes_\gamma$.
\item[ii)] Does ${\cal H}^\otimes_\gamma$ have a measure theoretic
interpretation as an $L_2$ space ? By the Kolmogorov theorem \cite{25}
the infinite product of probability measures is well defined and thus one
is tempted to identify ${\cal H}^\otimes_\gamma=\otimes_e L_2(SU(2),d\mu_H)$
with ${\cal H}^{0\prime}_\gamma:=L_2(\times_e SU(2),\otimes_e d\mu_H)$.
However, this cannot be
the case, the ITP Hilbert space is non-separable (as soon as $\dim({\cal
H}_e)>1$ for
almost all $e$ and $|E(\gamma)|=\infty$) while the latter Hilbert space
is separable, in fact, it is the subspace of ${\cal H}^0$ consisting
of the closed linear span of cylindrical functions over $\gamma'$
with $\gamma'\in \Gamma^\omega_0\cap E(\gamma)$.
\item[iii)] Yet, there is a relation between ${\cal H}^\otimes_\gamma$
and ${\cal H}^0$ through the inductive limit
of Hilbert spaces : We can find a directed sequence of elements $\gamma_n\in
\Gamma^\omega_0\cap E(\gamma)$, that is, $\gamma_m\subset\gamma_n$ for
$m\le n$, such that $\gamma$ is its limit in $\Gamma^\omega_\sigma$.
The subspaces ${\cal H}^0_{\gamma_n}\subset{\cal H}^0$ are isometric
isomorphic with the subspaces of ${\cal H}^\otimes_\gamma$ given by the
closed linear span of vectors of the form $\psi_{\gamma_n}\otimes[
\otimes_{e\in E(\gamma-\gamma_n)} 1]$ where $\psi_{\gamma_n}\in
{\cal H}^0_{\gamma_n}\equiv {\cal H}^\otimes_{\gamma_n}$ which provides
the
necessary isometric monomorphism to display ${\cal H}^\otimes_\gamma$
as the inductive limit of the ${\cal H}^0_{\gamma_n}$.
\item[vi)] So far we have looked only at a specific
$\gamma\in\Gamma^\omega_\sigma$. We now construct the total Hilbert space
$$
{\cal H}^\otimes:=\overline{\cup_{\gamma\in \Gamma^\omega_\sigma}
{\cal H}^\otimes_\gamma}
$$
equipped with the natural scalar product derived in \cite{113}.
This is to be compared with the
Hilbert space
$$
{\cal H}^0:=
\overline{\cup_{\gamma\in \Gamma^\omega_0}{\cal H}^0_\gamma}
=\overline{\cup_{\gamma\in \Gamma^\omega_\sigma}{\cal H}^{0\prime}_\gamma}
$$
The identity in the last line enables us to specify the precise sense in
which ${\cal H}^0\subset{\cal H}^\otimes$ : For any
$\gamma\in\Gamma^\omega_\sigma$ the space
${\cal H}^{0\prime}_\gamma$ is isometric isomorphic as specified in
iii) with the strong equivalence class Hilbert subspace
${\cal H}^\otimes_{\gamma,[1]}$ where $1_e=1$ is the constant function
equal to one. Thus, the Hilbert ${\cal H}^0$ space
describes the local excitations of the ``vacuum"
$\Omega^0$ with $\Omega^0_e=1$ for any possible analytic path $e$.

Notice that both Hilbert spaces are non-separable, but there are two
sources of non-separability : the Hilbert space ${\cal H}^0$ is
non-separable because $\Gamma^\omega_0$ has uncountable infinite
cardinality. This is also true for the ITP Hilbert space but
it has an additional character of non-separability : even for fixed $\gamma$
the Hilbert space
${\cal H}^\otimes_\gamma$ splits into an uncountably infinite number of
mutually orthogonal strong equivalence class Hilbert spaces and
${\cal H}^{0\prime}_\gamma$ is only one of them.
\item[v)] Recall that spin-network states \cite{9} form a basis for
${\cal H}^0$. The result of iv) states that they are no longer
a basis for the ITP. The spin-network basis is in fact the von Neumann
basis for the strong equivalence class Hilbert space determined
by $[\Omega^0]$ but for the others we need uncountably infinitely
many other bases, even for fixed $\gamma$. The technical reason for this is
that,
as remarked above, the unrestricted associativity law fails on the ITP.
\end{itemize}
We would now like to justify this huge blow up of the original
Hilbert space ${\cal H}^0$ from the point of view of physics. Clearly,
there is a blow up only when the initial data hypersurface is non-compact
as otherwise $\Gamma^\omega_0=\Gamma^\omega_\sigma$. Besides the
fact that like ${\cal H}^0$ it is another solution to implementing
the adjointness -- and canonical commutation relations, we have the
following:
\begin{itemize}
\item[a)] Let us fix $\gamma\in\Gamma^\omega_\sigma$ in order to describe
semi-classical physics on that graph as outlined above.
Given a classical initial data
set $m$ we can construct a coherent state $\psi_{\gamma,m}$ which
in fact is a $C_0$ vector $\otimes^\gamma_{\psi_m}$ for ${\cal
H}^\otimes_\gamma$
of unit norm. This coherent state can be considered as a ``vacuum"
or ``background state" for quantum field theory on the associated
spacetime.
As remarked above, the corresponding strong equivalence class Hilbert space
${\cal H}^\otimes_{\gamma,[\psi_m]}$ is obtained by acting on the
``vacuum" by local operators, resulting in a space isomorphic with
the familar Fock spaces and which is separable. In this sense, the
fact that ${\cal H}^\otimes_\gamma$ is non-separable, being an uncountably
infinite direct
sum of strong equivalence class Hilbert spaces, could simply account for the
fact that in quantum gravity {\it all vacua have to be considered
simultaneously, there is no distinguished vauum as we otherwise would
introduce a backgrond dependence into the theory}.
\item[b)] The Fock space structure of the strong equivalence classes
immediately suggests to try to identify suitable excitations of
$\psi_{\gamma,m}$ as graviton states propagating on a spacetime
fluctuating around the classical background deteremined by $m$
\cite{114a}. \\
Also, it is easy to check whether for different solutions
of Einstein's equations
the associated strong equivalence classes lie in different
weak classes and are thus physically different. For instance,
preliminary investigations indicate that Schwarzschild black hole spacetimes
with different masses lie in the {\it same} weak class. Thus, {\it unitary}
black hole evaportation and formation seems not to be excluded from the
outset.
\item[c)] From the point of view of ${\cal H}^{0\prime}_\gamma$
the Minkowski coherent state is an everywhere excited state like a
thermal state, the strong classes $[\Omega^0]$ and $[\psi_m]$ for
Minkowski data $m$ are orthogonal and lie in different weak classes.
The state $\Omega^0$ has no obvious semi-classical interpretation
in terms of coherent states for any classical spacetime.
\item[d)] It is easy to see that the GNS Hilbert space used
in \cite{109} is isometric isomorphic with a strong
equivalence class Hilbert space of our ITP construction.
Thus, our ITP framework collects a huge class of
representations in the ``folium" \cite{62a} of the
Hilbert space ${\cal H}^0$ and embeds them isometrically
into one huge Hilbert space ${\cal H}^\otimes$, thus
we have now {\it an inner product between different
GNS Hilbert spaces}! This demonstrates the power of
this framework because inner products between different
GNS Hilbert spaces are normally not easy to motivate.
\end{itemize}

\subsection{Photon Fock States on $\ab$}
\label{s8.3}

In \cite{117} Varadarajan investigated the
question in which sense the the techniques of $\ab,\mu_0$, which
in principle apply to any gauge field theory of connections for
compact gauge groups, can be used to describe the Fock states of
Maxwell theory. This is not at all an academic question because
presumably one wants to couple Maxwell theory to gravity also
in such a background independent representation as, in fact we
have indicated in section \ref{s7}. Moreover, linearized
gravity can be described in terms of connections as well
\cite{117a} where it becomes effectively a $U(1)^3$ Abelean
gauge theory just like Maxwell theory. Both theories are,
of course, ordinary free field theories on a Minkowski
background.

Varadarajan succeeded in displaying Fock states within
the framework of $\ab,\mu_0$ in a very precise way.
The crucial observation, unfortunately only valid if the
gauge group is Abelean,
is the following isomorphism between two different Poisson
subalgebras of the Poisson algebra on $\cal M$:
Consider a one-parameter family of
test functions of rapid decrease which are regularizations of
the $\delta-$distribution, for instance
\be \label{8.43}
f_r(x,y)=\frac{e^{-\frac{||x-y||^2}{2r^2}}}{(\sqrt{2\pi}r)^3}
\ee
where we have made use of the Euclidean spatial background
metric.
Given a path $p\in{\cal P}$ we denote its form factor
by
\be \label{8.44}
X^a_p(x):=\int_0^1 dt \dot{p}^a(t) \delta(x,p(t))
\ee
The smeared form factor is defined by
\be \label{8.45}
X^a_{p,r}(x):=\int d^3y f_r(x,y) X^a_p(y)=\int_0^1 dt \dot{p}^a(t)
f_r(x,p(t))
\ee
which is evidently a test function of rapid decrease.
Notice that a $U(1)$ holonomy maybe written as
\be \label{8.46}
h_p(A):=e^{i\int d^3x X^a_p(x) A_a(x)}
\ee
and we can define a smeared holonomy by
\be \label{8.47}
h_{p,r}(A):=e^{i\int d^3x X^a_{p,r}(x) A_a(x)}
\ee
Likewise we may define smeared electric fields as
\be \label{8.48}
E^a_r(x):=\int d^3y f_r(x,y) E^a(y)
\ee
If we denote by $q$ the electric charge (notice that in
our notation $\alpha=\hbar q^2$ is the fine structure
constant), then we obtain the
following Poisson subalgebras:
On the one hand we have
smeared holonomies but unsmeared electric fields with
\be \label{8.49}
\{h_{p,r},h_{p',r}\}=\{E^a(x),E^b(y\}=0,\;\;
\{E^a(x),h_{p,r}\}=iq^2 X^a_{p,r}(x) h_{p,r}
\ee
and on the other hand we have
unsmeared holonomies but smeared electric fields with
\be \label{8.50}
\{h_p,h_{p'}\}=\{E^a_r(x),E^b_r(y\}=0,\;\;
\{E^a_r(x),h_p\}=iq^2 X^a_{p,r}(x) h_p
\ee
Thus the two Poisson algebras are ismorphic and also
the $^\ast$ relations are isomorphic, both
$E^a(x),E^a_r(x)$ are real valued while
both $h_p,h_{P,r}$ are $U(1)$ valued. Thus, as abstract
$^\ast-$ Poisson algebras these two algebras are indistinguishable
and we may ask if we can find different representations
of it. Even better, notice that
$h_{p,r} h_{p',r}=h_{p\circ p',r},\;h_{p,r}^{-1}=h_{p^{-1},r}$
so the smeared holonomy algebra is also isomorphic to the
unsmeared one. It is crucial to point out that the right hand side
of both (\ref{8.49}), (\ref{8.50}) is a cylindrical function again
only in the Abelean case, see section \ref{s4.1}. Therefore
all that follows {\it is not true for $SU(2)$}.

Now we know that the unsmeared holonmy algebra is well represented
on the Hilbert space ${\cal H}^0=L_2(\ab,d\mu_0)$ while the
smeared holonomy algebra is well represented on the
Fock Hilbert space ${\cal H}_F=L_2({\cal S}',d\mu_F)$ where
${\cal S}'$ denotes the space of divergence free, tempered distributions
and $\mu_F$ is the Maxwell-Fock measure. These measures are
completely characterized by their generating functional
\be \label{8.51}
\omega_F(\hat{h}_{p,r}):=\mu_F(h_{p,r})=
e^{-\frac{1}{4\alpha}\int d^3x X^a_{p,r}(x)\sqrt{-\Delta}^{-1}
X^b_{p,r}\delta_{ab}}
\ee
since finite linear combinations of the $h_{p,r}$ are dense
in ${\cal H}_F$ \cite{117}. Here $\Delta=\delta^{ab}=\partial_a\partial_b$
denotes the Laplacian.
Here we have taken a loop $p$ rather than an open path so that
$X_{p,r}$ is transversal.
Also unsmeared electric fields are represented through the Fock
state $\omega_F$ by
\be \label{8.52}
\omega_F(\hat{h}_{p,r}\hat{E}^a(x)\hat{h}_{p',r})
=-\frac{\alpha}{2}[X^a_{p,r}(x)-X^a_{p',r}(x)]
\omega_F(\hat{h}_{p\circ p',r})
\ee
and any other expectation value follows from these and the
commutation relations.

Since $\omega_F$ defines a positive linear functional we may
define a new representation of the algebra $h_p,E^a_r$
by
\be \label{8.53}
\omega_r(\hat{h}_p):=\omega_F(\hat{h}_{p,r})
\mbox{ and }
\omega_r(\hat{h}_p\hat{E}^a_r(x)\hat{h}_{p'})
:=\omega_F(\hat{h}_{p,r}\hat{E}^a(x)\hat{h}_{p',r})
\ee
called the $r-$Fock representation. In order
to see whether there exists is a measure $\mu_r$ on $\ab$ that represents
$\omega_r$ in the sense of the Riesz representation theorem
we must check that $\omega_r$ is a positive linear functional on
$C(\ab)$. This can be done \cite{117}. In \cite{119} Velhinho
has computed explicitly the cylindrical projections of this
measure and showed that the one -- parameter family of measures
$\mu_r$ are expectedly mutually singular with respect to each
other and with respect to the uniform measure $\mu_0$.
Thus, none of these Hilbert spaces is contained in any other.
In fact, we have a natural map
\be \label{8.53a}
\Theta_r:\;{\cal S}'\to \agb;\;A\mapsto \Theta_r(A) \mbox{ where }
[\Theta_r(A)](p):=e^{i\int d^3x X_{p,r}^a A_a(x)}
\ee
and Velhinho showed that $\mu_r=(\Theta_r)_\ast\mu_F$ is just
the push-forward of the Fock measure.

Recall that the Fock vacuum $\Omega_F$ is defined to be the zero eigenvalue
coherent state, that is, it is annihilated by the annihilation operators
\be \label{8.54}
\hat{a}(f):=\frac{1}{\sqrt{2\alpha}}
\int d^3x f^a[\root 4\of{-\Delta} \hat{A}_a-i(\root 4\of{-\Delta})^{-1}\hat{E}^a]
\ee
where $f^a$ is any transversal smearing field. We then have in
fact that $\omega_F(.)=<\Omega_F,.\Omega_F>_{{\cal H}_F}$, that is
$\Omega_F$ is the cyclic vector that is determined by $\omega_F$ through
the GNS construction.
The idea is now the following: From (\ref{8.53}) we see that we can
easily answer any question in the $r-$Fock representation which has a
preimage in the Fock representation, we just have to replace
everywhere $h_{p,r},E^a(x)$ by $h_p,E^a_r(x)$.
Since in the $r-$Fock representations only exponentials of connections
are defined, we should exponentiate the annihilation operators
and select the Fock vacuum through the condition
\be \label{8.55}
e^{i\hat{a}(f)}\Omega_F=\Omega_F
\ee
In particular, choosing $f=\sqrt{2\alpha}(\root 4\of{-\Delta})^{-1} X_{p,r}$
for some loop $p$ we get
\be \label{8.56}
e^{\int d^3x X^a_{p,r}[i\hat{A}_a+(\sqrt{-\Delta})^{-1}\hat{E}^a]}
\Omega_F
=\Omega_F
\ee
Using the commutation relations and the Baker -- Campell -- Hausdorff
formula one can write (\ref{8.56}) in terms of $\hat{h}_{p,r}$
and the exponential of the electric field appearing in (\ref{8.56})
times a numerical factor. The resulting expression can then be
translated into the $r-$Fock representation.

This was Varadarajan's idea. He found that in fact there is no state
in ${\cal H}^0$ which satisfies the translated analogue of
(\ref{8.56}) but that there exists a distribution that does (we
must translate (\ref{8.56}) first into the dual action to
compute that distribution).
It is given (up to a constant) by
\be \label{8.57}
\Omega_r=\sum_s
e^{-\frac{\alpha}{2}\sum_{e,e'\in E(\gamma(s))} G^r_{e,e'} n_e(s) n_{e'}(s)}
T_s<T_s,.>_{{\cal H}^0}
\ee
where $s=(\gamma(s),\{n_e(s)\}_{e\in E(\gamma(s))})$ denotes
a charge network (the $U(1)$ analogue of a spin network) and
\be \label{8.58}
G^r_{e,e'}=\int d^3x X^a_{e,r}\sqrt{-\Delta}^{-1} X^b_{e',r}\delta^T_{ab}
\ee
where $\delta_{ab}^T=\delta_{ab}-\partial_a\Delta^{-1}\partial_b$
denotes the transverse projector.

Several remarks are in order concerning this result:
\begin{itemize}
\item[1)] {\it Distributional Fock States}\\
$n-$particle state excitations of the state $\Omega_F$ (and
also coherent states \cite{118}) can be easily translated into
distributional $n-$particle states (coherent states) by
using Varadarjan's prescription above. Thus, we get in fact a
{\bf Varadarajan map}
\be \label{8.59}
V:\;({\cal H}_F,{\cal L}({\cal H}_F)\mapsto
({\cal D}^\ast,{\cal L}'({\cal D}))
\ee
Of course, none
of the image states is normalizable with respect to $\mu_0$ and this
raises the question in which sense the kinematical
Hilbert space is useful at all in order to do semi-classical
analysis. One can {\it in this case define} a new scalar
product on these distributions simply by
\be \label{8.60}
<V\cdot\psi,V\cdot \psi'>_r:=<\psi,\psi'>_F
\ee
In particular we obtain $<\Omega_r,.\;\Omega_r>_r=\omega_r$
so $\Omega_r$ can be interpreted as the GNS cyclic vector
underlying $\omega_r$. With respect to this inner product
one can now perform semi-classical analysis. Of course,
in the non-Abelean case a Varadarjan map is not available
at this point.
\item[2)] {\it Electric Flux Operators}\\
In the non-Abelean theory it was crucial not to
work with electrical fields smeared in $D$ dimensions but
rather with those smeared in $D-1$ dimensions. However,
$(D-1)-$smeared electrical fields have no
pre-image under $V$ and in fact Velhinho showed that
there is no electric flux operator in the
$r-$Fock representation as to be expected. This seems
to be an obstruction to transfer the Varadarajan
map to the non -- Abelean case.
\item[3)] {\it Comparison with Heat Kernel Coherent States}\\
Formulas (\ref{8.22}) and (\ref{8.57}) look very similar to
each other (see also \cite{118}).
We can write (\ref{8.57}) more suggestively as
\be \label{8.61}
\Omega_r=\sum_s
e^{\frac{\alpha}{2}\sum_{e,e'\in E(\gamma(s))} G^r_{e,e'} R_e R_{e'}}
T_s<T_s,.>_{{\cal H}^0}
\ee
where $R_e$ are right invariant vector fields on $U(1)$. This
formula just asks to be analytically continued in order to arrive at
a coherent state because it looks like (\ref{8.22}).
The deeper origin of this apparent coincidence will be unravelled
in \cite{122} where it will be shown that the
Varadarajan coherent distributions are
a special case of the general formula (\ref{8.23}).

In \cite{118} it is speculated that one should generalize
(\ref{8.61}) in the obvious way to the non-Abelean case
by replacing charge nets by spin nets and $R_e R_{e'}$
by $R^j_e R^j_{e'}$ and to use the associated cut-off
states (called ``shadows" there) for semi-classical analysis.
However, it is unclear whether these shadows have
similarly nice properties as the cut-off states introduced in
\cite{111,112,113,114} because the metric $G^r_{e e'}$ is not
diagonal. Also it is unclear how one should then define
non-Abelean Fock states. Finally it is not clear
what the interpretation of the complexified group label
should be without which a semiclassical ineterpretation of those
states is out of reach.
\item[4)] {\it Other Operators}\\
One should not forget that important operators of Maxwell theory
such as the Hamiltonian operator are expressed as polynomials
of {\it un -- exponentiated} annihilation and creation operators.
However, such operators are not defined neither in the $r-$Fock
representation nor in ${\cal H}^0$. In \cite{114a} we will show
how to circumvent that problem.
\end{itemize}

\subsection{Applications}
\label{s8.4}

Beyond merely checking whether we have a quantum theory of the correct
classical theory, namely general relativity coupled to all known
matter, quantum gravity has certainly a huge impact on the whole
structure of physics. For instance, if the picture drawn in section
\ref{s7} is correct, then one must do quantum field theory on
one-dimensional polymer like structures rather than in a higher
dimensional manifold, presumably the ultraviolet divergences disappear
and while there are still bare and renormalized charges, masses etc.
the bare charges will presumably be finite while the renormalized charges
should better be called effective charges because they simply take into
account physical screening effects.

Quantum gravity effects are notoriously difficult to measure because the
Planck length is so incredibly tiny. It may therefore come as a surprise
that recently physicists have started to seriously discuss the possibility
to measure quantum gravity effects, mostly from astrophysical data
and gravitational wave detectors \cite{127}. See also the discussion in
the extremely beautiful review by Carlip \cite{128} and references therein.
The challenge is to compute these effects within quantum general
relativity. First pioneering steps towards the computation of the
so-called {\it $\gamma-$ray burst effect} have been
made, to date mostly at a phenomenological level, in \cite{129} for
photons and \cite{130} for neutrinos. A more detailed analysis based
on the coherent states proposed in \cite{111,112} will appear in \cite{114a}.

This is not the place to give a full-fledged account of these developments,
so we will restrict ourselves to presenting the main ideas for the
$\gamma-$ray burst effect.

A $\gamma-$ray burst is a light signal of extremely high energetic
photons (up to 1 TeV !) that travelled over cosmological distances
(say $10^9$ years). What is interesting about them is that the
signal is like a flash, that is, the intensity decays on the order of
$10^{-3}$s. The astrophysical origin of these bursts is still under
debate (see the references in \cite{130})
and we will have nothing to add on this debate here. What is important
though is that these photons probe the discrete (polymer) structure of
spacetime the more, the more energy they have which should lead to
an energy dependent velocity of light (dispersion) very similar
to the propagtion of light in cristals. More specifically, if one
plots the time signal of events as measured by a atmospherical
Cerenkov light detector \cite{131} within two disjoint energy channels
$[E_1-\Delta E,E_1+\Delta E]$ and
$[E_2-\Delta E,E_2+\Delta E]$ then one expects a time difference in the
peak of these signals given by\\
$t_2-t_1=\xi \frac{L}{c(0)}[(E_2/E_p)^\alpha-(E_1/E_p)^\alpha]$
where $L$ is the difference from the source (measured by the red shift
of the galaxy), $c(0)$ is the vacuum speed of light, $E_p$ is the effective
Planck scale energy of the order of $m_p$
and $\alpha,\xi$ are theory dependent constants of the order unity.
If $\alpha=\xi=1$, $E_p=m_p$ and $E_2-E_1=$1 TeV then for $L=10^9$
lightyears we get travel time differences of the order of $10^2$s
which is much larger than the duration of the peak. At present, the
sensitivity of available detectors is way below such a resoltion of
ms mainly because no detectors ahve been built for this specific purpose
but the construction of better detectors is on the way \cite{130}.

One may object that
1) quantum field theory effects from other interactions should be much
stronger than quantum gravity effects so that this effect would not
test so much quatum gravity but rather quantum field theory on
Minkowski space, 2) there are many possible astrophysical disturbances that
can cause dispersion such as interstellar dust and 3) it is not clear that
the photons of different energies have been emitted simultaneously.

The answer is as follows:\\
1) is ecluded by definition of quantum
field on Minkowski space: Such a theory is Poincar\'e invariant by
construction while an energy dependent dispersion breaks Lorentz invariance.
We see that the effect is {\it non-perturbative} because in any
perturbative approach to quantum gravity one treats gravity like the
other inetactions as a quantum field theory on a Minkowski background.\\
2) is excluded by the fact that the effect gets stronger with higher
energy while diffraction at dust gets weaker: The scale of dust or
gas molecules is transparent for such highly energetic photons.\\
3) is apparently excluded by model computations in astrophysics \cite{131}
for the known scenarios that lead to the $\gamma-$ray burst effect.

How would one then compute the effect within quantum general relativity ?
Basically, one would look at quantum Einstein-Maxwell theory and consider
states of the form $\psi_E\otimes \psi_M$ where $\psi_E$ is a
fixed coherent state
for the gravitational degrees of freedom, peaked at Minkowski initial data
and $\psi_M$ is a quantum state for the Maxwell-field. Given the
Einstein-Maxwell Hamiltonian
$$
H_{EM}=\frac{1}{2 e^2}\int d^3x \frac{q_{ab}}{\sqrt{\det(q)}}[E^a E^b+B^a B^b]
$$
one would quantize it as described in section \ref{s7} and then define
an effective Maxwell Hamiltonian by
$$
<\psi_M,\hat{H}^{eff}_M\psi'_M>_{{\cal H}_M}
:=<\psi_E\otimes\psi_M,
\hat{H}_{EM}\psi_E\otimes\psi_M>_{{\cal H}_E\otimes{\cal H}_M}
$$
At the moment we can do this computation only at the kinematical level
but as outlined in section \ref{s8.2} this should approximate the full
dynamical computation and at least gives an idea for the size of the effect.

Whatever technique is finally being used to carry out this computation
the mere existence of the effect is a {\it prediction} of any
background independent approach to quantum gravity. In fact,
the technical
reason for existence of the effect is a corollary from the Heisenberg
uncertainty relation: The quantum metric operators form a non-commuting set
of operators (they depend both on magnetic and electric degrees of freedom)
so that it is not possible to diagonalize them simultaneously. The best
one can do is to construct an approximate eigenstate for all of them
(namely a coherent state) but that state can then not be exactly Poincar\'e
invariant, only approximately.\\
\\
There are countless other applications of semiclassical
states such as an approach to quantum black holes from first principles
and a corresponding computation of the Hawking effect that takes full
account of the backreaction of the gravitational field towards infalling
matter which at the horizon becomes infinitely blue shifted so that
quantum gravity effects are no longer neglible.

\newpage

\section{Further Research Directions}
\label{s9}

In the second last section of this review we will describe briefly
three more major research directions within Canonical Quantum
General Relativity: Spin Foam Models, Quantum Black Holes and
Interfaces between Canonical Quantum General Relativity and
String Theory. To be sure, all three topics deserve to be treated in
a chapter of their own, however,
our presentation will be short since a thorough treatment would
require three additional reviews in their own right plus extra
background material in additional appendices which would
explode the already huge length of this review. Luckily, nice, pretty
self-contained, review articles, at least for the two first programmes,
already exist:\\
\\
For an introduction to spin-foam models we recommend the
really beautiful article by Baez \cite{132} which contains an almost
complete and up to date guide to the literature and the historical
development of the subject. See also the article by Barrett \cite{133} for
the closely related subject of state sum models. A summary of the
classical and quantum aspects of so-called {\it isolated horizons},
a local generalization of event horizons that is used in
black hole entropy calculations within quantum general relativity,
can be found in \cite{134}. The pivotal papers that describe the
details of the classical and quantum formulation respectively are
\cite{135} and \cite{136} respectively.

\subsection{Spin Foam Models}
\label{s9.1}

The prototype of spinfoam models are state sum models that had
been extensivley studied \cite{137} within the context of topological
quantum field theories \cite{138} long before spin foam models arose within
quantum gravity. The concrete connection of state sum models with
canonical quantum gravity was made by Reisenberger and Rovelli in their
seminal paper \cite{139} where they used the (Euclidean version of the)
Hamiltonian constraint described in section \ref{s6} in order to
write down a path integral formulation of the the theory.
Roughly speaking, this works as follows:\\
A heuristic method of how to solve the Hamiltonian constraint is
to take any kinematical state $\psi$ and to map it to
$\delta(\hat{H})\psi$ where $\delta(\hat{H})=\prod_{x\in \sigma}
\delta(\hat{H}(x)$. This is of course quite formal since neither
the $\hat{H}(x)$ are self-adjoint nor mutually commuting. It is
anyway a formal solution to the Hamiltonian constraint if we treat
the Diffeomorphism constraint similarly because the algebra of
deffeomorphisms and Hamiltonians is formally closed.
Proceeding formally, we may define a path integral formulation
of the $\delta-$distribution. Neglecting an (infinite) constant as usual
we obtain the functional integral
\be \label{9.1}
\delta(\hat{H})=\int [dN] e^{i\int_\sigma d^3x N(x) \hat{H}(x)}
\ee
This looks like a group averaging operation and we may try to define
a physical inner product between physical states
$\psi_{phys}:=\delta(\hat{H})\psi$ as
\be \label{9.2}
<\psi_{phys},\psi'_{phys}>_{phys}:=
<\psi,\delta(\hat{H})\psi'>=
\int_{{\cal N}} [dN] <\psi,e^{i\int_\sigma d^3x N(x) \hat{H}(x)}\psi'>
\ee
where $\cal N$ is the set of all lapse functions on $\sigma$.
In order to get time dependent lapse functions $\bar{N}(x,t)$ consider the
set of lapse functions $\overline{{\cal N}}_N$ on $M$ with
$\int_{-T}^T dt \bar{N}(x,t)=N(x)$ for some $T>0$.
Let also $\overline{{\cal N}}$ be the set of lapse functions over $M$.
Then
\ba \label{9.3}
&& \int_{\overline{{\cal N}}} [d\bar{N}]
<\psi,e^{i\int_M  d^4x N(x,t) \hat{H}(x)}\psi'>
\\
&=& \lim_{T\to \infty} \int_{\overline{{\cal N}}} [d\bar{N}]
<\psi,e^{i\int_{-T}^Tdt \int_\sigma  d^3x N(x,t) \hat{H}(x)}\psi'>
\nonumber\\
&=&\lim_{T\to\infty} \int_{{\cal N}} [dN]
<\psi,e^{i\int_\sigma  d^3x N(x) \hat{H}(x)}\psi'>
[\int_{\overline{{\cal N}}} [d\bar{N}]\delta(\int_{-T}^T dt
\bar{N}(x,t),N(x))]
\nonumber
\ea
Consider the integral
\be \label{9.4}
I^T_N:=\int_{\overline{{\cal N}}} [d\bar{N}] \delta(\int_{-T}^T dt
\bar{N}(x,t)=N(x))
\ee
appearing in the square bracket in the last line of (\ref{9.3}).
We claim that it is actually independent of $N(x)$. This can be verified by
introducing the constant shift
$\bar{N}(x,t)\mapsto \bar{N}(x,t)+\frac{N'(x)-N(x)}{2T}$
so that $I^T_N=I^T_{N'}=const.$. We conclude that (\ref{9.3}) and
(\ref{9.2}) are proportional to each other (by an infinite constant
$\lim_{T\to\infty} I^T_N$). The formula (\ref{9.3}) is then the starting
point for formulating a path integral through the usual skeletonization
process.

In any case we can now formally expand the exponent in (\ref{9.2})
and arrive at the following picture: Given two spin-network functions
$T_s,T_{s'}$ we have
\be \label{9.5}
<T_{s,phys},\psi'_{s',phys}>_{phys}:=
\sum_{n=0}^\infty \frac{i^n}{n!} \int_{{\cal N}} [dN]
<T_s,\hat{H}(N)^n T_{s'}>
\ee
Since $\hat{H}(N)$ is closed and densely defined on spin-network
functions, the matrix elements of powers of the Hamiltonian constraint
can be computed and since we integrate over all possible lapse functions
the result is manifestly spatially diffeomorphism invariant. Of course, the
result is badly divergent, but cutting off the integral over $N$ somehow
the following picture emerges: The power of $\hat{H}(N)^n$ corresponds
to a discrete $n$ time step evolution of an intial spin-net $s'$ to a
final one $s$. At each step $\hat{H}(N)$ changes the graph of the spin net
$s'$ according to the rules of section \ref{s6}. Let us associate a
hypersurface with each time step and let the respective spin nets
be embedded inside them. Connect the vertices of the spin-nets
in subsequent hypersurfaces by dotted lines. Since
$\hat{H}(N)$ adds edges to a graph, one of these dotted lines
branches up at some intermediate point into two additional dotted lines
which connect with the two newly created vertices.

We thus see that the quantum time evolution of edges become two-surfaces
(bounded by one or two edges and two dotted lines), that is, {\it
a spin foam}. Such kind of transition amplitudes are exactly of the form
as considered earlier by Reisenberger already \cite{140}.

Thus, the canonical theory seems to suggest a bubble evolution not unlike
the worldsheet formulation of string theory, although spin foams define
a background independent string theory in which the worldsheet is
not a smooth two-dimensional manifold but has necessarily (conical)
singularities due to the fact that the Hamiltonian constraint acts
non-trivially only at vertices in each time step.

In order to give mathematical meaning to these amplitudes
one obviously has to look for a better definition of the path
integral. One will therefore begin with stripping off all
the particulars of the specific theory that describes quantum gravity
and consider very general spin foam models and search for
criteria when they converge and when they do not. Then, in a second
step, one has to select among the converging ones the theory
which describes quantum gravity (if any).

It turns out that a systematic starting point are the so-called
BF topological field theories \cite{138}. In $D+1$ dimensions these are
described by an action ($D\ge 2$)
\be \label{9.6}
S_{BF}=\int_M \mbox{Tr}(B\wedge F)
\ee
where $B$ is a Lie$(G)$ valued $(D-1)-$form in a vector bundle
associated to a principal $G$ bundle $P$ under the adjoint
representation and $F$ is the curvature of a connection $A$
over $P$. The trace operation is with respect to the
the non-degenerate Cartan-Killing metric on Lie$(G)$
(assuming $G$ to be semi-simple), that is, basically the
Kronecker symbol (up to normalization). The equations of motion
are given by $F=DB=0$ where $D$ is the covariant differential
determined by $A$ (see section \ref{sa}). Thus $A$ is constrained
to be flat. The action has a huge symmetry, namely it is gauge invariant
and invariant under $A\mapsto A,\;B\mapsto B+Df$ for any $(D-2)-$form
$f$. Counting physical degrees of freedom it is easy to see that
almost nothing is left, the theory has only a finite number of degrees of
freedom, it is topological.

The connection with gravity is made through the Plebanski (first order)
action (in this section we set $\kappa=1$)
\be \label{9.7}
S_P=\int_M \mbox{Tr}((\ast[e\wedge e])\wedge F)
\ee
Here $e=(e_\mu^j)$ denotes the co-(D+1)-bein and
$\ast$ denotes the Hodge dual with respect to the
internal metric $\eta_{ij}$ which is just the
Minkowski (Euclidean) metric for Lorentzian
(Euclidean) general relativity with gauge group
$SO(D,1)$ ($SO(D+1)$). More specifically
\be \label{9.8}
(\ast[e\wedge e])_{ij}:=\frac{1}{(D-1)!}
\epsilon_{ij k_1..k_{D-1}}e^{k_1}\wedge..\wedge e^{k_{D-1}}
\ee
and plugging this into (\ref{9.7}) one easily sees that
(\ref{9.7}) equals the Einstein-Hilbert action for
orientable $M$ when $A$ is the spin-connection of
$e$ (which is one of the equations of motion that
one derives from (\ref{9.7})). Thus we see that
gravity is a BF theory modulo the constraint that
$B$ is in this case not an arbitrary $(D-1)-$form but
rather has to satisfy the so-called {\it simplicity}
constraint
\be \label{9.9}
B=\ast[e\wedge e]
\ee

The idea for writing a path integral for general relativity
is then the following: A lot is known about the path integral
quantization of BF theory in three and four dimensions \cite{137}.
Thus, it seems to be advisable to consider general relativity
as a BF theory in which the sum over histories is constrained
by (\ref{9.9}). One might wonder how it can happen that a
TQFT like BF theory with only a finite number of degrees of
freedom plus additional constraints can give rise to a field
theory like general relativity with an infinite number of
degrees of freedom. The answer is that (\ref{9.9}) breaks
a lot of the gauge invariance of BF theory so that gauge
degrees of freedom become physical degrees of freedom.
In order to sum over histories of $B$'s and $A$'s with
the constraint (\ref{9.9}) we must first write it in a form
in which only $B$'s appear. The algebraic condition on
$B$ such that there exists $e$ with (\ref{9.9}) satisfied has been
systematically analyzed by Freidel, Krasnov and Puzio in
\cite{141}. It can be written for $D\ge 3$ as
\be \label{9.10}
\epsilon^{ijklm_1..m_{D-3}} B^{\mu\nu}_{ij} B^{\rho\sigma}_{kl}
=\epsilon^{\mu\nu\rho\sigma\lambda_1..\lambda_{D-3}}
c^{m_1..m_{D-3}}_{\lambda_1..\lambda_{D-3}}
\ee
where $c$ is any totally skew (in both sets of indices)
tensor density and
\be \label{9.11}
B^{\mu\nu}_{ij}=\frac{1}{(D-1)!}\epsilon^{\mu\nu\rho_1..\rho_{D-1}}
\eta_{ik}\eta_{jl} B^{kl}_{\rho_1..\rho_{D-1}}
\ee
Actually for $D=3$ there is another solution to (\ref{9.10})
besides (\ref{9.9}) given by
\be \label{9.12}
B=\pm e\wedge e
\ee
but this solution gives rise again to a topological theory.
The constraint (\ref{9.10}) is enforced by adding to
the BF action a term of the form
\be \label{9.13}
\frac{1}{2}\int_M d^{D+1}x \Phi_{\mu\nu\rho\sigma}^{ijkl}
B^{\mu\nu}_{ij} B^{\rho\sigma}_{kl}
=:\frac{1}{2}\int_M \mbox{tr}(B\wedge \Phi(B))
=:\int_M \Phi\cdot C
\ee
where the Lagrange multiplier $\Phi$ is totally skew in both
index sets and we have denoted the simplicity constraint by
$C$.

Now the partition function for BF theory is given
by
\be \label{9.14}
Z_{BF}=\int [dA\;dB] e^{i\int_M \mbox{tr}(B\wedge F)}
\propto\int [dA] \delta(F)
\ee
where for {\it either signature} the factor of $i$ in front of the
action has to be there in order to enforce the flatness
constraint $\delta(F)$. That this defines the correct path integral
(up to proper regularization) has been verified by independent
methods, see \cite{137,138} and references therein. Since, from the point
of view of
BF theory, general relativity is a ``perturbation" (with the role
of the ``free" theory being played by BF theory) with
interaction term (\ref{9.13}) the partition function
for general relativity should be given by
\be \label{9.15}
Z_P=\int [dA\;dB\;d\Phi]
e^{i\int_M \mbox{tr}(B\wedge [F+\frac{1}{2}\Phi(B)])}
\propto \int [dA\;dB] \delta(C)
e^{i\int_M \mbox{tr}(B\wedge F)}
\ee
where the additional integral over the Lagrange
multiplier enforces the simplicity constraint.
Path integrals of the type (\ref{9.15}) were studied
by Freidel and Krasnov \cite{142} in terms of a generating
functional
\be \label{9.16}
Z[J]:=\int [dA\;dB]
e^{i\int_M \mbox{tr}(B\wedge [F+J])}
\ee
where $J$ is a two-form current. It is easy to see that formally
by a trick familiar from ordinary quantum field theory
\be \label{9.17}
Z_P=\int [d\Phi] \{e^{i\frac{1}{2}\int_M \mbox{tr}(\frac{\delta}{i\delta J}
\Phi(\frac{\delta}{i\delta J})])}Z[J]\}_{J=0}
\ee
which could then be the starting point for perturbative
expansions. Unfortunately, a truly systematic derivation
of spin foam models for general relativity starting
directly from (\ref{9.17}) is still missing.

We see that in order to define the partition function
for general relativity we must first define the one for
BF theory. Let us first consider the case that $G$ is
compact (Euclidean signature). Then the $\delta-$distribution
$\delta(F)$ in (\ref{9.14}) can be interpreted as the condition
that the holonomy of every contractible loop is trivial.
Furthermore, in order to regularize the functional integral,
we triangulate $M$, using some triangulation $T$ and interpret
the measure $[dA]$ as the uniform measure on $\ab$ restricted to $T$.
Then the condition $F=0$ amounts to saying that $h_\alpha=1_G$
where $\alpha$ is any contractible loop within $T$.
Let $\pi_1'(T)$ be the generators of the contractible subgroup
of the fundamental group of $T$. Hence the regulated BF
partition function becomes
\be \label{9.18}
Z_{BF}(T)=\int_{{\ab}_T} d\mu_{0T}(A)\prod_{\alpha\in \pi_1'(T)}
\delta(A(\alpha),1_G)
\ee
and we can use the Peter\&Weyl theorem in order to write
the $\delta-$distribution as
\be \label{9.19}
\delta(h,1_G)=\sum_{\pi\in\Pi} d_\pi \chi_\pi(h)
\ee
Now magically the integral (\ref{9.18}) is independent
of the choice of triangulation which can be traced back
to the fact that BF is a topological theory. The theory
defined by (\ref{9.18}) is known as the Turarev-Viro
state sum model for $D=2,G=SU(2)$ and as the Turarev-Ooguri-Crane-Yetter
model in $D=3,G=SO(4)$. Actually (\ref{9.18}) is still divergent
when one expands out the products of $\delta-$distributions
but this can be taken care of by using a quantum group regularization
at a root of unity which cuts off the sum over representations at those
of bounded dimension.

Let us now turn to Euclidean gravity for $D=3$. We somehow must
invoke the simplicity constraint into (\ref{9.18}). The idea is
to look at a canonical quantization of $BF$ theory with the
additional simplicity constraint imposed.
This analysis has been
started by by Barbieri \cite{143} leading to the consideration
of {\it quantum tetrahedra} and was completed by Baez and Barrett
\cite{144}. The result is as follows: Recall that $SO(4)$ is
homomorphic with $SU(2)\times SU(2)$, therefore its irreducible
representations can be labelled by two spin quantum numbers
$(j,j')$ (``left handed and right handed"). The simplicity
constraint now amounts to the constraint $j=j'$ explaining
the word ``simplicity". This motivates to define the
partition function for general relativity by restricting
the sum in
\be \label{9.20}
\delta(h,1_{SO(4)})=\sum_{j,j'} d_{\pi_{j,j'}} \chi_{\pi_{j,j'}}(h)
\ee
to
\be \label{9.21}
\delta'(h,1_{SO(4)})=\sum_j d_{\pi_{j,j}} \chi_{\pi_{j,j}}(h)
\ee
resulting in
\be \label{9.22}
Z_P(T)=\int_{{\ab}_T} d\mu_{0T}(A)\prod_{\alpha\in \pi_1'(T)}
\delta'(A(\alpha),1_G)
\ee
(Some version of) (\ref{9.22}) is referred to as the
Barrett-Crane model \cite{145}. The model has been
improved in its degree of uniqueness by Reisenberger
\cite{146} and also by Yetter, Barrett and Barrett
and Williams \cite{147}.

In contrast to (\ref{9.18}) the integral (\ref{9.22}) is
expectedly no longer independent of the triangulation $T$
so that one has to sum over all triangulations in order
to obtain triangulation independence. This amounts to
defining
\be \label{9.23}
Z_P=\sum_T w(T) Z_P(T)
\ee
Of course, the immediate question is how the weight factors
$w(T)$ should be chosen. Notice that for this section
we mean by a triangulation not an embedded triangulation but
a topological one, that is, in some sense four-dimensional
diffeomorphism invariance is {\it defined} to be taken care of.

A clue for how to do that comes from the matrix model approach
to two-dimensional quantum gravity, see e.g. \cite{148} and references
therein. Boulatov and Ooguri \cite{149} respectively
have shown that a Feynman like expansion
of a certain field theory over a group manifold (rather than a
space time) gives rise to all possible triangulations
of the Ponzano Regge (or the Turarev-Viro) model in three dimensions
with $G=SU(2)$
and the Crane-Yetter model in four dimensions respectively
\cite{137} with $G=SO(4)$.
In \cite{150} de Pietri, Freidel, Krasnov and
Rovelli applied these ideas in order to recover the
Barrett Crane model from a field theory formulation.
To see how this works, consider first the case of the
BF theory in $D=3$. Here one considers a real scalar field over
$SO(4)^4$ which is right invariant, that is
$\phi(h_1,h_2,h_3,h_4)=\phi(h_1 g,h_2 g,h_3 g,h_4 g)$ for any
$g\in SO(4)$. One can always obtain such a $\phi$
from a non-invariant field $\phi'$ by
$\phi=\int_{SU(2)} d\mu_H(g) R_g^\ast\phi'$.
The Boutalov -- Ooguri action is then given by
\ba \label{9.24}
S'_{BO}&=&\int_{SO(4)^4} d\mu_H(h_1)d\mu_H(h_2)d\mu_H(h_3)d\mu_H(h_4)
\phi^2(h_1,h_2,h_3,h_4)
\\
&& +\frac{\lambda}{5!}\int_{SO(4)^{10}}
d\mu_H(h_1)d\mu_H(h_2)d\mu_H(h_3)d\mu_H(h_4)d\mu_H(h_5)
\times\nonumber\\
&& \times d\mu_H(h_6)
d\mu_H(h_7)d\mu_H(h_8)d\mu_H(h_9)d\mu_H(h_{10})
\times \nonumber\\
&& \times \phi(h_1,h_2,h_3,h_4)\phi(h_5,h_6,h_7,h_8)\phi(h_7,h_3,h_8,h_9)
\phi(h_9,h_6,h_2,h_{10})\phi(h_{10},h_8,h_5,h_1)\nonumber
\ea
which looks almost like a $\lambda \phi^5$ theory. One can now develop
the usual Feynman rules for this field theory, giving rise to propagators
and vertex functions and construct the perturbation theory as an
expansion in powers of $\lambda$. The result is (for $\lambda=1$)
\be \label{9.25}
\int [d\phi] e^{-S_{BO}(\phi)}=\sum_T w(T) Z_{BF}(T)
\ee
with specific weight factors $w(T)$. Notice that the sum
over triangulations is redundant for BF theory but not
for general relativity.

Given the fact that the Barrett - Crane
model basically reduces the $SO(4)\cong SU(2)_L\times SU(2)_R$ of the BF
theory to $SU(2)$ it was natural to try to reduce the
Crane -- Yetter model to the Barrett -- Crane model by
requiring separate right invariance under $SU(2)$,
that is, $\phi(g_1,g_2,g_3,g_4)=\phi(g_1 h_1,g_2 h_2,g_3 h_3,g_4 h_4)$
for any $h_1,..,h_4\in SU(2)$.
Notice that such a field effectively only lives on $SU(2)^4$ precisely
as wanted (more precisely, its Peter\&Weyl expansion reduces to
simple representations).
This can be achieved by means of a projection
\be \label{9.26}
(P\phi)(g_1,..,g_4)=\int_{SU(2)^4} d\mu_H(h_1)d\mu_H(h_2)d\mu_H(h_3)d\mu_H(h_4)
\phi(g_1 h_1,g_2 h_2,g_3 h_3,g_4 h_4)
\ee
where we have chosen some internal direction in four dimensional Euclidean space
in order to write $SO(4)$ in terms of two copies of $SU(2)$ (to choose a
$SU(2)$ subgroup of $SO(4)$). The field
$P\phi$ is independent of that direction since it is invariant under simultaneous right
action by $SO(4)$ as well. The theory
considered in \cite{150} is given by (\ref{9.24}) just that
$\phi$ is replaced by $P\phi$, that is,
\ba \label{9.27}
S'_{BC} &=& \int_{SO(4)^4} d\mu_H(h_1)d\mu_H(h_2)d\mu_H(h_3)d\mu_H(h_4)
(P\phi)^2(h_1,h_2,h_3,h_4)
\\
&& +\frac{\lambda}{5!}\int_{SO(4)^{10}}
d\mu_H(h_1)d\mu_H(h_2)d\mu_H(h_3)d\mu_H(h_4)d\mu_H(h_5)
\nonumber\times \\
&& \times
d\mu_H(h_6)d\mu_H(h_7)d\mu_H(h_8)d\mu_H(h_9)d\mu_H(h_{10})
\times \nonumber\\
&\times &
(P\phi)(h_1,h_2,h_3,h_4)(P\phi)(h_5,h_6,h_7,h_8)(P\phi)(h_7,h_3,h_8,h_9)
\nonumber\times\\
&& \times(P\phi)(h_9,h_6,h_2,h_{10})(P\phi)(h_{10},h_8,h_5,h_1)\nonumber
\ea
It was shown that the resulting
Feynman expansion indeed gives rise to a sum over triangulations of the Barrett
Crane model.

The individual terms of the resulting series, however, are still
divergent. In \cite{151} Rovelli
and Perez suggested a slight modification of (\ref{9.27}) by removing the
projection in the quadratic term, that is,
\ba \label{9.28}
S'_{RP}&=&\int_{SO(4)^4} d\mu_H(h_1)d\mu_H(h_2)d\mu_H(h_3)d\mu_H(h_4)
\phi^2(h_1,h_2,h_3,h_4)
\\
&& +\frac{\lambda}{5!}\int_{SO(4)^{10}}
d\mu_H(h_1)d\mu_H(h_2)d\mu_H(h_3)d\mu_H(h_4)d\mu_H(h_5)
\times\nonumber\\
&& \times
d\mu_H(h_6)d\mu_H(h_7)d\mu_H(h_8)d\mu_H(h_9)d\mu_H(h_{10})
\nonumber\\
&\times&
(P\phi)(h_1,h_2,h_3,h_4)(P\phi)(h_5,h_6,h_7,h_8)(P\phi)(h_7,h_3,h_8,h_9)
\times \nonumber\\
&& \times
(P\phi)(h_9,h_6,h_2,h_{10})(P\phi)(h_{10},h_8,h_5,h_1)\nonumber
\ea
which is free of certain bubble divergences in its Feynman expansion.
In \cite{152} Perez proved that the resulting model, which
is only a slight variation of the
Barrett -- Crane model and which effectively only depends on simple
representations, is actually {\it finite} order by order in
perturbation theory (triangulation refinement). Of course, this
does not show that the series converges but it is anyway a remarkable
result that no renormalization is necessary. Besides, in \cite{153}
it was demonstrated that any Euclidean spin foam model can be written as a
field theory over a compact group manifold.

So far we have only discussed the Euclidean theory. Can we also
deal with the Lorentzian case ? In \cite{154} Barrett and Crane
modified their Euclidean model to the Lorentzian case. One obstacle
is that one now has to deal with the non-compact gauge group
$SO(1,3)$ for which all non-trivial unitary representations are
infinite dimensional. The unitary representations of the
universal covering group $SL(2,\Cl)$ are labelled by a pair
$(n,\rho)\in \Rl^+_0\times \Nl^+_0$, quite similar to the case of
the universal covering group $SU(2)\times SU(2)$ of $SO(4)$
which are labelled by a pair $(j,j')\in \Nl_0/2\times \Nl_0/2$.
For an exhaustive treatment see \cite{155}. Following an analogous
procedure that has lead to the constraint $j=j'$ in the Euclidean case
we now find that the simplicity constraint leads to $n\rho=0$,
that is, either $n=0$ or $\rho=0$. These representations pick
an $SL(2,\Rl)$ or $SU(2)$ subgroup within $SL(2,\Cl)$ for
$n=0$ or $\rho=0$ respectively. To see where this comes from,
one notices that the $B$ field of the BF theory
essentially becomes, upon canonical quantization, an angular momentum
operator and the Casimir
operators are given by $C_1=L_{ij} L^{ij},\;C_2=L_{ij}(\ast L)^{ij}$, the
simplicity constraint becomes $C_2=0$. In the Euclidean case the
spectra are $C_1=j(j+1)+j'(j'+1),\;C_2=j(j+1)-j'(j'+1)$ while
in the Lorentzian case they become
spectra are $C_1=[n^2-\rho^2-4]/4,\;C_2=n\rho/4$. We see that
in the Euclidean case the simple representations are
``spacelike" representations $C_1\ge 0$ while the simple representations
with $n=0,\rho=0$ for the Lorentzian theory are timelike and spacelike
respectively. The definition of the $\delta-$distribution becomes now
more complicated because there is no Peter\&Weyl basis any longer.
Rather one has direct integrals and sums respectively for the
simple continuous and discrete series of representations respectively
and in order to evaluate the state sum amplitudes one must now perform
also complicated integrals rather than only discrete sums. In \cite{156}
Baez and Barrett proved that nevertheless a large class of
these amplitudes are ``integrable".

In \cite{157} Perez and Rovelli
managed to show that also (a variant of) the Lorentzian Barrett -- Crane
model can be defined as a field theory on a group manifold including
the sum over triangulations again. Basically, what one does
is to replace in (\ref{9.28}) the group $SO(4)$ by $SL(2,\Cl)$
while the projection $P$ can now be performed with respect to any
of the two
subgroups $SL(2,\Rl)$ and $SU(2)$ respectively while
the field $\phi$ is now simultaneously
$SL(2,\Cl)$ right invariant. In \cite{157} the choice $SU(2)$ was made
in order to define $P$ which is therefore given by (\ref{9.26})
with $g_I\in SO(4)$ replaced by $g_I\in SL(2,\Cl),\;I=1,2,3,4$.
Finally, in \cite{158} Crane, Perez and Rovelli succeeded in proving,
using the results of \cite{156}, that the field theory \cite{157}
is finite order by order in parturbation theory at least on what
they call ``regular" triangulations.\\
\\
This concludes our brief report on the impressive progress that has been
made over the last few years in the ``spin foam model industry".
Let us conclude with a couple of remarks:
\begin{itemize}
\item[i)] {\it Spin Foams and Canonical Theory}\\
What is missing is an interpretation of these spin foam models.
Roughly speaking, what one should do is to impose boundary
conditions (boundary spin nets) on the partition function
and to sum over all
spin foam amplitudes and triangulations in between that are
compatible with the boundary spin-net. Provided that one can show
that the resulting object defines a positive semi-definite
sesqui-linear form one can compute its null space and complete
the corresponding factor space in order to obtain an inner product.
What one then would still have to show is that the theory
that one gets implements (some version of) the Hamiltonian
constraint. In other words, to be really convincing one
must make contact with the canonical theory somehow. An analysis
of this kind has been started in \cite{159}.

One can try to go the other way around and start from the
canonical theory and derive the path integral formulation through
some kind of Feynman-Kac formula. A natural starting point for such
an analysis would be by using coherent states as has often been stressed
by Klauder \cite{120}.
\item[ii)] {\it Semiclassical Analysis}\\
The Perez -- Rovelli variant of the Barrett -- Crane model
seems to be preferred at the moment but it is unclear whether
the modification they performed changes the physics significantly
or not. Moreover, in some sense there is always a jump in
passing from the BF theory to general relativity, in other
words, while it is extremely convincing that one should pass
to simple representations it would be nicer to start from
the constrained BF theory partition function (\ref{9.15})
and arrive at the Barrett -- Crane model by integrating
over the Freidel -- Krasnov -- Puzio Lagrange multiplicator.
Of course even then one has to make some guesses like the
choice of the measure $[dA\; dB\;d\Phi]$. So what one would
like to have are some independent arguments that the
models proposed have the correct classical limit for instance
by showing that they are a well-defined version of the
Reisenberger -- Rovelli projector (\ref{9.5}).
\item[iii)] {\it Sum over Triangulations}\\
While we seem to have finiteness proofs for the field theory
formulation order by order (``triangulation by triangulation"),
it would certainly be even better
if one could establish that the sum over triangulations
converges. ``But maybe this does not need to be the case at
all\footnote{Remark by the author to Alexandro Perez
at the ``Bleibtreu Meeting", 6th floor,
Bleibtreustrasse 12A, 10623 Berlin,
Germany, Feb. 16 -- 18, 2001, Fotini Markopoulou and Lee Smolin
(Organizers).}". The reason is that what we really
would like to show is that
\be \label{9.29}
<O>:=\frac{\int [d\phi] e^{-S[\phi]} O(\phi)}{Z}
\ee
converges for a sufficiently large set of observables
(how to express observables of general relativity in terms
of the field theory on the group manifold is another question).
This object should be regulated by cutting off the
sum over triangulations and then one takes the regulator away.
The objects (\ref{9.29}) possibly define the finite moments
of a rigorously defined measure on some field space on which
the the field $\phi$ lives. This is exactly how one
usually performs constructive quantum field theory, see
\cite{8,47n3,47n7,47n10}: Even in free scalar quantum field
theory none of the objects $[d\phi], e^{-S[\phi]},Z$ makes
sense separately, it is only the combination
$\frac{[d\phi] e^{-S[\phi]}}{Z}$ which can be given a
rigorous meaning.
\item[iv)] {\it Built in Causality and Appearance of
Renormalization Group}\\
In dealing with Lorentzian spin foams it is a valid question
in which sense the corresponding quantum evolution
is causal in any sense. These questions were first
addressed in \cite{160} by Markopoulou and Smolin.
One may even restrict the
class of spin foams to be considered by allowing only
those which are causal.

A different question related to the isssue of the classical limit
is whether there is some notion of a renormalization group
within spin foam models which then would answer the question
in which sense they depend on the class of triangulations that
we sum over or whether we are allowed to perform small
changes in the ``initial field theory action" without changing
the effective low energy (semiclassical) theory, in other words
whether there is a natural notion of universility classes and the
like. A first pioneering work has recently been published by
Markopoulou \cite{161} in which the Hopf algebra structure
underlying renormalization in ordinary field theory
discovered by Connes and Kreimer \cite{162}
was applied to coarsening processes of the triangulations
that underly of spin foams.
\end{itemize}

\subsection{Quantum Black Hole Physics}
\label{s9.2}

A first challenge of quantum black hole
physics is to give a microscopic explanation for the
Bekenstein -- Hawking entropy of a black hole \cite{163}
given by
\be \label{9.30}
S_{BH}=\frac{\mbox{Ar}(H)}{4\ell_p^2}
\ee
where Ar$(H)$ denotes the area of the event horizon $H$
as measured by the metric that describes the corresponding
black hole space time and in this section we set
$\ell_p^2=\hbar G_{Newton}$ instead of
$\hbar\kappa=8\pi\hbar G_{Newton}$.

In \cite{164} Krasnov performed a bold computation: Given
any surface $S$ with spherical topology, given some area $A$
and an interval
$[A-\Delta A,A+\Delta A]$, let us compute the number $N$ of
spin-network states $T_s$ such that
$<T_s,\widehat{\mbox{Ar}}(S) T_s>\in [A-\Delta A,A+\Delta A]$.
Of course, $N$ is infinite. But now let us mod out by the
gauge motions generated by the constraints: Most of the
divergence of $N$ stems from the fact that for given number
of punctures $S\cap \gamma(s)$ and fixed representations
$\vec{\pi}(s)$, there are uncountably many different spin
network states with the same area expectation value
because different positions of the punctures give different
spin-network states. This is no longer the case after moding
out by spatial diffemorphisms. There is, however, still
a source of divergence because what matters for the
area eigenvalue is more or less only the number of
punctures and the spins of the edges that intersect the
surfaces $S$, what happens outside or inside the surface
is irrelevant and certainly even after moding by spatial
diffeomorphisms one still had $N=\infty$ therefore.
Krasnov had to assume that this divergence would be taken care
of after moding out the action of the Hamiltonian constraint.
Hence, ignoring this final divergence his result for
$\Delta\approx \ell_p^2$ was very close to (\ref{9.30})
namely proportional to Ar$(S)/(4\ell_p^2)$. A similar
computation by Rovelli \cite{165} confirmed this value.

This result was promising enough in order to spend more
effort in making it water-tight: For instance, nothing
in \cite{164} could prevent one from performing the
computation for {\it any} surface, not necessarily
a black hole event horizon so that it was conceptually
unclear what the computation showed. Somehow one had
to invoke the information that $H$ {\it is} an event
horizon into the computation to get rid of the
divergences that were just mentioned. Also, given the
local nature of the area eigenvalue counting, it was desirable
to localize the notion of an event horizon which can be determined
only when one knows the entire spacetime (recall
that an event horizon \cite{14} is the external boundary of the
portion of
spacetime that does not lie in the past of null future infinity)
which is completely unphysical from an operational point of view because
one would never know if a horizon is really an event horizon since
the object under study could collide with a burnt out star in the
late period of the universe when all life has deceased.
Whether or not $H$ is a horizon one should be able
to determine by performing local measurements in spacetime.

These questions gave rise to a whole industry of its own,
called ``isolated horizons", which to a large extent
is a new beautiful chapter in classical general relativity.
In what follows we will try to summarize the main ingredients
of the framework, focussing on the quantum aspects.\\
\\
Notice that in canonical quantum gravity, as presently formulated,
we must specify a three manifold $\sigma$ of arbitary but fixed topology.
When $\sigma$ has a boundary, one must impose suitable boundary
conditions on the fields in order to obtain a well-defined action
principle. The idea is to {\it first} classically encode the presence of a
locally
defined horizon in the topology of $\sigma$ and the boundary conditions on
$(A,E)$ at the internal boundary (and the usual asymptotically flat
boundary conditions at spatial infinity $i^0$) and {\it then}
to quantize the system. Let us first give the abstract definition.
\begin{Definition} \label{def9.1} ~~~~~~\\
A submanifold $\Delta$ of a spacetime $(M,g)$ is said to be an isolated
horizon if\\
i)\\
$\Delta$ is topologically $\Rl\times S^2$, null with zero shear and
expansion. This conditions ensures that the covariant derivative
$\nabla$ on $M$ induces a unique covariant derivative on $\Delta$
via $D u=[\nabla\tilde{u}]_{|\Delta}$ where $\tilde{u}$ is any smooth
extension of the vector field $u$ on $\Delta$ to $M$.\\
ii)\\
There exists a null normal $l$ of $\Delta$ such that
$({\cal L}_l D-D{\cal L}_l)u\sim 0$ where $\sim$ denotes equality when
restricted to $\Delta$ (since $l$ is defined on $M$ we must use
an extension $\tilde{u}\tilde{Du}$ of $u,Du$ in order to act with
the Lie derivative ${\cal L}_l$).\\
iii)\\
The field equations hold at $\Delta$.
\end{Definition}
Notice that this definition is local to $\Delta$. We can think of
$l$ as ``time-direction" on $\Delta$ and so the first two conditions
imply that the geometry on $\Delta$ is stationary with respect to $l$.
These three conditions are very tight but less tight than those that
lead to event horizons although all known black hole families are
encompassed. For
instance it allows that there is
radiation within the bulk of $M$ which may even fall into the singularity
as long as it does not cross $\Delta$.
For a brief
discussion of all implications and an extension to matter
see \cite{134} and for a detailed derivation see \cite{136}.

For our limited
considerations concerning
the black hole entropy calculation it will be sufficient to describe the
consequences of definition \ref{def9.1} for the canonical quantization.
From now on we will restrict our attention to the portion of $M$ which
is bounded by two initial data hypersurfaces $\Sigma_1,\Sigma_2$,
spatial infinity $i^0$ and the isolated horizon $\Delta$.
As it is clear from the definition, the isolated horizon implies that
our initial data hypersurfaces $\Sigma$ that foliate $M$ are diffeomorphic
to $\sigma$ where $\sigma$ has an internal $S^2$ boundary. Then
definition \ref{def9.1} implies the following (we will
work with arbitrary Immirzi parameter $\beta$ but suppress
it in the canonical coordinates $(A=\Gamma+\beta K,E=E_1/\beta)$):\\
1)\\
There is a differentiable bijection $r^j:\; S^2\to S^2$
(meaning that $r^j r^k\delta_{jk}=1$). Given an $SU(2)$ principal fibre
bundle over $\sigma$ we obtain a principal $U(1)$ bundle over $S:=S^2$
by restricing the fibres over $s\in S^2\subset\sigma$ to those
$g\in SU(2)$ which preserve the internal vector $r^j$.
Consider the $U(1)$ connection on $S^2$ defined by
\be \label{9.30a}
W:=-\frac{1}{\sqrt{2}} [X^\ast \Gamma^j]r_j
\ee
where $X:\; S\to \sigma$ denotes the corresponding embedding
and $\Gamma$ the spin connection of the triad.\\
2)\\
The boundary conditions on the pull-backs $\underline{A},\underline{\ast E}$
to $S^2$ are that
\ba \label{9.31}
\underline{A}^j&:=&X^\ast A^j=W r^j
\nonumber\\
\underline{\ast E} &:=& X^\ast(\ast E^j) r_j=-\frac{a_0}{2\pi\beta} dW
\label{9.32}
\ea
where $a_0:=\mbox{Ar}_E(S)$ is constrained to be a constant
by the isolated horizon conditions (independent of $E$). Thus
$(\underline{A},\underline{\ast E})$ are completely determined by
$(W,dW)$ respectively.\\
3)\\
Finally, the symplectic structure of our classical system turns out to be
\be \label{9.33}
\Omega((\delta A,\delta E),(\delta A,\delta E))
=\frac{1}{\kappa}
[\int_{\sigma-S}\mbox{Tr}(\delta A\wedge \delta\ast E'
-\delta A'\wedge \delta\ast E)
+\frac{a_0}{\pi\beta} \int_S \delta W\wedge \delta W']
\ee
for arbitrary tangential vectors
$\delta A,\delta E,\delta A',\delta E'$ to the phase space $\cal M$.

Thus $\Omega=\Omega_\sigma+\Omega_S$ consists of a bulk and a
surface term. Clearly, classically the surface degrees of freedom are
determined by the bulk degrees of freedom by continuity but this
will change in quantum theory where the distributional nature of the
quantum configuration space excites additional degrees of freedom.
Interestingly, the surface symplectic structure is that of
a $U(1)$ Chern-Simons theory on $\Rl\times S$
with action
\be \label{9.34}
S_{CS}=\int_{\Rl\times S} W\wedge dW=\int_\Rl dt
\int_S \epsilon^{IJ}[\dot{W}_I W_J+W_t (dW)_{IJ}]
\ee
which displays $\epsilon^{IJ} W_J$ as the momentum canonically conjugate
to $W_I$ and the constraint is that $W$ be flat.

In order to quantize the system we will adopt the following strategy:
\begin{itemize}
\item[1.)] {\it Quantum Configuration Space}\\
Essentially we will make the bulk and surface configuration degrees of
freedom independent of each other, that is, $\ab=\ab_\sigma\times \ab_S$
with distributional $SU(2)$ and $U(1)$ connections respectively.
\item[2.)] {\it Kinematical Hilbert Space}\\
Accordingly the kinematical Hilbert space adopts a tensor product structure
${\cal H}^0={\cal H}^0_\sigma\otimes {\cal H}^0_S$ with
${\cal H}^0_\sigma=L_2(\ab_\sigma,d\mu_0),
{\cal H}^0_S=L_2(\ab_S,d\mu_0)$.
\item[3.)] {\it Quantum Boundary Conditions}\\
This structure suggests to solve (\ref{9.32}) in the symbolic form
\be \label{9.35}
[1_{{\cal H}^0_\sigma}\otimes e^{i\widehat{dW}}]\psi
=
[e^{-\frac{2\pi\beta i}{a_0}\widehat{\underline{\ast E}}}
\otimes 1_{{\cal H}^0_S}]\psi
\ee
The reason for this particular exponentiation of (\ref{9.32})
is required by the particulars of the quantization of
Chern-Simons theory.
\item[4.)] {\it Implementation of Quantum Dynamics}\\
Finally one has to impose the constraints at $S$. It turns
out that $\widehat{\underline{\ast E}}$ and
$\widehat{dW}$ generate $U(1)$ gauge transformations in the bulk
close to $S$ and on $S$ respectively so that through
(\ref{9.35}) {\it the Gauss constraint is already
satisfied}, in other words, the total state depending
on both bulk and surface degrees of freedom is gauge invariant!
Next, as already anticipated at the beginning of this
section, the diffeomorphism constraint restricted to $S$
basically tells us that what is important is the number
of punctures of the graph of a spin-network with the surface $S$
and not their position. Finally, the Hamiltonian constraint
vanishes identically at $S$ due to the definition of an
isolated horizon which, in particular, requires that
lapse functions that generate gauge motions induced by
$H$ must be identically zero at $S$ (This does not mean
that the lapse of a classical isolated horizon solution
must vanish at $S$, rather there is a subtle difference
between lapse functions that generate symmetries rather
than gauge transformations (see, e.g. \cite{30l} and references
therein) so that in this case lapse functions that do not
vanish at $S$ map between gauge inequivalent solutions).
\end{itemize}
Let us describe this in more detail. In order to solve
(\ref{9.35}) one makes a tensor product ansatz
$\psi=\psi_\sigma\otimes \psi_S$ implying
\be \label{9.36}
\psi_\sigma\otimes [e^{i\widehat{dW}}\psi_S]
=
[e^{-\frac{2\pi\beta i}{a_0}\widehat{\underline{\ast E}}}\psi_\sigma]
\otimes \psi_S
\ee
and one will try to look for eigenvectors of the exponentiated
operators with the same eigenvalues. Now, what we have
quantized in section \ref{s4} was not $\ast E^j$ bat rather
$\int_{S'} \ast E$ where $S'$ is any surface, however, the fact
that this function can be turned into an operator
means that there exists an operator valued distribution
which when restricted to $S$ is given on a function cylindrical
over $\gamma$ by
\be \label{9.37}
\widehat{\underline{\ast E}}(s)\psi_{\sigma,\gamma}=
4\pi i\ell_p^2\sum_{p\in S\cap \gamma} \eta \delta^{(2)}(s,X^{-1}(p))
\sum_{e\in E(\gamma);f(e)=p} L^j_e r_j(s) \psi_{\sigma,\gamma}
\ee
(the appearance of the $4\pi$ rather than $1/2$ as compared
to section \ref{s4} is due to our definition of
$\ell_p$ in this section).
Here $\eta=\frac{1}{2}\epsilon_{IJ} ds^I\wedge ds^J$ in a local
system of coordinates $s^I$ on $S$. Notice that only
left invariant vector fields appear because every edge that
intersects $S$ is of the ``down" type ($S$ carries outward
orientation and there is no interior of $\sigma$ with respect to
$S$ because $S$ is a boundary). No edge of the bulk graph $\gamma$
lies inside $S$ because edges inside $S$ label surface degrees of
freedom. Now the spectrum of (\ref{9.37}) can be computed by
inspection: The operator $i r_j L^j$ on $L_2(G,d\mu_H)$
is nothing else than the operator $2\hat{J}_3$ for the quantum
mechanics of the angular momentum whence it has spectrum
$2m$ with $m$ a half-integral quantum number. In fact, we can choose
a spin network basis in which $(L^j_e)^2,L^j_e r_j(s)$ are diagonal
and immediately obtain as ``distributional eigenvalues"
\be \label{9.37a}
\widehat{\underline{\ast E}}(s)\psi_{\sigma,s}=
8\pi \ell_p^2\sum_{p\in S\cap \gamma(s)} \eta \delta^{(2)}(s,X^{-1}(p))
\sum_{e\in E(\gamma(s));f(e)=p} m_e \psi_{\sigma,s}
\ee
where $|m_e|\le j_e$ is half-integral. The result (\ref{9.37})
motivates to split the bulk Hilbert space as
\be \label{9.38}
{\cal H}^0_\sigma=\oplus_{{\cal P},m} {\cal H}^{{\cal P},m}_\sigma
\ee
where ${\cal P}$ denotes the set of all punctures of $S$ (that is,
positions of points where a bulk graph intersects $S$)
and $m$ the possible eigenvalues (\ref{9.37}).

Next we turn to the operator $\widehat{dW}$. Since its eigenvalues
must match (\ref{9.37}) we conclude that the quantum curvature of
$W$ is flat everywhere except at the punctures. Consider the
spaces $\a^{{\cal P}},\g^{{\cal P}},{\cal D}^{{\cal P}}$ of
connections which are flat up to the punctures ${\cal P}$, gauge
transformations which equal the identity at ${\cal P}$ and
analytic diffeomorphisms which preserve ${\cal P}$. Consider the
moduli space
\be \label{9.39}
{\cal M}^{{\cal P}}:=\a^{{\cal P}}/(\g^{{\cal P}}\triangleleft
{\cal D}^{{\cal P}})
\ee
and turn it into a symplectic manifold by equipping it with
the Chern-Simons symplectic structure (the semi-direct
product in (\ref{9.39}) appears because diffeomorphisms
act non-trivially on gauge transformations).
The phase space ${\cal M}^{{\cal P}}$ is compact (one way to see
this is that it can be coordinatized by $U(1)$ holonomies,
see below)
and therefore does not admit the standard cotangent bundle
polarization. However, it can be quantized by the methods of
geometric quantization \cite{18} by choosing a (positive)
K\"ahler polarization. It would take us too far
to develop the necessary background for general geometric
quantization and quantization of Chern-Simons theory in
particular, see however the exhaustive treatment in
the beautiful thesis \cite{163}. The outcome of this
analysis is as follows:\\
1) {\it Phase Space}\\
First of all,
${\cal M}^{{\cal P}}$ can be identified with the torus
with $T^{2(n-1)}:=\Cl^{n-1}/(2\pi \Zl)^{2(n-1)}$ where
$n=|{\cal P}|$ is the number of punctures. To see at least
intuitively how this happens, notice that the holonomies around
loops and paths between the punctures separate the points
of $\a^{{\cal P}}/\g^{{\cal P}}$ since gauge transformations at
the punctures are trivial.
The punctured surface $S$ is homeomorphic
to a sphere with holes and the holonomy of a flat connection
along a loop depends only on its homotopy type. Thus, one might
think that the homotopy group of of the punctured sphere is
generated by $n$ elements $\alpha_p$ where $\alpha_p$ encloses
$p$ but not any other puncture, however, this is not true: Any
loop $\beta$ which encloses all punctures is contractible
``over the back of the sphere" and so is the loop
$\beta\circ (\circ_{p\in {\cal P}}\alpha_p^{-1})$.
Thus we may get rid of one of the $\alpha_p$, say $\alpha_{p_0}$
for some fixed puncture $p_0$. Next, consider paths
$e_p,\;p\not=p_0$ between $p$ and $p_0$ which do not intersect
any $\alpha_{p'},\;p'\not=p,p_0$. Obviously, any path
between $p,p'\in {\cal P}$ is homotop to $e_p\circ e_{p'}^{-1}$.
This explains already why the phase space should have
dimension $2(n-1)$. But each holonomy takes values in $U(1)$
which is diffeomorphic with $S^1=T^1=\Rl/(2\pi\Zl)$.
Finally, the diffeomorphisms in ${\cal D}^{{\cal P}}$ preserve the
homotopy type of the $\alpha_p,e_p;\;p\not=p_0$.\\
\\
2) {\it Geometric Quantization}\\
Let us introduce coordinates $x_p,y_p\in\Rl/(2\pi \Zl)$ with
$h_{\alpha_p}=e^{i x_p},h_{e_p}=e^{i y_p}$.
The numbers $x_p,y_p$ coordinatize
a point in ${\cal M}^{{\cal P}}$ and we can always
find a representative
\be \label{9.40a}
W=\sum_{p\not=p_0} [x_p X_p+y_p Y_p]
\ee
where the
one forms $X_p,Y_p$ are the Poincar\'e duals to
the $e_p,\alpha_p$, that is
\be \label{9.41}
\int_{e_p} X_{p'}=\int_{\alpha_p} Y_{p'}=\delta_{pp'}
\mbox{ and }
\int_{\alpha_p} X_{p'}=\int_{e_p} Y_{p'}=0
\ee
Pulling back the symplectic structure $\Omega_s$
by (\ref{9.41} we find
\be \label{9.42a}
\Omega=\frac{k}{2\pi} \sum_{p\not=p_0} dx_p\wedge y_p
\ee
Now the detailed framework of geometric quantization
reveals that there is an obstruction to quantization
which is spelled out in terms of Weil's integrality criterion
\cite{18}.
In our case it boils down to the condition
that the so-called level of the Chern-Simons theory
\be \label{9.40}
k:=\frac{a_0}{4\pi \beta \ell_p^2}
\ee
must be an integer.

Next consider
holomorphic wave functions of the $z_p=x_p+iy_p$ which defines
a so-called K\"ahler polarization (similar to the
Segal-Bargmann representation for the phase space
$\Rl^{2(n-1)}$). We can view functions on
$T^{2(n-1)}$ as functions on $\Cl^{2(n-1)}$ which
are invariant under translations within the
lattice $\Lambda=(2\pi \Zl)^{2(n-1)}$, that is,
periodic {\it holomorphic} functions. Now by Liouville's
theorem periodic holomorphic functions do not exist
so the best that one can achieve are
quasi-periodic functions which are also
called $\Theta-$functions \cite{166a}. These are holomorphic
functions which depend on the Teichm\"uller parameter
$\tau$ with $\Im(\tau)>0$ which determines the complex structure of the
torus and the positive level integer $k>0$. They satisfy the functional
equations (in one complex
dimension) $\Theta^k_\tau(z+2\pi)=\Theta^k_\tau(z),\;
\Theta^k_\tau(z+2\pi \tau)=\exp(- i k z+b) \Theta^k_\tau(z)$
where $b$ is an arbitrary complex number. It turns out that
the vector space of functions satisfying these functional
equations is real $k-$dimensional so that we get
$k$ solutions $\Theta^k_\tau(z,a)$ with $a=0,1,..,k-1\in \Zl_k$.
In our case we have $\tau=i$. The significance of the level $k$
is that the symplectic structure depends on it and that the
$\Theta-$ functions of level $k$ determine a $k-$dimensional
representation of the Heisenberg
group generated by the exponentials of the (pre-)quantum operators
$\hat{x}_p,\hat{y}_p$.
The final result is that in our case only
$\Theta$ functions $\psi_{S,a}$ labelled by $\vec{a}\in (\Zl_k)^{n-1}$ are
indistinguishable, in fact, they form a basis in the
prequantum Hilbert space of square integrable (with respect
to the Liouville measure times a damping
factor related to the K\"ahler potential) holomorphic
sections (of a complex
line bundle over the phase space). (It may come as
a surprise that therefore ${\cal H}^{{\cal P}}_S$ is
finite dimensional, namely $k^{n-1}$ but it really is not because the
number of quantum degrees of freedom is roughly given by the
Liouville volume of the phase space (which in our case is finite)
divided by the volume of a phase cell).
For the same reason the holonomy
operators $\hat{h}_{\alpha_p}$ have eigenvalues
$e^{2\pi i a_p/k}$. The operators $\hat{h}_{e_p}$ will
disappear from the final picture since we have to
take the quotient later on also with respect to gauge
transformations which are not trivial at the
punctures.\\
\\
We thus conclude that the
geometric quantization of
${\cal M}^{{\cal P}}$ leads to a Hilbert space
${\cal H}_S$ which is given by the inductive limit
of the Hilbert spaces ${\cal H}^{{\cal P}}_S$
where ${\cal P}$ ranges over all finite point subsets
of $S$ and where ${\cal H}^{{\cal P}}_S$ is isomorphic
with the geometric quantization of the corresponding
torus. On ${\cal H}^{{\cal P}}_S$ the
holonomy operators $\hat{h}_p$,
can be simultaneously diagonalized
these operators and their eigenvalues are given by
$\hat{h}_p\psi_{S,a}=e^{2\pi i a_p/k}$ where
$\sum_{p\in {\cal P}} a_p=0(\mbox{mod}\; k)$.

Let now $S_p\subset S$ be the interior of $\alpha_p$ then
the non-distributional way to state (\ref{9.36}) is
given by
\be \label{9.41a}
\psi_\sigma\otimes [\hat{h}_p\psi_S]
=
[e^{i-\frac{2\pi\beta i}{a_0}\int_{S_p}
\widehat{\underline{\ast E}}}\psi_\sigma]
\otimes \psi_S
\ee
for any $p\in {\cal P},\;\psi_S\in {\cal H}^{{\cal P}}_S,\;
\psi_\sigma\in {\cal H}^{{\cal P}}_\sigma$. This evidently
leads to the condition
\be \label
2 m_p=-a_p(\mbox{mod}\; k)
\ee
for the corresponding eigenvalues and we conclude that the
kinematical Hilbert space is given (modulo completion) by
\be \label{9.42}
{\cal H}^0=\oplus_{{\cal P},m,a;\;2m=-a(\mbox{mod}\; k)}
{\cal H}^{{\cal P},m}_\sigma\otimes{\cal H}^{{\cal P},a}_S
\ee
This Hilbert space is easily seen to solve the full
Gauss constraint already since we require states to be
gauge invariant in the bulk away from $S$ and the condition
(\ref{9.41}) is exactly the gauge invariance condition
for gauge transformations at the punctures (actually with
respect to a necessarily reduced gauge group).

Next we have to reduce with respect to the spatial
Diffeomorphism constraint which simply amounts to
replacing the spaces
${\cal H}^{{\cal P}}_\sigma,{\cal H}^{{\cal P}}_S$ by
${\cal H}^n_\sigma,{\cal H}^n_S$ where now only the
number of punctures is relevant.

Finally, with respect to the Hamiltonian constraint there
is nothing left to do for the reason already mentioned above
which is quite lucky because, as we have said before, there
is no proof that the proposed bulk Hamiltonian constraint
is the correct one but whatever it is it will vanish at $S$.\\
\\
Our final task will be to do the entropy counting.
The isolated horizon area operator $\widehat{\mbox{Ar}}(S)$
is a true Dirac observable in the present situation since
it is gauge invariant by construction and diffeomorphism
invariant under Diff$(S)$. Its eigenvalues on the
sector of the physical Hilbert space labelled by
$(n,\vec{j},\vec{m},\vec{a}),\;2\vec{m}+\vec{a}=0$ (mod $k$)
with $|m_l|\le j_l$ the spin
of the edge entering the $l$'th puncture (recall that
$(L^j_l)^2$ and $L^j_l r_j$ can be diagonalized simultaneously
with eigenvalues $4j_l(j_l+1)$ and $2m_l$ respectvely)
are given by
\be \label{9.43}
\mbox{Ar}(n,\vec{j})=8\pi \ell_p^2\beta \sum_{l=1}^n \sqrt{j_l(j_l+1)}
\ee
(the area operator acts on the bulk degrees of freedom
only). We are looking now for all those sectors for which
(\ref{9.43}) lies in the interval $[a_0-\delta,a_0+\delta]$.
Notice that when $n,\vec{j},\vec{m}$ are given then
$\vec{a}$ is completely fixed already. The crucial point is now
that
\be \label{9.44}
{\cal H}^{n,\vec{j},\vec{m}}_\sigma=\otimes
={\cal H}^{n,\vec{j},\vec{m}}_V\otimes
{\cal H}^{n,\vec{j},\vec{m}}_{bh}
\ee
where the first factor (corresponding to edges not intersecting $S$)
has infinite
dimension while the second has finite dimension (corresponding to edges
intersecting $S$).
We can summarize this in the formula (modulo completion)
\be \label{9.45}
{\cal H}_{phys}=\oplus_{n,\vec{j},\vec{m},\vec{a}=_k -2\vec{m}}
{\cal H}^{n,\vec{j},\vec{m}}_V\otimes
{\cal H}^{n,\vec{j},\vec{m}}_{bh}\otimes {\cal H}^{n,\vec{a}}_S
\ee
Next we form the corresponding microcanonical statistical ensemble
with the density matrix
\be \label{9.46}
\hat{\rho}_{bh}:=1_V\otimes
[\frac{1}{N_{a_0,\delta}}\sum_{n,\vec{j},\vec{m}}
|\psi^{n,\vec{j},\vec{m}}_{bh}><\psi^{n,\vec{j},\vec{m}}_{bh}|]
\otimes 1_S
\ee
where the sum is over all black hole sectors compatible with
$a_0,\delta$ and $N_{a_0,\delta}$ is their number.
Clearly, the quantum statistical entropy is given by
\be \label{9.47}
S=-\mbox{Tr}(\hat{\rho}_{bh}\ln(\hat{\rho}_{bh}))=\ln(N_{a_0,\delta})
\ee

Thus we just need to count states. The analysis is not entirely
straightforward but can be summarized as follows:\\
It turns out that expectedly most of the entropy comes from
those configurations with $j_l=1/2$ (maximum disorder).
Then $\mbox{Ar}(n,\vec{j})=4\pi \ell_p^2\beta n \sqrt{3}$.
If we choose $\delta>8\pi \ell_p^2\beta  \sqrt{3}$ then
we always find an even integer $n$ in order that the eigenvalue
lies in the required interval. Now the eigenvalue $j_l=1/2$
has degeneracy $2$ corresponding to the two possible
projections $m_l=\pm 1/2$ suggesting that there are $2^n$
such states, {\it one Boolean degree of freedom per puncture}. However, we
must satisfy
the constraint $2(m_1+..+m_n)=0$(mod $k$). Certainly for large $a_0$
we have $|2(m_1+..+m_n)|\le n\approx k/\sqrt{3}< k$ which means
that actually $m_1+..+m_n=0$, thus half of the spins must
be up the others are down. This brings the number of states
with $j_l=1/2$ down to
$\left( \begin{array}{c} n \\ n/2 \end{array} \right)$
which therefore is a lower bound for the number
$N_{a_0,\delta}$. The derivation of an upper bound
is more complicated but can be done with the result that
the leading order term is given by $S=\ln(2) n_0$ where
$n_0\approx k/\sqrt{3}$ which would already be the
leading order if we just had taken the lower bound and
applied Stirling's formula. We see that we {\it precisely}
reproduce the Bekenstein-Hawking entropy provided we choose the
Immirzi parameter to be
\be \label{9.48}
\beta=\frac{\ln(2)}{\pi\sqrt{3}}
\ee
Each of these three transcendent numbers has a well-understood
origin: $\pi=\kappa/(8 G_{Newton})$, $\sqrt{3}/2$ is the
lowest non-vanishing eigenvalue of $\sqrt{j(j+1)}$ and
$\ln(2)$ comes from $\ln(2^n)$.

The strategy to choose the Immirzi parameter according to
(\ref{9.48}) would be worthless if it would not be the {\it same}
value that one would have to match for various kinds of black
holes, not only the vacuum black holes that we have treated so
far. However, as one can show \cite{167} even for dilatonic and
Yang-Mills hair black holes the same value works. This relies
on the following facts: a) the presence of this bosonic matter
does not change the isolated horizon boundary conditions, b)
the matter fields are determined through $W$ at $S$ and therefore
c) matter has no independent surface degrees of freedom.
It should be pointed out that all of this works for
astrophysically realistic (Schwarzschild), four-dimensional,
non-supersymmetric black holes.

A couple of remarks are in order:
\begin{itemize}
\item[i)] {\it Non-Triviality}\\
The calculation is certainly very impressive because one could not
have expected from the outset that there would be a harmonic
interplay between classical general relativity (isolated
horizon boundary conditions), quantum gravity (discrete eigenvalues
of the area operator) and quantum Chern-Simons theory (horizon
degrees of freedom).
\item[ii)] {\it Other Results}\\
Recall that there has been established a precise dictionary
between the four laws of usual thermodynamics and black
hole thermodynamics for {\it event horizons}. It turns out
that one can write another dictionary for {\it isolated
horizons} \cite{168}. Also cosmological horizons
can be described by isolated horizon methods.

Next, we have pointed out before that the main series of the
area spectrum is by far not all of it. In particular, one can show
\cite{170b} that the number of eigenvalues in the interval
$[a_0-\ell_p^2,a_0+\ell_p^2]$
grows as $e^{\sqrt{a_0}/\ell_p}$ explicitly ruling out the naive
ansatz made in \cite{170a} that the area spectrum is evenly spaced
which seemed to be supported by the errornous computation \cite{70a}
and would have implied an even spacing of the spectrum. This has huge
observational consequences: The peak of the black body Hawking spectrum
from the black hole is at frequencies $\omega_0\approx 1/r_0$ where
$r_0\approx GM$ is the Schwarzschild radius of the black hole
(we neglect numerical constants and set $c=1$). Now $a_0=4\pi r_0^2$ and
since
energy emission of the black hole is due to ``area transitions" we obtain
spectral lines at $\hbar\omega\approx (\Delta M)
\approx \Delta (\sqrt{a}/G)\approx (\Delta a)/(g r_0)\approx
\omega_0 \Delta a/G$. We see that if the spectrum would be evenly spaced
at $\Delta a\approx \hbar G$ then $\omega\approx n \omega_0$ so we would
not get a black body spectrum at all, every line would be at a multiple
of the peak line.
\item[iii)] {\it Open Problems}\\
The case that we have treated above was for a static
isolated horizon. While rotating isolated
horizons can be treated classically \cite{169} so far
the quantum theory has not been developed. A related
question is whether one can also treat Hawking radiation
with the present framework and a pioneering ansatz was
made in \cite{170}.
Also, it has been conjectured that
the Bekenstein Hawking entropy is an inevitable,
universal property of any kind of quantum gravity theory
and a proof of that conjecture was begun in \cite{171}.
However, this calculation was shown not to apply in the
present context \cite{172}. Finally, a better understanding
about the role of the Immirzi parameter and whether or not
it should be fixed as displayed here would be desirable.
\item[iv)] {\it First Principle Calculation}\\
The isolated horizon description is an effective
one (not from first principles) because the presence of an
isolated horizon was put in at the classical level.
It would be far more desirable to begin with the
full quantum theory and to have {\it quantum criteria}
at one's disposal for when a given state represents
a quantum black hole. At this point the semi-classical
discussed in section \ref{s8.2} could be of some help.
\end{itemize}

\subsection{Connections Between Canonical Quantum General Relativity and
String -- (M) Theory}
\label{s9.3}

Smolin has conducted an ambitious programme, namely the investigation of
possible interfaces
between canonical quantum general relativity and M -- Theory, especially
in its Matrix theory incarnation \cite{190}. The ultimate goal of this
effort is to arrive at a background independent formulation of
M -- theory.

The possible links between these two major approaches to quantum gravity are
very complex and even a brief introduction would require
at least some background material for M -- Theory which would really
go much beyond what this review is intended to cover. We therefore
must unfortunately leave the reader with the literature cited and
just point out, as an example, the recent paper \cite{191} which seems to
indicate
that there is a conflict between Maldacena's conjecture \cite{192},
which says
that superstring theory on an Anti -- deSitter (AdS) background spacetime
is equivalent (``dual") to a conformal field theory (a super Yang
Mills theory) on the conformal boundary of the AdS space in the sense
that all scattering amplitudes in the bulk are completely determined
by the scattering amplitudes on the boundary, and Rehren's duality
\cite{193}, which says that there is a natural isomorphism between
nets of local algebras (in the sense of the Haag -- Kastler formulation
of quantum field theory) in
the bulk and on the conformal boundary whenever the AdS background is
available. This has actually already been observed earlier in
\cite{194}.
The conflict is the following: Rehren's duality maps a local theory
(spacelike separated
algebras commute), say the one on the boundary, to a theory which
is local, say in the bulk, however, not in the standard way: If
the former theory
is based on a Lagrangean principle so that fields can be indexed by
points on the boundary, the latter theory {\it does not admit
a Lagrangean formulation}, it is in
that
sense a non-local theory, in particular, there is no causal propagation in
the standard sense. Looking at the details of Rehren's map the technical
reason for this effect is that one has dropped one dimensional
information in this ``algebraic holography".

On the other
hand, the Maldacena conjecture seems to assume a map between two
Lagrangean theories (the effective low energy theory of string
theory in the bulk is a supergravity theory).
Thus, either the two dualities have nothing to do with each other
because (the low energy limit of) string theory on a given background
is not a theory to which the Haag-Kastler
framework applies (which would be extremely surprising) or
there is no Lagrangean origin for M -- Theory, not even for its low energy
limt (which is contrary to all what string theorists seem to assume for
decades).

\newpage

\section{Selection of Open Research Problems}
\label{s10}

Instead of summarizing the summary that we have given in this review,
we close with a (far from exhaustive) list
of open research problems which the author considers most
pressing to be solved.
\begin{itemize}
\item[1)] {\it Spectrum of the Volume Operator}\\
We have seen that the volume operator plays a prominent
role in the dynamical structure of the theory, it enters
the Hamiltonian constraint, matter Hamiltonians,
the generator of the Wick transform, the Quantum
Dirac Algebra, the asymptotic Poincar\'e algebra
and the (possible) complexifier of coherent
states. Despite this pivotal interaction of quantum dynamics
with the volume operator, very little is known about its
spectrum not even asymptotically (that is, in the limit
of large spins) and it would therefore be highly
desirable to gain more control over it.
\item[2)] {\it Rigorous Construction of the Wick Transform}\\
For various reasons it would be of benefit to return
to a complex connection formulation, if only because it is
closer to a manifestly covariant formulation of the theory
in terms of a path integral. It therefore seems to be mandatory
to construct the generator of that transform as a self-adjoint
operator. For the beginning of a corresponding analysis
in the simplified context of mini-superspace models see
\cite{196}. Related, although independent to this, is the question
whether there exist background independent measures on distributional
spaces of connections for non-compact gauge groups in analogy to the
structure provided by $\ab,\mu_0$.
\item[3)] {\it Correct Version of the Hamiltonian Constraint Operator}\\
We have tried to indicate that at this point we do not have
{\it the} Hamiltonian constraint but a huge class of consistent
proposals. None of them seems completely satisfactory though for
the reasons we have mentioned. It would be worthwhile to explore
which kind of generalizations are allowed which still lead to
anomaly-free constraint algebras while having the
correct classical limit.
\item[4)] {\it Proof of the Correct Classical Limit}\\
At this point the semi-classical analysis has just started.
The semi-classical states that are available all suffer from
one and the same desease: They are incapable to reproduce correct
expectation values for operators which map between spin network
states over different graphs. But this is precisely what
both the Hamiltonian and Diffeomorphism constraint do. An
appropriate improvement is therefore mandatory before we can even
seriously ask the question whether we have the correct theory.
\item[5)] {\it Contact with Quantum Field Theory on Curved Spacetimes}\\
We have indicated at various occasions how to make contact with
quantum field theory on curved spacetimes. Luckily, for bosonic
quantum field theories these questions can already be addressed
with the semi-classical states available. In its present form,
quantum field theory on curved spacetimes are most naturally
formulated in the language of algebraic quantum field theory
\cite{197}. It is almost clear that nets of local algebras
can have at most a semi-classical meaning since the axiom
of locality makes sense only when one has a background
spacetime available, in other words the fluctuations of the
quantum metric must be small compared to those of matter.
It would be crucial to make this correspondence manifest.
\item[6)] {\it Three -- and Four Dimensional Dirac Observables}\\
We have indicated in this article how one could use
matter in order to turn the area operator into a spatially
diffeomorphism invariant operator. We need something similar
with respect to the Hamiltonian constraint as well, at least
we should have a constructive procedure for how to arrive at them
by some algorithm which converges sufficiently fast so that one
has the notion of an approximate Dirac observable at least.
See \cite{198} for a first proposal.
\item[7)] {\it Avoidance of Classical Singularities}\\
As we mentioned, there are indications that the quantum
symmetry reduction of quantum general relativity to
certain Bianchi cosmologies predicts that there is
no big bang singularity at all. It would add faith to
these results if one could establish similar results within
the full theory without symmetry reduction.
%
\item[8)] {\it Hawking Effect from First Principles}\\
As mentioned in the previous section, the isolated
horizon approach to quantum black holes, as every
approach to quantum black holes that is presently available,
needs a classical input. It would be far more convincing
if one could develop pure quantum criteria for when a state
describes a black hole, thereby opening the possibility to
derive the Hawking effect from first principles.
\item[9)] {\it Proof of a Feynman -- Kac Formula for Canonical Quantum
Gravity}\\
What we need in order to connect the spin foam approach with
the canonical approach is some kind of Feynman -- Kac formula,
some possibility to derive the spin foam model out of canonical
quantum gravity or vice versa. This would be highly desirable
since path integrals and operator methods usually complement each
other. Again, coherent states could play a crucial role towards
this goal.
\item[10)] {\it Combinatorial Formulation of the Theory}\\
An ugly feature of the present framework is that it still depends
in a technically not too weak way on a background differential
(or even analytic) manifold and topology. If, as many suspect,
quantum gravity should allow for topology change then we must
get rid of these structures. Actually, the present framework
suggests its own way out of these limitations: Instead of
talking about embedded graphs we must learn how to formulate
the theory over algebraic graphs, see \cite{49,199}.
\item[11)] {\it Introduce Higher Form Variables in Higher Dimensions}\\
It is quite possible that supersymmetry does not play any role
in nature and that actually four spacetime dimensions are
sufficient. However, if M -- Theory is correct then quantum gravity
in four dimensions can be at most an effective theory. In order
to address this possibility one can start by trying to develop a
background independent quantization of 11D supergravity, the
low energy limit of M -- Theory. As we have seen, in order
to achieve background independence one must build the canonical
theory on $p-$form fields rather than metrics, similar as in
four dimensions. For a first ansatz see \cite{200}.

\item[12)] {\it Make contact with String (Membrane) Theory}
Related to this is the question whether methods of background
independent quantum gravity developed in three and four
dimensions cannot be used also in higher dimensions, for instance in
quantizing the super-membrabe non-perturbatively. The
super-membrane in 11D is one of the ``hot" candidates for
M -- Theory.
%
\item[13)] {\it Construction of Physical Inner Product}\\
We must develop an algorithm for how to arrive at a
physical inner product (automatically incorporating the correct
adjointness relations) for open constraint algebras, at least
in principle.
\item[14)] {\it Deparameterization, Reconstruction Problem}\\
Suppose somebody is going to find the complete space
of solutions to all quantum constraints, a consistent inner product
with all desired properties and a complete set of Dirac observables.
What is she/he going to do with it ? By definition
``nothing ever happens in quantum gravity" meaning that the
Dirac observables are ``constants of motion" (strictly speaking
only if $\sigma$ is compact). But although our world
is four-dimensionally diffeomorphism invariant (and therefore
all true observables should be highly non-local) we perform
local measurements every day. We must learn how to recover such
a local description from the frozen picture that we are confronted with
in quantum gravity. This is the reconstruction problem
\cite{128}. For a pretty proposal see \cite{28,29}.
\item[15)] {\it Representation Independent Formulation}\\
The lesson that we learn from algebraic quantum field theory is that
the important, primary ingredient are nets of algebras of operators (which
in our case, however, would presumably be rather non-local). Only in a
second step one studies representations of these algebras, of course
one must make sure that there exist physically interesting ones.
The advantage of this purely algebraic approach is that one can
perform a representation free structural analysis of the theory.
It would, for instance, be important to have an analog of
the DHR analysis (Doplicher, Haag, Roberts) for the
classification \cite{62a} of available representations for our theory at
one's disposal in order to know which features are tied to a specific
representation and which are not. Likewise it would be worthwhile
thinking about a suitable background free generalization of the
Haag -- Kastler axioms. For some first steps in that direction, although
within the constructive approach, see \cite{47n10}.
\end{itemize}
There is an endless chain of other problems in quantum general
relativity both on the technical and on the conceptual side which
we cannot possibly enumerate here. Hopefully, one day bright
students will figure them all out.



\newpage

\makeatletter
\@addtoreset{equation}{section}
\makeatother
\renewcommand{\theequation}{\thesection.\arabic{equation}}

\cleardoublepage

\part{Mathematical Tools}
\label{p3}

\section{The Dirac Algorithm for Field Theories with Constraints}
\label{sd}

It is a crime that the subsequent analysis is not a standard
ingredient of every course in theoretical mechanics. Every
interaction that we know today underlies a gauge theory, that is,
a field theory with constraints. However, constraints are generically 
at most mentioned and one usually 
finds out about the fact that one was truly betrayed in that beginning
theoretical mechanics course only much later. This is the more astonishing
as this really important topic can be taught at a truly elementary level.
Also quantum mechanics is not needed (at most for motivational
purposes), the theory can be formulated in purely classical terms.
We recommend the classic expositions by Dirac \cite{3} and by Hanson
et al \cite{3a}
as introductory texts. More advanced are the textbooks
\cite{107} and  \cite{3b}. For geometrical
quantization with constraints see \cite{3c} and for a more
mathematical formulation see \cite{3d}.\\
\\
We will consider only a finite number of degrees of freedom. The more general
case can be treated straightforwardly, at least at a formal level. We will 
also not consider 
the most general actions but only those which lead to phase spaces with 
a cotangential bundle topology. For the more general cases see the 
cited literature.
\begin{Definition} \label{defd.1}   ~~~~~~~~~\\
Consider a Lagrangean function $L:\;T_\ast({\cal C})\to 
\Cl;\;(q^a,v^a)\mapsto L(q,v)$ on the tangential bundle over 
the configuration manifold ${\cal C}$ where $v:=\dot{q}$ (velocity)
defines the corresponding action principle.\\
i)\\
The map 
\be \label{d.1}
\rho_L:\;T_\ast({\cal C})\to T^\ast({\cal C});\;(q,v)\mapsto
(q,p(q,v):=\frac{\partial L}{\partial v}(q,v))
\ee
is called Legendre transformation.\\
ii)\\
A Lagrangean is called singular provided that $\rho_L$ is not surjective,
that is,
\be \label{d.2}
\det((\frac{\partial^2 L}{\partial v^a\partial v^b})_{a,b=1}^m)=0
\ee
\end{Definition}
For singular Lagrangeans it is not possible to solve the velocities 
in terms of the momenta, the undelying reason being that the Lagrangean 
is invariant under certain symmetries.

Let $m=\dim({\cal C})$ and suppose that the rank of the matrix in 
(\ref{d.2}) is $m-r$ with $0<r\le m$. By the inverse function theorem
we can solve (at least locally) $m-r$ velocities for $m-r$ momenta
and the remaining velocities, that is w.l.g.
\be \label{d.3}
p_A=\frac{\partial L}{\partial v^A}(q,v)\;\Rightarrow\;
v^A=u^A(q^a,p_A,v^i)
\ee
where $a,b,..=1,..m,\;A,B,..=1,..,m-r,\;i,j,..=1,..,r$.
It follows that inserting (\ref{d.3}) into the remaining equations 
$p_i=\partial L/\partial v^i$ cannot depend on the $v^i$ any more as 
otherwise the rank would exceed $m-r$. We therefore obtain $r$ equations 
of the form
\be \label{d.4}
p_i=[\frac{\partial L}{\partial v^i}(q,v)]_{v^A=u^A(q^a,p_A,v^j}
=:\pi_i(q^a,p_A)
\ee
which show that the $p_a$ are not independent of each other.
\begin{Definition} \label{defd.2} ~~~~~~~~~~~\\
i)\\
The functions 
\be \label{d.5}
\phi_i(q^a,p_a):=p_i-\pi_i(q^a,p_A)
\ee
are called primary constraints.\\
ii)\\
The function 
\be \label{d.6}
H'(q^a,p_a,v^i):=[p_a v^a-L(q^a,p_a)]_{v^a=u^a(q^a,p_A,v^i)}
\ee
is called the primary Hamiltonian corresponding to $L$.
\end{Definition}
\begin{Lemma} \label{lad.1} ~~~~~~~~\\
The primary Hamiltonian is linear in $v^i$ with coefficients 
$\phi_i$.
\end{Lemma}
Proof of Lemma \ref{lad.1}:\\
Differentiating the expression 
\be \label{d.7}
H'(q^a,p_a,v^i)=p_A u^A(q^a,p_B,v^j)+p_i v^i-L(q^a,u^A(q^a,p_B,v^j),v^i)
\ee
by $v^i$ we obtain
\ba \label{d.7a}
\frac{\partial H'(q^a,p_a,v^j)}{\partial v^i}
&=& [p_A-(\frac{\partial L(q^a,v^a)}{\partial v^A})_{v^A=u^A}]
\frac{\partial u^A}{\partial v^i}+
[p_i-(\frac{\partial L(q^a,v^a)}{\partial v^i})_{v^A=u^A}]
\nonumber\\
&=& [p_i-\pi_i(q^a,p_A)]=\phi_i(q^a,p_a)
\ea
$\Box$\\
We conclude that we may write 
\be \label{d.8}
H'(q^a,p_a)=\tilde{H}(q^a,p_a)+v^i\phi_i(q^a,p_a)
\ee
where the new Hamiltonian $\tilde{H}$ is independent of the remaining
velocities $v^i$.
\begin{Theorem} \label{thd.1} ~~~~~~~~~\\
The Hamiltonian equations 
\be \label{d.9}
\dot{q}^a=\frac{\partial H'}{\partial p_a},\;
\dot{p}_a=-\frac{\partial H'}{\partial q^a},\;
0=\frac{\partial H'}{\partial v^i}
\ee
are equivalent to the Euler Lagrange equations 
\be \label{d.10}
\dot{q}^a=v^a,\;\;
\frac{\partial L}{\partial q^a}=
[\frac{d}{dt}\frac{\partial L}{\partial v^a}]_{v=\dot{q}}
\ee
\end{Theorem}
We leave the simple proof (just use carefully the definitions)
to the reader.

The phase space $\cal M$ of the constrained system is thus coordinatized by
the $q^a,p_a$ while the $v^i$ are {\it Lagrange multipliers}, they 
do not follow any prescribed dynamical trajectory and are completely
arbitrary. Our constrained phase space is equipped with the standard 
symplectic structure 
\be \label{d.11}
0=\{q^a,q^b\}=\{p_a,p_b\}=\{q^a,v^i\}=\{p_a,v^i\},\;
\{p_a,q^b\}=\delta^b_a
\ee
and the Hamiltonian $H'$.

The primary constraints force the system to the submanifold of the 
phase space defined by $\phi_i=0,\;i=1,..,r$ for which we use the short
hand notation $\phi=0$. 
This is consistent with 
the dynamics if and only if that submanifold is left invariant, that is,
\be \label{d.12}
\dot{\phi_i}=\{H',\phi_i\}=\{\tilde{H},\phi_i\}+v^j\{\phi_j,\phi_i\}
\ee
vanishes on the constraint surface $\bar{{\cal M}}:={\cal M}_{\phi=0}$
of the phase space. Now those primary constraints fall into the following 
three categories:\\
1)\\ 
$[\dot{\phi}_i]_{\phi=0}\equiv 0$ for $i=1,..,a$ is identically
satisfied for any $v^i$.\\
2i)\\
$[\dot{\phi}_i]_{\phi=0}\not=0$ and $\{\phi_j,\phi_i\}_{\phi=0}=0$
for all $j=1,..,r$ and $i=a+1,..,b$.\\
2ii)\\ 
$[\dot{\phi}_i]_{\phi=0}\not=0$ for generic $v^i$ but the 
matrix $\{\phi_j,\phi_i\}_{\phi=0}$ with $j=1,..,r;\;i=b+1,..,r$
has maximal rank $r-b$.\\
In case 2ii) we do not allow that the rank is smaller than $r-b$ 
since then we cannot find $v^i$ in order to set  
$[\dot{\phi}_i]_{\phi=0}=0$ and the theory would become inconsistent.
Inconsistent theories have to be ruled out anyway.

Let us now extend the set of primary constraints by the 
$\phi_i:=\dot{\phi}_{i-r+a}$ with $i=r+1,..,r+b-a$ and redefine $r$ 
by $r\to r':=r+b-a$. Now iterate the above case analysis 
(notice that $H'$ always only contains the first $r$ constraints while 
$\phi=0$ means $\phi_i=0,\;i=1,..,r'$) until 
case 2i) no longer appears ($b=a$). The iteration stops after at most 
$2m-r$ steps because in each step the number of (automatically functionally 
independent) constraints increases by at least one and $2m$ constraints 
constrain the phase space to a discrete set of points. 
\begin{Definition} \label{defd.3}
The constraints $\phi_i,\;i=r'-r$ are called secondary constraints.
Here $r'$ is the value of the redefined $r$ after the last iteration step. 
\end{Definition}
It follows that at the end of the procedure we have 
$[\dot{\phi}_i]_{\phi=0}\equiv 0$ identically for $i=1,..,a$ for any
choice of $v^i$ and some $0\le a\le r'$ and the  
matrix $\{\phi_j,\phi_i\}_{\phi=0}$ with $j=1,..,r;\;i=a+1,..,r'$
with $r'\ge r$ has maximal rank $r'-a\le r$.
Let now $v^j=v^j_0+\lambda^\mu v^j_\mu$ where $v^j_0(q^a,p_a)$ is a special 
solution of the inhomogeneous linear equation
\be \label{d.13}  
\{\tilde{H},\phi_i\}_{\phi=0}+v^j\{\phi_j,\phi_i\}_{\phi=0}=0
\ee
and $v^j_\mu(q^a,p_a),\;\mu=1,..,r-(r'-a)$ is a basis for the general 
solution of the homogeneous system. We define
\be \label{d.14}
H:=\tilde{H}+v_0^j\phi_j,\;\;\varphi_\mu:=v_\mu^j\phi_j
\ee
\begin{Definition} \label{defd.4}    ~~~~~~~~~\\
A function $f\in C^\infty({\cal M})$ is called of first class 
provided that $\{\phi_j,f\}_{\phi=0}=0$ for all $j=1,..,r'$,
otherwise of second class.
\end{Definition}
\begin{Lemma} \label{lad.2} ~~~~~~~~~~\\
i)\\
The functions $\varphi_\mu, H$ are of first class.\\
ii)\\
The first class functions form a subalgebra of the Poisson algebra on 
$\cal M$.
\end{Lemma}
Proof of Lemma \ref{lad.2}:\\
i) is clear from the construction. ii) follows by relaizing that 
if $f,f'$ are first class then there exist functions $f_{ij},f'_{ij}$
with $i,j=1,..,r'$ such that 
$\{\phi_i,f\}=f_{ij}\phi_j,\{\phi_i,f'\}=f'_{ij}\phi_j$. A short calculation
then reveals that $\{\phi_i,\{f,f'\}\}_{\phi=0}=0$.\\
$\Box$\\
Let now $H_\lambda:=H+\lambda^\mu \varphi_\mu$. Since at $\phi=0$
the finite time evolution of a function $f$ should be indpendent of the 
arbitrary parameters $\lambda^\mu$ we require that 
$\{H_{\lambda_1},..,\{H_{\lambda_N},f\},..\}_{\phi=0}$ is independent of the 
$\lambda_1,..\lambda_N$ for any $N=1,2,..$. It is easy to see 
from the above lemma that this is automatically the case if $f$ is of 
first class. However, since the multiple Poisson brackets 
contain only the first class constraints $\varphi_\mu$ it is actually
sufficient that $\{f,\varphi_\mu\}_{\phi=0}$ for all $\mu$. 

This motivates to extend the set of first class constraints 
$\varphi_\mu$ already found to a maximal set $C_\mu,\;\mu=1,..,k$ with 
$k\ge r-(r'-a)$ and to add them to the Hamiltonian with additional
lagrange multipliers.
Denote the subset of the constraints $\phi_i$ functionally independent
of the $C_\mu$, that is, the second class constraints, by 
$\phi_I,\;I=1,..,r'-k$. 
\begin{Definition} \label{defd.5} ~~~~~~~~~~~~~~\\
i)\\
The set $C_\mu$ is called the set of generators of gauge transformations.\\
ii)\\
A function $f\in C^\infty({\cal M})$ is called an observable 
provided that $\{f,C_\mu\}_{\phi=0}$ for all $\mu=1,..,k$.\\
iii)\\
The extended Hamiltonian is defined by
\be \label{d.15}
H_\lambda=H+\lambda^\mu C_\mu
\ee
\end{Definition}
The nomenclature stems from the fact that $\{C_\mu,f\}$ can be 
interpreted as an infinitesimal motion generated by the flow of the 
Hamiltonian vector field associated with $C_\mu$ and an obeservable is 
invariant under this flow at least on $\bar{{\cal M}}$. That all
first class constraints $C_\mu$ should be considered as generators of 
gauge transformations (so-called Dirac conjecture) and not only the 
$\varphi_\mu$ which appear 
in $H'$ is motivated by the fact that only the $C_\mu$ form a closed 
constraint algebra (see below), however, it does not follow strictly
from the formalism. That it is physically correct to proceed that way has
been confirmed in countless examples though and can even be proved 
under some restrictions \cite{3b}.
\begin{Lemma} \label{lad.3} ~~~~~~~~~~~\\
We have that $r'-k=2m'$ is even and that $(\{\phi_I,\phi_J\}_{\phi=0})$
is an invertible matrix.
\end{Lemma}
Proof of Lemma \ref{lad.3}:\\
Suppose that $(\{\phi_I,\phi_J\}_{\phi=0})$ is singular then there exist 
numbers
$x^J\in \Cl$ such that $\{\phi_I,C_0\}_{\phi=0}=0$ for all $I$ where 
$C_0=x^J \phi_J$. Since $\{C_\mu,C_0\}_{\phi=0}$ anyway we find 
$\{\phi_i,C_0\}_{\phi=0}=0$ for all $i$ so that $C_0$ is a 
first class constraint independent of the $C_\mu$. This 
is a contradiction to the assumed maximality. It follows that
$r'-k$ is even since $(\{\phi_I,\phi_J\}_{\phi=0})$ is an antisymmetric 
matrix.\\
$\Box$\\
\begin{Definition} \label{defd.6}  ~~~~~~~~~~~\\
Let $c^{IJ}:=((\{\phi_K,\phi_L\})^{-1})^{IJ}$. The Dirac bracket is defined
by
\be \label{d.16}
\{f,f'\}^\ast:=\{f,f'\}+\{\phi_I,f\} c^{IJ} \{\phi_J,f'\}
\ee
\end{Definition}
\begin{Theorem} \label{thd.2} ~~~~~~~\\
The Dirac bracket defines a closed but degenerate
two form on ${\cal M}$ with kernel spanned by $\chi_{\phi_I}$
where $\chi_f$ denotes the Hamiltonian vector field of
$f\in C^\infty({\cal M})$ with respect to the symplectic structure
determined by $\{.,.\}$.
\end{Theorem}
Proof of Theorem \ref{thd.2}:\\
Our conventions are $i_{\chi_f}\Omega+df=0$ and 
$\{f,f'\}=-i_{\chi_f} i_{\chi_{f'}}\Omega=\chi_f(f')=i_{\chi_f}(df')$
for the relation between a nondegenerate symplectic struture $\Omega$, 
Hamiltonian vector field $\chi_f$ and Poisson bracket $\{.,.\}$. Also
for a $p-$form $\omega=\omega_{\alpha_1..\alpha_p} 
dx^{\alpha_1}\wedge..\wedge dx^{\alpha_p}$
we define exterior derivative, contraction with vector fields $v$
and Lie derivative by
\ba \label{d.17}
d\omega &=&[\partial_{\alpha_1}\omega_{\alpha_2..\alpha_{p+1}}]
dx^{\alpha_1}\wedge..\wedge dx^{\alpha_{p+1}}
\nonumber\\
i_v\omega &=&p\; v^\alpha\omega_{\alpha \alpha_1..\alpha_{p-1}}
dx^{\alpha_1}\wedge..\wedge dx^{\alpha_{p-1}}
\nonumber\\
{\cal L}_v\omega &=& [i_v\cdot d+d\cdot i_v]\omega
\ea
Let $\Omega=\frac{1}{2}\Omega_{\alpha\beta} dx^\alpha\wedge dx^\beta$
(here $\alpha,\beta,..=1,..,2m$). Define the inverse of 
$\Omega_{\alpha\beta}$ by $\Omega^{\alpha\gamma}\Omega_{\gamma\beta}=
\delta^\alpha_\beta$. Then it is easy to verify that 
$\chi_f^\alpha=\Omega^{\alpha\beta}\partial_\beta f$ and therefore
$\Omega^{\alpha\beta}=-\{x^\alpha,x^\beta\}$.

We first of all verify that a nondegenerate two form is closed if and 
only if the associated Poisson bracket satisfies the Jacobi identity
\be \label{d.18}
\{f_{[1},\{f_2,f_{3]}\}\}=0
\ee
To see this we just need to use the formula $\{f,f'\}=-\Omega^{\alpha\beta}
(\partial_\alpha f)(\partial_\beta f')$ and the fact that
$\delta \Omega^{-1}=-\Omega^{-1}(\delta\Omega)\Omega^{-1}$ to conclude that
(\ref{d.18}) is equivalent with $\partial_{[\alpha}\Omega_{\beta\gamma]}=0$.

Next we verify directly from the definition for the Dirac bracket and by 
similar methods applied to $c_{IJ}$ 
that on all of ${\cal M}$ the Jacobi identity 
\be \label{d.19}
\{f_{[1},\{f_2,f_{3]}\}^\ast\}^\ast=0
\ee
holds. Moreover
\be \label{d.20}
\{f,\phi_I\}^\ast=-\{\phi_I,f\}^\ast=0
\ee
for any $I=1,..,2m'$ and and $f\in C^\infty({\cal M})$. We can therefore 
introduce local coordinates $x^\alpha=(x^a,x^I:=\phi_I)$ with 
$a=1,..2(m-m'),\; I=1,..,2m'$ such that $\{x^a,x^I\}=0$ (Darboux theorem
\cite{32}, replace the $\phi_I$ by equivalent constraints if necessary) and 
define 
$(\Omega^\ast)^{\alpha\beta}:=\{x^\alpha,x^\beta\}^\ast$. We 
then see that 
$(\Omega^\ast)^{aI}=(\Omega^\ast)^{IJ}=0$. Define 
$(\Omega^\ast)_{ab}$ to be the inverse of 
$(\Omega^\ast)^{ab}$ and $(\Omega^\ast)_{aI}=(\Omega^\ast)_{IJ}=0$. 
Then $\Omega^\ast=\pi^\ast\Omega$ where 
$\pi:\;{\cal M}\to{\cal M}'\;(x^a,x^I)\mapsto (x^a,0)$ is the
projection to the constraint manifold defined by second class 
constraints. That $\Omega^\ast$ is closed and has the anticipated kernel 
is now obvious.\\
$\Box$\\
Notice that for the first class constraints $C_\mu$ and the Hamiltonian
$H_\lambda$ we have for any $f\in C^\infty({\cal M})$ that
$\{C_\mu,f\}_{\phi=0}=\{C_\mu,f\}^\ast_{\phi=0}$ and 
$\{H_\lambda,f\}_{\phi=0}=\{H_\lambda,f\}^\ast_{\phi=0}$ (more generally
this holds for any first class function).
Thus, on the constraint surface the Dirac bracket defines the same 
equations of motion as the original bracket. Notice however that 
in general $\{f,f'\}_{\phi=0}\not=\{f,f'\}^\ast_{\phi=0}$ unless 
one uses a set of second class constraints $\phi_I$ which are themselves
Darboux coordinates which is always possible to achieve but generically 
difficult and even unpractical. However, the Dirac bracket is easily seen 
to have the important property
\be \label{d.21}
(\{f_{|{\cal M}'},f'_{|{\cal M}'}\}^\ast)_{|{\cal M}'}
=(\{f,f'\}^\ast)_{|{\cal M}'}
\ee
that is, with respect a Dirac bracket we may set the second class constraints
equal to zero before or after evaluating it.

Because of this and because the equations of motion and the gauge 
motions generated by the first class constraints are unaltered 
irrespective of whether we use the original Poisson bracket or the 
Dirac bracket we may just forget about the second class constraints 
for the rest of the analysis and work off the constraint surface 
defined by the sond class constraints while using the Dirac bracket.
The reason for treating the second class constraints differently from the 
first class constraints is as follows:\\
The cleanest way to treat a constrained Hamiltonian system is to 
compute the full constraint surface $\bar{{\cal M}}=\{m\in {\cal M};\;
\phi_i(m)=0\;\forall \;i=1,..,r'\}$. Since the Hamiltonian is a first
class function, its Hamiltonian flow preserves the constraint surface.
Since the Hamiltonian depends on arbitrary parameters, and physical 
observables must be independent of those, we have required that those 
oberservables must be independent of the Hamiltonian flow generated by 
the first class constraints, at least on the constraint surface. This
is, however, not possible to require for the second class constraints 
because their Hamiltonian flow does not preserve the constraint surface.
Thus, what one should do is to compute the gauge orbits $[m]$ of  
points $m$ on the constraint surface (gauge invariant quantities). The 
manifold so obtained 
is called the reduced phase space $\tilde{{\cal M}}$ and observables are
naturally functions on $\tilde{{\cal M}}$. The reduced phase space is 
automatically
equipped with a symplectic structure that one obtains locally by 
looking for a suitable set of first class constraints and conjugate Darboux 
coordinates (together with a suitable choice of second class 
constraints as Darboux coordinates). See \cite{18} for details.
One would then quantize the reduced system.

The reason for why that is not always done is that for non-linear systems it
is extremely difficult to compute $\bar{{\cal M}},\tilde{{\cal M}}$ 
even classically and the reduced symplectic structure on the observables 
might be so complicated that it is very hard to find a representation of 
the associated canonical commutation relations in the quantum theory.
Thus, in order to get started with the quantization Dirac has proposed to
solve the constraints not before but after the quantization. Roughly 
speaking, we turn the constraints into operators and impose that
physical states satisfy
\be \label{d.22}
\hat{C}_\mu\psi=0
\ee
(this equation must actually be read in a generalized sense, see
section \ref{si}). Notice that we impose this only for the first 
class constraints. To see why, notice that the first class constraints 
must satisfy a subalgebra of the Poisson algebra (we know that 
$\{C_\mu,C\nu\}_{\phi}=0$ therefore 
$\{C_\mu,C_\nu\}=f_{\mu\nu}\;^\rho C_\rho+f_{\mu\nu}\;^I \phi_I$
for some structure functions $f_{\mu\ni}^\rho,f_{\mu\nu}^I$
and since the Poisson bracket is first class again we know that
$f_{\mu\nu}^I=0$). Therefore upon suitable operator ordering  
for a solution of (\ref{d.22}) we have that
\be \label{d.23}
0=[\hat{C}_\mu,\hat{C}_\mu]\psi=\hat{f}_{\mu\nu}\;^\rho\hat{C}_\rho\psi
\ee
is a consistent equation. However, if we would extend (\ref{d.22})
to second class constraints we get the contradiction
\be \label{d.24}
0=[\hat{\phi}_I,\hat{\phi}_J]\psi\not=0
\ee
since the commutator is proportional to a quantization of $c_{IJ}$ which 
in the worst case is a constant (in general an operator which is 
not constrained to vanish). Thus, one solves the second class constraints
simply by restricting the argument of the wave function to the constraint
surface.

Two remarks are in order:\\
1)\\
Notice that every second class constraint classically removes one degree 
of freedom while every first class constraint removes two since not 
only we delete degrees of freedom but also compute gauge orbits. 
However,
since the number of second class constraints is always even, the reduced
phase space has always again an even number of physical degrees of freedom  
(otherwise it would not have a non-degenerate symplectic structure).
One may then wonder how it is possible that we just impose the constraint
on the state and do not compute its gauge orbit in addition. The answer 
is that the wave function already depends only on half of the number 
of kinematical degrees of freedom (configuration space). The imposition of 
the constraint is 
actually the condition that the state be gauge invariant and simultaneously
the constraint operator is deleted.\\
2)\\ 
One may also wonder why we do not simply remove the first class 
constraints as well. The procedure to do this is called {\it gauge fixing}.
Thus, we impose additional conditions $k_\mu=0$ which ideally pick
from each gauge orbit a unique representative and such that the 
matrix $(\{k_\mu,C_\nu\})$ is non-degenerate on the constraint surface.
One may then remove the constraints $C_\mu$ by considering the system
$k_\mu,C_\mu$ as second class constraints and by using the associated 
Dirac bracket. The reason for not doing this is that it is actually
very problematic: Usually functions with the required properties simply
do not exist, for instance gauge orbits can be cut more than once leading 
to the so-called Gribov copies \cite{107,3b}. Also, the geometric structure
of the system is very much veiled and different gauge conditions may 
lead to different physics.

Finally, let us display a trivial example:\\
Consider the phase space ${\cal M}=T^\ast(\Rl^3)$ with constraints
$\phi_1=p_1,\;\phi_2=q^2,\;\phi_3=p_2$ where $q^a,p_a,\;a=1,2,3$ are
canonically conjugate configuration and momentum coordinates.
It is easy to see that $C=\phi_1$ is the only first class constraint
and that $\phi_2,\phi_3$ is a pair of second class constraints.
For instance, functions which are independent of $q^1,q^2,p_2$ are 
first class but also the Hamiltonian
$H=-(q^1)^2+\sum_{a=1}^3 [(q^a)^2+(p_a)^2]$ and any function which 
is independent of $q^1$ is an observable but also the function 
$f=p_1 q^1$. The gauge motions generated by $C$ are translations
in the $q^1$ direction so that the value of $q^1$ is pure gauge.
Obviously then the only second class constraint reduced phase space
is ${\cal M}'=T^\ast(\Rl^2)$ while the fully reduced phase space
is $\tilde{{\cal M}}=T^\ast(\Rl^1)$.

\newpage

\section{Elements of Fibre Bundle Theory}
\label{sa}

This section recalls the most important structural elements of the theory of
connections
on principal fibre bundles and follows closely the excellent
exposition in \cite{50} to which the reader is referred for more details.
The reason for the inclusion of this section on standard material is the
pivotal role that the holonomy plays in canonical quantum gravity.

\begin{Definition} \label{defa.1}  ~~~~~~~~~~\\
A fibre bundle over a differential manifold $\sigma$ with atlas
$\{U_I,\varphi_I\}$ is a quintuple $(P,\sigma,\pi,F,G)$ consisting
of a differentiable manifod $P$ (called the total space), a differentiable
manifold $\sigma$ (called the base space), a differentiable surjection
$\pi:\;P\to\sigma$, a differentiable manifold $F$ (called the typical
fibre) which is diffeomorphic to every fibre $\pi^{-1}(x),\;x\in \sigma$
and a Lie group $G$ (called the structure group) which acts on
$F$ on the left, $\lambda:\; G\times F\to F;\;(h,f)\mapsto \lambda(h,f)=:
\lambda_h(f),\;\lambda_h\circ\lambda_{h'}=\lambda_{h h'},\;\lambda_{h^{-1}}=
(\lambda_h)^{-1}$. Furthermore, for every $U_I$ there exist diffeomorphisms
$\phi_I:\;U_I\times F\to \pi^{-1}(U_I)$, called local trivializations,
such that $\phi_{Ix}:F\to
F_x:=\pi^{-1}(x);\;f\mapsto \phi_{Ix}(f):=\phi_I(x,f)$ is a diffeomorphism
for every $x\in U_I$. Finally, we require that there exist maps
$h_{IJ}:\;U_I\cap U_J\not=\emptyset\to G$, called transition functions,
such that for every $x\in U_I\cap U_J\not=\emptyset$ we have
$\phi_{Jx}=\phi_{Ix}\circ\lambda_{h_{IJ}(x)}$.
\end{Definition}
Conversely, given $\sigma,F,G$ and the structure functions $h_{IJ}(x)$
with given left action $\lambda$ on $F$
we can reconstruct $P,\pi,\phi_I$ as follows: Define $P'=\cup_I U_I\times F$
and introduce an equivalence relation $\sim$
by saying that $(x,f)\in U_I\times F$
and $(x',f')\in U_J\times F$ for $U_I\cap U_J\not=\emptyset$ are equivalent
iff $x'=x$ and $f'=\lambda_{h_{IJ}(x)}(f)$. Then $P=P'/\sim$ is the set of
eqivalence classes $[(x,f)]$ with respect to this equivalence relation
with bundle projection $\pi([(x,f)]):=x$ and local trivializations
$\phi_I(x,f):=[(x,f)]$.\\
\begin{Definition} \label{defa.2} ~~~~~~~~~~~~ \\
Two bundles defined by the collections of tuples $\{(U_I,\phi_I)\}_I$ and
$\{(U'_J,\phi'_J)\}_J$ respectively
are said to be equivalent if the combined collection of tuples
$\{(U_I,\phi_I),(U'_J,\phi'_J)\}_{I,J}$ defines a bundle again.
A bundle automorphism is a diffeomorphism of $P$ that maps whole fibres to
whole fibres.
Equivalently then, two bundles are equivalent if there exists
a bundle automorphism which reduces to the identity on the base space.
A bundle is really an equivalence class of bundles.
\end{Definition}
Notice that the transition functions satisfy the cocycle condition
$h_{IJ} h_{JK} h_{KI}=1_G$
over $U_I\cap U_J\cap U_K$ and $h_{IJ}=h_{JI}^{-1}$ over $U_I\cap U_J$.
It is crucial to realize that in general $h_{IJ}$ is not a coboundary,
that is, there are in general no maps $h_I:\; U_I\to G$ such that
$h_{IJ}(x)=h_I(x)^{-1}h_J(x)$.
\begin{Definition} \label{defa.3} ~~~~~~~~~~\\
A fibre bundle is called trivial if its transition function cocycle
is a coboundary.
\end{Definition}
The reason for this notation is that trivial bundles are equivalent to
direct product bundles $\sigma\times F$: Given transition functions
$\phi_I$, it may be checked that the transition functions
$\phi'_I(x,f):=\phi_I(x,\lambda_{h_I(x)^{-1}}(f))$ are actually independent
of the label $I$ and thus there is only one of them. Therefore the bundle is
diffeomorphic with $\sigma\times F$.
\begin{Definition} \label{defa.4} ~~~~~~\\
A local section of $P$ is a smooth map $s_I:\; U_I\to P$ such that
$\pi\circ s_I=\mbox{id}_{U_I}$. A cross section is a global section, that is,
defined everywhere on $\sigma$.
\end{Definition}
\begin{Definition} \label{defa.5} ~~~~~~\\
A principal $G$ bundle is a fibre bundle where typical fibre and
structure group coincide with $G$. On a principal fibre bundle we may
define a right action $\rho:\;G\times P\to P;\rho_h(p):=
\phi_I(\pi(p),h_I(p)h)$ for $p\in \pi^{-1}(U_I)$ where $h_I:\;P\to G$
is uniquely defined by $(\pi(p)=x_I(p),h_I(p)):=\phi_I^{-1}(p)$.
Since $G$ acts transitively
on itself from the right, this right action is obviously transitive
in every fibre and fibre preserving. $s^\phi_I(x):=\phi_I(x,1_G)$ is called
the canonical local section. Conversely, given a system of
local sections $s_I$
one can construct local trivializations $\phi^s_I(x,h):=\rho_h(s_I(x))$,
called canonical local trivializations.
\end{Definition}
Notice the identity
$p=\rho_{h_I(p)}(s^\phi_I(\pi(p)))=\phi_I(\pi(p),h_I(p))=
\phi_{I\pi(p)}(h_I(p))$ for any
$p\in \pi^{-1}(U_I)$. If
$U_I\cap U_J\not=\emptyset$ and $p\in \pi^{-1}(U_I\cap U_J)$ this leads
to $\rho_{h_I(p)}(s^\phi_I(\pi(p)))=\rho^\phi_{h_J(p)}(s^\phi_J(\pi(p)))$.
Using the fact that $\rho$ is a right action we conclude
$s^\phi_J(\pi(p))=\rho_{h_I(p) h_J(p)^{-1}}(s^\phi_I(\pi(p))$. Since the
left hand
side does not depend any longer on the point $p$ in the fibre above
$x=\pi(p)$ we conclude that we have a $G-$valued functions
$h_{IJ}:\;U_I\cap U_J \to G,\;x\mapsto
[h_J(p)^{-1} h_I(p)]_{Ýp\in \pi^{-1}(x)}$ where the right hand side is
independent of the point in the fibre. The functions $h_{IJ}$ are actually
the structure functions of $P$: By definition we have $\phi_{Ix}(h_I(p))
=\phi_{Jx}(h_J(p))$, thus $h_I(p)=(\phi_{Ix}^{-1}\circ\phi_{Jx})(h_J(p))
\lambda_{h_{IJ}(x)}(h_J(p))=h_{IJ}(x)h_J(p)$ which also shows that the left
action in $P$ reduces to left translation in the fibre coordinate.

In a principal $G$ bundle it is easy to see, using transitivity of the right
action of $G$, that triviality
is equivalent with the existence of a global section. This is not
the case for vector bundles which always have the global section
$s_I(x)=\phi_I(x,0)$ but may have non-trivial transition functions.
\begin{Definition} \label{defa.6} ~~~~~~~~~~\\
A vector bundle $E$ is a fibre bundle whose typical fibre $F$
is a vector space. The vector bundle
associated with a principal $G$ bundle $P$ (where $G$ is the structure group
of $E$) under the left representation $\tau$ of $G$ on $F$,
denoted $E=P\times_\tau F$, is given by the set
of equivalence classes $[(p,f)]=\{(\rho_h(p),\tau(h^{-1})f);\;h\in G\}$
for $(p,f)\in P\times F$. The projection is given by
$\pi_E([(p,f)]):=\pi(p)$ and
local trivializations are given by
$\psi(x,f)=[(s_I(x),f)]$ since $[(\rho_h(s_I(x)),f)]=[(s_I(x),\tau(h)f]
=[s_I(x),f')]$. Transition functions result from
$u=[s_J(\pi(u)),f_J(u)]=[\rho_{h_{IJ}(\pi(u))}(s_I(\pi(u))),f_J(u))]=
[(s_I(\pi(u))),\tau(h_{IJ}(\pi(u)f_J(u))]=[(s_I(\pi(u)),f_I(u)]$
and are thus gven by $\tau(\rho_{IJ}(x))$.
\end{Definition}
Conversely, given any vector bundle $E$ we can construct a principal $G$
bundle $P$ such that $E$ is associated with it by going through the above
mentioned reconstruction process and by using the same structure
group (with $\tau$ as the defining representation) acting on the fibre $G$
by left translations and the same transition functions.
A vector bundle is then called trivial if its associated
principal fibre bundle is trivial.\\
\\
Every principal fibre bundle $P$ is naturally equipped with a vertical
distribution, that is, an assignment of a subspace $V_p(P)$ of the tangent
space $T_p(P)$ at each point $p$ of $P$ that is tangent to the fibre
above $\pi(p)$. (Notice that distributions are not necessarily
integrable, i.e they do not form the tangent spaces of a submanifold of
$P$). These vertical distributions are generated by the
fundamental vector fields $v_Y$ associated with an element $Y\in$Lie$(G)$
of the Lie algebra of $G$ which are defined through their action on
functions $f\in C^\infty(P)$:
\be \label{a.1}
(v_Y[f])(p):=(\frac{d}{dt})_{Ýt=0} f(\rho_{\exp(tY)}(p))
\ee
where $\exp:\;\mbox{Lie}(G)\to G$ denotes the exponential map.
The map $v:\;\mbox{Lie}(G)\to V_p(P);\;Y\to v_Y$ is a Lie agebra homomorphism
by construction.

The complement $H_p(P)$ of $V_p(P)$ in $T_p(P)$ is called the horizontal
distribution and is one way to define a connection on $P$. More precisely
\begin{Definition} \label{defa.7} ~~~~~~~~~\\
A connection on a principal $G$ bundle $P$ is a distribution of horizontal
subspaces $H_p(P)$ of $T_p(P)$ such that \\
a) $H_p(P)\oplus V_p(P)=T_p(P)$ (i.e. $H_p(P)\cap V_p(P)=\{0\},\;
H_p(P)\cup V_p(P)=T_p(P)$).\\
b) If $v(p)=v^H(p)+v^V(p)$ denotes the unique split of a smooth vector field
into
its horizontal and vertical components respectively, then the components
are smooth vector fields again.\\
c) $H_{\rho_h(p)}(P)=(\rho_h)_\ast H_p(P)$.
\end{Definition}
Condition c) tells us how horizontal supspaces in the same fibre are related.
Here $((\rho_h)_\ast v)[f]=v[(\rho_h)^\ast f]$ denotes the push-forward
of a vector field and $(\rho_h)^\ast f=f\circ\rho_h$ the pull-back
of a function.

A different, less geometrical definition of a connection consistent with
definition \ref{defa.7} is as follows:
\begin{Definition} \label{defa.8} ~~~~~~~~~\\
A connection on a principal $G$ bundle $P$ is a Lie algebra valued one
form $\omega$ on $P$ which projects $T_p(P)$ into $V_p(P)$, that is\\
a) $\omega(v_Y)=Y$\\
b) $(\rho_h)^\ast \omega=\mbox{ad}_{h^{-1}}(\omega)$\\
c) $H_p(P)=\{v\in T_p(P);\; i_v\omega=0\}$.
\end{Definition}
Here $ad:\;G\times\mbox{Lie}(G)\to Lie(G);(x,Y)\mapsto hYh^{-1}$ denotes the
adjoint action of $G$ on its own Lie algebra and $i_v$ denotes the
contraction of vector fields with forms. To see that both definitions
are consistent we notice that
\be \label{a.2}
((\rho_h)^\ast\omega)_p(v_p)=(\omega)_{\rho_h(p)}((\rho_h)_\ast v_p)
=(\mbox{ad}_{h^{-1}}\omega)_p(v_p)=h^{-1}\omega_p(v_p)h
\ee
so that $v_p\in H_p(P)$
implies $(\rho_h)_\ast v_p\in H_{\rho_h(p)}(P)$ indeed, demonstrating that
conditions b),c) of definition \ref{defa.8} imply condition c) of
definition \ref{defa.7}.
Condition a) is an additional requirement fixing an otherwise free constant
factor in $\omega$.\\
\\
For practical applications it is important to have a coordinate expression
for $\omega$. To that end, let us express $\omega$ in a local trivialization
$p=\phi_I(x,h)$.
Introducing matrix element indices $A,B,C,..$ for group elements
$h=(h_{AB})$ we have
\be \label{a.3}
v_Y^\mu(p)=(\frac{\partial\phi^\mu_I(x,h)}{\partial h_{AB}}
(hY)_{AB})_{Ý\phi_I(x,h)=p}
\ee
where $p^\mu$ denotes the coordinates of $p$. Recalling the definition
$(x_I(p)=\pi(p),h_I(p)):=\phi_I^{-1}(p)$ we claim that
\be \label{a.4}
(\omega_I(p))_{AB}=\mbox{ad}_{h_I(p)^{-1}}(\pi^\ast A_I)(p)_{AB}
+(h_I(p)^{-1})_{AC} dh_I(p)_{CB}
\ee
where $A_I(x)$ is a Lie$(G)$ valued one form on $U_I$. Let us check
that properties a), b) and c) are satisfied. \\
a)\\
We have
$(\pi^\ast A_I)(v_Y)_p=A_I(\pi_\ast v_Y)_{\pi(p)}$ but
$(\pi\_ast v_Y)^\mu(x)=[\partial \pi^\mu(\phi_I(x,h))/\partial h_{AB}]
(hY)_{AB}=0$ since $\pi(\phi_I(x,h))=x$ is independent of the fibre
coordinate $h$. On the other hand
\ba \label{a.5}
&& (h_I(p)^{-1} dh_I[v_Y]_p)_{AB}
=h_I(p)^{-1}_{AC} [\partial h_I(p)_{CB}/\partial p^\mu]
[\partial \phi^\mu(x,h)/\partial h_{DE} (hY)_{DE}]_{p=\phi_I(x,h)}
\nonumber\\
&=& h_I(p)^{-1}_{AD} (h_I(p)Y)_{DB}=Y_{AB}
\ea
where the $h_{AB},\;A,B=1,..,\dim(G)$ could be treated as independent
coordinates (although, depending on the group, this may not be the case)
because of the chain rule. More precisely,
\ba \label{a.6}
&& h_I(p)^{-1} dh_I[v_Y]_p
=h_I(p)^{-1} [\partial h_I(p)/\partial p^\mu]
[(\frac{d}{dt})_{t=0}\phi^\mu(x,h e^{t Y})]_{Ýp=\phi_I(x,h)}
\nonumber\\
&=& h_I(p)^{-1} (\frac{d}{dt})_{t=0}h_I(p) e^{t Y}=Y
\ea
b)\\
We have $\rho_h(p)=\phi_I(\pi(p),h_I(p)h)=\phi(\pi(p),h_I(\rho_h(p))$
since $\rho$ is fibre preserving whence
$h_I(\rho_h(p))=h_I(p)h$. Since $(\pi^\ast A_I)$ depends only on $\pi(p)$
we have $(\pi^\ast A_I)(\rho_h(p))=(\pi^\ast A_I)(p)$. Finally, since
$\rho^\ast d=d\rho^\ast$ we easily find
\be \label{a.7}
(\rho_h^\ast \omega)(p)=\mbox{ad}_{h_I(p)h}(\pi^\ast A_I)(p)+
(h_I(p) h)^{-1}d h_I(p) h=\omega(\rho_h(p))=\mbox{ad}_{h^{-1}}(\omega(p))
\ee
as claimed.\\
c)\\
Was already checked above.\\
\\
Consider the pull-back of $\omega$ to $\sigma$ by the canonical
local section $s^\phi_I(x)=\phi_I(x,1_G)$. Obviously $h_I(s^\phi(x))=1_G$
whence $((s^\phi_I)^\ast dh_I)(x)=d 1_G=0$ and
$((s^\phi_I)^\ast \pi^\ast A_I)(x)=
(\pi\circ s^\phi_I)^\ast A_I)(x)=A_I(x)$ since $\pi\circ s_I=
\mbox{id}_\sigma$ for any section. We conclude
\begin{Definition} \label{defa.8a} ~~~~~~~~~~~\\
The so-called connection potentials
\be \label{a.8}
A_I=(s^\phi_I)^\ast\omega
\ee
are nothing else than the pull-back of the connection by local sections.
\end{Definition}
By its very defintion, the connection
$\omega$ is globally defined therefore the above coordinate formula
must be independent of the trivialization. This implies the following
identity between the potentials $A_I(x)$
\be \label{a.9}
\pi^\ast A_I=\pi^\ast[\mbox{ad}_{h_{IJ}}\pi^\ast A_J-dh_{IJ}h_{IJ}^{-1}]
\ee
as one can easily verify using $(\pi^\ast h)_{IJ}(p)=h_I(p) h_J(p)^{-1}$.
We can also pull this identity back to $\sigma$ and obtain
\be \label{a.10}
A_I=\mbox{ad}_{h_{IJ}}(A_J)-dh_{IJ}h_{IJ}^{-1}
\ee
which is called the transformation behaviour of the connection potentials
under a change of section (or trivialization or gauge). Since the bundle
$P$ can be reconstructed from $G,\sigma$ and the transition functions
$h_{IJ}(x)$ we conclude that a connection can be defined uniquely by
a system of pairs consisting of connection potentials and local sections
$(A_I,s_I)$ respectvely, subject to the above transformation behaviour.
\begin{Definition} \label{defa.9} ~~~~~~~~~~\\
Given a principal $G$ bundle $P$ over $\sigma$ and a curve $c$ in $\sigma$
we define a curve $\tilde{c}$ to be the horizontal lift of $c$ provided
that\\
i) $\pi(\tilde{c})=c$\\
ii) $d\tilde{c}(t)/dt\in H_{\tilde{c}(t)}(P)$ for any $t$ in the domain
$[0,1]$ of the parametrization of $c$.
\end{Definition}
We now show that the lift is actually unique: We know that
$\tilde{c}(t)=\phi_I(c(t),h_{cI}(t)^{-1})
=\rho_{h_{cI}(t)^{-1}}(s^\phi_I(c(t))$ for some
function $h_{cI}(t)$ (to be solved for) when $c(t)$ lies in the chart $U_I$.
It follows that
\be \label{a.11}
d\tilde{c}(t)/dt=[\partial \phi_I/\partial x^a \dot{c}^a(t)
+\partial \phi_I/\partial h_{AB} (\dot{h}_{cI}(t)^{-1})_{AB}]_(Ý\phi_I(x,h)=
\tilde{c(t)}
\ee
That this vector is horizontal along $\tilde{c}(t)$ means that
$\omega[\dot{\tilde{c}}]_{\tilde{c}(t)}=0$. Using
$\omega=\mbox{ad}_{{h_I}^{-1}}(\pi^\ast A_I)+h_I^{-1} dh_I$ we find
\be \label{a.12}
\omega[\dot{\tilde{c}}]_{\tilde{c}(t)}=h_{cI}(t)
[A_{Ia}(c(t))h_{cI}(t)^{-1}\dot{c}^a(t)+\frac{d}{dt}({h}_{cI}(t)^{-1})]
\ee
implying the so-called parallel transport equation (dropping the index $I$)
\be \label{a.13}
\dot{h}_{cI}(t)=h_{cI}(t)A_{Ia}(c(t))\dot{c}^a(t)
\ee
which is an ordinary differential equation of first order and therefore
has a unique solution by the usual existence and uniqueness theorems
if we provide an initial datum $\tilde{c}(0)$. The point $\tilde{c}(1)$
is called the parallel transport of $\tilde{c}(0)$. Since the point $c(1)$
in the base is already known, the essential information is contained in the
group element $h_{cI}=h_{cI}(1)$ to which we will also refer to as the
holonomy of $A_I$ along $c$. It should be noted, however, that while
$\tilde{c(1})$ is globally defined, $h_{cI}$ depends on the choice of the
local trivialization. In fact, under a change
of trivialization $A_I(x)\mapsto A_J(x)=-dh_{JI}(x)h_{JI}^{-1}(x)+
\mbox{ad}_{h{JI}(x)}(A_I(x))$ we obtain $h_{cJ}=h_{JI}(c(0))h_{cI}
h_{JI}(c(1))^{-1}$ which maybe checked by inserting these formulas
into the parallel transport equation with $x,c(1)$ replaced by $c(t)$
and relying on the uniqueness property for solutions of ordinary
differential equations. It is easy to check that if $c$ is within
the domain of a chart, then an analytic formula for $h_c(A)$ is given by
\be \label{a.13a}
h_c(A)={\cal P}e^{\int_c A}=1+\sum_{n=1}^\infty
\int_0^1 dt_n\int_0^{t_n} dt_{n-1}..\int_0^{t_2} dt_1
A(t_1)..A(t_n)
\ee
where $A(t)=A_a^j(c(t))\dot{c}^a(t)\tau_j/2$, $\tau_j/2$ is a
Lie algebra basis and $\cal P$ denotes the path
ordering symbol (the smallest path parameter to the left).
\begin{Definition} \label{defa.10}  ~~~~~~~~~~\\
Let $V$ be a vector space and $\psi\in \bigwedge^n(P)\otimes V$ be a
vector valued $n-$form on $P$. The covariant derivative $\nabla\psi$
of $\psi$ is the element of $\bigwedge^{n+1}(P)\otimes V$ defined
uniquely by
\be \label{a.14}
(\nabla\psi)_p[v_1,..,v_{n+1}]:=d\psi_p[v_1^H,..,v_{n+1}^H]
\ee
where $v_k\in T_p(P),\;v_k^H$ is its horizontal component and $d$ is the
ordinary exteriour derivative.
\end{Definition}
This definition can be applied to the connection one form where the
vector space is given by $V=$Lie$(G)$.
\begin{Definition} \label{defa.11}  ~~~~~~~~~~~~~~~\\
The covariant derivative of the connection one-form
$\omega\in \bigwedge^1(P)\otimes \mbox{Lie}(P)$ is called the curvature
two-form $\Omega=\nabla\omega$ of $\omega$.
\end{Definition}
The curvature inherits from $\omega$ the property
\be \label{a.15}
\rho_h^\ast\Omega=\mbox{ad}_{h^{-1}}(\Omega)
\ee
To see this, notice that the property $(\rho_h )_\ast
H_p(P)=H_{\rho_h(p)}(P)$
of the horizontal suspaces means that $(\rho_h)_\ast v^H_p
\in H_{\rho_h(p)}(P)$ for any $v\in T_p(P)$. Since every element of
$H_{\rho_h(p)}(P)$ can be obtained this way and $(\rho_h)_\ast$ is a
bijection we conclude $[(\rho_h)_\ast v_p]^H=(\rho_h)_\ast v_p^H$.
Thus
\ba \label{a.16}
&&(\rho_h^\ast\Omega)_p(u_p,v_p)
=\Omega_{\rho_h(p)}((\rho_h)_\ast u_p,(\rho_h)_\ast v_p)
=d\omega_{\rho_h(p)}([(\rho_h)_\ast u_p]^H,[(\rho_h)_\ast v_p]^H)
\nonumber\\
&=&d\omega_{\rho_h(p)}((\rho_h)_\ast u_p^H,(\rho_h)_\ast v_p^H)
=(d\rho_h^\ast\omega)_p(u_p^H,v_p^H)
=\mbox{ad}_{h^{-1}}(d\omega_p)(u_p^H,v_p^H)
\nonumber\\
&=& \mbox{ad}_{h^{-1}}(\Omega_p)(u_p,v_p)
\ea
\begin{Definition} \label{defa.12}  ~~~~~~~~~~~\\
An element $\psi\in \bigwedge^n(P)\otimes F$ is said to be of type
$(\tau,F)$ (or equivariant under $\rho$) for some representation $\tau$
of $G$ on $F$ iff $\rho_h^\ast \psi=\tau(h)\psi$.
\end{Definition}
It follows that the curvature $\Omega$ is of type
$(\mbox{ad},\mbox{Lie}(G))$.
\begin{Definition}   \label{defa.13} ~~~~~~~~~~\\
Let $\psi\in \bigwedge^m(P)\otimes\mbox{Lie}(G),
\xi\in \bigwedge^n(P)\otimes\mbox{Lie}(G)$ then
\be \label{a.17}
[\psi,\xi]:=\psi\wedge\xi-(-1)^{mn}\xi\wedge\psi=\psi^j\wedge\xi^k
[\tau_j,\tau_k]\in\bigwedge^{m+n}(P)\otimes\mbox{Lie}(G)
\ee
where $\tau_j$ is some basis of the Lie algebra of $G$.
\end{Definition}
\begin{Theorem}[Cartan Structure Equation] \label{tha.1} ~~~~~~~~~\\
\be \label{a.18}
\Omega=d\omega+\omega\wedge\omega
\ee
\end{Theorem}
Proof of Theorem \ref{tha.1}:\\
Using the split $u=u^H+u^V$ it is clear that
$\omega\wedge\omega(u,v)=\omega\wedge\omega(u^V,v^V)$ because $\omega_p$
annihilates $H_p(P)$. Notice that $[\omega,\omega]=2\omega\wedge\omega$.

Likewise we write
\be \label{a.19}
d\omega(u,v)=d\omega(u^H,v^H)+d\omega(u^H,v^V)+d\omega(u^V,v^H)+
d\omega(u^V,v^V)
\ee
and use the differential geometric identity
$d\omega(u,v)=u[i_v\omega]-v[i_u\omega]-i_{[u,v]}\omega$ with
$(i_u\psi)(v_1,..,v_{n-1}):=\sum_{k=1}^n (-1)^{k+1}
\psi(v_1,,,v_{k-1},u,v_{k+1},..,v_n)$ for the contraction of an $n-$form
with a vector field (see, e.g. the second reference in \cite{32}).

To evaluate these four terms in (\ref{a.19}) we need two preliminary
results:\\
1)\\
We can always find $X,Y\in \mbox{Lie}(G)$ such that $u^V=v_X,v^V=v_Y$ are
displayed
as fundamental vector fields. It is easy to verify that
$[u^V,v^V]=[v_X,v_Y]=v_{[X,Y]}\in V_p(P)$ is a Lie algebra homomorphism.
We will exploit that $\omega(u^V)=X$ etc. is a constant.\\
2)\\
By definition of the Lie bracket of vector fields
$[u^V,v^H]=(d/dt)_{t=0} [\rho_{h^{u^V}(t)}]_\ast v^V\in H_p(P)$
since the push-forward by the right action preserves horizontal vector fields
($h^{u^V}(t)$ denotes the integral curve of $u^V$). We will exploit that
$\omega(w^H)=0$ for any horizontal vector field $w^H$.

Using these two properties it is immediate that
$d\omega(u^H,v^V)=d\omega(u^V,v^H)=0$ and that $d\omega(u^V,v^V)=
-\omega([v_X,v_Y])=-[X,Y]$. On the other hand
$$
\omega\wedge\omega(u^V,v^V)=i_{v^V} i_{u^V}\omega\wedge\omega
i_{v^V}[\omega(u_v)\omega-\omega\omega(u_v)]=[\omega(v_X),\omega(v_Y)]
=+[X,Y]
$$
Therefore we are left with
\be \label{a.19a}
[d\omega+\omega\wedge\omega](u,v)=d\omega(u^H,v^H)=\Omega(u,v)
\ee
$\Box$\\
\begin{Corollary}[Bianchi Identity]   \label{cola.1}  ~~~~~\\
\be \label{a.20}
\nabla\Omega=0
\ee
\end{Corollary}
To prove this, use the Cartan structure equation to infer
$d\omega=\omega\wedge\omega-\omega\wedge d\omega=[\omega\wedge\Omega]$
and use $\omega(u^H)=0$ again.
\begin{Definition} \label{defa.14} ~~~~~~~~~~~~\\
The local field strength $F_I:=2 s_I^\ast\Omega=2[dA_I+A_I\wedge A_I$ is
twice the pull-back by local sections of the curvature two-form.
\end{Definition}
Using the transformation behaviour of the connection potential under a
change of trivialization it is easy to verify the the corresponding change
of the field strength is given by
\be \label{a.21}
F_J(x)=\mbox{ad}_{g_{JI}(x)}(F_I(x))
\ee
whence traces of polynomials in the field strength, used in classical
action principles of gauge field theories are globally defined
(gauge invariant).
\begin{Definition} \label{defa.15} ~~~~~~~~~~~\\
Let $E=P\times_\tau F$ be a vector bundle associated to $P$, $c$ a curve in
$\sigma$ and $\tilde{c}$ its horizontal lift which we display as above as
$\tilde{c}(t)=\rho_{h_{cI}(t)^{-1}}(s^\phi_I(c(t))$. A local section
of $E$ is then given by $S_I(x)=[(s^\phi_I(x),f_I(x))]$ where $f_I(x)$
is called the fibre section, whence
\be \label{a.22}
S_I(c(t))=[(s^\phi_I(c(t)),f_I(c(t))]
=[\rho_{h_{cI}(t)^{-1}}(s^\phi_I(c(t)),\tau(h_{cI}(t))f_I(c(t))]
=[\tilde{c}(t),\tau(h_{cI}(t))f_I(c(t))]
\ee
The covariant differential of $S_I$ along $v:=\dot{c}(0)$ at $x=c(0)$ is
defined by
\be \label{a.23}
(\nabla_v S_I)_x:=[\tilde{c}(0),(\frac{d}{dt})_{t=0}\tau(h_{cI}(t))f_I(c(t))]
\ee
\end{Definition}
It is easy to see, using the equivalence relation in the definition of
$E$ and the definition of the horizontal lift that (\ref{a.23}) is actually
independent of the initial datum for $\tilde{c}$ or, equivalently,
the group element $h_0$ in $\tilde{c}(0)=\rho_{h_0}(s_I(x),1_G)$.
Notice that multiplication of sections by scalar functions is defined
by $f(x)S_I(x)=[(s^\phi_I(x),f(x)f_I(x)]$ so that the covariant differential
$\nabla$ satisfies the usual axioms for a covariant differential
(Leibniz rule).

As usual, one is interested for practical calculations in coordinate
expressions. To that end, consider a constant basis $e_\alpha$ in $F$ and
consider the special sections $S_{I\alpha}(x):=[(s^\phi_I(x),e_\alpha)]$.
From the differential equation for the holonomy (\ref{a.13}) with
initial condition $h_{cI}(0)=1_G$
we conclude
\ba \label{a.24}
(\nabla_v S_{I\alpha})(x)
&=&[(s^\phi_I(x),
(\frac{\partial\tau(h)}{\partial h_{AB}}]_{h=1_G} (A_{Ia}(x))_{AB}v^a
e^\alpha)] \\
&=&v^a A_{Ia}^j(x)[(s^\phi_I(x),
(\frac{d\tau(\exp(t\tau_j))}{dt)})_{t=0} e_\alpha)]
=v^a A_{Ia}^j(x)\tau^\tau_j S_{I\alpha}(x)
\nonumber
\ea
where we have abbreviated by $\tau^\tau_j=
(\frac{d\tau(\exp(t\tau_j))}{dt)})_{t=0}$ a basis of Lie$(G)$ in the
representation $\tau$ and have expanded $A_I=A_I^j\tau_j$ correspondingly.
Using the Leibniz rule and the fact that a general section may be
written as $S_I(x)=f_I^\alpha(x) S_{I\alpha}(x)$ we find
\be \label{a.25}
\nabla_v S_I=i_v[df_I^\alpha S_{I\alpha}+f_I^\alpha A_I^j \tau^\tau_j
S_{I\alpha}]
\ee
This expression becomes especially familiar if we use the standard basis
$(e_\alpha)^\beta=\delta_\alpha^\beta$ whence
$f_I^\alpha (M e_\alpha)=M^\beta_\alpha f_I^\alpha e_\beta=
(M f_I)^\alpha e_\alpha$ for any matrix $M$ so that
\be \label{a.26}
\nabla_v S_I=i_v[df_I+A_I^j \tau^\tau_j f_I]^\alpha S_{I\alpha}
=:[i_v(\nabla f_I)^\alpha] S_{I\alpha}
\ee
We now require that $S_I=S$ is actually globally defined which will require
a certain transformation behaviour of $f_I(x)$ under a change of section.
We have $p=\rho_{h_I(p)}(s^\phi_I(x))=\rho_{h_J(p)}(s^\phi_J(x))$ so that
$s^\phi_J(x)=\rho_{h_{IJ}(x)}(s^\phi_I(x))$, thus
$S_J(x)=[(s^\phi_I(x),\tau(h_{IJ}(x))f_J(x))]=S_I(x)$ requires that
the fibre section transforms as
\be \label{a.27}
f_J(x)=\tau(h_{JI}(x))f_I(x)
\ee
This leads to the following covariant transformation property of its
covariant derivative ($c(0)=x,\dot{c}(0)=v$):
\ba \label{a.28}
(\nabla_v f_J)(x)&=& i_v(df_J)_x+(\frac{d}{dt})_{t=0}\tau(h_{cJ}(t))f_J(x)
\nonumber\\
&=& \tau(h_{JI}(x))[
i_v(df_I)_x+\tau(h_{JI}(x))^{-1}[i_v d\tau(h_{JI})](x) f_I(x)
\nonumber\\
&& +\tau(h_{JI}(x))^{-1}(\frac{d}{dt})_{t=0}
\tau(h_{JI}(x)h_{cI}(t)h_{JI}(c(t))^{-1})
\tau(h_{JI}(x))f_I(x)]
\nonumber\\
&=& \tau(h_{JI}(x))[
i_v(df_I)_x+\tau(h_{JI}(x))^{-1}[i_v d\tau(h_{JI})](x) f_I(x)
\nonumber\\
&&
+(\frac{d}{dt})_{t=0}
\tau(h_{cI}(t)h_{JI}(c(t))^{-1})f_I(x)
\nonumber\\
&& +(\frac{d}{dt})_{t=0}\tau(h_{JI}(c(t))^{-1})
\tau(h_{JI}(x))f_I(x)]
\nonumber\\
&=& \tau(h_{JI}(x))[(\nabla_v f_I)(x)
+\{\tau(h_{JI}(x))^{-1}[i_v d\tau(h_{JI})](x) f_I(x)
\nonumber\\
&& +[i_v d\tau(h_{JI})^{-1}](x)\tau(h_{JI}(x))\}f_I(x)]
\nonumber\\
&=& \tau(h_{JI}(x))(\nabla_v f_I)(x)
\ea
which implies that the cross section $S$ has a globally defined covariant
differential.
\begin{Definition} \label{defa.16}
A cross section $S$ in $E=P\times\tau F$ is said to be parallel transported
along a curve $c$ in $\sigma$ iff $(\nabla_{\dot{c}(t)}S)(c(t))=0$
for all $t\in [0,1]$.
\end{Definition}
Notice that we may consider the covariant differential as a map
$\nabla:\; {\cal S}(E)\to {\cal S}(E)\otimes \bigwedge^1(\sigma)$ where
${\cal S}(E)$ denotes the space of sections of $E$.
We extend this definition to $\nabla:\;
{\cal S}(E)\otimes \bigwedge^n(\sigma)\to {\cal S}(E)\otimes
\bigwedge^{n+1}(\sigma)$ through the ``Leibniz rule"
\be \label{a.29}
\nabla (S\otimes \psi):=(\nabla S)\wedge\psi+S\otimes d\psi
\ee
This way we can rediscover the field strength through the square of
the covariant differential:
\ba \label{a.30}
\nabla^2 S&=&\nabla^2 S_\alpha\otimes f^\alpha
=\nabla [\nabla S_\alpha\otimes f^\alpha+S_\alpha \otimes
df^\alpha]
\nonumber\\
&=& \nabla S_\alpha\otimes [df^\alpha+A^\alpha_\beta f^\beta]
= S_\alpha\otimes \{A^\alpha_\gamma\wedge[df^\gamma+A^\gamma_\beta f^\beta]
+d(A^\alpha_\beta f^\beta)\}
\nonumber\\
&=&
=S_\alpha\otimes [dA^\alpha_\beta+A^\alpha_\gamma\wedge A^\gamma_\beta]
f^\beta
=\frac{1}{2}S_\alpha\otimes F^\alpha_\beta f^\beta
\ea

\newpage

\section{Tools from General Topology}
\label{sb}

We collect and prove here some important results from general topology
needed in the main text. For more details, see e.g. \cite{51}.
\begin{Definition} \label{defb.1} ~~~~~~~~~~~~\\
I)\\
i)\\
Let $X$ be a set and $\cal U$ a collection of subsets of $X$. We call
$X$ a topological space provided that\\
1) $\emptyset,X\in{\cal U}$\\
2) $\cal U$ is closed under finite intersections: $U_1,..,U_N\in{\cal U},
N\in \Nl
\;\Rightarrow \; \bigcap_{k=1}^N U_k\in{\cal U}$\\
3) $\cal U$ is closed under arbitrary (possibly uncountably infinite)
unions: $U_\alpha\in{\cal U},\alpha\in A
\;\Rightarrow \; \bigcup_{\alpha\in A} U_\alpha\in{\cal U}$\\
The sets $U\in{\cal U}$ are called open, their complements $X-U$ closed
in $X$. If $x\in X$ is a point and $U$ an open set containing it then
$U$ is called a neighbourhood of $x$ in $X$. A topology $\cal U$ is called
stronger (finer) then a topology $\cal U'$ which then is weaker (coarser)
if ${\cal U}'\subset {\cal U}$.\\
ii)\\
Let $(X,{\cal U}),(Y,{\cal V})$ be topological spaces such that $Y\subset X$.
The relative or subspace topology ${\cal U}_Y$ induced on $Y$ is given
by defining the sets $U\cap Y;\;U\in {\cal U}$ to be open. We say that
we have a topological inclusion, denoted $Y\hookrightarrow X$, provided
that the intrinsic topology is stronger than the relative one, that is,
${\cal U}_Y\subset {\cal V}$.\\
II)\\
i)\\
A function $f:\;X\to Y$ between topological spaces $X,Y$ is said to be
continuous provided that the preimage $f^{-1}(V)$ of any set $V\subset Y$
that is open in $Y$ is open in $X$. (The preimage is defined
by $f^{-1}(V)=\{x\in X;\;f(x)\in V\}$ and despite the notation does not
require $f$ to be either an injection or a surjection). One
easily shows that $f$ is continuous
if it is continuous at each point $x\in X$. Here $f$ is continuous at
$x\in X$ if for any open neighbourhood $V$ of $y=f(x)$ there exists an open
neighbourhood $U$ of $x$ such that $f(x')\in V$ for all $x'\in U$ (i.e.
$f(U)\subset V$).\\
ii)\\
If $f$ is a continuous bijection and also
$f^{-1}$ is continuous then $f$ is called a homeomorphism or a topological
isomorphism.
\end{Definition}
We see that a topology on a set $X$ is simply defined by saying which
sets are open, or equivalently, which functions are continuous.
The importance of homeomorphisms $f$ for topology is that not only the
spaces $X,Y$ can be identified set theoretically but also topologically,
that is, open sets can be identified with each other.

In order to get more topological spaces with more structure
one must add separation
and compactness properties. The one we need here is the following.
\begin{Definition} \label{defb.2} ~~~~~~~~~~\\
i)\\
A topological space $X$ is said to be Hausdorff iff for any two of its points
$x\not=y$ there exist neighbourhoods $U,V$ of $x,y$ respectively which are
disjoint.\\
ii)\\
A topological space $X$ is called compact if every open cover $\cal V$
of $X$ (a collection of open sets of $X$ whose union is all of $X$)
has a finite subcover.
\end{Definition}
\begin{Definition}  \label{defb.3}~~~~~~~~~~\\
i)\\
A net $(x^\alpha)$ in a topological space $X$ is a map
$\alpha\to x^\alpha$ from a
partially ordered and directed index set $A$ (relation $\ge$) to $X$.\\
ii)\\
A net $(x^\alpha)$
converges to $x$, denoted $\lim_\alpha x^\alpha=x$ if for every open
neighbourhood $U\subset X$ of $x$ there exists $\alpha(U)\in A$ such that
$x^\alpha\in U$ for every $\alpha\ge \alpha(U)$ (one says that $(x^\alpha)$
is eventually in $U$).\\
iii)\\
A subnet $(x^{\alpha(\beta)})$ of a net $(x^\alpha)$ is defined through
a map $B\to A;\;\beta\mapsto \alpha(\beta)$ between partially ordered and
directed index sets such that for any $\alpha_0\in A$ there exists
$\beta(\alpha_0)\in B$ with $\alpha(\beta)\ge \alpha_0$ for any
$\beta\ge \beta(\alpha_0)$ (one says that $B$ is cofinal for $A$). \\
iv) \\
A net $(x^\alpha)$ in a topological space $X$ is called universal if
for any subset $Y\in X$ the net $(x^\alpha)$ is eventually either only
in $Y$ or only in $X-Y$.
\end{Definition}
Notice that for a subnet there is no relation between the index sets
$A,B$ except that $\alpha(B)\subset A$ so that in particular the subnet
of a sequence ($A=\Nl$) may not be a sequence any longer.
The notions of closedness, continuity and compactness can be formulated in
terms of nets. The fact that
one uses nets instead of sequences is that lemma \ref{lab.1} is no longer
true when $A=\Nl$ unless we are dealing with metric spaces.
\begin{Lemma} \label{lab.1} ~~~~~~~~~~\\
i) A subset $Y$ of a toplogical space $X$ is closed if for every
convergent net $(x^\alpha)$ in $X$ with $x^\alpha\in Y\;\forall \alpha$
the limit actually lies in $Y$.\\
ii)\\
A function $f:\;X\to Y$ between topological spaces is
continuous if for every convergent net $(x^\alpha)$ in $X$, the net
$(f(x^\alpha))$ is convergent in $Y$. \\
iii) \\
A topological space $X$ is compact if every net has a convergent
subnet. The limit point of the convergent subnet is called a cluster
(accumulation) point of the original net.
\end{Lemma}
%
The proof is standard and will be omitted.
One easily sees that if a net converges (a function is continuous) in
a certain topology, then it does so in any weaker (stronger) topology.
In our applications direct products of topological spaces are of fundamental
importance.
\begin{Definition} \label{defb.4}   ~~~~~~~~~~~\\
The Tychonov topology on the direct product $X_\infty=\prod_{l\in{\cal L}}
X_l$
of topological spaces $X_l$, $\cal L$ any index set,
is the weakest topology such that all the projections
\be \label{b.1}
p_l:\;X_\infty\to X_l;\;(x_{l'})_{l'\in {\cal L}}\mapsto x_l
\ee
are continuous, that is, a net $x^\alpha=(x^\alpha_l)_{l\in {\cal L}}$
converges to $x=(x_l)_{l\in {\cal L}}$ iff $x^\alpha_l\to x_l$
for every $l\in {\cal L}$ pointwise (not necessarily uniformly) in
$\cal L$. Equivalently, the sets $p_l^{-1}(U_l)=
[\prod_{l'\not=l} X_{l'}]\times U_l$ are defined to be open and
form a base for the topology of $X_\infty$ (any open set can be obtained
from those by finite intersections and arbitrary unions).
\end{Definition}
The definition of this topology is motivated by the following theorem.
\begin{Theorem}[Tychonov] \label{thb.1} ~~~~~~~~~~~\\
Let $\cal L$ be an index set of arbitrary cardinality and suppose
that for each $l\in {\cal L}$ a compact topological space $X_l$ is given.
Then the direct product space $X_\infty=\prod_{l\in {\cal L}} X_l$
is a compact topological space in the Tychonov topology.
\end{Theorem}
We will give an elegant proof of the Tychonov theorem using the notion of
a universal net.
\begin{Lemma} \label{lab.2}   ~~~~~~~\\
i)\\
A universal net has at most one cluster point to which it then converges.\\
ii)\\
For any map $f:\;X\to Y$ between topological spaces the net $f(x^\alpha)$
in $Y$ is universal whenever $(x^\alpha)$ is universal in $x$ with no
restrictions on $f$.\\
iii)\\
Any net has a universal subnet.
\end{Lemma}
Proof of Lemma \ref{lab.2}:\\
i)\\
Suppose that $x$ is a cluster point of a universal net $(x^\alpha)$
and that the subnet $x^{\alpha(\beta)}$ converges to it. Thus for any
neighbourhood $U$ of $x$ the subnet is eventually in $U$, i.e. there exists
$\beta(U)$ such that $x^{\alpha(\beta)}\in U$ for any $\beta\ge \beta(U)$.
Since $(x^\alpha)$
is universal it must be eventually either in $U$ or $X-U$. Suppose there
was $\alpha_0$ such that $x^\alpha\in X-U$ for any $\alpha\ge  \alpha_0$.
By definition of a subnet we find $\beta(\alpha_0)$ such that
$\alpha(\beta)\ge \alpha_0$ for any $\beta\ge \beta(\alpha_0)$. Without
loss of generality we may choose $\beta(\alpha_0)\ge \beta(U)$. But then
we know already that the $x^{\alpha(\beta)},\;\beta\ge \beta(\alpha_0)$
are in $U$ which is a contradiction. Thus $x^\alpha$ is eventually in
$U$. Since $U$ was an arbitray neighbourhood of $x$, it follows that
$(x^\alpha)$ actually converges to $x$.\\
ii)\\
Obviously $f(x^\alpha)$ is eventually in $f(X)$ so we must show that
for any $V\subset f(X)$ we have $f(x^\alpha)$ eventually in $V$ or
$f(X)-V$. Let $U=f^{-1}(V)$ be the preimage of $V$, then $f(X-U)=f(X)-V$.
Since $(x^\alpha)$ is eventually in $U$ or $X-U$, the claim follows.\\
iii) \\
The proof can be found in exercise 2J d) together with theorem
2.5 in \cite{52}.\\
$\Box$\\
%
\begin{Corollary} \label{colb.1}   ~~~~~~~~~~~\\
A topological space $X$ is compact iff every universal net converges.
\end{Corollary}
Proof of Corollary \ref{colb.1}:\\
$\Rightarrow$:\\
Take any universal net $(x^\alpha)$. Since $X$ is compact it has a cluster
point to which it actually converges by lemma \ref{lab.1} i).\\
$\Leftarrow$:\\
Take any net $(x^\alpha)$. Then by lemma \ref{lab.1}iii) it has a universal
subnet $x^{\alpha(\beta)}$ which converges by assumtion.
Thus, $X$ is compact.\\
$\Box$\\
\\
Proof of Theorem \ref{thb.1}:\\
Let $(x^\alpha)=(x_l^\alpha)_{l\in {\cal L}}$ be any universal net in
$X_\infty=\prod_{l\in {\cal L}} X_l$. By lemma \ref{lab.1}ii) the
net $p_l((x^\alpha))=(x^\alpha_)$ is universal in $X_l$. Since $X_l$ is
compact, it converges to some $x_l$. Define $x:=(x_l)_{l\in {\cal L}}$.
By defintion of the Tychonov topology, $x^\alpha\to x$ iff
$x^\alpha_l\to x_l$ for any $l\in {\cal L}$ whence $(x^\alpha)$ converges.\\
$\Box$\\
This proof of the Tychonov theorem is shorter than the usual one in terms
of the (in)finite intersection property and technically clearer.
\begin{Definition} \label{defb.5} ~~~~\\
Let $Y$ be a subset of a topological space $X$. The subset topology induced
by $X$ on $Y$ is defined through the collection of open sets
${\cal V}:=\{U\cap Y;\;U\in {\cal U}\}$ where $\cal U$ defines the topology
of $X$.
\end{Definition}
\begin{Lemma} \label{lab.3}  ~~~~~~~\\
A closed subset $Y$ of a compact topological space $X$ is compact in the
subspace topology.
\end{Lemma}
Proof of Lemma \ref{lab.3}:\\
Let ${\cal V}$ be any open cover for $Y$. Since $Y$ is closed in $X$,
$X-Y$ is open in $X$ whence ${\cal U}={\cal V}\cup\{X-Y\}$ is an open cover
for $X$. Since $X$ is compact, it has a finite open subcover
$\{U_k\}_{k=1}^N\cup\{X-Y\}$ for some $N<\infty$ where $U_k$ is open in $X$.
By defintion of the supspace topology, $U_k\cap Y$ is open in $Y$ so that
$\{U_k\cap Y\}_{k=1}^N$ is a finite open subcover of $\cal V$.\\
$\Box$

In our discussion of the gauge orbit of connections we will deal with the
quotient of connections by the set of gauge transformations which is
a topological space again. The resulting quotient space carries a natural
topology, the quotient topology.
\begin{Definition} \label{defb.5a} ~~~~~~~\\
i)\\
Let $X,Y$ be topological spaces and $p:\;X\to Y$ a surjection. The map $p$
is said to be a quotient map provided that $V\subset Y$ is open in $Y$
if and only if $p^{-1}(V)$ is open in $X$.\\
ii)\\
If $X$ is a topological space, $Y$ a set and $p:\;X\to Y$ a surjection
then there exists a unique topology on $Y$ with respect to which $p$ is a
quotient map.\\
iii)\\
Let $X$ be a topological space and let $[X]$ be a partition of $X$ (i.e.
a collection of mutually disjoint subsets of $X$ whose union is $X$).
Denote by $[x],\;x\in X$ the subset of $X$ in that partition of $X$
which contains $x$. Equip $[X]$ with the quotient topology induced by
the map $[]:\;X\to [X];\;x\mapsto [x]$. Then $[X]$ is called the quotient
space of $X$.
\end{Definition}
Notice that the requirement for $p$ to be a quotient map is stronger than
that it be continuous which would only require that $p^{-1}(V)$ is open
in $X$ whenever $V$ is open in $Y$ (but not vice versa). Clearly in ii)
we define the topology on the set $Y$ to be those subsets $V$ for which
the preimage $p^{-1}(V)$ is open in $X$ and it is an elementary exercise
in the theory of mappings of sets to verify that the collection of
subsets of $Y$ so defined satisfies the axioms of a topology of
definition \ref{defb.1}.

Quotient spaces naturally arise if we have a
group action $\lambda:\;G\times X\to X;\;(g,x)\to
\lambda_g(x):=\lambda(g,x)$ on a topological space $xX$ and define
$[x]:=\{\lambda_g(x);\;g\in G\}$ to be the orbit of $x$. The orbits clearly
define a partition of $X$.
\begin{Lemma} \label{lab.4}     ~~~~~~~~~~\\
Let $X$ be a compact topological space, $Y$ a set and $p:\;X\to Y$
a surjection. Then $Y$ is compact in the quotient topology.
\end{Lemma}
Proof of Lemma \ref{lab.4}:\\
First of all, consider any subsets $V_1,V_2$ of $Y$.\\
On the one hand suppose
$x\in p^{-1}(V_1)\cap p^{-1}(V_2)$. Then there exist $y_1\in V_1,y_2\in
V_2$ such that $y_1=p(x)=y_2$, that is, $y_1=y_2\in V_1\cap V_2$ so
that actually $x\in p^{-1}(V_1\cap V_2)$. We conclude
$p^{-1}(V_1)\cap p^{-1}(V_2)\subset p^{-1}(V_1\cap V_2)$.\\
On the other hand,
let $x\in p^{-1}(V_1\cap V_2)$, then there exists $y\in V_1\cap V_2$
such that $x\in p^{-1}(y)$. Since $y\in V_1\cap V_2$ we have
$p^{-1}(y)\in p^{-1}(V_1)$ and $p^{-1}(y)\in p^{-1}(V_2)$,
thus $x\in p^{-1}(V_1)\cap P^{-1}(V_2)$. We conclude
$p^{-1}(V_1\cap V_2)\subset p^{-1}(V_1)\cap p^{-1}(V_2)$.\\
Thus, altogether
$p^{-1}(V_1)\cap p^{-1}(V_2)=p^{-1}(V_1\cap V_2)$ and
$p^{-1}(V_1)\cup p^{-1}(V_2)= p^{-1}(V_1\cup V_2)$ by taking complements.

Next, let $\cal V$ be an open cover of $Y$. Then, by definition of the
quotient topology, $p^{-1}(V)$ is open in $X$ and ${\cal U}:=\{p^{-1}(V);\;
V\in {\cal V}\}$ covers $X$ because $\bigcup_{U\in {\cal U}} U=
\bigcup_{V\in {\cal V}} p^{-1}(V)=
p^{-1}(\bigcup_{V\in {\cal V}} V)=p^{-1}(Y)=X$ since $p$ is a surjection and
$\cal V$ covers $Y$. We conclude that $\cal U$ is an open cover of $X$.

Since $X$ is compact, we find a finite, open subcover
$\{p^{-1}(V_k)\}_{k=1}^N$ of $X$ so that
$X=\bigcup_{k=1}^N p^{-1}(V_k)=p^{-1}(\bigcup_{k=1}^N V_k)=p^{-1}(Y)$
whence $Y=\bigcup_{k=1}^N V_k$, that is, $\{V_k\}_{k=1}^N$ is a finite
open subcover of $\cal V$ and $Y$ is compact.\\
$\Box$
\begin{Lemma} \label{lab.5} ~~~~~~~~~\\
Let $X$ be a Hausdorff space and $\lambda:\;G\times X\to X$ a continuous
group action on $X$ (i.e., $\lambda_g$ defined by
$\lambda_g(x):=\lambda(g,x)$ is continuous for any $g\in G$). Then
the quotient space $X/G:=\{[x];\;x\in X\}$ defined by the orbits
$[x]=\{\lambda_g(x);\;g\in G\}$ is Hausdorff in the quotient topology.
\end{Lemma}
Proof of Lemma \ref{lab.5}:\\
Let $[x]\not=[x']$ then certainly $x\not=x'$ since orbits are disjoint.
Since $X$ is Hausdorff we find disjoint open neighbourhoods $U,U'$ of
$x,x'$ respectively. We want to show that $U,U'$ can be chosen in such
a way that
\be \label{b.2}
[U]:=\{[y];\;y\in U\},\;[U']:=\{[y'];\;y'\in U'\}
\ee
are disjoint. First of all we notice that ($p$ the projection map)
\ba \label{b.3}
p^{-1}([U])&=&\bigcup_{y\in U} p^{-1}([y])=\{\lambda(g,y);\;y\in U,g\in G\}
=\bigcup_{g\in H} \lambda_g(U)=\bigcup_{g\in H} \lambda_{g^{-1}}(U)
\nonumber\\
&=& \bigcup_{g\in H} (\lambda_g)^{-1}(U)
\ea
where we have made use of $\lambda_{g^{-1}}=(\lambda_g)^{-1}$.
Since $U$ is open in $X$ and $\lambda_g$ is continuous by assumption, we
have that
$\lambda_g^{-1}(U)$ is open in $X$. Since arbitrary unions of open sets
are open it follows that $p^{-1}([U])$ is open in $X$, thus by the definition
of the quotient topology we have $[U],[U']$ open in $X/G$. Next,
obviously $[x]\in [U],[x']\in [U']$ whence $[U],[U']$ are open neighbourhoods
of $[x],[x']$ in $X/G$ respectively.

Let us now choose $V,V'$ to be open, disjoint
neighbourhoods of the orbits $p^{-1}([x])=\lambda_G(x),p^{-1}([x'])$
respectively. (This is certainly possible as otherwise there exists
$g\in G$ such that $\lambda_g(x),x'$ have no disjoint neighbourhoods
which is impossible because $\lambda_g(x)\not=x'$ (otherwise $[x]=[x']$)
and $X$ is Hausdorff). We claim that we can choose $U,U'$ in such a way
that $p^{-1}[U]:=\bigcup_{g\in G} \lambda_g(U)\subset V$ and
$p^{-1}[U']:=\bigcup_{g\in G} \lambda_g(U')\subset V'$.

Suppose that
were not the case. Then for any neighbourhood $U$ of $x$ we find
$z\in U$ and $g_0\in G$ such that $\lambda_{g_0}(z)\not\in V$.
Since by construction of $V$ we have that $V$ is a common open neighbourhood
of any $\lambda_g(x),\;g\in G$ we have in particular $y:=\lambda_{g_0}(x)\in
V$. It follows that we have found an open neighbourhood $V$ of
$y=\lambda_{g_0}(x)$ such that for any open neighbourhood $U$ of $x$
there exists $z\in U$ with $\lambda_{g_0}(z)\not\in V$. This means that
the map $\lambda_{g_0}$ is not continuous at $x$ in contradiction to
our assumption that $\lambda_g$ is everywhere continuous for any $g\in G$.

Therefore $p^{-1}([U])\cap p^{-1}([U'])=p^{-1}([U]\cap [U'])=\emptyset$
whence $[U]\cap [U']=\emptyset$, thus $X/G$ is Hausdorff.\\
$\Box$\\
\begin{Theorem} \label{thb.2}  ~~~~~~~~~~~~~\\
Let $X,Y$ be topological spaces and let $G$ be a group acting (not
necessarily continuously) on
them via $\lambda,\lambda'$ respectively. If $f:\; X\to Y$ is a
homeomorphism with respect to which the actions $\lambda,\lambda'$ are
equivariant then $f$ extends as a homeomorphism to the quotient spaces
$X/G,Y/G$ in their respective quotient topologies.
\end{Theorem}
Proof of Theorem \ref{thb.2}:\\
Equivariance means that $f\circ \lambda_g=\lambda'_g\circ f$ for all
$g\in G$ and since $f$ is a bijection, equivariance implies also
$\lambda_g\circ f^{-1}=f^{-1}\circ \lambda'_g$
Consider the corresponding quotient maps
\be \label{b.4}
p:\;X\to X/G;\;x\mapsto [x]_\lambda=\{\lambda_g(x);\;g\in G\} \mbox{ and }
p':\;Y\to Y/G;\;x\mapsto [x]_{\lambda'}=\{\lambda'_g(y);\;g\in G\}
\ee
Then due to equivariance
\be \label{b.5}
f([x]_\lambda)=\{f(\lambda_g(x));\;g\in G\}=\{\lambda'_g(f(x));\;g\in G\}
=[f(x)]_{\lambda'}
\ee
and similarly $f^{-1}([y]_{\lambda'})=[f^{-1}(y)]_{\lambda}$ so that
$f$ extends to a bijection between the corresponding equivalence classes.

Next we notice that
$p^{-1}([x]_\lambda)=\{\lambda_g(x);\;g\in G\}$ whence by (\ref{b.5})
we have $f(p^{-1}([x]_\lambda))=(p')^{-1}([f(x)]_{\lambda'})$ for all
$[x]_\lambda\in X/G$.
This shows that equivariance also implies
\be \label{b.6}
f\circ p^{-1}=(p')^{-1}\circ f\;\Rightarrow\;
f^{-1}\circ (p')^{-1}=p^{-1}\circ f^{-1}
\ee
Let then $B$ be open in $Y/G$, thus $(p')^{-1}(B)$ open in $Y$ by definition
of the quotient topology in $Y/G$, thus
$(f^{-1}\circ (p')^{-1})(B)=(p^{-1}\circ f^{-1})(B)$ open in $X$ since
$f$ is continuous, thus $f^{-1}(B)$ open in $X/G$ by definition of the
quotient topology in $X/G$. Likewise we see that $A$ open in $X/G$
implies $f(A)$ open in $Y/G$ since $f^{-1}$ is continuous.
It follows that $f,f^{-1}$ are continuous as maps between $X/G,Y/G$.\\
$\Box$

\newpage

\section{Elementary Introduction to Gel'fand Theory
for Abelean $C^\ast$ Algebras}
\label{se}

There are many good mathematical textbooks on operator algebra -- and
abstract $C^\ast-$algebra theory, see e.g. \cite{53,8b}.
The textbooks \cite{55} are more geared towards applications in
mathematical physics. For a pedagogical introduction with elegant proofs
the beautiful review \cite{56} is recommended.
\begin{Definition} \label{defe.1}  ~~~~~~~~~\\
i)\\
An algebra $\cal A$ is a vector space (taken over $\Cl$) together with
a multiplication map $\a\times\a\to\a;\;(a,a')\mapsto aa'$ which
is associative, $(ab)c=a(bc)$, and distributive,
$b(za+z'a')=zba+z'ba',\;(za+z'a')b=zab+z'a'b$ for all $a,a',b\in\a,\;
z,z'\in\Cl$. \\
ii) \\
An algebra $\a$ is called Abelean if all elements commute with each other
and unital if it has a (necessarily unique) unit element $1$ satisfying
$1a=a1=a$ for all $a\in\a$.\\
iii) \\
A vector subspace ${\cal B}$ of $\a$ is called a subalgebra if it
is closed under multiplication. A subalgebra $\cal I$ is called a left
(right) ideal if $ab\in {\cal I}$ ($ba\in {\cal I}$) for all $a\in \a,\;
b\in {\cal I}$ and a two-sided ideal (or simply ideal) if it is
simultaneously a left -- and
right ideal. An ideal of either kind is called maximal if there
is no other ideal containing it except for $\a$ itself.\\
iv)\\
An involution on an algebra $\a$ is a map $\ast:\;\a\to\a;a\mapsto a^\ast$
satisfying\\
1) $(za+z'b)^\ast=\bar{z} a^ast+\bar{z}' b^\ast$ (conjugate linear),\\
2) $(ab)^\ast=b^\ast a^\ast$ (reverses order) and\\
3) $(a^\ast)^\ast=a$ (squares to the identity)\\
for all $a,b \in \a,\;z,z'\in\Cl$. An algebra with involution is called
an $^\ast-$algebra.\\
v)\\
A homomorphism ($^\ast-$homomorphism) is a linear map $\Phi:\a\to{\cal B}$
between algebras ($^\ast-$algebras) that preserves the multiplicative
(and involutive) structure, that is, $\phi(ab)=\phi(a)\phi(b)$
(and $\phi(a^\ast)=(\phi(a))^\ast$).\\
vi)\\
A normed algebra $\a$ is equipped with a norm $||.||:\;\a\to\Rl^+$
(that is $||a+b||\le ||a||+||b||,\;||za||=|z|\;||a||,\;||a||=0\Leftrightarrow
a=0$, if the last property is drooped, then $||.||$ is only a seminorm)
whose compatibility with the mutiplicative structure is contained in
the submultiplicativity requirement $||ab||\le ||a||\;||b||$ for all
$a,b\in \a$. If $\a$ has an involution we require $||a^\ast||=||a||$ and
$\a$ is called a normed $^\ast-$algebra. If $\a$ is unital we require
$||1||=1$ (this is just a choice of normalization).\\
vii)\\
A norm induces a metric $d(a,b)=||a-b||$ and if the algebra $\a$ is complete
(every Cauchy sequence converges) then it is called a Banach algebra.\\
viii)\\
A $C^\ast-$algebra $\a$ is a Banach algebra with involution with the
following compatibility condition between the involutive and metrical
structure
\be \label{e.1}
||a^\ast a||=||a||^2
\ee
\end{Definition}
The innocent looking condition (\ref{e.1}) determines much of the structure
of $C^\ast-$algebras. If a $C^\ast-$algebra is not unital one can always
embed it isometrically into a larger unital $C^\ast-$algebra (see e.g.
\cite{56}). While this does not remove all problems with $C^\ast-$algebras
without identity in our applications only unital $C^\ast-$algebras will
appear and this is what we will assume form now on. If $\cal I$
is a two-sided ideal in an algebra $\a$ we can form the quotient algebra
$\a/{\cal I}$ which consists of the equivalence classes
$[a]:=\{a+b;\;b\in {\cal I}\}$ for any $a\in \a$ in which the rules
for addition, multiplication and scalar multiplication are given by
$[a]+[a']=[a+a'],\;[a][a']=[aa'],\;[za]=z[a]$ and it is easy to see that
the condition that $\cal I$ is an ideal is just sufficient for making these
rules independent of the representative.
Finally, if we think of $\a$ as an algebra of operators on a Hilbert
space and $||.||$ is the uniform operator norm then we see that we are
dealing with agebras of bounded operators only which trivializes
domain questions.
\begin{Definition} \label{defe.2}  ~~~~~~~~~~\\
The spectrum $\Delta(\a)$ of a unital Banach algebra $\a$ is the
set of all non-zero $^\ast-$homomorphisms $\chi:\;\a\to \Cl;\;
a\to \chi(a)$, called the characters.
\end{Definition}
Notice that $\Cl$ is itself a unital, Abelean $C^\ast-$algebra in the usual
metric topology of $\Rl^2$. Notice that $\chi(1)=1$ since
$\chi(a)=\chi(1a)=\chi(1)\chi(a)$ and if we choose $a\in \a$ such that
$\chi(a)\not=0$ the claim follows. Similarly $\chi(a^{-1})=\chi(a)^{-1}$
if $a$ has an inverse in $\a$, that is an element $a^{-1}$ with
$a a^{-1}=a^{-1} a=1$. Finally $\chi(0)=0$ since $1=\chi(1)=\chi(1+0)=
\chi(1)+\chi(0)=1+\chi(0)$.
\begin{Definition} \label{defe.3}  ~~~~~~~~~~~\\
For a character in a unital Banach algebra $\a$ define
$\mbox{ker}(\chi):=\{a\in \a;\;\chi(a)=0\}$ to be its kernel.
\end{Definition}
Clearly, $\mbox{ker}(\chi)$ is a two-sided ideal in $\a$ since
$\chi(ab)=\chi(ba)=\chi(a)\chi(b)=0$ for all
$a\in\a,\;b\in \mbox{ker}(\chi)$. Since $\chi$ is in particular
a linear functional on $\a$ considered as a vector space, it follows
that $\mbox{ker}(\chi)$ is a vector subspace of $\a$ of codimension one.
After taking its closure in $\a$ it is either still of codimension one
or of codimension zero, the latter being impossible since then
$\chi$ would be identically zero which we excluded in the definition
for a character.
It follows that there exist elements $a\in\a-\mbox{ker}(\chi)$ and that
$\a$ is the closure of the span of $a,\mbox{ker}(\chi)$. Thus, if there
is an ideal $\cal I$ of $\a$ properly containing $\mbox{ker}(\chi)$
then we may take such an $a\in {\cal I}-\mbox{ker}(\chi)$ from which we
conclude ${\cal I}=\a$. We conclude that the kernel of a character determines
a maximal ideal in $\a$.
\begin{Definition} \label{defe.4}  ~~~~~~~~~~~~~\\
Let $\a$ be a normed, unital algebra. The spectrum $\sigma(a)$ of $a\in\a$ is
defined to be the complement $\Cl-\rho(a)$ where
$\rho(a):=\{z\in\Cl;\;(a-z\cdot 1)^{-1}\in \a\}$ is called the resolvent
set of $a$. For $z\in\rho(a)$ one calls $r_z(a):=(a-z\cdot 1)^{-1}$ the
resolvent of $a$at $z$. The number
\be \label{e.1a}
r(a):=\sup(\{|z|;\;z\in \sigma(a)\}
\ee
is called the spectral radius of $a\in \a$.
\end{Definition}
Notice that the condition $a^{-1}\in\a$ implies that $||a^{-1}||$ exists,
that is, the inverse has a norm (``is bounded"). If we are dealing with an
algebra of
possibly unbounded operators on a Hilbert space then definition \ref{defe.4}
must be more precise: if $a$ is a densely defined, closed (the adjoint
$a^\ast\equiv a^\dagger$ is densely defined) linear operator on a Hilbert
space $\cal H$ with dense domain $D(a)$ then $z\in\rho(a)$ iff
$a-z\cdot 1$ is a bijection from $D(a)$ onto $\cal H$ with {\it bounded}
inverse.

We will need later the following technical result.
\begin{Lemma} \label{lae.0} ~~~~~~~~~~~\\
For the spectral radius the following identity holds
\be \label{e.1b}
r(a)=\lim_{n\to\infty} ||a^n||^{1/n}
\ee
\end{Lemma}
Proof of Lemma \ref{lae.0}:\\
First we show that the series of non-negative numbers
$x_n=||a^n||^{1/n}$ actually converges. For this purpose let $n\ge m\ge 1$
be any natural numbers and split $n$ uniquely as $n=km+r$ for natural
numbers $k,r$ with $0\le r<m$. By submultiplicativity of the norm we have
\be \label{e.1c}
||a^n||^{1/n}\le ||a^{km}||^{1/n}\; ||a^r||^{1/n}
\le ||a^m||^{k/n}\; ||a^r||^{1/n}
\ee
Fix $m$ and take $n\to\infty$ so that $k=(n-r)/m\to\infty$ while
$r\in\{0,..,m-1\}$ stays bounded. Thus the right hand side of (\ref{e.1c})
converges to $||a^m||^{1/m}$. It follows that the sequence $(x_n),\;
x_n=||a^n||^{1/n}$
is bounded and therefore must have an accumulation point each of which
must be smaller than $x_m$ for any $m\ge 1$. Let $\lim_n\sup(x_n)$ be the
largest accumulation point, then the inequality
$\lim_n\sup(x_n)\le x_m$ holds. Now take the infimum on the right hand side
which is also an accumulation point, then we get
\be \label{e.1d}
\lim_n\sup(x_n)\le \lim_m\inf(x_m)
\ee
which means that there is only one accumulation point, so the sequence
converges. Denote $x:=\lim_{n\to\infty} x_n$.

Now consider the geometrical (von Neumann) series for $z\not=0$
\be \label{e.1e}
r_z(a)=(a-z\cdot 1)^{-1}=-\frac{1}{z}\sum_{n=0}^\infty (\frac{a}{z})^n
\ee
which converges if there exists $0\le q<1$ with
$||(\frac{a}{z})^n||^{1/n}=||a^n||^{1/n}/|z|<q$ for all $n>n(q)$.
In other words, $z\in \rho(a)$ provided that $|z|>\lim_{n\to\infty} x_n$
or equivalently $z\in \sigma(a)$ provided that
\be \label{e.1f}
|z|\le x
\ee
Taking the supremum in $\sigma(a)$ on the left hand side of (\ref{e.1f})
we thus find
\be \label{e.1g}
r(a)\le x
\ee
Suppose now that $r(a)<x$. Then there exists a real
number $R$ with $r(a)<R<x$ and since obviously $R\in\rho(a)$ it is clear
that the resolvent $r_R(a)$ of $a$ at $R$ converges. Let $\phi$ be a
continuous linear functional on $\a$ then
\be \label{e.1k}
\phi(r_R(a))=-\frac{1}{R}\sum_{n=0}^\infty \phi((\frac{a}{z})^n)
\ee
exists which means that $\lim_{n\to\infty} \phi((\frac{a}{z})^n)=0$.
In other words, the function $n\mapsto \phi((\frac{a}{z})^n)$ is bounded
for all continuous linear functionals $\phi$.

Now the space $\a'$ of continuous linear forms on $\a$ is itself a Banach
space with norm $||\phi||:=\sup_{a\in \a}|\phi(a)|$. Consider the family
${\cal F}:=\{a^n/r^n;\;n\in\Nl\}$ then we have just shown that for
each $b\in{\cal F}$ the set $\{|\phi(b)|;\;\phi\in\a'\}$ is bounded.
Let us consider each $b\in{\cal F}$ as a map $b:\;\a'\to \Cl;\;\phi\to
\phi(b)$. We have $||b||':=\sup_{\phi\in\a'}|\phi(b)|/||\phi||=||b||$
where the norm in the last equality is the one in $\a$. By the principle
of uniform boundedness \cite{47n11} the set $\{||b||';\;b\in{\cal F}\}$
is bounded. Therefore we know that the set of norms $||a^n/r^n||$
is bounded. But
\be \label{e.1h}
||a^n/r^n||=(\frac{x}{r})^n(\frac{||a^n||^{1/n}}{x})^n
\ee
and the first fraction diverges while the second approaches $1$ as
$n\to\infty$.

Thus in fact $r(a)=\lim_{n\to\infty} ||a^n||^{1/n}$.\\
$\Box$\\
We will now start establishing the relation between characters and
maximal ideals.
\begin{Lemma} \label{lae.1} ~~~~~~~~~~~\\
If $\cal I$ is an ideal in a unital Banach algebra $\a$ then its closure
$\overline{{\cal I}}$ is still an ideal in $\a$. Every maximal ideal
is automatically closed.
\end{Lemma}
Proof of Lemma \ref{lae.1}:\\
Recall that the closure of a subset $Y$ in a topological space is $Y$
together with the limit points of convergent nets in $Y$. Let now
$\cal I$ be an ideal in $\a$ and let $(a^\alpha)$ be a net in $\cal I$
converging to $a\in\overline{{\cal I}}$. Then for any $b\in\a$ we have
$b a^\alpha\in {\cal I}$ since $\cal I$ is an ideal and
$\lim_\alpha ba ^\alpha=ba$ since $||b(a^\alpha-a)||\le
||b||\;||a^\alpha-a||\to 0$. Thus  $(ba^\alpha)$ is a net in $\cal I$
converging to $ba\in\a$ and since al limit points of converging nets in
$\cal I$ by definition lie in $\overline{{\cal I}}$
we actually have have $ba\in \overline{{\cal I}}$. Thus,
$\overline{{\cal I}}$ is an ideal.

Next we notice that every $a\in\a$ such that $||a-1||<1$ is invertible
(use $a^{-1}=-(1-(a-1))^{-1}$ and the geometric series representation
for the latter with convergence radius $1$). The set
$\{a\in\a;||a-1||\ge 1\}$ is a closed subset of $\a$ because if
$(a^\alpha)$ is a convergent net in it then the net of real
numbers $(||a^\alpha-1||)$ belongs to the set $\{x\in\Rl;\;x\ge 1\}$
and since it converges to $||a-1||$ it follows that $||a-1||\ge 1$ since
$\{x\in\Rl;\;x\ge 1\}$ is closed (that $b^\alpha\to b$ implies
$||b^\alpha||\to ||b||$ follows from the triangle inequality
$||a||\le ||a-b||+||b||,||b||\le ||a-b||+||a||$).
We conclude that every non-trivial (those not containing invertible
elements) ideal $\cal I$ must be contained in the closed set
$\{a\in\a;||a-1||\ge 1\}$ and so must its closure $\overline{{\cal I}}$.
Obviously $1\not\in \{a\in\a;||a-1||\ge 1\}$, hence, closures of
non-trivial ideals are non-trivial.

Finally a maximal ideal must be closed as otherwise its closure would be a
non-trivial ideal containing it.\\
$\Box$\\
\begin{Theorem}[Gel'fand] \label{the.1} ~~~~~~\\
If $\a$ is an Abelean, unital Banach algebra and $\cal I$ a two-sided,
maximal ideal in $\a$ then the quotient algebra $\a/{\cal I}$ is isomorphic
with $\Cl$.
\end{Theorem}
Proof of Theorem \ref{the.1}:\\
By lemma \ref{lae.1} ${\cal I}$ is closed in $\a$. We split the proof into
three parts.\\
$[i)]$ {\it If $\cal I$ is a maximal ideal in a unital Banach algebra $\a$ then
$\a/{\cal I}$ is a Banach algebra}\\
The norm on $\a/{\cal I}$ is given by
\be \label{e.2}
||[a]||:=\inf_{b\in [a]} ||b||
\ee
To see that this indeed defines a norm we check
\ba \label{e.3}
||[za]|| &=& ||z[a]||=\inf_{b\in [a]}||z b||=|z|\;||[a]||
\nonumber\\
||[a+a']||&=& ||[a]+[a']||=\inf_{b\in [a]+[a']}||b||=
\inf_{b\in [a],b'\in [a']} ||b+b'||
\nonumber\\
&\le&
\inf_{b\in [a],b'\in [a']} (||b||+||b'||)=||[a]||+||[a']||
\nonumber\\ ||[a]||&=&\inf_{b\in [a]} ||b||=0\Rightarrow [a]=[0]
\ea
In the second line we exploited that every representative of
$[a+a']$ can be written in the form $b+b'$ where $b,b'$ are representatives
of $[a],[a']$ and that the joint infimumm is the same as the infimum.
The conclusion in the last line means that $[a]$ contains elements
of arbitrarily small norm. (Consider a net of elements $(a+b^\alpha)$
in $[a]$ whose norm converges to zero. The net $(b^\alpha)$ is a net
in $\cal I$ and since $\cal I$ is closed it follows that the limit point
$a+b$ lies in $[a]$. Since $||a+b||=0$ and $||.||$ is a norm it follows
$a+b=0$, thus $0\in [a]$ and so $[a]=[0]$).

Suppose that $([a_n])$ is a Cauchy sequence in $\a/{\cal I}$.
We may assume $||[a_{n+1}]-[a_n]||=||[a_{n+1}-a_n]||<2^{-n}$ (pass
to a subsequence if necessary). Since
\be \label{e.4}
||[a_{n+1}]-[a_n]||=\inf_{b_{n+1}\in [a_{n+1}],b_n\in [a_n]}
||b_{n+1}-b_n||<2^{-n}
\ee
we certainly find representatives with $||c_{n+1}-c_n||<2^{-n+1}$.
Then for $n>m$
\be \label{e.5}
||c_n-c_m||=||\sum_{k=m+1}^{n-1} (c_{k+1}-c_k)||\le
\sum_{k=m+1}^{n-1} 2^{-k+1}=2^{-m}\sum_{k=0}^{m-n-1} 2^k\le 2^{-m+1}
\ee
which displays $(c_n)$ as a Cauchy sequence in $\a$. Since $\a$ is
complete this sequence converges to some $a\in \a$. But then
\be \label{e.6}
||[a_n]-[a]||=\inf_{b_n\in [a_n],b\in [a]}||b_n-b||\le ||c_n-a||
\ee
so $([a_n])$ converges to $[a]$. It follows that $\a/{\cal I}$ is
complete, that is, a Banach space with unit $[1]$.\\
$[ii)]$ {\it For an Abelean, unital algebra $\a$ an ideal $\cal I$ is maximal
in $\a$ iff $\a/{\cal I}-{[0]}$ consists of invertible elements only}\\
$\Rightarrow$:\\
Suppose we find $[0]\not=[a]\in\a/{\cal I}$ but that $[a]^{-1}$ does not
exist. This means that $a^{-1}$ does not exist since
$[a]^{-1}=[a^{-1}]$ as follows from $[a][a^{-1}]=[1]$. Consider now
the ideal $\a\cdot a=\{ba;\;b\in \a\}$ (this is a two-sided ideal because
$\a$ is Abelean). Since
${\cal I}\subset\a$ we certainly have ${\cal I}\cdot a\subset \a\cdot a$
and since ${\cal I}\cdot a={\cal I}$ because $\cal I$ is in particular a
right ideal we have ${\cal I}\subset \a\cdot a$. Now $a\in \a\cdot a$
since $1\in\a$ and $a\not\in{\cal I}$ because otherwise $[a]=[0]$
which we excluded. It follows that ${\cal I}$ is a proper subideal
of $\a\cdot a$. Finally, since $a^{-1}\not\in \a$, $\a\cdot a$ cannot
be all of $\a$, for instance $1\not\in \a\cdot a$ (an ideal that contains
$1$ or any invertible element is anyway the whole algebra). It follows that
$\cal I$ is not maximal.\\
$\Leftarrow$:\\
Suppose ${\cal I}$ is not a maximal ideal. Then we find a proper subideal
$\cal J$ of $\a$ of which $\cal I$ is a proper subideal. Since every
non-zero element of $\a/{\cal I}$ is invertible so is every element $[a]$
of ${\cal J}/{\cal I}$. But then ${\cal J}$ contains the invertible element
$a\in \a$ and thus $\cal J$ coincides with $\a$ which is a contradiction.\\
$[iii)]$ {\it A unital Banach algebra $\cal B$ in which every non-zero element
is invertible is isomorphic with $\Cl$}\\
Consider any $b\in{\cal B}$ then we claim that $\sigma(b)\not=\emptyset$.
Suppose that were not the case then $\rho(b)=\Cl$. Let $\phi$ be a continuous
linear functional on $\a$ considered as a vector space with metric.
Using linearity of $\phi$ and the expansion of $r_z(b)$ into an absolutely
geometric series we see that $z\mapsto\phi(r_z(b))$ is an entire analytic
function. Since $\phi$ is linear and continuous, it is bounded with
bound $||\phi||$. Thus $|\phi(r_z(b))|\le ||\phi||\;||r_z(b)||$.
Since $\lim_{z\to\infty} ||r_z(b)||=0$ (use the geometric series) and
$||r_z(a)||$ is everywhere
defined in $\Cl$ we conclude that $z\mapsto\phi(r_z(b))$ is an entire
bounded function which therefore, by Liouville's theorem, is a constant
$c_a=\phi(r_z(b))=\lim_{z\to\infty}\phi(r_z(b))=0$. Since $\phi$ was
arbitrary it follows that $r_z(a)=0$ implying that $b-z\cdot 1$ does not
exist which cannot be the case.

Thus we find $z_b\in\sigma(b)$, that is, $b-z_b\cdot 1$ is not invertible.
By assumption, only zero elements are not invertible, hence $b=z_b\cdot 1$
for some $z_b\in\Cl$ for any $b\in {\cal B}$. The map $b\mapsto z_b$
is then the searched for isomorphism ${\cal B}\to \Cl$. Notice that
$b=0$ iff $z_b=0$.\\

Let then ${\cal I}$ be a maximal ideal in a unital, Abelean Banach algebra
$\a$.
Then by i) ${\cal B}:=\a/{\cal I}$ is a unital Banach algebra and by ii)
each of its non-zero elements is invertible. Thus by iii) it is isomorphic
with $\Cl$.\\
$\Box$\\
\begin{Corollary} \label{col.1}  ~~~~~~~~~~\\
In an Abelean, unital Banach algebra $\a$ there is a one-to-one
correspondence between its spectrum $\Delta(\a)$ and the set $I(\a)$ of
maximal ideals in $\a$ via
\be \label{e.7}
\Delta(\a)\to I(\a);\;\chi\mapsto \mbox{ker}(\chi)
\ee
\end{Corollary}
Proof of Corollary \ref{col.1}:\\
That each character gives rise to a maximal ideal in $\a$ through its kernel
was already shown after definition \ref{defe.3}. Conversely, let $\cal I$
be a maximal ideal in a commutative unital
Banach algebra then we can apply theorem \ref{the.1} and
obtain a Banach algebra isomorphism $\chi:\a/{\cal I}\to \Cl;\;[a]\to
\chi([a])$. We can extend this to a homomorphism $\chi:\;\a\to \Cl$ by
$\chi(a):=\chi([a])$. By construction $\chi(a)=0$ iff $[a]=[0]$, that
is, iff $a\in {\cal I}$. In other words, the maximal ideal $\cal I$ is
the kernel of the character $\chi$.\\
$\Box$\\
The subsequent lemma explains the word ``spectrum".
\begin{Lemma} \label{lae.2}  ~~~~~~\\
Let $\a$ be a unital, commutative Banach algebra and $a\in \a$.
Then $z\in\sigma(a)$ iff there exists $\chi\in\Delta(\a)$ such that
$\chi(a)=z$.
\end{Lemma}
Proof of Lemma \ref{lae.2}:\\
The requirement $\chi(a)=z$ is equivalent with $\chi(a-z\cdot 1)=0$ so that
$a-z\cdot 1\in\mbox{ker}(\chi)$. Since $\mbox{ker}(\chi)$ is a maximal
ideal in $\a$ it cannot contain invertible elements, thus
$(a-z \cdot 1)^{-1}$ does not exist, hence $z\in \sigma(a)$.\\
$\Box$\\
We now equip the spectrum with a topology. We begin by showing
that the characters are in particular continuous linear functionals
on the topological vector space $\a$.
\begin{Definition} \label{defe.5}   ~~~~~~~~~\\
For a character $\chi$ in an Abelean, unital Banach algebra we define
its norm by
\be \label{e.8}
||\chi||:=\sup_{a\in \a} |\chi(a)|
\ee
\end{Definition}
\begin{Lemma} \label{lae.3}  ~~~~~~~~~~~~\\
The characters of an Abelean, unital Banach algebra form a subset
of the unit sphere
in $\a'$, the continuous linear functionals on $\a$ considered
as a topological vector space.
\end{Lemma} %
Proof of Theorem \ref{lae.3}:\\
By lemma \ref{lae.2} we showed that
$\sigma(a)=\{\chi(a);\;\chi\in\Delta(\a)\}$. It follows that
\be \label{e.9}
||\chi||=\sup_{a\in \a}\frac{|\chi(a)|}{||a||}
\le \sup_{a\in \a}\frac{\sup\{|\chi'(a)|;\;\chi'\in\Delta(\a)\}}{||a||}
=\sup_{a\in \a} \frac{\rho(a)}{||a||}\le 1
\ee
since by lemma \ref{lae.0} we have
$r(a)=\lim_{n\to\infty} ||a^n||^{1/n}\le ||a||$.
On the other hand $\chi(1)=1$, hence $||\chi||=1$ for every character
$\chi$. This shows that every character is a bounded linear functional
on $\a$, that is, $\Delta(\a)\subset\a'$.\\
$\Box$\\
Since we just showed that the characters are in particular bounded linear
functionals it is natural to equip the spectrum with the weak $^\ast$
topology of pointwise convergence induced from $\a'$.
\begin{Definition}  \label{defe.6}  ~~~~~~~~~\\
i)\\
The weak $^\ast$ topology on the topological dual $X'$ of a topological
vector space $X$ (the set of continuous (bounded) linear functionals)
is defined by pointwise convergence, that is, a net $(\phi^\alpha)$
in $X'$ converges to $\phi$ iff for any $x\in X$ the net of
complex numbers $(\phi^\alpha(x))$ converges to $\phi(x))$. Equivalently,
it is the weakest topology such that all the functions $x:\;X'\to \Cl;
\phi\to \phi(x)$ are continuous.\\
ii)\\
The Gel'fand topology on the spectrum of a unital, Abelean Banach algebra
is the weak $^\ast$ topology induced from $\a'$ on its subset
$\Delta(\a)$.\\
\end{Definition}
We now show that in the Gel'fand topology the spectrum becomes a compact
Hausdorff space. We need a preparational lemma.
\begin{Lemma} \label{lae.4}  ~~~\\
Let $X$ be a Banach space and $X'$ its topological dual. Then the
unit ball in $X'$ is closed and compact in the weak $^\ast$ topology.
\end{Lemma}
Proof of Lemma \ref{lae.4}:\\
The unit ball $B$ in $X'$ is defined as the subset of elements $\phi$
with norm smaller than or equal to unity, that is,
$||\phi||:=\sup_{x\in X}|\phi(x)|/||x||\le 1$.
By corollary \ref{colb.1} we must show that every universal net in $B$
converges. Let $\phi^\alpha$ be a universal net in $B$ and consider
for any given $x\in X$ the net of complex numbers $(\phi^\alpha(x))$
which are bounded by $||x||$. Our $x\in X$ defines a linear form
$X'\to\Cl;\; \phi\to \phi(x)$ whence by lemma \ref{lab.2}ii)
the net $(\phi^\alpha(x))$ is universal. It is contained in the
set $\{z\in\Cl;\;|z|\le ||x||\}$ which is compact in $\Cl$ and therefore
it converges. Define $\phi$ pointwise by the limit, that is,
$\phi(x):=\lim_\alpha \phi^\alpha(x)$. Then
\be \label{e.10}
||\phi||=\sup_{x\in X}\lim_\alpha |\phi^\alpha(x)|/||x||\le
\lim_\alpha ||\phi^\alpha||\le 1
\ee
Thus $\phi^\alpha$ converges pointwise to $\phi\in B$. In particular we have
shown that $B$ is closed.\\
$\Box$\\
\begin{Theorem} \label{the.2}  ~~~~~~~~~~\\
In the Gel'fand topology, the spectrum $\Delta(\a)$ of a unital, Abelean
Banach algebra is compact.
\end{Theorem}
Proof of Theorem \ref{the.2}:\\
Since we have shown 1) in lemma \ref{lae.3} that $\Delta(\a)$ is a subset
of the unit ball $B$ in $\a'$, 2) in lemma \ref{lae.4} that $B$ is
compact in the weak $^\ast$ topology and 3) in lemma \ref{lab.3} that closed
subspaces of compact spaces are are compact in the subspace topology it
will be sufficient to show that $\Delta(\a)$ is a closed in $B$
as the Gel'fand topology is the subspace topology induced from $B$.

Let then $(\chi^\alpha)$ be a net in $\delta(\a)$ converging to
$\chi\in B$. We have, e.g.,
$\chi(ab)=\lim_\alpha \chi^\alpha(ab)=\lim_\alpha \chi^\alpha(a)
\chi^\alpha(b)=\chi(a)\chi(b)$ and similar for pointwise addition, scalar
multiplication and involution in $\a$. It follows that $\chi$ is a
character, that is, $\chi\in \Delta(\a)$.\\
$\Box$\\
\begin{Definition}  \label{defe.7}  ~~~~~~~~\\
The Gel'fand transform is defined by
\be \label{e.11}
\bigvee:\;\a\to\Delta(\a)';\;a\mapsto\check{a} \mbox{ where }
\check{a}(\chi):=\chi(a)
\ee
Here $\Delta(\a)'$ denotes the continuous linear functionals on
$\Delta(\a)$ considered as a topological vector space.
\end{Definition}
It is clear that every $\check{a},\;a\in\a$ is a continuous linear functional
on the spectrum since for any net $(\chi^\alpha)$ in $\Delta(\a)$
which converges to $\chi$ we have $\lim_\alpha \check{a}(\chi^\alpha)
=\lim_\alpha \chi^\alpha(a)=\chi(a)=\check{a}(\chi)$ because convergence
of $(\chi^\alpha)$ means pointwise convergence on $\a$.
\begin{Theorem} \label{the.3}  ~~~~~~~~~~~~~\\
The Gel'fand transform extends to a homomorphism
\be \label{e.12}
\bigvee:\;\a\to C(\Delta(\a));\;a\to\check{a}
\ee
with the following additional properties:\\
1) range$(\check{a})=\sigma(a)$.\\
2) $||\check{a}||:=\sup_{\chi\in\Delta(\a)}|\check{a}(\chi)|=r(a)$.\\
3) The image $\bigvee(\a)$ separates the points of $\Delta(\a)$.
\end{Theorem}
Proof of Theorem \ref{the.3}:\\
0)\\
Morphism and Continuity:\\
We have for example
\be \label{e.13}
(ab)^{\bigvee}(\chi)=\chi(ab)=\chi(a)\chi(b)=
\check{a}(\chi)\check{b}(\chi)
\ee
for any $\chi\in \Delta(\a)$ and similar for $(a+b)^{\bigvee}$. Thus
multiplication and addition of functions are defined pointwise. That the
functions $\check{a}$ are continuous follows as after definition
\ref{defe.7} from the fact that the weak $^\ast$ topology on $\Delta(\a)$
{\it is defined} by asking that all the Gel'fand transforms $\check{a}$
be continuous and therefore is tautologous.\\
1)\\
We have
\be \label{e.14}
\mbox{range}(\check{a})=\{\check{a}(\chi);\;\chi\in\Delta(\a)\}
=\{\chi(a);\;\chi\in\Delta(\a)\}=\sigma(a)
\ee
as follows from lemma \ref{lae.2}.\\
2)\\
We have
\be \label{e.15}
||\check{a}||=\sup_{\chi\in\Delta(\a)} |\check{a}(\chi)|
=\sup_{\chi\in\Delta(\a)} |\chi(a)|
=\sup(\{|\chi(a)|;\;\chi\in\Delta(\a)\})=r(a)
\ee
by definition of the spectral radius. Notice that the sup-norm is a natural
norm on a space of continuous functions on a compact space.\\
3)\\
Recall that a collection of functions $\cal C$ on a topological space $X$
is said to separate its points iff for any $x_1\not= x_2$ we find
$f\in{\cal C}$ such that $f(x_1)\not=f(x_2)$. Consider then any
$\chi_1,\chi_2\in\Delta(\a)$ with $\chi_1\not=\chi_2$. By definition
of $\Delta(\a)$ there exists then $a\in\a$ such that
$\chi_1(a)=\check{a}(\chi_1)\not=\chi_2(a)=\check{a}(\chi_2)$.\\
$\Box$\\
To see that then $\Delta(\a)$ is a Hausdorff space recall the following
lemma.
\begin{Lemma} \label{lae.5}
Let $X$ be a topological space and ${\cal C}\subset C(X)$ a collection of
continuous functions on $X$ which separate the points of $X$. Then
the topology on $X$ is Hausdorff.
\end{Lemma}
Proof of Lemma \ref{lae.5}:\\
Let $x_1,x_2\in X$ with $x_1\not= x_2$ be any two distinct points. Since
$\cal C$ separates the points we find $f\in{\cal C}$ with
$f(x_1)\not=f(x_2)$. Let $d:=|f(x_2)-f(x_1)|$. Since $f$ is continuous
at $x_I$, for any $\epsilon>0$ we find a neighbourhood $U_I(\epsilon)$ of
$x_I,\; I=1,2$ such that $|f(x)-f(x_I)|<\epsilon$ for any $x\in
U_I(\epsilon)$. Now $d=|f(x_2)-f(x_1)|\le |f(x)-f(x_1)|+|f(x_2)-f(x)|$
for any $x\in X$.
Thus $d-\epsilon<|f(x_2)-f(x)|$ for any $x\in U_1(\epsilon)$ and
$d-\epsilon<|f(x_1)-f(x)|$ for any $x\in U_2(\epsilon)$.
Choose $\epsilon<d/2$. Then $U_1(\epsilon)\cap U_2(\epsilon)=\emptyset$.\\
$\Box$\\
\begin{Corollary} \label{cole.1}  ~~~~~~~~~~~\\
The Gel'fand topology on the spectrum of a unital, Abelean Banach algebra
is Hausdorff.
\end{Corollary}
Proof of Corollary \ref{cole.1}:\\
The proof follows trivially from the fact that by theorem \ref{the.3}
${\cal C}:=\{\check{a};\;a\in \a\}$ is a system of continuous functions
separating the points of $\Delta(\a)$ together with lemma \ref{lae.5}.\\
$\Box$\\
So far everything worked for an Abelean, unital Banach algebra $\a$.
We now invoke the further restriction that $\a$ be an Abelean, unital
$C^\ast$ algebra which makes the Gel'fand transform especially nice.
\begin{Theorem} \label{the.4} ~~~~~~~~\\
Let $\a$ be a unital, commutative $C^\ast-$algebra (not only a Banach
algebra). Then the Gel'fand
transform is an isometric isomorphism between $\a$ and the space of
continuous functions on its spectrum.
\end{Theorem}
Proof of Theorem \ref{the.4}:\\
First of all, using the fact that in a commutative $^\ast$ algebra every
element is normal (meaning that $[a,^\ast]=0$) we have, making
frequent use of the $C^\ast$ property (\ref{e.1})
\ba \label{e.16}
||a^{2^n}||^2 &=& ||a^{2^n} (a^{2^n})^\ast||=||(a a^\ast)^{2^n}||
\nonumber\\
&=&||(a a^\ast)^{2^{n-1}}((a a^\ast)^{2^{n-1}})^\ast||
=||(a a^\ast)^{2^{n-1}}||^2
\nonumber\\
&=& ||a a^\ast||^{2^n}=||a||^{2^{n+1}}
\ea
where in the third equality we exploited that $a a^\ast$ is self-adjoint
an in the fifth equality we iterated the equality between the expressions
at the end of the first and second line. We conclude that for any
natural number $n$
\be \label{e.17}
||a||=||a^{2^n}||^{1/2^n}
\ee
In lemma \ref{lae.0} we proved the formula $r(a)=\lim_{n\to\infty}
||a^n||^{1/n}$ meaning that every subsequence of the sequence
$(||a^n||^{1/n})$ has the same limit $r(a)$ including the one displayed
in (\ref{e.17}). Thus we have shown that for Abelian $C^\ast-$algebras
indeed
\be \label{e.18}
r(a)=||a||
\ee
and not only $r(a)\le ||a||$. By theorem \ref{the.3}2) we have
therefore
\be \label{e.19}
||\check{a}||=||a||
\ee
that is, isometry.

Consider now the system of complex valued functions on the spectrum given by
${\cal C}:=\{\check{a};\;a\in\a\}$. We claim that it has the following
properties:\\
i) ${\cal C}\subset C(\Delta(\a))$\\
ii) $\cal C$ separates the points of $\Delta(\a)$\\
iii) $\cal C$ is a closed (in the sup-norm topology) $^\ast$ subalgebra of
$C(\Delta(\a))$\\
iv) The constant functions belong to $\cal C$.\\
Property i), ii) are the assertions 0) and 3) of theorem \ref{the.3} while
iv) follows from the fact that $\a$ is unital, i.e.
$\check{1}(\chi)=\chi(1)=1$ so $\check{1}=1$. To show that iii) $\cal C$ is a
closed $^\ast$ algebra in $C(\Delta(\a))$ suppose that $(\check{a}^\alpha)$
is a net
in $\cal C$ converging to some $f\in C(\Delta(\a))$. Thus, $(\check{a}^\alpha)$
is in particular a Cauchy sequence, meaning that
$||\check{a}^\alpha-\check{a}^\beta||=||a^\alpha-a^\beta||$ becomes
arbitrarily small as $\alpha\,\beta$ grow, where we have used isometry.
It follows that $(a^\alpha)$ is a Cauchy seqence and therefore converges to
some $a\in \a$ since $\a$ is in particular a Banach algebra and therefore
complete. Therefore $f=\check{a}\in {\cal C}$, whence $\cal C$ is closed.
Clearly $\cal C$ is also a $^\ast$ subalgebra because $\a$ is an algebra
and $\bigvee$ a homomorphism.

Now reacll from theorem \ref{the.2} and corollary \ref{cole.1} that
$\Delta(\a)$ is a compact Hausdorff space. Then properties i),ii),iii)
of $\cal C$ enable us to apply the Stone-Weierstrass theorem (e.g. \cite{47n11})
which tells us that either ${\cal C}=C(\Delta(\a))$ or that there exists
$\chi_0\in\Delta(\a)$ such that $\check{a}(\chi_0)=0$ for all
$\check{a}\in {\cal C}$. By properety iv) the latter possibility is
excluded whence ${\cal C}=\bigvee(\a)$ is {\it all of} $C(\Delta(\a))$.
In other words, the Gel'fand transform is a surjection. Finally
it is an injection since $\check{a}=\check{a}'$ implies
$||\check{a}-\check{a}'||=||a-a'||=0$ by isometry, hence $a=a'$.\\
$\Box$\\
\begin{Corollary}  \label{cole.2}  ~~~~~~~~~~\\
Every compact Hausdorff space $X$ arises as the spectrum of an Abelean,
unital $C^\ast-$ algebra $\a$, specifically $\a=C(X),\;\Delta(\a)=X$.
\end{Corollary}
Proof of Corollary \ref{cole.2}:\\
Let $X$ be a compact Hausdorff space and define $\a:=C(X)$ equipped with
the sup-norm. Then
$X\subset\Delta(C(X))$ by the defintion $x(f):=f(x)=:\check{f}(x)$
for any $f\in \a$ so the Gel'fand transform is the identity map on $X$.
Thus, if $\Delta(C(X))-X\not=\emptyset$ then $\check{f}$ extends $f$
continuously to $\Delta(C(X))$.

Next let $(x^\alpha)$ be a net in $X$ which converges in $\Delta(C(X))$
then $\check{f}(x^\alpha)$ converges in $\Cl$ for any
$\check{f}\in C(\Delta(C(X)))$, i.e., $f(x^\alpha)$ converges
in $\Cl$ for any $f\in C(X)$. It follows that $(x^\alpha)$ converges in $X$,
that is, $X$ is closed in $\Delta(C(X))$.

Suppose now that $\Delta(C(X))-X\not=\emptyset$.
Thus we find $\chi_0\in \Delta(C(X))-X$. Now in a Hausdorff space
the one point sets are closed \cite{51}.
Therefore the sets $X,\{\chi_0\}$ are disjoint closed sets in the
compact Hausdorff space $\Delta(C(X))$. Since compact Hausdorff spaces
are normal spaces \cite{47n11} (i.e. one point sets are closed and
any two disjoint closed sets are contained in open disjoint sets) we may
apply Urysohns's lemma \cite{47n11} to conclude that there is a
continuous function $F:\;\Delta(C(X))\to \Rl$ with range in $[0,1]$
such that $F_{|X}=0$ and $F{|\{\chi_0\}}=F(\chi_0)=1$.

Consider then any $f\in C(X)$. Since $C(\Delta(C(X)))$ are {\it all}
continuous functions on $\Delta(C(X))$, there exist different continuous
extensions of $f$ to $\Delta(C(X))$, for instance the functions
$\check{f},\check{f}+F$ where $F$ is of the form just constructed.
However, this contradicts the fact that $\bigvee$ is an isomorphism
since it would not be surjective.\\
$\Box$\\
Corollary \ref{cole.2} tells us that a compact Hausdorff space can be
reconstructed from its Abelean, unital $C^\ast-$algebra of continuous
functions by constructing its spectrum. This is the starting point for
generalizations to non-commutative topological spaces \cite{8b}.

\newpage

\section{Tools from Measure Theory}
\label{sf}

For an introduction to general measure theory see e.g. the beautiful textbook
\cite{57} . For more advanced topics concerning the extension theory
of measures from self-consistent families of projections to
$\sigma-$additive ones, see e.g. \cite{25}.\\

Recall the notion of a topology and of continuous functions from
section \ref{sb}.
\begin{Definition} \label{deff.1}  ~~~~~~~~~\\
i)\\
Let $X$ be a set. Then a collection of subsets $\cal U$ of $X$ is called
a $\sigma-$algebra provided that \\
1) $X\in {\cal U}$,\\
2) $U\in {\cal U}$ implies $X-U\in {\cal U}$ and \\
3) $\cal U$ is closed under countabe unions, that is,
if $U_n\in {\cal U},\; n=1,2,..$ then also
$\cup_{n=1}^\infty U_n\in {\cal U}$.\\
The sets $U\in {\cal U}$ are called measurable and a space $X$ equipped with
a $\sigma-$algebra a measurable space.\\
ii)\\
Let $X$ be a measurable space and let $Y$ be a topological space. A function
$f:\;X\to Y$ is said to be measurable provided that the preimage
$f^{-1}(V)\subset X$ of any open set $V\subset Y$ is a measurable subset
in $X$.\\
iii)\\
Let $X$ be a topological space. The smallest $\sigma-$algebra on $X$ that
contains all open (and due to 2) therefore all closed) sets of $X$ is called
the Borel $\sigma-$algebra of $X$. The elements of the Borel $\sigma-$algebra
are called Borel sets.
\end{Definition}
Given a collection $\cal U$ of subsets of $X$ which is not yet a topology
($\sigma-$algebra) the weakest topology (smallest $\sigma-$algebra)
containing $\cal U$ is obtained by adding to the collection the
sets $X,\emptyset$ as well as arbitrary unions plus finite intersections
(countable unions and intersections).
Notice the similarity between a collection of sets $\cal U$
that qualify for a $\sigma-$algebra and a topology: In both cases the sets
$X,\emptyset$ belong to $\cal U$ but while open sets are closed under
arbitrary unions and finite intersections, measurable sets are closed
under countable unions and intersections. Note also that if
$X,Y$ are topological spaces and $f:\;X\to Y$ is continuous then $f$
is automatically measurable if $X$ is equipped with the Borel
$\sigma-$algebra.
\begin{Definition}  \label{deff.2}  ~~~~~~~~~\\
A complex measure $\mu$ on a measurable space $(X,{\cal U})$ is a function
$\mu:\; {\cal U}\to \Cl-{\infty};\;U\mapsto \mu(U)$ which is countably
(or $\sigma-$)additive, that is,
\be \label{f.1}
\mu(\bigcup_{n=1}^\infty U_n)= \sum_{n=1}^\infty \mu(U_n)
\ee
for any mutually disjoint measurable sets $U_n$. A positive measure
is also a $\sigma-$additive map $\mu:\; {\cal U}\to \Rl^+\cup{0,\infty}$
which however is postive semidefinite and may take the value $\infty$
with the convention $0\cdot \infty=0$ (which makes $[0,\infty]$ a set
in which commutative, distributive and associative law hold).
To avoid trivialities we assume that $\mu(U)<\infty$ for at least one
measurable set $U$. A measure is called a probability measure if
$\mu(X)=1$. The triple $(X,{\cal U},\mu)$ is called a measure space.
\end{Definition}
In what follows we will always assume that $\mu$ is a positive measure.\\
A very powerful tool in measure theory are characteristic functions of
subsets of $X$.
\begin{Definition} \label{deff.3}  ~~~~~~~~~~\\
A function $s:\;X\to \Cl$ in a measurable space $(X,{\cal U})$ is called
simple provided its range consists of finitely many points only.
If $z_k\in\Cl,\;k=1,..,N$ are these values and $S_k=s^{-1}(\{z_k\})$
then $s=\sum_{k=1}^N z_k \chi_{S_k}$ where $\chi_S$ with ($\chi_S(x)=1$ if
$x\in S$ and $\chi_S(x)=0$ otherwise) is called the characteristic
function of the subset $S\subset X$. Obviously, a simple function
is measurable if and only if the $S_k$ are measurable.
\end{Definition}
The justification for this definition lies in the following lemma.
\begin{Lemma} \label{laf.1}  ~~~~~~~~~~~\\
Let $f:X\to [0,\infty]$ be measurable. Then there exists a sequence
of measurable simple functions $s_n$ such that\\
a) $0\le s_1\le s_2\le...\le f$\\
b) $\lim_{n=1} s_n(x)=f(x)$ pointwise in $x\in X$.
\end{Lemma}
The proof can be found in \cite{57}, theorem 1.17.
%
\begin{Definition}  \label{deff.3a}
i)\\
For a simple measurable function $s=\sum_{k=1}^N z_k \chi_{S_k}$
with $z_k>0$ on a
measure space $(X,{\cal U},\mu)$ with positive measure $\mu$ we define
\be \label{f.2}
\mu(s):=\int_X d\mu(x) s(x):=\sum_{k=1}^N z_k \mu(S_k)
\ee
For a general measurable function $f:\;X\to [0,\infty]$ we define
\be \label{f.3}
\mu(f):=\sup_{0\le s\le f} \mu(s)
\ee
where the supremum is taken over the simple, positive measurable
functions that are nowhere larger than $f$.
The number $\mu(f)$ is called the Lebesgue integral of $f$.
For a general complex
valued, measurable function $f$ one can show that
we have a unique split as
$f=u+iv,\;u=u_+ - u_-,\;v=v_+ - v_-$ with non-negative measurable
functions $u_\pm,v_\pm$ and the integral is defined as
$\mu(f)=\mu(u_+)-\mu(u_-)+i[\mu(u_+)-\mu(u_-)]$. Also $|f|$ can be shown
to be measurable.\\
ii)\\
A measure $\mu$ is called positive definite if for every non-negative
measurable
function $f$ the condition $\mu(f)=0$ implies $f=0$ almost everywhere
(a.e., i.e. up to measure zero sets).
\end{Definition}
Of fundamental importance are conditions under which one is allowed to
echange integration and taking limits.
\begin{Theorem} \label{thf.1}  ~~~~~~~~~\\
Let $(X,{\cal U},\mu)$ be a measure space with positive measure $\mu$ and
let $(f_n)$ be a sequence of measurable functions that converges pointwise
on $X$ to the function $f$.\\
i) {\it Lebesgue Monotone Convergence Theorem}\\
Suppose that $0\le f_n(x)\le f_{n+1}(x)$ for all $x\in X$. Then $f$ is
measurable and $\lim_{n\to\infty} \mu(f_n)=\mu(f)$.\\
ii) {\it Lebesgue Dominated Convergence Theorem}\\
A function $F$ is said to be in $L_1(X,d\mu)$ if it is measurable and
$\mu(|F|)<\infty$. Suppose now that there exists $F\in L_1(X,s\mu)$
such that $|f_n(x)|\le |F(x)|$ for all $x\in X$. Then $f\in L_1(X,d\mu)$
and $\lim_{n\to\infty}\mu(|f-f_n|)=0$.
\end{Theorem}
It is easy to see that $\lim_{n\to\infty}\mu(|f-f_n|)=0$ implies
$\lim_{n\to\infty} \mu(f_n)=\mu(f)$.\\
Another convenient observation is the following.
\begin{Theorem} \label{thf.1a}  ~~~~~~~~~~\\
Let $(X,{\cal U},\mu)$ be a measure space. Let ${\cal U}'$ be the
collection of all $S\subset X$ such that there exist $U,V\in {\cal U}$
with $U\subset S\subset V$ and $\mu(V-U)=0$ (in particular
${\cal U}\subset {\cal U}'$). Define $\mu'(S)=\mu(U)$ in that case.
Then $(X,{\cal U}',\mu')$ is a measure space again, called the
completion of $(X,{\cal U},\mu)$.
\end{Theorem}
The theorem says that any measure can be completed. It means that if
we have a set which is not measurable but which can be sandwiched
between measurable sets whose difference has zero measure, then we can add
the set to the measurable sets and its measure is given by that of the
sandwiching sets.
\begin{Definition} \label{deff.3b}  ~~~~~~~~~~~~~~~\\
i)\\
A set $Y\subset X$ in a measure space $(X,{\cal U},\mu)$ is called thick
or a support for $\mu$ provided that for any measurable set $U\in {\cal U}$
the condition $U\cap Y=\emptyset$ implies $\mu(U)=0$. A support for
$\mu$ will be denoted by supp$(\mu)$.\\
ii) \\
For two measures $\mu_1,\mu_2$ on the same measurable space we say
that $\mu_1$ is regular (or absolutely continuous) with respect to
$\mu_2$ iff $\mu_2(U)=0$ for $U\in {\cal U}$ implies $\mu_1(U)=0$.
They are called mutually singular iff
$\mbox{supp}(\mu_1)\cap\mbox{supp}(\mu_2)=\emptyset$.
\end{Definition}
If $Y$ is a measurable support then $X-Y$ is measurable and since
$Y\cap (X-Y)=\emptyset$ we have $\mu(X-Y)=0$ explaining the word support.
If $Y$ is a support not measurable with respect to $\mu$ one can define
${\cal  U}'=[{\cal U}\cap Y]\cup Y,\;\mu'(U\cap Y)=\mu(U)$ and gets a measure
space $(Y,{\cal U}',\mu')$ for which $Y$ is measurable, called the trace. A
given support
does not mean that there are not smaller sets which are still thick.
If $\mu_2$ is a positive $\sigma-$finite (see below) measure and $\mu_1$
is a complex measure, then one can show (the Radon-Nikodym theorem) that
there is a unique (so-called Lebesgue) decomposition $\mu_1=\mu_1^a+\mu_1^s$
such
that $\mu_1^a,\mu_2^s$ are repectively absolutely continuous and singular
with respect to $\mu_2$ and that there exists $f\in L_1(X,d\mu_2)$,
called the Radon-Nikodym derivative, such that $d\mu_1^a=f\;d\mu_2$.

The following two definitions prepares to state the Riesz representation
(or Riesz -- Markov) theorem which will be of fundamental importance
for our applications.
\begin{Definition} \label{deff.4}  ~~~~~~~~~~\\
i)
A topological space is said to be locally compact if every point $x\in X$
has an open neighbourhood whose closure is compact.\\
ii)\\
A subset $S\subset X$ of a topological space $X$ is said to be
$\sigma-$compact if it is a countable union of compact sets.\\
iii)\\
A subset $S\subset X$ in a measure space $(X,{\cal U},\mu)$ with positive
measure $\mu$ is said to be $\sigma-$finite if $S$ is the countable union
of measurable sets $U_n$ with $\mu(U_n)<\infty$ for all $n\in\Nl$.
\end{Definition}
\begin{Definition} \label{deff.5} ~~~~~~~~~~~~\\
Let $X$ be a locally compact Hausdorff space and let $\cal U$ be its
naturally defined Borel $\sigma-$algebra.\\
i)\\
A measure $\mu$ defined on the Borel $\sigma-$algebra is called a Borel
measure.\\
ii)\\
A Borel set $S$ is said to be outer regular with respect to a positive Borel
measure $\mu$ provided that
\be \label{f.4}
\mu(S)=\inf\{\mu(O);\;S\subset O;\;O\in {\cal U}\mbox{ open}\}
\ee
iii)\\
A Borel set $S$ is said to be inner regular with respect to a positive Borel
measure $\mu$ provided that
\be \label{f.5}
\mu(S)=\sup\{\mu(K);\;S\supset K;\;K\in {\cal U}\mbox{ compact}\}
\ee
iv)
If $\mu$ is a positive Borel measure and every Borel set is both inner
and outer regular then $\mu$ is called regular.
\end{Definition}
\begin{Definition} \label{deff.6}  ~~~~~~~~~~~~\\
i)\\
Let X be a topological space. The support supp$(f)$ of a function $f:\; X\to
\Cl$ is the closure of the set $\{x\in;\;f(x)\not=0\}$. The vector
space of continuous functions of compact support is denoted by $C_0(X)$.\\
ii)\\
A linear functional $\Lambda:\;{\cal F}\to\Cl$ on the vector space
of functions $\cal F$ over a set $X$ is called positive if
$\Lambda(f)\in [0,\infty)$ for any $f\in {\cal F}$ such that $f(x)\in
[0,\infty)$ for all $x\in X$.
\end{Definition}
\begin{Theorem}[Riesz Representation Theorem]  \label{thf.2}  ~~~~~~\\
i)\\
Let $X$ be a locally compact Hausdorff space and let
$\Lambda:\;C_0(X)\to \Cl$ be a positive linear functional one the space
of continuous, complex-valued functions of compact support in $X$.
Then there exists a $\sigma-$algebra $\cal U$ on $X$ which contains the
Borel $\sigma-$algebra and a unique positive measure $\mu$ on $\cal U$
such that $\Lambda$ is represented by $\mu$, that is,
\be \label{f.6}
\Lambda(f)=\int_X d\mu(x) f(x)\;\;\;\forall\; f\in C_0(X)
\ee
Moreover, $\mu$ has the following properties:\\
1) $\mu(K)<\infty$ if $K\subset X$ is compact. \\
2) For every $S\in {\cal U}$ property (\ref{f.4}) holds.\\
3) For every open $S\in {\cal U}$ with $\mu(S)<\infty$
property (\ref{f.5}) holds.\\
4) If $S'\subset S\in {\cal U}$ and $\mu(S)=0$ then $S'\in {\cal U}$.\\
ii)\\
If, in addition to i), $X$ is $\sigma-$compact then $\mu$ has the
following additional properties:\\
5) $\mu$ is regular\\
6) For any $S\in {\cal U}$ and any $\epsilon>0$ there exists a closed set
$C$ and an open set $O$ such that $C\subset S\subset O$ and
$\mu(O-C)<\epsilon$.\\
7) For any $S\in {\cal U}$ there exist sets $C'$ and $O'$ which are
respectively countable
unions and intersections of closed and open sets respectively such that
$C'\subset S\subset O'$ and $\mu(O'-C')=0$.
\end{Theorem}
A very instructive proof of this theorem can be found in \cite{57}.
It is also worth pointing out the following theorem (see e.g. \cite{57})
which underlines the prominent role that continuous functions play for
Borel measures.
\begin{Theorem}[Lusin's Theorem] \label{thf.3}  ~~~~~~~~\\
Let $X$ be a locally compact Hausdorff
space $X$ with $\sigma-$algebra $\cal U$ and measure $\mu$ satisfying the
properties 1), 2), 3)
and 4) of theorem \ref{thf.2}. Let $f$ be a bounded measurable function with
support in a measurable set of finite measure. Then there exists a sequence
$(f_n)$ of {\it continous} functions of compact support, each of
which is bounded by the
same bound, such that $f(x)=\lim_{n\to \infty} f_n(x)$ almost everywhere
with respect to $\mu$ (i.e. they coincide pointwise up to sets of measure
zero).
\end{Theorem}
Let us also define the notion of faithfulness of measures:
\begin{Definition}  \label{deff.7}  ~~~~~~~~~~~\\
Let $X$ be a locally compact Hausdorff space and let ${\cal U},\mu$
have the properties of theorem \ref{thf.2}. Then $\mu$ is called faithful
if and only if the positive linear functional (\ref{f.6}) determined
by $\mu$ is positive definite, that is, if $f\in C_0(X)$ takes only
values in $[0,\infty)$ and $\Lambda(f)=0$ then $f=0$.
\end{Definition}
Notice that positive definiteness of a measure $\mu$ only allows us to
conclude that $f=0$ $\mu-$a.e. from $\mu(f)=0$ for positive measurable $f$.
Faithfulness
of the special kind of measures that come from positive definite linear
functionals alow us to conclude $f=0$ everywhere if $f$ is continuous
and of compact support. This means that every open set must have positive
measure for if a continuous function is positive at a point, it will be
bounded away from zero in a whole open neighbourhood of that point.

The application that we have in mind is that $X$ is not only locally
compact but actually compact so that the set $C_0(X)$ coincides with $C(X)$.
Hence, $C(X)$ contains the constant functions and we may w.l.g. assume
that $\Lambda(1)=1$ which is just a convenient choice of normalization.
(If $X$ is compact, so is every closed
subset, hence $X$ is locally compact). It is then trivially $\sigma-$compact
being its own cover by compact sets. Therefore the stronger version
ii) of theorem \ref{thf.2} applies and we see that by property 5) the
measure $\mu$ is regular. Furthermore, property 7) tells us that every
measurable set can be sandwiched between sets $C'\subset O'$ that belong
to the Borel $\sigma-$subalgebra such that $C'-O'$ is of measure zero.
In other words, every measurable set is a Borel set up to a set
of measure zero: Since $O'=S\cup (O'-S)$ we have from $\sigma-$additivity
$\mu(S)=\mu(O')$ since $0=\mu(O'-C')\ge \mu(O'-S)$ due to $O'-S\subset
O'-C'$. Thus effectively the measure $\mu$ in (\ref{f.6}) is a Borel measure
and in that sense we have the following corollary.
\begin{Corollary} \label{colf.1}  ~~~~~~~~\\
Let $X$ be a compact Hausdorff space and let $\Lambda:\;C(X)\to \Cl$
be a positive linear functional on the space of continuous functions
on $X$ with $\Lambda(1)=1$. Then there exists a unique, regular, Borel
probability measure
$\mu$ on the natural Borel $\sigma-$algebra $\cal U$ of $X$ such that
$\mu$ represents $\Lambda$, that is,
\be \label{f.7}
\Lambda(f)=\int_X d\mu(x) f(x)\;\;\;\forall\; f\in C(X)
\ee
\end{Corollary}
Notice that regularity of $\mu$ on a compact Hausdorff space $X$ reduces
to the fact that the measure of every measurable set can be approximated
arbitrarily well by open or compact (and hence closed since in a
Hausdorff space every compact subset is closed, see \cite{51}) sets
respectively. Also, Lusin's theorem simplifies to the statement that
every bounded measurable function can be approximated arbitarily
well be continuous functions with the same bound up to sets of measure
zero.

The notion of faithfulness actually comes from representation theory.
Indeed, the origin of positive linear functionals in physics are usually
states, that is, positive linear functionals $\omega$ on a unital
$C^\ast-$algebra $\a$ (see section \ref{se}), which is not necessarily
Abelean like the $C^\ast-$algebra $C(X)$ for a compact Hausdorff space
$X$, such that $\omega(1)=1$. Here a positive linear functional is a
map $\omega:\;\a\to \Cl;\;a\mapsto\omega(a)$ which satisfies
$\omega(a^\ast a)\ge0$ for any $a\in \a$. Elements $a$ of $\a$ of the form
$b^\ast b$ are called positive, denoted $a\ge 0$ (equivalently, $a\ge 0$ iff
for its spectrum $\sigma(a)\subset \Rl^+$ holds). One writes $a\ge a'$ if
$a-a'\ge 0$ which equips $\a$ with a partial order.
We will see in section \ref{sg} that positive linear
functionals give rise to a representation $\pi$
of the algebra on a Hilbert space via the GNS construction. If the
unital $C^\ast-$algebra is Abelean then we can always think of it as
an algebra of continuous functions on a compact Hausdorff space via
the Gel'fand isomorphism and if the associated measure is faithful, that is,
the
state is positive definite then the representation is faithful (or
non-degenerate), that is, $\pi(f)=0$ if and only if $f=0$.

Notice that every positive linear functional $\omega$ on a unital
$C^\ast$ algebra $\a$ is automatically bounded (continuous):\\
If $||.||$
denotes the norm on $\a$ and $^\ast$ the involution then for any
self-adjoint element $a=a^\ast$ we have $-||a||\cdot 1\le a\le ||a||\cdot 1$
since $||a||\ge r(a)$ (spectral radius). Hence
$\omega(||a||\cdot 1\pm a)=||a||\omega(1)\pm\omega(a)\ge 0$. Since
$\omega(1)\ge 0$ because $1=1^\ast 1$ is positive, it follows that in
particular $\omega(a)\in \Rl$ for self-adjoint $a$ so that
$|\omega(a)|/||a||\le \omega(1)$. If $a$ is arbitrary we can decompose it
uniquely into self-adjoint elements $a=a_+ +ia_-$ with $a_\pm=a_\pm^\ast$
and thus
$$
4||a_\pm^2||=||(a^\ast)^2+a^2\pm(a^\ast a+a a^\ast)||
\le ||(a^\ast)^2||+||a^2||+||a^\ast a||+||a a^\ast||=4||a||^2
$$
where we have made use twice of the $C^\ast-$algebra property
$||a^\ast a||=||a||^2$. It follows that
$$
|\omega(a)|^2=|\omega(a_+)+i\omega(a_-)|^2=
|\omega(a_+)|^2+|\omega(a_-)|^2\le \omega(1)[||a_+||^2+||a_-||^2]
\le 2\omega(1)||a||^2
$$
so a bound is given by $2\omega(1)$. One can actually show that a sharper
bound is given by $\omega(1)$ even for unital Banach algebras with
involution.\\
\\
We now turn to another direction within measure theory.
\begin{Definition} \label{deff.8} ~~~~~~~~~~~\\
Let $(X,{\cal U},\mu)$ be a measure space with a positive probability
measure $\mu$ on $X$. Let $\lambda:\;G\times X\to X;\;(g,x)\mapsto
\lambda_g(x)$ be a measure preserving group action, that is,
$(\lambda_g)_\ast\mu:=\mu\circ\lambda_g^{-1}=\mu$ for all $g\in G$, in
particular, $\lambda_g$ preserves $\cal U$. The group action is called
ergodic if the only invariant sets, that is, sets $S\in {\cal U}$ with
$\lambda_g(S)=S$ for all $g\in G$, have measure zero or one.
\end{Definition}
The definition captures exactly the intuitive idea of an ergodic
group action, namely that it spreads any set all over $X$ without changing
its measure.
It follows from the definition that a measure preserving group action
induces a unitary transformation on $L_2(X,d\mu)$ by the pull-back,
that is,
\be \label{f.8}
(\hat{U}(g)f)(x):=(\lambda_g^\ast f)(x)=f(\lambda_g(x))
\ee
Since the closed linear span of characteristic functions of measurable
sets is all of $L_2(X,d\mu)$ as we have seen above, it follows that
ergodicity is equivalent with the condition that
$\hat{U}(g)f=f$ $\mu-$a.e. for all $g\in G$ implies that
$f=$const. a.e. (Proof: If $\lambda$ is ergodic and
$f=\sum_k z_k \chi_{U_k}$
then $\hat{U}(g)f=\sum_k z_k \chi_{\lambda_{g^{-1}}(U_k)}=f$ a.e. for all
$g\in G$ implies that all $U_k$ must be invariant under $\lambda$, hence
that all of them have measure zero or one. If $U_k$ has measure zero
then $\chi_{U_k}=0$ a.e., if $U_k$ has measure one then $X-U_k$ has measure
zero so $\chi_{U_k}=\chi_X=1$ a.e. The converse implication is
similar).
\begin{Theorem}[von Neumann Mean Ergodic Theorem] \label{thf.4} ~~~~\\
Let $\Rl \to G;\;t\mapsto g_t$ be a one parameter group and
$\hat{U}:\; G\to {\cal B}(L_2(X,d\mu)$ be a unitray representation
of $G$. Let $\hat{P}$ be the projection on the closure of the
set of a.e. invariant vectors under $\hat{U}(g_t),\;t\in \Rl$. Then
\be \label{f.9}
(\hat{P}f)(x)=\lim_{T\to \infty}\frac{1}{2T}\int_{-T}^T dt
(\hat{U}(g_t)f)(x)\;\;\mu-\mbox{a.e.}
\ee
\end{Theorem}
For a proof see for instance \cite{47n11}. We conclude that $\lambda$
restricted to $t\to g_t$
is ergodic if and only if
\be \label{f.10}
\lim_{T\to \infty}\frac{1}{2T}\int_{-T}^T dt f(\lambda_{g_t}(x))
=[\int_X d\mu(x') f(x')]\cdot 1\;\;\mu-\mbox{a.e.}
\ee
Namely, if $t\to \lambda_{g_t}$ is ergodic, then the set of a.e. invariant
vectors is given by the constant functions whence $\hat{P}f\propto 1$,
that is,
\be \label{f.11}
\hat{P}f=<1,\hat{P}f>\cdot 1=<\hat{P} 1,f>1=<1,f>\cdot 1=[\int_X d\mu(x)
f(x)]\cdot 1 \ee
since $1(x)=1$ and the definition of the inner product. Comparing with
$\hat{P} f$ from (\ref{f.9}) gives the claimed result (\ref{f.10}).
Conversely, if (\ref{f.10}) holds then the right hand side is constant
almost everywhere and equals $\hat{P}f$ hence $t\to \lambda_{g_t}$
is ergodic by the above remark.

Criterion (\ref{f.10}) is interesting for the following reason:
Suppose that $\mu_1\not=\mu_2$ are different measures on the same
measurable space $(X,{\cal U})$, and that $t\to \lambda_{g_t}$ is a
measure preserving, ergodic group action with respect to both of them.
Then
\be \label{f.12}
[\int_X d\mu_1(x') f(x')]\cdot 1=_{\mu_1-\mbox{{\tiny a.e.}}}
\lim_{T\to \infty}\frac{1}{2T}\int_{-T}^T dt f(\lambda_{g_t}(x))
=_{\mu_2-\mbox{{\tiny a.e.}}}
[\int_X d\mu_2(x') f(x')]\cdot 1
\ee
for any $f\in L_1(X,d\mu_1)\cap L_1(X,d\mu_2)$. Now the left and right
hand side in (\ref{f.12}) do not depend at all on the point $x$
on which the middle term depends. Thus, if we can find
$f\in L_1(X,d\mu_1)\cap L_1(X,d\mu_2)\not=\emptyset$
such that the constants
$[\int_X d\mu_1(x') f(x')\not=[\int_X d\mu_2(x') f(x')]$ are different
from each other then the middle term must equal the left hand side
whenver $x\in\mbox{supp}(\mu_1)$ and it must equal the right hand side
whenver $x\in\mbox{supp}(\mu_2)$. This is no contradiction iff
$\mu_1,\mu_2$ are mutually singular with respect to each other.
Hence ergodicity gives a simple tool for investigating the singularity
structure of measures with respect to each other and one easily shows
that Gaussian measures with different covariances (e.g. scalar
fields with different masses) are built on mutually singular measures.
\begin{Definition} \label{deff.9} ~~~~~~~~~~~~\\
A one parameter-group of measure preserving transformations
$t\to \lambda_{g_t}$ is called mixing provided that
\be \label{f.13}
\lim_{t\to\infty} <f,\hat{U}(g_t)f'>=<f,1>\;<1,f'>
\ee
\end{Definition}
It is easy to see that mixing implies ergodicity: Suppose that $f'$ is
invariant a.e. under the one-parameter group. Then by (\ref{f.13})
and inserting the identity $1_{L_2}=\hat{P}\oplus[1_{L_2}-\hat{P}]$,
where $\hat{P}=|1><1|$ denotes the projection onto span$(\{1\})$, gives
\ba \label{f.14}
<f,f'>&=&<f,1>\;<1,f'>+<f,[1_{L_2}-\hat{P}]f'>=<f,1>\;<1,f'>
\nonumber\\
&& \Rightarrow\;<f,[1_{L_2}-\hat{P}]f'>=0\;\forall \;f\in L_2(X,d\mu)
\ea
hence $[1_{L_2}-\hat{P}]f'=0$ so that $f'=$const. a.e., that is,
ergodicity.

\newpage

\section{Spectral Theorem and GNS-Construction}
\label{sg}

As an application of appendices \ref{se} and \ref{sf} in addition to the
general theory of the main text we present an
elegant proof of the spectral theorem and sketch the GNS construction
due to Gel'fand, Naimark and Segal.

Let $\cal H$ be a Hilbert space and $a$ a bounded, linear, normal
operator on $\cal H$, that is
$||a||=\sup_{\psi\not=0}||a\psi||/||\psi||<\infty$ where
$||\psi||^2=<\psi,\psi>$ denotes the Hilbert space norm and
$[a,a^\dagger]=0$ where the bounded operator $a^\dagger$
is defined by $<a^\dagger \psi,\psi'>:=<\psi,a\psi'>$.
More precisely, consider the linear form on $\cal H$ defined by
\be \label{g.1}
l_\psi:{\cal H}\to \Cl;\psi'\to <\psi,a\psi'>
\ee
This linear form is continuous since
$|l_\psi(\psi')|\le ||\psi||\;||a||\;||\psi'||$ by the Schwarz inequality.
Hence, by the Riesz lemma there exists $\xi_\psi\in {\cal H}$ such that
$l_\psi=<\xi_\psi,.>$ and since $l_\psi$ is conjugate linear in $\psi$
it follows that $\psi\mapsto \xi_\psi:=a^\dagger\psi$ actually defines
a linear operator. Finally, $a^\dagger$ is bounded because
\be \label{g.2}
||a^\dagger \psi||^2=|<\psi,a a^\dagger\psi>|\le ||\psi||\;
||a a^\dagger\psi||
\le ||\psi||\; ||a||\;||a^\dagger\psi||
\ee
again by the Schwarz inequality.

Let $\a$ be the unital, Abelean $C^\ast-$algebra
generated by $1,a,a^\dagger$. It is Abelean since $a$ is normal and
the $C^\ast-$property follows from the following observation: Let
$b\in \a$, then $b$ is also normal and
$||b\psi||^2=<\psi,b^\dagger b\psi>=||b^\dagger \psi||^2$ so that
$||b||=||b^\dagger||$ for any $b\in\a$. Now by the Schwarz inequality
$||b\psi||^2=|<\psi,b^\dagger b\psi>|\le ||\psi||\;||b^\dagger b\psi||$
implying that $||b||^2=||b^\dagger||^2\le ||b^\dagger b||$. On the other
hand $||b^\dagger b||\le ||b||\;||b^\dagger||$ due to submultiplicativity.

Consider the spectrum $\Delta(\a)=\mbox{Hom}(\a,\Cl)$ and the
map $z:\;\Delta(\a)\to \Cl;\;\chi\mapsto \chi(a)$ which is continuous
by the definition of the Gel'fand topology on the spectrum. We have seen
already
that the range of this map coincides with $\sigma(a)$. Moreover,
$z$ is injective because $\chi(a)=\chi'(a)$ implies that $\chi,\chi'$
coincide on all polynomials of $a,a^\dagger$ since they are homomorphisms,
and thus on allof $\a$ by continuity whence $\chi=\chi'$. Thus, $z$ is a
continuous bijection between the spectra of $\a$ and $a$ respectively. Since
$a$ is bounded, both spectra are compact Hausdorff spaces. Now a continuous
bijection between compact Hausdorff spaces is automatically a
homeomorphism.
(Proof: Let $f:X\to Y$ be a continuous bijection and let $X$ be compact
and $Y$ Hausdorff. We must show that $f(U)$ is open in $Y$ for every
open subset $U\subset X$, or by taking complements, that images of
closed sets are closed. Now since $X$ is compact, it follows that
every closed set $U$ is also compact. Since $f$ is continuous, it follows
that $f(U)$ is compact. Since $Y$ is Hausdorff it follows that $f(U)$
is closed. See theorems 5.3 and 5.5 of \cite{51}). We conclude that
we can identify $\Delta(\a)$ topologically with $\sigma(a)$.
By defintion the polynomials $p$ in $a,a^\dagger$
lie dense in $\a$ and we have for $\chi\in \Delta(\a)$ that
\be \label{g.2a}
\chi(p(a,a^\dagger))=p(\chi(a),\overline{\chi(a)})=
p(z(\chi),\overline{z(\chi)})=[p\circ(z,\bar{z})](\chi)
=p(a,a^\dagger)^\bigvee(\chi)
\ee
so that the Gel'fand isometric isomorphism can be thought of as a map
$\bigvee:\;\a\to C(\sigma(a));\;b\mapsto\check{b}$ with
$\check{b}(z)=\chi(b)_{z=\chi(a)}$.

Now consider any state $\psi\in {\cal H}$ with $||\psi||=1$. Then
\be \label{g.3}
\omega_\psi:\;\a\to \Cl;\;b\mapsto <\psi,b\psi>
\ee
is obviously a state on $\a$. Via the Gel'fand transform we obtain
a positive linear functional on $C(\sigma(a))$ by
\be \label{g.4}
\Lambda_\psi:\;C(\sigma(a))\to \Cl;\;\check{b}\mapsto \omega_\psi(b)
\ee
and since $\sigma(a)$ is a compact Hausdorff space we can apply the
Riesz representation theorem in order to find a unique, regular Borel
measure $\mu_\psi$ on $\sigma(a)$ such that
\be \label{g.4a}
\omega_\psi(b)=\int_{\sigma(a)} d\mu_\psi(z) \check{b}(z)
\ee
The measure $\mu_\psi$ is caled a spectral measure.
The meaning of this formula is explained by the following definition.
\begin{Definition} \label{defg.1} ~~~~~~~~~\\
i)\\
A representation $\pi:\a\to {\cal B}({\cal H})$ of a $C^\ast-$algebra
is a $^\ast-$homomorphism into the $^\ast-$algebra of bounded operators
on a Hilbert space. The representation is said to be faithful if
$a\not=0$ implies $\pi(a)\not=0$. Two representations
$\pi_I;\a\to {\cal B}({\cal H}_I);\;I=1,2$ are called equivalent iff
there exists a Hilbert space isomorphism $U:\;{\cal H}_1\to {\cal H}_2$
such that $\pi_2(a)=U\pi_1(a) U^{-1}$ for all $a\in \a$. Finally,
a representation is called non-degenerate if
$\mbox{ker}(\pi):=\{\psi\in {\cal H};\;\pi(a)\psi=0\;\forall \; a\in \a\}$
is given by $\{0\}$.\\
ii)\\
Let $\omega$ be a state on a unital $C^\ast-algebra$ and define
the null space ${\cal N}_\omega:=\{a\in \a;\;\omega(a^\ast a)=0\}$.
The GNS representation with respect to $\omega$
\be \label{g.5}
\pi_\omega:\;\a\to {\cal B}({\cal H}_\omega) \mbox{ where }
{\cal H}_\omega:=\overline{\a/{\cal N}_\omega}:=\overline{\{[a];\;a\in \a\}}
\ee
where the overbar denotes completion and $[.]:\;\a\to \a/{\cal N}_\omega;\;
a\mapsto [a]:=\{a+b;\;b\in {\cal N}_\omega\}$ is the quotient map,
is densely defined by
\be \label{g.6}
\pi_\omega(a)[b]:=[ab]
\ee
and extended by continuity. The Hilbert space
${\cal H}_\omega$ is equipped with the inner product
\be \label{g.7}
<[a],[b]>_{{\cal H}_\omega}:=\omega(a^\ast b)
\ee
The Hilbert space state $\Omega_\omega:=[1]$ is cyclic for
${\cal H}_\omega$, that is, a dense set of Hilbert space space states
is obtained as $\{\pi_\omega(a)\Omega_\omega;\;a\in \a\}$. Moreover,
\be \label{g.8}
\omega(a)=<\Omega_\omega,\pi_\omega(a)\Omega_\omega>_{{\cal H}_\omega}
\ee
\end{Definition}
To see that this definition makes sense, one notices that
${\cal N}_\omega$ is a closed left ideal in $\a$ so that (\ref{g.6}),
(\ref{g.7})
are well-defined (using that the right hand side of (\ref{g.7}) defines a
positive
semidefinite sesquilinear form on ${\cal H}:=\a$ and hence the Schwarz
inequality applies) and that $||\pi_\omega(a)||=||a||$ is indeed bounded.
One can show that the triple $({\cal H}_\omega,\pi_\omega,\Omega_\omega)$
is fixed by condition (\ref{g.8}) up to unitary equivalence. Notice
that the state is not required to be pure (i.e. cannot be written as a
convex linear combination of other states) but if it is then one can show
that the representation is irreducible (does not contain invariant subspaces
different from itself and $\{0\}$).

It is almost clear that every non-degenerate representation is an orthogonal
sum of cyclic representations: Take an arbitrary element
$0\not=\psi\in{\cal H}$
and construct ${\cal H}_\psi:=\overline{\{\pi(a)\psi;\;a\in \a\}}$.
If ${\cal H}_\psi^\perp\not=\{0\}$ take $\psi'\in {\cal H}_\psi^\perp$ and
iterate. The rigrous proof makes use of the axiom of choice and will be
left to the reader.

Coming back to our concrete $C^\ast-$algebra $\a$ generated by a
normal, bounded operator $a\in {\cal B}((\cal H)$ on a given Hilbert space
$\cal H$ we see that it is represented as $\pi(b)=b$ on $\cal H$ and
that this representation is non-degenerate because $\a$ contains the
identity operator. We then find an index set $A$, vectors
$\psi_\alpha$ and closed, mutually orthogonal subspaces
${\cal H}_\alpha:=\overline{\{b\psi_\alpha;\;b\in \a\}}$ containing
$\psi_\alpha$ such that ${\cal H}=\oplus_{\alpha\in A} {\cal H}_\alpha$.
By construction, the subspaces ${\cal H}_\alpha$ are invariant for
$\a$. Then any vector $\psi\in {\cal H}$ is (in the closure of vectors
of) the form $\psi=\sum_{\alpha\in A} b_\alpha \psi_\alpha$ with
$b_\alpha\in \a$ and we have
\be \label{g.9}
<\psi,\psi'>=\sum_{\alpha\in A}
<\psi_\alpha,b_\alpha^\dagger b'_\alpha\psi_\alpha>
\ee
Using the result (\ref{g.4a}) we may write this as
\be \label{g.10}
<\psi,\psi'>=\sum_{\alpha\in A} \int_{\sigma(a)} d\mu_{\psi_\alpha}(z)
\overline{\check{b}_\alpha(z)}\check{b}'_\alpha(z)
\ee
where we have used that $(b^\dagger b')^{\bigvee}=\overline{\check{b}}
\check{b}'$.
This formula suggests to introduce the Hilbert spaces
$L_2(\sigma(a),d\mu_{\psi_\alpha})$ as well as the space
$\sigma:=\bigcup_{\alpha\in A} \sigma(a)_\alpha$ (disjoint union of copies of
$\sigma(a)$) and a measure $\mu$ on it defined by
$\mu_{|\sigma(a)_\alpha}:=\mu_{\psi_\alpha}$. Notice that
measurable sets are of the form $\bigcup_{\alpha\in B\subset A} U_\alpha$
where $U_\alpha$ is measurable in $\sigma(a)_\alpha$, $B$ can be any
subindex set and that
unions, intersections and differences of measurable sets are
performed componentwise. Let us now define the Hilbert space
$L_2(\sigma,d\mu)$. An element $\check{\psi}$ of $L_2(\sigma,d\mu)$
is a square integrable function on $\sigma$ with respect to the measure
$\mu$ and may be defined in terms of an array of functions
$\check{\psi}_\alpha \in L_2(\sigma(a)_\alpha,d\mu_{\psi\alpha)}$ through
$\check{\psi}_{|\sigma(a)_\alpha}=\check{\psi}_\alpha$. Notice that indeed
\ba \label{g.11}
<\check{\psi},\check{\psi}'>_{L_2(\sigma,d\mu)}
&=&\int_\sigma d\mu(z) \overline{\check{\psi}(z)}\check{\psi}'(z)
\nonumber\\
&=& \sum_{\alpha\in A}
\int_{\sigma(a)_\alpha} d\mu(z) \overline{\check{\psi}(z)}\check{\psi}'(z)
= \sum_{\alpha\in A}
\int_{\sigma(a)_\alpha} d\mu_{|\sigma(a)_\alpha}(z)
[\overline{\check{\psi}(z)}\check{\psi}'(z)]_{|\sigma(a)_\alpha}
\nonumber\\
&=& \sum_{\alpha\in A}
\int_{\sigma(a)} d\mu_{\psi_\alpha}(z)
\overline{\check{\psi}_\alpha(z)}\check{\psi}'_\alpha(z)
\ea
explaining the requirement that
$\check{\psi}_\alpha \in L_2(\sigma(a)_\alpha,d\mu_{\psi_\alpha})$.
Here we have made use of $\sigma-$additivity, that is,
$\mu(\bigcup_\alpha U_\alpha)=\sum_\alpha \mu(U_\alpha)=
\sum_\alpha \mu_{\psi_\alpha}(U_\alpha)$ for the mutually disjoint sets
$U_\alpha\subset U$.
Comparing (\ref{g.10}) and (\ref{g.11}) we see that we can identify
$L_2(\sigma, d\mu)$ with $\oplus_{\alpha\in A}
L_2(\sigma(a)_\alpha,d\mu_{\psi_\alpha})$ and obtain a unitary transformation
\be \label{g.12}
U:\;{\cal H}\to L_2(\sigma,d\mu);\;\;\psi=\sum_{\alpha\in A}
b_\alpha\psi_\alpha \mapsto\check{\psi} \mbox{ where }
\check{\psi}_{|\sigma(a)_\alpha}:=\check{b}_\alpha
\ee
Moreover, we have
\be \label{g.13}
Ub\psi=U\sum_\alpha bb_\alpha\psi_\alpha=\check{\psi}' \mbox{ where }
\check{\psi}'_{|\sigma(a)_\alpha}=(bb_\alpha)^\bigvee
=\check{b}\check{b}_\alpha
\ee
which means that on each supspace $L_(\sigma(a)_\alpha,d\mu_{\psi_\alpha})$
the operator $b$ is represented by multiplication by $\check{b}(z)$.
In particular, if $b=a$ or $b=a^\dagger$ it is represented by multiplication
by $z$ or $\bar{z}$ since $\chi(a)=z$.

This simple corollary from Gel'fand spectral theory and the Riesz
representation theorem is the spectral theorem for bounded operators.
It obviously generalizes to the case that we have a family
$(a_I)$ of bounded operators which together with their adjoints
mutually commute with each other. The only difference is that we now
get a homeomorphism between $\Delta(\a)$ and the joint spectrum
$\prod_I \sigma(a_I)$ via $\chi\mapsto (\chi(a_I))_I$. We can also strip
off the concrete Hilbert space context by considering an abstract
unital $C^\ast-$algebra $\a$ where instead of vector states $\psi_\alpha$
we use states $\omega_\alpha$ on $\a$ and apply the GNS construction.
That for given $a\in\a$ there is always a state $\omega$ with
$\omega(a^\ast a)>0$ follows from the Hahn-Banach theorem applied
to the vector space $X:=\a$ and its one-dimesional supspace
$Y:=\mbox{span}(a^\ast a)$ with the bounding function appearing in the theorem
given by the norm on $X$ and by defining $\omega(a^\ast a):=||a||^2$:
The Hahn-Banach theorem guarantees that then $\omega$ can be extended
as a positive linear functional to all of $\a$.
\begin{Theorem} \label{thg.1} ~~~~~~~~~~~~~\\
Let $(a_I)$ be a self-adjoint collection of mutually commuting elements of a
$C^\ast-$algebra ${\cal C}$. Then there exists a representation of
the sub$-C^\ast-$algebra $\a$ generated by this collection on a Hilbert
space $\cal H$ such that the $\pi(a_I)$ become multiplication operators.
\end{Theorem}
The extension of the spectral theorem to unbounded self-adjoint operators
operators on a Hilbert space can be traced back to the bounded
case by using the following trick. (Recall that a densely
defined operator $a$ with domain $D(A)$ is called self-adjoint
if $a^\dagger=a$ and $D(a^\dagger)=D(a)$ where
$$
D(a^\dagger):=\{\psi\in {\cal H};\;\sup_{0\not=\psi'\in D(a)}
|<\psi,a\psi'>|/||\psi'||<\infty\}
$$
and $a^\dagger$ is uniquely defined on $\psi\in D(a^\dagger)$ via
$<a^\dagger\psi,\psi'>=<\psi,a\psi'>$ for all $\psi\in D(a)$ through
the Riesz lemma):\\
The spectrum of $a$ will be an unbounded subset of the real line.
Let $f$ be a bijection $\Rl\to K$ where $K$ is a compact one-dimensional
subset of $\Cl$ and suppose that $f(a)$ is a bounded operator.
Then we can apply the spectral theorem for bounded normal operators
to $f(a)$ which then becomes a multiplication operator and if
$f^{-1}$ is a measurable function with respect to the spectral measure
$\mu$ then also $a$ itself is a multiplication operator. A popular
tool is the Caley transform $a\to u:=(a-i)(a+i)^{-1}$ which maps
$a$ to a unitary operator.

Finally, let us mention the spectral resolution. Let $a$ be a bounded
self-adjoint operator then by the spectral theorem there is a measure
$\mu$ and a representation such that
$<\psi,f(a)\psi>=\int_{\sigma(a)} d\mu_\psi(z) f(z)$ for any
measurable function $f$ and $\mu_\psi$ is the spectral measure of
$\psi$ in a cyclic representation. Let $S\subset \Rl$ be measurable and and
consider the operators
$p_S:=\chi_S(a)$ called the spectral projections where $\chi_S$ is the
characteristic function of $S$. Then
$<\psi,p_S\psi>=\int_{\sigma(a)} d\mu_\psi(z) \chi_S(z)$.
Let $p_z:=\chi_{(-\infty,z)}(a)$ for $z\in\Rl$ then we see that
we obtain the so-called projection valued measures
\be \label{g.14}
<\psi,dp_z\psi>=d<\psi,dp_z\psi>=d\mu_\psi(z)
\ee
whence
\be \label{g.15}
<\psi,f(a)\psi>=\int_\Rl <\psi,dp_z\psi> f(z)
\ee
for all $\psi\in {\cal H}$ or by the the polarization identity
\be \label{g.16}
f(a)=\int_\Rl dp_z f(z)
\ee
which is called the spectral resolution of $f(a)$.

\newpage


\section{Introduction to Refined Algebraic Quantization (RAQ)}
\label{si}

RAQ provides strong guidelines of how
to solve a given family of quantum constraints but unfortunately it is not
an algorithm
that one just has to apply in order to arrive at a satisfactory
end result. In particular, as presently formulated it has its limitations
since it does not cover the case that the constraints form an open algebra
with structure functions rather than structure constants as it would be the
case for a Lie algebra. Unfortunately, quantum gravity belongs to the
open algebra category of constrained systems.
We mainly follow Giulini and Marolf in \cite{47f1}.\\
\\
Let ${\cal H}_{kin}$ be a Hilbert space, referred to as the kinematical
Hilbert space because it is supposed to implement the adjointness -- and
canonical commutation relations of the elementary kinematical degrees
of freedom. However, these degrees of freedom are not observables
(classically they do not have vanishing Poisson brackets with
the constraints on the constraint surface) and the Hilbert space is
not the physical one on which the constraint operators would equal the
zero operators. The role of ${\cal H}_{kin}$ is to give the constraint
operators $(\hat{C}_I)_{I\in {\cal I}}$ a home, that is, there is a common
dense
domain ${\cal D}_{kin}\subset {\cal H}_{kin}$ which is supposed to be
invariant
under all the $\hat{C}_I$ and we also require that the $\hat{C}_I$
be closed operators (i.e. their adjoint is densely defined as well).
We do not require them to be bounded operators.
The label set $\cal I$ is rather arbitrary and usually is a combination of
direct products of finite and infinite sets (e.g. tensor or gauge group
indices times indices taking values in a separable space of smearing
functions).

We will further require that the constraints form a first class system
and that they actually form a Lie algebra, that is, there exist
complex valued structure constants $f_{IJ}\;^K$ such that
\be \label{i.1}
[\hat{C}_I,\hat{C}_J]=f_{IJ}\;^K\hat{C}_K
\ee
where the summation over
$K$ performed here will involve an integral for generic $\cal I$.
Notice that (\ref{i.1}) makes sense due to our requirement on
${\cal D}_{kin}$.
The case of an open algebra would correspond to the fact that the
structure constants become operator valued as well and then it becomes an
issue how to choose the operator ordering in (\ref{i.1}), in particular,
if constraint operators and structure constant operators are chosen
to be self-adjoint and anti-self-adjoint respectively (which would be
natural if their classical counterparts are classically real and imaginary
valued respectively) then one would have to order (\ref{i.1}) symmetrically
which would be a desaster for solving the constraints, see below, which
is why in the open case the constraints should not be chosen to be
self-adjoint operators. Notice that there is no contradiction because
self-adjointness usually is required to ensure that the spectrum
(measurement values) of the operator lies in the real line, however,
for constraint operators this requirement is void since we are only
interested in their kernel and the only requirement is that the
point zero belongs to the spectrum at all.

In order to allow for non-self-adjoint constraints, in what follows we will
assume that the set ${\cal C}:=\{\hat{C}_I;\;
I\in {\cal I}\}$ is self-adjoint (i.e. contains with $\hat{C}_I$ also
$\hat{C}_I^\dagger=\hat{C}_J$ for some $J$) which means that the dense domain
${\cal D}_{kin}$ is also a dense domain for the adjoints so that the
constraints are
explicitly closed operators. Let us now consider the self-adjoint set
of kinematical observables ${\cal O}_{kin}$, that is,
all operators on ${\cal H}_{kin}$ which have ${\cal D}_{kin}$ as
common dense domain together with their adjoints. Obviously,
${\cal O}_{kin}$ contains $\cal C$. Consider the
commutant of $\cal C$ within ${\cal O}_{kin}$, that is,
\be \label{i.2}
{\cal C}':=
\{O\in{\cal O}_{kin};\; [C,O]=0\;\forall \; C\in {\cal C}\}
\ee
It is clear that
${\cal C}'$ is a $^\ast-$subalgebra of ${\cal O}_{kin}$ since
$[O^\dagger,{\cal C}]=-([O,{\cal C}])^\dagger=0$ and
$[O O',{\cal C}]=O [O',{\cal C}]+[O ,{\cal C}]O'=0$ for any $O,O'\in
{\cal C}'$ since ${\cal C}^\dagger={\cal C}$ is a self-adjoint set.
Moreover, ${\cal C}$ might have a non-trivial center
\be \label{i.3}
{\cal Z}={\cal C}\cap {\cal C}'
\ee
which generates a two-sided ideal $I_{{\cal Z}}$ in ${\cal C}'$
corresponding to classical functions that vanish on the constraint
surface and is therfore physically uninteresting.
Hence we will define the algebra of physical observables to be
the quotient algebra
\be \label{i.4}
{\cal O}_{phys}:={\cal C}'/{\cal Z}
\ee

Usually the space ${\cal D}_{kin}$ comes with its own topology
$\tau$, different from
the subspace topology inherited from the Hilbert space topology $||.||$
on ${\cal H}_{kin}$, generically
a nuclear topology \cite{26} so that ${\cal D}_{kin}$ becomes a
Fr\'echet space (a space whose topology is generated by a countable family of
semi-norms that separates the points of ${\cal D}_{kin}$ and such that
${\cal D}_{kin}$
is complete in the associated norm; a general locally convex topological
vector space is not necessarily complete and the family of semi-norms
need not to be countable (it is then not metrizable)). The intrinsic
topology $\tau$ is then finer than $||.||$ since ${\cal D}_{kin}$
is complete but also dense in ${\cal H}_{kin}$ (if it would be coarser
then a Cauchy sequence in ${\cal D}_{kin}$ with respect to the intrinsic
topology
would also be one in the Hilbert space topology and since ${\cal D}_{kin}$
is dense
this completion would coincide with ${\cal H}_{kin}$). It follows that
the space of continuous linear functionals ${\cal D}'_{kin}$ (with respect
to the topology on ${\cal D}_{kin}$ contains
${\cal H}_{kin}$ since a Hilbert space is
reflexive, that is, ${\cal H}'_{kin}={\cal H}_{kin}$ by the Riesz lemma
so the elements of ${\cal H}_{kin}$ are in particular continuous
linear functionals on ${\cal D}_{kin}$ with respect to $||.||$ so that they
are also continuous with respect to $\tau$ (a function stays continuous if
one strengthens the topology on the domain space). Let
$(l^\alpha)$ be a net in ${\cal H}'_{kin}$ converging to $l$ then
\ba \label{i.5}
||l^\alpha-l||_{{\cal D}'_{kin}}
&=&\sup_{f\in {\cal D}_{kin}}\frac{|<l_\alpha-l,f>|}{||f||_{{\cal D}_{kin}}}
=\sup_{f\in {\cal D}_{kin}}
\frac{||f||_{{\cal H}_{kin}}}{||f||_{{cal D}_{kin}}}
\;\frac{|<l_\alpha-l,f>|}{||f||_{{\cal H}_{kin}}}
\nonumber\\
&\le& \sup_{f\in {\cal D}_{kin}}
\frac{|<l_\alpha-l,f>|}{||f||_{{\cal H}_{kin}}}
\le \sup_{f\in {\cal H}_{kin}}\frac{|<l_\alpha-l,f>|}{||f||_{{\cal H}_{kin}}}
=||l^\alpha-l||_{{\cal H}'_{kin}}
\ea
where we used $||f||_{{\cal H}_{kin}}/||f||_{{cal D}_{kin}}\ge 1$.
Thus it converges in ${\cal D}'_{kin}$ as well, that is, the topology on
${\cal D}'_{kin}$ is weaker than that of ${\cal H}_{kin}$. We thus
have topological inclusions
\be \label{i.6}
{\cal D}_{kin}\hookrightarrow{\cal H}_{kin}\hookrightarrow {\cal D}'_{kin}
\ee
sometimes called a Gel'fand triple.

Unfortunately the definition
of a Gel'fand triple requires a further input, the nuclear topology intrinsic
to ${\cal D}_{kin}$ which we want to avoid since there seems no physical
guiding principle (although then there are rather
strong theorems available concerning the completeness of generalized
eigenvectors \cite{26}). We thus equip ${\cal D}_{kin}$ simply with the
relative topology induced from ${\cal H}_{kin}$. The requirement that
${\cal D}_{kin}$ is dense is then no loss of generality since we may simply
replace
${\cal H}_{kin}$ by the completion of ${\cal D}_{kin}$. Instead of the
topological dual (which would coincide with ${\cal H}_{kin}$ we consider
the algebraic dual ${\cal D}^\ast_{kin}$ of all linear functionals on
${\cal D}_{kin}$. This space is naturally equipped with the weak $^\ast$
topology
of pointwise convergence, i.e. a net $(l^\alpha)$ converges to $l$ iff
the net of complex numbers $(l^\alpha(f))$ converges to $l(f)$ for any
$f\in {\cal D}_{kin}$ (but not uniformly).
Again we can consider ${\cal H}_{kin}$ as a subspace of ${\cal D}^\ast_{kin}$
and since a net converging in norm certainly converges pointwise we have
again topological inclusions
\be \label{i.7}
{\cal D}_{kin}\hookrightarrow{\cal H}_{kin}\hookrightarrow
{\cal D}^\ast_{kin}
\ee
which in abuse of notation we will still refer to as Gel'fand triple.
Thus, the only input left is the choice of ${\cal D}_{kin}$ for which,
however,
there are no general selection principles available at the moment
(see however \cite{47f1} for further discussion).

The reason for blowing up the structure beyond ${\cal H}_{kin}$ is that
generically the point zero does not lie in the discrete part of the
spectrum of ${\cal C}$, that is, if we look for solutions to the
constraints in the form $\hat{C}_I\psi=0$ for all $I\in {\cal I}$
for $\psi\in {\cal H}_{kin}$, then there are generically not enough
solutions
because $\psi$ would be an eigenvector with eigenvalue zero but since
zero does not lie in the discrete spectrum the eigenvectors do not
form the entire solution space.
This is precisely what happens with the diffeomorphism
constraint for the case of quantum gravity where the only eigenvectors are
the constant functions. We therefore look for generalized
eigenvectors $l\in {\cal D}^\ast_{kin}$ in the algebraic dual for which we
require
\be \label{i.8}
[(\hat{C}_I^\dagger)' l](f):=l(\hat{C}_I f)=0\;\forall \;I\in {\cal I},\;
f\in {\cal D}_{kin}
\ee
where the dual action of an operator $\hat{O}\in {\cal O}_{kin}$
on $l\in {\cal D}^\ast_{kin}$ is defined by
\be \label{i.9}
[\hat{O}' l](f):=l(\hat{O}^\dagger f)\;\forall \;f\in {\cal D}_{kin}
\ee
Notice that since we required $\cal C$ to be a self-adjoint can avoid
taking the adjoint in (\ref{i.8}) by passing to self-adjoint
representatives $\hat{C}_I$. Due to the adjoint operation
in (\ref{i.9}) we have an anti-linear representation of ${\cal O}_{kin}$
on ${\cal D}^\ast_{kin}$ which descends to an anti-linear representation of
${\cal O}_{phys}$ on
the space of solutions ${\cal D}^\ast_{phys}\subset {\cal D}^\ast_{kin}$
to (\ref{i.8}).

At this point, the space ${\cal D}^\ast_{phys}$ is just a subspace of
${\cal D}^\ast_{kin}$.
Ww would like to equip a subspace $H_{phys}$ of it with a Hilbert space
topology. The reason for not turning all of ${\cal D}^\ast_{phys}$ into
${\cal H}_{phys}$
is that then ${\cal O}_{phys}$ would be realized as an algebra of
bounded operators on ${\cal H}_{phys}$ since they are defined
everywhere on ${\cal D}^\ast_{phys}$ which would be unnatural if the
corresponding classical functions are unbounded.
In particular, the topology on ${\cal H}_{phys}$, as a complete norm
topology, should be finer than the relative topology induced from
${\cal D}^\ast_{kin}$.
The idea is then to consider ${\cal D}^\ast_{phys}$ as the algebraic dual
of a dense subspace ${\cal D}_{phys}\subset {\cal H}_{phys}$ so that
all of ${\cal O}_{phys}$ is densely defined there. In other words
we get a second Gel'fand triple
\be \label{i.10}
{\cal D}_{phys}\hookrightarrow{\cal H}_{phys}\hookrightarrow
{\cal D}^\ast_{phys}
\ee
with an anti-linear representation of ${\cal O}_{phys}$ on ${\cal H}_{phys}$
defined by (\ref{i.9}).

The choice of the inner product on ${\cal H}_{phys}$ is guided by the
requirement that the adjoint in the physical inner product, denoted by
$\star$, represents the adjoint in the kinematical one, that is,
\be \label{i.12}
<\psi,\hat{O}' \psi'>_{phys}=<(\hat{O}')^\star \psi, \psi'>_{phys}
=<(\hat{O}^\dagger)' \psi, \psi'>_{phys}
\ee
for all $\psi,\psi'\in {\cal D}_{phys}$. The canonical commutation relations
among observables are automatically implemented because by construction
${\cal H}_{phys}$ carries a representation of ${\cal O}_{phys}$ on which
the correct algebraic relations were already implemented as an abstract
algebra.

A systematic construction of the physical inner product is available
if we have an anti-linear (so-called) rigging map
\be \label{i.13}
\eta:\;{\cal D}_{kin}\to {\cal D}^\ast_{phys};\; f\mapsto \eta(f)
\ee
at our disposal which must be such that\\
1) the following is a positive
semi-definite sesquilinear form (linear in $f$, anti-linear in $f'$)
\be \label{i.14}
<\eta(f),\eta(f')>_{phys}:=[\eta(f')](f)\; \forall\; f,f'\in {\cal D}_{kin}
\ee
2) For any $\hat{O}\in {\cal O}_{phys}$ we have
\be \label{i.15}
\hat{O}'\eta(f)=\eta(\hat{O} f) \; \forall \; f\in {\cal D}_{kin}
\ee
which makes sure that the dual action preserves the space of solutions
since $\hat{C}'\hat{O}'\eta(f)=0$. Notice that bot the left and the
right hand side in (\ref{i.15}) are antilinear in $\hat{O}$.\\

We could then define
${\cal D}_{phys}:=\eta({\cal D}_{kin})/\mbox{ker}(\eta)$ (with the kernel
being understood with respect to $||.||_{phys}$) and complete it with respect
to (\ref{i.14}) to obtain ${\cal H}_{phys}$. Notice that (\ref{i.12})
is satisfied because for $\psi=\eta(f),\psi'=\eta(f')$ we have
\ba \label{i.16}
<\psi,\hat{O}'\psi'>_{phys}&=&<\eta(f),\eta(\hat{O}f')>_{phys}
=[\eta(\hat{O}f')](f)
=[\hat{O}'\eta(f')](f)=\eta(f')(\hat{O}^\dagger f)
\nonumber\\
&=&<\eta(\hat{O}^\dagger f),\eta(f')>_{phys}
=<(\hat{O}^\dagger)' \psi,\psi'>_{phys}
\ea
To see that ${\cal H}_{phys}$ is a subspace of ${\cal D}_{phys}^\ast$
with a finer topology, notice that the map
$J:\;{\cal H}_{phys}\to {\cal D}_{phys}^\ast$ defined by
$[J(\psi)](f):=<\psi, \eta(f)>_{phys}$ is an injection because
$J(\psi)$ vanishes iff $\psi$ is orthogonal to all $\eta(f)$ with respect
to $<.,.>_{phys}$ which means that $\psi=0$ because the image of $\eta$
is dense. Hence $J$ is an embedding (injective inclusion) of linear spaces.
Moreover, $J$ is evidently continuous: if
$||\psi^\alpha-\psi||_{phys}\to 0$ then $J(\psi^\alpha)\to J(\psi)$
in the weak $^\ast$ topology iff $[J(\psi^\alpha)](f)\to [J(\psi)](f)$
for any $f\in {\cal D}_{kin}$ which is clearly the case. So
convergence in ${\cal H}_{phys}$ implies convergence of
$J({\cal H}_{phys})$, hence the Hilbert space topology is stronger
than the relative topology on $J({\cal H}_{phys})$.

Thus, the existence of a rigging map solves the problem of defining
a suitable inner product. A heuristic idea of how to construct $\eta$ is
through
the group averaging proposal: Since ${\cal C}$ is a self-adjoint set
we may assume w.l.g that the $\hat{C}_I$ are self-adjoint, and since
they form a Lie algebra we may in principle exponentiate this Lie algebra
(using the spectral theorem)
and obtain a group of operators $t^I\to \exp(t^I\hat{C}_I)$ where
$t^I\in T$ is some set depending on the constraints.
Let then
\be \label{i.17}
\eta(f):=\overline{\int_T d\mu(t) \exp(t^I\hat{C}_I) f}
\ee
with a translation invariant measure $\mu$ on $T$.
One easily sees that with
\be \label{i.18}
[\eta(f)](f'):=\int_T d\mu(t) <\exp(t^I\hat{C}_I) f,f'>_{kin}
\ee
formally $[\eta(f)](\hat{C}_I f')=0$. Of course, one must check case by
case whether $T,\mu$ exist and that $\eta$ has the required
properties.

Let us make some short comments about the open algebra case:\\
Suppose that the classical constraint functions $C_I$ and the
structure functions $f_{IJ}\;^K$ are real and imaginary valued
respectively. As mentioned already, it is now excluded to choose the
corresponding operators to
be (anti)-self-adjoint opertors since this would require the
ordering
\be \label{i.19}
[\hat{C}_I,\hat{C}_J]=\frac{1}{2}(\hat{f}_{IJ}\;^K \hat{C}_K
+\hat{C}_K\hat{f}_{IJ}\;^K)
\ee
and would lead to the following quantum anomaly: If we impose
the condition (\ref{i.8}) then we would find for an element
$l\in {\cal D}^\ast_{phys}$ that
\be \label{i.20}
((\hat{f}_{IJ}\;^K)' \hat{C}_K'+\hat{C}_K'(\hat{f}_{IJ}\;^K)')l
=[\hat{C}_K',(\hat{f}_{IJ}\;^K)'] l=0
\ee
which means that $l$ is not only annihilated by the dual constraint operators
but also by (\ref{i.20}) which is not necessarily proportional
to a dual constraint operator any longer, implying that the physical
Hilbert space will be too small. If on the other hand we do not choose
the $\hat{C}_I$ to be self-adjoint, the anomaly problem is potentially
absent but now it is no longer true that $[\hat{C}_I' l](f)=l(\hat{C}_I f)$,
in other words, the question arises whether it is
$\hat{C}_I' l=0$ or $(\hat{C}_I^\dagger)' l=0$ that we should impose ?
The answer is that this just corresponds to a choice of operator
ordering since the classical limit of both $\hat{C}_I$ and
$\hat{C}_I^\dagger$ is given by the real valued function $C_I$ and thus the
answer is that the correct
ordering is the one in which the algebra is, besides being densely defined
and closed, also free of anomalies.\\
Thus, in the open algebra case we may proceed just as above with
the additional requirement of anomaly freeness. Of course, group
averaging does not work since we cannot eponentiate the algebra any longer.

We conclude this section with an example in order to illustrate
the procedure:\\
Suppose ${\cal H}_{kin}=L_2(\Rl^2,d^2x)$ and
$\hat{C}=\hat{p}_1=-i\partial/\partial x_1$. Obviously the kinematical
Hilbert space implements the adjointness and canonical commutation relations
among the basic variables $x_1,x_2,p_1,p_2$. A nuclear space choice would
be ${\cal D}_{kin}={\cal S}(\Rl^2)$ (test functions of rapid decrease).
The functions $l$ annihilated by $\hat{C}$ are those
that do not depend on $x^1$ and are thus not normalizable. However,
we can define them as elements of ${\cal D}^\ast_{kin}$ by
$l(f):=<l,f>_{kin}=\int_{\Rl^2} d^2x \overline{l(x)} f(x)$
which converges pointwise. Clearly $l(\hat{C}f)=0$ if $l_{,x^1}=0$.
The physical observable algebra consists of operators not involving
$\hat{x}_1$ and after taking the quotient with respect to the constraint
ideal they involve only $\hat{p}_2,\hat{x}_2$. Obviously they
leave the space ${\cal D}^\ast_{phys}$ invariant, consisting of those
elements of ${\cal D}^\ast_{kin}$ that are $x^1-$independent. The physical
Hilbert space that suggests itself (implementing the correct reality
condition) is therefore
${\cal H}_{phys}=L_2(\Rl,dx_2)$ which is a proper subspace of
${\cal D}^\ast_{phys}$ and we have $D_{phys}={\cal S}(\Rl)$. Now
an appropriate rigging map is obtained indeed by
$$
\eta(f)(x_1,x_2):=\overline{\int_{\Rl} dt \exp(it\hat{p}_1)f(x_1,x_2)}
=\overline{\int_{\Rl} dx_1 f(x_1,x_22)}=2\pi
\overline{\delta(\hat{C})f(x_1,x_2)}
$$
since $\hat{p}_1$ generates $x_1$ translations, produces
functions
independent of $x^1$ and $dt$ is an invariant measure on $T=\Rl$.
Notice that the integral converges because $f$ is of rapid decrease.
Notice also that we could define the delta distribution
of the constraint, using the spectral theorem. In the case of an Abelean
self-adjoint constraint algebra a reasonable ansatz for a rigging map is
always given by
\be \label{i.21}
\eta(f)=\overline{\prod_{I\in {\cal I}} \delta(\hat{C}_I) f}
\ee
We have
\ba
&& <\eta(f),\eta(f')>_{phys}:=\eta(f')[\eta(f)]=
\int_{\Rl} dt \int_{\Rl^2} d^2x \overline{f'(x_1+t,x_2)}f(x_1,x_2)
\nonumber\\
&=&\int_{\Rl} dx_2
\overline{[\int dx_1' f'(x_1',x_2)]}[\int dx_1 f(x_1,x_2)]
\ea
which is the same inner product as chosen above.



\end{document}